\documentclass[twoside,leqno]{report}

\pdfoutput=1

\usepackage{RR}
\usepackage[a4paper]{geometry}
\usepackage{a4wide}

\usepackage[utf8]{inputenc} 
\usepackage[T1]{fontenc} 

\usepackage[english]{babel}

\usepackage[numbers]{natbib}

\usepackage{xr-hyper}
\usepackage
  [pdftex, extension=pdf, colorlinks=true,
   citecolor=blue, filecolor=blue, linkcolor=blue, urlcolor=blue]{hyperref}

\usepackage{amssymb, amsmath, mathtools}
\usepackage{amsthm, thmtools} 
\usepackage{amsfonts, bbold, dsfont}

\usepackage[nottoc,other]{tocbibind}

\usepackage{minitoc} 

\usepackage[normalem]{ulem}

\usepackage{todonotes}

\usepackage[boxed]{algorithm2e}

\usepackage{enumitem}
\newenvironment{CompactList}{%
  \setlength{\parskip}{0pt}
  \begin{itemize}[nosep]}{%
  \end{itemize}}

\usepackage{rotating}
\usepackage{graphicx}

\graphicspath{{./}}

\usepackage{color}

\definecolor{darkred}{rgb}{0.7,0.1,0.1}
\definecolor{darkgreen}{rgb}{0.1,0.7,0.1}
\definecolor{darkblue}{rgb}{0.1,0.1,0.7}
\definecolor{orange}{rgb}{1,0.5,0}
\definecolor{lightred}{RGB}{255,232,232}
\definecolor{lightgreen}{RGB}{232,255,232}
\definecolor{lightblue}{RGB}{232,255,255}
\definecolor{lightorange}{rgb}{1,0.7,0.3}
\definecolor{lightgray}{gray}{0.9}

\newcommand{\assume}[1]{\textcolor{darkred}{\bf\boldmath{#1}}}

\newcommand{\defcolor}{lightgreen}
\newcommand{\lemcolor}{lightblue}
\newcommand{\thmcolor}{lightred}
\newcommand{\rmkcolor}{lightgray}

\newcommand{\coloredbox}[2]{\fcolorbox{black}{#1}{#2}}
\newcommand{\defbox}[1]{\coloredbox{\defcolor}{#1}}
\newcommand{\lembox}[1]{\coloredbox{\lemcolor}{#1}}
\newcommand{\thmbox}[1]{\coloredbox{\thmcolor}{#1}}
\newcommand{\rmkbox}[1]{\coloredbox{\rmkcolor}{#1}}

\usepackage{epic,eepic}

\definecolor{metriccolor}{RGB}{225,255,255}
\definecolor{completecolor}{RGB}{195,255,255}
\definecolor{vectorcolor}{RGB}{255,225,225}
\definecolor{normedcolor}{RGB}{255,195,195}
\definecolor{innercolor}{RGB}{255,135,135}
\definecolor{hilbertcolor}{RGB}{212,162,212}

\declaretheoremstyle[
  headfont=\normalfont\bfseries,
  notefont=\normalfont\bfseries\itshape,
  bodyfont=\normalfont]{mydef}
\declaretheoremstyle[
  headfont=\normalfont\bfseries,
  notefont=\normalfont\bfseries\itshape,
  bodyfont=\normalfont\itshape]{mythm}
\declaretheoremstyle[
  headfont=\normalfont\itshape,
  notefont=\normalfont\itshape,
  bodyfont=\normalfont]{myrmk}

\newcommand{\mydeclaretheoremshaded}[3]{
  \declaretheorem[
    #1,
    shaded={
      textwidth=0.99\textwidth,
      bgcolor=#2,
      rulecolor=black,
      rulewidth=0.5pt}]{#3}
}

\newcommand{\mydeclaretheoremstyle}[3]{
  \mydeclaretheoremshaded{numberlike=theorem, style=#1}{#2}{#3}
}

\newcommand{\thm}[1]{#1 theorem} 

\newcommand{\BNB}{Banach--Ne\v{c}as--Babu\v{s}ka}
\newcommand{\BNBT}{\thm{\BNB}}

\newcommand{\BL}{Beppo Levi}
\newcommand{\BLt}{\thm{{\BL} (monotone convergence)}}

\newcommand{\BoL}{(Borel--)Lebesgue}

\newcommand{\Cara}{Carathéodory}
\newcommand{\Carat}{\thm{{\Cara}'s extension}}
\newcommand{\CaratheodoryTh}{\Carat}

\newcommand{\CS}{Cauchy--Schwarz}

\newcommand{\Fl}{Fatou's lemma}

\newcommand{\FL}{Fatou--Lebesgue}
\newcommand{\FLt}{\thm{\FL}}

\newcommand{\Fubt}{\thm{Fubini}}

\newcommand{\FT}{Fubini--Tonelli}
\newcommand{\FTt}{\thm{\FT}}

\newcommand{\HK}{Henstock--Kurzweil}

\newcommand{\Ldcv}{Lebesgue's dominated convergence}
\newcommand{\Ldcvt}{\thm{\Ldcv}}

\newcommand{\Ledcv}{Lebesgue's extended dominated convergence}
\newcommand{\Ledcvt}{\thm{\Ledcv}}

\newcommand{\LM}{Lax--Milgram}
\newcommand{\LMT}{\thm{\LM}}

\newcommand{\LMC}{Lax--Milgram--C\'ea}
\newcommand{\LMCT}{\thm{\LMC}}

\newcommand{\LN}{Leibniz--Newton}

\newcommand{\RFr}{Riesz--Fr\'echet}
\newcommand{\RFrT}{\thm{{\RFr} representation}}

\newcommand{\RMK}{Riesz--Markov--Kakutani}
\newcommand{\RMKrt}{\thm{{\RMK} representation}}
\newcommand{\RieszMarkovKakutaniTh}{\RMKrt}

\newcommand{\Tont}{\thm{Tonelli}}

\mydeclaretheoremshaded{style=mythm}{\thmcolor}{theorem}
\mydeclaretheoremstyle{mythm}{\lemcolor}{lemma}
\mydeclaretheoremstyle{mydef}{\defcolor}{definition}
\mydeclaretheoremstyle{myrmk}{\rmkcolor}{remark}

\mydeclaretheoremshaded{numbered=no, name={\LMCT}}{\thmcolor}{lmcthm}
\mydeclaretheoremshaded{numbered=no, name={\LMT}}{\thmcolor}{lmthm}
\newcommand{\rLbbf}{Representation lemma for bounded bilinear forms}
\mydeclaretheoremshaded{numbered=no, name={\rLbbf}}{\lemcolor}{bilinlem}
\mydeclaretheoremshaded{numbered=no, name={\RFrT}}{\thmcolor}{rfrthm}
\newcommand{\opTcs}{\thm{Orthogonal projection} for a complete subspace}
\mydeclaretheoremshaded{numbered=no, name={\opTcs}}{\thmcolor}{projthm}
\newcommand{\fpT}{\thm{Fixed point}}
\mydeclaretheoremshaded{numbered=no, name={\fpT}}{\thmcolor}{fpthm}

\mydeclaretheoremshaded{numbered=no, name={\Ledcvt}}{\thmcolor}{ledcthm}
\mydeclaretheoremshaded{numbered=no, name={\Ldcvt}}{\thmcolor}{ldcthm}
\mydeclaretheoremshaded{numbered=no, name={\FLt}}{\thmcolor}{falthm}
\mydeclaretheoremshaded{numbered=no, name={\Fl}}{\thmcolor}{flem}
\mydeclaretheoremshaded{numbered=no, name={\Tont}}{\thmcolor}{ftthm}
\newcommand{\Utpm}{Uniqueness of tensor product measure}

\mydeclaretheoremshaded{numbered=no, name={\Utpm}}{\thmcolor}{utpmlem}
\mydeclaretheoremshaded{numbered=no, name={\BLt}}{\thmcolor}{blthm}
\mydeclaretheoremshaded{numbered=no, name={\Carat}}{\thmcolor}{cthm}
\newcommand{\Mct}{\thm{Monotone class}}
\newcommand{\mct}{\thm{monotone class}}
\newcommand{\Dplt}{%
  \thm{Dynkin \texorpdfstring{$\pi$}{pi}--\texorpdfstring{$\lambda$}{lambda}}}
\newcommand{\Abst}{{\Dplt} / {\mct}}
\mydeclaretheoremshaded{numbered=no, name={\Abst}}{\thmcolor}{absthm}

\newcommand{\thref}[1]{\ref{#1} ({\em \nameref{#1}})}
\newcommand{\threfc}[2]{\ref{#1} ({\em \nameref{#1}}, \uline{#2})}

\newcommand{\proofpar}[1]{{\bf\boldmath{#1}.}}
\newcommand{\proofparskip}[1]{\medskip\noindent\proofpar{#1}}

\newcommand{\FEM}{Finite Element Method}

\newcommand{\Vectorspace}{Vector space}
\newcommand{\vectorspace}{vector space}
\newcommand{\vectorsubspace}{vector subspace}

\newcommand{\software}[1]{{\sf #1}}
\newcommand{\coq}{\software{Coq}}
\newcommand{\isabelle}{\software{Isabelle/HOL}}

\newcommand{\st}{{\,|\,}}
\newcommand{\leftst}{\;\left|\;}
\newcommand{\rightst}{\;\right|\;}
\newcommand{\aka}{a.k.a.}

\newcommand{\eg}{e.g.}
\newcommand{\ie}{i.e.}

\newcommand{\perse}{{\em per se}}
\newcommand{\half}{{\frac{1}{2}}}

\newcommand{\oneovernplusone}{{\textstyle\frac{1}{n+1}}}

\newcommand{\lsrbra}{[\hspace*{-0.24em}(}
\newcommand{\rsrbra}{)\hspace*{-0.24em}]}

\newcommand{\makespace}[1]{\quad#1\quad}
\newcommand{\Equiv}{\Leftrightarrow}
\newcommand{\EQUIV}{\makespace{\Longleftrightarrow}}
\newcommand{\Implies}{\Rightarrow}
\newcommand{\IMPLIES}{\makespace{\Longrightarrow}}
\newcommand{\Conj}{\land}
\newcommand{\CONJ}{\makespace{\Conj}}
\newcommand{\Disj}{\lor}
\newcommand{\DISJ}{\makespace{\Disj}}

\newcommand{\AND}{\makespace{\mathrm{and}}}

\newcommand{\eqdef}{\stackrel{\mathrm{def.}}{=}}
\newcommand{\almev}{\mathrm{a.e.}}
\newcommand{\muae}[1]{#1\,\almev}
\newcommand{\opae}[2]{\stackrel{\muae{#2}}{#1}}
\newcommand{\eqae}[1]{\opae{=}{#1}}
\newcommand{\ltae}[1]{\opae{<}{#1}}
\newcommand{\leqae}[1]{\opae{\leq}{#1}}

\newcommand{\forallae}[1]{\forall_{\muae{#1}}}

\newcommand{\IMPLIESae}[1]{\opae{\IMPLIES}{#1}}
\newcommand{\plusae}[1]{\opae{+}{#1}}

\newcommand{\calA}{\mathcal{A}}
\newcommand{\calB}{\mathcal{B}}
\newcommand{\calBC}{\calB(\matC)}
\newcommand{\calBR}{\calB(\matR)}
\newcommand{\calBRbar}{\calB(\matRbar)}
\newcommand{\calBRbarplus}{\calB(\matRbarplus)}
\newcommand{\calBRn}{\calB(\matR^n)}
\newcommand{\calBRplus}{\calB(\matRplus)}
\newcommand{\calBRplusn}{\calB(\matRplus^n)}
\newcommand{\calBRtwo}{\calB(\matR^2)}
\newcommand{\calC}{\mathcal{C}}
\newcommand{\calCdiff}{\calC^\setminus}
\newcommand{\calD}{\mathcal{D}}

\newcommand{\calI}{\mathcal{I}}
\newcommand{\calIF}{\mathcal{IF}}
\newcommand{\calL}{\mathcal{L}}
\newcommand{\calM}{\mathcal{M}}
\newcommand{\calMC}{\calM_\matC}
\newcommand{\calMR}{\calM_\matR}

\newcommand{\calMplus}{\calM_+}

\newcommand{\calP}{\mathcal{P}}

\newcommand{\calR}{\mathcal{R}}
\newcommand{\calS}{\mathcal{S}}
\newcommand{\calSF}{\mathcal{SF}}
\newcommand{\calSFplus}{\calSF_+}
\newcommand{\calT}{\mathcal{T}}
\newcommand{\calU}{\mathcal{U}}

\newcommand{\Lambdainter}{\Lambda^\cap}

\newcommand{\matUN}{\mathds{1}}

\newcommand{\matC}{\mathbb{C}}
\newcommand{\matK}{\mathbb{K}}
\newcommand{\matN}{\mathbb{N}}

\newcommand{\matNbar}{\overline{\matN}}

\newcommand{\matQ}{\mathbb{Q}}
\newcommand{\matQplus}{\matQ_+}
\newcommand{\matQplusstar}{\matQplus^\star}
\newcommand{\matR}{\mathbb{R}}

\newcommand{\matRbar}{\overline{\matR}}
\newcommand{\matRbarminus}{\matRbar_-}

\newcommand{\matRbarplus}{\matRbar_+}
\newcommand{\matRbarplusstar}{\matRbarplus^\star}

\newcommand{\matRplus}{\matR_+}
\newcommand{\matRplusstar}{\matRplus^\star}
\newcommand{\matRd}{\matR^d}
\newcommand{\matZ}{\mathbb{Z}}
\newcommand{\eps}{\varepsilon}
\newcommand{\fhi}{\varphi}

\renewcommand{\emptyset}{\varnothing}
\renewcommand{\leq}{\leqslant}
\renewcommand{\geq}{\geqslant}

\newcommand{\Ap}{A^\prime}
\newcommand{\Bp}{{B^\prime}}

\newcommand{\Gp}{{G^\prime}}

\newcommand{\Ip}{I^\prime}
\newcommand{\Ipp}{{I^{\prime\prime}}}

\newcommand{\Kp}{{K^\prime}}
\newcommand{\Kpp}{{K^{\prime\prime}}}

\newcommand{\Op}{{O^\prime}}
\newcommand{\calTp}{{\calT^\prime}}
\newcommand{\Xp}{X^\prime}
\newcommand{\Xpp}{{X^{\prime\prime}}}
\newcommand{\Yp}{{Y^\prime}}

\newcommand{\cp}{{c^\prime}}
\newcommand{\fp}{{f^\prime}}

\newcommand{\gp}{{g^\prime}}

\newcommand{\up}{u^\prime}

\newcommand{\vp}{{v^\prime}}

\newcommand{\xp}{{x^\prime}}
\newcommand{\xpn}{{x^\prime_n}}
\newcommand{\xpnp}{{x^\prime_{n+1}}}
\newcommand{\xpz}{{x^\prime_0}}
\newcommand{\xpp}{{x^{\prime\prime}}}

\newcommand{\Sigmap}{\Sigma^\prime}
\newcommand{\Sigmapp}{\Sigma^{\prime\prime}}

\newcommand{\mup}{\mu^\prime}

\newcommand{\imalg}[2]{#2_#1}

\newcommand{\ialf}{\alpha}

\newcommand{\XxX}{{X \times X}}

\newcommand{\RbpxRb}{{\matRbarplus\times\matRbar}}
\newcommand{\RbpxRbp}{{\matRbarplus\times\matRbarplus}}

\newcommand{\olcap}{\:\overline{\cap}\:}
\newcommand{\oltimes}{\:\overline{\times}\:}
\newcommand{\olprod}{\overline{\prod}}

\newcommand{\supX}[1]{{\sup (#1 (X))}}
\newcommand{\supXf}{\supX{f}}

\newcommand{\maxX}[1]{{\max (#1 (X))}}
\newcommand{\maxXf}{\maxX{f}}

\newcommand{\infX}[1]{{\inf (#1 (X))}}
\newcommand{\infXf}{\infX{f}}
\newcommand{\infXtf}{\infX{\tf}}

\newcommand{\minX}[1]{{\min (#1 (X))}}
\newcommand{\minXf}{\minX{f}}

\newcommand{\floor}[1]{\left\lfloor #1 \right\rfloor}

\newcommand{\sgn}{\mathrm{sgn}}
\newcommand{\card}{\mathrm{card}}

\newcommand{\Span}[1]{{\mathrm{span} (#1)}}

\newcommand{\Ray}{\calR^o}
\newcommand{\Itvo}{\calI^o}
\newcommand{\Itvop}{\calI^{o,p}}
\newcommand{\FSp}[2]{{{#2}^{#1}}}

\newcommand{\FXE}{\FSp{X}{E}}
\newcommand{\FXExFXE}{{\FXE\times\FXE}}

\newcommand{\FXR}{\FSp{X}{\matR}}

\newcommand{\FXY}{\FSp{X}{Y}}
\newcommand{\FXYxFXY}{{\FXY\times\FXY}}

\newcommand{\FXYae}[1]{\left(\FXY\right)_{\muae{#1}}}
\newcommand{\FXRae}[1]{\left(\FXR\right)_{\muae{#1}}}

\newcommand{\Arrow}[2]{#1\to#2}
\newcommand{\ArER}{\Arrow{E}{\matR}}
\newcommand{\ArERxERER}{\Arrow{E/\calR\times E/\calR}{E/\calR}}

\newcommand{\ArNI}{\Arrow{\matN}{I}}
\newcommand{\ArNN}{\Arrow{\matN}{\matN}}
\newcommand{\ArNNxN}{\Arrow{\matN}{\matN^2}}

\newcommand{\ArNRb}{\Arrow{\matN}{\matRbar}}
\newcommand{\ArNRbp}{\Arrow{\matN}{\matRbarplus}}
\newcommand{\ArKxERER}{\Arrow{\matK\times E/\calR}{E/\calR}}

\newcommand{\ArRbpxRbRbp}{\Arrow{\RbpxRb}{\matRbarplus}}
\newcommand{\ArRbpxRbpRbp}{\Arrow{\RbpxRbp}{\matRbarplus}}

\newcommand{\ArPRRbp}{\Arrow{\calP(\matR)}{\matRbarplus}}
\newcommand{\ArFXRFXR}{\Arrow{\FXR}{\FXR}}
\newcommand{\ArXxXR}{\Arrow{\XxX}{\matR}}
\newcommand{\ArFXExFXEFXE}{\Arrow{\FXExFXE}{\FXE}}
\newcommand{\ArFXYFXY}{\Arrow{\FXY}{\FXY}}

\newcommand{\ArFXYxFXYFXY}{\Arrow{\FXYxFXY}{\FXY}}

\newcommand{\ArFXYIFXY}{\Arrow{(\FXY)^I}{\FXY}}

\newcommand{\ArXR}{\Arrow{X}{\matR}}
\newcommand{\ArXRp}{\Arrow{X}{\matRplus}}
\newcommand{\ArXRxR}{\Arrow{X}{\matR^2}}
\newcommand{\ArXRb}{\Arrow{X}{\matRbar}}
\newcommand{\ArXiRb}{\Arrow{X_i}{\matRbar}}
\newcommand{\ArXRbp}{\Arrow{X}{\matRbarplus}}
\newcommand{\ArXX}{\Arrow{X}{X}}
\newcommand{\ArXXp}{\Arrow{X}{\Xp}}
\newcommand{\ArXoxXtXp}{\Arrow{X_1\times X_2}{\Xp}}
\newcommand{\ArXpXpp}{\Arrow{\Xp}{\Xpp}}
\newcommand{\ArXY}{\Arrow{X}{Y}}
\newcommand{\ArYR}{\Arrow{Y}{\matR}}

\newcommand{\ArYRb}{\Arrow{Y}{\matRbar}}
\newcommand{\ArYRbp}{\Arrow{Y}{\matRbarplus}}
\newcommand{\ArYY}{\Arrow{Y}{Y}}

\newcommand{\ArYIY}{\Arrow{Y^I}{Y}}

\newcommand{\ArSigRb}{\Arrow{\Sigma}{\matRbar}}
\newcommand{\ArAB}{\Arrow{\calA}{\calB}}
\newcommand{\ArLoneR}{\Arrow{\calLone}{\matR}}
\newcommand{\ArXYae}[1]{\left(\ArXY\right)_{\muae{#1}}}

\newcommand{\Identity}{\mathrm{Id}}
\newcommand{\identity}[1]{\Identity_{#1}}

\newcommand{\idX}{\identity{X}}
\newcommand{\idmatN}{\identity{\matN}}

\newcommand{\vertiii}[1]{
  {\left\vert\kern-0.25ex\left\vert\kern-0.25ex\left\vert #1
    \right\vert\kern-0.25ex\right\vert\kern-0.25ex\right\vert}}

\newcommand{\nrm}[1]{\left\|#1\right\|}
\newcommand{\nrmdot}{\nrm{\cdot}}

\newcommand{\hP}{\hat{P}}
\newcommand{\hQ}{\hat{Q}}
\newcommand{\hR}{\hat{R}}
\newcommand{\hf}{\hat{f}}

\newcommand{\hfhi}{\hat{\fhi}}

\newcommand{\neglset}{{\bf N}}

\newcommand{\calLone}{\calL^1}
\newcommand{\callone}[1]{\calLone(#1)}

\newcommand{\tf}{\tilde{f}}
\newcommand{\tg}{\tilde{g}}
\newcommand{\tm}{\tilde{m}}
\newcommand{\tx}{\tilde{x}}
\newcommand{\tB}{\widetilde{B}}
\newcommand{\tC}{\widetilde{C}}

\newcommand{\cl}[1]{[#1]}
\newcommand{\clplus}{\cl{+}}
\newcommand{\cldot}{\cl{\cdot}}
\newcommand{\clf}{\cl{f}}
\newcommand{\clu}{\cl{u}}
\newcommand{\clup}{\cl{\up}}
\newcommand{\clv}{\cl{v}}
\newcommand{\clvp}{\cl{\vp}}

\newcommand{\Rintitle}{\texorpdfstring{$\matR$}{R}}
\newcommand{\Rbarintitle}{\texorpdfstring{$\matRbar$}{Rbar}}

\newcommand{\calLpintitle}[1]{\texorpdfstring{$\calL^{#1}$}{L#1}}

\newcommand{\Eqrel}{\calR}
\newcommand{\eqrel}[2]{#1 \; \Eqrel \; #2}
\newcommand{\quotplus}{\; [ + ] \;}
\newcommand{\quotdot}{\; [ \cdot ] \;}
\newcommand{\Eqrelae}[1]{\Eqrel_{\muae{#1}}}
\newcommand{\eqrelae}[3]{#2 \; \Eqrelae{#1} \; #3}
\newcommand{\Eqrelp}{\Eqrel^\prime}
\newcommand{\eqrelp}[2]{#1 \; \Eqrelp \; #2}
\newcommand{\Eqrelpae}[1]{\Eqrelp_{\muae{#1}}}
\newcommand{\eqrelpae}[3]{#2 \; \Eqrelpae{#1} \; #3}

\newcommand{\genrestr}[3]{{#2}#1{|#1{#3}}}
\newcommand{\srestr}[2]{\genrestr{_}{#1}{#2}}
\newcommand{\drestr}[2]{\genrestr{^}{#1}{#2}}
\newcommand{\restr}[2]{\srestr{#1}{#2}}

\newcommand{\argmin}{\mathrm{arg}\,\min}

\renewcommand{\Re}[1]{\mathrm{Re}\,#1}
\renewcommand{\Im}[1]{\mathrm{Im}\,#1}

\newcommand{\ulf}{\underline{f}}
\newcommand{\Gbar}{\overline{G}}
\newcommand{\Gbarp}{\Gbar^\prime}
\newcommand{\Xbar}{\overline{X}}
\newcommand{\fbar}{\overline{f}}
\newcommand{\calTbar}{\overline{\calT}}
\newcommand{\Sigmabar}{\overline{\Sigma}}
\newcommand{\Sigmabarp}{\Sigmabar^\prime}

\newcommand{\Dfsum}{\calD^+}

\newcounter{nexttheorem}
\stepcounter{nexttheorem}

\RRdate{January 2021}

\RRversion{2}
\RRdater{April 2021}

\RRauthor{
  François Clément%
  \thanks[serena]{%
    Inria \&
    CERMICS, École des Ponts, 77455 Marne-la-Vallée Cedex 2, France.
    \texttt{Francois.Clement@inria.fr}.}
  \and
  Vincent Martin%
  \thanks[lmac]{%
    Université de technologie de Compiègne,
    LMAC (Laboratory of Applied Mathematics of Compiègne),\protect\\
    CS 60319, 60203 Compiègne Cedex, France.
    \texttt{Vincent.Martin@utc.fr}.}
}
\authorhead{%
  F. Clément, \& V. Martin}

\RRtitle{%
  Intégrale de Lebesgue.\\
  Preuves détaillées en vue d'une formalisation en Coq}
\RRetitle{%
  Lebesgue integration.\\
  Detailed proofs to be formalized in Coq}
\titlehead{%
  Detailed proofs for Lebesgue integration}

\RRnote{%
  This work was partly supported by
  GT ELFIC from Labex DigiCosme - Paris-Saclay, and by
  the Paris Île-de-France Region (DIM RFSI MILC).}

\RRresume{%
  Pour obtenir la plus grande confiance en la correction de programmes de
  simulation numérique implémentant la méthode des éléments finis, il faut
  formaliser les notions et résultats mathématiques qui permettent d'établir la
  justesse de la méthode.
  Les espaces de Sobolev sont le cadre mathématique dans lequel la plupart des
  formulations faibles pour résoudre les équations aux dérivées partielles sont
  posées, et où les  solutions sont recherchées.
  La construction de ces espaces fonctionnels repose sur
  le calcul intégral et la théorie de la mesure.
  Ce chapitre de l'analyse fonctionnelle est donc un fondement
  théorique nécessaire à la définition de la méthode des éléments finis.
  L'objectif de ce document est de fournir à la communauté des chercheurs en
  preuve formelle des preuves papiers très détaillées des principaux résultats
  du calcul intégral et de la théorie de la mesure.}
\RRabstract{%
  To obtain the highest confidence on the correction of numerical simulation
  programs implementing the finite element method, one has to formalize the
  mathematical notions and results that allow to establish the soundness of the
  method.
  Sobolev spaces are the mathematical framework in which most weak formulations
  of partial derivative equations are stated, and where solutions are sought.
  These functional spaces are built on
  integration and measure theory.
  Hence, this chapter in functional analysis is a mandatory theoretical
  cornerstone for the definition of the finite element method.
  The purpose of this document is to provide the formal proof community with
  very detailed pen-and-paper proofs of the main results from
  integration and measure theory.}

\RRmotcle{%
  théorie de la mesure,
  intégrale de Lebesgue,
  preuve mathématique détaillée,
  preuve formelle en analyse réelle}
\RRkeyword{%
  measure theory,
  Lebesgue integration,
  detailed mathematical proof,
  formal proof in real analysis}

\RRprojet{Serena}

\RCParis

\begin{document}

\RRNo{9386}
\makeRR

\chapter*{Foreword}

\newcommand{\HALRR}{https://hal.inria.fr/hal-03105815}
\newcommand{\Version}[1]{%
  \href{https://hal.inria.fr/hal-03105815v#1}{Version~#1}}

\noindent
This document is intended to evolve over time.
Last version is release~1.1 ({\ie} version~2).\\
It is available at \url{\HALRR/}.

\bigskip\noindent
\Version{2} (release 1.1, 2021/04/01) is a minor revision.\\
Main changes are:
\begin{CompactList}
\item addition of this foreword;

\item reordering and sectioning:\\
  sections have become chapters and gathered into parts to ease the reading;\\
  all ``complements'' are gathered into a single chapter;\\
  Section~\ref{s:complements-on-algebraic-structures} about algebraic structures
  is moved right after Section~\ref{s:complements-on-set-theory} about
  set theory;\\
  Chapter~\ref{c:subset-systems} on subset systems is separated from
  Chapter~\ref{c:measurability} on measurability;

\item some color modifications to increase legibility after grayscale printing;

\item new contents in the introduction (Chapter~\ref{c:introduction-1}), and
  minor corrections in
  Chapters~\ref{c:notations}--\ref{c:statements-and-sketches-of-the-proofs}
  (including a bunch of hyperlinks in Chapters~\ref{c:proof-techniques}
  and~\ref{c:statements-and-sketches-of-the-proofs});

\item fix proofs of uniqueness in
  Lemma~\ref{l:uniq-of-meas-ext-from-p-syst} (statement changed too) and
  Theorem~\ref{t:caratheodory-lebesgue-meas-on-r};

\item fix statement and proof of Theorem~\ref{t:tonelli} and
    Lemma~\ref{l:tonelli-over-subset}, with new
    Definition~\ref{d:partial-fun-of-fun-from-prod-space};

\item fix statement of Theorem~\ref{t:fatou-lemma};

\item alternate proof of additivity for the integral of nonnegative simple
  functions in Lemma~\ref{l:int-in-sfplus-is-add} based on the new
  {\em disjoint} representation of simple functions of
  Lemma~\ref{l:sf-disj-repr};\\
  this involves new Definition~\ref{d:pseudopart},
  new Lemmas~\ref{l:if-is-meas},
  \ref{l:if-is-sigma-add},
  \ref{l:sf-disj-repr-is-subpart-of-can-repr},
  \ref{l:sf-is-alg-over-r},
  \ref{l:sfplus-disj-repr},
  \ref{l:sfplus-disj-repr-is-subpart-of-can-repr},
  \ref{l:sfplus-is-closed-under-pos-alg-ops},
  and~\ref{l:equiv-def-of-int-in-sfplus-disj},
  new Remarks~\ref{r:v2-new01}
  \ref{r:v2-new11},
  \ref{r:v2-new12},
  \ref{r:v2-new13},
  \ref{r:v2-new14},
  and~\ref{r:v2-new15},
  addition of uniqueness in
  Lemmas~\ref{l:sf-can-repr}
  and~\ref{l:sfplus-can-repr},
  thus impacting proofs of
  Lemmas~\ref{l:int-in-if-is-add},
  \ref{l:int-in-if-over-subset-is-add},
  \ref{l:int-in-sfplus-gen-int-in-if},
  \ref{l:int-in-sfplus-is-pos-lin},
  \ref{l:int-in-sfplus-over-subset-is-add},
  and~\ref{l:int-in-mplus-over-subset-is-sigma-add};

\item factor the proof on {\em layers} into new
  Lemmas~\ref{l:part-of-count-union-in-set-alg},
  \ref{l:part-of-count-union-in-sigma-alg},
  and~\ref{l:l-is-set-alg},
  thus impacting proofs of
  Lemmas~\ref{l:meas-is-cont-from-below},
  \ref{l:part-of-count-union-in-l},
  and~\ref{l:l-is-sigma-alg};

\item new
  Lemma~\ref{l:everywhere-implies-almost-everywhere-for-almost-the-same} and
  Remark~\ref{r:v2-new07} about reasoning with properties satisfied almost
  everywhere, now used in the proofs of
  Lemmas~\ref{l:compat-of-almost-bin-rel-with-refl},
  \ref{l:compat-of-almost-bin-rel-with-sym},
  \ref{l:compat-of-almost-bin-rel-with-antisym},
  \ref{l:compat-of-almost-bin-rel-with-trans},
  \ref{l:compat-of-almost-bin-rel-with-op},
  and~\ref{l:definiteness-implies-almost-definiteness}
  (statement changed too in the latter);

\item new
  Lemmas~\ref{l:inf-of-bounded-seq-is-bounded},
  \ref{l:sup-of-bounded-seq-is-bounded},
  \ref{l:mplus-is-closed-under-inf},
  and~\ref{l:mplus-is-closed-under-sup}
  shortening proofs of
  Lemmas~\ref{l:meas-of-meas-of-section-finite},
  and~\ref{l:meas-of-meas-of-section};

\item additional Remarks~\ref{r:v2-new02}, \ref{r:v2-new03}, \ref{r:v2-new04},
  \ref{r:v2-new05}, \ref{r:v2-new06}, \ref{r:v2-new08}, \ref{r:v2-new09},
  \ref{r:v2-new10}, \ref{r:v2-new16}, \ref{r:v2-new17}, \ref{r:v2-new18},
  \ref{r:v2-new19}, \ref{r:v2-new20}, \ref{r:v2-new21},
  and modification of Remarks~\ref{r:v2-mod1}, \ref{r:v2-mod2},
  \ref{r:v2-mod3}, and~\ref{r:v2-mod4};

\item Lemmas~\ref{l:count-union-of-sections-is-meas},
  and~\ref{l:count-inter-of-sections-is-meas} are moved from
  Section~\ref{s:measure-and-integration-over-product-space} to
  Section~\ref{s:product-of-measurable-spaces};

\item and possibly other slight modifications to improve the proof contents and
  their legibility.
\end{CompactList}

\bigskip\noindent
\Version{1} (release 1.0, 2021/01/14) is the first release.\\
It covers:
\begin{CompactList}
\item the main basic results in measure theory,
  including the construction of Lebesgue measure in~$\matR$ via {\Cara}'s
  extension scheme;

\item integration of nonnegative measurable functions,
  including the {\BLt} and {\Fl};

\item the integral over a product space, including the {\Tont};

\item integration of measurable functions with possibly changing sign,
  including the seminormed vector space~$\calL^1$, and {\Ldcvt}.
\end{CompactList}

\dominitoc
\setcounter{minitocdepth}{3}

\tableofcontents

\part{Overview}
\label{p:overview}

\chapter{Introduction}
\label{c:introduction-1}

\section{Formal proof}
\label{s:formal-proof}

A formal proof is conducted in a logical framework that provides dedicated
computer programs to mechanically check the validity of the proof, the
so-called formal proof assistants.
Such formal proofs may concern known mathematical theorems, but also properties
of some piece of other computer programs, {\eg} see~\cite{har:fpt:08},
and~\cite[Glossary p. 343]{bol:tcm:14}.
This field of computer science is extremely popular as it allows to certify
with no doubt the behavior of critical programs.

Interactive theorem provers are now known to be able to tackle real analysis.
For instance, in the field of Ordinary Differential Equations (ODEs) with
{\isabelle}~\cite{ih:nao:12,imm:fvc:14,it:fod:16}, and
{\coq}~\cite{ms:pao:13}, or in the field of Partial Differential Equations
(PDEs), again with {\isabelle}~\cite{ap:ihf:16}, and
{\coq}~\cite{bol:wen:13,bol:tcm:14}.
In the latter example, the salient aspect is that the round-off error due to
the use of IEEE-754 floating-point arithmetic can also be fully taken into
account.
But the price to pay is that every details of the proofs have to be dealt with,
and thus the availability of very detailed pen-and-paper proofs is a major
asset.

\section{Objective}
\label{s:objective}

Our long term purpose is to formally prove programs implementing the {\FEM}
(FEM).
The FEM is widely used to solve a broad class of PDEs, mainly because it has a
sound mathematical foundation,
{\eg} see~\cite{ztz:fem:13,cia:fem:78,qv:nap:94,bs:mtf:08,eg:tpf:04}.
The {\LMT}, one of the key ingredients to establish the FEM, was already
addressed in~\cite{cm:lmt:16} for a detailed pen-and-paper proof, and
in~\cite{bol:cfp:17} for a formal proof in~{\coq}.
The present document is a further contribution toward our ultimate goal.

The {\LMT} claims existence and uniqueness of the solution to the weak
formulation of a PDE problem, such as the Poisson problem, and its discrete
approximation;
it is stated on a Hilbert space, {\ie} a complete inner product space.
The simplest Hilbert functional spaces relevant for the resolution of PDEs
are the spaces~$L^2$ and~$H^1$.
More generally, when stronger results such as the {\BNBT} are involved, one may
be interested in the Sobolev spaces~$W^{m,p}$ that are Banach spaces, {\ie}
complete normed vector spaces.

Let~$d\in\{1,2,3\}$ be the spatial dimension.
Let~$\Omega$ be an open subset of~$\matRd$.
We consider real-valued functions defined almost everywhere over~$\Omega$.
For $0<p\leq\infty$, the Lebesgue space~$L^p(\Omega)$ is the space of
measurable functions for which the $p$-th power of the absolute value has a
finite integral over~$\Omega$.
For any natural number~$m$, the Sobolev space~$W^{m,p}(\Omega)$ is the
{\vectorsubspace} of~$L^p(\Omega)$ of functions for which all weak derivatives
up to order~$m$ also belong to~$L^p(\Omega)$.
Thus, Lebesgue spaces correspond to the case $m=0$,
$L^p(\Omega)=W^{0,p}(\Omega)$.
And the Hilbert spaces correspond to the case $p=2$,
$H^m(\Omega)\eqdef W^{m,2}(\Omega)$.

\section{Integration theories}
\label{s:integration-theories}

There is a huge variety of concepts of integral and integrability in the
literature, {\eg} see~\cite{cha:tic:02,bur:gi:07}, and one may wonder which one
to use.
Some are overridden by others, some are equivalent, and some have been
developed for specific situations, such as vector-valued functions or
functions defined on an infinite-dimensional domain.
But a few of them have gained popularity, be it for their appropriateness for
teaching, or their general-purpose nature: namely Riemann, Lebesgue, and {\HK}
integrations.

Lebesgue integral (with Lebesgue measure) is the traditional framework in which
the Sobolev functional spaces are defined.
Indeed, Riemann integral is disqualified because of its poor results on limit
and integral exchange making completeness unreachable, and so is {\HK}
integral, because of its not so obvious extension to the multidimensional case
and construction of a complete normed vector space of HK-integrable
functions~\cite{gw:faf:16,mt:dch:19}.

\section{Contents}
\label{s:contents}

The present version of this document covers all material up to the first
properties of the seminormed vector space~$\calL^1$ of integrable functions
(before taking the quotient to obtain the normed vector space~$L^1$).
Among the rich literature on Lebesgue integral theory, it was mostly derived
from the textbooks~\cite{mai:m2:14,gh:mip:13,rud:rca:87}.

It includes results on the general concepts of measurability, measure and
negligibility, and integration of nonnegative measurable functions culminating
with the {\BLt} allowing to exchange limit and integral for nondecreasing
sequences of measurable functions, and {\Fl} that only gives an upper bound
when the sequence is not monotone.
The formalization in {\coq} of all these aspects is presented
in~\cite{bol:cfp:21}.

In addition, this document also covers the building of the Lebesgue measure
in~$\matR$ using {\Cara}'s extension scheme, the integral over a product space
including the {\Tont} (for nonnegative measurable functions, whereas the
{\Fubt} deals with integrable functions), and the integral for measurable
functions with possibly changing sign, including {\Ldcvt}.

It is planned to add more results in a forthcoming version.

\section{Teaching}
\label{s:teaching}

This document is not primarily meant for teaching usage.
The objective was to be as comprehensive as possible in the proofs.
This led to very detailed demonstrations, and to a compact style of writing
that is not common, and may seem daunting to the uninformed reader.

However, the authors tried to give some insights on the integration theory and
on the proofs and theorems in the next introductory sections.
They also strove to give some indications in the proofs when they felt it
necessary.
They believe that this document could be useful for interested teachers, and
dedicated students.

\section{Disclaimer}
\label{s:disclaimer}

Note that the manuscript itself is not formally proved (and will never be).
Indeed, {\LaTeX} compilers are not formal proof tools.

Moreover, formalization is not just straightforward translation of mathematical
texts and formulas.
Some design choices have to be made and proof paths may differ, mainly to favor
usability of {\coq} theorems and ease formal developments.
Thus, there exist various differences between the mathematical setting
presented here, and the formal setting developed in {\coq}~\cite{bol:cfp:21}.
For instance, the {\coq} definition of measurability of subsets ({\ie} of
$\sigma$-algebra) takes the generators as parameter, and not only the basic
axioms.
The ubiquitous use of {\em total} functions (defined for {\em all} values of
their arguments) may be surprising at first: {\eg} the addition in~$\matRbar$
(``$\infty-\infty$'' is defined and takes the value~0), or the measure (one
can take the measure of any subset, even nonmeasurable ones).
Of course, just as in mathematical statements, unwanted behaviors are somehow
prevented, for instance by adding an hypothesis stating the legality of the
addition or the measurability of the subset (see~\cite{bol:cfp:21} for
details).

Hence, despite the care taken in its writing, this document might still be prone
to errors or holes in the demonstrations.
There could also exist simpler paths in the proofs.
Please, feel free to inform the authors of any such issue, and to share any
comments or suggestions\ldots

\section{Organization}
\label{s:organization}

Part~\ref{p:overview} of this document is organized as follows.
The notations are first collected in Chapter~\ref{c:notations}.
Then, the literature on the subject is briefly reviewed in
Chapter~\ref{c:state-of-the-art}.
Chapter~\ref{c:proof-techniques} gathers some proof techniques.
The chosen proof paths of the main results are then sketched in
Chapter~\ref{c:statements-and-sketches-of-the-proofs}.

Part~\ref{p:detailed-proofs} (Chapters~\ref{c:introduction-2}
to~\ref{c:integration-of-real-functions}) is the core of this document.
In this part, the definitions are presented, and the lemmas and theorems are
stated with their detailed proofs.
Its organization is described at the end of the introductory
Chapter~\ref{c:introduction-2}.

Chapter~\ref{c:conclusions-perspectives} concludes and gives some
perspectives.

Finally, an appendix gathers the list of statements in
Chapter~\ref{c:lists-of-statements}, and explicit dependencies (both ways) in
Chapters~\ref{c:the-proof-cites-explicitly}
and~\ref{c:is-explicitly-cited-in-the-proof-of}.
The appendix is not intended for printing!

\chapter{Notations}
\label{c:notations}

In this chapter (as in most of the document), we use the following
conventions:
\begin{CompactList}
\item
  capital letters~$X$ and~$Y$ denote ``surrounding'' sets, typically the
  domain and codomain of functions,
  $A$ and~$B$ are subsets of~$X$,
  and~$I$ denotes a set of indices;

\item
  calligraphic letter~$\calR$ denotes a binary relation on~$X$ ({\eg}
  equality or inequality);

\item
  the capital letter~$P$ denotes a predicate, {\ie} a function taking Boolean
  values;

\item
  small letters~$x$, $y$ and~$i$ denote elements of the set using the
  matching capital letter;

\item
  small letters~$a$ and~$b$ are (extended) real numbers (such that
  $a\leq b$),
  and~$n$ and~$p$ are (extended) natural numbers (such that $n\leq p$);

\item
  small letter~$f$ denotes a function, {\eg} from set~$X$ to set~$Y$;

\item
  the capital Greek letter~$\Sigma$ denotes a $\sigma$-algebra on~$X$, and
  the small Greek letter~$\mu$ denotes a measure on the measurable
  space~$(X,\Sigma)$.
\end{CompactList}

\medskip\noindent
The following notations and conventions are used throughout this document.
\begin{itemize}[itemsep=0pt,topsep=0pt]
\item Logic:
  \begin{CompactList}
  \item
    Using a compound (tuple of elements of~$X$, or subset of~$X$) in an
    expression at a location where only a single element makes sense is a
    shorthand for the same expression expanded for all elements of the
    compound;
    for instance, ``$\forall x,\xp\in X$'' means
    ``$\forall x\in X,\forall\xp\in X$'',
    ``$\forall(x_i)_{i\in I}\in X$'' means ``$\forall i\in I$, $x_i\in X$'',
    and ``$x,\xp\calR\xpp$'' means $x\calR\xpp$ and $\xp\calR\xpp$;

  \item
    $x_0\calR_1x_1\ldots\calR_mx_m$ is a shorthand for
    $x_0\calR_1x_1\Conj x_1\calR_2x_2\Conj\ldots\Conj x_{m-1}\calR_mx_m$;
  \item ``iff'' is a shorthand for ``if and only if''.
  \end{CompactList}

\item Set theory:
  \begin{CompactList}
  \item
    $\calP(X)$ denotes the power set of~$X$, {\ie} the set of its subsets;

  \item
    subsets are denoted using~$\subset$, and proper subsets by~$\subsetneq$;

  \item
    $n$-ary set operations, such as intersection and union, have precedence
    over binary set operations (no need for big parentheses);
    for instance, $\bigcup_{i\in I}A_i\cap B$ means
    $\left(\bigcup_{i\in I}A_i\right)\cap B$;

  \item
    superscript $^c$ denotes the absolute complement;

  \item
    $\setminus$ denotes the set difference (or relative complement):
    $A\setminus B\eqdef A\cap B^c$;\\
    when $B\subset A$, $A\setminus B$ is also called the
    {\em local complement of~$B$ (in~$A$)};

  \item
    $\uplus$ denotes the disjoint union:
    $A\uplus B$ means $A\cup B$ with the assumption that $A\cap B=\emptyset$;

  \item
    $\olcap$ is used to denote a set of traces of subsets,
    see Definition~\ref{d:trace-of-subsets-of-parties};

  \item
    $\oltimes$ is used to denote a set of Cartesian products of subsets,
    see Definition~\ref{d:prod-of-subsets-of-parties};

  \item
    $\card(X)$ is the cardinal of~$X$, {\ie} the number of its elements, with
    the convention $\card(X)\eqdef\infty$ when~$X$ is infinite;

  \item
    countable means either finite, or infinitely countable:
    $X$~is countable iff
    there exists an injection from~$X$ to~$\matN$ iff
    there exists $I\subset\matN$ and a bijection between~$I$ and~$X$;

  \item
    $\matUN_A$, or $\matUN_A^X$, denotes the indicator function of the
    subset~$A$ of~$X$.
    It is the function from~$X$ to~$\matR$, or~$\matRbar$, that takes the
    value~1 for all elements of~$A$, and the value~0 elsewhere;

  \item
    the set of functions from~$X$ to~$Y$ is either denoted~$\FXY$, or through
    the type annotation ``$\ArXY$''.
    Both compact expressions ``let~$f\in\FXY$'' and ``let~$f:\ArXY$'' mean
    ``let~$f$ be a function from~$X$ to~$Y$'';

  \item
    to avoid double parentheses/braces, the expressions involving inverse
    images of singletons are simplified: $f^{-1}(y)$ is a shorthand for
    $f^{-1}(\{y\})$;

  \item
    $\{P(f)\}$ is a shorthand for $\{x\in X\st P(f(x))\}$, or
    $f^{-1}(\{y\in Y\st P(y)\})$;
    for instance, $\{f<a\}$ means either $f^{-1}[-\infty,a)$ or
    $f^{-1}(-\infty,a)$,
    see Lemma~\ref{l:meas-of-num-fun-to-r}, and
    Lemma~\ref{l:meas-of-num-fun};

  \item
    $f^+$ and~$f^-$ denote the nonnegative and nonpositive parts of a
    numerical function~$f$, see Definition~\ref{d:nonneg-and-nonpos-parts}.
  \end{CompactList}

\item Totally ordered set:
  \begin{CompactList}
  \item
    when the nonempty set~$X$ is totally ordered, intervals follow the
    notations used for real numbers;
    for instance, $(x_1,x_2]$ represents the subset
    $\{x\in X\st x_1<x\leq x_2\}$;

  \item
    the square-and-round bracket notation is used for not specifying whether
    the bound is included or not: $(x_1,x_2\rsrbra$ means either $(x_1,x_2)$ or
    $(x_1,x_2]$;

  \item
    the lower and greater bounds of~$X$ may be denoted~$-\infty$ and~$+\infty$
    (as usual, the plus sign may be omitted);
    they may belong to~$X$ or not;
    $\Xbar$~denotes $X\cup\{-\infty,\infty\}$.

  \item
    rays, or half-lines, are denoted using the square-and-round notation:
    whether~$\infty$ belongs to~$X$ or not, $(x,\infty\rsrbra$ represents
    $\{\xp\in X\st x<\xp\}$, {\ie} either $(x,\infty]$ or $(x,\infty)$,
    see Definition~\ref{d:interval};

  \item
    $\Itvop_X$ denotes the set of open proper intervals of~$X$,
    see Definition~\ref{d:interval};

  \item
    $\Ray_X$ denotes the set of open rays of~$X$,
    see Definition~\ref{d:interval}.

  \item
    $\Itvo_X$ denotes the set of open intervals of~$X$,
    see Definition~\ref{d:interval}.
  \end{CompactList}

\item Numbers:
  \begin{CompactList}
  \item
    $\floor{\cdot}$ denotes the floor function from~$\matR$ to~$\matN$:
    $\floor{a}\leq a<\floor{a}+1$;

  \item
    $\lsrbra n..p\rsrbra$ denotes the integer interval
    $\lsrbra n,p\rsrbra\cap\matN$ (with the natural variants using
    parentheses and/or brackets when exclusion or inclusion of bounds is
    specified);

  \item
    $\ell(\lsrbra a,b\rsrbra)$ denotes the length of the
    interval~$\lsrbra a,b\rsrbra$,
    see Definition~\ref{d:len-of-int}.
  \end{CompactList}

\item General topology:
  \begin{CompactList}
  \item
    $\calT_X(G)$ denotes the topology generated by~$G\subset\calP(X)$,
    see Definition~\ref{d:gen-topo};
  \end{CompactList}

\item Measure theory:
  \begin{CompactList}
  \item
    $\Pi_X(G)$ denotes the $\pi$-system generated by~$G\subset\calP(X)$,
    see Definition~\ref{d:gen-p-syst};

  \item
    $\calA_X(G)$ denotes the set algebra generated by~$G\subset\calP(X)$,
    see Definition~\ref{d:gen-set-alg};

  \item
    $\calC_X(G)$ denotes the monotone class generated by~$G\subset\calP(X)$,
    see Definition~\ref{d:gen-monot-class};

  \item
    $\Lambda_X(G)$ denotes the $\lambda$-system generated
    by~$G\subset\calP(X)$,
    see Definition~\ref{d:gen-l-syst};

  \item
    $\Sigma_X(G)$ denotes the $\sigma$-algebra generated
    by~$G\subset\calP(X)$, see Definition~\ref{d:gen-sigma-alg};

  \item
    $\calB(X)$ denotes the Borel $\sigma$-algebra generated by the open
    subsets, see Definition~\ref{d:borel-sigma-alg};

  \item
    $\bigotimes_{i\in[1..m]}\Sigma_i$ denotes the tensor product of the
    $\sigma$-algebras~$(\Sigma_i)_{i\in[1..m]}$,
    see Definition~\ref{d:tensor-prod-of-sigma-algs};

  \item
    $\neglset$ denotes the set of negligible subsets,
    see Definition~\ref{d:negl-subset};

  \item
    the annotation~``$\muae{\mu}$'' specifies that the proposition is only
    considered almost everywhere,
    {\eg}~$\eqae{\mu}$, $\forallae{\mu}$, $\ArXYae{\mu}$, or~$\Eqrelae{\mu}$,
    see Definition~\ref{d:prop-almost-satisfied};

  \item
    $\delta_Y$ denotes the counting measure associated with~$Y$,
    $\delta_a$ denotes the Dirac measure at~$a$,
    see Lemma~\ref{l:count-meas} and Definition~\ref{d:dirac-meas};

  \item
    $\calL$ denotes the Lebesgue $\sigma$-algebra on~$\matR$ and
    $\lambda$ denotes the {\BoL} measure on~$\matR$,
    see Definitions~\ref{d:lambda-star-lebesgue-meas-cand}
    and~\ref{d:l-lebesgue-sigma-alg}, and
    Theorem~\ref{t:caratheodory-lebesgue-meas-on-r};

  \item
    $\mu_1\otimes\mu_2$ denotes the tensor product of measures~$\mu_1$
    and~$\mu_2$, see
    Definition~\ref{d:tensor-prod-meas},
    Definition~\ref{d:cand-tensor-prod-meas}, and
    Lemma~\ref{l:uniq-of-tensor-prod-meas};

  \item
    $\lambda^{\otimes 2}$ denotes the Lebesgue measure on~$\matR^2$,
    see Lemma~\ref{l:lebesgue-meas-on-r2};
  \end{CompactList}

\item Lebesgue integral:
  \begin{CompactList}
  \item
    $\calIF$ denotes the set of measurable indicator functions,
    see Definition~\ref{d:if-set-of-meas-indic-funs};

  \item
    $\calSF$ denotes the {\vectorspace} of simple functions,
    see Definition~\ref{d:sf-vector-space-of-simple-funs};

  \item
    $\calSFplus$ denotes the set of nonnegative simple functions,
    see Definition~\ref{d:sfplus-subset-of-nonneg-simple-funs};

  \item
    $\calMplus$ denotes the set of nonnegative measurable functions
    $\ArXRbp$, see Definition~\ref{d:mplus-subset-of-nonneg-meas-num-fun};

  \item
    $\calM$ denotes the set of measurable functions $\ArXRb$,
    see Definition~\ref{d:m-set-of-meas-num-funs};

  \item
    $\calMR$ denotes the {\vectorspace} of measurable functions $\ArXR$,
    see Definition~\ref{d:mr-vector-space-of-meas-num-fun-to-r};

  \item
    $\int f\,d\mu$ denotes the (Lebesgue) integral of~$f$ for the
    measure~$\mu$, whether the function belongs to~$\calIF$, $\calSFplus$,
    $\calMplus$, or~$\calM$,
    see Section~\ref{s:lebesgue-scheme}, Definition~\ref{d:int-in-if},
    Lemmas~\ref{l:int-in-sfplus} and~\ref{l:int-in-mplus}, and
    Definition~\ref{d:integral};

  \item
    $\int_A f\,d\mu$ denotes the integral of the restriction of~$f$ to~$A$,
    see Lemmas~\ref{l:int-in-if-over-subset},
    \ref{l:int-in-sfplus-over-subset}, \ref{l:int-in-mplus-over-subset},
    and~\ref{l:int-over-subset};

  \item
    $\int_a^bf(x)\,d\mu(x)$ denotes the integral of~$f$ over the
    interval~$(a,b)$, see Lemma~\ref{l:int-over-int};

  \item
    $\int f(x)\,dx$ denotes the integral of~$f$ for the Lebesgue measure,
    see Definition~\ref{d:int-for-lebesgue-meas-on-r};

  \item
    $(\calL^1,N_1)$ denotes the seminormed {\vectorspace} of integrable
    functions, see Lemma~\ref{l:seminorm-llone}, and
    Definition~\ref{d:llone-vector-space-of-int-fun};

  \item
    $\calI$ denotes the integral operator for integrable functions,
    see Lemma~\ref{l:int-is-pos-lin-form-on-llone}.
  \end{CompactList}
\end{itemize}

\chapter{State of the art}
\label{c:state-of-the-art}

After a brief survey of the wide variety of integrals, and of the means to
build Lebesgue integral and Lebesgue measure, we review some works of a few
authors, partly from the mathematical French school, that provide some details
about results in measure theory, Lebesgue integration, and basics of functional
analysis.

As usual, proofs provided in the literature are not comprehensive, and we
have to cover a series of sources to collect all the details necessary for a
formalization in a formal proof tool such as~{\coq}.
Usually, Lecture Notes in undergraduate mathematics are very helpful and
we selected~\cite{gos:cms1:93,gos:cms2:93,gos:cms3:93} among many other
possible choices.

\section{A zoology of integrals}
\label{s:a-zoology-of-integrals}

The history of integral calculus dates back at least from the Greeks, with the
method of exhaustion for the evaluation of areas, or volumes.
Since then, mathematicians have constantly endeavored to develop new
techniques.
The objective is mainly to be able to integrate a wider class of functions.
But also to establish more powerful results, such as convergence results, or
the Fundamental Theorems of Calculus (expressing that derivation and
integration are each other inverse operations).
And possibly to fit specific contexts, {\eg} driven by applications in
physics, or for teaching purposes.
For a wide panel, and pros and cons of different approaches, see for
instance~\cite{cuc:qia:98,cha:tic:02,bur:gi:07}, in which more than a hundred
named integrals are listed.
But, as pointed out in~\cite{tav:rgi:09}, building a universal integral ``will
probably be unnecessarily complicated when restricted to a simple setting''.

Considering the case of a numerical function defined on some interval, let us
review some of the most popular ones.

\subsection{{\LN} integral (late 17th century)}
\label{ss:leibniz-newton-integral-late-17th-century}

The {\LN} (LN-)integral is defined in a descriptive way as the difference of
any primitive evaluated at the bounds of the integration interval.
All continuous functions are LN-integrable.

This integral possesses interesting properties such as formulas for integration
by parts or by substitution, and a uniform convergence theorem.
But since the definition is not constructive, there is no regular process to
build the primitive of a function, even if it is continuous.
Moreover, other monotone or dominated convergence results must assume that the
limit is LN-integrable.
And it raises the question of integrability of functions with no primitives.

\subsection{Riemann integral (early 19th century)}
\label{ss:riemann-integral-early-19th-century}

The Riemann (R-)integral~\cite{rie:dft:68} consists in cutting the area
``under'' the graph of the function into vertical rectangular strips by
choosing a subdivision of the integration interval, and in increasing the
number of strips by making the step of the subdivisions go down to~0.
Equivalently, it might be more convenient to consider the Darboux
integral~\cite{dar:mfd:75}, which checks that both upper and lower so-called
Darboux sums have the same limit.

This simple approach has similar properties than the LN-integral, with for
instance a uniform convergence result, and R-integrable functions are the
piecewise continuous function which set of discontinuity points has null
measure.
But again, monotone convergence and dominated convergence theorems must assume
R-integrability of the limit.

\subsection{Lebesgue integral (early 20th century)}
\label{ss:lebesgue-integral-early-20th-century}

The key idea of the Lebesgue (L-) approach~\cite{leb:lir:04} is to consider
subdivisions of the codomain of the function, {\ie} cutting the area under the
graph into horizontal pieces, which are no longer continuous strips when the
function is not concave, and to define a measure for these pieces that
generalizes the length of intervals.

This has the great advantage of providing powerful monotone and dominated
convergence theorems, and to allow for an abstract setting in which one can
handle functions defined on more general spaces than the Euclidean
spaces~$\matR^n$.
And this paves the road for probability theory.
For instance, the Wiener measure used for the study of stochastic processes
such as Brownian motion is defined on the infinite-dimensional Banach space of
continuous functions from a compact interval to~$\matR^n$.
Moreover, the concept of property satisfied {\em almost everywhere} opens the
way to the~$L^p$ Lebesgue spaces as (complete) \underline{normed} vector
spaces (Banach spaces).

The main difficulty is the necessity to develop the measure theory to be able
to associate a length to the ``horizontal pieces''.
In general, this is not possible for all subsets of the domain of the function,
and leads to the concepts of {\em measurable subset} and
{\em measurable function}.

\medskip

Going back to the geometrical interpretation of the area under the graph of the
function, it is interesting to note that the opposition Riemann (subdivision of
the domain) versus Lebesgue (subdivision of the codomain) reflects in the field
of ordinary differential equations where the classical numerical schemes based
on a discretization of the time domain are now opposed to the recent quantized
state system solvers that are based on the quantization of the state variables,
see~\cite{ck:css:06,fk:sqs:14,fkb:pqs:17}.
Moreover, we may also cite~\cite{glu:nsl:10} in the field of signal processing
where the Lebesgue geometrical scheme is seen as a nonlinear sampling qualified
of ``noise-free quantization'', that could have impact on electronic design.

\subsection{{\HK} integral (mid-20th century)}
\label{ss:henstock-kurzweil-integral-mid-20th-century}

The {\HK} (HK-)integral, or gauge
integral~\cite{kur:god:57,hen:ti:63,bar:mti:01} is a generalization of the
R-integral for which the subdivision of the integration interval is no longer
uniform, but is driven by the variations of the function to integrate through a
so-called {\em gauge} function.

It is more powerful than the L-integral in the sense that a real-valued
function is L-integrable if and only if the function and its absolute value are
HK-integrable.
Unlike the L-integral, this leads to formulations of the fundamental theorem of
calculus without the need for the concept of improper integral.
The monotone and dominated convergence theorems are also valid.
Moreover, the HK-integral is usually considered more suited for teaching, for
it is hardly more complex than the R-integral.

However, unlike the L-integral, the HK-integral cannot be easily extended to
the Euclidean spaces~$\matR^n$, and the construction of a Banach space of
HK-integrable function is far less obvious than that of the $L^1$ Lebesgue
space~\cite{gw:faf:16,mt:dch:19}.

\subsection{Daniell integral (early 20th century)}
\label{ss:daniell-integral-early-20th-century}

The Daniell (D-) approach~\cite{dan:gfi:18,fol:ram:99} is a general scheme that
extends an {\em elementary} integral defined for {\em elementary} functions to
a much wider class of functions by checking that upper and lower elementary
integrals of elementary functions share the same limit, in a way that is
similar to the Darboux approach.
When applied to the integral of simple functions, this is somehow equivalent to
steps~3 and~4 of the Lebesgue scheme described later in
Section~\ref{s:lebesgue-scheme}.
But it is enough to consider the R-integral for the compactly supported
continuous functions ({\aka} the Cauchy integral) to get back the L-integral
for the Lebesgue measure on the Euclidean spaces~$\matR^n$.

This Daniell approach to the L-integral has the great advantage not to need the
concept of measure, which can be reconstructed afterwards by taking the
integral of indicator functions of measurable subsets.
Moreover, in the mid-20th century, this approach was shown in~\cite{bc:cmt:72}
to be compatible with constructive analysis.

\bigskip

In the present document, we choose the Lebesgue approach.

\section{Lebesgue integral and Lebesgue measure}
\label{s:lebesgue-integral-and-the-lebesgue-measure}

The Lebesgue integral was originally built using measure theory following the
so-called Lebesgue scheme, described in Section~\ref{s:lebesgue-scheme}.
We have seen that it may also be built in a constructive way by applying the
Daniell extension scheme to some elementary integral defined for elementary
functions, such as the Riemann integral for compactly supported continuous
functions.
Another alternative using the same basic ingredients consists in the completion
of the normed vector space of compactly supported continuous functions and the
extension of the Riemann integral which is uniformly
continuous~\cite{bou:int:65,die:ea2:68}.

The Lebesgue measure on~$\matR$ is a generalization of the length of bounded
intervals to a much wider class of subsets.
This extension is actually unique and complete.
It is defined on the Lebesgue $\sigma$-algebra which is generated by the Borel
subsets and the negligible subsets.
The restriction to the sole Borel subsets (the $\sigma$-algebra generated by
the open subsets, or the closed subsets) is sometimes called the
Borel(--Lebesgue) measure.
In the very same way, the Lebesgue measure on the Euclidean spaces~$\matR^n$ is
a generalization of the $n$-volume of rectangular boxes.
It is the tensor product of the Lebesgue measure on~$\matR$.

There are three main techniques for the construction of the Lebesgue measure
on~$\matR$.
The most popular one for teaching is through
{\CaratheodoryTh}~\cite{car:atm:63,dur:pte:19}.
The process builds progressively the Lebesgue $\sigma$-algebra and the Lebesgue
measure with their properties (see
Section~\ref{s:caratheodory-extension-scheme}).
A more abstract approach is based on the {\RieszMarkovKakutaniTh}, which
associates any positive linear form on the space of compactly supported
continuous functions with a unique measure, {\eg} see~\cite{rud:rca:87}.
The third way is not based on measure theory.
It follows the Daniell approach to integration, which is also based on the
Riemann integration of compactly supported continuous functions, and defines
the Lebesgue measure of any measurable subset as the integral of its indicator
function.

The construction of non-Lebesgue-measurable subsets requires the use the axiom
of choice, as for instance the Vitali subset~\cite{vit:spm:05}.

\medskip

In the present document, we build Lebesgue integral using the Lebesgue scheme
(see Section~\ref{s:lebesgue-scheme}), and Lebesgue measure through
{\CaratheodoryTh} (see Section~\ref{s:caratheodory-extension-scheme}).

\section{Our main sources}
\label{s:our-main-sources}

The main ingredients of measure theory and Lebesgue integration are presented
in an elementary manner in the textbook~\cite{mai:m2:14}, which is very well
suited for our purpose.
However, some results such as the construction of the Lebesgue measure and the
{\Tont} are admitted.

Details for the construction of the Lebesgue measure (through the {\Cara}
approach), and for the proofs of the {\FTt}s can be found in~\cite{gh:mip:13}.
This other textbook is quite comprehensive.
Many results are presented as exercises with their solution.

We may also cite~\cite{rud:rca:87}, \cite{ada:ss:75}, and~\cite{bre:af:83}.
The former for instance for a construction of the Lebesgue measure through the
{\RieszMarkovKakutaniTh}.
And the latter two for the~$L^p$ Lebesgue spaces and the~$W^{m,p}$ Sobolev
spaces.

\medskip

Note that some original forms of statements have been developed for the present
document, and even though the results are in general well known, the
formulations are new to our knowledge.
This includes:
\begin{CompactList}
\item
  the individual treatment of constitutive properties (closedness under set
  operations) of subset systems in Section~\ref{s:basic-properties};
\item
  the formalization of most concepts based on properties almost satisfied in
  Section~\ref{s:negligible-subset} with abstract results such as
  Lemmas~\ref{l:compat-of-almost-bin-rel-with-op},
  \ref{l:compat-of-almost-eq-with-op}, \ref{l:compat-of-almost-ineq-with-op},
  \ref{l:definiteness-implies-almost-definiteness},
  and~\ref{l:compat-of-int-in-mplus-with-almost-bin-rel} (the latter in
  Section~\ref{s:integration-of-nonnegative-measurable-functions});
\item
  the concept of almost sum in Section~\ref{s:negligibility-and-numbers} with
  Lemmas~\ref{l:almost-sum} and~\ref{l:compat-of-almost-sum-with-almost-eq},
  and Lemmas~\ref{l:minkowski-ineq-in-m} and~\ref{l:int-is-add} in
  Section~\ref{s:the-seminormed-vector-space-llone};
\item
  the concept of disjoint representation of simple functions in
  Section~\ref{s:integration-of-nonnegative-simple-functions} with
  Lemmas~\ref{l:sf-disj-repr}, \ref{l:sf-disj-repr-is-subpart-of-can-repr},
  and~\ref{l:sf-is-alg-over-r};
\item
  the fully detailed proof of the technical
  Lemma~\ref{l:change-of-variable-in-sum-in-sfplus} in
  Section~\ref{s:integration-of-nonnegative-simple-functions} to obtain
  additivity of the integral of nonnegative simple functions in
  Lemma~\ref{l:int-in-sfplus-is-add-alt-proof}.
\end{CompactList}

\chapter{Proof techniques}
\label{c:proof-techniques}

In the framework of integration theory, some proof techniques are used
repeatedly.
This chapter describes the most important ones:
the ``Lebesgue scheme'',
the ``{\Cara} extension scheme'', and
the ``{\Dplt} / {\mct} scheme''.

\section{Lebesgue scheme}
\label{s:lebesgue-scheme}

Consider a set~$X$ equipped with a $\sigma$-algebra~$\Sigma$ of measurable
subsets, and a measure~$\mu$ that associates a ``length'' (in~$\matRbarplus$)
to any measurable subset.
The set of extended real numbers~$\matRbar$ is equipped with the Borel
$\sigma$-algebra~$\calBRbar$ that is generated by all open subsets.
The purpose is to build, or prove properties of, the integral of
``measurable'' functions from~$X$ to~$\matRbar$, {\ie} for which the inverse
image of measurable subsets (of~$\matRbar$) are measurable subsets (of~$X$).

The Lebesgue scheme consists in establishing facts about the integral by
working successively inside three embedded functional subsets:
the set~$\calIF$ of indicator functions of measurable subsets,
the {\vectorspace}~$\calSF$ of simple functions (the linear span
of~$\calIF$),
and the set~$\calM$ of measurable functions.

The four steps of the scheme are the following:
\begin{enumerate}
\item
  establish the fact in the case of indicator functions (in~$\calIF$) for
  which the integral is simply the measure of the support;
\item
  then, generalize the fact to the case of nonnegative simple functions
  (in~$\calSFplus$) by nonnegative (finite) linear combination of indicator
  functions;
\item
  then again, generalize the fact to the case of nonnegative measurable
  functions (in~$\calMplus$) by taking the supremum for all lower nonnegative
  simple functions;
\item
  finally, generalize the fact to the case of all measurable functions
  (in~$\calM$) by taking the difference between the expressions involving
  nonnegative and nonpositive parts (integrable functions are those for which
  this difference is well-defined).
\end{enumerate}
Actually, a fifth step can be added to generalize the fact to the case of
numeric functions taking their values in~$\matR^n$, $\matC$, or~$\matC^n$ by
considering separately all components.

In the present document, the steps of Lebesgue scheme are mainly used to build
the Lebesgue integral (steps~1 to~4, see
Chapter~\ref{c:integration-of-nonnegative-functions} and
Section~\ref{s:definition-of-the-integral}), but also to establish some
properties such as the {\BLt} (step~3, see
Section~\ref{s:sketch-of-the-proof-of-the-beppo-levi-mono-cv-th} and
Theorem~\ref{t:beppo-levi-monot-conv}) and the {\Tont} (steps~1 to~3, see
Section~\ref{s:sketch-of-the-proof-of-the-tonelli-th} and
Theorem~\ref{t:tonelli}).

\section{{\Cara}'s extension scheme}
\label{s:caratheodory-extension-scheme}

The {\Cara} extension scheme consists in applying the eponymous theorem (see
Section~\ref{s:sketch-of-the-proof-of-caratheodory-ext-th}) to extend a
pre-measure ({\ie} defined on a ring of subsets, and not on a $\sigma$-algebra)
into a complete measure on a $\sigma$-algebra containing the initial ring of
subsets.

Let~$X$ be the ambient set, $\mu$ be the pre-measure and~$\calR$ be the ring
of subsets (of~$X$).
The two steps of the scheme are the following:
\begin{enumerate}
\item extend the pre-measure to any subset of~$X$ by taking the infimum of
  the sum of pre-measures of all coverings with elements of the ring~$\calR$
  (and the value~$\infty$ when there is no such covering):
  \begin{equation*}
    \mu^\star (A) \eqdef \inf \left\{
      \sum_{n \in \matN} \mu (A_n) \rightst \left.
      \vphantom{\sum_{n \in \matN} \mu (A_n)}
      (A_n)_{n \in \matN} \in \calR
      \Conj A \subset \bigcup_{n \in \matN} A_n \right\};
  \end{equation*}
  the theorem ensures that~$\mu^\star$ is an outer measure on~$X$, and that it
  is an extension of the pre-measure~$\mu$;

\item define the set of {\Cara}-measurable subsets as
  \begin{equation*}
    \Sigma \eqdef \{
      E \subset X \st
      \forall A \subset X,\;
      \mu^\star (A) = \mu^\star (A \cap E) + \mu^\star (A \setminus E) \};
  \end{equation*}
  the theorem ensures that~$\Sigma$ is a $\sigma$-algebra on~$X$
  containing the ring~$\calR$, and that the
  restriction~$\restr{\mu^\star}{\Sigma}$ is a complete measure
  on~$(X,\Sigma)$.
\end{enumerate}

The {\Cara} extension scheme is a popular tool to build measures.
The prominent example is the Lebesgue measure on~$\matR$ that extends the
length of open intervals (see Section~\ref{s:the-lebesgue-measure}).
But for instance, the extension scheme may also be used to build the
Lebesgue-Stieltjes measure associated with its cumulative distribution
function, and Loeb measures in a nonstandard analysis framework,
{\eg} see~\cite{loe:cns:75,cut:lmp:00}.

\section{{\Dplt} / {\mct} schemes}
\label{s:dynkin-p-l-th-monot-class-th-schemes}

A $\pi$-system, a set algebra, a monotone class, and a $\lambda$-system are all
different kinds of subset systems ({\ie} subsets of the power set), just like a
$\sigma$-algebra, but with less basic properties.

Roughly speaking, being a $\lambda$-system (resp. a monotone class) is what
is missing to a $\pi$-system (resp. an algebra of subsets) to be a
$\sigma$-algebra.
To prove that some property~$P$ holds on some $\sigma$-algebra~$\Sigma$
generated by some subset system~$G$, the idea is to proceed in two steps.
First, show that the property~$P$ holds on the simpler kind of subset
system ($\pi$-system, or set algebra) generated by~$G$, and then establish
that the subset of~$\Sigma$ where~$P$ holds is of the less simple kind of
subset system ($\lambda$-system, or monotone class).
Indeed, the {\Dplt} states that the $\lambda$-system generated by a
$\pi$-system is equal to the $\sigma$-algebra generated by the same
$\pi$-system, and similarly, the {\mct} states that the monotone class
generated by a set algebra is equal to the $\sigma$-algebra generated by the
same set algebra (see
Section~\ref{s:sketch-of-the-proof-of-the-dynkin-p-l-th-monot-class-th}).
Thus, $P$~is established on the whole $\sigma$-algebra.

The five steps of the scheme of the {\Dplt}, and of the {\mct} are the
following:
\begin{enumerate}
\item first define $\calS\eqdef\{A\in\Sigma\st P(A)\}$;
  obviously, we have $\calS\subset\Sigma$;

\item then, show that~$G\subset\calS$;

\item then, show that the $\pi$-system (resp. set algebra) $\calU^2_X(G)$ is a
  subset of~$\calS$;

\item then, show that~$\calS$ is a $\lambda$-system (resp. monotone class);

\item and finally, obtain the other inclusion $\Sigma\subset\calS$, because
  from monotonicity of subset system generation, and from the {\Dplt}
  (resp. {\mct}), {\ie} the equality in the middle, we have
  \begin{equation*}
    \Sigma
    = \Sigma_X (G)
    \subset \Sigma_X (\calU^2_X (G))
    = \calU^1_X (\calU^2_X (G))
    \subset \calU^1_X (\calS)
    = \calS
  \end{equation*}
\end{enumerate}
where $\calU^1_X=\Lambda_X$ (resp. $\calC_X$), {\ie} the generated
$\lambda$-system (resp. monotone class), and $\calU^2_X=\Pi_X$
(resp. $\calA_X$), {\ie} the generated $\pi$-system (resp. set algebra).

In the present document, the fifth step is embodied by
Lemma~\ref{l:usage-of-dynkin-pi-lambda-th}
for the {\Dplt}, and by
Lemma~\ref{l:usage-of-monot-class-th}
for the {\mct}.
The {\Dplt} scheme is used once when extending the equality of two measures on a
generator $\pi$-system to the whole $\sigma$-algebra in
Lemma~\ref{l:uniq-of-meas-ext-from-p-syst}, which is then used to establish
uniqueness of the Lebesgue measure on~$\matR$ in
Theorem~\ref{t:caratheodory-lebesgue-meas-on-r} (see also
Section~\ref{s:sketch-of-the-proof-of-caratheodory-ext-th}).
The {\mct} scheme is used twice when building the tensor product measure in the
context of the product of finite measure spaces:
to establish first the measurability of the measure of sections in
Lemma~\ref{l:meas-of-meas-of-section-finite}, and then the uniqueness of the
tensor product measure in Lemma~\ref{l:uniq-of-tensor-prod-meas-finite} (both
in the finite case).

\chapter{Statements and sketches of the proofs}
\label{c:statements-and-sketches-of-the-proofs}

This chapter gathers the sketches of the proofs of the main results that are
detailed in Part~\ref{p:detailed-proofs}.
Namely: {\Ldcvt} and its extended version, the {\Tont}, {\Fl}, the {\BLt},
{\Carat}, the {\Dplt}, and the {\mct}.

\section{Sketch of the proof of {\Ledcvt}}
\label{s:sketch-of-the-proof-of-lebesgue-ext-dom-cv-th} 

\begin{ledcthm}
  \mbox{}\\
  Let~$(X,\Sigma,\mu)$ be a measure space.
  Let~$(f_n)_{n\in\matN},f,g\in\calM$.
  Assume that the sequence is $\mu$-almost everywhere pointwise convergent
  towards~$f$.
  Assume that~$g$ is $\mu$-integrable, and that for all $n\in\matN$, we have
  $|f_n|\leqae{\mu}g$.
  Then, for all $n\in\matN$, $f_n$~is $\mu$-integrable, $f$~is
  $\mu$-integrable, and we have in~$\matR$
  \begin{equation}
    \label{e:ledcthm}
    \int f \, d\mu = \lim_{n \to \infty} \int f_n \, d\mu.
  \end{equation}
\end{ledcthm}

See Theorem~\ref{t:lebesgue-ext-dom-conv}.
The proof of {\Ledcvt} goes as follows (this proof uses the arithmetic
of~$\matRbar$, but with functions that are almost everywhere finite):
\begin{itemize}
\item
  $D\eqdef\{f=\liminf_{n\to\infty}f_n\}
  \cap\{f=\limsup_{n\to\infty}f_n\}
  \cap\bigcap_{n\in\matN}\{|f_n|\leq g\}\cap g^{-1}(\matRplus)\subset X$ is
  first shown to be measurable with $\mu(D^c)=0$;
  thus $f_n\matUN_D\eqae{\mu}f_n$, $f\matUN_D\eqae{\mu}f$, and
  $g\matUN_D\eqae{\mu}g$;
  moreover, $g\matUN_D$ belongs to~$\calLone$;

\item
  then, since $\lim_{n\to\infty}f_n\matUN_D=f\matUN_D$ and
  $|f_n\matUN_D|\leq g\matUN_D$, {\Ldcvt} (see
  Section~\ref{s:sketch-of-the-proof-of-lebesgue-dom-cv-th}) provides
  $f_n\matUN_D,f\matUN_D\in\calLone$, and the equality
  \begin{equation*}
    \int f \matUN_D \, d\mu
    = \lim_{n \to \infty} \int f_n \matUN_D \, d\mu;
  \end{equation*}

\item
  finally, the integrability of~$f_n$ and~$f$, and
  identity~\eqref{e:ledcthm} follow from the compatibility of the integral
  in~$\calM$ with almost equality.
\end{itemize}

\bigskip

For instance, {\Ledcvt} may be used to prove the Leibniz integral rule
(differentiation under the integral sign), {\eg} for the study of integrals
function of their upper bound.

\section{Sketch of the proof of {\Ldcvt}}
\label{s:sketch-of-the-proof-of-lebesgue-dom-cv-th} 

\begin{ldcthm}
  \mbox{}\\
  Let~$(X,\Sigma,\mu)$ be a measure space.
  Let~$(f_n)_{n\in\matN}\in\calM$.
  Assume that the sequence is pointwise convergent towards~$f$.
  Let~$g\in\calLone$.
  Assume that for all $n\in\matN$, $|f_n|\leq g$.
  Then, for all $n\in\matN$, $f_n\in\calLone$, $f\in\calLone$, the sequence
  is convergent towards~$f$ in~$\calLone$, and we have in~$\matR$
  \begin{equation}
    \label{e:ldcthm}
    \int \lim_{n \to \infty} f_n \, d\mu
    = \int f \, d\mu
    = \lim_{n \to \infty} \int f_n \, d\mu.
  \end{equation}
\end{ldcthm}

See Theorem~\ref{t:lebesgue-dom-conv}.
The proof of {\Ldcvt} goes as follows (this proof uses the arithmetic
of~$\matR$):
\begin{itemize}
\item
  monotonicity of integral in~$\calMplus$ provides $f_n,f\in\calLone$;

\item
  the sequence~$g_n\eqdef|f_n-f|$ is shown to be in~$\calLone\cap\calMplus$
  with limit~0;

\item
  the sequence~$2g-g_n$ is also shown to be in~$\calLone\cap\calMplus$, with
  limit inferior~$2g$;

\item
  then, {\Fl} (see Section~\ref{s:sketch-of-the-proof-of-fatou-lem}) and
  linearity of the integral in~$\calLone$ provide the inequality
  \begin{equation*}
    2 \int g \, d\mu
    \leq 2 \int g \, d\mu - \limsup_{n \to \infty} \int g_n \, d\mu;
  \end{equation*}
  thus, a nonnegativeness argument provides the nullity of the limit of the
  integral of~$g_n$'s, {\ie} the convergence of the~$f_n$'s towards~$f$
  in~$\calLone$;

\item
  finally, identity~\eqref{e:ldcthm} follows from nondecreasingness of the
  integral in~$\calLone$, the squeeze theorem, and linearity of the limit.
\end{itemize}

{\Ldcvt} is used in the present document to prove {\Ledcvt} (see
Section~\ref{s:sketch-of-the-proof-of-lebesgue-ext-dom-cv-th} and
Theorem~\ref{t:lebesgue-ext-dom-conv}).

\bigskip

{\Ldcvt} admits several variants: with lighter assumptions, or set in~$L^p$ for
$p\in[1,\infty)$.
For instance, it may be used to establish the Fourier inversion formula, or to
study the Gibbs phenomenon.

\section{Sketch of the proof of the {\Tont}}
\label{s:sketch-of-the-proof-of-the-tonelli-th} 

\begin{ftthm}
  \mbox{}\\
  Let~$(X_1,\Sigma_1,\mu_1)$ and~$(X_2,\Sigma_2,\mu_2)$ be $\sigma$-finite
  measure spaces.
  Let~$f\in\calMplus(X_1\times X_2,\Sigma_1\otimes\Sigma_2)$.
  Let~$i\in\{1,2\}$.
  Let~$j\eqdef3-i$.
  Let~$\psi$ be the permutation $((x_i,x_j)\mapsto(x_1,x_2))$.\\
  For all $x_i\in X_i$, let $f_{x_i}\eqdef(x_j\mapsto f\circ\psi(x_i,x_j))$.
  Let~$I_{f,i}\eqdef(x_i\mapsto\int f_{x_i}\,d\mu_j)$.\\
  Then, for all $x_i\in X_i$, $f_{x_i}\in\calMplus(X_j,\Sigma_j)$,
  $I_{f,i}\in\calMplus(X_i,\Sigma_i)$, and we have in~$\matRbarplus$
  \begin{equation}
    \int f \, d(\mu_1 \otimes \mu_2) = \int I_{f,i} \, d\mu_i.
  \end{equation}
\end{ftthm}

See Theorem~\ref{t:tonelli}.
The proof of the {\Tont} goes as follows (this proof uses the arithmetic
of~$\matRbarplus$ and the concepts of set algebra and monotone class;
it follows steps~1 to~3 of the Lebesgue scheme,
see Section~\ref{s:lebesgue-scheme}):

\begin{itemize}
\item the result is first established for indicator functions from properties
  of the measure (in particular the building of the tensor product measure
  that uses continuity of measures from below and from above);

\item then, the result is extended to nonnegative simple functions by taking
  (positive) linear combination of indicator functions and applying (positive)
  linearity of the integral in~$\calMplus$;

\item and finally, the result is extended to nonnegative measurable functions
  by taking the supremum of adapted sequences and applying the {\BLt} (see
  Section~\ref{s:sketch-of-the-proof-of-the-beppo-levi-mono-cv-th}).
\end{itemize}

The {\Tont} is used in the present document to establish identities:
for an integral over a subset in Lemma~\ref{l:tonelli-over-subset},
and for an integral of a tensor product function in
Lemma~\ref{l:tonelli-for-tensor-prod}.

\section{Sketch of the proof of {\Fl}}
\label{s:sketch-of-the-proof-of-fatou-lem} 

\begin{flem}
  \mbox{}\hfill
  Let~$(X,\Sigma,\mu)$ be a measure space.
  Let~$(f_n)_{n\in\matN}\in\calMplus$.\\
  Then, $\liminf_{n\to\infty}f_n\in\calMplus$, and we have
  in~$\matRbarplus$
  \begin{equation}
    \int \liminf_{n \to \infty} f_n \, d\mu
    \leq \liminf_{n \to \infty} \int f_n \, d\mu.
  \end{equation}
\end{flem}

See Theorem~\ref{t:fatou-lemma}.
The proof of {\Fl} goes as follows (this proof uses the arithmetic
of~$\matRbarplus$):
\begin{itemize}
\item
  the sequence $(\inf_{p\in\matN}f_{n+p})_{n\in\matN}$ is first shown to be
  pointwise nondecreasing in~$\calMplus$;

\item
  then, the {\BLt} (see
  Section~\ref{s:sketch-of-the-proof-of-the-beppo-levi-mono-cv-th}) provides
  the equality
  \begin{equation*}
    \int \liminf_{n \in \matN} f_n \, d\mu
    = \lim_{n \in \matN} \int \inf_{p \in \matN} f_{n + p} \, d\mu;
  \end{equation*}

\item
  and monotonicity of integral, and infimum and limit inferior definitions
  provide the inequality
  \begin{equation*}
    \lim_{n \in \matN} \int \inf_{p \in \matN} f_{n + p} \, d\mu
    \leq \liminf_{n \in \matN} \int f_n \, d\mu.
  \end{equation*}
\end{itemize}

{\Fl} is used in the present document to prove properties of the integral
in~$\calMplus$:
identity for pointwise convergent sequences in
Lemma~\ref{l:int-in-mplus-of-pointwise-conv-seq},
and {\Ldcvt} (see Section~\ref{s:sketch-of-the-proof-of-lebesgue-dom-cv-th} and
Theorem~\ref{t:lebesgue-dom-conv}).

\bigskip

The {\BLt} and {\Fl} can be established independently of one another, or
each one can be obtained as a consequence of the other.
Of course, those independent proofs do share a common ingredient to somehow
obtain an upper bound for the integral of the limit by the limit of the
integrals.
We chose to establish the {\BLt} first.

\section{Sketch of the proof of the {\BLt}}
\label{s:sketch-of-the-proof-of-the-beppo-levi-mono-cv-th} 

\begin{blthm}
  \mbox{}\\
  Let~$(X,\Sigma,\mu)$ be a measure space.
  Let~$(f_n)_{n\in\matN}\in\calMplus$.
  Assume that the sequence is pointwise nondecreasing.
  Then, $\lim_{n\to\infty}f_n\in\calMplus$, and we have
  in~$\matRbarplus$
  \begin{equation}
    \int \lim_{n \to \infty} f_n \, d\mu
    = \lim_{n \to \infty} \int f_n \, d\mu.
  \end{equation}
\end{blthm}

See Theorem~\ref{t:beppo-levi-monot-conv}.
The proof of the {\BLt} goes as follows (this proof uses the arithmetic
of~$\matRbarplus$, it follows step~3 of the Lebesgue scheme by taking
the ``supremum'' of quantities involving simple functions, see
Section~\ref{s:lebesgue-scheme}):
\begin{itemize}
\item
  let~$f\eqdef\lim_{n\to\infty}f_n$;
  monotonicity and completeness arguments provide that~$f$ is also
  $\sup_{n\to\infty}f_n$, that it belongs to~$\calMplus$, and the
  ``easy'' inequality
  \begin{equation*}
    \lim_{n \to \infty} \int f_n \, d\mu \leq \int f \, d\mu;
  \end{equation*}

\item
  let~$\fhi\in\calSFplus$ such that $\fhi\leq f$;
  using the nondecreasing sequence of measurable subsets
  $(\{a\fhi\leq f_n\})_{n\in\matN}$ for $a\in(0,1)$, continuity from below
  of~$\mu$, linearity, monotonicity and continuity arguments, allows to show
  the inequality
  \begin{equation*}
    \int \fhi \, d\mu \leq \lim_{n \to \infty} \int f_n \, d\mu;
  \end{equation*}

\item
  then, taking the supremum over~$\fhi$'s provides the other inequality
  \begin{equation*}
    \int f \, d\mu \leq \lim_{n \to \infty} \int f_n \, d\mu.
  \end{equation*}
\end{itemize}

The {\BLt} is used in the present document to prove properties of the
integral in~$\calMplus$ (possibly through Lemma~\ref{l:usage-of-adapted-seqs}):
homogeneity at~$\infty$ in Lemma~\ref{l:int-in-mplus-is-hom-at-infinity},
additivity in Lemma~\ref{l:int-in-mplus-is-add},
$\sigma$-additivity in Lemma~\ref{l:int-in-mplus-is-sigma-add},
identity for the integral over a subset in
Lemma~\ref{l:int-in-mplus-over-subset},
{\Fl} (see Section~\ref{s:sketch-of-the-proof-of-fatou-lem} and
Theorem~\ref{t:fatou-lemma}),
identity for the integral with the counting measure in
Lemma~\ref{l:int-in-mplus-for-count-meas}, and
the {\Tont} (see Section~\ref{s:sketch-of-the-proof-of-the-tonelli-th} and
Theorem~\ref{t:tonelli}).

\section{Sketch of the proof of {\Carat}}
\label{s:sketch-of-the-proof-of-caratheodory-ext-th} 

\begin{cthm}
  \mbox{}\\
  Let~$X$ be a set.
  Let~$\calR$ be a ring of subsets on~$X$
  ({\ie} closed under complement and union).
  Let~$\mu$ be a pre-measure defined on~$\calR$
  ({\ie} null on the empty set, and $\sigma$-additive).
  Then, $\mu$ can be extended into a complete measure on a $\sigma$-algebra
  containing~$\calR$ ({\ie} for which all negligible subsets are measurable,
  and of measure~0).
  Moreover, this extension is unique when~$\mu$ is $\sigma$-finite.
\end{cthm}

See Theorem~\ref{t:caratheodory-lebesgue-meas-on-r} (in the specific case of
$X\eqdef\matR$ and~$\calR$ is the ring generated by the intervals).
The proof of {\Carat} goes as follows (this proof uses the arithmetic
of~$\matRbarplus$ and the concepts of set algebra and monotone class;
the existence part may also be known as the {\Cara} extension scheme,
see Section~\ref{s:caratheodory-extension-scheme}):
\begin{itemize}
\item the function~$\mu^\star$ defined on~$\calP(X)$ by (with the convention
  $\inf\emptyset=\infty$)
  \begin{equation*}
    \forall A \subset X,\quad
    \mu^\star (A)
    \eqdef \inf \left\{ \sum_{n \in \matN} \mu (A_n) \rightst
    \left. \vphantom{\sum_{n \in \matN} \mu (A_n)}
      (A_n)_{n \in \matN} \in \calR
      \Conj A \subset \bigcup_{n \in \matN} A_n \right\}
  \end{equation*}
  is first shown to be an outer measure on~$X$ ({\ie} null on the empty set,
  nonnegative, monotone, and $\sigma$-subadditive) whose restriction
  to~$\calR$ is the pre-measure~$\mu$;

\item then, the set of subsets
  \begin{equation*}
    \Sigma \eqdef \left\{
      E \subset X \st \forall A \subset X,\;
      \mu^\star (A) = \mu^\star (A \cap E) + \mu^\star (A \setminus E) \right\}
  \end{equation*}
  is shown to be a $\sigma$-algebra on~$X$ containing the ring~$\calR$,
  while the restriction~$\restr{\mu^\star}{\Sigma}$ is shown to be a complete
  measure on~$(X,\Sigma)$ (closedness of~$\Sigma$ under countable union comes
  from additivity of~$\mu^\star$ on~$\Sigma$);

\item moreover, when~$\mu$ is $\sigma$-finite
  (with $X=\biguplus_{n\in\matN}X_n$ and $\mu(X_n)<\infty$), two {\Cara}
  extensions~$\mu_1$ and~$\mu_2$ are first restricted to~$X_n$,
  then the set $\{\mu_1(A\cap X_n)=\mu_2(A\cap X_n)\}$ is shown to be equal
  to~$\Sigma$, and finally uniqueness follows from $\sigma$-additivity
  of~$\mu_1$ and~$\mu_2$.
\end{itemize}

In the present document, {\Carat} is stated (and proved) in the specific case
of the building of the Lebesgue measure on~$\matR$ (see
Section~\ref{s:the-lebesgue-measure}).

\section{Sketch of the proof of the {\Dplt} / {\mct}}
\label{s:sketch-of-the-proof-of-the-dynkin-p-l-th-monot-class-th} 

Both theorems take advantage of the complementarity of being a
$\lambda$-system and being a $\pi$-system (for the {\Dplt}), or being a
monotone class and being a set algebra (for the {\mct}).
Using the notation of Section~\ref{s:dynkin-p-l-th-monot-class-th-schemes}
where~$\calU^1_X$ represents the generated $\lambda$-system~$\Lambda_X$
(resp. the generated monotone class~$\calC_X$), and~$\calU^2_X$ represents the
generated $\pi$-system~$\Pi_X$ (resp. the generated set algebra~$\calA_X$),
both theorems share the following abstract form:
\begin{absthm}
  \mbox{}\\
  Let~$X$ be a set.
  Let~$G\subset\calP(X)$.
  Assume that $\calU^2_X(G)=G$.
  Then, $\calU^1_X(G)=\Sigma_X(G)$.
\end{absthm}

See Theorems~\ref{t:dynkin-pi-lambda-th} and~\ref{t:monot-class}.
Their common proof goes as follows (these proofs use the concepts of
$\pi$-system and $\lambda$-system for the {\Dplt}, and of set algebra and
monotone class for the {\mct}):
\begin{itemize}
\item first, prove that for all~$\Gp$, $\calU^1_X(\Gp)=\Gp$ and
  $\calU^2_X(\Gp)=\Gp$ implies $\Sigma_X(\Gp)=\Gp$;

\item then, note that $\calU^1_X(\calU^1_X(G))=\calU^1_X(G)$
  ({\ie} idempotent law for subset system generation);

\item then, prove that $\calU^2_X(\calU^1_X(G))=\calU^1_X(G)$
  ({\ie} the property of~$G$ is transmitted to~$\calU^1_X(G)$);

\item then, apply the first result to $\Gp\eqdef\calU^1_X(G)$ and obtain
  $\Sigma_X(\calU^1_X(G))=\calU^1_X(G)$;

\item and finally, prove that $\Sigma_X(\calU^1_X(G))=\Sigma_X(G)$.
\end{itemize}
In both cases, the most technical part is the third point.

The {\Dplt} is used in the present document (through
Lemma~\ref{l:usage-of-dynkin-pi-lambda-th}) to extend the equality of two
measures on a generator $\pi$-system to the whole $\sigma$-algebra in
Lemma~\ref{l:uniq-of-meas-ext-from-p-syst}, which is used to establish
uniqueness of the Lebesgue measure on~$\matR$ in
Theorem~\ref{t:caratheodory-lebesgue-meas-on-r} (see also
Section~\ref{s:sketch-of-the-proof-of-caratheodory-ext-th}).
The {\mct} is used in the present document (through
Lemma~\ref{l:usage-of-monot-class-th}) to prove measurability of the measure of
sections (in the case of finite measure spaces) in
Lemma~\ref{l:meas-of-meas-of-section-finite}, and uniqueness of the tensor
product measure (also in the finite case) in
Lemma~\ref{l:uniq-of-tensor-prod-meas-finite}.
All these three proofs follow the {\Dplt} / {\mct} scheme (see
Section~\ref{s:dynkin-p-l-th-monot-class-th-schemes}).

\part{Detailed proofs}
\label{p:detailed-proofs}

\chapter{Introduction}
\label{c:introduction-2}

Statements are displayed inside colored boxes.
Their nature can be identified at a glance by using the following color code:
\begin{center}
  \rmkbox{light gray is for remarks}, \qquad
  \defbox{light green for definitions},\\
  \lembox{light blue for lemmas}, \quad and \quad
  \thmbox{light red for theorems}.
\end{center}
Definitions and results have a number and a name.
Inside the bodies of proofs, pertinent statements are referenced using both
their number and name.
When appropriate, some hints are given about the application, either to specify
arguments, or to provide justification or consequences;
they are \underline{underlined}.
Some useful definitions and results were already stated in~\cite{cm:lmt:16},
which was devoted to the detailed proof of the {\LMT}.
Those are numbered up to~{\thetheorem}, and the statements in the present
document are numbered starting from~{\thenexttheorem}.

Furthermore, as in~\cite{cm:lmt:16}, the most basic results are supposed to be
known and are not detailed further;
they are displayed in \assume{bold dark red}.
These include:
\begin{itemize}
\item Logic:
  tautologies from propositional calculus.

\item Set theory:
  \begin{CompactList}
  \item
    definition and properties of
    inclusion, intersection, (disjoint) union, complement, set difference,
    Cartesian product, power set, cardinality, and indicator function (that
    takes values~0 and~1),
    such as De~Morgan's laws,
    monotonicity of intersection,
    distributivity of intersection over (disjoint) union and set difference,
    distributivity of the Cartesian product over union,
    compatibility of intersection with Cartesian product,
    $\sigma$-additivity of the cardinality (with~$\infty$ absorbing element for
    addition in~$\matNbar$);

  \item
    definition and properties of
    equivalence and order binary relations;

  \item
    definition and properties of
    function, composition of functions,
    inverse image (compatibility with set operations),
    injective and bijective functions,
    restriction and extension;

  \item
    countability of finite Cartesian products of countable sets
    ($\matN^2$, $\matQ$, $\matQ\times\matQplusstar$).
  \end{CompactList}

\item Algebraic structures: results from group theory.

\item Topology: definition and properties of continuous functions.

\item Real analysis:
  \begin{CompactList}
  \item
    properties of the ordered and valued field~$\matR$ such as
    the Archimedean property,
    density of rational numbers, completeness,
    sum of the first terms of a geometric series;

  \item
    definition and properties of basic numeric analytic functions such as
    square root, power function,
    exponential function, natural logarithm function, and exponentiation;

  \item
    properties of the ordered set~$\matRbar$;

  \item
    properties of limits in~$\matR$/$\matRbar$ (cluster point),
    compactness of closed and bounded intervals,
    compatibility of limit with arithmetic operations,
    the squeeze theorem.
  \end{CompactList}
\end{itemize}

\bigskip

This part is organized as follows.
Chapter~\ref{c:complements} contains some results from various fields of
mathematics (set theory, algebraic structures, order theory, general topology,
real and extended real numbers including second-countability), that are needed
in the proofs of the integral theory.
We recall that the material stated in~\cite{cm:lmt:16} may also be used.

Then, Chapter~\ref{c:subset-systems} is devoted to subsets systems, from
$\pi$-system to $\sigma$-algebra.
Measurability and measurable space are presented in
Chapter~\ref{c:measurability}, and the specific case of sets of real and
extended numbers is treated in Chapter~\ref{c:measurability-and-numbers}.
Chapter~\ref{c:measure-space} is dedicated to measure and measure space, and
Chapter~\ref{c:measure-and-numbers} to the specific cases involving sets of
numbers, including the construction of the Lebesgue measure on~$\matR$ through
{\Carat}.
Finally, the integral of nonnegative functions is addressed in
Chapter~\ref{c:integration-of-nonnegative-functions} with for instance the
{\BLt}, {\Fl}, and the {\Tont}.
Chapter~\ref{c:integration-of-real-functions} is dedicated to the integral of
functions with arbitrary sign, with the seminormed {\vectorspace}~$\calLone$
and {\Ldcvt}.

\chapter{Complements}
\label{c:complements}

\minitoc

\section{Complements on set theory}
\label{s:complements-on-set-theory}

\begin{definition}[pseudopartition]
  \label{d:pseudopart}
  \mbox{}\\
  Let~$X$ be a set.
  Let~$I\subset\matN$.
  Subsets~$(X_i)_{i\in I}$ of~$X$ are said to form a
  {\em pseudopartition (of $X$)} iff
  \begin{equation}
    \label{e:pseudopart}
    \forall i, j \in I,\;
    i \not= j \Implies X_i \cap X_j = \emptyset
    \AND
    X = \biguplus_{i \in I} X_i.
  \end{equation}
\end{definition}

\begin{remark}
  \label{r:v2-new01}
  Note that, in contrast to proper partitions, parts of a pseudopartition may
  be empty.
\end{remark}

\begin{lemma}[compatibility of pseudopartition with intersection]
  \label{l:compat-of-pseudopart-with-inter}
  \mbox{}\hfill
  Let~$X$ be a set.\\
  Let~$I\subset\matN$.
  Let~$A,(X_i)_{i\in I}\subset X$.
  Assume that $(X_i)_{i\in I}$ form a pseudopartition of~$X$.\\
  Then, $(A\cap X_i)_{i\in I}$ form a pseudopartition of~$A$,
  {\ie} $A=\biguplus_{i \in I} (A\cap X_i)$.
\end{lemma}

\begin{proof}
  Direct consequence of
  Definition~\thref{d:pseudopart}, and
  \assume{distributivity of intersection over disjoint union}.
\end{proof}

\begin{lemma}[technical inclusion for countable union]
  \label{l:technical-inclusion-for-count-union}
  \mbox{}\\
  Let~$X$ be a set.
  Let~$(A_n)_{n\in\matN}\subset X$.
  Let~$\fhi:\ArNN$.
  Then, we have $\bigcup_{n\in\matN}A_{\fhi(n)}\subset\bigcup_{n\in\matN}A_n$.
\end{lemma}

\begin{proof}
  Let~$x\in\bigcup_{n\in\matN}A_{\fhi(n)}$.
  then, from
  \assume{the definition of union},
  there exists $n\in\matN$ such that $x\in A_{\fhi(n)}$.
  Thus, there exists $m=\fhi(n)\in\matN$ such that $x\in A_m$.
  Hence, from
  \assume{the definition of union},
  we have $x\in\bigcup_{m\in\matN}A_m$.

  Therefore, we have
  $\bigcup_{n\in\matN}A_{\fhi(n)}\subset\bigcup_{n\in\matN}A_n$.
\end{proof}

\begin{lemma}[order is meaningless in countable union]
  \label{l:order-is-meaningless-in-count-union}
  \mbox{}\hfill
  Let~$X$ be a set.
  Let~$(A_n)_{n\in\matN}\subset X$.
  Let~$\fhi:\ArNN$.
  Assume that~$\fhi$ is bijective.
  Then, we have $\bigcup_{n\in\matN}A_n=\bigcup_{n\in\matN}A_{\fhi(n)}$.
\end{lemma}

\begin{proof}
  Direct consequence of
  Lemma~\threfc{l:technical-inclusion-for-count-union}{%
    with $(A_n)_{n\in\matN}$ and $\fhi$,
    then $(A_{\fhi(n)})_{n\in\matN}$ and $\fhi^{-1}$ which satisfies
    $\fhi\circ\fhi^{-1}=\idmatN$}.
\end{proof}

\begin{lemma}[definition of double countable union]
  \label{l:def-of-double-count-union}
  \mbox{}\\
  Let~$X$ be a set.
  For all $n,m\in\matN$, let~$A_{n,m}\subset X$.
  Let~$\fhi,\psi:\ArNNxN$.
  Assume that $\fhi$ and $\psi$ are bijective.
  Then, $\bigcup_{p\in\matN}A_{\fhi(p)}=\bigcup_{p\in\matN}A_{\psi(p)}$.
  This union is denoted $\bigcup_{n,m\in\matN}A_{n,m}$
\end{lemma}

\begin{proof}
  Direct consequence of
  Lemma~\threfc{l:order-is-meaningless-in-count-union}{%
    with $(A_{\fhi(p)})_{p\in\matN}$ and $\fhi^{-1}\circ\psi$}.
\end{proof}

\begin{lemma}[double countable union]
  \label{l:double-count-union}
  \mbox{}\hfill
  Let~$X$ be a set.
  For all $n,m\in\matN$, let~$A_{n,m}\subset X$.
  Then, we have
  $\bigcup_{n,m\in\matN}A_{n,m}=
  \bigcup_{n\in\matN}\left(\bigcup_{m\in\matN}A_{n,m}\right)$.
\end{lemma}

\begin{proof}
  From
  \assume{countability of~$\matN^2$},
  let~$\fhi:\ArNNxN$ be a bijection.
  Then, from
  Lemma~\thref{l:def-of-double-count-union},
  we have $\bigcup_{n,m\in\matN}A_{n,m}=\bigcup_{p\in\matN}A_{\fhi(p)}$
  (and this does not depend on the choice for~$\fhi$).

  Let~$x\in\bigcup_{p\in\matN}A_{\fhi(p)}$.
  Then, from
  \assume{the definition of union},
  there exists $p_0\in\matN$ such that $x\in A_{\fhi(p_0)}$.
  Let~$(n_0,m_0)\eqdef\fhi(p_0)$.
  Thus, we have $x\in A_{n_0,m_0}$.
  Hence, from
  \assume{the definition of union},
  we have
  $x\in\bigcup_{m\in\matN}A_{n_0,m}\subset
  \bigcup_{n\in\matN}\left(\bigcup_{m\in\matN}A_{n,m}\right)$.

  Conversely, let
  $x\in\bigcup_{n\in\matN}\left(\bigcup_{m\in\matN}A_{n,m}\right)$.
  Then, from
  \assume{the definition of union},
  there exists $n_0\in\matN$ such that $x\in\bigcup_{m\in\matN}A_{n_0,m}$, and
  there exists $m_0\in\matN$ such that $x\in A_{n_0,m_0}$.\\
  Let~$p_0\eqdef\fhi^{-1}(n_0,m_0)$.
  Thus, we have $x\in A_{\fhi(p_0)}$.
  Hence, from
  \assume{the definition of union},
  we have $x\in\bigcup_{p\in\matN}A_{\fhi(p)}$.

  Therefore, we have
  $\bigcup_{n,m\in\matN}A_{n,m}=
  \bigcup_{n\in\matN}\left(\bigcup_{m\in\matN}A_{n,m}\right)$.
\end{proof}

\begin{remark}
  \label{r:v2-new02}
  The following lemma transforms a countable union of subsets into a countable
  disjoint union.
  It is achieved by adding the set differences of the subsets layer by layer.
  When the input sequence is nondecreasing, layers can be seen as onion peels.
\end{remark}

\begin{lemma}[partition of countable union]
  \label{l:part-of-count-union}
  \mbox{}\hfill
  Let~$X$ be a set.
  Let~$(A_n)_{n\in\matN}\subset X$.
  Let~$B_0\eqdef A_0$, and for all $n\in\matN$, let
  $B_{n+1}\eqdef A_{n+1}\setminus\bigcup_{i\in[0..n]}B_i$.
  Then, we have
  \begin{align}
    \label{e:part-of-count-union-1}
    \forall m, n \in \matN,\quad &
    m \not= n \IMPLIES B_m \cap B_n = \emptyset,\\
    \label{e:part-of-count-union-2}
    \forall n \in \matN,\quad &
    \bigcup_{i \in [0..n]} A_i = \biguplus_{i \in [0..n]} B_i,\\
    \label{e:part-of-count-union-3}
    & \bigcup_{i \in \matN} A_i = \biguplus_{i \in \matN} B_i.
  \end{align}
\end{lemma}

\begin{proof}
  Let~$p\in\matN$.
  Then, from
  \assume{the definition of set difference}, and
  \assume{De~Morgan's laws},
  \begin{equation*}
    B_{p + 1}
    = A_{p + 1} \cap \left( \bigcup_{i \in [0..p]} B_i \right)^c
    = A_{p + 1} \cap \bigcap_{i \in [0..p]} B_i^c.
  \end{equation*}
  Let~$i\in[0..p]$.
  Then, from
  \assume{properties of set operations},
  $B_{p+1}$ and~$B_i$ are disjoint.

  Let $m,n\in\matN$.
  Assume that $m\not=n$, and let $i\eqdef\min(m,n)$ and $p\eqdef\max(m,n)-1$.
  Then, we have $p\in\matN$ and $i\in[0..p]$, and thus
  $B_m\cap B_n=B_{p+1}\cap B_i=\emptyset$.

  For all $n\in\matN$, let $P(n)$ be the property:
  $\bigcup_{i\in[0..n]}A_i=\biguplus_{i\in[0..n]}B_i$.
  \proofpar{Induction: $P(0)$}
  Trivial.
  \proofpar{Induction: $P(n)$ implies $P(n+1)$}
  Let~$n\in\matN$.
  Assume that $P(n)$ holds.
  Let $B\eqdef\bigcup_{i\in[0..n]}B_i$.
  Then, from
  $P(n)$, and
  \assume{properties of set operations
    ({\eg}, $A\cup B=A\uplus(B\setminus A)$)},
  we have
  \begin{equation*}
    \bigcup_{i \in [0..n + 1]} A_i
    = B \cup A_{n + 1}
    = B \uplus \left( A_{n + 1} \setminus B \right)
    = B \uplus B_{n + 1}
    = \biguplus_{i \in [0..n + 1]} B_i.
  \end{equation*}
  Hence, for all $n\in\matN$, $P(n)$ holds.
  Moreover, from
  \assume{monotonicity of partial union}, and
  \assume{the definition of the limit of monotone sequence of subsets},
  we have $\bigcup_{i\in\matN}A_i=\biguplus_{i\in\matN}B_i$.

  Therefore, all three properties hold.
\end{proof}

\begin{definition}[trace of subsets of parties]
  \label{d:trace-of-subsets-of-parties}
  \mbox{}\hfill
  Let~$X$ be a set.
  Let~$Y\subset X$.
  Let~$i$ be the canonical injection from~$Y$ to~$X$.
  Let~$G\subset\calP(X)$.
  The notation $G\olcap Y$ denotes the set
  \begin{equation}
    \label{e:trace-of-subsets-of-parties}
    G \olcap Y \eqdef i^{-1} (G) = \left\{ A \cap Y \st A \in G \right\}.
  \end{equation}
\end{definition}

\begin{definition}[product of subsets of parties]
  \label{d:prod-of-subsets-of-parties}
  \mbox{}\\
  Let~$m\in[2..\infty)$.
  For all $i\in[1..m]$, let~$X_i$ be a set, and $G_i\subset\calP(X_i)$.
  Let~$X\eqdef\prod_{i\in[1..m]}X_i$.
  The notations $G_1\oltimes G_2$ and $\olprod_{i\in[1..m]}G_i$
  respectively denote the sets
  \begin{align}
    \label{e:prod-of-subsets-of-parties}
    G_1 \oltimes G_2
    & \eqdef \{ A_1 \times A_2 \in \calP (X_1 \times X_2) \st
      A_1 \in G_1 \Conj A_2 \in G_2 \}\\
    \olprod_{i \in [1..m]} G_i
    & \eqdef \left\{ \prod_{i \in [1..m]} A_i \in \calP (X) \rightst
      \left. \vphantom{\prod_{i \in [1..m]} A_i}
      \forall i \in [1..m],\;
      A_i \in G_i \right\}.
  \end{align}
\end{definition}

\begin{lemma}[restriction is masking]
  \label{l:restr-is-mask}
  \mbox{}\hfill
  Let~$X$ be a set.
  Let~$A\subset Y\subset X$.
  Let~$f:\ArYRb$.
  Let~$\hf:\ArXRb$.
  Assume that $\restr{\hf}{Y}=f$.
  Then, we have $\restr{\left(\hf\matUN_A\right)}{A}=\restr{f}{A}$ and
  $\restr{\left(\hf\matUN_A\right)}{A^c}=0$.
\end{lemma}

\begin{proof}
  Direct consequence of
  \assume{the definition of the indicator function}, and
  \assume{the definition of restriction of function}.
\end{proof}

\clearpage
\section{Complements on algebraic structures}
\label{s:complements-on-algebraic-structures}

\subsection{\Vectorspace}
\label{ss:vector-space}

\begin{definition}[relation compatible with vector operations]
  \label{d:relation-compatible-with-vector-ops}
  \mbox{}\hfill
  Let~$(E,+_E,\cdot_E)$ be a {\vectorspace}.
  An equivalence relation~$\calR$ on~$E$ is said
  {\em compatible with the vector operations} iff
  \begin{align}
    \label{e:relation-compatible-with-vector-ops-1}
    \forall u, \up, v, \vp \in E,\quad &
    \eqrel{u}{\up} \Conj \eqrel{v}{\vp}
    \Implies
    \eqrel{(u +_E v)}{(\up +_E \vp)},\\
    \label{e:relation-compatible-with-vector-ops-2}
    \forall \lambda \in \matK,\,
    \forall u, \up \in E,\quad &
    \eqrel{u}{\up} \Implies
    \eqrel{(\lambda \cdot_E u)}{(\lambda \cdot_E \up)}.
  \end{align}
\end{definition}

\begin{lemma}[quotient vector operations]
  \label{l:quotient-vector-ops}
  \mbox{}\hfill
  Let~$(E,+_E,\cdot_E)$ be a {\vectorspace}.\\
  Let~$\calR$ be an equivalence relation on~$E$.
  Assume that~$\calR$ is compatible with the vector operations.
  Then, the mappings $\clplus:\ArERxERER$ and $\cldot:\ArKxERER$ defined by
  \begin{align}
    \label{e:quotient-vector-ops-1}
    \forall u, v \in E,\quad &
    \clu \quotplus \clv \eqdef \cl{u +_E v},\\
    \label{e:quotient-vector-ops-2}
    \forall \lambda \in \matK,\,
    \forall u \in E,\quad &
    \lambda \quotdot \clu \eqdef \cl{\lambda \cdot_E u}.
  \end{align}
  are well-defined.
  These mappings are called
  {\em quotient vector operations induced on~$E/\calR$}.
\end{lemma}

\begin{proof}
  Let~$u,v\in E$.
  Let~$\lambda\in\matK$.
  Let~$\up,\vp\in E$ such that $\eqrel{u}{\up}$ and $\eqrel{v}{\vp}$, {\ie}
  such that $\clup=\clu$ and $\clvp=\clv$.
  Then, from
  Definition~\thref{d:relation-compatible-with-vector-ops},
  we have $\eqrel{(u+_Ev)}{(\up+_E\vp)}$ and
  $\eqrel{(\lambda\cdot_E u)}{(\lambda\cdot_E\up)}$,
  {\ie} $\cl{\up+_E\vp}=\cl{u+_Ev}$ and
  $\cl{\lambda\cdot_E\up}=\cl{\lambda\cdot_E u}$.
  Therefore, the quotient vector operations $\clplus$ and $\cldot$ defined
  by Equations~\eqref{e:quotient-vector-ops-1}
  and~\eqref{e:quotient-vector-ops-2} do not depend on the choice
  of the representative of classes, {\ie} they are well-defined.
\end{proof}

\begin{lemma}[quotient {\vectorspace}, equivalence relation]
  \label{l:quotient-vector-space-equiv-rel}
  \mbox{}\\
  Let~$(E,+_E,\cdot_E)$ be a {\vectorspace}.
  Let~$\calR$ be an equivalence relation on~$E$.
  Assume that~$\calR$ is compatible with the vector operations.
  Let~$\clplus$ and~$\cldot$ be the quotient vector operations induced on the
  quotient set $E/\calR$.
  Then, $(E/\calR,\clplus,\cldot)$ is a {\vectorspace}.
\end{lemma}

\begin{proof}
  From
  \assume{group theory},
  $(E/\calR,\clplus)$ is an abelian group with identity element $\cl{0_E}$
  Distributivity of the quotient scalar multiplication over quotient vector
  addition and field addition, compatibility of the quotient scalar
  multiplication with field multiplication, and 1 is the identity element for
  the quotient scalar multiplication are direct consequences of
  Definition~\thref{LM-d:space}, and
  Lemma~\thref{l:quotient-vector-ops}.
  Therefore, from
  Definition~\thref{LM-d:space},
  $(E/\calR,\clplus,\cldot)$ is a {\vectorspace}.
\end{proof}

\begin{lemma}[quotient {\vectorspace}]
  \label{l:quotient-vector-space}
  \mbox{}\hfill
  Let~$E$ be a {\vectorspace}.
  Let~$F$ be a {\vectorsubspace} of~$E$.
  Let~$\calR$ be the relation defined on~$E$ by for all $u,v\in E$,
  $\eqrel{u}{v}\Equiv v-u\in F$.\\
  Then, $\calR$ is an equivalence relation compatible with the vector
  operations of~$E$.

  The quotient set $E/\calR$ equipped with the quotient vector operations is
  thus a {\vectorspace} called {\em quotient {\vectorspace} of~$E$ by~$F$};
  it is denoted $E/F$.
  For all $u\in E$, the class of~$u$ is denoted $u+F$.
  The quotient vector operations are still denoted~$+$ and~$\cdot$ (the
  latter may be omitted).
\end{lemma}

\begin{proof}
  Let~$u,\up,v,\vp\in E$.
  Let~$\lambda\in\matK$.
  Assume that $\eqrel{u}{\up}$ and $\eqrel{v}{\vp}$, {\ie} $\up-u,\vp-v\in F$.
  Then, from
  Definition~\threfc{LM-d:space}{%
    vector addition is associative, and scalar multiplication is distributive
    over field addition}, and
  Lemma~\thref{LM-l:closed-under-vector-operations-is-subspace},
  we have $(\up+\vp)-(u+v)=(\up-u)+(\vp-v)\in F$ and
  $(\lambda\up)-(\lambda u)=\lambda(\up-u)\in F$,
  {\ie} $\eqrel{(u+v)}{(\up+\vp)}$ and $\eqrel{(\lambda u)}{(\lambda\up)}$.
  Hence, from
  Definition~\thref{d:relation-compatible-with-vector-ops},
  the equivalence relation~$\calR$ is compatible with the vector operations.
  Therefore, from
  Lemma~\thref{l:quotient-vector-space-equiv-rel},
  the quotient set $E/\calR$ equipped with the induced quotient vector
  operations is a {\vectorspace}.
\end{proof}

\begin{lemma}[linear map on quotient {\vectorspace}]
  \label{l:linear-map-on-quotient-vector-space}
  \mbox{}\hfill
  Let~$E,F$ be {\vectorspace}s.\\
  Let~$G$ be a {\vectorsubspace} of~$E$.
  Let~$f$ be a linear map from~$E$ to~$F$.
  Assume that $G\subset\ker(f)$.
  Then, the function $\clf\eqdef(u+G\mapsto f(u))$ is a linear map from~$E/G$
  to~$F$.
\end{lemma}

\begin{proof}
  Let~$u\in E$.
  Let~$\up\in u+G$.
  Then, from
  hypotheses, we have $\up-u\in G\subset\ker(f)$.
  Thus, from
  Definition~\thref{LM-d:linear-map}, and
  Definition~\thref{LM-d:kernel},
  we have $f(\up)=f(u)$.
  Hence, the function~$\clf$ does not depend on the choice of the
  representative of equivalence classes, it is well-defined.

  Let~$(u+G),(v+G)\in E/G$.
  Let~$\lambda\in\matK$.
  Then, from
  Lemma~\thref{l:quotient-vector-space},
  Lemma~\thref{l:quotient-vector-ops}, and
  Definition~\thref{LM-d:linear-map},
  we have
  \begin{gather*}
    \clf (a (u + G)) = \clf ((a u) + G) = f (a u)
    = a f (u) = a \clf (u + G),\\
    \clf ((u + G) + (v + G)) = \clf ((u + v) + G) = f (u + v)
    = f (u) + f (v) = \clf (u + G) + \clf (v + G).
  \end{gather*}

  Therefore, from
  Definition~\thref{LM-d:linear-map},
  $\clf$ is a linear map from~$E/G$ to~$F$.
\end{proof}

\subsection{Algebra over a field}
\label{ss:algebra-over-a-field}

\begin{remark}
  The following algebraic structure ``algebra over a field'' is not to be
  confused with the concept of ``set algebra'' defined in
  Section~\ref{s:set-alg}.
\end{remark}

\begin{remark}
  Most results on algebras over a field are valid on algebras over a ring, for
  which the property of {\vectorspace} over the field is replaced by that of
  module over the ring.
\end{remark}

\begin{definition}[algebra over a field]
  \label{d:alg-over-a-field}
  \mbox{}\\
  Let~$\matK$ be a field.
  A set~$E$ equipped with three algebra operations (a vector addition~$+$, a
  scalar multiplication~$\cdot$, and a vector multiplication~$\times$), is
  called {\em algebra (over field~$\matK$)}, or {\em $\matK$-algebra}, iff
  $(E,+,\cdot)$ is a $\matK$-{\vectorspace}, and
  vector multiplication is bilinear (or left and right distributive over
  vector addition, and compatible with scalars):
  \begin{align}
    \label{e:alg-over-a-field-1}
    \forall u, v, w \in E,\quad &
    (u + v) \times w
    = (u \times w) + (v \times w),\\
    \label{e:alg-over-a-field-2}
    \forall u, v, w \in E,\quad &
    u \times (v + w)
    = (u \times v) + (u \times w),\\
    \label{e:alg-over-a-field-3}
    \forall \lambda, \mu \in \matK,\;
    \forall u, v \in E,\quad &
    (\lambda \cdot u) \times (\mu \cdot v)
    = (\lambda \mu) \cdot (u \times v).
  \end{align}
\end{definition}

\begin{remark}
  The~$\cdot$ and~$\times$ infix signs in the scalar and vector
  multiplications may be omitted.
\end{remark}

\begin{lemma}[$\matK$~is $\matK$-algebra]
  \label{l:k-is-k-alg}
  \mbox{}\hfill
  Let~$\matK$ be a field.
  It is an algebra over itself.
\end{lemma}

\begin{proof}
  Direct consequence of
  Definition~\thref{d:alg-over-a-field}, and
  \assume{field properties of~$\matK$}.
\end{proof}

\begin{definition}[inherited algebra operations]
  \label{d:inherited-alg-ops}
  \mbox{}\\
  Let~$X$ be a nonempty set.
  Let~$\matK$ be a field.
  Let~$(E,+_E,\cdot_E,\times_E)$ be an algebra over field~$\matK$.
  The {\em algebra operations inherited on~$\FXE$} are the mappings~$+_\FXE$
  and~$\cdot_\FXE$ of Definition~\thref{LM-d:inherited-vector-operations},
  and the mapping $\times_\FXE:\ArFXExFXEFXE$ defined by
  \begin{equation}
    \label{e:inherited-alg-ops}
    \forall f, g \in \FXE,\,
    \forall x \in X,\quad
    (f \times_\FXE g) (x) \eqdef f (x) \times_E g (x).
  \end{equation}
\end{definition}

\begin{remark}
  Usually, inherited algebra operations are denoted the same way as the
  algebra operations of the target algebra.
\end{remark}

\begin{lemma}[algebra of functions to algebra]
  \label{l:alg-of-funs-to-alg}
  \mbox{}\\
  Let~$\matK$ be a field.
  Let~$X$ be a nonempty set.
  Let~$(E,+_E,\cdot_E,\times_E)$ be a $\matK$-algebra.
  Let~$+_\FXE$, $\cdot_\FXE$ and~$\times_\FXE$ be the algebra operations
  inherited on~$\FXE$.
  Then, $(\FXE,+_\FXE,\cdot_\FXE,\times_\FXE)$ is a $\matK$-algebra.
\end{lemma}

\begin{proof}
  From
  Lemma~\thref{LM-l:space-of-functions-to-space},
  $(\FXE,+_\FXE,\cdot_\FXE)$ is a $\matK$-{\vectorspace}.
  Then, bilinearity of the inherited vector multiplication is a direct
  consequence of
  Definition~\thref{d:alg-over-a-field}, and
  Definition~\thref{d:inherited-alg-ops}.
  Therefore, from
  Definition~\thref{d:alg-over-a-field},
  $\FXE$ equipped with~$+_\FXE$, $\cdot_\FXE$ and~$\times_\FXE$ is a
  $\matK$-algebra.
\end{proof}

\begin{lemma}[$\matK^X$~is algebra]
  \label{l:maps-to-k-is-alg}
  \mbox{}\\
  Let~$\matK$ be a field.
  Let~$X$ be a nonempty set.
  Then, $\matK^X$ is a $\matK$-algebra.
\end{lemma}

\begin{proof}
  Direct consequence of
  Lemma~\thref{l:k-is-k-alg}, and
  Lemma~\thref{l:alg-of-funs-to-alg}.
\end{proof}

\begin{definition}[subalgebra]
  \label{d:subalg}
  \mbox{}\hfill
  Let~$\matK$ be a field.
  Let~$(E,+,\cdot,\times)$ be a $\matK$-algebra.\\
  A subset~$F$ of~$E$ equipped with the restrictions~$\restr{+}{F}$,
  $\restr{\cdot}{F}$ and~$\restr{\times}{F}$ of the algebra operations to~$F$
  is called {\em ($\matK$-)subalgebra of~$E$} iff
  $(F,\restr{+}{F},\restr{\cdot}{F},\restr{\times}{F})$~is a $\matK$-algebra.
\end{definition}

\begin{remark}
  Usually, restrictions~$\restr{+}{F}$, $\restr{\cdot}{F}$,
  and~$\restr{\times}{F}$ are still denoted~$+$, $\cdot$ and~$\times$.
\end{remark}

\begin{lemma}[{\vectorsubspace} and closed under multiplication is subalgebra]
  \label{l:subspace-and-closed-under-mult-is-subalg}
  \mbox{}\\
  Let~$\matK$ be a field.
  Let~$(E,+,\cdot,\times)$ be a $\matK$-algebra.
  Let~$F\subset E$.
  Then, $F$~is a $\matK$-subalgebra of~$E$ iff
  $F$~is a $\matK$-{\vectorsubspace} of~$E$, and
  $F$~is closed under vector multiplication:
  \begin{equation}
    \label{e:subspace-and-closed-under-mult-is-subalg}
    \forall u, v \in F,\quad u \times v \in F.
  \end{equation}
\end{lemma}

\begin{proof}
  \proofpar{``Left'' implies ``right''}
  Assume first that~$F$ is a $\matK$-subalgebra of~$E$.
  Then, from
  Definition~\threfc{d:subalg}{$F$~is a $\matK$-algebra},
  Definition~\thref{d:alg-over-a-field},
  $(F,\restr{+}{F},\restr{\cdot}{F})$ is a $\matK$-{\vectorspace}, and~$F$ is
  closed under the restriction to~$F$ of the three operations.
  Thus, from
  Definition~\thref{LM-d:subspace},
  $F$~is a $\matK$-{\vectorsubspace} of~$E$, and~$F$ is closed under vector
  multiplication.

  \proofparskip{``Right'' implies ``left''}
  Conversely, assume now that~$F$ is a $\matK$-{\vectorsubspace} of~$E$, and
  that it is closed under vector multiplication.
  Then, from
  Definition~\thref{LM-d:subspace},
  $(F,\restr{+}{F},\restr{\cdot}{F})$ is a $\matK$-{\vectorspace}.
  Moreover, from
  Lemma~\thref{LM-l:closed-under-vector-operations-is-subspace},
  $F$~is also closed under vector addition and scalar multiplication.
  Thus, since~$F$ is a subset of~$E$, and~$E$ is a $\matK$-algebra,
  Equations~\eqref{e:alg-over-a-field-1} to~\eqref{e:alg-over-a-field-3} are
  trivially satisfied over~$F$ with the restrictions of the three operations.
  Hence, from
  Definition~\thref{d:alg-over-a-field}, and
  Definition~\thref{d:subalg},
  $F$~is a $\matK$-subalgebra of~$E$.

  \medskip\noindent
  Therefore, we have the equivalence.
\end{proof}

\begin{lemma}[closed under algebra operations is subalgebra]
  \label{l:closed-under-alg-ops-is-subalg}
  \mbox{}\\
  Let~$\matK$ be a field.
  Let~$E$ be a $\matK$-algebra.
  Let~$F\subset E$.
  Then, $F$~is a $\matK$-subalgebra of~$E$ iff
  $0_E\in F$, and
  $F$~is closed under vector addition, scalar and vector multiplications:
  \begin{align}
    \label{e:closed-under-alg-ops-is-subalg-1}
    \forall u, v \in F,\quad & u + v \in F,\\
    \label{e:closed-under-alg-ops-is-subalg-2}
    \forall \lambda \in \matK,\; \forall u \in F,\quad & \lambda u \in F,\\
    \label{e:closed-under-alg-ops-is-subalg-3}
    \forall u, v \in F,\quad & u \times v \in F.
  \end{align}
\end{lemma}

\begin{proof}
  Direct consequence of
  Lemma~\thref{l:subspace-and-closed-under-mult-is-subalg}, and
  Lemma~\thref{LM-l:closed-under-vector-operations-is-subspace}.
\end{proof}

\subsection{Seminormed {\vectorspace}}
\label{ss:seminormed-vector-spaces}

\begin{definition}[seminorm]
  \label{d:seminorm}
  \mbox{}\\
  Let~$\matK$ be a valued field.
  Let~$E$ be a $\matK$-{\vectorspace}.
  A function $\nrmdot:\ArER$ is called {\em seminorm over~$E$} iff
  it is absolutely homogeneous of degree~1,
  and it satisfies the triangle inequality:
  \begin{align}
    \label{e:seminorm-1}
    \forall \lambda \in \matK,\,
    \forall u \in E,\quad &
    \nrm{\lambda u} = | \lambda | \, \nrm{u},\\
    \label{e:seminorm-2}
    \forall u, v \in E,\quad &
    \nrm{u + v} \leq \nrm{u} + \nrm{v}.
  \end{align}

  If so, $(E,\nrmdot)$ (or simply~$E$) is called
  {\em seminormed ($\matK$-){\vectorspace}}.
\end{definition}

\begin{remark}
  Most results from~\cite{cm:lmt:16} on normed {\vectorspace}s can be
  generalized to the case of seminormed {\vectorspace}, sometimes with
  slight modifications of the statement.
  In particular, the associated distance becomes a pseudometric.
\end{remark}

\begin{lemma}[definite seminorm is norm]
  \label{l:definite-seminorm-is-norm}
  \mbox{}\hfill
  Let~$(E,\nrmdot)$ be a seminormed {\vectorspace}.
  Then, $(E,\nrmdot)$ is a normed {\vectorspace} iff
  $\nrmdot$~is definite, {\ie} for all $u\in E$, $\nrm{u}=0\Equiv u=0$.
\end{lemma}

\begin{proof}
  Direct consequence of
  Definition~\thref{d:seminorm},
  Definition~\thref{LM-d:normed-space}, and
  Definition~\thref{LM-d:norm}.
\end{proof}

\clearpage
\section{Complements on order theory}
\label{s:complements-on-order-theory}

\begin{remark}
  In the following definition, the strict inequality ``$<$'' naturally means
  ``$\leq$ and not equal'' (equality is related to the set object).

  We recall the notations~$\pm\infty$ to represent the extreme bounds of a
  totally ordered set;
  they may belong to the set, or not.
\end{remark}

\begin{definition}[interval]
  \label{d:interval}
  \mbox{}\hfill
  Let~$(X,\leq)$ be a totally ordered nonempty set.
  Let~$a,b\in X$.

  The subset $\{x\in X\st a<x<b\}$ is called the
  {\em open proper interval from~$a$ to~$b$} (with excluded bounds~$a$
  and~$b$);
  it is denoted $(a,b)$.
  The set of open proper intervals for all $a,b\in X$ is denoted~$\Itvop_X$.

  The subsets $\{x\in X\st x<b\}$ and $\{x\in X\st a<x\}$ are called
  {\em open left ray and open right ray} (with excluded bounds~$a$ and~$b$)
  (or {\em half-lines});
  they are denoted $(a,\infty\rsrbra$ and $\lsrbra-\infty,b)$.
  The set of open rays for all $a,b\in X$ is denoted~$\Ray_X$.

  The notation $\lsrbra-\infty,\infty\rsrbra$ may be used to represent the
  whole set~$X$.

  In all cases, square brackets ``$[,]$'' are used to specify that bounds are
  included, and square-and-round brackets ``$\lsrbra,\rsrbra$'' are used to
  avoid specifying inclusion or exclusion of the bounds.

  Let~$a,b\in X\cup\{\pm\infty\}$.
  The subset $\lsrbra a,b\rsrbra$ is either a proper interval, a ray, or the
  whole set;
  it is called {\em interval}.
  The set of open intervals for all $a,b\in X\cup\{\pm\infty\}$ is
  denoted~$\Itvo_X\eqdef\Itvop_X\cup\Ray_X\cup\{X\}$.
\end{definition}

\begin{remark}
  Note that if~$X$ contains at least two elements, $\Itvop_X$, $\Ray_X$
  and~$\{X\}$ are pairwise disjoint.
  In particular, $(a,\infty)$ is an open ray when $\infty\not\in X$, and an
  open proper interval when $\infty\in X$.
  More generally, open rays are never open proper intervals.
  For instance, when~$\infty$ belongs to the set, $(a,\infty\rsrbra$ is
  actually $(a,\infty]$ (an open subset of~$\matRbar$), which is distinct
  from the open proper interval $(a,\infty)$.
\end{remark}

\begin{lemma}[empty open interval]
  \label{l:empty-open-int}
  \mbox{}\\
  Let~$(X,\leq)$ be a totally ordered nonempty set.
  Assume that~$X$ is dense-in-itself:
  \begin{equation}
    \label{e:empty-open-int}
    \forall x, y \in X,\quad
    x < y
    \IMPLIES \exists z \in X,\;
    x < z < y.
  \end{equation}
  Let~$a,b\in X\cup\{\pm\infty\}$.
  Then, the open interval $(a,b)$ is empty iff $b\leq a$.
\end{lemma}

\begin{proof}
  Then, from
  Definition~\thref{d:interval},
  Equation~\eqref{e:empty-open-int},
  \assume{transitivity of order (which provides the other implication)}, and
  \assume{the definition of strict inequality},
  we have
  \begin{align*}
    (a, b) = \emptyset
    & \EQUIV \forall x \in X,\; \neg (a < x < b)\\
    & \EQUIV \neg (\exists x \in X,\; a < x < b)
    \EQUIV \neg (a < b)
    \EQUIV b \leq a.
  \end{align*}
\end{proof}

\begin{remark}
  Of course, the previous statement is wrong in the presence of isolated
  points.
  For instance, $(0,1)$ is empty in the discrete sets~$\matN$ and~$\matZ$.
\end{remark}

\begin{remark}
  In the following lemma, the left and right square-and-round brackets must
  remain the same on the sides of each interval.
  For instance ``$\lsrbra_1$'' denotes either ``$[$'' or ``$($'', but it
  remains identical in the three intervals of
  Equation~\eqref{e:inter-of-ints-and-rays}.
\end{remark}

\begin{lemma}[intervals are closed under finite intersection]
  \label{l:int-are-closed-under-finite-inter}
  \mbox{}\\
  Let~$(X,\leq)$ be a totally ordered nonempty set.
  Then, the intersection of two proper intervals is a proper interval, the
  intersection of two rays is either a ray (if they both point towards the
  same direction), or a proper interval, and the intersection of a proper
  interval and a ray is a proper interval.
  In particular, for all~$a,b,c,d\in X\cup\{\pm\infty\}$, we have (see remark
  above)
  \begin{equation}
    \label{e:inter-of-ints-and-rays}
    \lsrbra_1 a, b \rsrbra_2 \cap \lsrbra_1 c, d \rsrbra_2
    = \lsrbra_1 \max (a, c), \min (b, d) \rsrbra_2.
  \end{equation}
  Hence, the closure of~$\Ray_X\cup\{X\}$ under finite intersection
  is~$\Itvo_X$, and~$\Itvop_X$ and~$\Itvo_X$ are closed under finite
  intersection.
\end{lemma}

\begin{proof}
  Direct consequence of
  Definition~\thref{d:interval},
  \assume{totally ordered set properties of~$X$}, and
  \assume{induction on the number of operands of the finite intersection}.
\end{proof}

\begin{lemma}[empty intersection of open intervals]
  \label{l:empty-inter-of-open-ints}
  \mbox{}\\
  Let~$(X,\leq)$ be a totally ordered nonempty set.
  Assume that~$X$ is dense-in-itself:
  \begin{equation}
    \label{e:empty-inter-of-open-ints}
    \forall x, y \in X,\quad
    x < y
    \IMPLIES \exists z \in X,\;
    x < z < y.
  \end{equation}
  Let~$a,b,c,d\in X\cup\{\pm\infty\}$.
  Then, $(a,b)\cap(c,d)$ is empty iff
  $b\leq a$, $d\leq c$, $d\leq a$ or $b\leq c$.
\end{lemma}

\begin{proof}
  Direct consequence of
  Lemma~\thref{l:int-are-closed-under-finite-inter},
  Lem\-ma~\thref{l:empty-open-int}, and
  \assume{totally ordered set properties of~$X$}.
\end{proof}

\clearpage
\section{Complements on general topology}
\label{s:complements-on-general-topology}

\begin{remark}
  In~\cite{cm:lmt:16}, we have only covered topology for metric spaces in
  which subsets are open when they contain a ball centered in each of their
  points.
  In the present document, we deal with the general case of topological spaces
  for which the collection of open subsets is given, {\eg} via a topological
  basis.
\end{remark}

\begin{definition}[topological space]
  \label{d:topological-space}
  \mbox{}\hfill
  Let~$X$ be a set.
  A subset~$\calT$ of~$\calP(X)$ is called {\em topology of~$X$} iff
  $\emptyset,X\in\calT$, and
  $\calT$~is closed under (infinite) union and finite intersection.

  If so, $(X,\calT)$ (or simply~$X$) is called {\em topological space},
  elements of~$\calT$ are called {\em open subsets of~$X$}, and
  the complement of elements of~$\calT$ are called
  {\em closed subsets of~$X$}.
\end{definition}

\begin{lemma}[intersection of topologies]
  \label{l:inter-of-topo}
  \mbox{}\\
  Let~$X$ and~$I$ be sets.
  Let~$(\calT_i)_{i\in I}$ be topologies on~$X$.
  Then, $\bigcap_{i\in I}\calT_i$ is a topology on~$X$.
\end{lemma}

\begin{proof}
  Direct consequence of
  Definition~\thref{d:topological-space}, and
  \assume{the definition and properties of intersection and union of
    subsets}.
\end{proof}

\begin{definition}[generated topology]
  \label{d:gen-topo}
  \mbox{}\hfill
  Let~$X$ be a set.
  Let~$G\subset\calP(X)$.\\
  The {\em topology on~$X$ generated by~$G$} is the intersection of all
  topologies on~$X$ containing~$G$;
  it is denoted~$\calT_X(G)$.
  The generator~$G$ is also called {\em subbase of the topology}.
\end{definition}

\begin{lemma}[generated topology is minimum]
  \label{l:gen-topo-is-min}
  \mbox{}\\
  Let~$X$ be a set.
  Let~$G\subset\calP(X)$.
  Then, $\calT_X(G)$~is the smallest topology on~$X$ containing~$G$.
\end{lemma}

\begin{proof}
  Direct consequence of
  Definition~\thref{d:gen-topo},
  Lemma~\thref{l:inter-of-topo}, and
  \assume{properties of the intersection}.
\end{proof}

\begin{lemma}[equivalent definition of generated topology]
  \label{l:equiv-def-of-gen-topo}
  \mbox{}\\
  Let~$X$ be a set.
  Let~$G\subset\calP(X)$.
  Let~$O\subset X$.
  Then, $O\in\calT_X(G)$ iff
  $O$~is the union of finite intersections of elements of~$G\cup\{X\}$.
\end{lemma}

\begin{proof}
  Direct consequence of
  Definition~\thref{d:topological-space},
  \assume{associativity and commutativity of union},
  \assume{distributivity of intersection and union (both ways)}
  (thus, the set of unions of finite intersections of elements of~$G$ is a
  topology of~$X$ containing~$G$, and contained in all topologies of~$X$
  containing~$G$), and
  Lemma~\thref{l:gen-topo-is-min}.
\end{proof}

\begin{definition}[topological basis]
  \label{d:topological-basis}
  \mbox{}\\
  Let~$(X,\calT)$ be a topological space.
  Let~$I$ be a set.
  A set $\{B_i\in\calT\st i\in I\}$ is called
  {\em topological basis of~$(X,\calT)$} iff
  for all $O\in\calT$, there exists $J\subset I$ such that
  $O=\bigcup_{j\in J}B_j$.
\end{definition}

\begin{lemma}[augmented topological basis]
  \label{l:augmented-topo-basis}
  \mbox{}\hfill
  Let~$(X,\calT)$ be a topological space.\\
  Let~$\calB$ be a topological basis of~$(X,\calT)$.
  Let~$O\in\calT$.
  Then, $\calB\cup\{O\}$ is a topological basis of~$(X,\calT)$.
\end{lemma}

\begin{proof}
  Direct consequence of
  Definition~\threfc{d:topological-basis}{with~$O$ open}.
\end{proof}

\begin{definition}[order topology]
  \label{d:order-topo}
  \mbox{}\\
  Let~$(X,\leq)$ be a totally ordered set.
  The topology~$\calT_X(\Ray_X)$ is called {\em order topology on~$X$}.
\end{definition}

\begin{remark}
  See Definition~\thref{d:interval} for the definition of~$\Ray_X$.
  Note that the order topology is the standard topology on the totally
  ordered sets of numbers~$\matN$, $\matZ$, $\matQ$, and~$\matR$.
\end{remark}

\begin{lemma}[topological basis of order topology]
  \label{l:topological-basis-of-order-topo}
  \mbox{}\\
  Let~$(X,\leq)$ be a totally ordered set.
  Then, $\Itvo_X$ is a topological basis for the order topology on~$X$.
\end{lemma}

\begin{proof}
  Direct consequence of
  Definition~\thref{d:order-topo},
  Lemma~\thref{l:equiv-def-of-gen-topo},
  Lemma~\thref{l:int-are-closed-under-finite-inter}, and
  Definition~\thref{d:topological-basis}.
\end{proof}

\begin{remark}
  See Definition~\thref{d:interval} for the definition of~$\Itvo_X$.
\end{remark}

\begin{lemma}[trace topology on subset]
  \label{l:trace-topo-on-subset}
  \mbox{}\\
  Let~$(X,\calT)$ be a topological space, and~$\calB$ be a topological basis
  of $(X,\calT)$.
  Let~$Y\subset X$.\\
  Then, $\calT_Y\eqdef\calT\olcap Y$~is a topology of~$Y$, and
  $\calB\olcap Y$ is a topological basis of~$(Y,\calT_Y)$.

  $\calT_Y$ is called {\em trace topology}, and $(Y,\calT_Y)$ is said
  {\em topological subspace of~$(X,\calT)$}.
\end{lemma}

\begin{proof}
  Direct consequence of
  Definition~\thref{d:topological-space},
  Definition~\thref{d:topological-basis},
  Definition~\thref{d:trace-of-subsets-of-parties},
  \assume{distributivity of intersection over union}, and
  \assume{commutativity of intersection}.
\end{proof}

\begin{lemma}[box topology on Cartesian product]
  \label{l:box-topo-on-cartesian-prod}
  \mbox{}\hfill
  Let~$I$ be a set.
  For all $i\in I$, let~$(X_i,\calT_i)$ be a topological space and~$\calB_i$
  be a topological basis of $(X_i,\calT_i)$.
  Let~$X\eqdef\prod_{i\in I}X_i$.
  Then, $\calT\eqdef\olprod_{i\in I}\calT_i$ is a topology of~$X$, and
  $\olprod_{i\in I}\calB_i$~is a topological basis of~$(X,\calT)$.

  $\calT$~is called the
  {\em box topology of~$X$ (induced by the~$\calT_i$'s)}.
\end{lemma}

\begin{proof}
  Direct consequence of
  Definition~\thref{d:topological-space},
  Definition~\thref{d:topological-basis},
  Definition~\thref{d:prod-of-subsets-of-parties}, and
  \assume{distributivity of the Cartesian product over union}.
\end{proof}

\subsection{Second axiom of countability}
\label{ss:second-axiom-of-countability}

\begin{definition}[second-countability]
  \label{d:second-count}
  \mbox{}\hfill
  A topological space $(X,\calT)$ is said
  {\em second-countable}, or {\em completely separable}, iff
  it admits a countable topological basis,
\end{definition}

\begin{remark}
  This corresponds to the existence of $I\subset\matN$ in
  Definition~\thref{d:topological-basis}.
\end{remark}

\begin{lemma}[complete countable topological basis]
  \label{l:complete-count-topo-basis}
  \mbox{}\\
  Let~$(X,\calT)$ be a second-countable topological space.
  Let~$\calB$ be a countable topological basis of~$(X,\calT)$.
  Let~$O\in\calT$.
  Then, $\calB\cup\{O\}$~is a countable topological basis of~$(X,\calT)$.
\end{lemma}

\begin{proof}
  Direct consequence of
  Definition~\thref{d:second-count}, and
  Lemma~\thref{l:augmented-topo-basis}.
\end{proof}

\begin{lemma}[compatibility of second-countability with Cartesian product]
  \label{l:compat-of-second-count-with-cartesian-prod}
  \mbox{}\\
  Let~$I$ be a set.
  For all $i\in I$, let~$(X_i,\calT_i)$ be a second-countable topological
  space, and let~$\calB_i$ be a countable topological basis of
  $(X_i,\calT_i)$.
  Let~$X\eqdef\prod_{i\in I}X_i$.
  Assume that~$X$ is equipped with the box
  topology~$\calT\eqdef\olprod_{i\in I}\calT_i$.
  Then, $\olprod_{i\in I}\calB_i$~is a countable topological basis
  of~$(X,\calT)$.
  Hence, $(X,\calT)$~is second-countable.
\end{lemma}

\begin{proof}
  Direct consequence of
  Lemma~\thref{l:box-topo-on-cartesian-prod},
  Definition~\thref{d:prod-of-subsets-of-parties}, and
  \assume{compatibility of finite Cartesian product with countability}.
\end{proof}

\begin{lemma}[complete countable topological basis of product space]
  \label{l:complete-count-topo-basis-of-prod-space}
  \mbox{}\hfill
  Let~$I$ be a set.\\
  For all $i\in I$, let~$(X_i,\calT_i)$ be a second-countable topological
  space, let~$\calB_i$ be a countable topological basis of it, and
  let~$O_i\in\calT_i$.
  Let~$X\eqdef\prod_{i\in I}X_i$.
  Assume that~$X$ is equipped with the box
  topology~$\calT\eqdef\olprod_{i\in I}\calT_i$.
  Then, $\olprod_{i\in I]}\calB_i\cup\{O_i\}$~is a countable topological
  basis of~$(X,\calT)$.
\end{lemma}

\begin{proof}
  Direct consequence of
  Definition~\thref{d:prod-of-subsets-of-parties},
  Lemma~\thref{l:complete-count-topo-basis}, and
  Lemma~\thref{l:box-topo-on-cartesian-prod}.
\end{proof}

\subsection{Complements on metric space}
\label{ss:complements-on-metric-space}

\begin{definition}[pseudometric]
  \label{d:pseudometric}
  \mbox{}\\
  Let~$X$ be a nonempty set.
  A function~$d:\ArXxXR$ is called {\em pseudometric over~$X$} iff
  it is nonnegative,
  symmetric,
  it is zero on the diagonal,
  and it satisfies the triangle inequality:
  \begin{align}
    \label{e:pseudometric-1}
    \forall x, y \in X,\quad & d (x, y) \geq 0,\\
    \label{e:pseudometric-2}
    \forall x, y \in X,\quad & d (y, x) = d (x, y),\\
    \label{e:pseudometric-3}
    \forall x \in X, \quad & d (x, x) = 0,\\
    \label{e:pseudometric-4}
    \forall x, y, z \in X,\quad & d (x, z) \leq d (x, y) + d (y, z).
  \end{align}

  If so, $(X,d)$ (or simply~$X$) is called {\em pseudometric space}.
\end{definition}

\begin{remark}
  A pseudometric becomes a metric when it is also definite.
  Hence, most results from~\cite{cm:lmt:16} on metric spaces are still valid
  on pseudometric spaces.
\end{remark}

\begin{lemma}[equivalent definition of convergent sequence]
  \label{l:equiv-def-of-conv-seq}
  \mbox{}\\
  Let~$(X,d)$ be a metric space.
  Let~$(x_n)_{n\in\matN},l\in X$.
  Then, $(x_n)_{n\in\matN}$ is convergent with limit~$l$ iff
  \begin{equation}
    \label{e:equiv-def-of-conv-seq}
    \forall k \in \matN,\;
    \exists N \in \matN,\;
    \forall n \in [N..\infty),\quad
    d (x_n, l) \leq \frac{1}{k + 1}.
  \end{equation}
\end{lemma}

\begin{proof}
  \proofpar{``Left'' implies ``right''}\\
  Direct consequence of
  Definition~\threfc{LM-d:convergent-sequence}{%
    with $\eps\eqdef\frac{1}{k+1}$}.

  \proofparskip{``Right'' implies ``left''}
  Assume that Equation~\eqref{e:equiv-def-of-conv-seq} holds.
  Let~$\eps>0$.
  Then, from
  \assume{the Archimedean property of~$\matR$}, and
  \assume{ordered field properties of~$\matR$},
  let $k\in\matN$ such that $k\geq\frac{1}{\eps}-1$, {\ie}
  $\frac{1}{k+1}\leq\eps$.
  Thus, from assumption, there exists $N\in\matN$ such that for all
  $n\in[N..\infty)$, we have $d(x_n,l)\leq\frac{1}{k+1}\leq\eps$.
  Hence, from
  Definition~\thref{LM-d:convergent-sequence},
  the sequence $(x_n)_{n\in\matN}$ is convergent with limit~$l$.

  \medskip\noindent
  Therefore, we have the equivalence.
\end{proof}

\begin{lemma}[convergent subsequence of Cauchy sequence]
  \label{l:conv-subseq-of-cauchy-seq}
  \mbox{}\\
  Let~$(X,d)$ be a metric space.
  Let~$(x_n)_{n\in\matN}\in X$.
  Assume that $(x_n)_{n\in\matN}$ is a Cauchy sequence.
  Let~$(n_k)_{k\in\matN}\in\matN$.
  Assume that~$(n_k)_{k\in\matN}$ is nondecreasing and that the
  subsequence~$(x_{n_k})_{k\in\matN}$ is convergent.
  Then, $(x_n)_{n\in\matN}$ is convergent with the same limit.
\end{lemma}

\begin{proof}
  From
  Lemma~\thref{LM-l:limit-is-unique},
  let~$x\in X$ be the limit of the subsequence.
  Let~$\eps>0$.
  Then, from
  Definition~\thref{LM-d:cauchy-sequence},
  let $N\in\matN$ such that for all $p,q\geq N$, we have
  $d(x_p,x_q)\leq\frac{\eps}{2}$.
  Moreover, as $(n_k)_{k\in\matN}$ is increasing,
  let $\Kp\in\matN$ such that for all $k\geq \Kp$, we have $n_k>N$,
  and from
  Definition~\thref{LM-d:convergent-sequence},
  let $\Kpp\in\matN$ such that for all $k\geq
  \Kpp$, we have $d(x_{n_k},x)\leq\frac{\eps}{2}$.
  Let~$K\eqdef\max(\Kp,\Kpp)$.
  Then, from
  Definition~\threfc{LM-d:distance}{triangle inequality},
  we have for all $n\geq N$,
  \begin{equation*}
    d (x_n, x)
    \leq d (x_n, x_{n_K}) + d (x_{n_K}, x)
    \leq \frac{\eps}{2} + \frac{\eps}{2}
    = \eps.
  \end{equation*}
  Therefore, from
  Definition~\thref{LM-d:convergent-sequence},
  $(x_n)_{n\in\matN}$ is convergent with limit~$x$.
\end{proof}

\begin{definition}[cluster point]
  \label{d:cluster-point}
  \mbox{}\hfill
  Let~$(X,d)$ be a metric space.
  Let~$(x_n)_{n\in\matN}\in X$.\\
  A {\em cluster point} of the sequence is the limit~$x$ of any convergent
  subsequence of~$(x_n)_{n\in\matN}$:
  \begin{equation}
    \label{e:cluster-point}
    \forall \eps > 0,\;
    \forall N \in \matN,\;
    \exists n \geq N,\quad
    d (x_n, x) \leq \eps.
  \end{equation}
\end{definition}

\clearpage
\section{Complements on real numbers}
\label{s:complements-on-real-numbers}

\subsection{Real numbers}
\label{ss:real-numbers}

\begin{lemma}[finite cover of compact interval]
  \label{l:finite-cover-of-compact-int}
  \mbox{}\\
  Let~$a,b,(a_n)_{n\in\matN},(b_n)_{n\in\matN}\in\matR$.
  Assume that $a\leq b$ and $[a,b]\subset\bigcup_{n\in\matN}(a_n,b_n)$.
  Then, there exists $q\in\matN$ and $(i_p)_{p\in[0..q]}\in\matN$ pairwise
  distinct such that $[a,b]\subset\bigcup_{p\in[0..q]}(a_{i_p},b_{i_p})$ with
  $a_{i_0}<a$, $b<b_{i_q}$, and for all $p\in[0..q-1]$, $a_{i_{p+1}}<b_{i_p}$.
\end{lemma}

\begin{proof}
  From
  \assume{the definition of compactness}, and
  \assume{compactness of~$[a,b]$},
  there exists $n\in\matN$ such that
  $[a,b]\subset\bigcup_{j\in[0..n]}(a_j,b_j)$.

  Let $J\subset[0..n]$ and $x\in\matR$.
  Assume that $[x,b]\subset\bigcup_{j\in J}(a_j,b_j)$ and $x\leq b$.

  \proofparskip{(1). Next index:
    $\exists i\eqdef i_J(x)\in J,\;a_i<x<b_i$}\\
  Direct consequence of
  \assume{the definition of union (with $x\in[x,b]$)}.

  \proofparskip{(2). Next cover:
    $[b_i,b]\subset\bigcup_{j\in J\setminus\{i\}}(a_j,b_j)$}\\
  \proofpar{Case $b<b_i$} Trivial.
  \proofpar{Case $b_i\leq b$}
  Then, from~(1), we have $b_i\in(x,b]$.
  Let $y\in[b_i,b]\subset[x,b]$.
  Then, from
  \assume{the definition of union},
  there exists $j\in J$ such that $a_j<y<b_j$.
  Hence, from
  \assume{transitivity of order in~$\matR$},
  we have $b_i<b_j$, {\ie} $j\not=i$, and
  $[b_i,b]\subset\bigcup_{j\in J\setminus\{i\}}(a_j,b_j)$.

  \medskip
  Let~$i_{-1}\eqdef-1$, $J_{-1}\eqdef[0..n]$, and~$b_{i_{-1}}=b_{-1}\eqdef a$.
  Let~$(i_p)_{p\in[0..q]}$ be the sequence of integers computed by the
  following algorithm:
  \begin{center}
    \begin{minipage}{0.5\textwidth}
      \begin{algorithm}[H]
        $p \eqdef -1$\;
        \Repeat{$b < b_{i_p}$}{
          $i_{p + 1} \eqdef i_{J_p} (b_{i_p})$\;
          $J_{p + 1} \eqdef J_p \setminus \{ i_{p + 1} \}$\;
          $p \eqdef p + 1$\;
        }
        \Return $(i_0, \ldots, i_p)$\;
      \end{algorithm}
    \end{minipage}
  \end{center}

  For all $p\in\{-1\}\cup\matN$, let $I_p\eqdef\{i_0,\ldots,i_p\}$ (with
  $I_{-1}\eqdef\emptyset$), and~$P(p)$ be the property:
  \begin{align*}
    & \card (I_p) = p + 1
      \CONJ [0..n] = I_p \uplus J_p
      \CONJ (\forall m \in [0..p],\; a_{i_m} < b_{i_{m - 1}})\quad\Conj\\
    & [a, b_{i_p}) \subset \bigcup_{j \in I_p} (a_j, b_j)
    \CONJ [b_{i_p}, b] \subset \bigcup_{j \in J_p} (a_j, b_j).
  \end{align*}
  Let~us show that there exists~$q\in[0..n]$ such that~$P(q)\Conj b<b_{i_q}$
  holds.

  \proofparskip{(3). Initialization: $P(-1)\Conj b_{i_{-1}}\leq b$}
  Trivial.

  \proofparskip{(4). Iterations:
    $\forall p\in\{-1\}\cup\matN,\;
    P(p)\Conj b_{i_p}\leq b$ implies $P(p+1)$}\\
  Let~$p\in\{-1\}\cup\matN$.
  Assume that~$P(p)\Conj b_{i_p}\leq b$ holds.\\
  Let $J\eqdef J_p$ and $x\eqdef b_{i_p}$.
  Then, we have $[x,b]\subset\bigcup_{j\in J}(a_j,b_i)$ and $x\leq b$.
  Thus, from~(1), there exists $i_{p+1}=i_J(x)\in J=J_p$ ({\ie}
  $i_{p+1}\not\in I_p$), such that $a_{i_{p+1}}<b_{i_p}<b_{i_{p+1}}$.
  Moreover, from~(2), we have
  $[b_{i_{p+1}},b]\subset\bigcup_{j\in J_{p+1}}(a_j,b_j)$.
  Then, we have
  \begin{gather*}
    \card (I_{p + 1}) = \card (I_p) + 1 = p + 2,\\
    [0..n] = I_p \uplus \{ i_{p+1} \} \uplus (J_p \setminus \{ i_{p+1} \})
    = I_{p+1} \uplus J_{p+1}.\\
    a_{i_{p+1}} < b_{i_p},\\
    [a, b_{i_{p+1}}) = [a, b_{i_p}) \uplus [b_{i_p}, b_{i_{p+1}})
    \subset \bigcup_{j \in I_p} (a_j, b_j) \cup (a_{i_{p+1}}, b_{i_{p+1}})
    = \bigcup_{j \in I_{p+1}} (a_j, b_j),\\
    [b_{i_{p + 1}}, b] \subset \bigcup_{j \in J_{p + 1}} (a_j, b_j).
  \end{gather*}
  Hence, $P(p+1)$~holds.

  \proofparskip{(5). Termination: $\exists q\in[0..n],\;P(q)\Conj b<b_{i_q}$}
  Assume that $P(p)\Conj b_{i_p}\leq b$ holds for all~$p$ in~$[0..n]$.
  Then, from~(4), we have $P(n+1)$.
  Thus, from
  \assume{additivity of the cardinality},
  we have
  \begin{equation*}
    n + 1 = \card ([0..n])
    = \card (I_{n + 1}) + \card (J_{p + 1})
    \geq \card (I_{n + 1})
    = n + 2.
  \end{equation*}
  Which is impossible.
  Hence, there exists $q\in[0..n]$ such that $P(q)\Conj b<b_{i_q}$ holds.\\
  Moreover, we have
  $[a,b]\subset[a,b_{i_q})\subset\bigcup_{j\in I_q}(a_j,b_j)$.

  \medskip
  Therefore, there exists $q\in\matN$ and $(i_p)_{p\in[0..q]}\in\matN$ pairwise
  distinct such that $[a,b]$ is included in
  $\bigcup_{p\in[0..q]}(a_{i_p},b_{i_p})$ with $a_{i_0}<a$, $b<b_{i_q}$, and
  for all $p\in[0..q-1]$, $a_{i_{p+1}}<b_{i_p}$.
\end{proof}

\begin{definition}[H\"older conjugates in~$\matR$]
  \label{d:holder-conjugates-in-r}
  \mbox{}\\
  Real numbers $p,q\in(1,\infty)$ are said
  {\em H\"older conjugates in~$\matR$} iff
  $\frac{1}{p}+\frac{1}{q}=1$.
\end{definition}

\begin{lemma}[2 is self-H\"older conjugate in~$\matR$]
  \label{l:two-is-self-holder-conjugate-in-r}
  \mbox{}\\
  The real number~2 is H\"older conjugate in~$\matR$ with itself.
\end{lemma}

\begin{proof}
  Direct consequence of
  Definition~\thref{d:holder-conjugates-in-r}, and
  since $\half+\half=1$.
\end{proof}

\begin{lemma}[Young's inequality for products in~$\matR$]
  \label{l:youngs-ineq-for-prod-in-r}
  \mbox{}\hfill
  Let~$p,q\in(1,\infty)$.
  Assume that~$p$ and~$q$ are H\"older conjugates.
  Let~$a,b\in\matRplus$.
  Then, we have $ab\leq\frac{a^p}{p}+\frac{b^q}{q}$.
\end{lemma}

\begin{proof}
  \proofpar{Case $ab=0$}
  Then, from
  \assume{nonnegativeness of exponentiation},
  \assume{closedness of multiplicative inverse in~$\matRplusstar$,
    and of the multiplication and addition in~$\matRplus$},
  The right-hand side is nonnegative.
  hence, the inequality holds.

  \proofparskip{Case $ab>0$}
  Then, from
  \assume{the zero-product property in~$\matRplus$ (contrapositive)},
  we have $a,b>0$.
  Thus, from
  \assume{algebraic properties of the natural logarithm function},
  Definition~\thref{d:holder-conjugates-in-r}, and
  \assume{concavity of the natural logarithm function (with
    $\frac{1}{p}+\frac{1}{q}=1$)},
  we have
  \begin{equation*}
    \ln (a b)
    = \ln a + \ln b
    = \frac{1}{p} \ln (a^p) + \frac{1}{q} \ln (b^q)
    \leq \ln \left( \frac{a^p}{p} + \frac{b^q}{q} \right).
  \end{equation*}
  Hence, from
  \assume{monotonicity of the exponential function}, and since
  \assume{the exponential and logarithm functions in~$\matR$ are each other
    inverse},
  we have $ab\leq\frac{a^p}{p}+\frac{b^q}{q}$.

  Therefore, the inequality always holds.
\end{proof}

\begin{lemma}[Young's inequality for products in~$\matR$, case $p=2$]
  \label{l:youngs-ineq-for-prod-in-r-case-p-two}
  \mbox{}\\
  Let~$a,b\in\matRplus$.
  Let~$\eps>0$.
  Then, we have $ab\leq\frac{a^2}{2\eps}+\frac{\eps b^2}{2}$.
\end{lemma}

\begin{proof}
  Direct consequence of
  Lemma~\thref{l:two-is-self-holder-conjugate-in-r},
  Lemma~\threfc{l:youngs-ineq-for-prod-in-r}{%
    with $p=q\eqdef2$ and
    $\frac{a}{\sqrt{\eps}},\sqrt{\eps}b\in\matRplus$}, and
  \assume{properties of square root and multiplicative inverse
    in~$\matRplusstar$, and of multiplication in~$\matRplus$}.
\end{proof}

\begin{remark}
  Note that a similar result also holds in the general case of H\"older
  conjugate numbers $p,q\in(1,\infty)$:
  for all $a,b\in\matRplus$, for all $\eps>0$, we have
  $ab\leq\frac{a^p}{p\eps^\frac{2}{q}}+\frac{\eps^\frac{2}{p}b^q}{q}$.
\end{remark}

\clearpage
\subsection{Extended real numbers}
\label{ss:ext-real-numbers}

\begin{definition}[extended real numbers, $\matRbar$]
  \label{d:ext-real-nums-rbar}
  \mbox{}\hfill
  The set of {\em extended real numbers} is
  $\matRbar\eqdef\matR\uplus\{-\infty,\infty\}$, and the order in~$\matR$ is
  extended to~$\matRbar$ with the following rule:
  \begin{equation}
    \label{e:total-order-on-rbar}
    1.\quad
    \forall a \in \matR,\quad
    -\infty < a < \infty.
  \end{equation}
\end{definition}

\begin{lemma}[order in~$\matRbar$ is total]
  \label{l:order-in-rbar-is-total}
  \mbox{}\hfill
  $(\matRbar,\leq)$ is a totally ordered set.
\end{lemma}

\begin{proof}
  Direct consequence of
  \assume{the definition of total order},
  Definition~\thref{d:ext-real-nums-rbar}, and
  \assume{totality of order in~$\matR$}.
\end{proof}

\begin{remark}
  The goal of this section is to clarify some properties of basic operations
  such as addition, multiplication, and exponentiation, when extended
  to~$\matRbar$.
\end{remark}

\begin{remark}
  Note that, as with regular real numbers, all results on operations on
  extended real numbers can be lifted into similar results on functions taking
  their values in~$\matRbar$.
\end{remark}

\begin{definition}[addition in~$\matRbar$]
  \label{d:add-in-rbar}
  \mbox{}\\
  Addition and subtraction in~$\matR$ are extended to~$\matRbar$ with the
  following rules:
  \begin{equation}
    \label{e:add-in-rbar}
    \left\{
      \begin{array}{l}
        1.\quad
        \forall a > -\infty,\quad
        a + \infty = \infty + a \eqdef \infty,\\
        2.\quad
        \forall a < \infty,\quad
        a + (-\infty) = -\infty + a \eqdef -\infty,\\
        3.\quad
        \infty + (-\infty)
        \mbox{ and } -\infty + \infty \mbox{ are undefined},\\
        4.\quad
        - (\pm \infty) = \mp \infty,\\
        5.\quad
        \forall a, b \in \matRbar,\quad
        a + (-b) \mbox{ defined}
        \IMPLIES
        a - b \eqdef a + (-b).
      \end{array}
    \right.
  \end{equation}
\end{definition}

\begin{lemma}[zero is identity element for addition in~$\matRbar$]
  \label{l:zero-is-identity-element-for-add-in-rbar}
  \mbox{}\\
  Let~$a\in\matRbar$.
  Then, we have $a+0=0+a=a$.
\end{lemma}

\begin{proof}
  Direct consequence of
  \assume{abelian group properties of~$(\matR,+)$}, and
  Definition~\threfc{d:add-in-rbar}{%
    new rules~1, and~2 are compatible with the property}.
\end{proof}

\begin{lemma}[addition in~$\matRbar$ is associative when defined]
  \label{l:add-in-rbar-is-assoc-when-defined}
  \mbox{}\\
  Let~$a,b,c\in\matRbar$.
  Then, $a+(b+c)$ and $(a+b)+c$ are either equal or both undefined.
\end{lemma}

\begin{proof}
  Direct consequence of
  \assume{associativity of addition in~$\matR$}, and
  Definition~\threfc{d:add-in-rbar}{%
    new rules~1, 2, and~3 are compatible with associativity}.
\end{proof}

\begin{lemma}[addition in~$\matRbar$ is commutative when defined]
  \label{l:add-in-rbar-is-comm-when-defined}
  \mbox{}\\
  Let~$a,b\in\matRbar$.
  Then, $a+b$ and $b+a$ are either equal or both undefined.
\end{lemma}

\begin{proof}
  Direct consequence of
  \assume{commutativity of addition in~$\matR$}, and
  Definition~\threfc{d:add-in-rbar}{%
    new rules~1, 2, and~3 are compatible with commutativity}.
\end{proof}

\begin{lemma}[infinity-sum property in~$\matRbar$]
  \label{l:infinity-sum-prop-in-rbar}
  \mbox{}\hfill
  Let~$a,b\in\matRbar$.
  Then, we have
  \begin{align}
    \label{e:infinity-sum-prop-in-rbar-1}
    a + b = \infty \EQUIV &
    (a = \infty \CONJ b > -\infty) \DISJ (a > -\infty \CONJ b = \infty)\\
    \label{e:infinity-sum-prop-in-rbar-2}
    a + b = -\infty \EQUIV &
    (a = -\infty \CONJ b < \infty) \DISJ (a < \infty \CONJ b = -\infty).
  \end{align}

  Moreover, if we assume that the sum $a+b$ is well-defined, then we have
  \begin{equation}
    \label{e:infinity-sum-prop-in-rbar-3}
    a + b = \pm\infty \EQUIV a = \pm\infty \DISJ b = \pm\infty.
  \end{equation}
\end{lemma}

\begin{proof}
  Direct consequence of
  Definition~\threfc{d:add-in-rbar}{rules~1, 2 and~3}.
\end{proof}

\begin{lemma}[additive inverse in~$\matRbar$ is monotone]
  \label{l:additive-inverse-in-rbar-is-monot}
  \mbox{}\\
  Let~$a,b\in\matRbar$.
  Then, we have $a\leq b$ iff $-b\leq-a$.
\end{lemma}

\begin{proof}
  Direct consequence of
  Definition~\thref{d:ext-real-nums-rbar},
  \assume{monotonicity of additive inverse in~$\matR$}, and
  Definition~\threfc{d:add-in-rbar}{rule~4}.
\end{proof}

\begin{definition}[multiplication in~$\matRbar$]
  \label{d:mult-in-rbar}
  \mbox{}\\
  Multiplication and division by nonzero in~$\matR$ are extended
  to~$\matRbar$ with the following rules:
  \begin{equation}
    \label{e:mult-in-rbar}
    \left\{
      \begin{array}{l}
        1.\quad
        \forall a > 0,\quad
        a \times (\pm \infty) = \pm \infty \times a \eqdef \pm \infty,\\
        2.\quad
        \forall a < 0,\quad
        a \times (\pm \infty) = \pm \infty \times a \eqdef \mp \infty,\\
        3.\quad
        0 \times (\pm \infty)
        \mbox{ and } \pm \infty \times 0 \mbox{ are undefined},\\
        4.\quad \displaystyle
        \frac{1}{0} \mbox{ is undefined, and }
        \frac{1}{\pm \infty} \eqdef 0,\\
        5.\quad \displaystyle
        \forall a, b \in \matRbar,\quad
        \frac{1}{b} \mbox{ and } a \times \frac{1}{b} \mbox{ defined}
        \IMPLIES
        \frac{a}{b} \eqdef a \times \frac{1}{b}.
      \end{array}
    \right.
  \end{equation}
\end{definition}

\begin{remark}
  The 5th rule in the previous definition implies
  that~$\frac{\infty}{\pm\infty}$, $\frac{-\infty}{\pm\infty}$,
  and~$\frac{a}{0}$ (for all $a\in\matRbar$) are undefined.
  Note that rule~3 is modified in the context of measure theory in
  Definition~\ref{d:mult-in-rbar-mt}.
\end{remark}

\begin{lemma}[multiplication in~$\matRbar$ is associative when defined]
  \label{l:mult-in-rbar-is-assoc-when-defined}
  \mbox{}\\
  Let~$a,b,c\in\matRbar$.
  Then, $a\times(b\times c)$ and $(a\times b)\times c$ are either equal or
  both undefined.
\end{lemma}

\begin{proof}
  Direct consequence of
  \assume{associativity of multiplication in~$\matR$}, and
  Definition~\threfc{d:mult-in-rbar}{%
    new rules~1, 2, and~3 are compatible with associativity}.
\end{proof}

\begin{lemma}[multiplication in~$\matRbar$ is commutative when defined]
  \label{l:mult-in-rbar-is-comm-when-defined}
  \mbox{}\\
  Let~$a,b\in\matRbar$.
  Then, $a\times b$ and $b\times a$ are either equal or both undefined.
\end{lemma}

\begin{proof}
  Direct consequence of
  \assume{commutativity of multiplication in~$\matR$}, and
  Definition~\threfc{d:mult-in-rbar}{%
    new rules~1, 2, and~3 are compatible with commutativity}.
\end{proof}

\begin{lemma}[multiplication in~$\matRbar$ is left distributive over addition
  when defined]
  \label{l:mult-in-rbar-is-left-distr-over-add-when-defined}
  \mbox{}\\
  Let~$a,b,c\in\matRbar$.
  Then, $a\times(b+c)$ and $(a\times b)+(a\times c)$ are either equal or both
  undefined.
\end{lemma}

\begin{proof}
  Direct consequence of
  \assume{left distributivity of multiplication over addition in~$\matR$},
  Definition~\thref{d:add-in-rbar}, and
  Definition~\thref{d:mult-in-rbar}
  (new rules are compatible with left distributivity of multiplication over
  addition).
\end{proof}

\begin{lemma}[multiplication in~$\matRbar$ is right distributive over addition
  when defined]
  \label{l:mult-in-rbar-is-right-distr-over-add-when-defined}
  \mbox{}\\
  Let~$a,b,c\in\matRbar$.
  Then, $(a+b)\times c$ and $(a\times c)+(b\times c)$ are either equal or both
  undefined.
\end{lemma}

\begin{proof}
  Direct consequence of
  Lemma~\threfc{l:mult-in-rbar-is-comm-when-defined}{%
    used twice}, and
  Lemma~\thref{l:mult-in-rbar-is-left-distr-over-add-when-defined}.
\end{proof}

\begin{lemma}[zero-product property in~$\matRbar$]
  \label{l:zero-prod-prop-in-rbar}
  \mbox{}\hfill
  Let~$a,b\in\matRbar$.
  Then, we have
  \begin{equation}
    \label{e:zero-prod-prop-in-rbar}
    a b = 0
    \EQUIV
    a b \mbox{ is defined}
    \CONJ
    (a = 0 \DISJ b = 0).
  \end{equation}
\end{lemma}

\begin{proof}
  Direct consequence of
  \assume{the zero-product property in~$\matR$}, and
  Definition~\threfc{d:mult-in-rbar}{rule~3}.
\end{proof}

\begin{lemma}[infinity-product property in~$\matRbar$]
  \label{l:infinity-prod-prop-in-rbar}
  \mbox{}\hfill
  Let~$a,b\in\matRbar$.
  Then, we have
  \begin{align}
    \label{e:infinity-prod-prop-in-rbar-1}
    a b = \infty \EQUIV
    & (a = \infty \CONJ b > 0) \DISJ (a = -\infty \CONJ b < 0) \quad\Disj\\
    \nonumber
    & (a > 0 \CONJ b = \infty) \DISJ (a < 0 \CONJ b = -\infty),\\
    \label{e:infinity-prod-prop-in-rbar-2}
    a b = -\infty \EQUIV
    & (a = -\infty \CONJ b > 0) \DISJ (a = \infty \CONJ b < 0) \quad\Disj\\
    \nonumber
    & (a > 0 \CONJ b = -\infty) \DISJ (a < 0 \CONJ b = \infty).
  \end{align}
\end{lemma}

\begin{proof}
  \proofpar{``Left'' implies ``right''}
  Assume that $ab=\pm\infty$.
  Then, from
  Definition~\threfc{d:mult-in-rbar}{rules~1 and~2},
  we have $a=\pm\infty$ or $b=\pm\infty$.
  Assume that the other operand is zero.
  Then, from
  Definition~\threfc{d:mult-in-rbar}{rule~3},
  the product~$ab$ is undefined.
  Which is impossible.

  Assume that $ab=\infty$.
  Then, from
  Definition~\threfc{d:mult-in-rbar}{rules~1 and~2},
  we have either $a=\infty$ and $b>0$, $a=-\infty$ and $b<0$, $b=\infty$ and
  $a>0$, or $b=-\infty$ and $a<0$.

  Similarly, assume now that $ab=-\infty$.
  Then, again from
  Definition~\threfc{d:mult-in-rbar}{rules~1 and~2},
  we have either $a=-\infty$ and $b>0$, $a=\infty$ and $b<0$, $b=-\infty$ and
  $a>0$, or $b=\infty$ and $a<0$.

  \proofparskip{``Right'' implies ``left''}
  Direct consequence of
  Definition~\threfc{d:mult-in-rbar}{rules~1 and~2}.

  \medskip\noindent
  Therefore, we have the equivalences.
\end{proof}

\begin{lemma}[finite-product property in~$\matRbar$]
  \label{l:finite-prod-prop-in-rbar}
  \mbox{}\hfill
  Let $a,b\in\matRbar$.
  Then, we have
  \begin{equation}
    \label{e:finite-prod-prop-in-rbar}
    a b \mbox{ is defined} \CONJ a b \in \matR \EQUIV a, b \in \matR.
  \end{equation}
\end{lemma}

\begin{proof}
  Direct consequence of
  Definition~\threfc{d:mult-in-rbar}{%
    new rules are not finite-product rules}.
\end{proof}

\begin{definition}[absolute value in~$\matRbar$]
  \label{d:abs-in-rbar}
  \mbox{}\\
  The absolute value in~$\matR$ is extended to~$\matRbar$ with the following
  rule:
  \begin{equation}
    | \pm \infty | = \infty.
  \end{equation}
\end{definition}

\begin{lemma}[equivalent definition of absolute value in~$\matRbar$]
  \label{l:equiv-def-of-abs-in-rbar}
  \mbox{}\\
  Let~$a\in\matRbar$.
  Then, we have $|a|=\max(-a,a)$.
\end{lemma}

\begin{proof}
  Direct consequence of
  \assume{the definition of absolute value in~$\matR$}, and
  Definition~\thref{d:abs-in-rbar}.
\end{proof}

\begin{lemma}[bounded absolute value in~$\matRbar$]
  \label{l:bounded-abs-in-rbar}
  \mbox{}\\
  Let~$a,b\in\matRbar$.
  Then, we have $|a|\leq b$ iff $-b\leq a\leq b$.
\end{lemma}

\begin{proof}
  Direct consequence of
  Lemma~\threfc{l:equiv-def-of-abs-in-rbar}{$-a\leq b$ and $a\leq b$}, and
  Lemma~\threfc{l:additive-inverse-in-rbar-is-monot}{$-b\leq a$}.
\end{proof}

\begin{lemma}[bounded absolute value in~$\matRbar$ (strict)]
  \label{l:bounded-abs-in-rbar-strict}
  \mbox{}\\
  Let~$a,b\in\matRbar$.
  Then, we have $|a|<b$ iff $-b<a<b$.
\end{lemma}

\begin{proof}
  Direct consequence of
  Lemma~\threfc{l:equiv-def-of-abs-in-rbar}{%
    $-a<b$ and $a<b$}, and
  Lemma~\threfc{l:additive-inverse-in-rbar-is-monot}{%
    contrapositive both ways, $-b<a$}.
\end{proof}

\begin{lemma}[finite absolute value in~$\matRbar$]
  \label{l:finite-abs-in-rbar}
  \mbox{}\hfill
  Let~$a\in\matRbar$.
  Then, we have $|a|$ finite iff $a$ finite.
\end{lemma}

\begin{proof}
  Direct consequence of
  Lemma~\threfc{l:bounded-abs-in-rbar-strict}{with $b\eqdef\infty$}.
\end{proof}

\begin{lemma}[absolute value in~$\matRbar$ is nonnegative]
  \label{l:abs-in-rbar-is-nonneg}
  \mbox{}\hfill
  The absolute value in~$\matRbar$ is nonnegative.
\end{lemma}

\begin{proof}
  Direct consequence of
  Definition~\thref{d:abs-in-rbar},
  \assume{nonnegativeness of the absolute value in~$\matR$}. and
  \assume{nonnegativeness of~$\infty$}.
\end{proof}

\begin{lemma}[absolute value in~$\matRbar$ is even]
  \label{l:abs-in-rbar-is-even}
  \mbox{}\hfill
  The absolute value in~$\matRbar$ is even.
\end{lemma}

\begin{proof}
  Direct consequence of
  \assume{evenness of the absolute value in~$\matR$}. and
  Definition~\threfc{d:abs-in-rbar}{%
    new rule is compatible with evenness},
\end{proof}

\begin{lemma}[absolute value in~$\matRbar$ is definite]
  \label{l:abs-in-rbar-is-definite}
  \mbox{}\hfill
  The absolute value in~$\matRbar$ is definite.
\end{lemma}

\begin{proof}
  Direct consequence of
  \assume{definiteness of the absolute value in~$\matR$}, and
  Definition~\threfc{d:abs-in-rbar}{%
    new rule is not a zero-absolute-value rule}.
\end{proof}

\begin{lemma}[absolute value in~$\matRbar$ satisfies triangle inequality]
  \label{l:abs-in-rbar-satisfies-triangle-ineq}
  \mbox{}\\
  Let~$a,b\in\matRbar$.
  Assume that~$a+b$ is well-defined.
  Then, we have $|a+b|\leq|a|+|b|$.
\end{lemma}

\begin{proof}
  Direct consequence of
  Definition~\thref{d:abs-in-rbar},
  \assume{the triangle inequality for the absolute value in~$\matR$},
  Definition~\threfc{d:add-in-rbar}{%
    $\infty$ is absorbing for addition in~$\matRbarplus$}, and
  Definition~\threfc{d:ext-real-nums-rbar}{%
    $\infty$ is the maximal element}.
\end{proof}

\begin{definition}[exponential and logarithm in~$\matRbar$]
  \label{d:exp-and-log-in-rbar}
  \mbox{}\\
  The exponential function in~$\matR$ and the natural logarithm function
  in~$\matRplusstar$ are respectively extended to~$\matRbar$
  and~$\matRbarplus$ with the following rules:
  \begin{equation}
    \exp (-\infty) = 0,\quad
    \exp \infty = \infty,\quad
    \ln 0 = -\infty,
    \AND
    \ln \infty = \infty.
  \end{equation}
\end{definition}

\begin{lemma}[exponential and logarithm in~$\matRbar$ are inverse]
  \label{l:exp-and-log-in-rbar-are-inverse}
  \mbox{}\\
  The exponential function in~$\matRbar$ and the natural logarithm function
  in~$\matRbarplus$ are each other inverse.
\end{lemma}

\begin{proof}
  Direct consequence of
  \assume{properties of the exponential and natural logarithm functions
    in~$\matR$}, and
  Definition~\threfc{d:exp-and-log-in-rbar}{%
    new rules are compatible with the property}.
\end{proof}

\begin{definition}[exponentiation in~$\matRbar$]
  \label{d:exp-in-rbar}
  \mbox{}\hfill
  Exponentiation in~$\matR$ (with either positive base and any exponent, or
  zero base and positive exponent) is extended to~$\matRbar$ with the
  following rule:
  \begin{equation}
    \label{e:exp-in-rbar-1}
    \forall a \in \matRbarplus,\;
    \forall b \in \matRbar,\quad
    b \ln a \mbox{ defined}
    \IMPLIES
    a^b \eqdef \exp (b \ln a).
  \end{equation}
\end{definition}

\begin{lemma}[exponentiation in~$\matRbar$]
  \label{l:exp-in-rbar}
  \mbox{}\\
  Exponentiation in~$\matRbar$ is the function
  $\exp:\ArRbpxRbRbp$ defined by
  \begin{equation}
    \label{e:exp-in-rbar-2}
    \left\{
      \begin{array}{l}
        1.\quad
        \forall a \in \matRplusstar,\;
        \forall b \in \matR,\quad
        a^b \eqdef \exp (b \ln a) \in \matRplusstar,\\
        2.\quad
        \forall a \in [0, 1),\;
        \forall b > 0,\quad
        a^\infty = 0^b \eqdef 0 \CONJ a^{-\infty} = \infty^b \eqdef \infty,\\
        3.\quad
        \forall a \in (1, \infty],\;
        \forall b < 0,\quad
        a^\infty = 0^b \eqdef \infty \CONJ a^{-\infty} = \infty^b \eqdef 0,\\
        4.\quad
        0^0, \infty^0, \mbox{ and } 1^{\pm\infty} \mbox{ are undefined}.
      \end{array}
    \right.
  \end{equation}
\end{lemma}

\begin{proof}
  Direct consequence of
  \assume{the definition of exponentiation in~$\matR$},
  Definition~\thref{d:exp-in-rbar},
  Definition~\thref{d:exp-and-log-in-rbar}, and
  Definition~\thref{d:mult-in-rbar}.
\end{proof}

\begin{lemma}[topology of~$\matRbar$]
  \label{l:topo-of-rbar}
  \mbox{}\\
  The totally ordered set~$(\matRbar,\leq)$ is equipped with the order
  topology~$\calT_{\matRbar}(\Ray_{\matRbar})$ generated by the open rays,
  for which the open intervals~$\Itvo_{\matRbar}$ constitute a topological
  basis.
\end{lemma}

\begin{proof}
  Direct consequence of
  Definition~\thref{d:ext-real-nums-rbar},
  Definition~\thref{d:order-topo}, and
  Lemma~\thref{l:topological-basis-of-order-topo}.
\end{proof}

\begin{remark}
  Note that~$\matRbar$ is actually metrizable since it is homeomorphic to a
  bounded segment, {\eg}~$[-\pi/2,\pi/2]$ or~$[-1,1]$.
  Indeed, metrics can be defined on~$\matRbar$, for instance by using the
  arctangent function, or the hyperbolic tangent function.
\end{remark}

\begin{lemma}[trace topology on~$\matR$]
  \label{l:trace-topo-on-r}
  \mbox{}\\
  Let~$\calTbar$ be the order topology on~$\matRbar$.
  Then, the trace topology~$\calTbar_\matR$ is equal to the order topology
  on~$\matR$.
\end{lemma}

\begin{proof}
  Direct consequence of
  Lemma~\thref{l:trace-topo-on-subset},
  Lemma~\threfc{l:topological-basis-of-order-topo}{%
    since $\matR\in\Itvo_\matR\subset\Itvo_{\matRbar}$}, and
  Definition~\thref{d:topological-basis}.
\end{proof}

\begin{remark}
  As a consequence, open subsets of~$\matR$ are also open subsets
  of~$\matRbar$.
\end{remark}

\begin{lemma}[convergence towards~$-\infty$]
  \label{l:conv-towards-minus-infinity}
  \mbox{}\\
  Let~$(x_n)_{n\in\matN}\in\matRbar$.
  Then, we have $\lim_{n\to\infty}x_n=-\infty$ iff
  \begin{equation}
    \label{e:conv-towards-minus-infinity}
    \forall k \in \matN,\;
    \exists N \in \matN,\;
    \forall n \in [N..\infty),\quad
    x_n \leq -k.
  \end{equation}
  If so, the sequence is said convergent towards~$-\infty$.
\end{lemma}

\begin{proof}
  Direct consequence of
  Lemma~\thref{l:topo-of-rbar},
  \assume{the definition of the limit using neighborhoods},
  \assume{the Archimedean property of~$\matRbar$}, and
  \assume{totally ordered set properties of~$\matRbar$}.
\end{proof}

\begin{lemma}[continuity of addition in~$\matRbar$]
  \label{l:continuity-of-add-in-rbar}
  \mbox{}\hfill
  Addition in~$\matRbar$ is continuous when defined.
\end{lemma}

\begin{proof}
  Direct consequence of
  \assume{continuity of addition in~$\matR$}.
  \assume{unboundedness of addition when an operand tends
    towards~$\pm\infty$}, and
  Definition~\threfc{d:add-in-rbar}{rules~1 and~2}.
\end{proof}

\begin{lemma}[continuity of multiplication in~$\matRbar$]
  \label{l:continuity-of-mult-in-rbar}
  \mbox{}\\
  Multiplication in~$\matRbar$ is continuous when defined.
\end{lemma}

\begin{proof}
  Direct consequence of
  \assume{continuity of multiplication in~$\matR$}.
  \assume{unboundedness of multiplication when an operand tends
    towards~$\pm\infty$ and the other is nonzero}, and
  Definition~\threfc{d:mult-in-rbar}{rules~1 and~2}.
\end{proof}

\begin{lemma}[absolute value in~$\matRbar$ is continuous]
  \label{l:abs-in-rbar-is-cont}
  \mbox{}\hfill
  The absolute value in~$\matRbar$ is continuous.
\end{lemma}

\begin{proof}
  Direct consequence of
  \assume{continuity of the absolute value in~$\matR$}, and
  Definition~\threfc{d:abs-in-rbar}{
    absolute value is closed in $\{\pm\infty\}$}.
\end{proof}

\subsubsection{Nonnegative extended real numbers}
\label{sss:nonnegative-ext-real-numbers}

\begin{lemma}[addition in~$\matRbarplus$ is closed]
  \label{l:add-in-rbarplus-is-closed}
  \mbox{}\hfill
  Addition in~$\matRbarplus$ is closed.
\end{lemma}

\begin{proof}
  Direct consequence of
  \assume{closedness of addition in~$\matRplus$}, and
  Definition~\threfc{d:add-in-rbar}{%
    undefined forms of rule~3 cannot occur}.
\end{proof}

\begin{lemma}[addition in~$\matRbarplus$ is associative]
  \label{l:add-in-rbarplus-is-assoc}
  \mbox{}\hfill
  Addition in~$\matRbarplus$ is associative.
\end{lemma}

\begin{proof}
  Direct consequence of
  Lemma~\thref{l:add-in-rbar-is-assoc-when-defined}, and
  Definition~\threfc{d:add-in-rbar}{%
    undefined forms of rule~3 cannot occur}.
\end{proof}

\begin{lemma}[addition in~$\matRbarplus$ is commutative]
  \label{l:add-in-rbarplus-is-comm}
  \mbox{}\hfill
  Addition in~$\matRbarplus$ is commutative.
\end{lemma}

\begin{proof}
  Direct consequence of
  Lemma~\thref{l:add-in-rbar-is-comm-when-defined}, and
  Definition~\threfc{d:add-in-rbar}{%
    undefined forms of rule~3 cannot occur}.
\end{proof}

\begin{lemma}[infinity-sum property in~$\matRbarplus$]
  \label{l:infinity-sum-prop-in-rbarplus}
  \mbox{}\hfill
  Let~$a,b\in\matRbarplus$.
  Then, we have
  \begin{equation}
    \label{e:infinity-sum-prop-in-rbarplus}
    a + b = \infty \EQUIV a = \infty \DISJ b = \infty.
  \end{equation}
\end{lemma}

\begin{proof}
  Direct consequence of
  Lemma~\thref{l:infinity-sum-prop-in-rbar}.
\end{proof}

\begin{lemma}[series are convergent in~$\matRbarplus$]
  \label{l:series-are-conv-in-rbarplus}
  \mbox{}\\
  Let~$(a_n)_{n\in\matN}\in\matRbarplus$.
  Then, we have $\sum_{n\in\matN}a_n\in\matRbarplus$.
\end{lemma}

\begin{proof}
  Direct consequence of
  \assume{completeness of~$\matRbarplus$}.
\end{proof}

\begin{lemma}[technical upper bound in series in~$\matRbarplus$]
  \label{l:technical-upper-bound-in-series-in-rbarplus}
  \mbox{}\hfill
  Let~$(a_p)_{p\in\matN}\in\matRbarplus$.\\
  Let~$\fhi:\ArNN$.
  Assume that~$\fhi$ is injective.
  Then, we have $\sum_{j\in\matN}a_{\fhi(j)}\leq\sum_{p\in\matN}a_p$.
\end{lemma}

\begin{proof}
  Let
  \begin{equation*}
    A = \sum_{j \in \matN} a_{\fhi (j)}
    \eqdef \lim_{i \to \infty} \sum_{j \in [0..i]} a_{\fhi (j)}
    \AND
    B = \sum_{p \in \matN} a_p
    \eqdef \lim_{n \to \infty} \sum_{p \in [0..n]} a_p.
  \end{equation*}
  Then, from
  Lemma~\thref{l:series-are-conv-in-rbarplus},
  we have $A,B\in\matRbarplus$.
  Let~$i\in\matN$.
  Let
  \begin{equation*}
    n \eqdef \max \{ \fhi (j) \st j \in [0..i] \}.
  \end{equation*}
  Then, from
  \assume{the definition of the maximum},
  we have $\fhi([0..i])\subset[0..n]\subset\matN$.
  Thus, from
  \assume{totally ordered set properties of~$\matRbarplus$},
  we have
  \begin{equation*}
    \sum_{j \in [0..i]} a_{\fhi (j)} \leq \sum_{p \in [0..n]} a_p \leq B.
  \end{equation*}
  Hence, from
  \assume{monotonicity of the limit (when $n\to\infty$)},
  we have $A\leq B$.

  Therefore, we have $\sum_{j\in\matN}a_{\fhi(j)}\leq\sum_{p\in\matN}a_p$.
\end{proof}

\begin{lemma}[order is meaningless in series in~$\matRbarplus$]
  \label{l:order-is-meaningless-in-series-in-rbarplus}
  \mbox{}\\
  Let~$(a_p)_{p\in\matN}\in\matRbarplus$.
  Let~$\fhi:\ArNN$.
  Assume that~$\fhi$ is bijective.
  Then, $\sum_{p\in\matN}a_p=\sum_{j\in\matN}a_{\fhi(j)}$.
\end{lemma}

\begin{proof}
  Direct consequence of
  Lemma~\threfc{l:technical-upper-bound-in-series-in-rbarplus}{%
    first used with the sequence~$(a_p)_{n\in\matN}$ and the function~$\fhi$,
    and then with the sequence
    $(b_j)_{j\in\matN}\eqdef(a_{\fhi(j)})_{j\in\matN}$ and the
    function~$\fhi^{-1}$ which satisfies $\fhi\circ\fhi^{-1}=\idmatN$}.
\end{proof}

\begin{lemma}[definition of double series in~$\matRbarplus$]
  \label{l:def-of-double-series-in-rbarplus}
  \mbox{}\\
  Let~$(a_{p,q})_{p,q\in\matN}\in\matRbarplus$.
  Let~$\fhi,\psi:\ArNNxN$.
  Assume that~$\fhi$ and~$\psi$ are bijections.\\
  Then, we have $\sum_{j\in\matN}a_{\fhi(j)}=\sum_{j\in\matN}a_{\psi(j)}$.
  This sum is denoted $\sum_{p,q\in\matN}a_{p,q}$.
\end{lemma}

\begin{proof}
  Direct consequence of
  Lemma~\threfc{l:order-is-meaningless-in-series-in-rbarplus}{%
    with $(a_{\fhi(j)})_{j\in\matN}$ and $\fhi^{-1}\circ\psi$}.
\end{proof}

\begin{lemma}[double series in~$\matRbarplus$]
  \label{l:double-series-in-rbarplus}
  \mbox{}\\
  Let~$(a_{p,q})_{p,q\in\matN}\in\matRbarplus$.
  Then, we have
  $\sum_{p,q\in\matN}a_{p,q}=
  \sum_{p\in\matN}\left(\sum_{q\in\matN}a_{p,q}\right)$.
\end{lemma}

\begin{proof}
  From
  \assume{countability of~$\matN^2$},
  let~$\fhi:\ArNNxN$ be a bijection.
  Then, from
  Lemma~\thref{l:def-of-double-series-in-rbarplus},
  let $A\eqdef\sum_{p,q\in\matN}=\sum_{j\in\matN}a_{\fhi(j)}$ (the sum does
  not depend on the choice for~$\fhi$).
  Let~$B\eqdef\sum_{p\in\matN}\left(\sum_{q\in\matN}a_{p,q}\right)$.
  Then, from
  Lemma~\thref{l:series-are-conv-in-rbarplus},
  we have $A,B\in\matRbarplus$.

  Let~$i\in\matN$.
  Let~$n\eqdef\max\{\pi_1(\fhi(j))\st j\in[0..i]\}$ and
  $m\eqdef\max\{\pi_2(\fhi(j))\st j\in[0..i]\}$ where the functions~$\pi_1$
  and~$\pi_2$ are defined by $\forall p,q\in\matN$, $\pi_1(p,q)=p$ and
  $\pi_2(p,q)=q$.
  Then, from
  \assume{the definition of the maximum},
  we have $\fhi([0..i])\subset[0..n]\times[0..m]\subset\matN^2$.
  Thus, from
  \assume{totally ordered set properties of~$\matRbarplus$},
  we have
  \begin{equation*}
    \sum_{j \in [0..i]} a_{\fhi (j)} \leq
    \sum_{p \in [0..n]} \left( \sum_{p \in [0..m]} a_{p, q} \right) \leq B.
  \end{equation*}
  Hence, from
  \assume{monotonicity of the limit (when $i\to\infty$)},
  we have $A\leq B$.

  Let~$n,m\in\matN$.
  Let~$i\eqdef\max\{\fhi^{-1}(p,q)\st(p,q)\in[0..n]\times[0..m]\}$.
  Then, from
  \assume{the definition of the maximum},
  we have $[0..n]\times[0..m]\subset\fhi([0..i])=\matN^2$.
  Thus, from
  \assume{totally ordered set properties of~$\matRbarplus$},
  we have
  \begin{equation*}
    \sum_{p \in [0..n]} \left( \sum_{q \in [0..m]} a_{p, q} \right) \leq
    \sum_{j \in [0..i]} a_{\fhi (j)} \leq A.
  \end{equation*}
  Hence, from
  \assume{monotonicity of the limit (when $m\to\infty$, then
    $n\to\infty$)},
  we have $B\leq A$.

  Therefore, we have
  $\sum_{p,q\in\matN}=\sum_{p\in\matN}\left(\sum_{q\in\matN}a_{p,q}\right)$.
\end{proof}

\begin{definition}[multiplication in~$\matRbarplus$]
  \label{d:mult-in-rbarplus}
  \mbox{}\\
  Multiplication and division in~$\matRbar$ are restricted to~$\matRbarplus$
  by replacing the fourth rule in~\eqref{e:mult-in-rbar} by
  \begin{equation}
    \label{e:mult-in-rbarplus}
    4^\prime.\quad
    \frac{1}{0} \eqdef \infty, \mbox{ and }
    \frac{1}{\infty} \eqdef 0.
  \end{equation}
\end{definition}

\begin{remark}
  When restricting to nonnegative extended numbers, making multiplicative
  inverse a bijection from~$\matRbarplus$ onto itself through
  Definition~\ref{d:mult-in-rbarplus} implies that for all $a>0$,
  $\frac{a}{0}\eqdef\infty$.
  The expressions $0\times\infty$, $\frac{\infty}{\infty}$ and~$\frac{0}{0}$
  remain undefined.
\end{remark}

\begin{lemma}[multiplication in~$\matRbarplus$ is closed when defined]
  \label{l:mult-in-rbarplus-is-closed-when-defined}
  \mbox{}\\
  When they are defined, multiplication and division in~$\matRbarplus$ are
  closed.
\end{lemma}

\begin{proof}
  Direct consequence of
  \assume{closedness of multiplication and division by nonzero
    in~$\matRplus$},
  Definition~\thref{d:mult-in-rbar}, and
  Definition~\thref{d:mult-in-rbarplus}.
\end{proof}

\begin{lemma}[zero-product property in~$\matRbarplus$]
  \label{l:zero-prod-prop-in-rbarplus}
  \mbox{}\hfill
  Let~$a,b\in\matRbarplus$.
  Then, we have
  \begin{equation}
    \label{e:zero-prod-prop-in-rbarplus}
    a b = 0
    \EQUIV
    a b \mbox{ is defined}
    \CONJ
    (a = 0 \DISJ b = 0).
  \end{equation}
\end{lemma}

\begin{proof}
  Direct consequence of
  Lemma~\thref{l:zero-prod-prop-in-rbar}, and
  Definition~\threfc{d:mult-in-rbarplus}{%
    new rule is not a zero-product rule}.
\end{proof}

\begin{lemma}[infinity-product property in~$\matRbarplus$]
  \label{l:infinity-prod-prop-in-rbarplus}
  \mbox{}\hfill
  Let~$a,b\in\matRbarplus$.
  Then, we have
  \begin{equation}
    \label{e:infinity-prod-prop-in-rbarplus}
    a b = \infty
    \EQUIV (a = \infty \CONJ b \not= 0)
    \DISJ (a \not= 0 \CONJ b = \infty).
  \end{equation}
\end{lemma}

\begin{proof}
  Direct consequence of
  Lemma~\thref{l:infinity-prod-prop-in-rbar}, and
  Definition~\threfc{d:mult-in-rbarplus}{%
    new rule is not an infinity-product rule}.
\end{proof}

\begin{lemma}[finite-product property in~$\matRbarplus$]
  \label{l:finite-prod-prop-in-rbarplus}
  \mbox{}\hfill
  Let~$a,b\in\matRbarplus$.
  Then, we have
  \begin{equation}
    \label{e:finite-prod-prop-in-rbarplus}
    a b \mbox{ is defined } \CONJ a b \in \matRplus
    \EQUIV a, b \in \matRplus.
  \end{equation}
\end{lemma}

\begin{proof}
  Direct consequence of
  Lemma~\thref{l:finite-prod-prop-in-rbar},
  \assume{closedness of multiplication in~$\matRplus$}, and
  Definition~\threfc{d:mult-in-rbarplus}{%
    new rule is not a finite-product rule}.
\end{proof}

\subsubsection{In the context of measure theory}
\label{sss:in-the-context-of-measure-theory}

\begin{definition}[multiplication in~$\matRbar$ (measure theory)]
  \label{d:mult-in-rbar-mt}
  \mbox{}\\
  In the context of measure theory (and probability), the third rule
  in~\eqref{e:mult-in-rbar} is replaced by
  \begin{equation}
    \label{e:mult-in-rbar-mt}
    3^\prime.\quad 0 \times (\pm \infty) = \pm \infty \times 0 \eqdef 0.
  \end{equation}
\end{definition}

\begin{remark}
  In the context of measure theory, making~$0$ an absorbing element for
  multiplication in~$\matRbar$ through the previous definition implies that
  $\frac{\infty}{\pm\infty}=\frac{-\infty}{\pm\infty}\eqdef0$.
  The expression~$\frac{a}{0}$ (for all $a\in\matRbar$) remains undefined.
  Note that multiplication in~$\matRbar$ is always well-defined in this
  context.
\end{remark}

\begin{lemma}[zero-product property in~$\matRbar$ (measure theory)]
  \label{l:zero-prod-prop-in-rbar-mt}
  \mbox{}\\
  In the context of measure theory, let $a,b\in\matRbar$.
  Then, we have
  \begin{equation}
    \label{e:zero-prod-prop-in-rbar-mt}
    a b = 0 \EQUIV a = 0 \DISJ b = 0.
  \end{equation}
\end{lemma}

\begin{proof}
  Direct consequence of
  Lemma~\thref{l:zero-prod-prop-in-rbar}, and
  Definition~\threfc{d:mult-in-rbar-mt}{%
    new rule~3$^\prime$ is compatible with the property}.
\end{proof}

\begin{lemma}[infinity-product property in~$\matRbar$ (measure theory)]
  \label{l:infinity-prod-prop-in-rbar-mt}
  \mbox{}\\
  In the context of measure theory, let $a,b\in\matRbar$.
  Then, we have
  \begin{align}
    \nonumber
    a b = \infty \EQUIV
    & (a = \infty \CONJ b > 0) \DISJ (a = -\infty \CONJ b < 0) \quad\Disj\\
    \label{e:infinity-prod-prop-in-rbar-mt-1}
    & (a > 0 \CONJ b = \infty) \DISJ (a < 0 \CONJ b = -\infty),\\
    \nonumber
    a b = -\infty \EQUIV
    & (a = -\infty \CONJ b > 0) \DISJ (a = \infty \CONJ b < 0) \quad\Disj\\
    \label{e:infinity-prod-prop-in-rbar-mt-2}
    & (a > 0 \CONJ b = -\infty) \DISJ (a < 0 \CONJ b = \infty).
  \end{align}
\end{lemma}

\begin{proof}
  Direct consequence of
  Lemma~\thref{l:infinity-prod-prop-in-rbar}, and
  Definition~\threfc{d:mult-in-rbar-mt}{%
    new rule is not an infinity-product rule}.
\end{proof}

\begin{lemma}[finite-product property in~$\matRbar$ (measure theory)]
  \label{l:finite-prod-prop-in-rbar-mt}
  \mbox{}\\
  In the context of measure theory, let $a,b\in\matRbar$.
  Then, we have
  \begin{equation}
    \label{e:finite-prod-prop-in-rbar-mt}
    a b \in \matR
    \EQUIV a, b \in \matR
    \DISJ a = 0
    \DISJ b = 0.
  \end{equation}
\end{lemma}

\begin{proof}
  Direct consequence of
  Lemma~\threfc{l:infinity-prod-prop-in-rbar-mt}{contra\-positive},
  \assume{De~Morgan's laws}, and
  \assume{distributivity of logical conjunction over logical disjunction}.
\end{proof}

\begin{lemma}[multiplication in~$\matRbarplus$ is closed (measure theory)]
  \label{l:mult-in-rbarplus-is-closed-mt}
  \mbox{}\\
  In the context of measure theory, multiplication and division are total
  functions $\ArRbpxRbpRbp$.
  In particular, we have $\frac{\infty}{\infty}=\frac{0}{0}\eqdef0$, and for
  all $a>0$, $\frac{a}{0}\eqdef\infty$.
\end{lemma}

\begin{proof}
  Direct consequence of
  Definition~\thref{d:mult-in-rbar},
  Definition~\thref{d:mult-in-rbarplus}, and
  Definition~\thref{d:mult-in-rbar-mt}.
\end{proof}

\begin{remark}
  Of course, in the context of measure theory, multiplication and division
  are no longer continuous on the whole boundary of
  $\matRbarplus\times\matRbarplus$.
\end{remark}

\begin{lemma}[multiplication in~$\matRbarplus$ is associative (measure theory)]
  \label{l:mult-in-rbarplus-is-assoc-mt}
  \mbox{}\\
  In the context of measure theory, multiplication in~$\matRbarplus$ is
  associative.
\end{lemma}

\begin{proof}
  Direct consequence of
  Lemma~\thref{l:mult-in-rbar-is-assoc-when-defined}, and
  Definition~\threfc{d:mult-in-rbar-mt}{%
    no longer undefined forms}.
\end{proof}

\begin{lemma}[multiplication in~$\matRbarplus$ is commutative (measure
  theory)]
  \label{l:mult-in-rbarplus-is-comm-mt}
  \mbox{}\\
  In the context of measure theory, multiplication in~$\matRbarplus$ is
  commutative.
\end{lemma}

\begin{proof}
  Direct consequence of
  Lemma~\thref{l:mult-in-rbar-is-comm-when-defined}, and
  Definition~\threfc{d:mult-in-rbar-mt}{%
    no longer undefined forms}.
\end{proof}

\begin{lemma}[multiplication in~$\matRbarplus$ is distributive over addition
  (measure theory)]
  \label{l:mult-in-rbarplus-is-distr-over-add-mt}
  \mbox{}\\
  In the context of measure theory, multiplication is (left and right)
  distributive over addition in~$\matRbarplus$.
\end{lemma}

\begin{proof}
  Direct consequence of
  Lemma~\thref{l:mult-in-rbar-is-left-distr-over-add-when-defined},
  Lemma~\thref{l:mult-in-rbar-is-right-distr-over-add-when-defined}, and
  Definition~\threfc{d:mult-in-rbar-mt}{%
    no longer undefined forms}.
\end{proof}

\begin{lemma}[zero-product property in~$\matRbarplus$ (measure theory)]
  \label{l:zero-prod-prop-in-rbarplus-mt}
  \mbox{}\\
  In the context of measure theory, let $a,b\in\matRbarplus$.
  Then, we have
  \begin{equation}
    \label{e:zero-prod-prop-in-rbarplus-mt}
    a b = 0 \EQUIV a = 0 \DISJ b = 0.
  \end{equation}
\end{lemma}

\begin{proof}
  Direct consequence of
  Lemma~\thref{l:zero-prod-prop-in-rbarplus}, and
  Definition~\threfc{d:mult-in-rbar-mt}{%
    new rule~3$^\prime$ is compatible with the property}.
\end{proof}

\begin{lemma}[infinity-product property in~$\matRbarplus$ (measure theory)]
  \label{l:infinity-prod-prop-in-rbarplus-mt}
  \mbox{}\\
  In the context of measure theory, let $a,b\in\matRbarplus$.
  Then, we have
  \begin{equation}
    \label{e:infinity-prod-prop-in-rbarplus-mt}
    a b = \infty
    \EQUIV (a = \infty \CONJ b > 0)
    \DISJ (b = \infty \CONJ a > 0).
  \end{equation}
\end{lemma}

\begin{proof}
  Direct consequence of
  Lemma~\thref{l:infinity-prod-prop-in-rbarplus}, and
  Definition~\threfc{d:mult-in-rbar-mt}{%
    new rule is not an infinity-product rule}.
\end{proof}

\begin{lemma}[finite-product property in~$\matRbarplus$ (measure theory)]
  \label{l:finite-prod-prop-in-rbarplus-mt}
  \mbox{}\\
  In the context of measure theory, let $a,b\in\matRbarplus$.
  Then, we have
  \begin{equation}
    \label{e:finite-prod-prop-in-rbarplus-mt}
    a b \in \matRplus
    \EQUIV a, b \in \matRplus
    \DISJ a = 0
    \DISJ b = 0.
  \end{equation}
\end{lemma}

\begin{proof}
  Direct consequence of
  Lemma~\thref{l:finite-prod-prop-in-rbar-mt}, and
  \assume{closedness of multiplication in~$\matRplus$}.
\end{proof}

\begin{lemma}[exponentiation in~$\matRbar$ (measure theory)]
  \label{l:exp-in-rbar-mt}
  \mbox{}\\
  In the context of measure theory, the fourth rule in~\eqref{e:exp-in-rbar-2}
  is replaced by
  \begin{equation}
    \label{e:exp-in-rbar-mt-1}
    4^\prime.\quad 0^0 = \infty^0 = 1^{\pm \infty} \eqdef 1.
  \end{equation}
  Thus, we have
  \begin{equation}
    \label{e:exp-in-rbar-mt-2}
    \forall a \in \matRbarplus,\;
    a^0 = 1
    \AND
    \forall b \in \matRbar,\;
    1^b = 1.
  \end{equation}
\end{lemma}

\begin{proof}
  Direct consequence of
  Definition~\thref{d:exp-in-rbar},
  Definition~\thref{d:mult-in-rbar-mt}, and
  Lemma~\threfc{l:exp-in-rbar}{%
    new rule only affects rule~4}.
\end{proof}

\begin{definition}[H\"older conjugates]
  \label{d:holder-conjugates}
  \mbox{}\\
  Extended numbers $p,q\in[1,\infty]$ are said {\em H\"older conjugates} iff
  $\frac{1}{p}+\frac{1}{q}=1$.
\end{definition}

\begin{remark}
  The previous definition implies that extended numbers~1 and~$\infty$ are
  H\"older conjugate, since from rule~4$^\prime$ of
  Definition~\ref{d:mult-in-rbarplus}, we have $\frac{1}{\infty}=0$.
\end{remark}

\begin{lemma}[Young's inequality for products (measure theory)]
  \label{l:youngs-ineq-for-prod-mt}
  \mbox{}\\
  In the context of measure theory, let $p,q\in(1,\infty)$.
  Assume that~$p$ and~$q$ are H\"older conjugates.
  Let~$a,b\in\matRbarplus$.
  Then, we have $ab\leq\frac{a^p}{p}+\frac{b^q}{q}$.
\end{lemma}

\begin{proof}
  From
  Lemma~\thref{l:mult-in-rbarplus-is-closed-mt},
  Lemma~\threfc{l:exp-in-rbar}{%
    with $p,q\not=0,\pm\infty$},
  \assume{closedness of multiplicative inverse in~$\matRplusstar$},
  Lemma~\thref{l:mult-in-rbarplus-is-closed-when-defined},
  Definition~\threfc{d:mult-in-rbar}{%
    rule~5 with $b=p,q\in\matRplusstar$}, and
  Lemma~\thref{l:add-in-rbarplus-is-closed},
  $ab$, $a^p$, $b^q$, $\frac{a^p}{p}$, $\frac{b^q}{q}$, and
  $\frac{a^p}{p}+\frac{b^q}{q}$ are well-defined, and belong
  to~$\matRbarplus$.

  \proofparskip{Case $ab=\infty$}
  Then, from
  Lemma~\threfc{l:infinity-prod-prop-in-rbarplus-mt}{%
    $a=\infty$ or $b=\infty$},
  Lemma~\threfc{l:exp-in-rbar}{%
    rule~2 with $b=p,q>0$},
  Definition~\threfc{d:mult-in-rbar}{%
    rule~1 with $a\eqdef\frac{1}{p},\frac{1}{q}>0$}, and
  Definition~\threfc{d:add-in-rbar}{%
    rule~1, the other operand is not~$-\infty$},
  the right-hand side equals~$\infty$.
  Thus, the (in)equality holds.

  \proofparskip{Case $ab=0$}
  Then, the inequality holds since the right-hand side is nonnegative.

  \proofparskip{Case $ab\in\matRplusstar$}
  Direct consequence of
  Lemma~\threfc{l:zero-prod-prop-in-rbarplus-mt}{%
    contrapositive},
  Lemma~\threfc{l:infinity-prod-prop-in-rbarplus}{%
    contrapositive}, and
  Lemma~\thref{l:youngs-ineq-for-prod-in-r}.

  Therefore, the inequality always holds.
\end{proof}

\begin{lemma}[Young's inequality for products, case $p=2$ (measure theory)]
  \label{l:youngs-ineq-for-prod-case-p-two-mt}
  \mbox{}\\
  In the context of measure theory, let $a,b\in\matRbarplus$.
  Let~$\eps\in\matRplusstar$.
  Then, we have $ab\leq\frac{a^2}{2\eps}+\frac{\eps b^2}{2}$.
\end{lemma}

\begin{proof}
  Direct consequence of
  Lemma~\thref{l:two-is-self-holder-conjugate-in-r},
  Lemma~\threfc{l:youngs-ineq-for-prod-mt}{%
    with $p=q\eqdef2$ and
    $\frac{a}{\sqrt{\eps}},\sqrt{\eps}b\in\matRbarplus$},
  \assume{properties of square root and multiplicative inverse
    in~$\matRplusstar$},
  Lemma~\thref{l:mult-in-rbarplus-is-closed-mt},
  Lemma~\thref{l:mult-in-rbarplus-is-assoc-mt}, and
  Lemma~\thref{l:mult-in-rbarplus-is-comm-mt}.
\end{proof}

\clearpage
\subsection{Second countability and real numbers}
\label{ss:second-countability-and-real-numbers}

\begin{definition}[connected component in~$\matR$]
  \label{d:conn-comp-in-r}
  \mbox{}\\
  Let~$A\subset\matR$.
  Let~$x\in A$.
  The {\em connected component of~$A$ containing~$x$} is the union of all
  open intervals~$I$ containing~$x$ and included in~$A$;
  it is denoted~$I^A_x$.
\end{definition}

\begin{lemma}[connected component of open subset of~$\matR$ is open interval]
  \label{l:conn-comp-of-open-subset-of-r-is-open-int}
  \mbox{}\\
  Let~$O\subset\matR$ be open.
  Let~$x\in O$.
  Then, $I^O_x$ is an open interval contained in~$O$.
\end{lemma}

\begin{proof}
  Note that from
  Definition~\thref{LM-d:open-subset},
  there exists~$\eps>0$ such that $(x-\eps,x+\eps)\subset O$.
  Then, from
  Definition~\thref{d:conn-comp-in-r},
  $(x-\eps,x+\eps)\subset I^O_x$.
  Thus, $I^O_x$ is nonempty, contains~$x$.
  Moreover, from
  Definition~\threfc{d:topological-space}{closedness under union},
  $I^O_x$ is open.

  Let~$a,b,c\in\matR$ such that $a<c<b$ and $a,b\in I^O_x$.
  Then, from
  Definition~\thref{d:conn-comp-in-r}, and
  \assume{the definition of union},
  there exists intervals~$I_a(x)$ and~$I_b(x)$ such that
  $x,a\in I_a(x)\subset I^O_x\subset O$ and
  $x,b\in I_b(x)\subset I^O_x\subset O$.
  \proofpar{Case $c=x$}
  Then, we have $c=x\in I^O_x$.
  \proofpar{Case $c<x$}
  Then, we have $a<c<x$.
  Thus, since $a,x\in I_a(x)$, and $I_a(x)$~is an interval,
  we have successively $c\in(a,x)\subset I_a(x)\subset I^O_x$.
  \proofpar{Case $x<c$}
  Then, we have $x<c<b$.
  Thus, since $x,b\in I_b(x)$, and $I_b(x)$~is an interval,
  we have successively $c\in(x,b)\subset I_b(x)\subset I^O_x$.
  Hence, we always have $c\in I^O_x$.
  Thus, from
  Definition~\threfc{d:interval}{with $X\eqdef\matR$},
  $I^O_x$ is an interval.

  Therefore, $I^O_x$ is an open interval contained in~$O$.
\end{proof}

\begin{lemma}[connected component of open subset of~$\matR$ is maximal]
  \label{l:conn-comp-of-open-subset-of-r-is-maximal}
  \mbox{}\\
  Let~$O\subset\matR$ be open.
  Let~$x,y\in O$.
  Assume that $y\in I^O_x$.
  Then, we have $I^O_y=I^O_x$.
\end{lemma}

\begin{proof}
  From
  Definition~\threfc{d:conn-comp-in-r}{union},
  there exists an open interval~$I$ such that $y\in I$ and $x\in I\subset O$.
  Thus, we also have $x\in I^O_y$.
  But, from
  Lemma~\thref{l:conn-comp-of-open-subset-of-r-is-open-int},
  $I^O_x$ is an open interval such that $y\in I^O_x\subset O$, and $I^O_y$ is
  an open interval such that $x\in I^O_y\subset O$.
  Thus, from
  Definition~\thref{d:conn-comp-in-r},
  we have $I^O_x\subset I^O_y$ and $I^O_y\subset I^O_x$.
  Therefore, we have $I^O_x=I^O_y$.
\end{proof}

\begin{lemma}[connected components of open subset of~$\matR$ are equal or
  disjoint]
  \label{l:conn-comps-of-open-subset-of-r-equal-or-disj}
  \mbox{}\\
  Let~$O\subset\matR$ be open.
  Let~$x,y\in O$.
  Then, we have either $I^O_x=I^O_y$, or $I^O_x\cap I^O_y=\emptyset$.
\end{lemma}

\begin{proof}
  Assume that $I^O_x\cap I^O_y\not=\emptyset$.
  Then, there exists $z\in I^O_x\cap I^O_y$.
  Thus, from
  Lemma~\thref{l:conn-comp-of-open-subset-of-r-is-maximal},
  we have $I^O_z=I^O_x$ and $I^O_z=I^O_y$.
  Therefore, $I^O_x$ and~$I^O_y$ are equal or disjoint.
\end{proof}

\begin{theorem}[countable connected components of open subsets of~$\matR$]
  \label{t:count-conn-comps-of-open-subsets-of-r}
  \mbox{}\\
  Let~$O\subset\matR$ be open.
  Then, $O$~is a countable union of disjoint open intervals.
\end{theorem}

\begin{proof}
  Let~$x\in O$.
  Then, from
  Definition~\thref{LM-d:open-subset},
  there exists $\eps>0$ such that the open interval $(x-\eps,x+\eps)$ is
  included in~$O$.
  From
  \assume{density of rational numbers in~$\matR$},
  there exists $q\in\matQ\cap(x-\eps,x+\eps)$.
  Then, by construction, we have $x\in(x-\eps,x+\eps)\subset I^O_q$.
  Thus, we have the inclusion $O\subset\bigcup_{q\in\matQ\cap O}I^O_q$.

  Conversely, let $q\in\matQ\cap O$.
  Then, from
  Definition~\thref{d:conn-comp-in-r},
  we have $I^O_q\subset O$.
  Thus, we also have the other inclusion
  $\bigcup_{q\in\matQ\cap O}I^O_q\subset O$.
  Hence, the equality.

  Therefore, from
  \assume{countability of~$\matQ$},
  Lemma~\thref{l:conn-comps-of-open-subset-of-r-equal-or-disj}, and
  after eliminating doubles in the union,
  $O$~is a countable union of disjoint open intervals.
\end{proof}

\begin{lemma}[rational approximation of lower bound of open interval]
  \label{l:rat-approx-of-lower-bound-of-open-int}
  \mbox{}\\
  Let~$a,b\in\matRbar$.
  Assume that $a<b$.
  Then, there exists a sequence $(a_n)_{n\in\matN}\in\matQ\cap(a,b)$ that is
  nonincreasing with limit~$a$.
\end{lemma}

\begin{proof}
  Note that $-\infty\leq a<b\leq\infty$.
  From
  \assume{density of rational numbers in~$\matR$},
  let $a_0\in(a,b)\cap\matQ$.

  \proofparskip{Case $a=-\infty$}
  Then, from
  \assume{density of rational numbers in~$\matR$},
  for all $n\in\matN$, let $a_{n+1}$ be in~$(-\infty,-2|a_n|)\cap\matQ$.
  Hence, from
  \assume{ordered field properties of~$\matR$}, and
  \assume{the Archimedean property of~$\matR$},
  the sequence $(a_n)_{n\in\matN}$ belongs to~$(a,b)$, and is nonincreasing
  with limit~$a=-\infty$.

  \proofparskip{Case~$a$ finite}
  Then, from
  \assume{density of rational numbers in~$\matR$},
  for all $n\in\matN$, let $a_{n+1}$ be
  in~$\left(a,\frac{a+a_n}{2}\right)\cap\matQ\subset(a,b)\cap\matQ$.
  Let~$n\in\matN$.
  Then, from
  \assume{ordered field properties of~$\matR$},
  we have $a_{n+1}<a_n$ and $0<a_{n+1}-a<\frac{a_n-a}{2}$.
  Thus, a trivial finite induction on index~$i$ shows that
  \begin{equation*}
    \forall i \in [0..n + 1],\quad
    0
    < a_{n + 1} - a
    < \frac{a_i - a}{2^{n + 1 - i}}
    < \frac{a_0 - a}{2^{n + 1}}.
  \end{equation*}
  Hence, from
  \assume{the squeeze theorem},
  the sequence $(a_n)_{n\in\matN}$ is nonincreasing with limit~$a$.
\end{proof}

\begin{lemma}[rational approximation of upper bound of open interval]
  \label{l:rat-approx-of-upper-bound-of-open-int}
  \mbox{}\\
  Let~$a,b\in\matRbar$.
  Assume that $a<b$.
  Then, there exists a sequence $(b_n)_{n\in\matN}\in\matQ\cap(a,b)$ that is
  nonincreasing with limit~$b$.
\end{lemma}

\begin{proof}
  Direct consequence of
  Lemma~\threfc{l:rat-approx-of-lower-bound-of-open-int}{%
    with $a\eqdef-b$ and $b\eqdef-a$}.
\end{proof}

\begin{lemma}[open intervals with rational bounds cover open interval]
  \label{l:open-int-with-rat-bounds-cover-open-int}
  \mbox{}\hfill
  Let~$a,b\in\matRbar$.
  Assume that $a<b$.
  Then, there exists $(a_n)_{n\in\matN},(b_n)_{n\in\matN}\in\matQ$ such that
  $(a,b)=\bigcup_{n\in\matN}(a_n,b_n)$.
\end{lemma}

\begin{proof}
  Direct consequence of
  Lemma~\threfc{l:rat-approx-of-lower-bound-of-open-int}{%
    there exists $(a_n)_{n\in\matN}\in\matQ\cap(a,b)$},
  Lemma~\threfc{l:rat-approx-of-upper-bound-of-open-int}{%
    there exists $(b_n)_{n\in\matN}\in\matQ\cap(a,b)$},
  Definition~\threfc{LM-d:convergent-sequence}{%
    with $\eps\eqdef x-a$, then $\eps\eqdef b-x$ for all $x\in(a,b)$
    (hence, $(a,b)\subset\bigcup_{n\in\matN}(a_n,b_n)$)}, and
  Definition~\threfc{d:topological-space}{%
    closedness under union ($\bigcup_{n\in\matN}(a_n,b_n)\subset(a,b)$)}.
\end{proof}

\begin{theorem}[$\matR$~is second-countable]
  \label{t:r-is-second-countable}
  \mbox{}\hfill
  Let~$d$ be the Euclidean distance on~$\matR$.\\
  Then, $\{(a,b)\st a,b\in\matQ\Conj a<b\}$ is a topological basis
  of~$(\matR,d)$.
  Hence, $(\matR,d)$ is second-countable.
\end{theorem}

\begin{proof}
  Direct consequence of
  Theorem~\thref{t:count-conn-comps-of-open-subsets-of-r},
  Lemma~\thref{l:open-int-with-rat-bounds-cover-open-int},
  Definition~\thref{d:topological-basis},
  \assume{countability of~$\matQ\times\matQplusstar$}, and
  Definition~\thref{d:second-count}.
\end{proof}

\begin{lemma}[$\matR^n$~is second-countable]
  \label{l:rn-is-second-countable}
  \mbox{}\hfill
  Let~$n\in[2..\infty)$.
  Let~$d$ be the Euclidean distance on~$\matR^n$.
  Then, $\left\{\left.\prod_{i\in[1..n]}(a_i,b_i)\rightst
  \forall i\in[1..n],a_i,b_i\in\matQ\Conj a_i<b_i\right\}$ is a topological
  basis of~$(\matR^n,d)$.
  Hence, $(\matR^n,d)$ is second-countable.
\end{lemma}

\begin{proof}
  Direct consequence of
  Theorem~\thref{t:r-is-second-countable},
  Lemma~\thref{l:compat-of-second-count-with-cartesian-prod},
  Lemma~\thref{l:box-topo-on-cartesian-prod} and
  Definition~\thref{d:second-count}.
\end{proof}

\begin{lemma}[open intervals with rational bounds cover open interval
  of~$\matRbar$]
  \label{l:open-int-with-rat-bounds-cover-open-int-of-rbar}
  \mbox{}\\
  Let~$a,b\in\matRbar$.
  Assume that $a<b$.
  Then, there exists $(a_n)_{n\in\matN},(b_n)_{n\in\matN}\in\matQ$ such that
  \begin{equation}
    \label{e:open-int-with-rat-bounds-cover-open-int-of-rbar}
    (a, b) = \bigcup_{n \in \matN} (a_n, b_n),\quad
    [-\infty, b) = \bigcup_{n \in \matN} [-\infty, b_n)
    \AND
    (a, \infty] = \bigcup_{n \in \matN} (a_n, \infty].
  \end{equation}
\end{lemma}

\begin{proof}
  Direct consequence of
  Lemma~\thref{l:open-int-with-rat-bounds-cover-open-int},
  Lemma~\thref{l:rat-approx-of-lower-bound-of-open-int}, and
  Lemma~\thref{l:rat-approx-of-upper-bound-of-open-int}.
\end{proof}

\begin{lemma}[$\matRbar$~is second-countable]
  \label{l:rbar-is-second-countable}
  \mbox{}\\
  Let~$\calTbar\eqdef\calT_{\matRbar}(\Ray_{\matRbar})$ be the order topology
  on~$\matRbar$.
  Then, the open intervals with rational bounds constitute a topological
  basis of~$(\matRbar,\calTbar)$.
  Hence, $(\matRbar,\calTbar)$ is second-countable.
\end{lemma}

\begin{proof}
  Direct consequence of
  Lemma~\thref{l:topo-of-rbar},
  Theorem~\threfc{t:count-conn-comps-of-open-subsets-of-r}{%
    similar proof for $\matRbar$},
  Lemma~\thref{l:open-int-with-rat-bounds-cover-open-int-of-rbar},
  Definition~\thref{d:topological-basis},
  \assume{countability of~$\matQ^2$},
  \assume{compatibility of union with countability}, and
  Definition~\thref{d:second-count}.
\end{proof}

\clearpage
\subsection{Infimum, supremum}
\label{ss:inf-sup}

\begin{lemma}[extrema of constant function]
  \label{l:extr-of-const-fun}
  \mbox{}\hfill
  Let~$X$ be a nonempty set.
  Let~$f:\ArXRb$.
  Let~$a\in\matRbar$.
  Assume that~$f$ is constant of value~$a$.
  Then,
  \begin{equation}
    \label{e:extr-of-const-fun}
    \supXf = \maxXf = \infXf = \minXf = a.
  \end{equation}
\end{lemma}

\begin{proof}
  Direct consequence of
  \assume{the definition of constant function},
  Lemma~\thref{LM-l:equivalent-definition-of-maximum},
  Definition~\thref{LM-d:maximum},
  Lemma~\thref{LM-l:equivalent-definition-of-minimum}, and
  Definition~\thref{LM-d:minimum}.
\end{proof}

\begin{lemma}[equivalent definition of finite infimum]
  \label{l:equiv-def-of-finite-inf}
  \mbox{}\hfill
  Let~$X$ be a nonempty set.\\
  Let~$f:\ArXRb$.
  Let~$l\in\matR$.
  Then, we have $l=\infXf$ iff
  $l$~is a lower bound of~$f(X)$, and
  there exists $(x_n)_{n\in\matN}\in X$ such that
  $(f(x_n))_{n\in\matN}\in\matR$ is nonincreasing with limit~$l$.
\end{lemma}

\begin{proof}
  \proofpar{``Left'' implies ``right''}
  Assume first that $l=\infXf$.
  Then, from
  Definition~\thref{LM-d:infimum},
  $l$~is a (finite) lower bound for~$f(X)$.
  Thus, from
  Lemma~\thref{LM-l:finite-infimum-discrete},
  for all $n\in\matN$, there exists $\xpn\in X$ such that
  $f(\xpn)<l+\frac{1}{n+1}$.
  Let~$x_0\eqdef\xpz$.
  For all $n\in\matN$, let $x_{n+1}\eqdef\argmin(f(x_n),f(\xpnp))$.
  Then, for all $n\in\matN$, we have
  \begin{equation*}
    f (x_{n + 1})
    \leq f (x_n)
    \leq f (\xpn)
    < l + \frac{1}{n + 1}.
  \end{equation*}
  Thus, from
  \assume{the definition of monotone sequence},
  the sequence $(f(x_n))_{n\in\matN}$ is nonincreasing.
  Moreover, let $k\in\matN$.
  Let~$N\eqdef k$.
  Let~$n\in\matN$ such that $n\geq N$.
  Then, from
  \assume{ordered field properties of~$\matR$},
  we have $f(x_n)\leq f(x_N)=f(x_k)<l+\frac{1}{k+1}$.
  Hence, from
  Lemma~\thref{l:equiv-def-of-conv-seq},
  the sequence $(f(x_n))_{n\in\matN}$ is convergent with limit~$l$.

  \proofparskip{``Right'' implies ``left''}
  Conversely, assume now that~$l$ is a (finite) lower bound of~$f(X)$, and
  that there exists $(x_n)_{n\in\matN}\in X$ such that $(f(x_n))_{n\in\matN}$
  is nonincreasing with limit~$l$.
  Let~$n\in\matN$, and let $\eps_n\eqdef\frac{1}{2(n+1)}>0$.
  Then, from
  Definition~\thref{LM-d:convergent-sequence},
  let $N\in\matN$ such that for all $p\geq N$, $|f(x_p)-l|<\eps_n$.
  Let~$\xpn\eqdef x_N$.
  Then, from
  \assume{ordered field properties of~$\matR$},
  we have $f(\xpn)\leq l+\eps_n<l+\frac{1}{n+1}$.
  Hence, from
  Lemma~\thref{LM-l:finite-infimum-discrete},
  we have $l=\infXf$.

  \medskip\noindent
  Therefore, we have the equivalence.
\end{proof}

\begin{lemma}[equivalent definition of finite infimum in~$\matRbar$]
  \label{l:equiv-def-of-finite-inf-in-rbar}
  \mbox{}\hfill
  Let~$X$ be a nonempty set.\\
  Let~$f:\ArXRb$.
  Let~$l\in\matR$.
  Then, we have $l=\infXf$ iff
  $l$~is a lower bound of~$f(X)$, and
  there exists $(x_n)_{n\in\matN}\in X$ such that
  $(f(x_n))_{n\in\matN}\in\matR$ is nonincreasing with limit~$l$.
\end{lemma}

\begin{proof}
  \proofpar{Case $-\infty\in f(X)$}
  Then, from
  \assume{the definition of~$-\infty$}, and
  Definition~\threfc{LM-d:infimum}{lower bound},
  $-\infty$ ~is the only lower bound of~$f(X)$.
  Thus, since $l\in\matR$, both propositions ``$l=\infXf$'' and
  ``$l$~is a lower bound of~$f(X)$'' are wrong.
  Hence, we have the equivalence.

  \proofparskip{Case $f(X)\subset\matR$}
  Direct consequence of
  Lemma~\thref{l:equiv-def-of-finite-inf}.

  \proofparskip{Case $-\infty\not\in f(X)$, and $\infty\in f(X)$}
  Then, from
  Lemma~\threfc{l:extr-of-const-fun}{%
    contrapositive with $\infXf\not=\infty$},
  the function cannot be constant of value~$\infty$.
  Thus, since $-\infty\not\in f(X)$, let $\tx\in X$ such that
  $f(\tx)\in\matR$.
  Let~$\pi_f:\ArXX$ be the projection defined by
  \begin{equation*}
    \pi_f (x) \eqdef \left\{
      \begin{array}{ll}
        x & \mbox{when } f (x) \not= \infty,\\
        \tx & \mbox{otherwise}.
      \end{array}
    \right.
  \end{equation*}
  Let~$\tf\eqdef f\circ\pi_f$.
  Then, by construction,
  we have $\tf(X)\subset\matR$ and $\tf\leq f$.

  Let~$l$ be a lower bound of~$\tf(X)$.
  Then, $l$~is also a lower bound of~$f(X)$.
  Conversely, let~$l$ be a lower bound of~$f(X)$.
  Let~$x\in X$.
  Then, from
  the definitions of~$\pi_f$ and~$\tf$,
  we have
  \begin{equation*}
    l \leq f (\pi_f (x)) = \tf (x).
  \end{equation*}
  Thus, $l$~is also a lower bound of~$\tf(X)$.
  Hence, $f(X)$ and~$\tf(X)$ have the same lower bounds.

  Moreover, from
  Definition~\threfc{LM-d:infimum}{lower bound},
  $\infXf$ is also a lower bound of~$\tf(X)$, and $\infXtf$ is also a lower
  bound of~$f(X)$.
  Thus, from
  Definition~\threfc{LM-d:infimum}{greatest lower bound},
  we have $\infXf\leq\infXtf$, and $\infXtf\leq\infXf$.
  Hence, from
  \assume{totally ordered set properties of~$\matRbar$},
  we have $\infXtf=\infXf$.

  Furthermore, let $(\tx_n)_{n\in\matN}$ be a sequence in~$X$.
  Then, from
  the definition of~$\pi_f$ and~$\tf$,
  the sequence $(x_n\eqdef\pi_f(\tx_n))_{n\in\matN}\in X$ is such that both
  sequences $(f(x_n))_{n\in\matN}$ and $(\tf(\tx_n))_{n\in\matN}$ are
  identical.
  Conversely, assume that $(x_n)_{n\in\matN}\in X$ is such that
  $(f(x_n))_{n\in\matN}$ is in~$\matR$.
  Then, from
  the definition of~$\pi_f$ and~$\tf$,
  both sequences $(f(x_n))_{n\in\matN}$ and $(\tf(x_n))_{n\in\matN}$ are
  identical.

  Therefore, from what precedes, we have the equivalences
  $l=\infXf$ iff
  $l=\infXtf$ iff
  $l$~is a lower bound of~$\tf(X)$, and there exists a sequence
  $(\tx_n)_{n\in\matN}\in X$ such that $(\tf(\tx_n))_{n\in\matN}$ is
  nonincreasing with limit~$l$ iff
  $l$~is a lower bound of~$f(X)$, and there exists a sequence
  $(x_n)_{n\in\matN}\in X$ such that $(f(x_n))_{n\in\matN}\in\matR$ is
  nonincreasing with limit~$l$
\end{proof}

\begin{lemma}[equivalent definition of infimum]
  \label{l:equiv-def-of-inf}
  \mbox{}\hfill
  Let~$X$ be a nonempty set.\\
  Let~$f:\ArXRb$.
  Let~$l\in\matRbar$.
  Then, we have $l=\infXf$ iff
  $l$~is a lower bound of~$f(X)$, and
  there exists $(x_n)_{n\in\matN}\in X$ such that
  $(f(x_n))_{n\in\matN}\in\matRbar$ is nonincreasing with limit~$l$.
\end{lemma}

\begin{proof}
  \proofpar{Case $\infXf=-\infty$}

  \proofparskip{``Left'' implies ``right''}
  Assume first that $l=\infXf=-\infty$.
  Then, from
  \assume{the definition of~$-\infty$},
  $l$~is a lower bound of~$f(X)$.
  Moreover, from
  Lemma~\threfc{LM-l:finite-infimum}{contrapositive},
  for all $m\in\matR$, there exists $x\in X$ such that $f(x)<m$.
  For all $n\in\matN$, let $\xpn\in X$ such that $f(\xpn)<-n$.
  Let~$x_0\eqdef\xpz$.
  For all $n\in\matN$, let $x_{n+1}\eqdef\argmin(f(x_n),f(\xpnp))$.
  Then, for all $n\in\matN$, we have $f(x_{n+1})\leq f(x_n)\leq f(\xpn)<-n$.
  Thus, from
  \assume{the definition of monotone sequence},
  the sequence $(f(x_n))_{n\in\matN}$ is nonincreasing.
  Moreover, let $k\in\matN$.
  Let~$N\eqdef k$.
  Let~$n\in\matN$ such that $n\geq N$.
  Then, from
  \assume{totally ordered set properties of~$\matRbar$},
  we have
  \begin{equation*}
    f (x_n) \leq f (x_N) = f (x_k) < -k.
  \end{equation*}
  Hence, from
  Lemma~\thref{l:conv-towards-minus-infinity},
  the sequence $(f(x_n))_{n\in\matN}$ is convergent in~$\matRbar$ with
  limit~$-\infty=l$.

  \proofparskip{``Right'' implies ``left''}
  Conversely, assume now that~$l$ is a lower bound of~$f(X)$.
  Then, from
  Definition~\threfc{LM-d:infimum}{greatest lower bound}, and
  \assume{the definition of~$-\infty$},
  we have
  \begin{equation*}
    -\infty \leq l \leq \infXf = -\infty.
  \end{equation*}
  Hence, we have $l=-\infty=\infXf$.

  \proofparskip{Case $\infXf=\infty$}

  Then, from
  Definition~\threfc{LM-d:infimum}{lower bound}, and
  \assume{the definition of~$\infty$},
  we have for all $x\in X$, $\infty=\infXf\leq f(x)\leq\infty$.
  Which means that~$f$ is a constant function of value~$\infty$.
  Thus, from
  \assume{the definition of~$\infty$},
  every extended number is a lower bound of~$f(X)=\{\infty\}$.
  Moreover, for all sequence $(x_n)_{n\in\matN}\in X$, the sequence
  $(f(x_n))_{n\in\matN}$ is stationary of value~$\infty$.
  Thus, from
  \assume{the definition of monotone sequence}, and
  Lemma~\thref{LM-l:stationary-sequence-is-convergent},
  $(f(x_n))_{n\in\matN}\in\matRbar$ is nonincreasing with
  limit~$\infty$.
  For instance, the stationary sequence $(\tx)_{n\in\matN}$ for some
  $\tx\in X\not=\emptyset$ is such a sequence.
  Hence, we have the equivalence $l=\infty(=\infXf)$ iff
  there exists a sequence $(x_n)_{n\in\matN}\in X$ such that the sequence
  $(f(x_n))_{n\in\matN}\in\matRbar$ is nonincreasing with limit~$l$.

  \proofparskip{Case $\infXf\in\matR$}

  Direct consequence of
  Lemma~\thref{l:equiv-def-of-finite-inf-in-rbar}.

  \medskip\noindent
  Therefore, we always have the equivalence.
\end{proof}

\begin{lemma}[infimum is smaller than supremum]
  \label{l:inf-is-smaller-than-sup}
  \mbox{}\\
  Let~$X$ be a nonempty set.
  Let~$f:\ArXRb$.
  Then, we have $\inf(f(X))\leq\sup(f(X))$.
\end{lemma}

\begin{proof}
  Direct consequence of
  Definition~\threfc{LM-d:infimum}{lower bound},
  Definition~\threfc{LM-d:supremum}{upper bound}, and
  Lemma~\threfc{l:order-in-rbar-is-total}{transitivity}.
\end{proof}

\begin{lemma}[infimum is monotone]
  \label{l:inf-is-monot}
  \mbox{}\hfill
  Let~$X$ and~$Y$ be nonempty sets.\\
  Assume that $Y\subset X$.
  Let~$f:\ArXRb$.
  Then, we have $\inf(f(X))\leq\inf(f(Y))$.
\end{lemma}

\begin{proof}
  Direct consequence of
  Definition~\threfc{LM-d:infimum}{%
    lower bound with $X$, then greatest lower bound with $Y$}, and
  Lemma~\threfc{l:order-in-rbar-is-total}{transitivity}.
\end{proof}

\begin{lemma}[supremum is monotone]
  \label{l:sup-is-monot}
  \mbox{}\hfill
  Let~$X$ and~$Y$ be nonempty sets.\\
  Assume that $Y\subset X$.
  Let~$f:\ArXRb$.
  Then, we have $\sup(f(Y))\leq\sup(f(X))$.
\end{lemma}

\begin{proof}
  Direct consequence of
  Lemma~\thref{LM-l:duality-infimum-supremum},
  Lemma~\thref{l:inf-is-monot}, and
  \assume{monotonicity of additive inverse}.
\end{proof}

\begin{lemma}[compatibility of infimum with absolute value]
  \label{l:compat-of-inf-with-abs}
  \mbox{}\\
  Let~$X$ be a nonempty set.
  Let~$f:\ArXRb$.
  Then, we have $|\inf(f(X))|\leq\sup(|f(X)|)$.
\end{lemma}

\begin{proof}
  From
  Lemma~\threfc{l:equiv-def-of-abs-in-rbar}{$f\leq|f|$},
  Lemma~\thref{l:inf-is-monot},
  Lemma~\thref{l:inf-is-smaller-than-sup},
  Lemma~\thref{l:abs-in-rbar-is-even}, and
  Lemma~\thref{l:sup-is-monot},
  we have
  \begin{equation*}
    \inf (f (X)) \leq \inf | f (X) | \leq \sup | f (X) |
    \AND
    \sup (-f (X)) \leq \sup | f (X) |.
  \end{equation*}
  \proofpar{Case $\inf(f(X))\geq0$}
  Then, from
  Lemma~\thref{l:equiv-def-of-abs-in-rbar},
  we have $|\inf(f(X))|=\inf(f(X))\leq\sup|f(X)|$.
  \proofpar{Case $\inf(f(X))<0$}
  Then, from
  Lemma~\thref{LM-l:duality-infimum-supremum},
  we have $|\inf(f(X))=-\inf(f(X))=\sup(-f(X))\leq\sup|f(X)|$.
\end{proof}

\begin{lemma}[compatibility of supremum with absolute value]
  \label{l:compat-of-sup-with-abs}
  \mbox{}\\
  Let~$X$ be a nonempty set.
  Let~$f:\ArXRb$.
  Then, we have $|\sup(f(X))|\leq\sup(|f(X)|)$.
\end{lemma}

\begin{proof}
  Direct consequence of
  Lemma~\thref{LM-l:duality-infimum-supremum},
  Lemma~\threfc{l:compat-of-inf-with-abs}{with $(-f_n)_{n\in\matN}$}, and
  Lemma~\thref{l:abs-in-rbar-is-even}.
\end{proof}

\begin{lemma}[compatibility of translation with infimum]
  \label{l:compat-of-translation-with-inf}
  \mbox{}\hfill
  Let~$X$ be a nonempty set.\\
  Let~$(f_n)_{n\in\matN}:\ArXRb$.
  Then, for all $p\in\matN$, we have
  $\inf_{n\in\matN}f_n\leq\inf_{n\in\matN}f_{n+p}$.
\end{lemma}

\begin{proof}
  Direct consequence of
  Lemma~\threfc{l:inf-is-monot}{with $[p..\infty)\subset\matN$}.
\end{proof}

\begin{lemma}[compatibility of translation with supremum]
  \label{l:compat-of-translation-with-sup}
  \mbox{}\hfill
  Let~$X$ be a nonempty set.\\
  Let~$(f_n)_{n\in\matN}:\ArXRb$.
  Then, for all $p\in\matN$, we have
  $\sup_{n\in\matN}f_{n+p}\leq\sup_{n\in\matN}f_n$.
\end{lemma}

\begin{proof}
  Direct consequence of
  Lemma~\threfc{l:sup-is-monot}{with $[p..\infty)\subset\matN$}.
\end{proof}

\begin{lemma}[infimum of sequence is monotone]
  \label{l:inf-of-seq-is-monot}
  \mbox{}\\
  Let~$X$ be a nonempty set.
  Let~$(f_n)_{n\in\matN},(g_n)_{n\in\matN}:\ArXRb$.
  Assume that for all $n\in\matN$, $f_n\leq g_n$
  Then, we have $\inf_{n\in\matN}f_n\leq\inf_{n\in\matN}g_n$.
\end{lemma}

\begin{proof}
  Direct consequence of
  Definition~\threfc{LM-d:infimum}{%
    with $X\eqdef\matN$,
    lower bound for $\inf_{n\in\matN}f_n$ and
    greatest lower bound for $\inf_{n\in\matN}g_n$}, and
  Lemma~\threfc{l:order-in-rbar-is-total}{transitivity}.
\end{proof}

\begin{lemma}[supremum of sequence is monotone]
  \label{l:sup-of-seq-is-monot}
  \mbox{}\\
  Let~$X$ be a nonempty set.
  Let~$(f_n)_{n\in\matN},(g_n)_{n\in\matN}:\ArXRb$.
  Assume that for all $n\in\matN$, $f_n\leq g_n$
  Then, we have $\sup_{n\in\matN}f_n\leq\sup_{n\in\matN}g_n$.
\end{lemma}

\begin{proof}
  Direct consequence of
  Lemma~\thref{LM-l:duality-infimum-supremum},
  Lemma~\threfc{l:inf-of-seq-is-monot}{with $-g_n\leq-f_n$}, and
  \assume{monotonicity of additive inverse}.
\end{proof}

\begin{lemma}[infimum of bounded sequence is bounded]
  \label{l:inf-of-bounded-seq-is-bounded}
  \mbox{}\\
  Let~$X$ be a nonempty set. Let $a,b\in\matRbar$.
  Let~$(f_n)_{n\in\matN}:\ArXRb$.\\
  If for all $n\in\matN$, $f_n\leq b$, then
  we have $\inf_{n\in\matN}f_n\leq b$.\\
  If for all $n\in\matN$, $a\leq f_n$, then
  we have $a\leq\inf_{n\in\matN}f_n$.
\end{lemma}

\begin{proof}
  Direct consequence of
  Lemma~\threfc{l:extr-of-const-fun}{with $X\eqdef\matN$}, and
  Lemma~\thref{l:inf-of-seq-is-monot}.
\end{proof}

\begin{lemma}[supremum of bounded sequence is bounded]
  \label{l:sup-of-bounded-seq-is-bounded}
  \mbox{}\\
  Let~$X$ be a nonempty set. Let $a,b\in\matRbar$.
  Let~$(f_n)_{n\in\matN}:\ArXRb$.\\
  If for all $n\in\matN$, $f_n\leq b$, then
  we have $\sup_{n\in\matN}f_n\leq b$.\\
  If for all $n\in\matN$, $a\leq f_n$, then
  we have $a\leq\sup_{n\in\matN}f_n$.
\end{lemma}

\begin{proof}
  Direct consequence of
  Lemma~\threfc{l:extr-of-const-fun}{with $X\eqdef\matN$}, and
  Lemma~\thref{l:sup-of-seq-is-monot}.
\end{proof}

\clearpage
\subsection{Limit inferior, limit superior}
\label{ss:liminf-limsup}

\begin{lemma}[limit inferior]
  \label{l:liminf}
  \mbox{}\hfill
  Let~$X$ be a nonempty set.
  Let~$(f_n)_{n\in\matN}:\ArXRb$.\\
  Then, for all $x\in X$,
  $(\inf_{p\in\matN}f_{n+p}(x))_{n\in\matN}$ is nondecreasing, and we have
  \begin{equation}
    \label{e:liminf-1}
    \forall x \in X,\quad
    \lim_{n \to \infty} \left(
      \inf_{p \in \matN} f_{n + p} (x) \right)
    = \sup_{n \in \matN} \left(
      \inf_{p \in \matN} f_{n + p} (x) \right)
    \quad \in \matRbar.
  \end{equation}

  The {\em limit inferior} of the sequence is the function
  $\ArXRb$ defined by
  \begin{equation}
    \label{e:liminf-2}
    \forall x \in X,\quad
    \liminf_{n \to \infty} f_n (x)
    \eqdef \lim_{n \to \infty} \left(
      \inf_{p \in \matN} f_{n + p} (x) \right).
  \end{equation}
\end{lemma}

\begin{proof}
  Direct consequence of
  Lemma~\thref{l:inf-is-monot}, and
  \assume{completeness of~$\matRbar$
    (a nondecreasing sequence is convergent and its limit is its least upper
    bound)}.
\end{proof}

\begin{lemma}[limit inferior is~$\infty$]
  \label{l:liminf-is-inf}
  \mbox{}\hfill
  Let~$X$ be a nonempty set.
  Let~$(f_n)_{n\in\matN}:\ArXRb$.
  Let~$x\in X$.
  Assume that $\liminf_{n\to\infty}f_n(x)=\infty$.
  Then, we have $\lim_{n\to\infty}f_n(x)=\infty$
\end{lemma}

\begin{proof}
  Let~$a\in\matR$.
  Then, from
  \assume{the definition of~$\infty$},
  we have $a<\liminf_{n\to\infty}f_n(x)$.
  Thus, from
  Lemma~\thref{l:liminf}, and
  \assume{the definition of the limit},
  there exists $N\in\matN$ such that $a<\inf_{p\in\matN}f_{N+p}(x)$.
  Hence, from
  Definition~\threfc{LM-d:infimum}{lower bound},
  we have for all $p\in\matN$,
  $a<\inf_{p\in\matN}f_{N+p}(x)\leq f_{N+p}(x)$.
  Therefore, from
  \assume{the definition of the limit},
  we have $\lim_{n\to\infty}f_n(x)=\infty$.
\end{proof}

\begin{lemma}[equivalent definition of the limit inferior]
  \label{l:equiv-def-of-liminf}
  \mbox{}\\
  Let~$X$ be a nonempty set.
  Let~$(f_n)_{n\in\matN}:\ArXRb$.
  Then, for all $x\in X$, $\liminf_{n\to\infty}f_n(x)$ is the
  smallest cluster point of the sequence $(f_n(x))_{n\in\matN}$.
\end{lemma}

\begin{proof}
  Let~$x\in X$.
  For all $n\in\matN$, let $F_n^-(x)\eqdef\inf_{p\in\matN}f_{n+p}(x)$.\\
  Let~$\ulf(x)\eqdef\liminf_{n\to\infty}f_n(x)
  =\lim_{n\to\infty}F_n^-(x) \in \matRbar$.

  \medskip
  Let~us first show that $\ulf(x)$ is a cluster point of the sequence
  $(f_n(x))_{n\in\matN}$.

  \proofparskip{Case $\ulf(x)$ is finite}
  Let~$\eps>0$.
  Let~$M\in\matN$.
  Then, from
  Lemma~\threfc{l:liminf}{%
    $(F_n^-(x))_{n\in\matN}$ is nondecreasing}, and
  Definition~\threfc{LM-d:convergent-sequence}{with $(F_n^-(x))_{n\in\matN}$},
  there exists $N\in\matN$ such that we have
  \begin{equation*}
    \forall k \in[N..\infty),\quad
    \ulf (x) - \eps
    \leq F_k^- (x)
    \leq \ulf (x).
  \end{equation*}
  Let~$K\eqdef\max(M,N)$.
  Then, from
  Lemma~\threfc{LM-l:finite-infimum}{for~$F_K^-$},
  there exists $k\geq K\geq M$ such that
  \begin{equation*}
    \ulf (x) - \eps
    \leq F_K^- (x)
    \leq f_k (x)
    < F_K^- (x) + \eps
    \leq \ulf (x) + \eps.
  \end{equation*}
  Hence, from
  Definition~\thref{d:cluster-point},
  $\ulf(x)$ is a cluster point of the sequence $(f_n(x))_{n\in\matN}$.

  \proofparskip{Case $\ulf(x)=-\infty$}
  From
  Lemma~\thref{l:liminf}, and
  Definition~\threfc{LM-d:supremum}{upper bound},
  we have for all $n\in\matN$, $F_n^-(x)=-\infty$.
  Let~$n\in\matN$.
  Let~$a\in\matR$.
  Then, $F_n^-(x)<a$.
  Thus, from
  Definition~\threfc{LM-d:infimum}{greatest lower bound, contrapositive},
  there exists $P\in\matN$ such that $f_{n+P}(x)<a$.
  Hence, from
  \assume{the definition of cluster point in~$-\infty$},
  $\ulf(x)=-\infty$ is a cluster point of the sequence $(f_n(x))_{n\in\matN}$.

  \proofparskip{Case $\ulf(x)=\infty$}
  Then, from
  Lemma~\thref{l:liminf-is-inf},
  we have $\lim_{n\to\infty}f_n(x)=\infty$, and from
  \assume{the fact that the limit of a convergent sequence is its only
    cluster point},
  $\ulf(x)=\infty$ is a cluster point of the sequence $(f_n(x))_{n\in\matN}$.

  \medskip
  Now, let~$f(x)$ be a cluster point of the sequence $(f_n(x))_{n\in\matN}$.
  Let~us show that $\ulf(x)\leq f(x)$.

  \proofparskip{Case $f(x)=\infty$}
  Then, from
  \assume{the definition of~$\infty$},
  we have $\ulf(x)\leq f(x)$.

  \proofparskip{Case $f(x)<\infty$}
  Let~$a\in\matR$ such that $f(x)<a$.
  Let~$\eps\eqdef\frac{a-f(x)}{2}>0$.
  Let~$n\in\matN$.
  Then, from
  Definition~\thref{d:cluster-point},
  there exists $k\geq n$ such that $f_k(x)\leq f(x)+\eps<a$.
  Thus, from
  Definition~\threfc{LM-d:infimum}{lower bound},
  $F_n^-(x)<a$.
  Hence, from
  \assume{monotonicity of the limit}, and
  Lemma~\thref{l:liminf},
  we have $\ulf(x)=\lim_{n\to\infty}F_n^-(x)\leq a$.
  Since this is true for all $a>f(x)$, we also have $\ulf(x)\leq f(x)$.

  \medskip\noindent
  Therefore, for all $x\in X$, $\ulf(x)$ is the smallest cluster point of the
  sequence $(f_n(x))_{n\in\matN}.$
\end{proof}

\begin{lemma}[limit inferior is invariant by translation]
  \label{l:liminf-is-invariant-by-translation}
  \mbox{}\\
  Let~$X$ be a nonempty set.
  Let~$(f_n)_{n\in\matN}:\ArXRb$.
  Then, we have
  \begin{equation}
    \label{e:liminf-is-invariant-by-translation}
    \forall k \in \matN,\quad
    \forall x \in X,\quad
    \liminf_{n \to \infty} f_{k + n} (x)
    = \liminf_{n \to \infty} f_n (x).
  \end{equation}
\end{lemma}

\begin{proof}
  Direct consequence of
  Lemma~\thref{l:equiv-def-of-liminf},
  Definition~\thref{d:cluster-point}, and
  \assume{compatibility of translation with limit}.
\end{proof}

\begin{lemma}[limit inferior is monotone]
  \label{l:liminf-is-monot}
  \mbox{}\hfill
  Let~$X$ be a nonempty set.\\
  Let~$(f_n)_{n\in\matN},(g_n)_{n\in\matN}:\ArXRb$.
  Assume that~$(f_n)_{n\in\matN}\leq(g_n)_{n\in\matN}$ from some rank:
  \begin{equation}
    \label{e:liminf-is-monot-1}
    \exists N \in \matN,\;
    \forall n \in [N..\infty),\;
    \forall x \in X,\quad
    f_n (x) \leq g_n (x).
  \end{equation}
  Then, we have
  \begin{equation}
    \label{e:liminf-is-monot-2}
    \forall x \in X,\quad
    \liminf_{n \to \infty} f_n (x)
    \leq \liminf_{n \to \infty} g_n (x).
  \end{equation}
\end{lemma}

\begin{proof}
  Direct consequence of
  Lemma~\threfc{l:liminf-is-invariant-by-translation}{%
    with $k\eqdef N$},
  Lemma~\thref{l:liminf},
  Lemma~\thref{l:inf-of-seq-is-monot}, and
  Lemma~\thref{l:sup-of-seq-is-monot}.
\end{proof}

\begin{lemma}[limit superior]
  \label{l:limsup}
  \mbox{}\hfill
  Let~$X$ be a nonempty set.
  Let~$(f_n)_{n\in\matN}:\ArXRb$.\\
  Then, for all $x\in X$,
  $(\sup_{p\in\matN}f_{n+p}(x))_{n\in\matN}$ is nonincreasing, and we have
  \begin{equation}
    \label{e:limsup-1}
    \forall x \in X,\quad
    \lim_{n \to \infty} \left(
      \sup_{p \in \matN} f_{n + p} (x) \right)
    = \inf_{n \in \matN} \left(
      \sup_{p \in \matN} f_{n + p} (x) \right)
    \quad \mbox{in } \matRbar.
  \end{equation}

  The {\em limit superior} of the sequence is the function
  $\ArXRb$ defined by
  \begin{equation}
    \label{e:limsup-2}
    \forall x \in X,\quad
    \limsup_{n \to \infty} f_n (x)
    \eqdef \lim_{n \in \matN} \left(
      \sup_{p \in \matN} f_{n + p} (x) \right).
  \end{equation}
\end{lemma}

\begin{proof}
  Let~$x\in X$.
  For all $n\in\matN$, let $F_n(x)\eqdef\{f_{n+p}(x)\st p\in\matN\}$.
  Then, we have
  \begin{equation*}
    \forall n \in \matN,\quad
    \sup F_n (x)
    = \sup_{p \in \matN} f_{n+p} (x)
    \CONJ
    F_n (x) \subset F_{n+1} (x) \subset \matRbar.
  \end{equation*}
  Thus, the sequence $(\sup F_n(x))_{n\in\matN}$ is nonincreasing
  in~$\matRbar$.
  Therefore, from
  \assume{completeness of~$\matRbar$},
  the sequence is convergent and its limit is its greatest lower bound
  $\inf_{n\in\matN}(\sup F_n(x))$.
\end{proof}

\begin{lemma}[duality limit inferior-limit superior]
  \label{l:duality-liminf-limsup}
  \mbox{}\\
  Let~$X$ be a nonempty set.
  Let~$(f_n)_{n\in\matN}:\ArXRb$.
  Then, we have
  \begin{equation}
    \label{e:duality-liminf-limsup}
    \forall x \in X,\quad
    \limsup_{n \to \infty} f_n (x)
    = - \liminf_{n \to \infty} (- f_n (x)).
  \end{equation}
\end{lemma}

\begin{proof}
  Direct consequence of
  Lemma~\thref{l:limsup},
  Lemma~\thref{l:liminf},
  \assume{linearity and uniqueness of the limit}, and
  Lemma~\threfc{LM-l:duality-infimum-supremum}{with $X\eqdef\matN$}.
\end{proof}

\begin{lemma}[equivalent definition of limit superior]
  \label{l:equiv-def-of-limsup}
  \mbox{}\\
  Let~$X$ be a nonempty set.
  Let~$(f_n)_{n\in\matN}:\ArXRb$.
  Then, for all $x\in X$, $\limsup_{n\to\infty}f_n(x)$ is the
  largest cluster point of the sequence $(f_n(x))_{n\in\matN}$.
\end{lemma}

\begin{proof}
  Let~$x\in X$.
  Let~$\fbar(x)\eqdef\limsup_{n\to\infty}f_n(x)$.
  Then, from
  Lemma~\thref{l:duality-liminf-limsup}, and
  Lemma~\thref{l:equiv-def-of-liminf},
  $-\fbar(x)$~is the smallest cluster point of the sequence
  $(-f_n(x))_{n\in\matN}$.
  Therefore, from
  \assume{linearity of the limit}, and
  \assume{totally ordered set properties of~$\matRbar$},
  $\fbar(x)$~is the largest cluster point of the sequence
  $(f_n(x))_{n\in\matN}$.
\end{proof}

\begin{lemma}[limit inferior is smaller than limit superior]
  \label{l:liminf-is-smaller-than-limsup}
  \mbox{}\\
  Let~$X$ be a nonempty set.
  Let~$(f_n)_{n\in\matN}:\ArXRb$.
  Then, we have
  \begin{equation}
    \label{e:liminf-is-smaller-than-limsup}
    \forall x \in X,\quad
    \liminf_{n \to \infty} f_n (x)
    \leq \limsup_{n \to \infty} f_n (x).
  \end{equation}
\end{lemma}

\begin{proof}
  Direct consequence of
  Lemma~\thref{l:equiv-def-of-liminf},
  Lemma~\thref{l:equiv-def-of-limsup}, and
  Lemma~\thref{l:order-in-rbar-is-total}{transitivity}.
\end{proof}

\begin{lemma}[limit superior is monotone]
  \label{l:limsup-is-monot}
  \mbox{}\hfill
  Let~$X$ be a nonempty set.\\
  Let~$(f_n)_{n\in\matN},(g_n)_{n\in\matN}:\ArXRb$.
  Assume that~$(f_n)_{n\in\matN}\leq(g_n)_{n\in\matN}$ from some rank:
  \begin{equation}
    \label{e:limsup-is-monot-1}
    \exists N \in \matN,\;
    \forall n \in [N..\infty),\;
    \forall x \in X,\quad
    f_n (x) \leq g_n (x).
  \end{equation}
  Then, we have
  \begin{equation}
    \label{e:limsup-is-monot-2}
    \forall x \in X,\quad
    \limsup_{n \to \infty} f_n (x)
    \leq \limsup_{n \to \infty} g_n (x).
  \end{equation}
\end{lemma}

\begin{proof}
  Direct consequence of
  \assume{monotonicity of additive inverse in~$\matRbar$},
  Lemma~\threfc{l:liminf-is-monot}{with $-g_n\leq-f_n$}, and
  Lemma~\thref{l:duality-liminf-limsup}.
\end{proof}

\begin{lemma}[compatibility of limit inferior with absolute value]
  \label{l:compat-liminf-with-abs}
  \mbox{}\\
  Let~$X$ be a nonempty set.
  Let~$(f_n)_{n\in\matN}:\ArXRbp$.
  Then, we have
  \begin{equation}
    \label{e:compat-liminf-with-abs}
    \forall x \in X,\quad
    \left| \liminf_{n \to \infty} f_n (x) \right|
    \leq \limsup_{n \to \infty} | f_n (x) |.
  \end{equation}
\end{lemma}

\begin{proof}
  From
  Lemma~\threfc{l:compat-of-inf-with-abs}{%
    with $X\eqdef[n..\infty)\times X$},
  for all $x\in X$, for all $n\in\matN$, we have
  \begin{equation*}
    \left| \inf_{p \in \matN} f_{n + p} (x) \right|
    \leq \sup_{p \in \matN} | f_{n + p} (x) |.
  \end{equation*}
  Therefore, from
  Lemma~\thref{l:liminf},
  \assume{compatibility of the limit with the absolute value},
  \assume{monotonicity of the limit (when $n\to\infty$)}, and
  Lemma~\thref{l:limsup},
  we have
  \begin{equation*}
    \left| \liminf_{n \to \infty} f_n (x) \right|
    = \left|
      \lim_{n \to \infty} \inf_{p \in \matN} f_{n + p} (x) \right|
    = \lim_{n \to \infty} \left|
      \inf_{p \in \matN} f_{n + p} (x) \right|
    \leq \lim_{n \to \infty} \sup_{p \in \matN} | f_{n + p} (x) |
    = \limsup_{n \to \infty} | f_n (x) |.
  \end{equation*}
\end{proof}

\begin{lemma}[compatibility of limit superior with absolute value]
  \label{l:compat-limsup-with-abs}
  \mbox{}\\
  Let~$X$ be a nonempty set.
  Let~$(f_n)_{n\in\matN}:\ArXRbp$.
  Then, we have
  \begin{equation}
    \label{e:compat-limsup-with-abs}
    \forall x \in X,\quad
    \left| \limsup_{n \to \infty} f_n (x) \right|
    \leq \limsup_{n \to \infty} | f_n (x) |.
  \end{equation}
\end{lemma}

\begin{proof}
  Direct consequence of
  Lemma~\thref{l:duality-liminf-limsup},
  Lemma~\threfc{l:compat-liminf-with-abs}{with $-f_n$}, and
  Lemma~\thref{l:abs-in-rbar-is-even}.
\end{proof}

\begin{definition}[pointwise convergence]
  \label{d:pointwise-conv}
  \mbox{}\\
  Let~$X$ be a nonempty set.
  A sequence $(f_n)_{n\in\matN}$ of functions $\ArXRb$ is said
  {\em pointwise convergent} iff
  for all $x\in X$, $(f_n(x))_{n\in\matN}$ is convergent in~$\matRbar$.
\end{definition}

\begin{lemma}[limit inferior and limit superior of pointwise convergent]
  \label{l:liminf-and-limsup-of-pointwise-conv}
  \mbox{}\\
  Let~$X$ be a nonempty set.
  Let~$(f_n)_{n\in\matN}:\ArXRb$.
  Assume that the sequence is pointwise convergent.
  Then, we have
  \begin{equation}
    \label{e:liminf-and-limsup-of-pointwise-conv}
    \forall x \in X,\quad
    \liminf_{n \to \infty} f_n (x)
    = \limsup_{n \to \infty} f_n (x)
    = \lim_{n \to \infty} f_n (x).
  \end{equation}
\end{lemma}

\begin{proof}
  Direct consequence of
  Definition~\thref{d:pointwise-conv},
  Lemma~\thref{l:equiv-def-of-liminf},
  Lemma~\thref{l:equiv-def-of-limsup}, and
  \assume{the fact that a convergent sequence has its limit as the unique
    cluster point}.
\end{proof}

\begin{lemma}[limit inferior bounded from below]
  \label{l:liminf-bounded-from-below}
  \mbox{}\\
  Let~$X$ be a nonempty set.
  Let~$(f_n)_{n\in\matN}:\ArXRb$.
  Let~$m\in\matRbar$.
  Assume that the sequence is bounded by~$m$ from below from some rank:
  \begin{equation}
    \label{e:liminf-bounded-from-below-1}
    \exists N \in \matN,\;
    \forall n \in [N..\infty),\;
    \forall x \in X,\quad
    m \leq f_n (x).
  \end{equation}
  Then, we have
  \begin{equation}
    \label{e:liminf-bounded-from-below-2}
    \forall x \in X,\quad
    m \leq \liminf_{n \to \infty} f_n (x).
  \end{equation}
\end{lemma}

\begin{proof}
  Direct consequence of
  Lemma~\thref{l:liminf-is-monot},
  Lemma~\thref{LM-l:stationary-sequence-is-convergent},
  Definition~\threfc{LM-d:stationary-sequence}{%
    constant sequence is stationary ($N=0$)}, and
  Lemma~\threfc{l:liminf-and-limsup-of-pointwise-conv}{%
    with the constant sequence of value~$m$}.
\end{proof}

\begin{lemma}[limit inferior bounded from above]
  \label{l:liminf-bounded-from-above}
  \mbox{}\\
  Let~$X$ be a nonempty set.
  Let~$(f_n)_{n\in\matN}:\ArXRb$.
  Let~$M\in\matRbar$.
  Assume that the sequence is bounded by~$M$ from above from some rank:
  \begin{equation}
    \label{e:liminf-bounded-from-above-1}
    \exists N \in \matN,\;
    \forall n \in [N..\infty),\;
    \forall x \in X,\quad
    f_n (x) \leq M.
  \end{equation}
  Then, we have
  \begin{equation}
    \label{e:liminf-bounded-from-above-2}
    \forall x \in X,\quad
    \liminf_{n \to \infty} f_n (x) \leq M.
  \end{equation}
\end{lemma}

\begin{proof}
  Direct consequence of
  Lemma~\thref{l:liminf-is-monot},
  Lemma~\thref{LM-l:stationary-sequence-is-convergent},
  Definition~\threfc{LM-d:stationary-sequence}{%
    constant sequence is stationary ($N=0$)}, and
  Lemma~\threfc{l:liminf-and-limsup-of-pointwise-conv}{%
    with the constant sequence of value~$M$}.
\end{proof}

\begin{lemma}[limit superior bounded from below]
  \label{l:limsup-bounded-from-below}
  \mbox{}\\
  Let~$X$ be a nonempty set.
  Let~$(f_n)_{n\in\matN}:\ArXRb$.
  Let~$m\in\matRbar$.
  Assume that the sequence is bounded by~$m$ from below from some rank:
  \begin{equation}
    \label{e:limsup-bounded-from-below-1}
    \exists N \in \matN,\;
    \forall n \in [N..\infty),\;
    \forall x \in X,\quad
    m \leq f_n (x).
  \end{equation}
  Then, we have
  \begin{equation}
    \label{e:limsup-bounded-from-below-2}
    \forall x \in X,\quad
    m \leq \limsup_{n \to \infty} f_n (x).
  \end{equation}
\end{lemma}

\begin{proof}
  Direct consequence of
  Lemma~\threfc{l:liminf-bounded-from-above}{%
    with $f_n\eqdef-f_n$ and $M\eqdef-m$}, and
  Lemma~\thref{l:duality-liminf-limsup}.
\end{proof}

\begin{lemma}[limit superior bounded from above]
  \label{l:limsup-bounded-from-above}
  \mbox{}\\
  Let~$X$ be a nonempty set.
  Let~$(f_n)_{n\in\matN}:\ArXRb$.
  Let~$M\in\matRbar$.
  Assume that the sequence is bounded by~$M$ from above from some rank:
  \begin{equation}
    \label{e:limsup-bounded-from-above-1}
    \exists N \in \matN,\;
    \forall n \in [N..\infty),\;
    \forall x \in X,\quad
    f_n (x) \leq M.
  \end{equation}
  Then, we have
  \begin{equation}
    \label{e:limsup-bounded-from-above-2}
    \forall x \in X,\quad
    \limsup_{n \to \infty} f_n (x) \leq M.
  \end{equation}
\end{lemma}

\begin{proof}
  Direct consequence of
  Lemma~\threfc{l:liminf-bounded-from-below}{%
    with $f_n\eqdef-f_n$ and $m\eqdef-M$}, and
  Lemma~\thref{l:duality-liminf-limsup}.
\end{proof}

\begin{lemma}[limit inferior, limit superior and pointwise convergence]
  \label{l:liminf-limsup-and-pointwise-conv}
  \mbox{}\\
  Let~$X$ be a nonempty set.
  Let~$(f_n)_{n\in\matN}:\ArXRb$.
  Assume that
  \begin{equation}
    \label{e:liminf-limsup-and-pointwise-conv-1}
    \forall x \in X,\quad
    \limsup_{n \to \infty} f_n (x)
    \leq \liminf_{n \to \infty} f_n (x).
  \end{equation}
  Then, the sequence is pointwise convergent and
  \begin{equation}
    \label{e:liminf-limsup-and-pointwise-conv-2}
    \forall x \in X,\quad
    \liminf_{n \to \infty} f_n (x)
    = \limsup_{n \to \infty} f_n (x)
    = \lim_{n \to \infty} f_n (x).
  \end{equation}
\end{lemma}

\begin{proof}
  Let~$x\in X$.
  Then, from
  Lemma~\thref{l:liminf-is-smaller-than-limsup},
  we have
  $\liminf_{n\to\infty}f_n(x)
  =\limsup_{n\to\infty}f_n(x)
  \eqdef f(x)\in\matRbar$.

  \proofparskip{Case $f(x)=\infty$}
  Then, from
  Lemma~\thref{l:liminf-is-inf},
  we have
  \begin{equation*}
    \lim_{n \to \infty} f_n (x) = \infty = f (x).
  \end{equation*}

  \proofparskip{Case $f(x)=-\infty$}
  Then, from
  Lemma~\thref{l:duality-liminf-limsup}, and
  Lemma~\thref{l:liminf-is-inf},
  we have $\lim_{n\to\infty}f_n(x)=-\infty=f(x)$.

  \proofparskip{Case $f(x)\in\matR$}
  Let~$\eps>0$.
  Then, we have
  \begin{equation*}
    f (x) - \eps < \liminf_{n \to \infty} f_n (x)
    = f (x)
    = \limsup_{n \to \infty} f_n (x) < f (x) + \eps.
  \end{equation*}
  Thus, from
  Lemma~\thref{l:equiv-def-of-liminf},
  there exists $N^-\in\matN$ such that for all $n\geq N^-$, we have
  $f(x)-\eps<f_n(x)$, and from
  Lemma~\thref{l:equiv-def-of-limsup},
  there exists $N^+\in\matN$ such that for all $n\geq N^+$, we have
  $f_n(x)<f(x)+\eps$.\\
  Let~$N\eqdef\max(N^-,N^+)$.
  Then, for all $n\geq N$, we have $f(x)-\eps<f_n(x)<f(x)+\eps$.
  Therefore, from
  Definition~\thref{LM-d:convergent-sequence},
  the sequence $(f_n(x))_{n\in\matN}$ is convergent with limit~$f(x)$.
\end{proof}

\clearpage
\subsection{Truncating a function}
\label{ss:truncating-a-function}

\begin{definition}[finite part]
  \label{d:finite-part}
  \mbox{}\\
  Let~$X$ be a set.
  Let~$f:\ArXRb$.
  The {\em finite part of~$f$} is the function $f\matUN_{f^{-1}(\matR)}$.
\end{definition}

\begin{lemma}[finite part is finite]
  \label{l:finite-part-is-finite}
  \mbox{}\\
  Let~$X$ be a set.
  Let~$f:\ArXRb$.
  Then, the finite part of~$f$ is finite.
\end{lemma}

\begin{proof}
  Direct consequence of
  Definition~\thref{d:finite-part},
  Definition~\threfc{d:ext-real-nums-rbar}{%
    $\matRbar=\{\pm\infty\}\uplus\matR$},
  \assume{properties of inverse image
  ($X=f^{-1}(\matRbar)=f^{-1}(\pm\infty)\uplus f^{-1}(\matR)$)}, and
  \assume{the definition of the indicator function}.
\end{proof}

\begin{definition}[nonnegative and nonpositive parts]
  \label{d:nonneg-and-nonpos-parts}
  \mbox{}\\
  Let~$X$ be a set.
  Let~$f:\ArXRb$.
  The functions $f^+\eqdef\max(f,0)$ and $f^-\eqdef\max(-f,0)$ are
  respectively called the {\em nonnegative part of~$f$} and the
  {\em nonpositive part of~$f$}.
\end{definition}

\begin{lemma}[equivalent definition of nonnegative and nonpositive parts]
  \mbox{}\\
  \label{l:equiv-def-of-nonneg-and-nonpos-parts}
  Let~$X$ be a set.
  Let~$f:\ArXRb$.
  Then, we have $f^+=f\matUN_{f^{-1}(\matRbarplus)}$ and
  $f^-=-f\matUN_{f^{-1}(\matRbarminus)}$.
\end{lemma}

\begin{proof}
  Direct consequence of
  Definition~\thref{d:nonneg-and-nonpos-parts},
  \assume{the definition of the maximum}, and
  \assume{the definition of the indicator function}.
\end{proof}

\begin{lemma}[nonnegative and nonpositive parts are nonnegative]
  \label{l:nonneg-and-nonpos-parts-are-nonneg}
  \mbox{}\\
  Let~$X$ be a set.
  Let~$f:\ArXRb$.
  Then, we have $f^+,f^-\geq0$.
\end{lemma}

\begin{proof}
  Direct consequence of
  Definition~\thref{d:nonneg-and-nonpos-parts}, and
  \assume{the definition of the maximum}.
\end{proof}

\begin{lemma}[nonnegative and nonpositive parts are orthogonal]
  \label{l:nonneg-and-nonpos-parts-are-orthogonal}
  \mbox{}\\
  Let~$X$ be a set.
  Let~$f:\ArXRb$.
  Let~$x\in X$.
  Then, we have
  \begin{equation}
    \label{e:nonneg-and-nonpos-parts-are-orthogonal}
    f^+ (x) = 0 \ (\mbox{and } f^- (x) = - f (x)) \DISJ
    f^- (x) = 0 \ (\mbox{and } f^+ (x) = f (x)).
  \end{equation}
\end{lemma}

\begin{proof}
  Direct consequence of
  \assume{the definition of the maximum},
  \assume{the partition of extended real numbers into negative, zero, and
    positive extended numbers}.
\end{proof}

\begin{lemma}[decomposition into nonnegative and nonpositive parts]
  \label{l:decomp-into-nonneg-and-nonpos-parts}
  \mbox{}\\
  Let~$X$ be a set.
  Let~$f:\ArXRb$.
  Then, we have $f=f^+-f^-$ and $|f|=f^++f^-$.
\end{lemma}

\begin{proof}
  Direct consequence of
  Lemma~\threfc{l:nonneg-and-nonpos-parts-are-orthogonal}{%
    $f^+$ and~$f^-$ cannot take value~$\infty$ at the same point},
  Definition~\thref{d:add-in-rbar}, and
  Definition~\thref{d:abs-in-rbar}.
\end{proof}

\begin{lemma}[compatibility of nonpositive and nonnegative parts with addition]
  \label{l:compat-of-nonpos-and-nonneg-parts-with-add}
  \mbox{}\\
  Let~$X$ be a set.
  Let~$f,g:\ArXRb$ such that their sum is well-defined.
  Then, we have
  \begin{equation}
    \label{e:compat-of-nonpos-and-nonneg-parts-with-add}
    (f + g)^+ + f^- + g^- = (f + g)^- + f^+ + g^+.
  \end{equation}
\end{lemma}

\begin{proof}
  Let~$x\in X$.

  \proofparskip{Case~$f(x)$ and~$g(x)$ finite}
  From
  Lemma~\threfc{l:decomp-into-nonneg-and-nonpos-parts}{%
    with $f+g$, then $f$ and $g$},
  we have
  \begin{equation*}
    (f + g)^+ (x) - (f + g)^- (x)
    = f (x) + g (x)
    = (f^+ (x) - f^- (x)) + (g^+ (x) - g^- (x)).
  \end{equation*}
  Then, from
  Definition~\thref{d:nonneg-and-nonpos-parts},
  $(f+g)^+(x)$, $(f+g)^-(x)$, $f^+(x)$, $f^-(x)$, $g^+(x)$ and~$g^-(x)$, are
  finite.
  Hence, from
  \assume{abelian group properties of~$\matR$},
  we have
  \begin{equation*}
    (f + g)^+ (x) + f^- (x) + g^- (x) = (f + g)^- (x) + f^+ (x) + g^+ (x).
  \end{equation*}

  \proofparskip{Case~$f(x)$ or~$g(x)$ is~$\infty$}
  Then, from
  Definition~\threfc{d:add-in-rbar}{rule~3 cannot occur}, and
  Lemma~\thref{l:nonneg-and-nonpos-parts-are-orthogonal},
  we have
  \begin{equation*}
    (f + g)^+ (x) = f (x) + g (x) = \infty
    \AND
    ( f^+ (x) = \infty \DISJ g^+ (x) = \infty).
  \end{equation*}
  Thus, from
  Lemma~\thref{l:nonneg-and-nonpos-parts-are-nonneg},
  Lemma~\thref{l:infinity-sum-prop-in-rbarplus},
  Lemma~\thref{l:add-in-rbarplus-is-assoc}, and
  Lemma~\thref{l:add-in-rbarplus-is-comm},
  we have
  \begin{equation*}
    (f + g)^+ (x) + f^- (x) + g^- (x) =\infty
    \AND
    (f + g)^- (x) + f^+ (x) + g^+ (x)=\infty.
  \end{equation*}
  Hence,
  Equation~\eqref{e:compat-of-nonpos-and-nonneg-parts-with-add}
  is satisfied.

  \proofparskip{Case~$f(x)$ or~$g(x)$ is~$-\infty$}
  Same reasoning with the nonpositive parts.

  \medskip\noindent
  Therefore, we always have the identity.
\end{proof}

\begin{lemma}[compatibility of nonpositive and nonnegative parts with mask]
  \label{l:compat-of-nonpos-and-nonneg-parts-with-mask}
  \mbox{}\\
  Let~$X$ be a set.
  Let~$A\subset X$.
  Let~$f:\ArXRb$.
  Then, we have $(f\matUN_A)^\pm=f^\pm\matUN_A$
\end{lemma}

\begin{proof}
  Direct consequence of
  Definition~\thref{d:nonneg-and-nonpos-parts}, and
  \assume{nonnegativeness of the indicator function}.
\end{proof}

\begin{lemma}[compatibility of nonpositive and non\-negative parts with
  restriction]
  \label{l:compat-of-nonpos-and-nonneg-parts-with-restr}
  \mbox{}\\
  Let~$X$ be a set.
  Let~$A\subset X$.
  Let~$f:\ArXRb$.
  Then, we have $(\restr{f}{A})^\pm=\restr{f^\pm}{A}$.
\end{lemma}

\begin{proof}
  Direct consequence of
  Definition~\thref{d:nonneg-and-nonpos-parts}, and
  \assume{compatibility of min/max with restriction of function}.
\end{proof}

\chapter{Subset systems}
\label{c:subset-systems}

\minitoc

\begin{remark}
  \label{r:v2-mod1}
  Subset systems of a set~$X$ are subsets of its power set~$\calP(X)$.

  Interesting subset systems are those closed under a family of set
  operations such as complement, union or intersection.
  The simplest one is $\pi$-system that is nonempty and only closed under
  intersection.
  The most elaborate one is $\sigma$-algebra that is somehow closed under all
  set operations, including countable union and intersection.

  Measure theory and Lebesgue integration are usually based on
  $\sigma$-algebra that is the most general subset system concept with which
  we can build desirable properties such as $\sigma$-additivity for measures,
  and linearity and powerful convergence theorems for the integral.
  Thus, the last two sections of this chapter are dedicated to ways to extend
  weaker subset systems, such as set algebra or $\lambda$-system, into a
  $\sigma$-algebra.
\end{remark}

\section{Basic properties}
\label{s:basic-properties}

\begin{lemma}[nonempty and with empty or full]
  \label{l:nonempty-and-empty-or-full}
  \mbox{}\hfill
  Let~$X$ be a set.
  Let~$\calS\subset\calP(X)$.\\
  If~$\calS$ contains either the empty set or the full set,
  then it is nonempty.\\
  Assume that~$\calS$ is nonempty and closed under complement.
  Then, it contains the empty set if it is closed under intersection,
  and it contains the full set if it is closed under union.
\end{lemma}

\begin{proof}
  Direct consequence of
  \assume{the identities $A\setminus A=\emptyset$, and $A\cup A^c=X$}.
\end{proof}

\begin{lemma}[with empty and full]
  \label{l:empty-and-full}
  \mbox{}\\
  Let~$X$ be a set.
  Let~$\calS\subset\calP(X)$.
  Assume that~$\calS$ is closed under complement.\\
  Then, it contains the empty set iff it contains the full set.
\end{lemma}

\begin{proof}
  Direct consequence of
  \assume{the identities $\emptyset^c=X$, and $X^c=\emptyset$}.
\end{proof}

\begin{remark}
  \mbox{}\\
  The local complement is the set difference when the first operand contains
  the second one.
\end{remark}

\begin{lemma}[closedness under local complement and complement]
  \label{l:local-compl-and-compl}
  \mbox{}\hfill
  Let~$X$ be a set.\\
  Let~$\calS\subset\calP(X)$.
  Assume that~$\calS$ contains the full set and is closed under local
  complement.\\
  Then, it is closed under complement.
\end{lemma}

\begin{proof}
  Direct consequence of
  \assume{the identity $A^c=X\setminus A$}.
\end{proof}

\begin{lemma}[closedness under disjoint union and local complement]
  \label{l:union-disj-local-compl-equiv}
  \mbox{}\\
  Let~$X$ be a set.
  Let~$\calS\subset\calP(X)$.
  Assume that~$\calS$ is closed under complement.\\
  Then, it is closed under disjoint union iff
  it is closed under local complement.
\end{lemma}

\begin{proof}
  Direct consequence of
  \assume{the identity $A\setminus B=(A^c\cup B)^c$},
  \assume{monotonicity of complement
    (then $A\cap B=\emptyset\Equiv B\subset A^c$)}, and
  \assume{the identity $A\cup B=(A^c\setminus B)^c$}.
\end{proof}

\begin{lemma}[closedness under set difference and local complement]
  \label{l:set-diff-and-local-compl}
  \mbox{}\\
  Let~$X$ be a set.
  Let~$\calS\subset\calP(X)$.\\
  If~$\calS$ is closed under set difference,
  then it is closed under local complement.\\
  If~$\calS$ is closed under intersection and local complement,
  then it is closed under set difference.
\end{lemma}

\begin{proof}
  Direct consequence of
  \assume{the identity $A\setminus B=A\setminus(A\cap B)$
    with $A\cap B\subset A$}.
\end{proof}

\begin{lemma}[closedness under intersection and set difference]
  \label{l:inter-set-diff-equiv}
  \mbox{}\\
  Let~$X$ be a set.
  Let~$\calS\subset\calP(X)$.\\
  If~$\calS$ is closed under complement and intersection,
  then it is closed under set difference.\\
  If~$\calS$ is closed under set difference,
  then it is closed under intersection.
\end{lemma}

\begin{proof}
  Direct consequence of
  \assume{the identity $A\cap B=A\setminus(A\setminus B)$}.
\end{proof}

\begin{lemma}[closedness under union and intersection]
  \label{l:union-inter-equiv}
  \mbox{}\\
  Let~$X$ be a set.
  Let~$\calS\subset\calP(X)$.
  Assume that~$\calS$ is closed under complement.\\
  Then, it is closed under union iff it is closed under intersection.
\end{lemma}

\begin{proof}
  Direct consequence of
  \assume{De Morgan's laws}.
\end{proof}

\begin{lemma}[closedness under union and set difference]
  \label{l:union-set-diff-equiv}
  \mbox{}\\
  Let~$X$ be a set.
  Let~$\calS\subset\calP(X)$.
  Assume that~$\calS$ is closed under complement.\\
  Then, it is closed under union iff it is closed under set difference.
\end{lemma}

\begin{proof}
  Direct consequence of
  Lemma~\thref{l:union-inter-equiv}, and
  Lem\-ma~\thref{l:inter-set-diff-equiv}.
\end{proof}

\begin{lemma}[closedness under finite operations]
  \label{l:finite-ops-equiv}
  \mbox{}\hfill
  Let~$X$ be a set.
  Let~$\calS\subset\calP(X)$.\\
  $\calS$ is closed under finite intersection iff
  it is closed under intersection.\\
  $\calS$ is closed under finite union iff
  it is closed under union.\\
  $\calS$ is closed under finite disjoint union iff
  it is closed under disjoint union.
\end{lemma}

\begin{proof}
  Direct consequence of
  \assume{induction on the number of operands}.
\end{proof}

\begin{lemma}[closedness under finite union and intersection]
  \label{l:finite-union-inter-equiv}
  \mbox{}\\
  Let~$X$ be a set.
  Let~$\calS\subset\calP(X)$.
  Assume that~$\calS$ is closed under complement.\\
  Then, it is closed under finite union iff
  it is closed under finite intersection, and \\
  it is closed under finite monotone union iff
  it is closed under finite monotone intersection.
\end{lemma}

\begin{proof}
  Direct consequence of
  Lemma~\thref{l:finite-ops-equiv},
  Lemma~\thref{l:union-inter-equiv},
  \assume{monotonicity of complement}, and
  \assume{De Morgan's laws}.
\end{proof}

\begin{remark}
  \mbox{}\\
  Note that obviously, closedness under (finite or countable) union implies
  closedness under (finite or countable) disjoint union.
  In the same way, closedness under finite or countable intersection
  (resp. union) implies closedness under finite or countable monotone
  intersection (resp. union).
\end{remark}

\begin{remark}
  Note that closedness under a countable subset operation actually means that
  the subset system is closed under this operation with at most a countable
  number of operands, {\ie} it is also valid for a finite number ({\eg}~two).
  Except for countable disjoint union (see next lemma).
\end{remark}

\begin{lemma}[closedness under countable and finite disjoint union]
  \label{l:count-and-finite-union-disj}
  \mbox{}\\
  Let~$X$ be a set.
  Let~$\calS\subset\calP(X)$.
  Assume that~$\calS$ contains the empty set and is closed under countable
  disjoint union.
  Then, it is closed under finite disjoint union.
\end{lemma}

\begin{proof}
  Direct consequence of
  \assume{extension of finite disjoint union into countable disjoint union
    using the empty set}.
\end{proof}

\begin{lemma}[closedness under countable disjoint union and local complement]
  \label{l:count-union-disj-local-compl}
  \mbox{}\\
  Let~$X$ be a set.
  Let~$\calS\subset\calP(X)$.
  Assume that~$\calS$ contains the full set and is closed under complement and
  countable disjoint union.
  Then, it is closed under local complement.
\end{lemma}

\begin{proof}
  Direct consequence of
  Lemma~\threfc{l:empty-and-full}{$\emptyset\in\calS$},
  Lemma~\threfc{l:count-and-finite-union-disj}{%
    $\calS$ closed under finite disjoint union},
  Lemma~\threfc{l:finite-ops-equiv}{$\calS$ closed under disjoint union}, and
  Lemma~\thref{l:union-disj-local-compl-equiv}.
\end{proof}

\begin{lemma}[closedness under countable union and intersection]
  \label{l:count-union-inter-equiv}
  \mbox{}\\
  Let~$X$ be a set.
  Let~$\calS\subset\calP(X)$.
  Assume that~$\calS$ is closed under complement.\\
  Then, it is closed under countable union iff
  it is closed under countable intersection,
  and it is closed under countable monotone union iff
  it is closed under countable monotone intersection.
\end{lemma}

\begin{proof}
  Direct consequence of
  \assume{De Morgan's laws}, and
  \assume{monotonicity of complement}.
\end{proof}

\begin{lemma}[closedness under countable disjoint and monotone union]
  \label{l:count-union-disj-and-monot}
  \mbox{}\\
  Let~$X$ be a set.
  Let~$\calS\subset\calP(X)$.
  Assume that~$\calS$ is closed under local complement and countable disjoint
  union.
  Then, it is closed under countable monotone union.
\end{lemma}

\begin{proof}
  Let~$(A_n)_{n\in\matN}\in\calS$.
  Assume that for all $n\in\matN$, $A_n\subset A_{n+1}$.\\
  Let~$B_0\eqdef A_0$, and for all $n\in\matN$, let
  $B_{n+1}\eqdef A_{n+1}\setminus\bigcup_{p\in[0..n]}B_p$.
  Then, from
  Lemma~\thref{l:part-of-count-union},
  the sequence~$(B_n)_{n\in\matN}$ is pairwise disjoint,
  for all $n\in\matN$, we have
  $A_n=\bigcup_{i\in[0..n]}A_i=\biguplus_{i\in[0..n]}B_i$, and
  $\bigcup_{n\in\matN}A_n=\biguplus_{n\in\matN}B_n$.
  Moreover, $B_0=A_0\in\calS$, and for all $n\in\matN$, we have
  $B_{n+1}=A_{n+1}\setminus A_n\in\calS$.
  Hence, $\bigcup_{n\in\matN}A_n\in\calS$.

  Therefore, $\calS$ is closed under countable monotone union.
\end{proof}

\begin{lemma}[closedness under countable monotone and disjoint union]
  \label{l:count-union-monot-and-disj}
  \mbox{}\\
  Let~$X$ be a set.
  Let~$\calS\subset\calP(X)$.
  Assume that~$\calS$ is closed under complement, local complement and
  countable monotone union.
  Then, it is closed under countable disjoint union.
\end{lemma}

\begin{proof}
  Let~$(A_n)_{n\in\matN}\in\calS$.
  Assume that for all $p,q\in\matN$, $p\not=q$ implies
  $A_p\cap A_q=\emptyset$.\\
  For all $n\in\matN$, let $B_n\eqdef\biguplus_{p\in[0..n]}A_p$.
  Then, from
  \assume{properties of union},
  the sequence~$(B_n)_{n\in\matN}$ is nondecreasing, for all $n\in\matN$,
  we have $\biguplus_{p\in[0..n]}A_p=\bigcup_{p\in[0..n]}B_p=B_n$, and
  $\biguplus_{n\in\matN}A_n=\bigcup_{n\in\matN}B_n$.
  Moreover, $B_0=A_0\in\calS$, and from
  \assume{De Morgan's laws}, and
  \assume{the definition of set difference},
  for all $n\in\matN$, we have
  $B_n\cap A_{n+1}=\biguplus_{p\in[0..n]}A_p\cap A_{n+1}=\emptyset$
  ({\ie} $B_n\subset A_{n+1}^c$), and
  \begin{equation*}
    B_{n + 1}
    = A_{n + 1} \uplus \biguplus_{p \in [0..n]} A_p
    = A_{n + 1} \uplus B_n
    = (A_{n + 1}^c \cap B_n^c)^c
    = (A_{n + 1}^c \setminus B_n)^c.
  \end{equation*}
  Then, from a trivial induction, for all $n\in\matN$, we have $B_n\in\calS$.
  Hence, $\biguplus_{n\in\matN}A_n\in\calS$.

  Therefore, $\calS$ is closed under countable disjoint union.
\end{proof}

\begin{lemma}[closedness under countable disjoint union and countable union]
  \label{l:count-union-disj-and-union}
  \mbox{}\\
  Let~$X$ be a set.
  Let~$\calS\subset\calP(X)$.
  Assume that~$\calS$ is closed under complement, intersection and
  countable disjoint union.
  Then, it is closed under countable union.
\end{lemma}

\begin{proof}
  From
  Lemma~\thref{l:union-inter-equiv}, and
  Lemma~\thref{l:finite-ops-equiv},
  $\calS$ is closed under finite union.

  Let~$(A_n)_{n\in\matN}\in\calS$.
  Let~$B_0\eqdef A_0$, and for all $n\in\matN$, let
  $B_{n+1}\eqdef A_{n+1}\setminus\bigcup_{p\in[0..n]}B_p$.
  Then, from
  Lemma~\thref{l:part-of-count-union},
  the sequence~$(B_n)_{n\in\matN}$ is pairwise disjoint,
  for all $n\in\matN$, we have
  $\bigcup_{p\in[0..n]}A_p=\biguplus_{p\in[0..n]}B_p$, and
  $\bigcup_{n\in\matN}A_n=\biguplus_{n\in\matN}B_n$.
  Moreover, $B_0\in\calS$, and from
  Lemma~\thref{l:inter-set-diff-equiv},
  for all $n\in\matN$, we have
  $B_{n+1}=A_{n+1}\setminus\bigcup_{p\in[0..n]}A_p\in\calS$.
  Hence, $\bigcup_{n\in\matN}A_n\in\calS$.

  Therefore, $\calS$ is closed under countable union.
\end{proof}

\begin{lemma}[closedness under countable monotone union and countable union]
  \label{l:count-union-monot-and-union}
  \mbox{}\\
  Let~$X$ be a set.
  Let~$\calS\subset\calP(X)$.
  Assume that~$\calS$ is closed under complement, union and countable monotone
  union.
  Then, it is closed under countable union.
\end{lemma}

\begin{proof}
  Direct consequence of
  Lemma~\threfc{l:union-inter-equiv}{$\calS$ is closed under intersection},
  Lemma~\threfc{l:inter-set-diff-equiv}{%
    $\calS$ is closed under set difference},
  Lemma~\threfc{l:set-diff-and-local-compl}{%
    $\calS$ is closed under local set difference},
  Lemma~\threfc{l:count-union-monot-and-disj}{%
    $\calS$ is closed under disjoint union}, and
  Lemma~\thref{l:count-union-disj-and-union}.
\end{proof}

\clearpage
\section{Pi-system}
\label{s:p-syst}

\begin{definition}[$\pi$-system]
  \label{d:p-syst}
  \mbox{}\hfill
  Let~$X$ be a set.\\
  A subset~$\Pi$ of~$\calP(X)$ is called {\em $\pi$-system on~$X$} iff
  it is nonempty and closed under finite intersection:
  \begin{align}
    \label{e:p-syst-1}
    & \Pi \not= \emptyset,\\
    \label{e:p-syst-2}
    \forall n \in \matN,\;
    \forall (A_p)_{p \in [0..n]} \in \Pi,\quad &
    \bigcap_{p \in [0..n]} A_p \in \Pi.
  \end{align}
\end{definition}

\begin{remark}
  Note that adding the empty set to a $\pi$-system keeps it a $\pi$-system.
  We may replace in the previous definition the nonempty
  condition~\eqref{e:p-syst-1} by the ``contain-the-empty-set'' condition, and
  still maintain the {\Dplt} (see Theorem~\ref{t:dynkin-pi-lambda-th}).
\end{remark}

\begin{remark}
  \label{r:v2-new03}
  The following set of four lemmas and a definition is written in an almost
  identical manner for $\pi$-systems
  (statements~\ref{l:inter-of-p-systs}--\ref{l:p-syst-gen-is-idem}),
  set algebras (\ref{l:inter-of-set-algs}--\ref{l:set-alg-gen-is-idem}),
  monotone classes
  (\ref{l:inter-of-monot-classes}--\ref{l:monot-class-gen-is-idem}),
  $\lambda$-systems (\ref{l:inter-of-l-systs}--\ref{l:l-syst-gen-is-idem}),
  and $\sigma$-algebras
  (\ref{l:inter-of-sigma-algs}--\ref{l:sigma-alg-gen-is-idem}).
  They all derive from the definition and properties of intersection, and
  reflexivity of inclusion, thus allowing for generated systems that are
  minimum, satisfy monotonicity and idempotence.
  The very short proofs are almost identical.

  Note that, unlike subsequent subset systems, the $\pi$-systems require
  nonemptiness, that shows up in all statements.
\end{remark}

\begin{lemma}[intersection of $\pi$-systems]
  \label{l:inter-of-p-systs}
  \mbox{}\hfill
  Let~$X$ and~$I$ be sets.
  Let~$(\Pi_i)_{i\in I}$ be $\pi$-systems on~$X$.
  Then, $\bigcap_{i\in I}\Pi_i$ is closed under intersection, {\ie} it is a
  $\pi$-system iff it is nonempty.
\end{lemma}

\begin{proof}
  Direct consequence of
  Definition~\thref{d:p-syst}, and
  \assume{the definition of intersection}.
\end{proof}

\begin{definition}[generated $\pi$-system]
  \label{d:gen-p-syst}
  \mbox{}\\
  Let~$X$ be a set.
  Let~$G\subset\calP(X)$.
  Assume that~$G\not=\emptyset$.
  The {\em $\pi$-system generated by~$G$} is the intersection of all
  $\pi$-systems on~$X$ containing~$G$;
  it is denoted~$\Pi_X(G)$.
\end{definition}

\begin{lemma}[generated $\pi$-system is minimum]
  \label{l:gen-p-syst-is-min}
  \mbox{}\hfill
  Let~$X$ be a set.
  Let~$G\subset\calP(X)$.
  Assume that~$G\not=\emptyset$.
  Then, $\Pi_X(G)$~is the smallest $\pi$-system on~$X$ containing~$G$.
\end{lemma}

\begin{proof}
  Direct consequence of
  Definition~\thref{d:gen-p-syst},
  Lemma~\thref{l:inter-of-p-systs}, and
  \assume{properties of the intersection}.
\end{proof}

\begin{lemma}[$\pi$-system generation is monotone]
  \label{l:p-syst-gen-is-monot}
  \mbox{}\hfill
  Let~$X$ be a set.
  Let~$G_1,G_2\subset\calP(X)$.\\
  Assume that $\emptyset\not=G_1\subset G_2$.
  Then, we have $\Pi_X(G_1)\subset\Pi_X(G_2)$.
\end{lemma}

\begin{proof}
  Direct consequence of
  \assume{the definition of inclusion ($G_2\not=\emptyset$)}, and
  Lemma~\threfc{l:gen-p-syst-is-min}{%
    with $G\eqdef G_2$, then $G\eqdef G_1$}.
\end{proof}

\begin{lemma}[$\pi$-system generation is idempotent]
  \label{l:p-syst-gen-is-idem}
  \mbox{}\\
  Let~$X$ be a set.
  Let~$\calS\subset\calP(X)$.
  Assume that $\calS\not=\emptyset$.
  Then, $\calS$ is a $\pi$-system on~$X$ iff $\Pi_X(\calS)=\calS$.
\end{lemma}

\begin{proof}
  Direct consequence of
  \assume{reflexivity of inclusion}, and
  Lemma~\threfc{l:gen-p-syst-is-min}{%
    $\calS$ and $\Pi_X(\calS)$ are both $\pi$-systems containing $\calS$}.
\end{proof}

\clearpage
\section{Set algebra}
\label{s:set-alg}

\begin{remark}
  \label{r:v2-mod2}
  The following concept of ``set algebra'' is not to be confused with the
  algebraic structure ``algebra over a field'' defined in
  Section~\ref{ss:algebra-over-a-field}.

  Note that the same concept is sometimes called ``field (of sets)'', which is
  of course not to be confused with the algebraic structure either.
\end{remark}

\begin{definition}[set algebra]
  \label{d:set-alg}
  \mbox{}\hfill
  Let~$X$ be a set.
  A subset~$\calA$ of~$\calP(X)$ is called {\em set algebra on~$X$} iff
  it contains the empty set,
  it is closed under complement, and
  under finite union:
  \begin{align}
    \label{e:set-alg-1}
    & \emptyset \in \calA,\\
    \label{e:set-alg-2}
    \forall A \in \calA,\quad &
    A^c \in \calA,\\
    \label{e:set-alg-3}
    \forall n \in \matN,\;
    \forall (A_i)_{i \in [0..n]} \in \calA,\quad &
    \bigcup_{i \in [0..n]} A_i \in \calA.
  \end{align}
\end{definition}

\begin{lemma}[equivalent definition of set algebra]
  \label{l:equiv-def-of-set-alg}
  \mbox{}\\
  Let~$X$ be a set.
  Let~$\calA\subset\calP(X)$.
  Then, $\calA$~is a set algebra on~$X$ iff
  \eqref{e:set-alg-2} holds and
  \begin{gather}
    \label{e:equiv-def-of-set-alg-1}
    \emptyset \in \calA \DISJ X \in \calA \DISJ \calA \not= \emptyset,\\
    \label{e:equiv-def-of-set-alg-2}
    \forall A, B \in \calA,\;
    A \cup B \in \calA \DISJ
    \forall A, B \in \calA,\;
    A \cap B \in \calA.
  \end{gather}
\end{lemma}

\begin{proof}
  Direct consequence of
  Definition~\thref{d:set-alg},
  Lemma~\thref{l:nonempty-and-empty-or-full},
  Lemma~\thref{l:empty-and-full},
  Lemma~\thref{l:finite-ops-equiv}, and
  Lemma~\thref{l:finite-union-inter-equiv}.
\end{proof}

\begin{lemma}[other equivalent definition of set algebra]
  \label{l:other-equiv-def-of-set-alg}
  \mbox{}\hfill
  Let~$X$ be a set.
  Let~$\calA\subset\calP(X)$.
  Then, $\calA$~is a set algebra on~$X$ iff
  it contains the full set~$X$, and
  it is closed under set difference:
  \begin{align}
    \label{e:other-equiv-def-of-set-alg-1}
    & X \in \calA.\\
    \label{e:other-equiv-def-of-set-alg-2}
    \forall A, B \in \calA,\quad & A \setminus B \in \calA.
  \end{align}
\end{lemma}

\begin{proof}
  Direct consequence of
  Lemma~\thref{l:equiv-def-of-set-alg},
  Lemma~\threfc{l:inter-set-diff-equiv}{both ways},
  Lemma~\thref{l:set-diff-and-local-compl}, and
  Lemma~\thref{l:local-compl-and-compl}.
\end{proof}

\begin{lemma}[set algebra is closed under local complement]
  \label{l:set-alg-is-closed-under-local-compl}
  \mbox{}\\
  Let~$X$ be a set.
  Let~$\calA$ be a set algebra on~$X$.
  Then, $\calA$~is closed under local complement.
\end{lemma}

\begin{proof}
  Direct consequence of
  Lemma~\thref{l:other-equiv-def-of-set-alg}, and
  Lem\-ma~\thref{l:set-diff-and-local-compl}.
\end{proof}

\begin{lemma}[intersection of set algebras]
  \label{l:inter-of-set-algs}
  \mbox{}\\
  Let~$X$ and~$I$ be sets.
  Let~$(\calA_i)_{i\in I}$ be set algebras on~$X$.
  Then, $\bigcap_{i\in I}\calA_i$ is a set algebra on~$X$.
\end{lemma}

\begin{proof}
  Direct consequence of
  Definition~\thref{d:set-alg}, and
  \assume{the definition of intersection}.
\end{proof}

\begin{definition}[generated set algebra]
  \label{d:gen-set-alg}
  \mbox{}\hfill
  Let~$X$ be a set.
  Let~$G\subset\calP(X)$.
  The {\em set algebra generated by~$G$} is the intersection of all set
  algebras on~$X$ containing~$G$;
  it is denoted~$\calA_X(G)$.
\end{definition}

\begin{lemma}[generated set algebra is minimum]
  \label{l:gen-set-alg-is-min}
  \mbox{}\\
  Let~$X$ be a set.
  Let~$G\subset\calP(X)$.
  Then, $\calA_X(G)$~is the smallest set algebra on~$X$ containing~$G$.
\end{lemma}

\begin{proof}
  Direct consequence of
  Definition~\thref{d:gen-set-alg},
  Lemma~\thref{l:inter-of-set-algs}, and
  \assume{properties of the intersection}.
\end{proof}

\begin{lemma}[set algebra generation is monotone]
  \label{l:set-alg-gen-is-monot}
  \mbox{}\\
  Let~$X$ be a set.
  Let~$G_1,G_2\subset\calP(X)$.
  Assume that $G_1\subset G_2$.
  Then, we have $\calA_X(G_1)\subset\calA_X(G_2)$.
\end{lemma}

\begin{proof}
  Lemma~\threfc{l:gen-set-alg-is-min}{%
    with $G\eqdef G_2$, then $G\eqdef G_1$}.
\end{proof}

\begin{lemma}[set algebra generation is idempotent]
  \label{l:set-alg-gen-is-idem}
  \mbox{}\\
  Let~$X$ be a set.
  Let~$\calS\subset\calP(X)$.
  Then, $\calS$ is a set algebra on~$X$ iff $\calA_X(\calS)=\calS$.
\end{lemma}

\begin{proof}
  Direct consequence of
  \assume{reflexivity of inclusion},
  Lemma~\threfc{l:gen-set-alg-is-min}{%
    $\calS$ and $\calA_X(\calS)$ are both set algebras containing $\calS$}.
\end{proof}

\begin{lemma}[partition of countable union in set algebra]
  \label{l:part-of-count-union-in-set-alg}
  \mbox{}\\
  Let~$X$ be a set.
  Let~$\calA$ be a set algebra on~$X$.
  Let~$(A_n)_{n\in\matN}\subset\calA$.
  Let~$B_0\eqdef A_0$, and for all $n\in\matN$, let
  $B_{n+1}\eqdef A_{n+1}\setminus\bigcup_{i\in[0..n]}B_i$.
  Then, we have
  \begin{align}
    \label{e:part-of-count-union-in-set-alg-1}
    \forall n \in \matN,\quad & B_n \in \calA,\\
    \label{e:part-of-count-union-in-set-alg-2}
    \forall m, n \in \matN,\quad &
    m \not= n \IMPLIES B_m \cap B_n = \emptyset,\\
    \label{e:part-of-count-union-in-set-alg-3}
    \forall n \in \matN,\quad &
    \bigcup_{i \in [0..n]} A_i = \biguplus_{i \in [0..n]} B_i \in \calA.
  \end{align}
\end{lemma}

\begin{proof}
  For all $n\in\matN$, let $P(n)$ be the property:
  $(\forall i\in\matN,\;i\leq n\Implies B_i\in\calA)
  \Conj\bigcup_{i\in[0..n]}B_i\in\calA$.\\
  \proofpar{Induction: $P(0)$}
  Trivial.\\
  \proofpar{Induction: $P(n)$ implies $P(n+1)$}
  Direct consequence of
  Lemma~\threfc{l:other-equiv-def-of-set-alg}{closedness under set difference},
  \assume{associativity of intersection}, and
  Lemma~\threfc{l:equiv-def-of-set-alg}{closedness under union}.\\
  Hence, $P(n)$~holds for all~$n\in\matN$.

  Therefore, from
  Lemma~\thref{l:part-of-count-union},
  all three properties hold.
\end{proof}

\begin{lemma}[explicit set algebra]
  \label{l:explicit-set-alg}
  \mbox{}\hfill
  Let~$X$ be a set.
  Let~$G\subset\calP(X)$.\\
  Assume that~$G$ contains the full set, is closed under intersection, and
  satisfies the property:
  \begin{equation}
    \label{e:explicit-set-alg-1}
    \forall A \in G,\;
    \exists B_1, B_2 \in G,\quad
    B_1 \cap B_2 = \emptyset
    \CONJ
    A^c = B_1 \uplus B_2.
  \end{equation}
  Then, the set algebra generated by~$G$ is the set of finite disjoint unions
  of elements of~$G$:
  \begin{align}
    \nonumber
    \calA_X (G)
    = \left\{ \biguplus_{p \in [0..n]} A_p \rightst
    & n \in \matN
      \CONJ \forall p \in [0..n],\;
      A_p \in G\\
    \label{e:explicit-set-alg-2}
    &
      \left.
      \vphantom{\biguplus_{p \in [o..n]} A_p}
      \Conj\quad \forall p, q \in [0..n],\;
      p \not= q
      \Implies A_p \cap A_q = \emptyset \right\}.
  \end{align}
\end{lemma}

\begin{proof}
  Let~$\calA\eqdef\calA_X(G)$.
  Let~$\calU$ be the set of finite disjoint unions of elements of~$G$.

  \proofparskip{(0). $\calU\subset\calA$}
  From
  Lemma~\threfc{l:gen-set-alg-is-min}{$G\subset\calA$}, and
  Definition~\threfc{d:set-alg}{$\calA$~is closed under finite union},
  we have $\calU\subset\calA$.

  \proofparskip{(1a). $G\subset\calU$}
  Direct consequence of
  the definition of~$\calU$ (with $n\eqdef 0$).

  \proofparskip{(1b). $X\in\calU$}
  Direct consequence of~(1a).

  \proofparskip{(1c). $\calU$ is closed under complement}
  Let~$A\in\calU$.

  From the definition of~$\calU$, let $n\in\matN$ and
  $(A_p)_{p\in[0..n]}\in G$ such that for all $p,q\in[0..n]$, $p\not=q$
  implies $A_p\cap A_q=\emptyset$, and $A=\biguplus_{p\in[0..n]}A_p$.
  Let~$p\in[0..n]$.
  Then, from~\eqref{e:explicit-set-alg-1},  let $B_p^0,B_p^1\in G$ such that
  $B_p^0\cap B_p^1=\emptyset$ and $A_p^c=B_p^0\uplus B_p^1$.

  Let~$I\eqdef\{0,1\}^{[0..n]}$ (its cardinality is~$2^{n+1}$).
  For all $\fhi\in I$, let $C^\fhi\eqdef\bigcap_{p\in[0..n]}B_p^{\fhi(p)}$.\\
  Let~$\fhi\in I$.
  Then, from
  Lemma~\threfc{l:finite-ops-equiv}{$G$ is closed under finite intersection},
  we have $C^\fhi\in G$.
  Let~$\psi\in I$.
  Assume that $\fhi\not=\psi$, {\ie} there exists $q\in[0..n]$ such that
  $\fhi(q)\not=\psi(q)$.
  Then, from
  \assume{the definition of intersection},
  we have $C^\fhi\subset B_q^{\fhi(q)}$ and $C^\psi\subset B_q^{\psi(q)}$.
  Hence, from
  \assume{monotonicity of intersection},
  we have
  \begin{equation*}
    C^\fhi \cap C^\psi
    \subset B_q^{\fhi (q)} \cap B_q^{\psi (q)}
    = \emptyset.
  \end{equation*}
  Moreover, from
  \assume{De~Morgan's laws}, and
  \assume{distributivity of intersection over union},
  we have
  \begin{equation*}
    A^c
    = \bigcap_{p \in [0..n]} \left( B_p^0 \uplus B_p^1 \right)
    = \bigcup_{\fhi \in I} \left(
      \bigcap_{p \in [0..n]} B_p^{\fhi (p)} \right)
    = \biguplus_{\fhi \in I} C^\fhi.
  \end{equation*}
  Thus, $A^c\in\calU$.
  Hence, $\calU$~is closed under complement.

  \proofparskip{(1d). $\calU$ is closed under intersection}
  Let~$A^0,A^1\in\calU$.

  Let~$\ialf\in\{0,1\}$.
  From the definition of~$\calU$, let $n^\ialf\in\matN$, let
  $(A^\ialf_{p^\ialf})_{p^\ialf\in[0..n^\ialf]}\in G$ such that for all
  $p^\ialf,q^\ialf\in[0..n^\ialf]$, $p^\ialf\not=q^\ialf$ implies
  $A^\ialf_{p^\ialf}\cap A^\ialf_{q^\ialf}=\emptyset$, and
  $A^\ialf=\biguplus_{p^\ialf\in[0..n^\ialf]}A^\ialf_{p^\ialf}$.

  For all $p^0\in[0..n^0]$, for all $p^1\in[0..n^1]$, let
  $B_{p^0,p^1}\eqdef A^0_{p^0}\cap A^1_{p^1}\in G$.
  Let~$n\eqdef n^0n^1+n^0+n^1$.
  Let~$\fhi:[0..n]\to[0..n^0]\times[0..n^1]$ be a bijection (their
  common cardinality is $n+1=(n^0+1)(n^1+1)$).
  Let~$p,q\in[0..n]$.
  Assume that $p\not=q$.
  Let~$p^0,q^0\in[0..n^0]$ and $p^1,q^1\in[0..n^1]$ such that
  \begin{equation*}
    (p^0, p^1) = \fhi (p)
    \AND
    (q^0, q^1) = \fhi (q).
  \end{equation*}
  Then, from
  \assume{the definition of bijection and injection},
  we have $p^0\not=q^0$ or $p^1\not=q^1$, {\ie}
  $A^0_{p^0}\cap A^0_{q^0}=\emptyset$ or $A^1_{p^1}\cap A^1_{q^1}=\emptyset$.
  Thus, from
  \assume{associativity and commutativity of intersection}, and since
  \assume{$\emptyset$ is absorbing for intersection},
  we have
  \begin{equation*}
    B_{\fhi (p)} \cap B_{\fhi (q)}
    = (A^0_{p^0} \cap A^1_{p^1}) \cap (A^0_{q^0} \cap A^1_{q^1})
    = (A^0_{p^0} \cap A^0_{q^0}) \cap (A^1_{p^1} \cap A^1_{q^1})
    = \emptyset.
  \end{equation*}
  Moreover, from
  \assume{left and right distributivity of intersection over union}, and
  \assume{associativity and commutativity of union},
  we have
  \begin{equation*}
    A^0 \cap A^1
    = \left( \bigcup_{p^0 \in [0..n^0]} A^0_{p^0} \right)
    \cap \left( \bigcup_{p^1 \in [0..n^1]} A^1_{p^1} \right)
    = \bigcup_{(p^0,p^1) \in [0..n^0] \times [0..n^1]} B_{p^0, p^1}
    = \biguplus_{p \in [0..n]} B_{\fhi (p)}.
  \end{equation*}
  Thus, $A^0\cap A^1\in\calU$.
  Hence, $\calU$~is closed under intersection.

  \proofparskip{(2). $\calA\subset\calU$}
  From~(1b), (1c), (1d),
  Lemma~\threfc{l:equiv-def-of-set-alg}{$\calU$ is set algebra},
  (1a), and
  Lemma~\thref{l:gen-set-alg-is-min},
  we have $\calA\subset\calU$.

  \medskip\noindent
  Therefore, from~(0) and~(2), we have $\calA=\calU$.
\end{proof}

\clearpage
\section{Monotone class}
\label{s:monot-class}

\begin{definition}[monotone class]
  \label{d:monot-class}
  \mbox{}\hfill
  Let~$X$ be a set.
  A subset~$\calC$ of~$\calP(X)$ is called\\ {\em monotone class on~$X$} iff
  it is closed under countable monotone union and intersection:
  \begin{align}
    \label{e:monot-class-1}
    \forall (A_n)_{n \in \matN} \in \calC,\quad &
    (\forall n \in \matN,\; A_n \subset A_{n + 1})
    \IMPLIES
    \bigcup_{n \in \matN} A_n \in \calC,\\
    \label{e:monot-class-2}
    \forall (A_n)_{n \in \matN} \in \calC,\quad &
    (\forall n \in \matN,\; A_n \supset A_{n + 1})
    \IMPLIES
    \bigcap_{n \in \matN} A_n \in \calC.
  \end{align}
\end{definition}

\begin{lemma}[intersection of monotone classes]
  \label{l:inter-of-monot-classes}
  \mbox{}\hfill
  Let~$X$ and~$I$ be sets.\\
  Let~$(\calC_i)_{i\in I}$ be a family of monotone classes on~$X$.
  Then, $\bigcap_{i\in I}\calC_i$ is a monotone class on~$X$.
\end{lemma}

\begin{proof}
  \mbox{}\\
  Direct consequence of
  Definition~\thref{d:monot-class}, and
  \assume{the definition of intersection}.
\end{proof}

\begin{definition}[generated monotone class]
  \label{d:gen-monot-class}
  \mbox{}\\
  Let~$X$ be a set.
  Let~$G\subset\calP(X)$.
  The {\em monotone class generated by~$G$} is the intersection of all
  monotone classes on~$X$ containing~$G$;
  it is denoted by $\calC_X(G)$.
\end{definition}

\begin{lemma}[generated monotone class is minimum]
  \label{l:gen-monot-class-is-min}
  \mbox{}\\
  Let~$X$ be a set.
  Let~$G\subset\calP(X)$.
  Then, $\calC_X(G)$ is the smallest monotone class on~$X$ containing~$G$.
\end{lemma}

\begin{proof}
  Direct consequence of
  Definition~\thref{d:gen-monot-class},
  Lemma~\thref{l:inter-of-monot-classes}, and
  \assume{properties of the intersection}.
\end{proof}

\begin{lemma}[monotone class generation is monotone]
  \label{l:monot-class-gen-is-monot}
  \mbox{}\\
  Let~$X$ be a set.
  Let~$G_1,G_2\subset\calP(X)$.
  Assume that $G_1\subset G_2$.
  Then, we have $\calC_X(G_1)\subset\calC_X(G_2)$.
\end{lemma}

\begin{proof}
  Lemma~\threfc{l:gen-monot-class-is-min}{%
    with $G\eqdef G_2$, then $G\eqdef G_1$}.
\end{proof}

\begin{lemma}[monotone class generation is idempotent]
  \label{l:monot-class-gen-is-idem}
  \mbox{}\\
  Let~$X$ be a set.
  Let~$\calS\subset\calP(X)$.
  Then, $\calS$ is a monotone class on~$X$ iff $\calC_X(\calS)=\calS$.
\end{lemma}

\begin{proof}
  Direct consequence of
  \assume{reflexivity of inclusion},
  Lemma~\threfc{l:gen-monot-class-is-min}{%
    $\calS$ and $\calC_X(\calS)$ are both monotone classes containing $\calS$}.
\end{proof}

\begin{definition}[monotone class and symmetric set difference]
  \label{d:monot-class-and-symm-set-diff}
  \mbox{}\hfill
  Let~$X$ be a set.\\
  Let~$\calC$ be a monotone class on~$X$.
  For all $A\subset X$, let~$\calCdiff_A$ be the subset system defined by
  \begin{equation}
    \label{e:monot-class-and-set-diff}
    \calCdiff_A \eqdef \{
      B \subset X \leftst
      A \setminus B \in \calC
      \Conj B \setminus A \in \calC \right.
  \}.
  \end{equation}
\end{definition}

\begin{lemma}[$\calCdiff$ is symmetric]
  \label{l:c-diff-is-symmetric}
  \mbox{}\hfill
  Let~$X$ be a set.\\
  Let~$\calC$ be a monotone class on~$X$.
  Then, for all $A,B\subset X$, we have $B\in\calCdiff_A$ iff
  $A\in\calCdiff_B$.
\end{lemma}

\begin{proof}
  Direct consequence of
  Definition~\thref{d:monot-class-and-symm-set-diff}.
\end{proof}

\begin{lemma}[$\calCdiff$ is monotone class]
  \label{l:c-diff-is-monot-class}
  \mbox{}\hfill
  Let~$X$ be a set.\\
  Let~$\calC$ be a monotone class on~$X$.
  Then, for all $A\subset X$, $\calCdiff_A$ is a monotone class on~$X$.
\end{lemma}

\begin{proof}
  Let~$A\subset X$.
  Let~$(B_n)_{n\in\matN}\in\calCdiff_A$.
  Let~$n\in\matN$.
  Then, from
  Definition~\thref{d:monot-class-and-symm-set-diff},
  we have $A\setminus B_n\in\calC$ and $B_n\setminus A\in\calC$.

  Assume first that, for all $n\in\matN$, $B_n\subset B_{n+1}$.
  Let~$n\in\matN$.
  Then, from
  \assume{monotonicity of set difference},
  we have $A\setminus B_n\supset A\setminus B_{n+1}$ and
  $B_n\setminus A\subset B_{n+1}\setminus A$.
  Let~$B\eqdef\bigcup_{n\in\matN}B_n$.
  Then, from
  \assume{the definition of set difference},
  \assume{De~Morgan's laws},
  Definition~\thref{d:monot-class-and-symm-set-diff}, and
  Definition~\thref{d:monot-class},
  we have
  $A\setminus B=\bigcap_{n\in\matN}(A\setminus B_n)\in\calC$ and
  $B\setminus A=\bigcup_{n\in\matN}(B_n\setminus A)\in\calC$.
  Thus, from
  Definition~\thref{d:monot-class-and-symm-set-diff},
  we have $B\in\calCdiff_A$.
  Hence, $\calCdiff_A$~is closed under nondecreasing union.

  Assume now that, for all $n\in\matN$, $B_n\supset B_{n+1}$.
  Let~$n\in\matN$.
  Then, from
  \assume{monotonicity of set difference},
  we have $A\setminus B_n\subset A\setminus B_{n+1}$ and
  $B_n\setminus A\supset B_{n+1}\setminus A$.
  Let~$B\eqdef\bigcap_{n\in\matN}B_n$.
  Then, from
  \assume{the definition of set difference},
  \assume{De~Morgan's laws},
  Definition~\thref{d:monot-class-and-symm-set-diff}, and
  Definition~\thref{d:monot-class},
  we have
  $A\setminus B=\bigcup_{n\in\matN}(A\setminus B_n)\in\calC$ and
  $B\setminus A=\bigcap_{n\in\matN}(B_n\setminus A)\in\calC$.
  Thus, from
  Definition~\thref{d:monot-class-and-symm-set-diff},
  we have $B\in\calCdiff_A$.
  Hence, $\calCdiff_A$~is closed under nonincreasing intersection.

  Therefore, from
  Definition~\thref{d:monot-class},
  $\calCdiff_A$~is a monotone class on~$X$.
\end{proof}

\begin{lemma}[monotone class is closed under set difference]
  \label{l:monot-class-is-closed-under-set-diff}
  \mbox{}\\
  Let~$X$ be a set.
  Let~$G\subset\calP(X)$.
  Assume that~$G$ is closed under set difference.\\
  Then, $\calC_X(G)$ is closed under set difference.
\end{lemma}

\begin{proof}
  For all~$A\subset X$, we use the simplified notation
  $\calCdiff_A\eqdef(\calC_X(G))_A^{\setminus}$.

  \proofparskip{(1). $\forall A\in G,\;\calC_X(G)\subset\calCdiff_A$}
  Let~$A\in G$.\\
  Let~$B\in G$.
  Then, we have $A\setminus B,\ B\setminus A\in G$.
  Thus, from
  Lemma~\threfc{l:gen-monot-class-is-min}{$G\subset\calC_X(G)$}, and
  Definition~\thref{d:monot-class-and-symm-set-diff},
  we have $B\in\calCdiff_A$, {\ie} $G\subset\calCdiff_A$.
  Hence, from
  Lemma~\thref{l:c-diff-is-monot-class}, and
  Lemma~\thref{l:gen-monot-class-is-min},
  we have $\calC_X(G)\subset\calCdiff_A$.

  \proofparskip{(2).
    $\forall B\in\calC_X(G),\;\calC_X(G)\subset\calCdiff_B$}
  Let~$B\in\calC_X(G)$.\\
  Let~$A\in G$.
  Then, from~(1), we have $B\in\calCdiff_A$, and from
  Lemma~\thref{l:c-diff-is-symmetric},
  we have $A\in\calCdiff_B$, {\ie} $G\subset\calCdiff_B$.
  Hence, from
  Lemma~\thref{l:c-diff-is-monot-class}, and
  Lemma~\thref{l:gen-monot-class-is-min},
  we have $\calC_X(G)\subset\calCdiff_B$.

  \medskip\noindent
  Let~$A,B\in\calC_X(G)$.
  Then, from~(2), we have $B\in\calCdiff_A$, and from
  Definition~\thref{d:monot-class-and-symm-set-diff},
  we have $A\setminus B\in\calC_X(G)$.

  \medskip\noindent
  Therefore, $\calC_X(G)$~is closed under set difference.
\end{proof}

\begin{lemma}[monotone class generated by set algebra]
  \label{l:monot-class-gen-by-set-alg}
  \mbox{}\hfill
  Let~$X$ be a set.
  Let~$G\subset\calP(X)$.\\
  Assume that~$G$ is a set algebra on~$X$.
  Then, $\calC_X(G)$ is a set algebra on~$X$.
\end{lemma}

\begin{proof}
  Direct consequence of
  Lemma~\threfc{l:other-equiv-def-of-set-alg}{%
    $G$ contains the full set, and is closed under set difference},
  Lemma~\threfc{l:gen-monot-class-is-min}{%
    $X\in G\subset\calC_X(G)$},
  Lemma~\threfc{l:monot-class-is-closed-under-set-diff}{%
    $\calC_X(G)$ is closed under set difference}, and
  Lemma~\thref{l:other-equiv-def-of-set-alg}.
\end{proof}

\clearpage
\section{Lambda-system}
\label{s:l-syst}

\begin{definition}[$\lambda$-system]
  \label{d:l-syst}
  \mbox{}\\
  Let~$X$ be a set.
  A subset~$\Lambda$ of~$\calP(X)$ is called {\em $\lambda$-system on~$X$}, or
  {\em Dynkin-system on~$X$} iff
  it contains the full set,
  it is closed under complement, and
  under countable disjoint union:
  \begin{align}
    \label{e:l-syst-1}
    & X \in \Lambda,\\
    \label{e:l-syst-2}
    \forall A \in \Lambda,\quad &
    A^c \in \Lambda,\\
    \label{e:l-syst-3}
    \forall (A_n)_{n \in \matN} \in \Lambda,\quad &
    (\forall p, q \in \matN,\;
      p \not= q \Implies A_p \cap A_q = \emptyset)
    \IMPLIES
    \biguplus_{n \in \matN} A_n \in \Lambda.
  \end{align}
\end{definition}

\begin{lemma}[equivalent definition of $\lambda$-system]
  \label{l:equiv-def-of-l-syst}
  \mbox{}\\
  Let~$X$ be a set.
  Let~$\Lambda\subset\calP(X)$.
  Then, $\Lambda$ is a $\lambda$-system on~$X$ iff
  \eqref{e:l-syst-1} holds, and
  it is closed under local complement, and
  under countable monotone union:
  \begin{align}
    \label{e:equiv-def-of-l-syst-1}
    \forall A, B \in \Lambda,\quad &
    B \subset A \Rightarrow A \setminus B \in \Lambda,\\
    \label{e:equiv-def-of-l-syst-2}
    \forall (A_n)_{n \in \matN} \in \Lambda,\quad &
    (\forall n \in \matN,\; A_n \subset A_{n + 1})
    \IMPLIES
    \bigcup_{n \in \matN} A_n \in \Lambda.
  \end{align}
\end{lemma}

\begin{proof}
  \proofpar{``Left'' implies ``right''}\\
  Direct consequence of
  Definition~\thref{d:l-syst},
  Lemma~\threfc{l:count-union-disj-local-compl}{%
    $\Lambda$ is closed under local complement}, and
  Lemma~\threfc{l:count-union-disj-and-monot}{%
    $\Lambda$ is closed under countable monotone union}.

  \proofparskip{``Right'' implies ``left''}
  Direct consequence of
  Lemma~\threfc{l:local-compl-and-compl}{%
    $\Lambda$ is closed under complement}, and
  Lemma~\threfc{l:count-union-monot-and-disj}{%
    $\Lambda$ is closed under countable disjoint union}.
\end{proof}

\begin{lemma}[other properties of $\lambda$-system]
  \label{l:other-prop-of-l-syst}
  \mbox{}\\
  Let~$X$ be a set.
  Let~$\Lambda$ be a $\lambda$-system on~$X$.
  Then, $\Lambda$ is nonempty, contains the empty set, and is closed under
  countable monotone intersection.
\end{lemma}

\begin{proof}
  Direct consequence of
  Definition~\threfc{d:l-syst}{closedness under complement},
  Lemma~\thref{l:nonempty-and-empty-or-full},
  Lemma~\thref{l:empty-and-full},
  Lemma~\thref{l:equiv-def-of-l-syst}, and
  Lemma~\thref{l:count-union-inter-equiv}.
\end{proof}

\begin{lemma}[intersection of $\lambda$-systems]
  \label{l:inter-of-l-systs}
  \mbox{}\\
  Let~$X$ and~$I$ be sets.
  Let~$(\Lambda_i)_{i\in I}$ be $\lambda$-systems on~$X$.
  Then, if $\bigcap_{i\in I}\Lambda_i$ is a $\lambda$-system on~$X$.
\end{lemma}

\begin{proof}
  Direct consequence of
  Definition~\thref{d:l-syst}, and
  \assume{the definition of intersection}.
\end{proof}

\begin{definition}[generated $\lambda$-system]
  \label{d:gen-l-syst}
  \mbox{}\hfill
  Let~$X$ be a set.
  Let~$G\subset\calP(X)$.
  The {\em $\lambda$-system generated by~$G$} is the intersection of all
  $\lambda$-systems on~$X$ containing~$G$;
  it is denoted~$\Lambda_X(G)$.
\end{definition}

\begin{lemma}[generated $\lambda$-system is minimum]
  \label{l:gen-l-syst-is-min}
  \mbox{}\\
  Let~$X$ be a set.
  Let~$G\subset\calP(X)$.
  Then, $\Lambda_X(G)$~is the smallest $\lambda$-system on~$X$ containing~$G$.
\end{lemma}

\begin{proof}
  Direct consequence of
  Definition~\thref{d:gen-l-syst},
  Lemma~\thref{l:inter-of-l-systs}, and
  \assume{properties of the intersection}.
\end{proof}

\begin{lemma}[$\lambda$-system generation is monotone]
  \label{l:l-syst-gen-is-monot}
  \mbox{}\\
  Let~$X$ be a set.
  Let~$G_1,G_2\subset\calP(X)$.
  Assume that $G_1\subset G_2$.
  Then, we have $\Lambda_X(G_1)\subset\Lambda_X(G_2)$.
\end{lemma}

\begin{proof}
  Lemma~\threfc{l:gen-l-syst-is-min}{%
    with $G\eqdef G_2$, then $G\eqdef G_1$}.
\end{proof}

\begin{lemma}[$\lambda$-system generation is idempotent]
  \label{l:l-syst-gen-is-idem}
  \mbox{}\\
  Let~$X$ be a set.
  Let~$\calS\subset\calP(X)$.
  Then, $\calS$ is a $\lambda$-system on~$X$ iff $\Lambda_X(\calS)=\calS$.
\end{lemma}

\begin{proof}
  Direct consequence of
  \assume{reflexivity of inclusion},
  Lemma~\threfc{l:gen-l-syst-is-min}{%
    $\calS$ and $\Lambda_X(\calS)$ are both $\lambda$-systems containing
    $\calS$}.
\end{proof}

\begin{definition}[$\lambda$-system and intersection]
  \label{d:l-syst-and-inter}
  \mbox{}\hfill
  Let~$X$ be a set.\\
  Let~$\Lambda$ be a $\lambda$-system on~$X$.
  For all $A\subset X$, let~$\Lambdainter_A$ be the subset system defined by
  \begin{equation}
    \label{e:inter-inv-image-of-l-syst}
    \Lambdainter_A \eqdef \{
      B \in \Lambda \leftst
      A \cap B \in \Lambda \right.
  \}.
  \end{equation}
\end{definition}

\begin{lemma}[$\Lambdainter$ is symmetric]
  \label{l:l-inter-is-symmetric}
  \mbox{}\\
  Let~$X$ be a set.
  Let~$\Lambda$ be a $\lambda$-system on~$X$.
  Then, for all $A,B\in\Lambda$, we have $B\in\Lambdainter_A$ iff
  $A\in\Lambdainter_B$.
\end{lemma}

\begin{proof}
  Direct consequence of
  Definition~\thref{d:l-syst-and-inter}.
\end{proof}

\begin{lemma}[$\Lambdainter$ is $\lambda$-system]
  \label{l:l-inter-is-l-syst}
  \mbox{}\\
  Let~$X$ be a set.
  Let~$\Lambda$ be a $\lambda$-system on~$X$.
  Then, for all $A\in\Lambda$, $\Lambdainter_A$ is a $\lambda$-system on~$X$.
\end{lemma}

\begin{proof}
  Let~$A\in \Lambda$.

  From
  Definition~\threfc{d:l-syst}{$X\in\Lambda$}, and
  \assume{the identity $A\cap X=A\in\Lambda$},
  we have $X\in\Lambdainter_A$.

  Let $B,C\in\Lambdainter_A$, {\ie} $B,C,A\cap B,A\cap C\in\Lambda$.
  Assume that $C\subset B$.\\
  Then, from
  Lemma~\threfc{l:equiv-def-of-l-syst}{closedness under local complement},
  we have $B\setminus C\in\Lambda$.
  Moreover, from
  \assume{distributivity of intersection over set difference},
  \assume{monotonicity of intersection}, and
  Lemma~\threfc{l:equiv-def-of-l-syst}{closedness under local complement},
  we have $(A\cap C)\subset(A\cap B)$ and
  $A\cap(B\setminus C)=(A\cap B)\setminus(A\cap C)\in\Lambda$.
  Hence, $\Lambdainter_A$ is closed under local complement.

  Let~$(B_n)_{n\in\matN}\in\Lambdainter_A$, {\ie} for all $n\in\matN$,
  $B_n,A\cap B_n\in\Lambda$.
  Assume that for all $n\in\matN$, $B_n\subset B_{n+1}$.\\
  Then, from
  Lemma~\threfc{l:equiv-def-of-l-syst}{%
    closedness under countable monotone union},
  we have $\bigcup_{n\in\matN}B_n\in\Lambda$.
  Moreover, from
  \assume{distributivity of intersection over union},
  \assume{monotonicity of intersection}, and
  Lemma~\threfc{l:equiv-def-of-l-syst}{%
    closedness under countable monotone union},
  we have
  \begin{equation*}
    (A \cap B_n) \subset (A \cap B_{n+1})
    \AND
    A \cap \bigcup_{n \in \matN} B_n
    = \bigcup_{n \in \matN} (A \cap B_n) \in \Lambda.
  \end{equation*}
  Hence, $\Lambdainter_A$ is closed under countable monotone union.

  Therefore, from
  Lemma~\thref{l:equiv-def-of-l-syst},
  $\Lambdainter_A$ is a $\lambda$-system on~$X$.
\end{proof}

\begin{lemma}[$\lambda$-system with intersection]
  \label{l:l-syst-with-inter}
  \mbox{}\hfill
  Let~$X$ be a set.
  Let~$G\subset\calP(X)$.\\
  Assume that~$G$ is closed under intersection.
  Then,  $\forall A\in\Lambda_X(G)$, $(\Lambda_X(G))_A^\cap=\Lambda_X(G)$.
\end{lemma}

\begin{proof}
  For all~$A\subset X$, we use the simplified notation
  $\Lambdainter_A\eqdef(\Lambda_X(G))_A^\cap$.

  \proofparskip{(0)}
  From
  Definition~\thref{d:l-syst-and-inter},
  we have $\forall A\in\Lambda_X(G)$, $\Lambdainter_A\subset\Lambda_X(G)$.

  \proofparskip{(1). $\forall A\in G,\;\Lambda_X(G)\subset\Lambdainter_A$}
  Let~$A\in G$.\\
  Let~$B\in G$.
  Then, from
  Lemma~\threfc{l:gen-l-syst-is-min}{$G\subset\Lambda_X(G)$}, and
  Definition~\thref{d:l-syst-and-inter},
  we have $B\in\Lambdainter_A$, {\ie} $G\subset\Lambdainter_A$.
  Hence, from
  Lemma~\threfc{l:l-inter-is-l-syst}{with $A\in G\subset\Lambda_X(G)$}, and
  Lemma~\thref{l:gen-l-syst-is-min},
  we have $\Lambda_X(G)\subset\Lambdainter_A$.

  \proofparskip{(2).
    $\forall B\in\Lambda_X(G),\;\Lambda_X(G)\subset\Lambdainter_B$}
  Let~$B\in\Lambda_X(G)$.\\
  Let~$A\in G$.
  Then, from~(1), we have $B\in\Lambda_X(G)\subset\Lambdainter_A$.
  Thus, from
  Lemma~\threfc{l:gen-l-syst-is-min}{$G\subset\Lambda_X(G)$}, and
  Lemma~\threfc{l:l-inter-is-symmetric}{with $A,B\in\Lambda_X(G)$},
  we have $A\in\Lambdainter_B$, {\ie} $G\subset\Lambdainter_B$.
  Hence, from
  Lemma~\thref{l:l-inter-is-l-syst}, and
  Lemma~\thref{l:gen-l-syst-is-min},
  we have $\Lambda_X(G)\subset\Lambdainter_B$.

  \medskip\noindent
  Therefore, from~(0) and~(2), we have the equality.
\end{proof}

\begin{lemma}[$\lambda$-system is closed under intersection]
  \label{l:l-syst-is-closed-under-inter}
  \mbox{}\hfill
  Let~$X$ be a set.
  Let~$G\subset\calP(X)$.\\
  Assume that~$G$ is closed under intersection.
  Then, $\Lambda_X(G)$ is closed under intersection.
\end{lemma}

\begin{proof}
  Let~$A,B\in\Lambda_X(G)$.
  Then, from
  Lemma~\threfc{l:l-syst-with-inter}{%
    $B$ belongs to $\Lambda_X(G)=(\Lambda_X(G))_A^\cap$}, and
  Definition~\thref{d:l-syst-and-inter},
  we have $A\cap B\in\Lambda_X(G)$.

  \medskip\noindent
  Therefore, $\Lambda_X(G)$ is closed under intersection.
\end{proof}

\begin{lemma}[$\lambda$-system generated by $\pi$-system]
  \label{l:l-syst-gen-by-p-syst}
  \mbox{}\hfill
  Let~$X$ be a set.
  Let~$G\subset\calP(X)$.\\
  Assume that~$G$ is a $\pi$-system on~$X$.
  Then, $\Lambda_X(G)$ is a $\pi$-system on~$X$.
\end{lemma}

\begin{proof}
  Direct consequence of
  Definition~\threfc{d:p-syst}{%
    $G\not=\emptyset$ and $G$ is closed under finite intersec\-tion},
  Lemma~\threfc{l:finite-ops-equiv}{$G$ is closed under intersection},
  Lemma~\threfc{l:gen-l-syst-is-min}{%
    $\emptyset\not=G\subset\Lambda_X(G)$},
  Lemma~\threfc{l:l-syst-is-closed-under-inter}{%
    $\Lambda_X(G)$ is closed under intersection},
  Lemma~\threfc{l:finite-ops-equiv}{%
    $\Lambda_X(G)$ is closed under finite intersection}, and
  Definition~\thref{d:p-syst}.
\end{proof}

\clearpage
\section{Sigma-algebra}
\label{s:sigma-alg}

\begin{remark}
  \label{r:v2-new04}
  Note that the following concept of ``$\sigma$-algebra'' is sometimes called
  ``$\sigma$-field''.
\end{remark}

\begin{definition}[$\sigma$-algebra]
  \label{d:sigma-alg}
  \mbox{}\hfill
  Let~$X$ be a set.
  A subset~$\Sigma$ of~$\calP(X)$ is called {\em $\sigma$-algebra on~$X$} iff
  it contains the empty set,
  it is closed under complement, and
  under countable union:
  \begin{align}
    \label{e:sigma-alg-1}
    & \emptyset \in \Sigma,\\
    \label{e:sigma-alg-2}
    \forall A \in \Sigma,\quad &
    A^c \in \Sigma,\\
    \label{e:sigma-alg-3}
    \forall I \subset \matN,\;
    \forall (A_i)_{i \in I} \in \Sigma,\quad &
    \bigcup_{i \in I} A_i \in \Sigma.
  \end{align}
\end{definition}

\begin{lemma}[equivalent definition of $\sigma$-algebra]
  \label{l:equiv-def-of-sigma-alg}
  \mbox{}\\
  Let~$X$ be a set.
  Let~$\Sigma\subset\calP(X)$.
  Then, $\Sigma$~is a $\sigma$-algebra on~$X$ iff
  \eqref{e:sigma-alg-2} holds and
  \begin{gather}
    \label{e:equiv-def-of-sigma-alg-1}
    \emptyset \in \Sigma \DISJ X \in \Sigma \DISJ \Sigma \not= \emptyset,\\
    \label{e:equiv-def-of-sigma-alg-2}
    \forall I \subset \matN,\;
    \forall (A_i)_{i \in I} \in \Sigma,\;
    \bigcup_{i \in I} A_i \in \Sigma \DISJ
    \forall I \subset \matN,\;
    \forall (A_i)_{i \in I} \in \Sigma,\;
    \bigcap_{i \in I} A_i \in \Sigma.
  \end{gather}
\end{lemma}

\begin{proof}
  Direct consequence of
  Definition~\thref{d:sigma-alg},
  Lemma~\thref{l:nonempty-and-empty-or-full},
  Lemma~\thref{l:empty-and-full}, and
  Lemma~\thref{l:count-union-inter-equiv}.
\end{proof}

\begin{remark}
  Note that from the previous lemma, we may define $\sigma$-algebras as subset
  systems that satisfies~\eqref{e:sigma-alg-2}, and any term in each of the
  disjunctions~\eqref{e:equiv-def-of-sigma-alg-1}
  and~\eqref{e:equiv-def-of-sigma-alg-2}.
\end{remark}

\begin{lemma}[$\sigma$-algebra is set algebra]
  \label{l:sigma-alg-is-set-alg}
  \mbox{}\\
  Let~$X$ be a set.
  Let~$\Sigma$ be a $\sigma$-algebra on~$X$.
  Then, $\Sigma$~is a set algebra on~$X$.
\end{lemma}

\begin{proof}
  Direct consequence of
  Definition~\thref{d:set-alg}, and
  Definition~\thref{d:sigma-alg}.
\end{proof}

\begin{lemma}[$\sigma$-algebra is closed under set difference]
  \label{l:sigma-alg-is-closed-under-set-diff}
  \mbox{}\hfill
  Let~$X$ be a set.\\
  Let~$\Sigma$ be a $\sigma$-algebra on~$X$.
  Then, $\Sigma$~is closed under set difference and local complement.
\end{lemma}

\begin{proof}
  Direct consequence of
  Lemma~\thref{l:sigma-alg-is-set-alg},
  Lemma~\thref{l:other-equiv-def-of-set-alg}, and
  Lemma~\thref{l:set-alg-is-closed-under-local-compl}.
\end{proof}

\begin{lemma}[other properties of $\sigma$-algebra]
  \label{l:other-prop-of-sigma-alg}
  \mbox{}\hfill
  Let~$X$ be a set.
  Let~$\Sigma$ be a $\sigma$-algebra on~$X$.\\
  Then, $\Sigma$ is closed under countable monotone intersection and union, and
  countable disjoint union.
\end{lemma}

\begin{proof}
  Direct consequence of
  Lemma~\thref{l:equiv-def-of-sigma-alg}.
\end{proof}

\begin{lemma}[partition of countable union in $\sigma$-algebra]
  \label{l:part-of-count-union-in-sigma-alg}
  \mbox{}\\
  Let~$X$ be a set.
  Let~$\Sigma$ be a $\sigma$-algebra on~$X$.
  Let~$(A_n)_{n\in\matN}\subset\Sigma$.
  Let~$B_0\eqdef A_0$, and for all $n\in\matN$, let
  $B_{n+1}\eqdef A_{n+1}\setminus\bigcup_{i\in[0..n]}B_i$.
  Then, we have
  \begin{align}
    \label{e:part-of-count-union-in-sigma-alg-1}
    \forall n \in \matN,\quad & B_n \in \Sigma,\\
    \label{e:part-of-count-union-in-sigma-alg-2}
    \forall m, n \in \matN,\quad &
    m \not= n \IMPLIES B_m \cap B_n = \emptyset,\\
    \label{e:part-of-count-union-in-sigma-alg-3}
    \forall n \in \matN,\quad &
    \bigcup_{i \in [0..n]} A_i = \biguplus_{i \in [0..n]} B_i \in \Sigma,\\
    \label{e:part-of-count-union-in-sigma-alg-4}
    & \bigcup_{i \in \matN} A_i = \biguplus_{i \in \matN} B_i \in \Sigma.
  \end{align}
\end{lemma}

\begin{proof}
  Direct consequence of
  Lemma~\thref{l:sigma-alg-is-set-alg},
  Lemma~\thref{l:part-of-count-union-in-set-alg},
  Lemma~\thref{l:part-of-count-union}, and
  Definition~\threfc{d:sigma-alg}{%
    closedness under countable union with $I=\matN$}.
\end{proof}

\begin{lemma}[intersection of $\sigma$-algebras]
  \label{l:inter-of-sigma-algs}
  \mbox{}\\
  Let~$X$ and~$I$ be sets.
  Let~$(\Sigma_i)_{i\in I}$ be $\sigma$-algebras on~$X$.
  Then, $\bigcap_{i\in I}\Sigma_i$ is a $\sigma$-algebra on~$X$.
\end{lemma}

\begin{proof}
  Direct consequence of
  Definition~\thref{d:sigma-alg}, and
  \assume{the definition of intersection}.
\end{proof}

\begin{definition}[generated $\sigma$-algebra]
  \label{d:gen-sigma-alg}
  \mbox{}\hfill
  Let~$X$ be a set.
  Let~$G\subset\calP(X)$.
  The {\em $\sigma$-algebra generated by~$G$} is the intersection of all
  $\sigma$-algebras on~$X$ containing~$G$;
  it is denoted~$\Sigma_X(G)$.
\end{definition}

\begin{lemma}[generated $\sigma$-algebra is minimum]
  \label{l:gen-sigma-alg-is-min}
  \mbox{}\\
  Let~$X$ be a set.
  Let~$G\subset\calP(X)$.
  Then, $\Sigma_X(G)$~is the smallest $\sigma$-algebra on~$X$
  containing~$G$.
\end{lemma}

\begin{proof}
  Direct consequence of
  Definition~\thref{d:gen-sigma-alg},
  Lemma~\thref{l:inter-of-sigma-algs}, and
  \assume{properties of the intersection}.
\end{proof}

\begin{lemma}[$\sigma$-algebra generation is monotone]
  \label{l:sigma-alg-gen-is-monot}
  \mbox{}\\
  Let~$X$ be a set.
  Let~$G_1,G_2\subset\calP(X)$.
  Assume that $G_1\subset G_2$.
  Then, we have $\Sigma_X(G_1)\subset\Sigma_X(G_2)$.
\end{lemma}

\begin{proof}
  Lemma~\threfc{l:gen-sigma-alg-is-min}{%
    with $G\eqdef G_2$, then $G\eqdef G_1$}.
\end{proof}

\begin{lemma}[$\sigma$-algebra generation is idempotent]
  \label{l:sigma-alg-gen-is-idem}
  \mbox{}\\
  Let~$X$ be a set.
  Let~$\calS\subset\calP(X)$.
  Then, $\calS$ is a $\sigma$-algebra on~$X$ iff $\Sigma_X(\calS)=\calS$.
\end{lemma}

\begin{proof}
  Direct consequence of
  \assume{reflexivity of inclusion},
  Lemma~\threfc{l:gen-sigma-alg-is-min}{%
    $\calS$ and $\Sigma_X(\calS)$ are both $\sigma$-algebras containing
    $\calS$}.
\end{proof}

\begin{lemma}[$\sigma$-algebra is $\pi$-system]
  \label{l:sigma-alg-is-p-syst}
  \mbox{}\\
  Let~$X$ be a set.
  Let~$\Sigma$ be a $\sigma$-algebra on~$X$.
  Then, $\Sigma$~is a $\pi$-system on~$X$.
\end{lemma}

\begin{proof}
  Direct consequence of
  Definition~\thref{d:p-syst}, and
  Lemma~\thref{l:equiv-def-of-sigma-alg}.
\end{proof}

\begin{lemma}[$\sigma$-algebra contains $\pi$-system]
  \label{l:sigma-alg-contains-p-syst}
  \mbox{}\\
  Let~$X$ be a set.
  Let~$G\subset\calP(X)$.
  Assume that~$G\neq\emptyset$.
  Then, we have $\Pi_X(G)\subset\Sigma_X(G)$.
\end{lemma}

\begin{proof}
  Direct consequence of
  Lemma~\threfc{l:gen-sigma-alg-is-min}{%
    $\Sigma_X(G)$ is a $\sigma$-algebra, and $G\subset\Sigma_X(G)$},
  Lemma~\thref{l:sigma-alg-is-p-syst}, and
  Lemma~\thref{l:gen-p-syst-is-min}.
\end{proof}

\begin{lemma}[$\pi$-system contains $\sigma$-algebra]
  \label{l:p-syst-contains-sigma-alg}
  \mbox{}\\
  Let~$X$ be a set.
  Let~$G\subset\calP(X)$.
  Assume that~$G\neq\emptyset$, and that~$\Pi_X(G)$ is closed under complement
  and countable disjoint union.
  Then, we have $\Sigma_X(G)\subset\Pi_X(G)$.
\end{lemma}

\begin{proof}
  Direct consequence of
  Lemma~\threfc{l:gen-p-syst-is-min}{%
    $\Pi_X(G)$ is a $\pi$-system, and $G\subset\Pi_X(G)$},
  Definition~\threfc{d:p-syst}{%
    $\Pi_X(G)$ is nonempty, and closed under intersection},
  Lemma~\threfc{l:nonempty-and-empty-or-full}{$\emptyset\in\Pi_X(G)$},
  Lemma~\threfc{l:count-union-disj-and-union}{%
    $\Pi_X(G)$ is closed under countable union},
  Definition~\threfc{d:sigma-alg}{$\Pi_X(G)$ is a $\sigma$-algebra}, and
  Lemma~\thref{l:gen-sigma-alg-is-min}.
\end{proof}

\begin{lemma}[$\sigma$-algebra generated by $\pi$-system]
  \label{l:sigma-alg-gen-by-p-syst}
  \mbox{}\\
  Let~$X$ be a set.
  Let~$G\subset\calP(X)$.
  Assume that~$G\neq\emptyset$.
  Then, we have $\Sigma_X(\Pi_X(G))=\Sigma_X(G)$.
\end{lemma}

\begin{proof}
  Direct consequence of
  Lemma~\threfc{l:sigma-alg-contains-p-syst}{$\Pi_X(G)\subset\Sigma_X(G)$},
  Lem\-ma~\threfc{l:gen-sigma-alg-is-min}{%
    $\Sigma_X(G)$ is a $\sigma$-algebra, and
    $\Sigma_X(\Pi_X(G))\subset\Sigma_X(G)$},
  Lemma~\threfc{l:gen-p-syst-is-min}{$G\subset\Pi_X(G)$}, and
  Lemma~\threfc{l:sigma-alg-gen-is-monot}{%
    $\Sigma_X(G)$ is included in $\Sigma_X(\Pi_X(G))$}.
\end{proof}

\begin{lemma}[$\sigma$-algebra contains set algebra]
  \label{l:sigma-alg-contains-set-alg}
  \mbox{}\\
  Let~$X$ be a set.
  Let~$G\subset\calP(X)$.
  Then, we have $\calA_X(G)\subset\Sigma_X(G)$.
\end{lemma}

\begin{proof}
  Direct consequence of
  Lemma~\threfc{l:gen-sigma-alg-is-min}{%
    $\Sigma_X(G)$ is a $\sigma$-algebra, and $G\subset\Sigma_X(G)$},
  Lemma~\thref{l:sigma-alg-is-set-alg}, and
  Lemma~\thref{l:gen-set-alg-is-min}.
\end{proof}

\begin{lemma}[set algebra contains $\sigma$-algebra]
  \label{l:set-alg-contains-sigma-alg}
  \mbox{}\hfill
  Let~$X$ be a set.
  Let~$G\subset\calP(X)$.\\
  Assume that~$\calA_X(G)$ is closed under countable monotone union.
  Then, we have $\Sigma_X(G)\subset\calA_X(G)$.
\end{lemma}

\begin{proof}
  Direct consequence of
  Lemma~\threfc{l:gen-set-alg-is-min}{%
    $\calA_X(G)$ is a set algebra, and $G\subset\calA_X(G)$},
  Definition~\threfc{d:set-alg}{%
    $\calA_X(G)$ contains~$\emptyset$, is closed under comple\-ment and union},
  Lemma~\threfc{l:count-union-monot-and-union}{%
    $\calA_X(G)$ is closed under countable union},
  Definition~\threfc{d:sigma-alg}{$\calA_X(G)$ is a $\sigma$-algebra}, and
  Lemma~\thref{l:gen-sigma-alg-is-min}.
\end{proof}

\begin{lemma}[$\sigma$-algebra generated by set algebra]
  \label{l:sigma-alg-gen-by-set-alg}
  \mbox{}\\
  Let~$X$ be a set.
  Let~$G\subset\calP(X)$.
  Then, we have $\Sigma_X(\calA_X(G))=\Sigma_X(G)$.
\end{lemma}

\begin{proof}
  Direct consequence of
  Lemma~\threfc{l:sigma-alg-contains-set-alg}{$\calA_X(G)\subset\Sigma_X(G)$},
  Lem\-ma~\threfc{l:gen-sigma-alg-is-min}{%
     $\Sigma_X(G)$ is a $\sigma$-algebra, and
     $\Sigma_X(\calA_X(G))\subset\Sigma_X(G)$},
  Lemma~\threfc{l:gen-set-alg-is-min}{$G\subset\calA_X(G)$}, and
  Lemma~\threfc{l:sigma-alg-gen-is-monot}{%
    $\Sigma_X(G)$ is included in $\Sigma_X(\calA_X(G))$}.
\end{proof}

\begin{lemma}[$\sigma$-algebra is monotone class]
  \label{l:sigma-alg-is-monot-class}
  \mbox{}\\
  Let~$X$ be a set.
  Let~$\Sigma$ be a $\sigma$-algebra on~$X$.
  Then, $\Sigma$~is a monotone class on~$X$.
\end{lemma}

\begin{proof}
  Direct consequence of
  Definition~\thref{d:monot-class}, and
  Lemma~\thref{l:other-prop-of-sigma-alg}.
\end{proof}

\begin{lemma}[$\sigma$-algebra contains monotone class]
  \label{l:sigma-alg-contains-monot-class}
  \mbox{}\\
  Let~$X$ be a set.
  Let~$G\subset\calP(X)$.
  Then, we have $\calC_X(G)\subset\Sigma_X(G)$.
\end{lemma}

\begin{proof}
  Direct consequence of
  Lemma~\threfc{l:gen-sigma-alg-is-min}{%
    $\Sigma_X(G)$ is a $\sigma$-algebra, and $G\subset\Sigma_X(G)$},
  Lemma~\thref{l:sigma-alg-is-monot-class}, and
  Lemma~\thref{l:gen-monot-class-is-min}.
\end{proof}

\begin{lemma}[monotone class contains $\sigma$-algebra]
  \label{l:monot-class-contains-sigma-alg}
  \mbox{}\\
  Let~$X$ be a set.
  Let~$G\subset\calP(X)$.
  Assume that~$\calC_X(G)$ contains the empty set, and is closed under
  complement and union.
  Then, we have $\Sigma_X(G)\subset\calC_X(G)$.
\end{lemma}

\begin{proof}
  Direct consequence of
  Lemma~\threfc{l:gen-monot-class-is-min}{%
    $\calC_X(G)$ is a monotone class, and $G\subset\calC_X(G)$},
  Definition~\threfc{d:monot-class}{%
    $\calC_X(G)$ is closed under count\-able monotone union},
  Lemma~\threfc{l:count-union-monot-and-union}{%
    $\calC_X(G)$ is closed under countable union},
  Definition~\threfc{d:sigma-alg}{$\calC_X(G)$ is a $\sigma$-algebra}, and
  Lemma~\thref{l:gen-sigma-alg-is-min}.
\end{proof}

\begin{lemma}[$\sigma$-algebra generated by monotone class]
  \label{l:sigma-alg-gen-by-monot-class}
  \mbox{}\\
  Let~$X$ be a set.
  Let~$G\subset\calP(X)$.
  Then, we have $\Sigma_X(\calC_X(G))=\Sigma_X(G)$.
\end{lemma}

\begin{proof}
  Direct consequence of
  Lemma~\threfc{l:sigma-alg-contains-monot-class}{%
    $\calC_X(G)\subset\Sigma_X(G)$},
  Lemma~\threfc{l:gen-sigma-alg-is-min}{%
     $\Sigma_X(G)$ is a $\sigma$-algebra, and
     $\Sigma_X(\calC_X(G))\subset\Sigma_X(G)$},
  Lemma~\threfc{l:gen-monot-class-is-min}{$G\subset\calC_X(G)$}, and
  Lemma~\threfc{l:sigma-alg-gen-is-monot}{%
    $\Sigma_X(G)\subset\Sigma_X(\calC_X(G))$}.
\end{proof}

\begin{lemma}[$\sigma$-algebra is $\lambda$-system]
  \label{l:sigma-alg-is-l-syst}
  \mbox{}\\
  Let~$X$ be a set.
  Let~$\Sigma$ be a $\sigma$-algebra on~$X$.
  Then, $\Sigma$~is a $\lambda$-system on~$X$.
\end{lemma}

\begin{proof}
  Direct consequence of
  Definition~\thref{d:l-syst},
  Lemma~\thref{l:equiv-def-of-sigma-alg}, and
  Lemma~\thref{l:other-prop-of-sigma-alg}.
\end{proof}

\begin{lemma}[$\sigma$-algebra contains $\lambda$-system]
  \label{l:sigma-alg-contains-l-syst}
  \mbox{}\\
  Let~$X$ be a set.
  Let~$G\subset\calP(X)$.
  Then, we have $\Lambda_X(G)\subset\Sigma_X(G)$.
\end{lemma}

\begin{proof}
  Direct consequence of
  Lemma~\threfc{l:gen-sigma-alg-is-min}{%
    $\Sigma_X(G)$ is a $\sigma$-algebra, and $G\subset\Sigma_X(G)$},
  Lemma~\thref{l:sigma-alg-is-l-syst}, and
  Lemma~\thref{l:gen-l-syst-is-min}.
\end{proof}

\begin{lemma}[$\lambda$-system contains $\sigma$-algebra]
  \label{l:l-syst-contains-sigma-alg}
  \mbox{}\hfill
  Let~$X$ be a set.
  Let~$G\subset\calP(X)$.\\
  Assume that~$\Lambda_X(G)$ is closed under intersection.
  Then, we have $\Sigma_X(G)\subset\Lambda_X(G)$.
\end{lemma}

\begin{proof}
  Direct consequence of
  Lemma~\threfc{l:gen-l-syst-is-min}{%
    $\Lambda_X(G)$ is a $\lambda$-system, and $G\subset\Lambda_X(G)$},
  Definition~\threfc{d:l-syst}{%
    $\Lambda_X(G)$ contains the full set, is closed under comple\-ment and
    countable disjoint union},
  Lemma~\threfc{l:count-union-disj-and-union}{%
    $\Lambda_X(G)$ is closed under countable union},
  Lemma~\threfc{l:equiv-def-of-sigma-alg}{%
    $\Lambda_X(G)$ is a $\sigma$-algebra}, and
  Lemma~\thref{l:gen-sigma-alg-is-min}.
\end{proof}

\begin{lemma}[$\sigma$-algebra generated by $\lambda$-system]
  \label{l:sigma-alg-gen-by-l-syst}
  \mbox{}\\
  Let~$X$ be a set.
  Let~$G\subset\calP(X)$.
  Then, we have $\Sigma_X(\Lambda_X(G))=\Sigma_X(G)$.
\end{lemma}

\begin{proof}
  Direct consequence of
  Lemma~\threfc{l:sigma-alg-contains-l-syst}{$\Lambda_X(G)\subset\Sigma_X(G)$},
  Lem\-ma~\threfc{l:gen-sigma-alg-is-min}{%
     $\Sigma_X(G)$ is a $\sigma$-algebra, and
     $\Sigma_X(\Lambda_X(G))\subset\Sigma_X(G)$},
  Lemma~\threfc{l:gen-l-syst-is-min}{$G\subset\Lambda_X(G)$}, and
  Lemma~\threfc{l:sigma-alg-gen-is-monot}{%
    $\Sigma_X(G)$ is included in $\Sigma_X(\Lambda_X(G))$}.
\end{proof}

\begin{lemma}[other $\sigma$-algebra generator]
  \label{l:other-sigma-alg-gen}
  \mbox{}\hfill
  Let~$X$ be a set.
  Let~$G_1,G_2\subset\calP(X)$.\\
  Assume that $G_1\subset\Sigma_X(G_2)$ and $G_2\subset\Sigma_X(G_1)$.
  Then, we have $\Sigma_X(G_1)=\Sigma_X(G_2)$.
\end{lemma}

\begin{proof}
  From
  Lemma~\thref{l:sigma-alg-gen-is-monot}, and
  Lemma~\thref{l:sigma-alg-gen-is-idem},
  we have
  \begin{equation*}
    \Sigma_X (G_1)
    \subset \Sigma_X \left( \Sigma_X (G_2) \right)
    = \Sigma_X (G_2)
    \AND
    \Sigma_X (G_2)
    \subset \Sigma_X \left( \Sigma_X (G_1) \right)
    = \Sigma_X (G_1).
  \end{equation*}
  Therefore, we have $\Sigma_X(G_1)=\Sigma_X(G_2)$.
\end{proof}

\begin{lemma}[complete generated $\sigma$-algebra]
  \label{l:complete-gen-sigma-alg}
  \mbox{}\hfill
  Let~$X$ be a set.
  Let~$G_1,G_2\subset\calP(X)$.\\
  Assume that $G_2\subset\Sigma_X(G_1)$.
  Then, we have $\Sigma_X(G_1\cup G_2)=\Sigma_X(G_1)$.
\end{lemma}

\begin{proof}
  From
  Lemma~\thref{l:gen-sigma-alg-is-min}, and
  Lemma~\thref{l:sigma-alg-gen-is-monot},
  we have $G_1\subset\Sigma_X(G_1)\subset\Sigma_X(G_1\cup G_2)$.
  From
  Lemma~\threfc{l:gen-sigma-alg-is-min}{$G_1\subset\Sigma_X(G_1)$}, and
  \assume{monotonicity of union},
  we have $G_1\cup G_2\subset\Sigma_X(G_1)$.

  Therefore, from
  Lemma~\threfc{l:other-sigma-alg-gen}{with $G_1$ and $G_1\cup G_2$},
  we have $\Sigma_X(G_1\cup G_2)=\Sigma_X(G_1)$.
\end{proof}

\begin{lemma}[countable $\sigma$-algebra generator]
  \label{l:count-sigma-alg-gen}
  \mbox{}\\
  Let~$X$ be a set.
  Let $G_1\subset G_2\subset\calP(X)$.
  Assume that all elements of~$G_2$ are countable unions of elements of~$G_1$.
  Then, we have $\Sigma_X(G_1)=\Sigma_X(G_2)$.
\end{lemma}

\begin{proof}
  Direct consequence of
  Lemma~\threfc{l:gen-sigma-alg-is-min}{%
    $G_1\subset G_2\subset\Sigma_X(G_2)$},
  Definition~\threfc{d:sigma-alg}{%
    closedness under countable union, thus $G_2\subset\Sigma_X(G_1)$}, and
  Lemma~\thref{l:other-sigma-alg-gen}.
\end{proof}

\begin{remark}
  \label{r:v2-mod3}
  Note that in the following proof, the point~(1) does not depend on the
  hypotheses.
  It could be an independent lemma that characterizes
  $\Sigma_1\oltimes\Sigma_2$.

  Note also that $\Sigma_1\oltimes\Sigma_2$ may not be a $\sigma$-algebra.
  See Section~\ref{s:product-of-measurable-spaces} for a definition of a
  $\sigma$-algebra on the product of measurable spaces.
\end{remark}

\begin{lemma}[set algebra generated by product of $\sigma$-algebras]
  \label{l:set-alg-gen-by-prod-of-sigma-algs}
  \mbox{}\hfill
  Let~$X_1$ and~$X_2$ be sets.\\
  For all $i\in\{1,2\}$, let~$\Sigma_i$ be a $\sigma$-algebra on~$X_i$.
  Let~$X\eqdef X_1\times X_2$, and~$\Sigmabar\eqdef\Sigma_1\oltimes\Sigma_2$.\\
  Then, $\calA_X(\Sigmabar)$ is the set of finite disjoint unions of elements
  of~$\Sigmabar$.
\end{lemma}

\begin{proof}
  \proofpar{(1). $X\in\Sigmabar$}
  Direct consequence of
  Definition~\thref{d:prod-of-subsets-of-parties}, and
  Lemma~\threfc{l:equiv-def-of-sigma-alg}{%
    $\Sigma_1$ and $\Sigma_2$ contain full set.}.

  \proofparskip{(2). $\Sigmabar$ is closed under intersection}
  Let~$A,B\in\Sigmabar$.\\
  From
  Definition~\thref{d:prod-of-subsets-of-parties},
  let $A_1,B_1\in\Sigma_1$ and $A_2,B_2\in\Sigma_2$ such that
  $A=A_1\times A_2$ and $B=B_1\times B_2$.
  Then, from
  \assume{compatibility of intersection with Cartesian product}, and
  Lemma~\threfc{l:equiv-def-of-sigma-alg}{%
    closedness under intersection for $\Sigma_1$ and $\Sigma_2$},
  we have $A\cap B=(A_1\cap B_1)\times(A_2\cap B_2)$ with
  $A_1\cap B_1\in\Sigma_1$ and $A_2\cap B_2\in\Sigma_2$.
  Thus, $A\cap B\in\Sigmabar$.
  Hence, $\Sigmabar$~is closed under intersection.

  \proofparskip{(3).
    $\forall A\in\Sigmabar,\;
    \exists B,C\in\Sigmabar,\;
    B\cap C=\emptyset\Conj A^c=B\uplus C$}
  Let~$A\in\Sigmabar$.\\
  From
  Definition~\thref{d:prod-of-subsets-of-parties},
  let $A_1\in\Sigma_1$ and $A_2\in\Sigma_2$ such that $A=A_1\times A_2$.
  Let~$B\eqdef X_1\times A_2^c$ and $C\eqdef A_1^c\times A_2$.
  Then, from
  Definition~\thref{d:prod-of-subsets-of-parties},
  Lemma~\threfc{l:equiv-def-of-sigma-alg}{%
    $\Sigma_1$ contains full set, and closedness under complement
  for $\Sigma_1$ and $\Sigma_2$}, and
  \assume{set operations properties},
  we have $B,C\in\Sigmabar$, $B\cap C=\emptyset$, and $A^c=B\uplus C$.

  \medskip\noindent
  Therefore, from~(1), (2), (3), and
  Lemma~\threfc{l:explicit-set-alg}{with $G\eqdef\Sigmabar$},
  $\calA_X(\Sigmabar)$~is the set of finite disjoint unions of elements
  of~$\Sigmabar$.
\end{proof}

\clearpage
\section{{\Dplt}}
\label{s:dynkin-pi-lambda-th}

\begin{lemma}[$\pi$-system and $\lambda$-system is $\sigma$-algebra]
  \label{l:p-syst-and-l-syst-is-sigma-alg}
  \mbox{}\hfill
  Let~$X$ be a set.
  Let~$\calS\subset\calP(X)$.
  Assume that~$\calS$ is a $\pi$-system on~$X$ and a $\lambda$-system on~$X$.
  Then, $\calS$~is a $\sigma$-algebra on~$X$.
\end{lemma}

\begin{proof}
  Direct consequence of
  Lemma~\threfc{l:l-syst-gen-is-idem}{%
    $\Lambda_X(\calS)=\calS$},
  Definition~\threfc{d:p-syst}{%
    $\Lambda_X(\calS)$ is closed under intersection},
  Lemma~\threfc{l:l-syst-contains-sigma-alg}{%
    $\Sigma_X(\calS)$ is included in $\Lambda_X(\calS)$},
  Lemma~\threfc{l:sigma-alg-contains-l-syst}{%
    $\Lambda_X(\calS)\subset\Sigma_X(\calS)$}, and
  Lemma~\threfc{l:sigma-alg-gen-is-idem}{%
    $\Sigma_X(\calS)=\Lambda_X(\calS)=\calS$,
    thus $\calS$~is a $\sigma$-algebra}.
\end{proof}

\begin{remark}
  \label{r:v2-new05}
  See the sketch of next proof in
  Section~\ref{s:sketch-of-the-proof-of-the-dynkin-p-l-th-monot-class-th}.
\end{remark}

\begin{theorem}[{\Dplt}]
  \label{t:dynkin-pi-lambda-th}
  \mbox{}\\
  Let~$X$ be a set.
  Let~$\Pi$ be a $\pi$-system on~$X$.
  Then, we have $\Lambda_X(\Pi)=\Sigma_X(\Pi)$.
\end{theorem}

\begin{proof}
  Direct consequence of
  Lemma~\threfc{l:l-syst-gen-by-p-syst}{%
    $\Lambda_X(\Pi)$ is $\pi$-system},
  Lemma~\threfc{l:gen-l-syst-is-min}{%
    $\Lambda_X(\Pi)$ is $\lambda$-system},
  Lemma~\threfc{l:p-syst-and-l-syst-is-sigma-alg}{%
    $\Lambda_X(\Pi)$ is $\sigma$-algebra},
  Lemma~\threfc{l:sigma-alg-gen-is-idem}{%
    $\Sigma_X(\Lambda_X(\Pi))=\Lambda_X(\Pi)$}, and
  Lemma~\threfc{l:sigma-alg-gen-by-l-syst}{%
    $\Sigma_X(\Lambda_X(\Pi))=\Sigma_X(\Pi)$}.
\end{proof}

\begin{remark}
  Note that the {\Dplt} may take the following form:
  if a $\lambda$-system contains a $\pi$-system, then it also contains the
  $\sigma$-algebra generated by the $\pi$-system.

  Similarly, the next statement is an application lemma for the previous
  theorem.
  It is used to prove Lemma~\ref{l:uniq-of-meas-ext-from-p-syst} in
  Section~\ref{s:uniqueness-condition}, itself later used to prove uniqueness
  of the Lebesgue measure.
\end{remark}

\begin{lemma}[usage of {\Dplt}]
  \label{l:usage-of-dynkin-pi-lambda-th}
  \mbox{}\hfill
  Let~$X$ be a set.
  Let~$\Sigma$ be a $\sigma$-algebra on~$X$.\\
  Let~$P$ be a predicate over~$\calP(X)$, and
  let $\calS\eqdef\{A\in\Sigma\st P(A)\}$.
  Let~$G\subset\calP(X)$ be a nonempty generator of~$\Sigma$.
  Assume that~$\Pi_X(G)\subset\calS$, and that~$\calS$ is a $\lambda$-system.\\
  Then, we have $\calS=\Sigma$, {\ie}~$P$ holds for all subsets in~$\Sigma$.
\end{lemma}

\begin{proof}
  From
  the definition of~$\calS$,
  we have $\calS\subset\Sigma$.

  Let~$\Pi\eqdef\Pi_X(G)$.
  Then, from
  Definition~\threfc{d:gen-sigma-alg}{$\Sigma=\Sigma_X(G)$},
  Lemma~\threfc{l:gen-p-syst-is-min}{%
    $G\subset\Pi$, and $\Pi$ is a $\pi$-system},
  Lemma~\threfc{l:sigma-alg-gen-is-monot}{%
    $\Sigma_X(G)$ is included in $\Sigma_X(\Pi)$},
  Theorem~\threfc{t:dynkin-pi-lambda-th}{$\Sigma_X(\Pi)=\Lambda_X(\Pi)$},
  Lemma~\threfc{l:l-syst-gen-is-monot}{%
    $\Lambda_X(\Pi)\subset\Lambda_X(\calS)$}, and
  Lemma~\threfc{l:l-syst-gen-is-idem}{%
    $\Lambda_X(\calS)=\calS$},
  we have $\Sigma\subset\calS$.

  Therefore, we have the equality, {\ie}~$P$ holds for all subsets
  in~$\Sigma$.
\end{proof}

\clearpage
\section{{\Mct}}
\label{s:monot-class-th}

\begin{lemma}[algebra and monotone class is $\sigma$-algebra]
  \label{l:set-alg-and-monot-class-is-sigma-alg}
  \mbox{}\hfill
  Let~$X$ be a set.
  Let~$\calS\subset\calP(X)$.
  Assume that~$\calS$ is a set algebra on~$X$ and a monotone class on~$X$.
  Then, $\calS$~is a $\sigma$-algebra on~$X$.
\end{lemma}

\begin{proof}
  Direct consequence of
  Lemma~\threfc{l:monot-class-gen-is-idem}{%
    $\calC_X(\calS)=\calS$},
  Definition~\threfc{d:set-alg}{%
    $\calC_X(\calS)$ contains the empty set, and is closed under complement and
    union},
  Lemma~\threfc{l:monot-class-contains-sigma-alg}{%
    $\Sigma_X(\calS)\subset\calC_X(\calS)$},
  Lemma~\threfc{l:sigma-alg-contains-monot-class}{%
    $\calC_X(\calS)\subset\Sigma_X(\calS)$}, and
  Lemma~\threfc{l:sigma-alg-gen-is-idem}{%
    $\Sigma_X(\calS)=\calC_X(\calS)=\calS$,
    thus $\calS$ is $\sigma$-algebra.}.
\end{proof}

\begin{remark}
  \label{r:v2-new06}
  See the sketch of next proof in
  Section~\ref{s:sketch-of-the-proof-of-the-dynkin-p-l-th-monot-class-th}.
\end{remark}

\begin{theorem}[monotone class]
  \label{t:monot-class}
  \mbox{}\\
  Let~$X$ be a set.
  Let~$\calA$ be a set algebra on~$X$.
  Then, $\calC_X(\calA)=\Sigma_X(\calA)$.
\end{theorem}

\begin{proof}
  Direct consequence of
  Lemma~\threfc{l:monot-class-gen-by-set-alg}{%
    $\calC_X(\calA)$ is set algebra},
  Lemma~\threfc{l:gen-monot-class-is-min}{%
    $\calC_X(\calA)$ is monotone class},
  Lemma~\threfc{l:set-alg-and-monot-class-is-sigma-alg}{%
    $\calC_X(\calA)$ is $\sigma$-algebra},
  Lemma~\threfc{l:sigma-alg-gen-is-idem}{%
    thus $\Sigma_X(\calC_X(\calA))=\calC_X(\calA)$}, and
  Lemma~\threfc{l:sigma-alg-gen-by-monot-class}{%
    $\Sigma_X(\calC_X(\calA))=\Sigma_X(\calA)$}.
\end{proof}

\begin{remark}
  Note that the {\mct} may take the following form:
  if a monotone class contains a set algebra, then it also contains the
  $\sigma$-algebra generated by the algebra.

  Similarly, the next statement is an application lemma for the previous
  theorem.
  It is used to prove
  Lemmas~\ref{l:meas-of-meas-of-section-finite}
  and~\ref{l:uniq-of-tensor-prod-meas} in
  Section~\ref{s:measure-and-integration-over-product-space} in the context of
  product spaces.
\end{remark}

\begin{lemma}[usage of {\mct}]
  \label{l:usage-of-monot-class-th}
  \mbox{}\hfill
  Let~$X$ be a set.
  Let~$\Sigma$ be a $\sigma$-algebra on~$X$.
  Let~$P$ be a predicate over~$\calP(X)$, and
  let $\calS\eqdef\{A\in\Sigma\st P(A)\}$.
  Let~$G\subset\calP(X)$ be a generator of~$\Sigma$.
  Assume that $\calA_X(G)\subset\calS$, and that~$\calS$ is a monotone
  class.\\
  Then, we have~$\calS=\Sigma$, {\ie}~$P$ holds for all subsets in~$\Sigma$.
\end{lemma}

\begin{proof}
  From
  the definition of~$\calS$,
  we have $\calS\subset\Sigma$.

  Let~$\calA\eqdef\calA_X(G)$.
  Then, from
  Definition~\threfc{d:gen-sigma-alg}{$\Sigma=\Sigma_X(G)$},
  Lemma~\threfc{l:gen-set-alg-is-min}{%
    $G\subset\calA$, and $\calA$ is set algebra},
  Lemma~\threfc{l:sigma-alg-gen-is-monot}{%
    $\Sigma_X(G)\subset\Sigma_X(\calA)$},
  Theorem~\threfc{t:monot-class}{$\Sigma_X(\calA)=\calC_X(\calA)$},
  Lem\-ma~\threfc{l:monot-class-gen-is-monot}{%
    $\calC_X(\calA)\subset\calC_X(\calS)$}, and
  Lemma~\threfc{l:monot-class-gen-is-idem}{%
    $\calC_X(\calS)=\calS$},
  we have $\Sigma\subset\calS$.

  Therefore, we have the equality, {\ie}~$P$ holds for all subsets
  in~$\Sigma$.
\end{proof}

\chapter{Measurability}
\label{c:measurability}

\minitoc

\section{Measurable space and Borel subsets}
\label{s:measurable-space-and-borel-subsets}

\begin{definition}[measurable space]
  \label{d:measurable-space}
  \mbox{}\hfill
  Let~$X$ be a set.
  Let~$\Sigma$ be a $\sigma$-algebra on~$X$.\\
  Then, $(X,\Sigma)$ is called {\em measurable space}, and elements
  of~$\Sigma$ are said {\em ($\Sigma$-)measurable}.
\end{definition}

\begin{definition}[Borel $\sigma$-algebra]
  \label{d:borel-sigma-alg}
  \mbox{}\\
  Let~$(X,\calT)$ be a topological space.
  The {\em Borel $\sigma$-algebra of~$X$} is the $\sigma$-algebra generated
  by the open subsets;
  it is denoted $\calB(X)\eqdef\Sigma_X(\calT)$.

  $\calB(X)$-measurable subsets are called {\em Borel subsets of~$X$}.
\end{definition}

\begin{lemma}[some Borel subsets]
  \label{l:some-borel-subsets}
  \mbox{}\\
  Let~$(X,\calT)$ be a topological space.
  Then, open and closed subsets are Borel subsets of~$X$.

  Moreover, if the space~$X$ is separable, then countable subsets of~$X$ are
  Borel subsets of~$X$.
\end{lemma}

\begin{proof}
  From
  Definition~\thref{d:borel-sigma-alg}, and
  Definition~\thref{d:sigma-alg},
  the Borel subsets are the elements of the Borel $\sigma$-algebra.

  Let~$Y$ be an open subset of~$X$.
  Then, from
  Definition~\thref{d:borel-sigma-alg}, and
  Lemma~\thref{l:gen-sigma-alg-is-min},
  we have $Y\in\calT\subset\calB(X)$.

  Let~$Y$ be a closed subset of~$X$.
  Then, from
  Definition~\threfc{d:topological-space}{closed subset},
  Definition~\thref{d:borel-sigma-alg}, and
  Lemma~\thref{l:gen-sigma-alg-is-min},
  we have $Y^c\in\calT\subset\calB(X)$.
  Hence, from
  Definition~\threfc{d:sigma-alg}{closedness under complement}, and from
  \assume{involutiveness of complement},
  we have $Y=(Y^c)^c\in\calB(X)$.

  Let~$Y$ be a countable subset of~$X$.
  Then, from
  \assume{the definition of countability},
  there exists $I\subset\matN$, and $(x_i)_{i\in I}\in X$ such that
  $Y=\bigcup_{i\in I}\{x_i\}$.
  Let $i\in I$.
  Then, from
  \assume{closedness of singletons in a separable space},
  $\{x_i\}$~is closed, thus it belongs to~$\calB(X)$.
  Hence, from
  Definition~\threfc{d:sigma-alg}{closedness under countable union},
  we have $Y\in\calB(X)$.

  Therefore, open, closed, and countable subsets are Borel subsets of~$X$.
\end{proof}

\begin{lemma}[countable Borel $\sigma$-algebra generator]
  \label{l:count-borel-sigma-alg-gen}
  \mbox{}\\
  Let~$(X,\calT)$ be a topological space.
  Let~$G\subset\calT$.
  Assume that all open subsets of~$X$ are countable unions of elements
  of~$G$.
  Then, we have $\calB(X)=\Sigma_X(G)$.
\end{lemma}

\begin{proof}
  Direct consequence of
  Definition~\threfc{d:borel-sigma-alg}{$\calB(X)=\Sigma_X(\calT)$}, and
  Lemma~\threfc{l:count-sigma-alg-gen}{%
    with $G_1\eqdef G$ and $G_2\eqdef\calT$}.
\end{proof}

\begin{remark}
  The previous lemma means that if~$G$ contains a countable topological basis
  of~$(X,\calT)$ (and possibly other open subsets) (see
  Definitions~\ref{d:topological-basis} and~\ref{d:second-count}), then the
  Borel $\sigma$-algebra of~$X$ is generated by~$G$.
\end{remark}

\begin{remark}
  Note that metric spaces are separable topological spaces, thus both previous
  lemmas apply in metric spaces.
\end{remark}

\clearpage
\section{Measurable function}
\label{s:measurable-function}

\begin{definition}[measurable function]
  \label{d:meas-fun}
  \mbox{}\hfill
  Let~$(X,\Sigma)$ and $(\Xp,\Sigmap)$ be measurable spaces.\\
  A function~$f:\ArXXp$ is said
  {\em measurable (for~$\Sigma$ and~$\Sigmap$)} iff
  \begin{equation}
    \label{e:meas-fun}
    f^{-1} (\Sigmap) \subset \Sigma
    \qquad \mbox{{\ie} }
    \forall \Ap \in \Sigmap,\;
    f^{-1} (\Ap) \in \Sigma.
  \end{equation}

  If $\Sigma$ and $\Sigmap$ are the Borel $\sigma$-algebras of~$X$
  and~$\Xp$, then~$f$ is called {\em Borel function}.
\end{definition}

\begin{lemma}[inverse $\sigma$-algebra]
  \label{l:inverse-sigma-alg}
  \mbox{}\\
  Let~$X$ be a set.
  Let~$(\Xp,\Sigmap)$ be a measurable space.
  Let~$f:\ArXXp$.
  Then, $\Sigma=f^{-1}(\Sigmap)$ is the smallest $\sigma$-algebra
  of~$f^{-1}(\Xp)\subset X$ such that~$f$ is measurable for~$\Sigma$
  and~$\Sigmap$.

  It is called the {\em inverse $\sigma$-algebra of~$\Sigmap$ by~$f$}.
\end{lemma}

\begin{proof}
  Direct consequence of
  Definition~\thref{d:measurable-space},
  Definition~\thref{d:sigma-alg},
  \assume{homogeneity of inverse image},
  \assume{compatibility of inverse image with complement, union and
    intersection}, and
  Definition~\thref{d:meas-fun}.
\end{proof}

\begin{lemma}[image $\sigma$-algebra]
  \label{l:image-sigma-alg}
  \mbox{}\hfill
  Let~$(X,\Sigma)$ be a measurable space.
  Let~$\Xp$ be a set.\\
  Let~$f:\ArXXp$.
  Then, $\Sigmap=\imalg{f}{\Sigma}\eqdef
  \{\Ap\subset\Xp\st f^{-1}(\Ap)\in\Sigma\}$ is the largest
  $\sigma$-algebra of~$f(X)$ such that~$f$ is measurable for~$\Sigma$
  and~$\Sigmap$.
  It is called the {\em image $\sigma$-algebra of~$\Sigma$ by~$f$}.
\end{lemma}

\begin{proof}
  Direct consequence of
  Definition~\thref{d:measurable-space},
  Definition~\thref{d:sigma-alg},
  \assume{compatibility of inverse image with complement, union and
    intersection},
  \assume{homogeneity of inverse image ($f^{-1}(\emptyset)=\emptyset$)}, and
  Definition~\thref{d:meas-fun}.
\end{proof}

\begin{lemma}[identity function is measurable]
  \label{l:identity-fun-is-meas}
  \mbox{}\\
  Let~$(X,\Sigma)$ be a measurable space.
  Then, the identity function is measurable for~$\Sigma$ (and~$\Sigma$).
\end{lemma}

\begin{proof}
  Direct consequence of
  \assume{the definition of the identity function}, and
  Definition~\thref{d:meas-fun}.
\end{proof}

\begin{lemma}[constant function is measurable]
  \label{l:const-fun-is-meas}
  \mbox{}\hfill
  Let~$(X,\Sigma)$ and $(\Xp,\Sigmap)$ be measurable spaces.
  Let~$f:\ArXXp$.
  Assume that~$f$ is constant.
  Then, $f$~is measurable for~$\Sigma$ and~$\Sigmap$.
\end{lemma}

\begin{proof}
  Let~$\cp\in \Xp$ be the constant value taken by the function~$f$.
  Let~$\Ap\in\Sigmap$.\\
  \proofpar{Case $\cp\in\Ap$}
  Then, we have $f^{-1}(\Ap)=X$.
  \proofpar{Case $\cp\not\in\Ap$}
  Then, we have $f^{-1}(\Ap)=\emptyset$.
  Thus, from
  Lemma~\thref{l:equiv-def-of-sigma-alg},
  $f^{-1}(\Ap)$ always belongs to the $\sigma$-algebra~$\Sigma$.
  Therefore, $f$~is measurable for~$\Sigma$ and~$\Sigmap$.
\end{proof}

\begin{lemma}[inverse image of generating family]
  \label{l:inverse-image-of-gen-family}
  \mbox{}\\
  Let~$(X,\Sigma)$ and $(\Xp,\Sigmap)$ be measurable spaces.
  Let~$f:\ArXXp$.
  Then,  we have
  \begin{equation}
    \label{e:inverse-image-of-gen-family}
    \forall \Gp \subset \calP (\Xp),\quad
    \Sigma_X (f^{-1} (\Gp)) = f^{-1} (\Sigma_{\Xp} (\Gp)).
  \end{equation}
\end{lemma}

\begin{proof}
  \proofpar{``Left'' included in ``right''}
  Let~$\Gp\subset\calP(\Xp)$.
  Then, from
  Lemma~\thref{l:gen-sigma-alg-is-min},
  we have $\Gp\subset\Sigma_{\Xp}(\Gp)$.
  Thus, from
  \assume{monotonicity of inverse image},
  $f^{-1}(\Gp)$ is a subset of $f^{-1}(\Sigma_{\Xp}(\Gp))$.
  Hence, from
  Lemma~\thref{l:sigma-alg-gen-is-monot},
  Lemma~\threfc{l:inverse-sigma-alg}{%
    $f^{-1}(\Sigma_{\Xp}(\Gp))$ is $\sigma$-algebra}, and
  Lemma~\thref{l:sigma-alg-gen-is-idem},
  we have
  $\Sigma_X(f^{-1}(\Gp))\subset\Sigma_X(f^{-1}(\Sigma_{\Xp}(\Gp)))
  =f^{-1}(\Sigma_{\Xp}(\Gp))$.

  \proofparskip{``Right'' included in ``left''}
  Conversely, let $\Ap\in\Gp\subset\calP(\Xp)$.
  Then, from
  \assume{monotonicity of inverse image}, and
  Lemma~\thref{l:gen-sigma-alg-is-min},
  we have
  \begin{equation*}
    f^{-1} (\Ap) \in f^{-1} (\Gp) \subset \Sigma_X (f^{-1} (\Gp)).
  \end{equation*}
  Thus, from
  Lemma~\thref{l:image-sigma-alg},
  we have $\Gp\subset\imalg{f}{(\Sigma_X(f^{-1}(\Gp)))}$, and from
  Lemma~\thref{l:gen-sigma-alg-is-min},
  we have $\Sigma_{\Xp}(\Gp)\subset\imalg{f}{(\Sigma_X(f^{-1}(\Gp)))}$.
  Hence, from
  Lemma~\thref{l:image-sigma-alg}, and
  \assume{the definition of inverse image},
  we have
  \begin{equation*}
    f^{-1} (\Sigma_{\Xp} (\Gp)) \subset \Sigma_X (f^{-1} (\Gp)).
  \end{equation*}

  Therefore, we have $\Sigma_X(f^{-1}(\Gp))=f^{-1}(\Sigma_{\Xp}(\Gp))$.
\end{proof}

\begin{lemma}[equivalent definition of measurable function]
  \label{l:equiv-def-of-meas-fun}
  \mbox{}\hfill
  Let~$(X,\Sigma)$ and $(\Xp,\Sigmap)$ be measurable spaces.
  Let~$f:\ArXXp$.
  Then, $f$~is measurable for~$\Sigma$ and~$\Sigmap$ iff
  \begin{equation}
    \label{e:equiv-def-of-meas-fun}
    \exists \Gp \subset \calP (\Xp),\quad
    \Sigma_{\Xp} (\Gp) = \Sigmap
    \IMPLIES
    f^{-1} (\Gp) \subset \Sigma.
  \end{equation}
\end{lemma}

\begin{proof}
  \proofpar{``Left'' implies ``right''}
  Assume first that~$f$ is measurable for~$\Sigma$ and~$\Sigmap$.
  Let~$\Gp\eqdef\Sigmap$.
  Then, from
  Lemma~\thref{l:sigma-alg-gen-is-idem},
  we have $\Sigma_{\Xp}(\Sigmap)=\Sigmap$.
  Hence, from
  Definition~\thref{d:meas-fun},
  we have $f^{-1}(\Gp)=f^{-1}(\Sigmap)\subset\Sigma$.

  \proofparskip{``Right'' implies ``left''}
  Conversely, assume now that there exists $\Gp\subset\calP(\Xp)$ such that
  $\Sigma_{\Xp}(\Gp)=\Sigmap$ and $f^{-1}(\Gp)\subset\Sigma$.
  Then, from
  Lemma~\thref{l:inverse-image-of-gen-family},
  Lemma~\thref{l:sigma-alg-gen-is-monot}, and
  Lemma~\thref{l:sigma-alg-gen-is-idem},
  we have $f^{-1}(\Sigmap)=f^{-1}(\Sigma_{\Xp}(\Gp))
  =\Sigma_X(f^{-1}(\Gp))\subset\Sigma_X(\Sigma)=\Sigma$.

  \medskip\noindent
  Therefore, we have the equivalence.
\end{proof}

\begin{lemma}[continuous is measurable]
  \label{l:cont-is-meas}
  \mbox{}\\
  Let~$(X,\calT)$ and $(\Xp,\calTp)$ be topological spaces.
  Assume that~$X$ and~$\Xp$ are equipped with their Borel $\sigma$-algebras.
  Let~$f:\ArXXp$.
  Assume that~$f$ is continuous.
  Then, $f$~is a Borel function.
\end{lemma}

\begin{proof}
  Let~$\Op\in\calTp$.
  Then, from
  \assume{the definition of continuity},
  $f^{-1}(\Op)\in\calT$.
  Thus, from
  Lemma~\thref{l:gen-sigma-alg-is-min}, and
  Definition~\thref{d:borel-sigma-alg},
  we have $\calB(\Xp)=\Sigma_{\Xp}(\calTp)$ and
  $f^{-1}(\calTp)\subset\calT\subset\Sigma_X(\calT)=\calB(X)$.
  Therefore, from
  Lemma~\thref{l:equiv-def-of-meas-fun}, and
  Definition~\thref{d:meas-fun},
  $f$~is a Borel function from~$X$ to~$\Xp$.
\end{proof}

\begin{lemma}[compatibility of measurability with composition]
  \label{l:compat-of-meas-with-comp}
  \mbox{}\\
  Let~$(X,\Sigma)$, $(\Xp,\Sigmap)$,
  and $(\Xpp,\Sigmapp)$ be measurable spaces.
  Let~$f:\ArXXp$ and~$g:\ArXpXpp$.
  Assume that~$f$ is measurable for~$\Sigma$ and~$\Sigmap$, and
  that~$g$ is measurable for~$\Sigmap$ and~$\Sigmapp$.
  Then, $g\circ f$ is measurable for~$\Sigma$ and~$\Sigmapp$.
\end{lemma}

\begin{proof}
  Direct consequence of
  \assume{the definition of composition},
  Definition~\threfc{d:meas-fun}{with~$f$ and~$g$}, and
  \assume{monotonicity of inverse image}.
\end{proof}

\clearpage
\section{Measurable subspace}
\label{s:measurable-subspace}

\begin{remark}
  We recall that~$\olcap$ denotes a set of traces of subsets, see
  Definition~\ref{d:trace-of-subsets-of-parties}.
\end{remark}

\begin{lemma}[trace $\sigma$-algebra]
  \label{l:trace-sigma-alg}
  \mbox{}\\
  Let~$(X,\Sigma)$ be a measurable space.
  Let~$Y\subset X$.
  Let~$i$ be the  canonical injection from~$Y$ to~$X$.
  Then, $\Sigma\olcap Y$ is a $\sigma$-algebra of~$Y$,
  and~$i$ is measurable for~$\Sigma\olcap Y$ and~$\Sigma$.

  $\Sigma\olcap Y$ is called {\em trace $\sigma$-algebra of~$\Sigma$ on~$Y$}.
  The measurable space $(Y,\Sigma\olcap Y)$ is said
  {\em measurable subspace of~$(X,\Sigma)$}.
\end{lemma}

\begin{proof}
  From
  Definition~\threfc{d:measurable-space}{$\Sigma$ is a $\sigma$-algebra},
  Definition~\threfc{d:sigma-alg}{$\emptyset\in\Sigma$}, and since
  \assume{$\emptyset$ is absorbing for intersection},
  we have $\emptyset=\emptyset\cap Y\in\Sigma\olcap Y$.

  Let~$A\in\Sigma\olcap Y$.
  Then, from
  Definition~\thref{d:trace-of-subsets-of-parties},
  let~$B\in\Sigma$ such that $A=B\cap Y$.
  Then, from
  \assume{the definition of set difference},
  \assume{De~Morgan's laws},
  \assume{distributivity of intersection over union},
  \assume{the definition of complement}, and
  \assume{commutativity of intersection},
  we have
  \begin{equation*}
    Y \setminus A
    = Y \setminus (B \cap Y)
    = Y \cap (B \cap Y)^c
    = Y \cap (B^c \cup Y^c)
    = (Y \cap B^c) \cup (Y \cap Y^c)
    = B^c \cap Y.
  \end{equation*}
  Hence, from
  Definition~\threfc{d:sigma-alg}{closedness under complement},
  we have $Y\setminus A\in\Sigma\olcap Y$.

  Let~$I$ be a set.
  Let~$(A_i)_{i\in I}\in\Sigma\olcap Y$.
  Then, from
  Definition~\thref{d:trace-of-subsets-of-parties},
  for all $i\in I$, let~$B_i\in\Sigma$ such that $A_i=B_i\cap Y$.
  Then, from
  \assume{distributivity of intersection over union},
  we have
  \begin{equation*}
    \bigcup_{i \in I} A_i
    = \bigcup_{i \in I} (B_i \cap Y)
    = \bigcup_{i \in I} B_i \cap Y.
  \end{equation*}
  Hence, from
  Definition~\threfc{d:sigma-alg}{closedness under union},
  we have $\bigcup_{i\in I}A_i\in\Sigma\olcap Y$.

  Therefore, from
  Definition~\thref{d:sigma-alg},
  $\Sigma\olcap Y$~is a $\sigma$-algebra of~$Y$.
  Moreover, from
  \assume{the definition of the canonical injection}, and
  Definition~\thref{d:meas-fun},
  $i$~is measurable for~$\Sigma\olcap Y$ and~$\Sigma$.
\end{proof}

\begin{lemma}[measurability of measurable subspace]
  \label{l:meas-of-meas-subspace}
  \mbox{}\\
  Let~$(X,\Sigma)$ be a measurable space.
  Let~$Y\subset X$.
  Then, we have $Y\in\Sigma$ iff
  $\Sigma\olcap Y=\{A\in\Sigma\st A\subset Y\}$.
\end{lemma}

\begin{proof}
  Let~$A\in\Sigma$ such that $A\subset Y$.
  Then, from
  Definition~\thref{d:trace-of-subsets-of-parties},
  $A=A\cap Y$ belongs to $\Sigma\olcap Y$.
  Thus, we always have $\{A\in\Sigma\st A\subset Y\}\subset\Sigma\olcap Y$.

  \proofparskip{``Left'' implies ``right''}
  Assume first that $Y\in\Sigma$.
  Let~$A\in\Sigma$, and let $\Ap\eqdef A\cap Y$.
  Then, from
  Lemma~\threfc{l:equiv-def-of-sigma-alg}{%
    closedness under countable intersection},
  we have $\Ap\in\Sigma$ and $\Ap\subset Y$.
  Thus, from
  Definition~\thref{d:trace-of-subsets-of-parties},
  $\Sigma\olcap Y$ is included in $\{A\in\Sigma\st A\subset Y\}$.
  Hence, $\Sigma\olcap Y=\{A\in\Sigma\st A\subset Y\}$.

  \proofparskip{``Right'' implies ``left''}
  Conversely, assume now that $\Sigma\olcap Y=\{A\in\Sigma\st A\subset Y\}$.
  Then, from
  Lemma~\thref{l:trace-sigma-alg}, and
  Lemma~\threfc{l:equiv-def-of-sigma-alg}{contains full set},
  we have $Y\in\Sigma\olcap Y\subset\Sigma$.

  \medskip\noindent
  Therefore, we have the equivalence.
\end{proof}

\begin{lemma}[generating measurable subspace]
  \label{l:gen-meas-subspace}
  \mbox{}\\
  Let~$X$ be a set.
  Let~$Y\subset X$.
  Let~$G\subset\calP(X)$.
  Then, we have $\Sigma_Y(G\olcap Y)=\Sigma_X(G)\olcap Y$.
\end{lemma}

\begin{proof}
  Let~$i$ be the canonical injection from~$Y$ to~$X$.
  Then, from
  Definition~\thref{d:trace-of-subsets-of-parties},
  we have $i^{-1}(G)=G\olcap Y$ and $i^{-1}(\Sigma_X(G))=\Sigma_X(G)\olcap Y$.
  Therefore, from
  Lemma~\thref{l:inverse-image-of-gen-family},
  we have
  \begin{equation*}
    \Sigma_Y (G \olcap Y)
    = \Sigma_Y (i^{-1} (G))
    = i^{-1} (\Sigma_X (G))
    = \Sigma_X (G) \olcap Y.
  \end{equation*}
\end{proof}

\begin{lemma}[Borel sub-$\sigma$-algebra]
  \label{l:borel-sub-sigma-alg}
  \mbox{}\hfill
  Let~$(X,\calT)$ be a topological space.\\
  Let~$Y\subset X$.
  Then, we have $\calB(Y)=\calB(X)\olcap Y$, and $Y\in\calB(X)$ iff
  $\calB(Y)=\{A\in\calB(X)\st A\subset Y\}$.
\end{lemma}

\begin{proof}
  Direct consequence of
  Definition~\thref{d:borel-sigma-alg},
  \assume{the definition of trace topology on~$Y$},
  Lemma~\thref{l:gen-meas-subspace}, and
  Lemma~\thref{l:meas-of-meas-subspace}.
\end{proof}

\begin{lemma}[characterization of Borel subsets]
  \label{l:characterization-of-borel-subsets}
  \mbox{}\hfill
  Let~$(X,\calT)$ be a topology space.\\
  Let $(Y_n)_{n\in\matN}\in\calB(X)$.
  Assume that $X=\biguplus_{n\in\matN}Y_n$.
  Let~$A\subset X$.
  Then, we have
  \begin{equation}
    \label{e:characterization-of-borel-subsets}
    A \in \calB (X)
    \EQUIV
    \forall n \in \matN,\quad
    A \cap Y_n \in \calB (Y_n).
  \end{equation}
\end{lemma}

\begin{proof}
  \proofpar{``Left'' implies ``right''}
  Assume first that $A\in\calB(X)$.
  Let~$n\in\matN$.
  Then, $A\cap Y_n\subset Y_n$, and from
  Lemma~\threfc{l:equiv-def-of-sigma-alg}{%
    closedness under countable intersection},
  we have $A\cap Y_n\in\calB(X)$.
  Hence, from
  Lemma~\thref{l:borel-sub-sigma-alg},
  we have $A\cap Y_n\in\calB(Y_n)$.

  \proofparskip{``Right'' implies ``left''}
  Conversely, assume now that for all $n\in\matN$, $A\cap Y_n\in\calB(Y_n)$.
  Let~$n\in\matN$.
  Then, since $Y_n\in\calB(X)$, from
  Lemma~\thref{l:borel-sub-sigma-alg},
  we have $A\cap Y_n\in\calB(X)$.
  Hence, from
  Lemma~\thref{l:compat-of-pseudopart-with-inter},
  Definition~\thref{d:borel-sigma-alg}, and
  Definition~\threfc{d:sigma-alg}{closedness under countable union},
  we have
  \begin{equation*}
    A = \biguplus_{n \in \matN} (A \cap Y_n) \in \calB(X).
  \end{equation*}

  \medskip\noindent
  Therefore, we have the equivalence.
\end{proof}

\begin{lemma}[source restriction of measurable function]
  \label{l:source-restr-of-meas-fun}
  \mbox{}\hfill
  Let~$(X,\Sigma)$ and $(\Xp,\Sigmap)$ be measurable spaces.
  Let~$f:\ArXXp$.
  Let~$Y\subset X$.
  Let~$\restr{f}{Y}$ be the restriction of~$f$ to the source~$Y$.
  Assume that $f$ is measurable for~$\Sigma$ and~$\Sigmap$.
  Then, $\restr{f}{Y}$ is measurable for $\Sigma\olcap Y$ and~$\Sigmap$.
\end{lemma}

\begin{proof}
  Let~$i$ be the canonical injection from~$Y$ to~$X$.
  Then, from
  Definition~\thref{d:trace-of-subsets-of-parties}, and
  Lemma~\thref{l:trace-sigma-alg},
  $i^{-1}(\Sigma)=\Sigma\olcap Y$ is a $\sigma$-algebra of~$Y$.
  Thus, from
  Definition~\thref{d:meas-fun},
  $i$~is measurable for~$\Sigma\olcap Y$ and~$\Sigma$.
  Therefore, from
  Lemma~\thref{l:compat-of-meas-with-comp},
  $\restr{f}{Y}=f\circ i$ is measurable for $\Sigma\olcap Y$ and~$\Sigmap$.
\end{proof}

\begin{lemma}[destination restriction of measurable function]
  \label{l:destination-restr-of-meas-fun}
  \mbox{}\hfill
  Let~$(X,\Sigma)$ and $(\Xp,\Sigmap)$ be measurable spaces.
  Let~$f:\ArXXp$.
  Let~$\Yp\subset\Xp$.
  Assume that $f(X)\subset\Yp$.
  Let~$\drestr{f}{\Yp}$ be the restriction of~$f$ to the
  destination~$\Yp$.
  Then, $f$~is measurable for~$\Sigma$ and~$\Sigmap$ iff
  $\drestr{f}{\Yp}$~is measurable for~$\Sigma$ and~$\Sigmap\olcap\Yp$.
\end{lemma}

\begin{proof}
  Let~$\Ap\in\Sigmap$.
  Then, from
  \assume{the definition of set difference}, and
  \assume{monotonicity of inverse image and complement},
  we have
  \begin{equation*}
    f^{-1} (\Ap \setminus \Yp)
    = f^{-1} (\Ap \cap {\Yp}^c)
    \subset f^{-1} ({\Yp}^c)
    \subset f^{-1} (f(X)^c)
    = \emptyset.
  \end{equation*}
  Then, from
  \assume{compatibility of inverse image with disjoint union}, and
  \assume{the definition of destination restriction},
  we have
  \begin{equation*}
    f^{-1} (\Ap)
    = f^{-1} ((\Ap \cap \Yp) \uplus (\Ap \setminus \Yp))
    = f^{-1} (\Ap \cap \Yp)
    = \left( \drestr{f}{\Yp} \right)^{-1} (\Ap \cap \Yp).
  \end{equation*}
  Hence,
  $f^{-1}(\Sigmap)=\left(\drestr{f}{\Yp}\right)^{-1}(\Sigmap\olcap\Yp)$.
  Therefore, from
  Definition~\thref{d:meas-fun},
  $\drestr{f}{\Yp}$ is measurable for $\Sigma$ and~$\Sigmap\olcap\Yp$.
\end{proof}

\begin{lemma}[measurability of function defined on a pseudopartition]
  \label{l:meas-of-fun-def-on-pseudopart}
  \mbox{}\hfill
  Let~$(X,\Sigma)$ and $(\Xp,\Sigmap)$ be measurable spaces.
  Let~$I\subset\matN$.
  For all $i\in I$, let~$X_i\in\Sigma$ and~$f_i:\ArXXp$.
  Assume that for all $i\in I$, $f_i$~is measurable
  for~$\Sigma$ and~$\Sigmap$, and that $X=\biguplus_{i\in I}X_i$.
  Then, the function defined by~$f_i$ on~$X_i$ for all $i\in I$ is measurable
  for~$\Sigma$ and~$\Sigmap$.
\end{lemma}

\begin{proof}
  Let~$f$ be the function defined by for all $i\in I$, for all $x\in X_i$,
  $f(x)\eqdef f_i(x)$.
  Let~$\Ap\in\Sigmap$.\\
  Let~$i\in I$.
  Then, from
  Definition~\threfc{d:meas-fun}{with~$f_i$},
  we have $f_i^{-1}(\Ap)\in\Sigma$.
  Thus, from
  Lemma~\threfc{l:equiv-def-of-sigma-alg}{%
    closedness under countable intersection and union},
  we have
  $f^{-1}(\Ap)=\bigcup_{i\in I}\left(X_i\cap f_i^{-1}(\Ap)\right)\in\Sigma$.
  Hence, $f^{-1}(\Sigmap)\subset\Sigma$.

  Therefore, from
  Definition~\thref{d:meas-fun},
  $f$~is measurable for~$\Sigma$ and~$\Sigmap$.
\end{proof}

\clearpage
\section{Product of measurable spaces}
\label{s:product-of-measurable-spaces}

\begin{remark}
  We recall that~$[n..p]$ denotes an interval of integers, and that~$\olprod$
  and~$\oltimes$ denote a set of Cartesian products of subsets, see
  Definition~\ref{d:prod-of-subsets-of-parties}.

  The concepts presented in this section are used mainly in
  Section~\ref{s:measure-and-integration-over-product-space}.
\end{remark}

\begin{definition}[tensor product of $\sigma$-algebras]
  \label{d:tensor-prod-of-sigma-algs}
  \mbox{}\\
  Let~$m\in[2..\infty)$.
  For all $i\in[1..m]$, let~$(X_i,\Sigma_i)$ be a measurable space.
  Let~$X\eqdef\prod_{i\in[1..m]}X_i$
  and~$\Sigmabar\eqdef\olprod_{i\in[1..m]}\Sigma_i$.
  The {\em (tensor) product of the
    $\sigma$-algebras~$(\Sigma_i)_{i\in[1..m]}$} on~$X$ is the
  $\sigma$-algebra generated by~$\Sigmabar$;
  it is denoted $\bigotimes_{i\in[1..m]}\Sigma_i\eqdef\Sigma_X(\Sigmabar)$.
\end{definition}

\begin{lemma}[product of measurable subsets is measurable]
  \label{l:prod-of-meas-subsets-is-meas}
  \mbox{}\\
  Let~$m\in[2..\infty)$.
  For all $i\in[1..m]$, let~$(X_i,\Sigma_i)$ be a measurable space,
  and~$A_i\in \Sigma_i$.\\
  Let~$\Sigmabar\eqdef\olprod_{i\in[1..m]}\Sigma_i$
  and~$\Sigma\eqdef\bigotimes_{i\in[1..m]}\Sigma_i$.
  Then, $\prod_{i\in[1..m]}A_i\in\Sigmabar\subset\Sigma$.
\end{lemma}

\begin{proof}
  Direct consequence of
  Definition~\thref{d:prod-of-subsets-of-parties},
  Definition~\thref{d:tensor-prod-of-sigma-algs}, and
  Lemma~\threfc{l:gen-sigma-alg-is-min}{%
    $\Sigmabar\subset\Sigma_X(\Sigmabar)$}.
\end{proof}

\begin{lemma}[measurability of function to product space]
  \label{l:meas-of-fun-to-prod-space}
  \mbox{}\\
  Let~$m\in[2..\infty)$.
  Let~$(X,\Sigma)$ be a measurable space.
  For all $i\in[1..m]$, let~$(\Xp_i,\Sigmap_i)$ be a measurable space,
  and~$f_i:\ArXXp_i$.
  Then, $(f_i)_{i\in[1..m]}$~is measurable for~$\Sigma$
  and~$\bigotimes_{i\in[1..m]}\Sigmap_i$ iff
  for all $i\in[1..m]$, $f_i$ is measurable for~$\Sigma$ and~$\Sigmap_i$.
\end{lemma}

\begin{proof}
  Let~$\Sigmap\eqdef\bigotimes_{i\in[1..m]}\Sigmap_i$.

  \proofparskip{``Left'' implies ``right''}
  Assume first that~$(f_i)_{i\in[1..m]}$ is measurable for~$\Sigma$
  and~$\Sigmap$.\\
  Let~$i\in[1..m]$.
  Let~$\Ap_i\in\Sigmap_i$.
  Let~$\Ap\eqdef\Xp_1\times\ldots\times\Ap_i\times\ldots\times\Xp_m$.
  Then, from
  Lemma~\threfc{l:equiv-def-of-sigma-alg}{$\Xp_i\in\Sigmap_i$},
  Definition~\thref{d:tensor-prod-of-sigma-algs}, and
  Lemma~\thref{l:prod-of-meas-subsets-is-meas},
  we have $\Ap\in\Sigmap$.
  Thus, from
  Definition~\threfc{d:meas-fun}{with~$f$},
  we have $f_i^{-1}(\Ap_i)=f^{-1}(\Ap)\in\Sigma$.
  Hence, from
  Definition~\thref{d:meas-fun},
  $f_i$~is measurable for~$\Sigma$ and~$\Sigmap_i$.

  \proofparskip{``Right'' implies ``left''}
  Assume now that for all~$i\in[1..m]$, $f_i$~is measurable
  for~$\Sigma$ and~$\Sigmap_i$.\\
  Let~$\Sigmabarp\eqdef\olprod_{i\in[1..m]}\Sigmap_i$
  and~$f\eqdef(f_i)_{i\in[1..p]}$.
  Let~$\Ap\eqdef\prod_{i\in[1..m]}\Ap_i\in\Sigmabarp$.
  From
  Definition~\thref{d:prod-of-subsets-of-parties},
  for all $i\in[1..m]$, we have $\Ap_i\in\Sigmap_i$.
  Let~$x\in X$.
  Then, $f(x)\in\Ap$ iff
  for all $i\in[1..m]$, $f_i(x)\in\Ap_i$.
  Thus, we have $f^{-1}(\Ap)=\bigcap_{i\in[1..m]}f_i^{-1}(\Ap_i)$.
  Hence, from
  Definition~\threfc{d:meas-fun}{with~$f_i$}, and
  Lemma~\threfc{l:equiv-def-of-sigma-alg}{%
    closedness under countable intersection},
  we have $f^{-1}(\Sigmabarp)
  =\bigcap_{i\in[1..m]}f_i^{-1}(\Sigmap_i)\subset\Sigma$.
  Thus, from
  Lemma~\threfc{l:equiv-def-of-meas-fun}{%
    with $\Gp\eqdef\Sigmabarp$},
  $f$~is measurable for~$\Sigma$ and~$\Sigmap$.

  \medskip\noindent
  Therefore, we have the equivalence.
\end{proof}

\begin{lemma}[canonical projection is measurable]
  \label{l:can-proj-is-meas}
  \mbox{}\\
  Let~$m\in[2..\infty)$.
  For all $i\in[1..m]$, let~$(X_i,\Sigma_i)$ be a measurable space,
  and let~$\pi_i$ be the canonical projection
  from~$X\eqdef\prod_{i\in[1..m]}X_i$ onto~$X_i$.
  Let~$\Sigma\eqdef\bigotimes_{i\in[1..m]}\Sigma_i$.\\
  Then, $\Sigma$~is the smallest $\sigma$-algebra on~$X$ such that for all
  $i\in[1..m]$, $\pi_i$~is measurable for~$\Sigma$ and~$\Sigma_i$.
\end{lemma}

\begin{proof}
  Let~$\Sigmap$ be a $\sigma$-algebra on~$X$.
  Then, from
  Lemma~\threfc{l:meas-of-fun-to-prod-space}{%
    with $\Sigma\eqdef\Sigmap$, $\Xp_i\eqdef X_i$,
    $\Sigmap_i\eqdef\Sigma_i$, and $f\eqdef\idX=(\pi_i)_{i\in[1..m]}$},
  Definition~\thref{d:meas-fun}, and
  \assume{reflexivity of inclusion}
  we have
  \begin{align*}
    & \forall i \in [1..m],\;
    \pi_i \mbox{ is measurable for } \Sigmap \mbox{ and } \Sigma_i\\
    & \EQUIV \idX \mbox{ is measurable for } \Sigmap \mbox{ and } \Sigma\\
    & \EQUIV \idX^{-1} (\Sigma) = \Sigma \subset \Sigmap.
  \end{align*}
\end{proof}

\begin{lemma}[permutation is measurable]
  \label{l:perm-is-meas}
  \mbox{}\hfill
  Let~$m\in[2..\infty)$.
  Let~$\psi$ be a permutation of~$[1..m]$.
  For all $i\in[1..m]$, let~$(X_i,\Sigma_i)$ be a measurable space,
  and let~$\pi_i$ be the canonical projection
  from~$X\eqdef\prod_{i\in[1..m]}X_i$ onto~$X_i$.
  Let~$\Sigma\eqdef\bigotimes_{i\in[1..m]}\Sigma_i$,
  and~$\Sigma^\psi\eqdef\bigotimes_{i\in[1..m]}\Sigma_{\psi(i)}$.\\
  Then, the permutation of coordinates $(\pi_{\psi(i)})_{i\in[1..m]}$ is
  measurable for~$\Sigma$ and~$\Sigma^\psi$.
\end{lemma}

\begin{proof}
  Direct consequence of
  Lemma~\thref{l:meas-of-fun-to-prod-space}, and
  Lem\-ma~\threfc{l:can-proj-is-meas}{%
    with $\Xp_i\eqdef X_{\psi(i)}$ and $f_i\eqdef\pi_{\psi(i)}$}.
\end{proof}

\begin{lemma}[generating product measurable space]
  \label{l:gen-prod-meas-space}
  \mbox{}\hfill
  Let~$m\in[2..\infty)$.
  For all $i\in[1..m]$, let~$(X_i,\Sigma_i)$ be a measurable space,
  $G_i\subset\calP(X_i)$, and assume that $X_i\in G_i$ and
  $\Sigma_i=\Sigma_{X_i}(G_i)$.
  Let~$X\eqdef\prod_{i\in[1..m]}X_i$
  and~$\Sigma\eqdef\bigotimes_{i\in[1..m]}\Sigma_i$.
  Then, we have $\Sigma=\Sigma_X\left(\olprod_{i\in[1..m]}G_i\right)$.
\end{lemma}

\begin{proof}
  Let~$\Gbar\eqdef\olprod_{i\in[1..m]}G_i$.

  \proofparskip{$\Sigma\subset\Sigma_X(\Gbar)$}
  Let $i\in[1..m]$.
  Let~$\pi_i$ be the canonical projection from~$X$ onto~$X_i$.
  Then, from
  Lemma~\thref{l:inverse-image-of-gen-family},
  we have
  \begin{equation*}
    \pi_i^{-1} (\Sigma_i)
    = \pi_i^{-1} (\Sigma_{X_i} (G_i))
    = \Sigma_X (\pi_i^{-1} (G_i)).
  \end{equation*}
  Moreover, since $X_i\in G_i$, and from
  Definition~\thref{d:prod-of-subsets-of-parties},
  we have
  \begin{equation*}
    \pi_i^{-1} (G_i)
    = \{ X_1 \times \ldots \times A_i \times \ldots \times X_m
    \st A_i \in G_i \}
    \subset \Gbar.
  \end{equation*}

  Then, from
  Lemma~\thref{l:sigma-alg-gen-is-monot},
  we have $\pi_i^{-1}(\Sigma_i)\subset\Sigma_X(\Gbar)$.
  Thus, from
  Definition~\thref{d:meas-fun},
  $\pi_i$~is measurable for~$\Sigma_X(\Gbar)$ and~$\Sigma_i$.
  Hence, from
  Lemma~\threfc{l:can-proj-is-meas}{%
    smallest $\sigma$-algebra},
  we have $\Sigma\subset\Sigma_X(\Gbar)$.

  \proofparskip{$\Sigma_X(\Gbar)\subset\Sigma$}
  Let~$\Sigmabar\eqdef\olprod_{i\in[1..m]}\Sigma_i$.\\
  Direct consequence of
  Lemma~\threfc{l:gen-sigma-alg-is-min}{%
    $G_i\subset\Sigma_i$},
  Definition~\threfc{d:prod-of-subsets-of-parties}{%
    $\Gbar\subset\Sigmabar$},
  Lemma~\threfc{l:sigma-alg-gen-is-monot}{%
    $\Sigma_X(\Gbar)$ is included in $\Sigma_X(\Sigmabar)$}, and
  Definition~\threfc{d:tensor-prod-of-sigma-algs}{%
    $\Sigma=\Sigma_X(\Sigmabar)$},

  \medskip\noindent
  Therefore, we have the equality.
\end{proof}

\begin{remark}
  For the sake of simplicity, we only present the remainder of this section
  in the case of the product of two measure spaces.
  When $i\in\{1,2\}$, the complement $\{1,2\}\setminus\{i\}$ is~$\{3-i\}$.
\end{remark}

\begin{definition}[section in Cartesian product]
  \label{d:section-in-cartesian-prod}
  \mbox{}\hfill
  Let~$X_1$ and~$X_2$ be sets.
  Let~$A\subset X_1\times X_2$.
  Let~$i\in\{1,2\}$.
  Let~$j\eqdef3-i$.
  Let~$x_i\in X_i$.
  The {\em $i$-th section of~$A$ in~$x_i$} is the subset
  \begin{equation}
    \label{e:section-in-cartesian-prod}
    s_i (x_i, A) \eqdef \{ x_j \in X_j \st (x_1, x_2) \in A \}.
  \end{equation}
\end{definition}

\begin{lemma}[section of product]
  \label{l:section-of-prod}
  \mbox{}\hfill
  Let~$X_1$ and~$X_2$ be sets.
  Let~$A_1\in X_1$ and $A_2\in X_2$.
  Let~$i\in\{1,2\}$.
  Let~$j\eqdef3-i$.
  Let~$x_i\in X_i$.
  Then, we have
  \begin{equation}
    \label{e:section-of-prod}
    s_i (x_i, A_1 \times A_2)
    = \left\{
      \begin{array}{lc}
        A_j & \mbox{when } x_i \in A_i,\\
        \emptyset & \mbox{otherwise}.
      \end{array}
    \right.
  \end{equation}
\end{lemma}

\begin{proof}
  Direct consequence of
  Definition~\thref{d:section-in-cartesian-prod}.
\end{proof}

\begin{lemma}[compatibility of section with set operations]
  \label{l:compat-of-section-with-set-ops}
  \mbox{}\hfill
  Let~$X_1$ and~$X_2$ be sets.\\
  Let~$X\eqdef X_1\times X_2$.
  Let~$i\in\{1,2\}$.
  Let~$x_i\in X_i$.
  Then, the sections are compatible with the empty set, the complement,
  countable union and intersection, and are monotone:
  \begin{align}
    \label{e:compat-of-section-with-set-ops-1}
    & s_i (x_i, \emptyset) = \emptyset,\\
    \label{e:compat-of-section-with-set-ops-2}
    \forall A \subset X,\quad &
    s_i (x_i, A^c) = s_i (x_i, A)^c,\\
    \label{e:compat-of-section-with-set-ops-3}
    \forall I \subset \matN,\;
    \forall (A_n)_{n \in I} \subset X,\quad &
    s_i \left( x_i, \bigcup_{n \in I} A_n \right)
    = \bigcup_{n \in I} s_i (x_i, A_n),\\
    \label{e:compat-of-section-with-set-ops-4}
    \forall I \subset \matN,\;
    \forall (A_n)_{n \in I} \subset X,\quad &
    s_i \left( x_i, \bigcap_{n \in I} A_n \right)
    = \bigcap_{n \in I} s_i (x_i, A_n),\\
    \label{e:compat-of-section-with-set-ops-5}
    \forall A, B \subset X,\quad &
    A \subset B \IMPLIES
    s_i (x_i, A) \subset s_i (x_i, B).
  \end{align}
\end{lemma}

\begin{proof}
  Direct consequences of
  Definition~\thref{d:section-in-cartesian-prod}.
\end{proof}

\begin{lemma}[measurability of section]
  \label{l:measurability-of-section}
  \mbox{}\hfill
  Let~$(X_1,\Sigma_1)$ and~$(X_2,\Sigma_2)$ be measurable spaces.
  Let~$A\in\Sigma_1\otimes\Sigma_2$.
  Let~$i\in\{1,2\}$.
  Let~$j\eqdef3-i$.
  Let~$x_i\in X_i$.
  Then, we have $s_i(x_i,A)\in\Sigma_j$.
\end{lemma}

\begin{proof}
  Let~$\Sigmabar\eqdef\Sigma_1\oltimes\Sigma_2$
  and~$\Sigma\eqdef\Sigma_1\otimes\Sigma_2$.
  Let~$\calS_i\eqdef\{A\subset X\st s_i(x_i,A)\in\Sigma_j\}$.

  From
  Definition~\thref{d:measurable-space},
  $\Sigma_j$ is a $\sigma$-algebras.

  Let~$A_1\in\Sigma_1$, and $A_2\in\Sigma_2$.
  Then, from
  Lemma~\threfc{l:section-of-prod}{%
    $s_i(x_i,A_1\times A_2)$ belongs to $\{\emptyset,A_j\}$},
  Definition~\threfc{d:sigma-alg}{%
    $\emptyset,A_j\in\Sigma_j$}, and
  the definition of~$\calS_i$,
  we have $A_1\times A_2\in\calS_i$.
  Hence, from
  Definition~\thref{d:prod-of-subsets-of-parties},
  we have $\Sigmabar\subset\calS_i$.

  From
  Lemma~\thref{l:compat-of-section-with-set-ops}, and
  Definition~\thref{d:sigma-alg},
  $\calS_i$~contains~$\emptyset$, and is closed under complement and
  countable union.
  Thus, from
  Definition~\thref{d:sigma-alg},
  $\calS_i$~is a $\sigma$-algebra on~$X$.
  Hence, from
  Definition~\threfc{d:tensor-prod-of-sigma-algs}{%
    $\Sigma$~is a generated by $\Sigmabar$},
  Lemma~\thref{l:gen-sigma-alg-is-min},
  we have $\Sigma\subset\calS_i$.

  Therefore, for all $A\in\Sigma$ we have $s_i(x_i,A)\in\Sigma_j$.
\end{proof}

\begin{lemma}[countable union of sections is measurable]
  \label{l:count-union-of-sections-is-meas}
  \mbox{}\\
  Let~$(X_1,\Sigma_1)$ and~$(X_2,\Sigma_2)$ be measurable spaces.
  Let $I\subset\matN$.
  Let~$(A_n)_{n\in I}\in\Sigma_1\otimes\Sigma_2$.
  Let~$i$ be in~$\{1,2\}$.
  Let~$j\eqdef3-i$.
  Let~$x_i\in X_i$.
  Then, we have
  \begin{equation}
    \label{e:count-union-of-sections-is-meas}
    s_i \left( x_i, \bigcup_{n \in I} A_n \right)
    = \bigcup_{n \in I} s_i (x_i, A_n)  \in \Sigma_j.
  \end{equation}
\end{lemma}

\begin{proof}
  Direct consequence of
  Lemma~\threfc{l:compat-of-section-with-set-ops}{%
    with count\-able union},
  Definition~\threfc{d:measurable-space}{$\Sigma_j$ is a $\sigma$-algebra},
  Definition~\threfc{d:sigma-alg}{closedness under countable union}, and
  Lemma~\thref{l:measurability-of-section}.
\end{proof}

\begin{lemma}[countable intersection of sections is measurable]
  \label{l:count-inter-of-sections-is-meas}
  \mbox{}\\
  Let~$(X_1,\Sigma_1)$ and~$(X_2,\Sigma_2)$ be measurable spaces.
  Let~$(A_n)_{n\in\matN}\in\Sigma_1\otimes\Sigma_2$.
  Let~$i$ be in~$\{1,2\}$.
  Let~$j\eqdef3-i$.
  Let~$x_i\in X_i$.
  Then, we have
  \begin{equation}
    \label{e:count-inter-of-sections-is-meas}
    s_i \left( x_i, \bigcap_{n \in \matN} A_n \right)
    = \bigcap_{n \in \matN} s_i (x_i, A_n)  \in \Sigma_j.
  \end{equation}
\end{lemma}

\begin{proof}
  Direct consequence of
  Lemma~\threfc{l:compat-of-section-with-set-ops}{%
    with count\-able intersection},
  Lemma~\threfc{l:equiv-def-of-sigma-alg}{%
    closedness under countable intersection}, and
  Lemma~\thref{l:measurability-of-section}.
\end{proof}

\begin{lemma}[indicator of section]
  \label{l:indic-of-section}
  \mbox{}\hfill
  Let~$(X_1,\Sigma_1)$ and~$(X_2,\Sigma_2)$ be measurable spaces.
  Let~$A\subset X_1\times X_2$.
  Let~$i\in\{1,2\}$.
  Let~$j\eqdef3-i$.
  Then, we have
  \begin{equation}
    \label{e:indic-of-section}
    \forall x_1 \in X_1,\quad
    \forall x_2 \in X_2,\quad
    \matUN_A (x_1, x_2) = \matUN_{s_i (x_i, A)} (x_j).
  \end{equation}
\end{lemma}

\begin{proof}
  Direct consequence of
  Definition~\thref{d:section-in-cartesian-prod}, and
  \assume{the definition of the indicator function}.
\end{proof}

\begin{lemma}[measurability of function from product space]
  \label{l:meas-of-fun-from-prod-space}
  \mbox{}\\
  Let~$(X_1,\Sigma_1)$, $(X_2,\Sigma_2)$ and~$(\Xp,\Sigmap)$ be measurable
  spaces.
  Let~$i\in\{1,2\}$.
  Let~$j\eqdef3-i$.
  Let~$f:\ArXoxXtXp$.
  Assume that~$f$ is measurable for $\Sigma_1\otimes\Sigma_2$ and~$\Sigmap$.\\
  Then, for all $x_i\in X_i$, the function $(x_j\mapsto f(x_1,x_2))$ is
  measurable for~$\Sigma_j$ and~$\Sigmap$.
\end{lemma}

\begin{proof}
  Direct consequence of
  \assume{properties of inverse image
    ($(f^j_{x_i})^{-1}(\Ap)=s_i(x_i,f^{-1}(\Ap))$ where
    $f^j_{x_i}\eqdef(x_j\mapsto f(x_1,x_2))$ and $\Ap\in\Sigmap$)},
  Lemma~\threfc{l:measurability-of-section}{%
    $(f^j_{x_i})^{-1}(\Ap)\in\Sigma_j$}, and
  Definition~\thref{d:meas-fun}.
\end{proof}

\begin{remark}
  Note that the reciprocal of the previous lemma is false.

  Indeed, let us build a counter-example.
  Let $X_1=X_2=\Xp\eqdef\matR$,
  $\Sigma_1=\Sigma_2\eqdef\Sigma_{\matR}(\{\{x\}\}_{x\in\matR})$, and
  $\Sigmap\eqdef\calBR$.
  Let~$D\eqdef\{(x,x)\st x\in\matR\}\subset X_1\times X_2$.
  Then, the tensor product $\sigma$-algebra
  $\Sigma\eqdef\Sigma_1\otimes\Sigma_2$ is also generated by all singletons
  of~$\matR^2$, and~$D$ is not measurable.
  Let~$f=\matUN_D$.
  Then, $f$~is not measurable for~$\Sigma$ and~$\Sigmap$, but for all
  $x\in X$, $(y\mapsto f(x,y))=(y\mapsto f(y,x))=\matUN_{\{x\}}$ is obviously
  measurable for~$\Sigma$ and~$\Sigmap$.
\end{remark}

\chapter{Measurability and numbers}
\label{c:measurability-and-numbers}

\minitoc

\begin{remark}
  We address in this chapter the specific case of the measurable spaces of
  numbers~$(\matRbar,\calBRbar)$ and~$(\matR,\calBR)$, and of measurable
  functions to them.
\end{remark}

\section{Borel subset of numbers}
\label{s:borel-subset-of-numbers}

\subsection{Borel subset of real numbers}
\label{ss:borel-subset-of-real-numbers}

\begin{lemma}[Borel $\sigma$-algebra of~$\matR$]
  \label{l:borel-sigma-alg-of-r}
  \mbox{}\\
  $\calBR$~is generated by any of the following sets of intervals:
  \begin{gather}
    \label{e:borel-sigma-alg-of-r-1}
    \{ (a, b) \}_{a < b},\quad
    \{ [a, b) \}_{a < b},\quad
    \{ [a, b] \}_{a < b},\quad
    \{ (a, b] \}_{a < b},\\
    \label{e:borel-sigma-alg-of-r-2}
    \{ (-\infty, b) \}_{b \in \matR},\quad
    \{ (-\infty, b] \}_{b \in \matR},\quad
    \{ (a, \infty) \}_{a \in \matR},\quad
    \{ [a, \infty) \}_{a \in \matR}.
  \end{gather}
\end{lemma}

\begin{proof}
  From
  Theorem~\thref{t:count-conn-comps-of-open-subsets-of-r}, and
  Lemma~\thref{l:count-borel-sigma-alg-gen},
  we have $\calBR=\Sigma_\matR(\{(a,b)\}_{a<b})$.

  Let~$a,b\in\matR$ such that $a<b$.
  Then, we have the following countable unions:
  \begin{gather*}
    [a, b) = \{ a \} \cup (a, b),\quad
    [a, b] = \{ a \} \cup (a, b) \cup \{ b \},\quad
    (a, b] = (a, b) \cup \{ b \},\\
    (a, b)
    = \bigcup_{n \in \matN} \left[ a + \oneovernplusone, b \right)
    = \bigcup_{n \in \matN} \left[
      a + \oneovernplusone, b - \oneovernplusone \right]
    = \bigcup_{n \in \matN} \left( a, b - \oneovernplusone \right].
  \end{gather*}
  Thus, from
  Definition~\thref{d:gen-sigma-alg},
  Definition~\threfc{d:sigma-alg}{%
    closedness under count\-able union}, and
  Lemma~\threfc{l:some-borel-subsets}{singletons are Borel subsets},
  we have
  \begin{gather*}
    \{ [a, b) \}_{a < b},\;
    \{ [a, b] \}_{a < b},\;
    \{ (a, b] \}_{a < b}
    \;\subset\; \calBR = \Sigma_\matR (\{ (a, b) \}_{a < b}),\\
    \{ (a, b) \}_{a < b}
    \;\subset\; \Sigma_\matR (\{ [a, b) \}_{a < b}),\;
    \Sigma_\matR (\{ [a, b] \}_{a < b}),\;
    \Sigma_\matR (\{ (a, b] \}_{a < b}).
  \end{gather*}
  Hence, from
  Lemma~\thref{l:other-sigma-alg-gen},
  we have
  \begin{equation*}
    \calBR
    = \Sigma_\matR (\{ (a, b) \}_{a < b})
    = \Sigma_\matR (\{ [a, b) \}_{a < b})
    = \Sigma_\matR (\{ [a, b] \}_{a < b})
    = \Sigma_\matR (\{ (a, b] \}_{a < b}).
  \end{equation*}

  Let~$a,b\in\matR$ such that $a<b$.
  Then, we have the following (finite) unions and intersections:
  \begin{gather*}
    (-\infty, b] = (-\infty, b) \cup \{ b \},\quad
    [a, \infty) = \{ a \} \cup (a, \infty),\\
    [a, b)
    = (-\infty, a)^c \cap (-\infty, b)
    = [a, \infty) \cap [b, \infty)^c,\\
    (a, b]
    = (-\infty, a]^c \cap (-\infty, b]
    = (a, \infty) \cap (b, \infty)^c.
  \end{gather*}
  Thus, since $(-\infty,b)$ and $(a,\infty)$ are open, and from
  Definition~\thref{d:gen-sigma-alg},
  Lem\-ma~\threfc{l:equiv-def-of-sigma-alg}{%
    closedness under complement, countable union and intersection}, and
  Lemma~\threfc{l:some-borel-subsets}{singletons are Borel subsets},
  we have
  \begin{gather*}
    \{ (-\infty, b) \}_{b \in \matR},\;
    \{ [a, \infty) \}_{a \in \matR}
    \;\subset\; \calBR
    = \Sigma_\matR (\{ [a, b) \}_{a < b}),\\
    \{ (-\infty, b] \}_{b \in \matR},\;
    \{ (a, \infty) \}_{a \in \matR}
    \;\subset\; \calBR
    = \Sigma_\matR (\{ (a, b] \}_{a < b}),\\
    \{ [a, b) \}_{a < b}
    \;\subset\; \Sigma_\matR (\{ (-\infty, b) \}_{b \in \matR}),\;
    \Sigma_\matR (\{ [a, \infty) \}_{a \in \matR}),\\
    \{ (a, b] \}_{a < b}
    \;\subset\; \Sigma_\matR (\{ (-\infty, b] \}_{b \in \matR}),\;
    \Sigma_\matR (\{ (a, \infty) \}_{a \in \matR}).
  \end{gather*}
  Hence, from
  Lemma~\thref{l:other-sigma-alg-gen},
  we have
  \begin{gather*}
    \calBR
    = \Sigma_\matR (\{ [a, b) \}_{a < b})
    = \Sigma_\matR (\{ (-\infty, b) \}_{b \in \matR})
    = \Sigma_\matR (\{ [a, \infty) \}_{a \in \matR}),\\
    \calBR
    = \Sigma_\matR (\{ (a, b] \}_{a < b})
    = \Sigma_\matR (\{ (-\infty, b] \}_{b \in \matR})
    = \Sigma_\matR (\{ (a, \infty) \}_{a \in \matR}).
  \end{gather*}

  Therefore, all eight sets of intervals generate the Borel $\sigma$-algebra
  of~$\matR$.
\end{proof}

\begin{lemma}[countable generator of Borel $\sigma$-algebra of~$\matR$]
  \label{l:count-gen-of-borel-sigma-alg-of-r}
  \mbox{}\\
  In
  Lemma~\thref{l:borel-sigma-alg-of-r},
  finite bounds of intervals, $a$ and~$b$, may be taken rational.
\end{lemma}

\begin{proof}
  Same proof as for
  Lemma~\thref{l:borel-sigma-alg-of-r},
  but with using the countability of the topological basis of~$\matR$
  equipped with the Euclidean distance through
  Theorem~\thref{t:r-is-second-countable} instead of
  Theorem~\thref{t:count-conn-comps-of-open-subsets-of-r}.
\end{proof}

\subsection{Borel subset of extended real numbers}
\label{ss:borel-subset-of-ext-real-numbers}

\begin{lemma}[Borel $\sigma$-algebra of~$\matRbar$]
  \label{l:borel-sigma-alg-of-rbar}
  \mbox{}\\
  $\calBRbar$~is generated by any of the following sets of intervals:
  \begin{equation}
    \label{e:borel-sigma-alg-of-rbar}
    \{ [-\infty, b) \}_{b \in \matR},\quad
    \{ [-\infty, b] \}_{b \in \matR},\quad
    \{ (a, \infty] \}_{a \in \matR},\quad
    \{ [a, \infty] \}_{a \in \matR}.
  \end{equation}
\end{lemma}

\begin{proof}
  From
  Theorem~\threfc{t:count-conn-comps-of-open-subsets-of-r}{%
    similar proof in $\matRbar$},
  and
  Lemma~\thref{l:count-borel-sigma-alg-gen},
  we have
  \begin{equation*}
    \calBRbar
    = \Sigma_{\matRbar} (\{ (a, b) \}_{a < b}
      \cup \{ [-\infty, b) \}_{b \in \matR}
      \cup \{ (a, \infty] \}_{a \in \matR}).
  \end{equation*}

  Let~$a,b\in\matR$ such that $a<b$.
  Then, we have the following countable unions and (finite) intersections:
  \begin{gather*}
    [-\infty, b)
    = \bigcup_{n \in \matN} \left(
      b - \oneovernplusone, \infty \right]^c, \quad
    (a, \infty]
    = \bigcup_{n \in \matN} \left[ -\infty, a + \oneovernplusone \right)^c,\\
    (a, b)
    = \bigcup_{n \in \matN} \left[ -\infty, a + \oneovernplusone \right)^c
    \cap [-\infty, b)
    = (a, \infty] \cap
    \bigcup_{n \in \matN} \left( b - \oneovernplusone, \infty \right]^c.
  \end{gather*}
  Thus, from
  Definition~\thref{d:gen-sigma-alg}, and
  Lemma~\threfc{l:equiv-def-of-sigma-alg}{%
    closedness under complement, countable union and intersection},
  we have
  \begin{gather*}
    \{ [-\infty, b) \}_{b \in \matR},\;
    \{ [a, \infty] \}_{a \in \matR}
    \;\subset\; \calBRbar
    = \Sigma_{\matRbar}
    \left( \{ (a, b) \}_{a < b}
      \cup \{ [-\infty, b) \}_{b \in \matR}
      \cup \{ (a, \infty] \}_{a \in \matR} \right),\\
    \{ (a, b) \}_{a < b},\;
    \{ [-\infty, b) \}_{b \in \matR},\;
    \{ (a, \infty] \}_{a \in \matR}
    \;\subset\; \Sigma_{\matRbar} (\{ [-\infty, b) \}_{b \in \matR}),\;
    \Sigma_{\matRbar} (\{ (a, \infty] \}_{a \in \matR}).
  \end{gather*}
  Hence, from
  Lemma~\thref{l:other-sigma-alg-gen},
  we have
  \begin{equation*}
    \calBRbar
    = \Sigma_{\matRbar} (\{ (a, b) \}_{a < b}
      \cup \{ [-\infty, b) \}_{b \in \matR}
      \cup \{ (a, \infty] \}_{a \in \matR})
    = \Sigma_{\matRbar} (\{ [-\infty, b) \}_{b \in \matR})
    = \Sigma_{\matRbar} (\{ (a, \infty] \}_{a \in \matR}).
  \end{equation*}

  Let~$a,b\in\matR$ such that $a<b$.
  Then, we have the following complements:
  \begin{gather*}
    [-\infty, b] = (b, \infty]^c,\quad [a, \infty] = [-\infty, a)^c,\\
    (a, \infty] = [-\infty, a]^c,\quad [-\infty, b) = [b, \infty]^c.
  \end{gather*}
  Thus, from
  Definition~\thref{d:gen-sigma-alg},
  Lemma~\threfc{l:equiv-def-of-sigma-alg}{%
    closedness under complement, countable union and intersection}, and
  Lemma~\thref{l:other-sigma-alg-gen},
  we have
  \begin{gather*}
    \{ [-\infty, b] \}_{b \in \matR}
    \;\subset\; \calBRbar
    = \Sigma_{\matRbar} (\{ (a, \infty] \}_{a \in \matR}),\\
    \{ [a, \infty] \}_{a \in \matR}
    \;\subset\; \calBRbar
    = \Sigma_{\matRbar} (\{ [-\infty, b) \}_{b \in \matR}),\\
    \{ (a, \infty] \}_{a \in \matR}
    \;\subset\; \Sigma_{\matRbar} (\{ [-\infty, b] \}_{b \in \matR}),\\
    \{ [-\infty, b) \}_{b \in \matR}
    \;\subset\; \Sigma_{\matRbar} (\{ [a, \infty] \}_{a \in \matR}).
  \end{gather*}
  Hence, from
  Lemma~\thref{l:other-sigma-alg-gen},
  we have
  \begin{gather*}
    \calBRbar
    = \Sigma_{\matRbar} (\{ (a, \infty] \}_{a \in \matR})
    = \Sigma_{\matRbar} (\{ [-\infty, b] \}_{b \in \matR}),\\
    \calBRbar
    = \Sigma_{\matRbar} (\{ [-\infty, b) \}_{b \in \matR})
    = \Sigma_{\matRbar} (\{ [a, \infty] \}_{a \in \matR}).
  \end{gather*}

  Therefore, all four sets of intervals generate the Borel $\sigma$-algebra
  of~$\matRbar$.
\end{proof}

\begin{lemma}[Borel subsets of~$\matRbar$ and~$\matR$]
  \label{l:borel-subsets-of-rbar-and-r}
  \mbox{}\hfill
  Let~$A\subset\matRbar$.
  Then, $A\in\calBRbar$ iff $A\cap\matR\in\calBR$.
\end{lemma}

\begin{proof}
  Direct consequence of
  Definition~\threfc{d:borel-sigma-alg}{%
      $\matR$ (open subset) belongs to~$\calBRbar$},
  Definition~\threfc{d:sigma-alg}{%
    $\{-\infty,\infty\}$ = $\matR^c\in\calBRbar$},
  Lemma~\threfc{l:characterization-of-borel-subsets}{%
    with $X\eqdef\matRbar$, $n\eqdef2$, $Y_1\eqdef\matR$, and
    $Y_2=\{-\infty,\infty\}$},
  Lemma~\threfc{l:borel-sub-sigma-alg}{%
    with $X\eqdef\matRbar$ and $Y=\{-\infty,\infty\}$, {\ie}
    $\calB(\{-\infty,\infty\})=\calP(\{-\infty,\infty\})$, and thus
    $A\cap\{-\infty,\infty\}$ always belongs to~$\calB(\{-\infty,\infty\})$}.
\end{proof}

\begin{remark}
  In other words, Borel subsets of~$\matRbar$ are Borel subsets of~$\matR$,
  or the union of Borel subsets of~$\matR$ with~$\{-\infty\}$, $\{\infty\}$,
  or $\{-\infty,\infty\}$.
\end{remark}

\subsection{Borel subset of nonnegative numbers}
\label{ss:borel-subset-of-nonnegative-numbers}

\begin{lemma}[Borel $\sigma$-algebra of~$\matRplus$]
  \label{l:borel-sigma-alg-of-rplus}
  \mbox{}\hfill
  We have $\calBRplus=\{A\in\calBR\st A\subset\matRplus\}$.
\end{lemma}

\begin{proof}
  In~$\matR$, we have $\matRplus=[0,\infty)=(-\infty,0)^c$.
  Thus, from
  Definition~\threfc{d:sigma-alg}{closedness under complement}, and
  Lemma~\thref{l:borel-sigma-alg-of-r},
  $\matRplus$ belongs to~$\calBR$.
  Therefore, from
  Lemma~\thref{l:borel-sub-sigma-alg},
  we have $\calBRplus=\{A\in\calBR\st A\subset\matRplus\}$.
\end{proof}

\begin{lemma}[Borel $\sigma$-algebra of~$\matRbarplus$]
  \label{l:borel-sigma-alg-of-rbarplus}
  \mbox{}\\
  $\calBRbarplus$ is generated by any of the
  following sets of intervals:
  \begin{equation}
    \label{e:borel-sigma-alg-of-rbarplus}
    \{ [0, b) \}_{b \in \matRplus},\quad
    \{ [0, b] \}_{b \in \matRplus},\quad
    \{ (a, \infty] \}_{a \in \matRplus},\quad
    \{ [a, \infty] \}_{a \in \matRplus}.
  \end{equation}
\end{lemma}

\begin{proof}
  Direct consequence of
  Lemma~\thref{l:borel-sigma-alg-of-rbar}, and
  Lemma~\thref{l:gen-meas-subspace}.
\end{proof}

\subsection{Product of Borel subsets of  numbers}
\label{ss:product-borel-subsets-of-numbers}

\begin{lemma}[Borel $\sigma$-algebra of~$\matR^n$]
  \label{l:borel-sigma-alg-of-rm}
  \mbox{}\\
  Let~$n\in[2..\infty)$.
  For all $i\in[1..n]$, let~$\calB_i\eqdef\calBR$.
  Then, we have
  \begin{equation}
    \label{e:borel-sigma-alg-of-rm}
    \calBRn = \calBR^{n \otimes} \eqdef \bigotimes_{i \in [1..n]} \calB_i.
  \end{equation}
\end{lemma}

\begin{proof}
  Let~$G\eqdef\{(a,b)\st a,b\in\matQ\Conj a<b\}$, $\Gp\eqdef G\cup\{\matR\}$,
  and~$\Gbarp\eqdef\olprod_{i\in[1..n]}\Gp$.
  Then, from
  Lemma~\thref{l:count-gen-of-borel-sigma-alg-of-r},
  Lemma~\threfc{l:equiv-def-of-sigma-alg}{contains full set}, and
  Lemma~\threfc{l:complete-gen-sigma-alg}{%
    with $\Gp\eqdef\{\matR\}\subset\calBR$},
  we have
  \begin{equation*}
    \calBR = \Sigma_\matR (G) = \Sigma_\matR (\Gp).
  \end{equation*}
  Hence, from
  Lemma~\thref{l:gen-prod-meas-space},
  we have $\calBR^{n\otimes}=\Sigma_{\matR^n}(\Gbarp)$.

  Let~$d$ be the Euclidean distance on~$\matR^n$.
  Then, from
  Lemma~\threfc{l:rn-is-second-countable}{%
    $\Gbar$ countable topological basis of~$(\matR^n,d)$}, and
  Lemma~\threfc{l:complete-count-topo-basis-of-prod-space}{%
    with $A_i\eqdef\matR$ open in~$\matR$},
  $\Gbarp$~is a countable topological basis of~$(\matR^n,d)$.
  Hence, from
  Lemma~\threfc{l:count-borel-sigma-alg-gen}{%
    with $X\eqdef\matR^n$ and $G\eqdef\Gbarp$},
  we have $\calBRn=\Sigma_{\matR^n}(\Gbarp)$.

  Therefore, we have the equality $\calBRn=\calBR^{m\otimes}$.
\end{proof}

\begin{remark}
  Note that the previous lemma also holds for the product of subsets
  of~$\matR$.
  In particular, this more general result could be used to prove that
  $\calBRplusn=\left(\calBRplus\right)^{n\otimes}$.

  Note also that identifying~$\matR^2$ with~$\matC$ via the canonical
  isometry $((x,y)\mapsto x+iy)$, allows to identify open subsets
  of~$\matR^2$ to those of~$\matC$.
  Hence, a function from some measurable space $(X,\Sigma)$ to
  $(\matC,\calBC)$ is measurable iff its real and imaginary parts ($\Re{f}$
  and $\Im{f}$) are measurable for~$\Sigma$ and~$\calBR$.
\end{remark}

\clearpage
\section{Measurable numeric function}
\label{s:measurable-numeric-function}

\subsection{Measurable numeric function to~\Rintitle}
\label{ss:measurable-numeric-function-to-r}

\begin{definition}[$\calMR$, {\vectorspace} of measurable numeric functions
  to~$\matR$]
  \label{d:mr-vector-space-of-meas-num-fun-to-r}
  \mbox{}\\
  Let~$(X,\Sigma)$ be a measurable space.
  The set of functions $\ArXR$ that are measurable for~$\Sigma$ and~$\calBR$
  is called the
  {\em {\vectorspace} of finite-valued measurable functions (over~$X$)};
  it is denoted $\calMR(X,\Sigma)$ (or simply~$\calMR$).
\end{definition}

\begin{remark}
  The set~$\calMR$ is shown below to be a {\vectorspace};
  hence its name.
\end{remark}

\begin{lemma}[measurability of indicator function]
  \label{l:meas-of-indic-fun}
  \mbox{}\\
  Let~$(X,\Sigma)$ be a measurable space.
  Let~$A\subset X$.
  Then, we have $\matUN_A\in\calMR$ iff $A\in\Sigma$.
\end{lemma}

\begin{proof}
  $\matUN_A$ takes the values~0 and~1.
  Let~$\Ap\in\calBR$.\\
  \proofpar{Case $0,1\in\Ap$}
  Then, $\matUN_A^{-1}(\Ap)=X$.
  \proofpar{Case $0\in\Ap$ and $1\not\in\Ap$}
  Then, $\matUN_A^{-1}(\Ap)=A^c$.\\
  \proofpar{Case $0\not\in\Ap$ and $1\in\Ap$}
  Then, $\matUN_A^{-1}(\Ap)=A$.
  \proofpar{Case $0,1\not\in\Ap$}
  Then, $\matUN_A^{-1}(\Ap)=\emptyset$.

  Hence, from
  Definition~\threfc{d:measurable-space}{$\Sigma$ is a $\sigma$-algebra}, and
  Lemma~\thref{l:equiv-def-of-sigma-alg},
  $\matUN_A^{-1}(\Ap)$ belongs to the $\sigma$-algebra~$\Sigma$ iff
  $A\in\Sigma$.
  Therefore, from
  Definition~\thref{d:sigma-alg}, and
  Definition~\thref{d:mr-vector-space-of-meas-num-fun-to-r},
  $\matUN_A\in\calMR$ iff $A\in\Sigma$.
\end{proof}

\begin{lemma}[measurability of numeric function to~$\matR$]
  \label{l:meas-of-num-fun-to-r}
  \mbox{}\hfill
  Let~$(X,\Sigma)$ be a measurable space.
  Let~$f:\ArXR$.
  Then, $f\in\calMR$ iff
  one of the following conditions is satisfied:
  \begin{align}
    \label{e:meas-of-num-fun-to-r-1}
    \forall a \in \matR,\quad &
    \{ f < a \} \eqdef f^{-1} (-\infty, a) \in \Sigma,\\
    \label{e:meas-of-num-fun-to-r-2}
    \forall a \in \matR,\quad &
    \{ f \leq a \} \eqdef f^{-1} (-\infty, a] \in \Sigma,\\
    \label{e:meas-of-num-fun-to-r-3}
    \forall a \in \matR,\quad &
    \{ f > a \} \eqdef f^{-1} (a, \infty) \in \Sigma,\\
    \label{e:meas-of-num-fun-to-r-4}
    \forall a \in \matR,\quad &
    \{ f \geq a \} \eqdef f^{-1} [a, \infty) \in \Sigma.
  \end{align}
\end{lemma}

\begin{proof}
  \proofpar{``Left'' implies ``right''}
  Direct consequence of
  Definition~\thref{d:mr-vector-space-of-meas-num-fun-to-r},
  Definition~\thref{d:meas-fun},
  Definition~\thref{d:borel-sigma-alg},
  Lemma~\thref{l:some-borel-subsets},
  \assume{open intervals of~$\matR$ are open subsets}, and
  \assume{closed intervals of~$\matR$, are closed subsets}.

  \proofparskip{``Right'' implies ``left''}
  Direct consequence of
  Lemma~\thref{l:borel-sigma-alg-of-r},
  Lemma~\thref{l:equiv-def-of-meas-fun}, and
  Definition~\thref{d:mr-vector-space-of-meas-num-fun-to-r}.

  \medskip\noindent
  Therefore, we have the equivalence.
\end{proof}

\begin{lemma}[inverse image is measurable in~$\matR$]
  \label{l:inverse-image-is-meas-in-r}
  \mbox{}\\
  Let~$(X,\Sigma)$ be a measurable space.
  Let~$f\in\calMR$.
  Then, we have
  \begin{equation}
    \label{e:inverse-image-is-meas-in-r}
    \forall a \in \matR,\quad
    \{ f = a \} \eqdef f^{-1} (a) \in \Sigma.
  \end{equation}
\end{lemma}

\begin{proof}
  Direct consequence of
  Definition~\thref{d:mr-vector-space-of-meas-num-fun-to-r},
  Definition~\thref{d:meas-fun},
  Lemma~\threfc{l:some-borel-subsets}{singletons are measurable}.
\end{proof}

\begin{lemma}[$\calMR$~is algebra]
  \label{l:mr-is-alg}
  \mbox{}\\
  Let~$(X,\Sigma)$ be a measurable space.
  Then, $\calMR$~is a subalgebra of $(\matR^X,+,\cdot,\times)$.
\end{lemma}

\begin{proof}
  Let~$f,g\in\calMR$.
  Then, from
  Definition~\thref{d:mr-vector-space-of-meas-num-fun-to-r},
  Lemma~\thref{l:meas-of-fun-to-prod-space}, and
  Lemma~\threfc{l:borel-sigma-alg-of-rm}{with $m=2$},
  the function $(f,g):\ArXRxR$ is measurable for~$\Sigma$
  and~$\calBRtwo$.
  Moreover, from
  \assume{continuity of addition and multiplication from~$\matR^2$
    to~$\matR$},
  Lemma~\thref{l:cont-is-meas}, and
  Lemma~\thref{l:compat-of-meas-with-comp},
  $f+g$ and $fg$ are measurable for~$\Sigma$ and~$\calBR$.

  Let~$f:\ArXR$.
  Let~$a\in\matR$.
  Then, from
  Lemma~\thref{l:const-fun-is-meas}, and
  Definition~\thref{d:mr-vector-space-of-meas-num-fun-to-r},
  we have $a\in\calMR$.
  Thus, from the previous result, we have $af\in\calMR$.

  Therefore, from
  Lemma~\threfc{l:k-is-k-alg}{with $\matK\eqdef\matR$},
  Lemma~\threfc{l:alg-of-funs-to-alg}{with $\matK\eqdef\matR$},
  and
  Lemma~\thref{l:closed-under-alg-ops-is-subalg},
  $\calMR$~is a subalgebra of $(\matR^X,+,\cdot,\times)$.
\end{proof}

\begin{remark}
  Note that we may also show that~$\calMC$, the space of measurable functions
  to~$(\matC,\calBC)$, is a subalgebra of $(\matC,+,\cdot,\times)$.
\end{remark}

\begin{lemma}[$\calMR$~is {\vectorspace}]
  \label{l:mr-is-vector-space}
  \mbox{}\\
  Let~$(X,\Sigma)$ be a measurable space.
  Then, $\calMR$~is a {\vectorspace} with the zero function as zero.
\end{lemma}

\begin{proof}
  Direct consequence of
  Lemma~\thref{l:mr-is-alg},
  Definition~\thref{d:subalg},
  Definition~\threfc{d:alg-over-a-field}{$\calMR$~is a {\vectorspace}}, and
  Lemma~\threfc{l:closed-under-alg-ops-is-subalg}{%
    $0_{\calMR}=0_{\matR^X}$ is the zero function}.
\end{proof}

\subsection{Measurable numeric function to~\Rbarintitle}
\label{ss:measurable-numeric-function-to-rbar}

\begin{definition}[$\calM$, set of measurable numeric functions]
  \label{d:m-set-of-meas-num-funs}
  \mbox{}\\
  Let~$(X,\Sigma)$ be a measurable space.
  The set of functions $\ArXRb$ that are measurable for~$\Sigma$
  and~$\calBRbar$ is called the {\em set of measurable functions (over~$X$)};
  it is denoted $\calM(X,\Sigma)$ (or simply~$\calM$).
\end{definition}

\begin{remark}
  \mbox{}\\
  We use the convention of keeping the name of functions $\ArXR$ when they
  are ``extended'' as functions $\ArXRb$.
  This allows us to consider~$\calMR$ as a subset of~$\calM$ in the next
  lemma.
\end{remark}

\begin{lemma}[$\calM$ and finite is~$\calMR$]
  \label{l:m-and-finite-is-mr}
  \mbox{}\hfill
  Let~$(X,\Sigma)$ be a measurable space.\\
  Let~$\FXR$ be the set of functions from~$X$ to~$\matR$.
  Then, we have $\calMR=\calM\cap\FXR$.
\end{lemma}

\begin{proof}
  \proofpar{$\calMR\subset\calM\cap\FXR$}
  Let~$f\in\calMR$.
  Let~$A\in\calBRbar$.
  Then, from
  Definition~\thref{d:mr-vector-space-of-meas-num-fun-to-r},
  Definition~\threfc{d:ext-real-nums-rbar}{%
    $\matRbar=\matR\uplus\{\pm\infty\}$},
  \assume{compatibility of inverse image with union},
  Lemma~\threfc{l:borel-subsets-of-rbar-and-r}{%
    $A\cap\matR\in\calBR$}, and
  Definition~\thref{d:meas-fun},
  we have $f\in\FXR$, and
  \begin{equation*}
    f^{-1} (A)
    = f^{-1} (A \cap \matR) \cup f^{-1} (A \cap \{ \pm\infty \})
    = f^{-1} (A \cap \matR)
    \in \Sigma.
  \end{equation*}
  Hence, from
  Definition~\thref{d:meas-fun}, and
  Definition~\thref{d:m-set-of-meas-num-funs},
  $f\in\calM\cap\FXR$.

  \proofparskip{$\calM\cap\FXR\subset\calMR$}
  Let~$f\in\calM\cap\FXR$.
  Let~$A\in\calBR$.
  Then, from
  Definition~\thref{d:m-set-of-meas-num-funs},
  Lemma~\threfc{l:borel-subsets-of-rbar-and-r}{$A\cap\matR=A\in\calBR$}, and
  Definition~\thref{d:meas-fun},
  we have $A\in\calBRbar$, and $f^{-1}(A)\in\Sigma$.
  Hence, from
  Definition~\thref{d:meas-fun}, and
  Definition~\thref{d:mr-vector-space-of-meas-num-fun-to-r},
  we have $f\in\calMR$.

  \medskip\noindent
  Therefore, we have the equality.
\end{proof}

\begin{lemma}[measurability of numeric function]
  \label{l:meas-of-num-fun}
  \mbox{}\hfill
  Let~$(X,\Sigma)$ be a measurable space.
  Let~$f:\ArXRb$.
  Then, $f\in\calM$ iff
  one of the following conditions is satisfied:
  \begin{align}
    \label{e:meas-of-num-fun-1}
    \forall a \in \matR,\quad &
    \{ f < a \} \eqdef f^{-1} [-\infty, a) \in \Sigma,\\
    \label{e:meas-of-num-fun-2}
    \forall a \in \matR,\quad &
    \{ f \leq a \} \eqdef f^{-1} [-\infty, a] \in \Sigma,\\
    \label{e:meas-of-num-fun-3}
    \forall a \in \matR,\quad &
    \{ f > a \} \eqdef f^{-1} (a, \infty] \in \Sigma,\\
    \label{e:meas-of-num-fun-4}
    \forall a \in \matR,\quad &
    \{ f \geq a \} \eqdef f^{-1} [a, \infty] \in \Sigma.
  \end{align}
\end{lemma}

\begin{proof}
  \proofpar{``Left'' implies ``right''}
  Direct consequence of
  Definition~\thref{d:m-set-of-meas-num-funs},
  Definition~\thref{d:meas-fun},
  Definition~\thref{d:borel-sigma-alg},
  Lem\-ma~\thref{l:some-borel-subsets},
  \assume{open intervals of~$\matRbar$ are open subsets}, and
  \assume{closed intervals of~$\matRbar$, are closed subsets}.

  \proofparskip{``Right'' implies ``left''}
  Direct consequence of
  Lemma~\thref{l:borel-sigma-alg-of-rbar},
  Lemma~\thref{l:equiv-def-of-meas-fun}, and
  Definition~\thref{d:m-set-of-meas-num-funs}.

  \medskip\noindent
  Therefore, we have the equivalence.
\end{proof}

\begin{lemma}[inverse image is measurable]
  \label{l:inverse-image-is-meas}
  \mbox{}\\
  Let~$(X,\Sigma)$ be a measurable space.
  Let~$f\in\calM$.
  Then, we have
  \begin{equation}
    \label{e:inverse-image-is-meas}
    \forall a \in \matRbar,\quad
    \{ f = a \} \eqdef f^{-1} (a) \in \Sigma.
  \end{equation}
\end{lemma}

\begin{proof}
  Direct consequence of
  Definition~\thref{d:m-set-of-meas-num-funs},
  Definition~\thref{d:meas-fun},
  Lemma~\threfc{l:some-borel-subsets}{singletons are measurable}.
\end{proof}

\begin{lemma}[$\calM$ is closed under finite part]
  \label{l:m-is-closed-under-finite-part}
  \mbox{}\hfill
  Let~$(X,\Sigma)$ be a measurable space.
  Let~$f\in\calM$.
  Let~$A\eqdef f^{-1}(\matR)$.
  Then, $A\in\Sigma$ and (the finite part of~$f$)
  $f\matUN_A\in\calMR(\subset\calM)$.
\end{lemma}

\begin{proof}
  From
  Definition~\thref{d:ext-real-nums-rbar},
  we have $\matR=\{\pm\infty\}^c$.
  Let
  \begin{equation*}
    B
    \eqdef f^{-1} (\{ \pm\infty \})
    = f^{-1} (-\infty) \uplus f^{-1} (\infty).
  \end{equation*}
  Then, from
  \assume{compatibility of inverse image with disjoint union},
  Lemma~\thref{l:inverse-image-is-meas},
  Definition~\threfc{d:measurable-space}{$\Sigma$ is a $\sigma$-algebra}, and
  Definition~\threfc{d:sigma-alg}{closedness under union},
  we have $B\in\Sigma$.
  Hence, from
  \assume{compatibility of inverse image with complement},
  Definition~\thref{d:pseudopart}, and
  Definition~\threfc{d:sigma-alg}{closedness under complement},
  we have $A=B^c\in\Sigma$, and $(A,B)$ form a pseudopartition of $X$.

  From
  Definition~\thref{d:finite-part},
  let~$\hf\eqdef f\matUN_A$ be the finite part of~$f$.
  Therefore, from
  Lemma~\thref{l:finite-part-is-finite},
  \assume{the definition of the indicator function},
  Lemma~\threfc{l:const-fun-is-meas}{$0\in\calM$},
  Lemma~\threfc{l:meas-of-fun-def-on-pseudopart}{%
    with $(\Xp,\Sigmap)\eqdef(\matRbar,\calBRbar)$,
    and the pseudopartition $X=A\uplus B$}, and
  Lemma~\thref{l:m-and-finite-is-mr},
  we have $\hf\in\calM\cap\FXR=\calMR$.
\end{proof}

\begin{lemma}[$\calM$ is closed under addition when defined]
  \label{l:m-is-closed-under-add-when-defined}
  \mbox{}\\
  Let~$(X,\Sigma)$ be a measurable space.
  Let~$f,g\in\calM$.
  Assume that~$f$ and~$g$ never take opposite infinite values at the same
  point.
  Then, $f+g$~is well-defined, and belongs to~$\calM$.
\end{lemma}

\begin{proof}
  By hypothesis, the subsets $\{f=\infty\}\cap\{g=-\infty\}$ and
  $\{f=-\infty\}\cap\{g=\infty\}$ are empty.
  Thus, from
  Definition~\thref{d:pseudopart},
  the whole space reduces to the pseudopartition
  \begin{equation*}
    X = X_{\pm\infty}^c \uplus X_\infty \uplus X_{-\infty}
  \end{equation*}
  (we recall that the notation~$\uplus$ is for disjoint union) where
  \begin{align*}
    X_{\pm\infty}^c & \eqdef f^{-1} (\matR) \cap g^{-1} (\matR),\\
    X_\infty & \eqdef \{ f = \infty \} \cup \{ g = \infty \},\\
    X_{-\infty} & \eqdef \{ f = -\infty \} \cup \{ g = -\infty \}.
  \end{align*}
  Then, from
  Definition~\thref{d:m-set-of-meas-num-funs},
  Definition~\threfc{d:meas-fun}{for~$f$ and~$g$},
  Definition~\threfc{d:borel-sigma-alg}{%
    $\matR$ open in~$\matRbar$ belongs to~$\calBRbar$},
  Lemma~\threfc{l:equiv-def-of-sigma-alg}{%
    closedness under countable union and intersection}, and
  Lemma~\threfc{l:inverse-image-is-meas}{%
    with $a\eqdef\pm\infty$},
  $X_{\pm\infty}^c$, $X_\infty$ and~$X_{-\infty}$ belong to~$\Sigma$.

  Let~$\tilde{f}$ and~$\tilde{g}$ be the finite parts of~$f$ and~$g$.
  Then, from
  Definition~\thref{d:finite-part}, and
  Definition~\thref{d:add-in-rbar},
  the sum $f+g$ can be defined on the pseudopartition by
  \begin{equation*}
    f + g \eqdef \left\{
      \begin{array}{ll}
        \tilde{f} + \tilde{g} & \mbox{on } X_{\pm\infty}^c,\\
        \infty & \mbox{on } X_{\infty},\\
        -\infty & \mbox{on } X_{-\infty}.
      \end{array}
    \right.
  \end{equation*}
  Therefore, from
  Lemma~\threfc{l:m-is-closed-under-finite-part}{%
    with $f$ and $g$},
  Lemma~\threfc{l:mr-is-alg}{sum},
  Lemma~\thref{l:const-fun-is-meas}, and
  Lemma~\thref{l:meas-of-fun-def-on-pseudopart},
  we have $f+g\in\calM$.
\end{proof}

\begin{lemma}[$\calM$ is closed under finite sum when defined]
  \label{l:m-is-closed-under-finite-sum-when-defined}
  \mbox{}\\
  Let~$(X,\Sigma)$ be a measurable space.
  Let~$I$ be a finite set.
  Let~$(f_i)_{i\in I}\in\calM$.
  Assume that the $f_i$'s never take opposite infinite values at the same
  point, {\ie} for all $i,j\in I$, $i\not=j$ implies
  $\{f_i=\infty\}\cap\{f_j=-\infty\}=\emptyset$.
  Then, $\sum_{i\in I}f_i$ is well-defined, and belongs to~$\calM$.
\end{lemma}

\begin{proof}
  Direct consequence of
  Lemma~\thref{l:m-is-closed-under-add-when-defined}, and
  induction on the cardinality of~$I$.
\end{proof}

\begin{lemma}[$\calM$ is closed under multiplication]
  \label{l:m-is-closed-under-mult}
  \mbox{}\\
  Let~$(X,\Sigma)$ be a measurable space.
  Let~$f,g\in\calM$.
  Then, $fg$~is well-defined, and belongs to~$\calM$.
\end{lemma}

\begin{proof}
  From
  Lemma~\thref{l:zero-prod-prop-in-rbarplus-mt}, and
  Lemma~\thref{l:infinity-prod-prop-in-rbarplus},
  the whole space can be pseudopartitioned into
  $X=X_{\pm\infty}^c\uplus X_0\uplus X_{\infty}\uplus X_{-\infty}$ (the
  notation~$\uplus$ is for disjoint union) where
  \begin{align*}
    X_{\pm\infty}^c \eqdef & f^{-1} (\matR) \cap g^{-1} (\matR),\\
    X_0 \eqdef &
      \left( \{ | f | = \infty \} \cap \{ g = 0 \} \right)
      \cup \left( \{ f = 0\} \cap \{ | g | = \infty\} \right),\\
    X_{\infty} \eqdef &
      \left( \{ f = \infty \} \cap \{ g > 0 \} \right)
      \cup \left( \{ f = -\infty \} \cap \{ g < 0 \} \right)\\
    &
      \cup \left( \{ f > 0\} \cap \{ g = \infty\} \right)
      \cup \left( \{ f < 0\} \cap \{ g = -\infty\} \right),\\
    X_{-\infty} \eqdef &
      \left( \{ f = -\infty \} \cap \{ g > 0 \} \right)
      \cup \left( \{ f = \infty \} \cap \{ g < 0 \} \right)\\
    &
      \cup \left( \{ f > 0\} \cap \{ g = -\infty\} \right)
      \cup \left( \{ f < 0\} \cap \{ g = \infty\} \right).
  \end{align*}
  Then, from
  Definition~\thref{d:m-set-of-meas-num-funs},
  Definition~\threfc{d:meas-fun}{for~$f$ and~$g$},
  Definition~\threfc{d:borel-sigma-alg}{%
    $\matR$ open in~$\matRbar$ belongs to~$\calBRbar$},
  Lemma~\threfc{l:inverse-image-is-meas}{%
    with $a\eqdef\pm\infty,0$}
  Lemma~\threfc{l:meas-of-num-fun}{%
    with $a\eqdef0$}, and
  Lemma~\thref{l:equiv-def-of-sigma-alg}{%
    closedness under countable union and intersection},
  we have $X_{\pm\infty}^c,X_0,X_\infty,X_{-\infty}\in\Sigma$.

  Let~$\tilde{f}$ and~$\tilde{g}$ be the finite parts of~$f$ and~$g$.
  Then, from
  Definition~\thref{d:finite-part},
  Definition~\thref{d:mult-in-rbar}, and
  Definition~\thref{d:mult-in-rbar-mt},
  the product $fg$ can be defined on the pseudopartition by
  \begin{equation*}
    f g \eqdef \left\{
      \begin{array}{ll}
        \tilde{f} \tilde{g} & \mbox{on } X_{\pm\infty}^c,\\
        0 & \mbox{on } X_0,\\
        \infty & \mbox{on } X_{\infty},\\
        -\infty & \mbox{on } X_{-\infty}.
      \end{array}
    \right.
  \end{equation*}
  Therefore, from
  Lemma~\threfc{l:m-is-closed-under-finite-part}{%
    with $f$ and $g$},
  Lemma~\threfc{l:mr-is-alg}{product},
  Lemma~\thref{l:const-fun-is-meas}, and
  Lemma~\thref{l:meas-of-fun-def-on-pseudopart},
  we have $fg\in\calM$.
\end{proof}

\begin{lemma}[$\calM$ is closed under finite product]
  \label{l:m-is-closed-under-finite-prod}
  \mbox{}\\
  Let~$(X,\Sigma)$ be a measurable space.
  Let~$I$ be a finite set.
  Let~$(f_i)_{i\in I}\in\calM$.
  Then, $\prod_{i\in I}f_i\in\calM$.
\end{lemma}

\begin{proof}
  Direct consequence of
  Lemma~\thref{l:m-is-closed-under-mult}, and
  induction on the cardinality of~$I$.
\end{proof}

\begin{lemma}[$\calM$ is closed under scalar multiplication]
  \label{l:m-is-closed-under-scalar-mult}
  \mbox{}\\
  Let~$(X,\Sigma)$ be a measurable space.
  Let~$a\in\matRbar$.
  Let~$f\in\calM$.
  Then, we have $af\in\calM$.
\end{lemma}

\begin{proof}
  Direct consequence of
  Lemma~\thref{l:m-is-closed-under-mult}, and
  Lemma~\thref{l:const-fun-is-meas}.
\end{proof}

\begin{lemma}[$\calM$ is closed under infimum]
  \label{l:m-is-closed-under-inf}
  \mbox{}\\
  Let~$(X,\Sigma)$ be a measurable space.
  Let~$I\subset\matN$.
  Let~$(f_i)_{i\in I}\in\calM$.
  Then, we have $\inf_{i\in I}f_i\in\calM$.
\end{lemma}

\begin{proof}
  Let~$a\in\matR$.
  Then, from
  Definition~\thref{LM-d:infimum},
  Lemma~\threfc{l:meas-of-num-fun}{with~$f_i$}, and
  Lemma~\threfc{l:equiv-def-of-sigma-alg}{%
    closedness under countable intersection},
  we have
  \begin{equation*}
    \left\{ \inf_{i \in I} f_i \geq a \right\}
    = \bigcap_{i \in I} \{ f_i \geq a \} \in \Sigma.
  \end{equation*}

  Therefore, from
  Lemma~\threfc{l:meas-of-num-fun}{%
    with $\inf_{i\in I}f_i$},
  $\inf_{i\in I}f_i$ belongs to~$\calM$.
\end{proof}

\begin{lemma}[$\calM$ is closed under supremum]
  \label{l:m-is-closed-under-sup}
  \mbox{}\\
  Let~$(X,\Sigma)$ be a measurable space.
  Let~$I\subset\matN$.
  Let~$(f_i)_{i\in I}\in\calM$.
  Then, we have $\sup_{i\in I}f_i\in\calM$.
\end{lemma}

\begin{proof}
  Let~$a\in\matR$.
  Then, from
  Definition~\thref{LM-d:supremum},
  Lemma~\threfc{l:meas-of-num-fun}{%
    with all the~$f_i$'s}, and
  Lemma~\threfc{l:equiv-def-of-sigma-alg}{%
    closedness under countable intersection},
  \begin{equation*}
    \left\{ \sup_{i \in I} f_i \leq a \right\}
    = \bigcap_{i \in I} \{ f_i \leq a \} \in \Sigma.
  \end{equation*}

  Therefore, from
  Lemma~\threfc{l:meas-of-num-fun}{%
    with $\sup_{i\in I}f_i$},
  $\sup_{i\in I}f_i$ belongs to~$\calM$.
\end{proof}

\begin{lemma}[$\calM$ is closed under limit inferior]
  \label{l:m-is-closed-under-liminf}
  \mbox{}\\
  Let~$(X,\Sigma)$ be a measurable space.
  Let~$(f_n)_{n\in\matN}\in\calM$.
  Then, we have $\liminf_{n\to\infty}f_n\in\calM$.
\end{lemma}

\begin{proof}
  Direct consequence of
  Lemma~\threfc{l:m-is-closed-under-inf}{%
    $F_n^-\eqdef\inf_{p\in\matN}f_{n+p}\in\calM$},
  Lemma~\threfc{l:m-is-closed-under-sup}{%
    $\ulf\eqdef\sup_{n\in\matN}F_n^-\in\calM$}, and
  Lemma~\threfc{l:liminf}{$\liminf_{n\to\infty}f_n=\ulf$}.
\end{proof}

\begin{lemma}[$\calM$ is closed under limit superior]
  \label{l:m-is-closed-under-limsup}
  \mbox{}\\
  Let~$(X,\Sigma)$ be a measurable space.
  Let~$(f_n)_{n\in\matN}\in\calM$.
  Then, we have $\limsup_{n\to\infty}f_n\in\calM$.
\end{lemma}

\begin{proof}
  Direct consequence of
  Lemma~\threfc{l:m-is-closed-under-sup}{%
    $F_n^+\eqdef\sup_{p\in\matN}f_{n+p}$ belongs to~$\calM$},
  Lemma~\threfc{l:m-is-closed-under-inf}{%
    $\fbar\eqdef\inf_{n\in\matN}F_n^+\in\calM$}, and
  Lemma~\threfc{l:limsup}{$\limsup_{n\to\infty}f_n=\fbar$}.
\end{proof}

\begin{lemma}[$\calM$ is closed under limit when pointwise convergent]
  \label{l:m-is-closed-under-limit-when-pointwise-conv}
  \mbox{}\\
  Let~$(X,\Sigma)$ be a measurable space.
  Let~$(f_n)_{n\in\matN}\in\calM$.
  Assume that~$(f_n(x))_{n\in\matN}$ is pointwise convergent in~$\matRbar$.
  Then, we have $\lim_{n\to\infty}f_n\in\calM$.
\end{lemma}

\begin{proof}
  Direct consequence of
  Lemma~\thref{l:liminf-limsup-and-pointwise-conv},
  Lemma~\thref{l:m-is-closed-under-liminf}, or
  Lemma~\thref{l:m-is-closed-under-limsup}.
\end{proof}

\begin{lemma}[measurability and masking]
  \label{l:meas-and-masking}
  \mbox{}\\
  Let~$(X,\Sigma)$ be a measurable space.
  Let~$f:\ArXRb$.
  Then, $f\in\calM$ iff
  for all $A\in\Sigma$, $f\matUN_A\in\calM$.
\end{lemma}

\begin{proof}
  \proofpar{``Left'' implies ``right''}
  Direct consequence of
  Lemma~\thref{l:meas-of-indic-fun},
  Lemma~\threfc{l:m-and-finite-is-mr}{$\calMR\subset\calM$}, and
  Lemma~\thref{l:m-is-closed-under-mult}.

  \proofparskip{``Right'' implies ``left''}
  Direct consequence of
  Definition~\threfc{d:measurable-space}{$\Sigma$ is a $\sigma$-algebra},
  Definition~\threfc{d:sigma-alg}{$X\in\Sigma$},
  \assume{the definition of the indicator function ($\matUN_X\equiv1$)}, and
  Definition~\threfc{d:mult-in-rbar}{1 is unity}.

  \medskip\noindent
  Therefore, we have the equivalence.
\end{proof}

\begin{lemma}[measurability of restriction]
  \label{l:meas-of-restr}
  \mbox{}\\
  Let~$(X,\Sigma)$ be a measurable space.
  Let~$A\in\Sigma$.
  Let~$Y\subset X$ such that $A\subset Y$.
  Let~$f:\ArYRb$.
  Let~$\hf:\ArXRb$.
  Assume that $\restr{\hf}{Y}=f$.
  Then, we have $\restr{f}{A}\in\calM(A,\Sigma\olcap A)$ iff
  $\hf\matUN_A\in\calM(X,\Sigma)$.
\end{lemma}

\begin{proof}
  Let~$\Bp\in\calBRbar$.
  \proofpar{Case $0\not\in\Bp$}
  Then, from
  \assume{the definition of the indicator function},
  we have $(\hf\matUN_A)^{-1}(\Bp)=f^{-1}(\Bp)\cap A$.
  \proofpar{Case $0\in\Bp$}
  Then, from
  \assume{compatibility of inverse image with disjoint union},
  \assume{the definition of the indicator function}, and
  \assume{associativity of disjoint union},
  we have
  \begin{align*}
    (\hf \matUN_A)^{-1} (\Bp)
    & = (\hf \matUN_A)^{-1} \left( \Bp \setminus \{ 0 \} \right)
    \;\uplus\; (\hf \matUN_A)^{-1} (0)\\
    & = \left( f^{-1} \left( \Bp \setminus \{ 0 \} \right) \cap A \right)
    \;\uplus\; \left(
      \left( f^{-1} (0) \cap A \right)
      \;\uplus\; A^c \right)\\
    & = \left( f^{-1} (\Bp) \cap A \right) \;\uplus\; A^c.
  \end{align*}
  Moreover, from
  \assume{the definition of restriction of function},
  we have $\restr{f}{A}^{-1}(\Bp)=f^{-1}(\Bp)\cap A$.
  Hence, we have $(\hf\matUN_A)^{-1}(\Bp)=\restr{f}{A}^{-1}(\Bp)\uplus C$
  with $C\in\{\emptyset,A^c\}$.

  Let~$B\subset A$.
  Assume first that $B\in\Sigma$.
  Then, from
  Definition~\threfc{d:measurable-space}{$\Sigma$ is a $\sigma$-algebra}, and
  Definition~\threfc{d:sigma-alg}{%
    closedness under complement and countable union},
  we have $B\uplus A^c\in\Sigma$.
  Conversely, assume now that $B\uplus A^c\in\Sigma$.
  Then, from
  Lemma~\threfc{l:equiv-def-of-sigma-alg}{%
    closedness under countable intersection},
  we have $(B\uplus A^c)\cap A=B\in\Sigma$.
  Hence, we have the equivalence $B\in\Sigma$ iff $B\uplus A^c\in\Sigma$.

  Therefore, from
  Definition~\thref{d:meas-fun}, and
  Definition~\thref{d:m-set-of-meas-num-funs},
  we have the equivalence~$\restr{f}{A}\in\calM(A,\Sigma\olcap A)$ iff
  $\hf\matUN_A\in\calM(X,\Sigma)$.
\end{proof}

\subsection{Nonnegative measurable numeric function}
\label{ss:nonnegative-measurable-numeric-function}

\begin{definition}[$\calMplus$, subset of nonnegative measurable numeric
  functions]
  \label{d:mplus-subset-of-nonneg-meas-num-fun}
  \mbox{}\\
  Let~$(X,\Sigma)$ be a measurable space.
  The {\em subset of nonnegative measurable functions (over~$X$)} is denoted
  $\calMplus(X,\Sigma)$ (or simply $\calMplus$);
  it is defined by
  $\calMplus(X,\Sigma)\eqdef\{f\in\calM\st f(X)\subset\matRbarplus\}$.
\end{definition}

\begin{lemma}[measurability of nonnegative and nonpositive parts]
  \label{l:meas-of-nonneg-and-nonpos-parts}
  \mbox{}\\
  Let~$(X,\Sigma)$ be a measurable space.
  Let~$f:\ArXRb$.
  Then, we have $f\in\calM$ iff
  $f^+,f^-\in\calMplus$.
\end{lemma}

\begin{proof}
  \proofpar{``Left'' implies ``right''}
  Direct consequence of
  Definition~\thref{d:nonneg-and-nonpos-parts},
  Lemma~\threfc{l:m-is-closed-under-scalar-mult}{$-f\in\calM$},
  Lemma~\threfc{l:const-fun-is-meas}{$0\in\calM$},
  Lemma~\threfc{l:m-is-closed-under-sup}{%
    maximum is supremum},
  Lemma~\thref{l:nonneg-and-nonpos-parts-are-nonneg}, and
  Definition~\thref{d:mplus-subset-of-nonneg-meas-num-fun}.

  \proofparskip{``Right'' implies ``left''}
  Direct consequence of
  Definition~\threfc{d:mplus-subset-of-nonneg-meas-num-fun}{%
    $\calMplus\subset\calM$},
  Lemma~\threfc{l:m-is-closed-under-scalar-mult}{$-f^-\!\in\!\calM$},
  Lemma~\threfc{l:decomp-into-nonneg-and-nonpos-parts}{%
    $f^+-f^-$ is well-defined}, and
  Lemma~\thref{l:m-is-closed-under-add-when-defined}.

  \medskip\noindent
  Therefore, we have the equivalence.
\end{proof}

\begin{lemma}[$\calMplus$ is closed under finite part]
  \label{l:mplus-is-closed-under-finite-part}
  \mbox{}\hfill
  Let~$(X,\Sigma)$ be a measurable space.
  Let~$f\in\calMplus$.
  Let~$A\eqdef f^{-1}(\matR)$.
  Then, $A\in\Sigma$ and (the finite part of~$f$)
  $f\matUN_A\in\calMR\cap\calMplus$.
\end{lemma}

\begin{proof}
  Direct consequence of
  Definition~\thref{d:mplus-subset-of-nonneg-meas-num-fun},
  Lemma~\thref{l:m-is-closed-under-finite-part},
  \assume{nonnegativeness of the indicator function}, and
  \assume{closedness of multiplication in~$\matRplus$}.
\end{proof}

\begin{lemma}[$\calM$ is closed under absolute value]
  \label{l:m-is-closed-under-abs}
  \mbox{}\\
  Let~$(X,\Sigma)$ be a measurable space.
  Let~$f\in\calM$.
  Then, we have $|f|\in\calMplus\subset\calM$.

   Moreover, if $f\in\calMR$, then we have $|f|\in\calMR\cap\calMplus$.
\end{lemma}

\begin{proof}
  Direct consequence of
  Lemma~\thref{l:abs-in-rbar-is-cont},
  Lemma~\thref{l:cont-is-meas},
  Lemma~\thref{l:compat-of-meas-with-comp},
  Definition~\threfc{d:m-set-of-meas-num-funs}{%
    $|f|\in\calM$},
  Lemma~\thref{l:abs-in-rbar-is-nonneg},
  Definition~\threfc{d:mplus-subset-of-nonneg-meas-num-fun}{%
    $|f|\in\calMplus$},
  \assume{closedness of absolute value in~$\matR$}, and
  Definition~\threfc{d:mr-vector-space-of-meas-num-fun-to-r}{$|f|\in\calMR$}.
\end{proof}

\begin{lemma}[$\calMplus$ is closed under addition]
  \label{l:mplus-is-closed-under-add}
  \mbox{}\hfill
  Let~$(X,\Sigma)$ be a measurable space.\\
  Let~$f,g\in\calMplus$.
  Then, $f+g$ is well-defined, and belongs to~$\calMplus$.
\end{lemma}

\begin{proof}
  Direct consequence of
  Definition~\thref{d:mplus-subset-of-nonneg-meas-num-fun},
  Lemma~\threfc{l:m-is-closed-under-add-when-defined}{%
    $f$~and~$g$ cannot take opposite infinite values}, and
  Lemma~\thref{l:add-in-rbarplus-is-closed}.
\end{proof}

\begin{lemma}[$\calMplus$ is closed under multiplication]
  \label{l:mplus-is-closed-under-mult}
  \mbox{}\\
  Let~$(X,\Sigma)$ be a measurable space.
  Let~$f,g\in\calMplus$.
  Then, we have $fg\in\calMplus$.
\end{lemma}

\begin{proof}
  Direct consequence of
  Definition~\thref{d:mplus-subset-of-nonneg-meas-num-fun},
  Lemma~\thref{l:m-is-closed-under-mult}, and
  Lemma~\thref{l:mult-in-rbarplus-is-closed-mt}.
\end{proof}

\begin{lemma}[$\calMplus$ is closed under nonnegative scalar multiplication]
  \label{l:mplus-is-closed-under-nonneg-scalar-mult}
  \mbox{}\\
  Let~$(X,\Sigma)$ be a measurable space.
  Let~$a\in\matRbarplus$.
  Let~$f\in\calMplus$.
  Then, we have $af\in\calMplus$.
\end{lemma}

\begin{proof}
  Direct consequence of
  Definition~\thref{d:mplus-subset-of-nonneg-meas-num-fun},
  Lemma~\thref{l:m-is-closed-under-scalar-mult}, and
  Lemma~\thref{l:mult-in-rbarplus-is-closed-mt}.
\end{proof}

\begin{lemma}[$\calMplus$ is closed under infimum]
  \label{l:mplus-is-closed-under-inf}
  \mbox{}\\
  Let~$(X,\Sigma)$ be a measurable space.
  Let~$I\subset\matN$.
  Let~$(f_i)_{i\in I}\in\calMplus$.
  Then, we have $\inf_{i\in I}f_i\in\calMplus$.
\end{lemma}

\begin{proof}
  Direct consequence of
  Lemma~\thref{l:m-is-closed-under-inf}, and
  Lemma~\threfc{l:inf-of-bounded-seq-is-bounded}{with $a\eqdef 0$}.
\end{proof}

\begin{lemma}[$\calMplus$ is closed under supremum]
  \label{l:mplus-is-closed-under-sup}
  \mbox{}\\
  Let~$(X,\Sigma)$ be a measurable space.
  Let~$I\subset\matN$.
  Let~$(f_i)_{i\in I}\in\calMplus$.
  Then, we have $\sup_{i\in I}f_i\in\calMplus$.
\end{lemma}

\begin{proof}
  Direct consequence of
  Lemma~\thref{l:m-is-closed-under-sup}, and
  Lemma~\threfc{l:sup-of-bounded-seq-is-bounded}{with $a\eqdef 0$}.
\end{proof}

\begin{lemma}[$\calMplus$ is closed under limit when pointwise convergent]
  \label{l:mplus-is-closed-under-limit-when-pointwise-conv}
  \mbox{}\\
  Let~$(X,\Sigma)$ be a measurable space.
  Let~$(f_n)_{n\in\matN}\in\calMplus$.
  Assume that~$(f_n)_{n\in\matN}$ is pointwise convergent in~$\matRbarplus$.
  Then, we have $\lim_{n\to\infty}f_n\in\calMplus$.
\end{lemma}

\begin{proof}
  Direct consequence of
  Definition~\thref{d:mplus-subset-of-nonneg-meas-num-fun},
  Lemma~\thref{l:m-is-closed-under-limit-when-pointwise-conv}, and
  \assume{completeness of~$\matRbarplus$}.
\end{proof}

\begin{lemma}[$\calMplus$ is closed under countable sum]
  \label{l:mplus-is-closed-under-count-sum}
  \mbox{}\hfill
  Let~$(X,\Sigma)$ be a measurable space.\\
  Let~$I\subset\matN$.
  Let~$(f_i)_{i\in I}\in\calMplus$.
  Then, $\sum_{i\in I}f_i$ is well-defined, and belongs to~$\calMplus$.
\end{lemma}

\begin{proof}
  Direct consequence of
  Lemma~\thref{l:series-are-conv-in-rbarplus},
  Lemma~\thref{l:mplus-is-closed-under-add},
  induction on~$n\in\matN$ (with
    $g_n\eqdef\sum_{i\in I\cap[0..n]}f_i\in\calMplus$,
    with the convention that a sum indexed by~$\emptyset$ is~0), and
  Lemma~\threfc{l:mplus-is-closed-under-limit-when-pointwise-conv}{%
    with $f_n\eqdef g_n$}.
\end{proof}

\subsection{Tensor product of measurable numeric functions}
\label{ss:tensor-product-of-measurable-numeric-functions}

\begin{definition}[tensor product of numeric functions]
  \label{d:tensor-prod-of-num-funs}
  \mbox{}\\
  Let~$m\in[2..\infty)$.
  For all $i\in[1..m]$, let~$X_i$ be a set, and~$f_i:\ArXiRb$.
  Let~$X\eqdef\prod_{i\in[1..m]}X_i$.
  The {\em tensor product of~$(f_i)_{i\in[1..m]}$} is the function
  $\bigotimes_{i\in[1..m]}f_i:\ArXRb$ defined by
  \begin{equation}
    \label{e:tensor-prod-of-num-funs}
    \forall x \eqdef (x_i)_{i \in [1..m]} \in X,\quad
    \left( \bigotimes_{i \in [1..m]} f_i \right) (x)
    \eqdef \prod_{i \in [1..m]} f_i (x_i).
  \end{equation}
\end{definition}

\begin{lemma}[measurability of tensor product of numeric functions]
  \label{l:meas-of-tensor-prod-of-num-funs}
  \mbox{}\\
  Let~$m\in[2..\infty)$.
  For all $i\in[1..m]$, let~$(X_i,\Sigma_i)$ be a measurable space,
  and~$f_i\in\calM(X_i,\Sigma_i)$.\\
  Then, we have
  $\bigotimes_{i\in[1..m]}f_i\in\calM\left(
    \prod_{i\in[1..m]}X_i,\bigotimes_{i\in[1..m]}\Sigma_i\right)$.
\end{lemma}

\begin{proof}
  Let~$X\eqdef\prod_{i\in[1..m]}X_i$
  and~$\Sigma\eqdef\bigotimes_{i\in[1..m]}\Sigma_i$.
  Let~$i\in[1..m]$.
  Let~$\pi_i$ be the canonical projection from~$X$ onto~$X_i$.
  Then, from
  Definition~\thref{d:m-set-of-meas-num-funs},
  Lemma~\thref{l:can-proj-is-meas}, and
  Lemma~\thref{l:compat-of-meas-with-comp},
  $f_i\circ\pi_i$ belongs to~$\calM(\!X,\!\Sigma)$.
  Therefore, from
  Definition~\thref{d:tensor-prod-of-num-funs}, and
  Lemma~\threfc{l:m-is-closed-under-finite-prod}{with $I\eqdef[1..m]$},
  we have
  \begin{equation*}
    \bigotimes_{i \in [1..m]} f_i
    = \prod_{i \in [1..m]} (f_i \circ \pi_i) \in \calM (X, \Sigma).
  \end{equation*}
\end{proof}

\chapter{Measure space}
\label{c:measure-space}

\minitoc

\section{Measure}
\label{s:measure}

\begin{remark}
  We recall that~$\uplus$ denotes disjoint union.
\end{remark}

\begin{definition}[additivity over measurable space]
  \label{d:add-over-meas-space}
  \mbox{}\hfill
  Let~$(X,\Sigma)$ be a measurable space.\\
  A function $\mu:\ArSigRb$ is said {\em additive} iff
  for all $n\in\matN$, for all $(A_i)_{i\in[0..n]}\in\Sigma$,
  \begin{equation}
    \label{e:add-over-meas-space}
    (\forall p, q \in [0..n],\;
    p \not= q \Implies A_p \cap A_q = \emptyset)
    \IMPLIES
    \mu \left( \biguplus_{i \in [0..n]} A_i \right)
    = \sum_{i \in [0..n]} \mu (A_i).
  \end{equation}
\end{definition}

\begin{definition}[$\sigma$-additivity over measurable space]
  \label{d:sigma-add-over-meas-space}
  \mbox{}\hfill
  Let~$(X,\Sigma)$ be a measurable space.
  A function $\mu:\ArSigRb$ is said {\em $\sigma$-additive} iff
  for all $I\subset\matN$, for all $(A_i)_{i\in I}\in\Sigma$,
  \begin{equation}
    \label{e:sigma-add-over-meas-space}
    (\forall p, q \in I,\;
    p \not= q \Implies A_p \cap A_q = \emptyset)
    \IMPLIES
    \mu \left( \biguplus_{i \in I} A_i \right)
    = \sum_{i \in I} \mu (A_i).
  \end{equation}
\end{definition}

\begin{remark}
  Note that from
  Definitions~\ref{d:measurable-space} and~\ref{d:sigma-alg} (closedness under
  countable union), both previous definitions are well-defined.
\end{remark}

\begin{lemma}[$\sigma$-additivity implies additivity]
  \label{l:sigma-add-implies-add}
  \mbox{}\hfill
  Let~$(X,\Sigma)$ be a measurable space.
  Let~$\mu:\ArSigRb$.
  Assume that~$\mu$ is $\sigma$-additive.
  Then, $\mu$~is additive.
\end{lemma}

\begin{proof}
  Direct consequence of
  Definition~\threfc{d:sigma-add-over-meas-space}{%
    with $I\eqdef[0..n]$}, and
  Definition~\thref{d:add-over-meas-space}.
\end{proof}

\begin{definition}[measure]
  \label{d:meas}
  \mbox{}\\
  Let~$(X,\Sigma)$ be a measurable space.
  A function $\mu:\ArSigRb$ is called {\em measure on~$(X,\Sigma)$} iff
  it is nonnegative,
  $\mu(\emptyset)=0$, and
  it is $\sigma$-additive.
  If so, $(X,\Sigma,\mu)$ is called {\em measure space}.
\end{definition}

\begin{remark}
  The previous definition is actually that of {\em nonnegative} measure, but as
  we do not consider ``signed'' measures in this document, we omit the
  qualifier ``nonnegative''.
\end{remark}

\begin{lemma}[measure over countable pseudopartition]
  \label{l:meas-over-count-pseudopart}
  \mbox{}\\
  Let~$(X,\Sigma,\mu)$ be a measure space.
  Let~$I\subset\matN$.
  Let~$A,(B_i)_{i\in I}\in\Sigma$.
  Assume that $X=\biguplus_{i\in I}B_i$.
  Then, for all $i\in I$, $A\cap B_i=B_i\cap A\in\Sigma$, and we have
  \begin{equation}
    \label{e:meas-over-count-partition}
    \mu (A) = \sum_{i \in I} \mu (A \cap B_i) = \sum_{i \in I} \mu (B_i \cap A).
  \end{equation}
\end{lemma}

\begin{proof}
  Direct consequence of
  Lemma~\thref{l:compat-of-pseudopart-with-inter},
  Lemma~\threfc{l:equiv-def-of-sigma-alg}{%
    closedness under countable intersection with $\card(I)$ equals 2},
  Definition~\threfc{d:meas}{$\sigma$-additive},
  Definition~\thref{d:sigma-add-over-meas-space}, and
  \assume{commutativity of intersection}.
\end{proof}

\begin{lemma}[measure is monotone]
  \label{l:meas-is-monot}
  \mbox{}\\
  Let~$(X,\Sigma,\mu)$ be a measure space.
  Then, $\mu$ is nondecreasing over~$\Sigma$:
  \begin{equation}
    \label{e:meas-is-monot}
    \forall A, B \in \Sigma,\quad A \subset B \IMPLIES \mu (A) \leq \mu (B).
  \end{equation}

  Moreover, if $\mu(A)$ is finite, then we have
  $\mu(B\setminus A)=\mu(B)-\mu(A)$.
\end{lemma}

\begin{proof}
  Let~$A,B\in\Sigma$.
  Assume that $A\subset B$.
  Then, from
  \assume{the definition of set difference}, and
  Lemma~\thref{l:sigma-alg-is-closed-under-set-diff},
  we have $A\cap(B\setminus A)=\emptyset$, $B=A\uplus(B\setminus A)$ and
  $B\setminus A\in\Sigma$.
  Therefore, from
  Definition~\threfc{d:meas}{nonnegativeness}, and
  Definition~\thref{d:sigma-add-over-meas-space}
  we have $\mu(B)=\mu(A)+\mu(B\setminus A)\geq\mu(A)$.

  Assume now that $\mu(A)<\infty$.
  Then, from
  Definition~\threfc{d:add-in-rbar}{rule~5 applies},
  we have $\mu(B\setminus A)=\mu(B)-\mu(A)$.
\end{proof}

\begin{lemma}[measure satisfies the finite Boole inequality]
  \label{l:meas-satisfies-finite-boole-ineq}
  \mbox{}\\
  Let~$(X,\Sigma,\mu)$ be a measure space.
  Then, $\mu$ satisfies the finite Boole inequality:
  \begin{equation}
    \label{e:meas-satisfies-finite-boole-ineq}
    \forall n \in \matN,\;
    \forall (A_i)_{i \in [0..n]} \in \Sigma,\quad
    \mu \left( \bigcup_{i \in [0..n]} A_i \right)
    \leq \sum_{i \in [0..n]} \mu (A_i).
  \end{equation}
\end{lemma}

\begin{proof}
  For~$n\in\matN$, let~$P(n)$ be the property
  $\forall(A_i)_{i\in[0..n]}\in\Sigma$,
  $\mu\left(\bigcup_{i\in[0..n]}A_i\right)\leq\sum_{i\in[0..n]}\mu(A_i)$.

  \proofparskip{Induction: $P(0)$}
  Trivial.

  \proofparskip{Induction: $P(n)$ implies $P(n+1)$}
  Let~$n\in\matN$.
  Assume that $P(n)$ holds.\\
  For all $i\in[0..(n+1)]$, let $A_i\in\Sigma$.
  From
  Definition~\thref{d:meas},
  Definition~\threfc{d:measurable-space}{$\Sigma$ is a $\sigma$-algebra}, and
  Definition~\threfc{d:sigma-alg}{closedness under countable union},
  let $B\eqdef\bigcup_{i\in[0..n]}A_i\in\Sigma$.
  Then, from
  \assume{the definition of set difference},
  \assume{associativity of union}, and
  Lemma~\thref{l:sigma-alg-is-closed-under-set-diff},
  we have
  \begin{equation*}
    B \cap (A_{n + 1} \setminus B) = \emptyset,\quad
    \bigcup_{i \in [0..(n + 1)]} A_i = B \uplus (A_{n + 1} \setminus B),
    \AND
    A_{n + 1} \setminus B \in \Sigma.
  \end{equation*}
  Thus, from
  Definition~\threfc{d:meas}{$\sigma$-additive},
  Definition~\threfc{d:sigma-add-over-meas-space}{%
    with $\card(I)=2$},
  $P(n)$,
  Lemma~\threfc{l:meas-is-monot}{%
    with $A_{n+1}\setminus B\subset A_{n+1}$}, and
  \assume{monotonicity of addition},
  we have
  \begin{equation*}
    \mu \left( \bigcup_{i \in [0..(n + 1)]} A_i \right)
    = \mu (B \uplus (A_{n + 1} \setminus B))
    = \mu (B) + \mu (A_{n + 1} \setminus B)
    \leq \sum_{i \in [0..(n + 1)]} \mu (A_i).
  \end{equation*}
  Hence, $P(n+1)$ holds.

  Therefore, by induction, we have $P(n)$ for all $n\in\matN$.
\end{proof}

\begin{definition}[continuity from below]
  \label{d:continuity-from-below}
  \mbox{}\\
  Let~$(X,\Sigma)$ be a measurable space.
  A function $\mu:\ArSigRb$ is said {\em continuous from below} iff
  \begin{equation}
    \label{e:continuity-from-below}
    \forall (A_n)_{n \in \matN} \in \Sigma,\quad
    (\forall n \in \matN,\;
    A_n \subset A_{n+1})
    \IMPLIES
    \mu \left( \bigcup_{n \in \matN} A_n \right)
    = \lim_{n \to \infty} \mu (A_n).
  \end{equation}
\end{definition}

\begin{lemma}[measure is continuous from below]
  \label{l:meas-is-cont-from-below}
  \mbox{}\\
  Let~$(X,\Sigma,\mu)$ be a measure space.
  Then, $\mu$ is continuous from below.
  Moreover, we have
  \begin{equation}
    \label{e:meas-is-cont-from-below}
    \forall (A_n)_{n \in \matN} \in \Sigma,\quad
    (\forall n \in \matN,\;
    A_n \subset A_{n+1})
    \IMPLIES
    \mu \left( \bigcup_{n \in \matN} A_n \right)
    = \sup_{n \in \matN} \mu (A_n).
  \end{equation}
\end{lemma}

\begin{proof}
  Let~$(A_n)_{n\in\matN}\in\Sigma$.
  Assume that the sequence $(A_n)_{n\in\matN}$ is nondecreasing.

  Let~$B_0\eqdef A_0\in\Sigma$.
  for all $n\in\matN$,
  let $B_{n+1}\eqdef A_{n+1}\setminus\bigcup_{i\in[0..n]}B_i$.
  Then, from
  Lemma~\thref{l:part-of-count-union-in-sigma-alg}, and
  \assume{partial union law for nondecreasing sequence},
  the sequence~$(B_n)_{n\in\matN}$ is pairwise disjoint, and we have
  \begin{align*}
    \forall n \in \matN,\quad &
    B_n \in \Sigma
    \CONJ
    \biguplus_{i \in [0..n]} B_i = \bigcup_{i \in [0..n]} A_i = A_n
    \in \Sigma,\\
    & \bigcup_{n \in \matN} A_n = \biguplus_{n \in \matN} B_n \in \Sigma.
  \end{align*}
  Hence, from
  Definition~\threfc{d:meas}{$\sigma$-additivity},
  Definition~\thref{d:sigma-add-over-meas-space}, and
  \assume{the definition of the sum of a sequence of positive numbers},
  we have
  \begin{align*}
    \mu \left( \bigcup_{n \in \matN} A_n \right)
    = \mu \left( \biguplus_{n \in \matN} B_n \right)
    & = \sum_{n \in \matN} \mu (B_n)\\
    & = \lim_{n \to \infty} \sum_{i \in [0..n]} \mu (B_i)
      = \lim_{n \to \infty}
      \mu \left( \biguplus_{i \in [0..n]} B_n \right)
      = \lim_{n \to \infty} \mu (A_n).
  \end{align*}
  Therefore, from
  Definition~\thref{d:continuity-from-below},
  $\mu$~is continuous from below.

  Moreover, Equation~\eqref{e:meas-is-cont-from-below} is a
  direct consequence of
  Lemma~\thref{l:meas-is-monot}, and
  \assume{properties of nondecreasing sequences in~$\matRbarplus$}.
\end{proof}

\begin{definition}[continuity from above]
  \label{d:continuity-from-above}
  \mbox{}\\
  Let~$(X,\Sigma)$ be a measurable space.
  A function $\mu:\ArSigRb$ is said {\em continuous from above} iff
  \begin{align}
    \label{e:continuity-from-above}
    \forall (A_n)_{n \in \matN} \in \Sigma,\quad
    & (\forall n \in \matN,\;
      A_n \supset A_{n+1} \CONJ
      \exists n_0 \in \matN,\;
      \mu (A_{n_0}) < \infty)\\
    & \Longrightarrow\quad
      \mu \left( \bigcap_{n \in \matN} A_n \right)
      = \lim_{n \to \infty} \mu (A_n)
      < \infty.
  \end{align}
\end{definition}

\begin{lemma}[measure is continuous from above]
  \label{l:meas-is-cont-from-above}
  \mbox{}\\
  Let~$(X,\Sigma,\mu)$ be a measure space.
  Then, $\mu$ is continuous from above.
  Moreover, we have
  \begin{align}
    \label{e:meas-is-cont-from-above}
    \forall (A_n)_{n \in \matN} \in \Sigma,\quad
    & (\forall n \in \matN,\;
      A_n \supset A_{n+1} \CONJ
      \exists n_0 \in \matN,\;
      \mu (A_{n_0}) < \infty)\\
    & \Longrightarrow\quad
      \mu \left( \bigcap_{n \in \matN} A_n \right)
      = \inf_{n \in \matN} \mu (A_n)
      < \infty.
  \end{align}
\end{lemma}

\begin{proof}
  Let~$(A_n)_{n\in\matN}\in\Sigma$.
  Assume that the sequence $(A_n)_{n\in\matN}$ is nonincreasing.
  Then, from
  \assume{properties of the intersection and of the limit},
  we have
  \begin{equation*}
    \bigcap_{n \in \matN} A_n
    = \bigcap_{n \geq n_0} A_n
    \AND
    \lim_{n \to \infty} \mu (A_n)
    = \lim_{n \to \infty} \mu (A_{n_0 + n}).
  \end{equation*}

  For all $n\in\matN$, let $B_n\eqdef A_{n_0}\setminus A_{n_0+n}$
  (thus $B_0=\emptyset$).
  Then, from
  \assume{the definition and properties of set difference},
  Lemma~\thref{l:sigma-alg-is-closed-under-set-diff}, and
  Lemma~\threfc{l:equiv-def-of-sigma-alg}{%
    closedness under countable intersection},
  $(B_n)_{n\in\matN}$ is a nondecreasing sequence of measurable subsets such
  that $\bigcup_{n\in\matN}B_n
  =A_{n_0}\setminus\bigcap_{n\geq n_0}A_{n}\in\Sigma$.
  Thus, from
  Lemma~\thref{l:meas-is-cont-from-below},
  Definition~\thref{d:continuity-from-below}, and
  Lemma~\threfc{l:meas-is-monot}{with $\mu(A_{n_0})<\infty$},
  we have
  \begin{equation*}
    \mu (A_{n_0}) - \mu \left( \bigcap_{n \geq n_0} A_n \right)
    = \mu \left( \bigcup_{n \in \matN} B_n \right)
    = \lim_{n \to \infty} \mu (B_n)
    = \lim_{n \to \infty}
    \left( \mu (A_{n_0}) - \mu (A_{n_0 + n}) \right).
  \end{equation*}
  Hence, from
  \assume{linearity of the limit},
  \assume{additive group properties of~$\matR$}, and
  Lemma~\thref{l:meas-is-monot},
  we have
  \begin{equation*}
    \mu \left( \bigcap_{n \in \matN} A_n \right)
    = \mu \left( \bigcap_{n \geq n_0} A_n \right)
    = \lim_{n \to \infty} \mu (A_{n_0 + n})
    = \lim_{n \to \infty} \mu (A_n)
    \leq \mu (A_{n_0})
    < \infty.
  \end{equation*}
  Therefore, from
  Definition~\thref{d:continuity-from-above},
  $\mu$~is continuous from above.

  Moreover, Equation~\eqref{e:meas-is-cont-from-above} is a
  direct consequence of
  Lemma~\thref{l:meas-is-monot}, and
  \assume{properties of nonincreasing sequences in~$\matRbarplus$}.
\end{proof}

\begin{lemma}[measure satisfies the Boole inequality]
  \label{l:meas-satisfies-boole-ineq}
  \mbox{}\\
  Let~$(X,\Sigma,\mu)$ be a measure space.
  Then, $\mu$ satisfies the Boole inequality:
  \begin{equation}
    \label{e:meas-satisfies-boole-ineq}
    \forall (A_n)_{n \in \matN} \in \Sigma,\quad
    \mu \left( \bigcup_{n \in \matN} A_n \right)
    \leq \sum_{n \in \matN} \mu (A_n).
  \end{equation}
\end{lemma}

\begin{proof}
  Let~$(A_n)_{n\in\matN}\in\Sigma$.
  For all $n\in\matN$, from
  Definition~\thref{d:meas},
  Definition~\threfc{d:measurable-space}{$\Sigma$ is a $\sigma$-algebra}, and
  Definition~\threfc{d:sigma-alg}{closedness under countable union},
  let $B_n\eqdef\bigcup_{i\in[0..n]}A_i\in\Sigma$.
  Then, from
  \assume{properties of union}, and
  Definition~\threfc{d:sigma-alg}{closedness under countable union},
  the sequence $(B_n)_{n\in\matN}$ is nondecreasing and
  \begin{equation*}
    \bigcup_{n \in \matN} B_n = \bigcup_{n \in \matN} A_n \in \Sigma.
  \end{equation*}
  Thus, from
  Lemma~\thref{l:meas-is-cont-from-below},
  we have
  \begin{equation*}
    \mu \left( \bigcup_{n \in \matN} A_n \right)
    = \mu \left( \bigcup_{n \in \matN} B_n \right)
    = \sup_{n \in \matN} \mu (B_n).
  \end{equation*}
  Let~$n\in\matN$.
  Then, from
  Lemma~\thref{l:meas-satisfies-finite-boole-ineq}, and
  Definition~\threfc{d:meas}{nonnegativeness},
  we have
  \begin{equation*}
    \mu (B_n)
    = \mu \left( \bigcup_{i \in [0..n]} A_i \right)
    \leq \sum_{i \in [0..n]} \mu (A_i)
    \leq \sum_{n \in \matN} \mu (A_n).
  \end{equation*}
  Therefore, from
  Definition~\threfc{LM-d:supremum}{least upper bound},
  we have
  \begin{equation*}
    \mu \left( \bigcup_{n \in \matN} A_n \right)
    \leq \sum_{n \in \matN} \mu (A_n).
  \end{equation*}
\end{proof}

\begin{lemma}[equivalent definition of measure]
  \label{l:equiv-def-of-meas}
  \mbox{}\\
  Let~$(X,\Sigma)$ be a measurable space.
  Let~$\mu:\ArSigRb$.
  Assume that~$\mu$ is nonnegative, and that $\mu(\emptyset)=0$.
  Then, $\mu$~is a measure on~$(X,\Sigma)$ iff
  it is additive and continuous from below.
\end{lemma}

\begin{proof}
  \proofpar{``Left'' implies ``right''}
  Direct consequence of
  Definition~\thref{d:meas},
  Lemma~\thref{l:sigma-add-implies-add}, and
  Lemma~\thref{l:meas-is-cont-from-below}.

  \proofparskip{``Right'' implies ``left''}
  Assume that~$\mu$ is additive and continuous from below.\\
  Let~$I\subset\matN$.
  Let~$(A_i)_{i\in I}\in\Sigma$.
  Assume that for all $p,q\in I$, $p\not=q\Implies A_p\cap A_q=\emptyset$.
  For all $n\in\matN$, let $B_n\eqdef\biguplus_{i\in I, i\in[0..n]}A_i$.
  Then, from
  Definition~\thref{d:meas},
  Definition~\threfc{d:measurable-space}{$\Sigma$ is a $\sigma$-algebra}, and
  Definition~\threfc{d:sigma-alg}{closedness under countable union},
  $(B_n)_{n\in\matN}$ is a nondecreasing sequence of measurable subsets such
  that $\bigcup_{n\in\matN}B_n=\biguplus_{i\in I}A_i\in\Sigma$.
  Thus, from
  Definition~\thref{d:continuity-from-below}, and
  Definition~\thref{d:add-over-meas-space},
  \begin{align*}
    \mu \left( \biguplus_{i \in I} A_i \right)
    = \mu \left( \bigcup_{n \in \matN} B_n \right)
    & = \lim_{n \to \infty} \mu (B_n)\\
    & = \lim_{n \to \infty}
      \mu \left( \biguplus_{i \in I \cap [0..n]} A_i \right)
      = \lim_{n \to \infty} \sum_{i \in I \cap [0..n]} \mu (A_i)
      = \sum_{i \in I} \mu (A_i).
  \end{align*}
  Hence, from
  Definition~\thref{d:sigma-add-over-meas-space}, and
  Definition~\thref{d:meas},
  $\mu$~is a measure on~$(X,\Sigma)$.

  \medskip\noindent
  Therefore, we have the equivalence.
\end{proof}

\begin{definition}[finite measure]
  \label{d:finite-meas}
  \mbox{}\hfill
  Let~$(X,\Sigma,\mu)$ be a measure space.\\
  The measure~$\mu$ is said {\em finite} iff
  $\mu(X)<\infty$.
  If so, the measure space $(X,\Sigma,\mu)$ is also said {\em finite}.
\end{definition}

\begin{lemma}[finite measure is bounded]
  \label{l:finite-meas-is-bounded}
  \mbox{}\\
  Let~$(X,\Sigma,\mu)$ be a finite measure space.
  Then, $\mu$~is bounded.
\end{lemma}

\begin{proof}
  Direct consequence of
  Lemma~\threfc{l:equiv-def-of-sigma-alg}{contains full set}, and
  Lemma~\threfc{l:meas-is-monot}{$\mu(A)\leq\mu(X)<\infty$}.
\end{proof}

\begin{definition}[$\sigma$-finite measure]
  \label{d:sigma-finite-meas}
  \mbox{}\\
  Let~$(X,\Sigma,\mu)$ be a measure space.
  The measure~$\mu$ is said {\em $\sigma$-finite} iff
  \begin{equation}
    \label{e:sigma-finite-meas}
    \exists (A_n)_{n \in \matN} \in \Sigma,\quad
    (\forall n \in \matN,\quad
    \mu (A_n) < \infty)
    \CONJ
    X = \bigcup_{n \in \matN} A_n.
  \end{equation}

  If so, the measure space $(X,\Sigma,\mu)$ is also said
  {\em $\sigma$-finite}.
\end{definition}

\begin{lemma}[equivalent definition of $\sigma$-finite measure]
  \label{l:equiv-def-of-sigma-finite-meas}
  \mbox{}\\
   Let~$(X,\Sigma,\mu)$ be a measure space.
  Then, the measure~$\mu$ is $\sigma$-finite iff
  \begin{equation}
    \label{e:equiv-def-of-sigma-finite-meas}
    \exists (B_n)_{n \in \matN} \in \Sigma,\quad
    (\forall n \in \matN,\quad
    B_n \subset B_{n + 1}
    \CONJ
    \mu (B_n) < \infty)
    \CONJ
    X = \bigcup_{n \in \matN} B_n.
  \end{equation}
\end{lemma}

\begin{proof}
  \proofparskip{``Left'' implies ``right''}
  From
  Definition~\thref{d:sigma-finite-meas},
  there exists $(A_n)_{n\in\matN}\in\Sigma$ such that, for all $n\in\matN$,
  $\mu(A_n)<\infty$, and $X=\bigcup_{n\in\matN}A_n$.
  For all $n\in\matN$, from
  Definition~\thref{d:meas},
  Definition~\threfc{d:measurable-space}{$\Sigma$ is a $\sigma$-algebra}, and
  Definition~\threfc{d:sigma-alg}{closedness under countable union},
  let $B_n\eqdef\bigcup_{p\in[0..n]}A_p\in\Sigma$.

  Let~$n\in\matN$.
  Then, from
  \assume{associativity of union},
  we have $B_{n+1}=B_n\cup A_{n+1}$.
  Thus, we have $B_n\subset B_{n+1}$, and
  $\bigcup_{p\in[0..n]}B_p=B_n=\bigcup_{p\in[0..n]}A_n$.
  Hence, from
  Lemma~\thref{l:meas-satisfies-finite-boole-ineq},
  \assume{closedness of addition in~$\matRplus$}, and
  \assume{the definition of countable union},
  we have, for all $n\in\matN$,
  $\mu(B_n)\leq\sum_{p\in[0..n]}\mu(A_p)<\infty$, and
  $X=\bigcup_{n\in\matN}B_n$.

  \proofparskip{``Right'' implies ``left''}
  Trivial.

  \medskip\noindent
  Therefore, we have the equivalence.
\end{proof}

\begin{definition}[diffuse measure]
  \label{d:diffuse-meas}
  \mbox{}\\
  Let~$(X,\Sigma,\mu)$ be a measure space.
  Assume that~$\Sigma$ contains all singletons of~$X$.
  The measure~$\mu$ is said {\em diffuse} iff
  for all $x\in X$, $\mu(\{x\})=0$.
  If so, the measure space~$(X,\Sigma,\mu)$ is also said {\em diffuse}.
\end{definition}

\begin{lemma}[finite measure is $\sigma$-finite]
  \label{l:finite-meas-is-sigma-finite}
  \mbox{}\\
  Let~$(X,\Sigma,\mu)$ be a finite measure space.
  Then, $\mu$~is $\sigma$-finite.
\end{lemma}

\begin{proof}
  Direct consequence of
  Definition~\threfc{d:sigma-finite-meas}{with $A_n=X$}, and
  Definition~\thref{d:finite-meas}.
\end{proof}

\begin{lemma}[trace measure]
  \label{l:trace-meas}
  \mbox{}\\
  Let~$(X,\Sigma,\mu)$ be a measure space.
  Let~$Y\in\Sigma$.
  Then, $\restr{\mu}{\Sigma\olcap Y}$ is a measure on~$(Y,\Sigma\olcap Y)$.

  The measure~$\mu_Y\eqdef\restr{\mu}{\Sigma\olcap Y}$ is called
  {\em trace measure on~$Y$}.
  The measure space $(Y,\Sigma\olcap Y,\mu_Y)$ is called
  {\em trace measure space on~$Y$}.
\end{lemma}

\begin{proof}
  Direct consequence of
  Lemma~\thref{l:trace-sigma-alg},
  Lemma~\threfc{l:meas-of-meas-subspace}{%
    $\Sigma\olcap Y\subset\Sigma$}, and
  Definition~\thref{d:meas}.
\end{proof}

\begin{lemma}[restricted measure]
  \label{l:restr-meas}
  \mbox{}\\
  Let~$(X,\Sigma,\mu)$ be a measure space.
  Let~$Y\in\Sigma$.
  Then, the function~$\mup_Y$ defined on~$\Sigma$ by
  \begin{equation}
    \label{e:restr-meas}
    \forall A \in \Sigma,\quad
    \mup_Y (A) \eqdef \mu (A \cap Y)
  \end{equation}
  is a measure on~$(X,\Sigma)$.
\end{lemma}

\begin{proof}
  Direct consequence of
  Lemma~\threfc{l:equiv-def-of-sigma-alg}{%
    closedness under countable intersection (with $\card(I)=2$)},
  \assume{properties of intersection}, and
  Definition~\thref{d:meas}.
\end{proof}

\begin{remark}
  Note that measures~$\mu_Y$ and~$\mup_Y$ from the two previous lemmas are
  distinct since they are not defined on the same $\sigma$-algebra.
  But they coincide on the trace $\sigma$-algebra $\Sigma\olcap Y$.
\end{remark}

\clearpage
\section{Negligible subset}
\label{s:negligible-subset}

\begin{definition}[negligible subset]
  \label{d:negl-subset}
  \mbox{}\hfill
  Let~$(X,\Sigma,\mu)$ be a measure space.\\
  A subset~$A$ of~$X$ is said {\em ($\mu$-)negligible} iff
  there exists $B\in\Sigma$ such that $A\subset B$ and $\mu(B)=0$.

  The set of $\mu$-negligible subsets is denoted $\neglset(X,\Sigma,\mu)$
  (or simply~$\neglset$).
\end{definition}

\begin{definition}[complete measure]
  \label{d:complete-meas}
  \mbox{}\hfill
  Let~$(X,\Sigma,\mu)$ be a measure space.
  The measure~$\mu$ is said {\em complete} iff
  $\neglset(X,\Sigma,\mu)\subset\Sigma$.
  If so, the measure space~$(X,\Sigma,\mu)$ is also said {\em complete}.
\end{definition}

\begin{definition}[considerable subset]
  \label{d:considerable-subset}
  \mbox{}\\
  Let~$(X,\Sigma,\mu)$ be a measure space.
  A subset~$A$ of~$X$ is said {\em ($\mu$-)considerable} iff
  $A\not\in\neglset(X,\Sigma,\mu)$.
\end{definition}

\begin{lemma}[equivalent definition of considerable subset]
  \label{l:equiv-def-of-considerable-subset}
  \mbox{}\hfill
  Let~$(X,\Sigma,\mu)$ be a measure space.
  Let~$A\subset X$.
  Then, $A$ is $\mu$-considerable iff
  for all $B\in\Sigma$, $A\subset B$ implies $\mu(B)>0$.
\end{lemma}

\begin{proof}
  Direct consequence of
  Definition~\thref{d:considerable-subset},
  Definition~\thref{d:negl-subset},
  Definition~\threfc{d:meas}{nonnegativeness}, and
  \assume{the tautology $\neg(P\Conj Q)\Equiv(P\Implies\neg Q)$}.
\end{proof}

\begin{remark}
  In the previous lemma, ``considerable'' naturally means ``non-negligible''.
\end{remark}

\begin{lemma}[negligibility of measurable subset]
  \label{l:negl-of-meas-subset}
  \mbox{}\\
  Let~$(X,\Sigma,\mu)$ be a measure space.
  Let~$A\in\Sigma$.
  Then, we have $A\in\neglset$ iff
  $\mu(A)=0$.
\end{lemma}

\begin{proof}
  Direct consequence of
  Definition~\threfc{d:negl-subset}{with $B=A$}.
\end{proof}

\begin{lemma}[empty set is negligible]
  \label{l:empty-set-is-negl}
  \mbox{}\hfill
  Let~$(X,\Sigma,\mu)$ be a measure space.
  Then, $\emptyset\in\neglset$.
\end{lemma}

\begin{proof}
  Direct consequence of
  Lemma~\thref{l:negl-of-meas-subset},
  Definition~\thref{d:meas},
  Definition~\threfc{d:measurable-space}{$\Sigma$ is a $\sigma$-algebra},
  Definition~\threfc{d:sigma-alg}{$\emptyset\in\Sigma$}, and
  Definition~\threfc{d:meas}{$\mu(\emptyset)=0$}.
\end{proof}

\begin{lemma}[compatibility of null measure with countable union]
  \label{l:compat-of-null-meas-with-count-union}
  \mbox{}\\
  Let~$(X,\Sigma,\mu)$ be a measure space.
  Let~$I\subset\matN$. 
  For all $i\in I$, let~$A_i\in\Sigma$ such that $\mu(A_i)=0$.
  Then, we have $\mu\left(\bigcup_{i\in I}A_i\right)=0$.
\end{lemma}

\begin{proof}
  Direct consequence of
  Definition~\thref{d:meas},
  Definition~\threfc{d:measurable-space}{$\Sigma$ is a $\sigma$-algebra},
  Definition~\threfc{d:sigma-alg}{%
    closedness under countable union},
  Definition~\threfc{d:meas}{nonnegativeness},
  Lemma~\thref{l:meas-satisfies-finite-boole-ineq},
  Lemma~\thref{l:meas-satisfies-boole-ineq},
  \assume{additive group properties of~$\matR$}, and
  Lem\-ma~\threfc{LM-l:stationary-sequence-is-convergent}{%
    countable sum of zero terms is zero}.
\end{proof}

\begin{lemma}[$\neglset$~is closed under countable union]
  \label{l:n-is-closed-under-count-union}
  \mbox{}\\
  Let~$(X,\Sigma,\mu)$ be a measure space.
  Let~$I\subset\matN$. 
  Let~$(A_i)_{i\in I}\in\neglset$.
  Then, we have $\bigcup_{i\in I}A_i\in\neglset$.
\end{lemma}

\begin{proof}
  Direct consequence of
  Definition~\thref{d:negl-subset},
  \assume{monotonicity of union}, and
  Lemma~\thref{l:compat-of-null-meas-with-count-union}.
\end{proof}

\begin{lemma}[subset of negligible is negligible]
  \label{l:subset-of-negl-is-negl}
  \mbox{}\\
  Let~$(X,\Sigma,\mu)$ be a measure space.
  Let~$A\in\neglset$.
  Then, we have $\calP(A)\subset\neglset$.
\end{lemma}

\begin{proof}
  Direct consequence of
  Definition~\thref{d:negl-subset},
  \assume{the definition of the power set}, and
  \assume{transitivity of the inclusion}.
\end{proof}

\begin{definition}[property almost satisfied]
  \label{d:prop-almost-satisfied}
  \mbox{}\\
  Let~$(X,\Sigma,\mu)$ be a measure space.
  A predicate~$P$ defined on~$X$ is said
  {\em satisfied ($\mu$-)almost everywhere} iff
  $\{\neg P\}\eqdef\{x\in X\st\neg P(x)\}\in\neglset$;
  this is denoted either
  ``$P\,\muae{\mu}$'',
  ``$\forall x\in X$, $P(x)\,\muae{\mu}$'',
  ``$\forallae{\mu}x\in X$, $P(x)$'', or
  ``for $\mu$-almost all $x\in X$, $P(x)$''
  (or simply without the mention of the measure~$\mu$).
\end{definition}

\begin{remark}
  When a single binary relation (using infix notation) is involved in~$P$,
  the annotation ``$\muae{\mu}$'' may be put above the infix operator, as
  in~``$\eqae{\mu}$'' (or simply as in~``$\eqae{}$'').
\end{remark}

\begin{lemma}[everywhere implies almost everywhere]
  \label{l:everywhere-implies-almost-everywhere}
  \mbox{}\\
  Let~$(X,\Sigma,\mu)$ be a measure space.
  Let~$P$ be a predicate on~$X$.
  Then, we have
  \begin{equation}
    \label{e:everywhere-implies-almost-everywhere}
    (\forall x \in X,\; P(x))
    \IMPLIES
    P \, \muae{\mu}.
  \end{equation}
\end{lemma}

\begin{proof}
  Direct consequence of
  Definition~\thref{d:prop-almost-satisfied}, and
  Lemma~\thref{l:empty-set-is-negl}.
\end{proof}

\begin{lemma}[everywhere implies almost everywhere for almost the same]
  \label{l:everywhere-implies-almost-everywhere-for-almost-the-same}
  \mbox{}\\
  Let~$(X,\Sigma,\mu)$ be a measure space.
  Let~$Y$ be a set.
  Let~$P$ be a predicate on~$Y$, lifted into a predicate on~$\FXY$.
  Let~$f$ and~$g$ be functions from~$X$ to~$Y$.
  Assume that $f\eqae{\mu}g$.
  Then, we have
  \begin{equation}
    \label{e:everywhere-implies-almost-everywhere-for-almost-the-same}
    (\forall x \in X,\; P (f (x)))
    \IMPLIES
    P (g) \, \muae{\mu}
  \end{equation}
\end{lemma}

\begin{proof}
  Direct consequence of
  \assume{monotonicity of complement ($\{P(g)\}^c\subset\{f=g\}^c$)},
  Definition~\threfc{d:prop-almost-satisfied}{$\{f=g\}^c\in\neglset$}, and
  Lemma~\threfc{l:subset-of-negl-is-negl}{%
    $\{P(g)\}^c\in\neglset$}.
\end{proof}

\begin{remark}
  \label{r:v2-new07}
  The previous lemma allows the use in most statements of functions defined
  almost everywhere rather than regular total functions.
\end{remark}

\begin{lemma}[extended almost modus ponens]
  \label{l:ext-almost-modus-ponens}
  \mbox{}\\
  Let~$(X,\Sigma,\mu)$ be a measure space.
  Let~$P$ and~$Q$ be predicates on~$X$.
  Then, we have
  \begin{equation}
    \label{e:ext-almost-modus-ponens}
    (P \IMPLIESae{\mu} Q)
    \CONJ
    P \, \muae{\mu}
    \IMPLIES
    Q \, \muae{\mu}
  \end{equation}
\end{lemma}

\begin{proof}
  Assume that $P\Implies Q$ and~$P$ hold almost everywhere.

  From
  Definition~\thref{d:prop-almost-satisfied}, and
  \assume{modus ponens},
  we have
  \begin{equation*}
    \{ P \Implies Q \}^c, \{ P \}^c \in \neglset
    \AND
    B \eqdef \{ P \Implies Q \} \cap \{ P \} \subset \{ Q \}.
  \end{equation*}
  Hence, from
  \assume{monotonicity of complement},
  \assume{De~Morgan's laws}, and
  Lemma~\threfc{l:n-is-closed-under-count-union}{%
    with $\card(I)=2$},
  we have
  \begin{equation*}
    \{ Q \}^c \subset B^c
    \AND
    B^c = \{ P \Implies Q \}^c \cup \{ P \}^c \in \neglset.
  \end{equation*}
  Therefore, from
  Lemma~\thref{l:subset-of-negl-is-negl}, and
  Definition~\thref{d:prop-almost-satisfied},
  we have for $\mu$-almost all $x\in X$, $Q(x)$.
\end{proof}

\begin{lemma}[almost modus ponens]
  \label{l:almost-modus-ponens}
  \mbox{}\\
  Let~$(X,\Sigma,\mu)$ be a measure space.
  Let~$P$ and~$Q$ be predicates on~$X$.
  Then, we have
  \begin{equation}
    \label{e:almost-modus-ponens}
    (\forall x \in X,\; P (x) \Implies Q (x))
    \CONJ
    P \, \muae{\mu}
    \IMPLIES
    Q \, \muae{\mu}
  \end{equation}
\end{lemma}

\begin{proof}
  Direct consequence of
  Lemma~\threfc{l:everywhere-implies-almost-everywhere}{%
    with predicate $P\Implies Q$}, and
  Lemma~\thref{l:ext-almost-modus-ponens}.
\end{proof}

\begin{remark}
  The two previous lemmas allow to still use modus ponens in reasoning when
  predicates are only valid almost everywhere.
\end{remark}

\begin{definition}[almost definition]
  \label{d:almost-definition}
  \mbox{}\\
  Let~$(X,\Sigma,\mu)$ be a measure space.
  Let~$Y$ be a set. 
  A function~$f:\ArXY$ is said
  {\em defined ($\mu$-)almost everywhere in~$X$} iff
  the property ``$f(x)$ is defined'' is satisfied $\mu$-almost everywhere.

  The
  {\em set of functions $\ArXY$ defined ($\mu$-)almost everywhere}
  is denoted~$\FXYae{\mu}$, or through the type annotation $\ArXYae{\mu}$
  (or simply without the mention of the measure~$\mu$).
\end{definition}

\begin{definition}[almost binary relation]
  \label{d:almost-bin-rel}
  \mbox{}\hfill
  Let~$(X,\Sigma,\mu)$ be a measure space.\\
  Let~$Y$ be a set.
  Let~$\calR$ be a binary relation over~$Y$, lifted into a binary relation
  over~$\FXY$.
  The functions~$f,g:\ArXYae{\mu}$ are said
  {\em in relation ($\mu$-)almost everywhere through~$\Eqrel$} iff
  the property ``$\eqrel{f(x)}{g(x)}$'' is satisfied $\mu$-almost everywhere;
  this is denoted $\eqrelae{\mu}{f}{g}$ (or simply $\eqrelae{}{f}{g}$).
\end{definition}

\begin{lemma}[compatibility of almost binary relation with reflexivity]
  \label{l:compat-of-almost-bin-rel-with-refl}
  \mbox{}\\
  Let~$(X,\Sigma,\mu)$ be a measure space.
  Let~$Y$ be a set.
  Let~$\Eqrel$ be a binary relation over~$Y$, lifted into a binary relation
  over~$\FXY$.
  Assume that~$\Eqrel$ is reflexive.
  Then, $\Eqrelae{\mu}$ is also reflexive.
\end{lemma}

\begin{proof}
  Direct consequence of
  Definition~\thref{d:almost-bin-rel},
  Definition~\thref{d:almost-definition},
  Definition~\thref{d:prop-almost-satisfied},
  \assume{the definition of reflexivity},
  Lemma~\thref{l:everywhere-implies-almost-everywhere}, and
  Lemma~\thref{l:everywhere-implies-almost-everywhere-for-almost-the-same}.
\end{proof}

\begin{lemma}[compatibility of almost binary relation with symmetry]
  \label{l:compat-of-almost-bin-rel-with-sym}
  \mbox{}\\
  Let~$(X,\Sigma,\mu)$ be a measure space.
  Let~$Y$ be a set.
  Let~$\Eqrel$ be a binary relation over~$Y$, lifted into a binary relation
  over~$\FXY$.
  Assume that~$\Eqrel$ is symmetric.
  Then, $\Eqrelae{\mu}$ is also symmetric.
\end{lemma}

\begin{proof}
  Direct consequence of
  Definition~\thref{d:almost-bin-rel},
  Definition~\thref{d:almost-definition},
  Definition~\thref{d:prop-almost-satisfied},
  \assume{the definition of symmetry},
  Lemma~\thref{l:almost-modus-ponens}, and
  Lemma~\thref{l:everywhere-implies-almost-everywhere-for-almost-the-same}.
\end{proof}

\begin{lemma}[compatibility of almost binary relation with antisymmetry]
  \label{l:compat-of-almost-bin-rel-with-antisym}
  \mbox{}\\
  Let~$(X,\Sigma,\mu)$ be a measure space.
  Let~$Y$ be a set.
  Let~$\Eqrel$ be a binary relation over~$Y$, lifted into a binary relation
  over~$\FXY$.
  Assume that~$\Eqrel$ is antisymmetric.
  Then, $\Eqrelae{\mu}$ is ``almost'' antisymmetric (where equality is replaced
  by almost equality).
\end{lemma}

\begin{proof}
  Let~$f,g:\ArXYae{\mu}$.
  Assume that $\eqrelae{\mu}{f}{g}$ and $\eqrelae{\mu}{g}{f}$.
  Let us show that $f\eqae{\mu}g$.

  From
  Definition~\thref{d:almost-bin-rel},
  Definition~\thref{d:almost-definition},
  Definition~\thref{d:prop-almost-satisfied}, and
  Lemma~\thref{l:everywhere-implies-almost-everywhere-for-almost-the-same},
  we have $\{\eqrel{f}{g}\}^c,\{\eqrel{g}{f}\}^c\in\neglset$.
  Moreover, from
  \assume{the definition of antisymmetry},
  we have $B\eqdef\{\eqrel{f}{g}\}\cap\{\eqrel{g}{f}\}\subset\{f=g\}$.
  Hence, from
  \assume{monotonicity of complement},
  \assume{De~Morgan's laws}, and
  Lemma~\threfc{l:n-is-closed-under-count-union}{with $\card(I)=2$},
  we have
  $\{f=g\}^c\subset B^c=\{\eqrel{f}{g}\}^c\cup\{\eqrel{g}{f}\}^c\in\neglset$.

  Therefore, from
  Lemma~\thref{l:subset-of-negl-is-negl}, and
  Definition~\thref{d:prop-almost-satisfied},
  we have $f\eqae{\mu}g$.
\end{proof}

\begin{lemma}[compatibility of almost binary relation with transitivity]
  \label{l:compat-of-almost-bin-rel-with-trans}
  \mbox{}\\
  Let~$(X,\Sigma,\mu)$ be a measure space.
  Let~$Y$ be a set.
  Let~$\Eqrel$ be a binary relation over~$Y$, lifted into a binary relation
  over~$\FXY$.
  Assume that~$\Eqrel$ is transitive.
  Then, $\Eqrelae{\mu}$ is also transitive.
\end{lemma}

\begin{proof}
  Let~$f,g,h:\ArXYae{\mu}$.
  Assume that $\eqrelae{\mu}{f}{g}$ and $\eqrelae{\mu}{g}{h}$.
  Let us show that $\eqrelae{\mu}{f}{h}$.

  From
  Definition~\thref{d:almost-bin-rel},
  Definition~\thref{d:almost-definition},
  Definition~\thref{d:prop-almost-satisfied}, and
  Lemma~\thref{l:everywhere-implies-almost-everywhere-for-almost-the-same},
  we have $\{\eqrel{f}{g}\}^c,\{\eqrel{g}{h}\}^c\in\neglset$.
  Moreover, from
  \assume{the definition of transitivity},
  we have
  $B\eqdef\{\eqrel{f}{g}\}\cap\{\eqrel{g}{h}\}\subset\{\eqrel{f}{h}\}$.
  Hence, from
  \assume{monotonicity of complement},
  \assume{De~Morgan's laws}, and
  Lemma~\threfc{l:n-is-closed-under-count-union}{with $\card(I)=2$},
  we have
  $\{\eqrel{f}{h}\}^c\subset
  B^c=\{\eqrel{f}{g}\}^c\cup\{\eqrel{g}{h}\}^c\in\neglset$.

  Therefore, from
  Lemma~\thref{l:subset-of-negl-is-negl},
  Definition~\thref{d:prop-almost-satisfied}, and
  Definition~\thref{d:almost-bin-rel},
  we have $\eqrelae{\mu}{f}{h}$.
\end{proof}

\begin{remark}
  From the similarity of their formal expressions, antisymmetry and
  transitivity can be abstracted under the form
  \begin{equation*}
    \forall f, g, h : \ArXY,\quad
    \eqrel{f}{g} \CONJ \eqrel{g}{F(f, h)} \IMPLIES \eqrelp{f}{F (g, h)}
  \end{equation*}
  where the binary relation~$\Eqrelp$ is the equality for antisymmetry
  and~$\Eqrel$ for transitivity, and where the
  function~$F:\ArFXYxFXYFXY$ is the first projection for antisymmetry and the
  second projection for transitivity.
  Moreover, the proofs of the two previous lemmas only differ from the
  expression of their instances of~$\Eqrelp$ and~$F$.
  Thus, both statements and proofs of these lemmas can be abstracted under
  more general forms.
\end{remark}

\begin{lemma}[almost equivalence is equivalence relation]
  \label{l:almost-equiv-is-equiv-rel}
  \mbox{}\hfill
  Let~$(X,\Sigma,\mu)$ be a measure space.
  Let~$Y$ be a set.
  Let~$\Eqrel$ be a binary relation over~$Y$, lifted into a binary relation
  over~$\FXY$.
  Assume that~$\Eqrel$ is an equivalence relation.
  Then, $\Eqrelae{\mu}$ is also an equivalence relation.
\end{lemma}

\begin{proof}
  Direct consequence of
  \assume{the definition of equivalence relation},
  Lemma~\thref{l:compat-of-almost-bin-rel-with-refl},
  Lemma~\thref{l:compat-of-almost-bin-rel-with-sym}, and
  Lemma~\thref{l:compat-of-almost-bin-rel-with-trans}.
\end{proof}

\begin{lemma}[almost equality is equivalence relation]
  \label{l:almost-eq-is-equiv-rel}
  \mbox{}\\
  Let~$(X,\Sigma,\mu)$ be a measure space.
  Let~$Y$ be a set.
  Then, $\eqae{\mu}$ is an equivalence relation over~$Y^X$.
\end{lemma}

\begin{proof}
  Direct consequence of
  Lemma~\threfc{l:almost-equiv-is-equiv-rel}{%
    with equality over~$Y$}.
\end{proof}

\begin{lemma}[almost order is order relation]
  \label{l:almost-order-is-order-rel}
  \mbox{}\\
  Let~$(X,\Sigma,\mu)$ be a measure space.
  Let~$Y$ be a set. 
  Let~$\Eqrel$ be a binary relation over~$Y$, lifted into a binary relation
  over~$\FXY$.
  Assume that~$\Eqrel$ is an order relation.
  Then, $\Eqrelae{\mu}$ is an ``almost'' order relation (where equality is
  replaced by almost equality in antisymmetry).
\end{lemma}

\begin{proof}
  Direct consequence of
  \assume{the definition of order relation},
  Lemma~\thref{l:compat-of-almost-bin-rel-with-refl},
  Lemma~\thref{l:compat-of-almost-bin-rel-with-antisym}, and
  Lemma~\thref{l:compat-of-almost-bin-rel-with-trans}.
\end{proof}

\begin{lemma}[compatibility of almost binary relation with operator]
  \label{l:compat-of-almost-bin-rel-with-op}
  \mbox{}\\
  Let~$(X,\Sigma,\mu)$ be a measure space.
  Let~$Y$ be a nonempty set.
  Let~$y_0\in Y$.
  Let~$\Eqrel$ and~$\Eqrelp$ be binary relations over~$Y$, lifted into binary
  relations over~$\FXY$.
  Assume that~$\eqrel{y_0}{y_0}$.
  Let~$I$ be a nonempty subset of~$\matN$.
  Let~$\Diamond:\ArYIY$, lifted into an operator $\ArFXYIFXY$.
  Assume that for all $(f_i)_{i\in I},(g_i)_{i\in I}:\ArXY$, we have
  \begin{equation}
    \label{e:compat-of-almost-bin-rel-with-op-1}
    (\forall i \in I,\; \eqrel{f_i}{g_i})
    \IMPLIES
    \eqrelp{\Diamond (f_i)_{i \in I}}{\Diamond (g_i)_{i \in I}}.
  \end{equation}
  Then, for all $(f_i)_{i\in I},(g_i)_{i\in I}:\ArXYae{\mu}$, we have
  \begin{equation}
    \label{e:compat-of-almost-bin-rel-with-op-2}
    (\forall i \in I,\; \eqrelae{\mu}{f_i}{g_i})
    \IMPLIES
    \eqrelpae{\mu}{\Diamond (f_i)_{i \in I}}{\Diamond (g_i)_{i \in I}}.
  \end{equation}
\end{lemma}

\begin{proof}
  For all $i\in I$, let $f_i,g_i:\ArXYae{\mu}$, and assume that
  $\eqrelae{\mu}{f_i}{g_i}$ holds.

  For all $i\in I$, let $B_i\eqdef\{\eqrel{f_i}{g_i}\}$.
  For all $i\in I$, let $\tf_i,\tg_i:\ArXY$ defined by
  \begin{equation*}
    \left.
      \begin{array}{l}
        \tf_i (x) \eqdef f_i (x)\\
        \tg_i (x) \eqdef g_i (x)
      \end{array}
    \right\}
    \mbox{ when } x \in \bigcap_{i \in I} B_i,
    \AND
    \tf_i (x) = \tg_i (x) \eqdef y_0 \mbox{ otherwise}.
  \end{equation*}
  Let~$i\in I$.
  Then, from
  Definition~\thref{d:almost-bin-rel},
  Definition~\thref{d:almost-definition},
  Definition~\thref{d:prop-almost-satisfied}, and
  Lemma~\thref{l:everywhere-implies-almost-everywhere-for-almost-the-same},
  we have $B_i^c\in\neglset$.
  Let~$x\in X$.
  Then, by construction, and from reflexivity of~$\Eqrel$, we have
  $\eqrel{\tf_i(x)}{\tg_i(x)}$.
  Thus, from assumption,
  $\eqrelp{\Diamond(\tf_i)_{i\in I}(x)}{\Diamond(\tg_i)_{i\in I}(x)}$ also
  holds.
  Moreover, by construction, we have
  \begin{equation*}
    \forall x \in \bigcap_{i \in I} B_i,\quad
    \Diamond (\tf_i)_{i \in I} (x)
    = \Diamond (f_i)_{i \in I} (x)
    \AND
    \Diamond (\tg_i)_{i \in I} (x)
    = \Diamond (g_i)_{i \in I} (x).
  \end{equation*}
  Thus, we have
  $\bigcap_{i\in I}B_i
  \subset\{\eqrelp{\Diamond(f_i)_{i\in I}}{\Diamond(g_i)_{i\in I}}\}$.
  Hence, from
  \assume{monotonicity of complement},
  \assume{De~Morgan's laws}, and
  Lemma~\thref{l:n-is-closed-under-count-union},
  we have
  \begin{equation*}
    \{ \eqrelp{\Diamond (f_i)_{i \in I}}{\Diamond (g_i)_{i \in I}} \}^c
    \subset \bigcup_{i \in I} \{ \eqrel{f_i}{g_i} \}^c
    \in \neglset.
  \end{equation*}

  Therefore, from
  Definition~\thref{d:prop-almost-satisfied},
  Definition~\thref{d:almost-bin-rel}, and
  Definition~\thref{d:almost-definition},
  we have $\eqrelpae{\mu}{\Diamond(f_i)_{i\in I}}{\Diamond(g_i)_{i\in I}}$.
\end{proof}

\begin{lemma}[compatibility of almost equality with operator]
  \label{l:compat-of-almost-eq-with-op}
  \mbox{}\\
  Let~$(X,\Sigma,\mu)$ be a measure space.
  Let~$Y$ be a set. 
  Let~$I\subset\matN$. 
  Let~$\Diamond:\ArYIY$, lifted into an operator $\ArFXYIFXY$.
  Let~$(f_i)_{i\in I},(g_i)_{i\in I}:\ArXYae{\mu}$.
  Assume that for all $i\in I$, $f_i\eqae{\mu}g_i$.
  Then, we have $\Diamond(f_i)_{i\in I}\eqae{\mu}\Diamond(g_i)_{i\in I}$.
\end{lemma}

\begin{proof}
  Direct consequence of
  Lemma~\threfc{l:compat-of-almost-bin-rel-with-op}{%
    with $\Eqrel=\Eqrelp\eqdef$ equality},
  \assume{reflexivity of equality}, and
  \assume{the definition of function ((for all $i\in I$, $f_i=g_i$) implies
    $\Diamond(f_i)_{i\in I}=\Diamond(g_i)_{i\in I}$)}.
\end{proof}

\begin{lemma}[compatibility of almost inequality with operator]
  \label{l:compat-of-almost-ineq-with-op}
  \mbox{}\\
  Let~$(X,\Sigma,\mu)$ be a measure space.
  Let~$Y$ be an ordered set. 
  Let~$\Eqrelp$ be a binary relation on~$Y$, lifted into a binary relation
  on~$\FXY$.
  Let~$I\subset\matN$. 
  Let~$\Diamond:\ArYIY$, lifted into an operator $\ArFXYIFXY$.
  Assume that for all $(f_i)_{i\in I},(g_i)_{i\in I}:\ArXY$, we have
  \begin{equation}
    \label{e:compat-of-almost-ineq-with-op-1}
    (\forall i \in I,\; f_i \leq g_i)
    \IMPLIES
    \eqrelp{\Diamond (f_i)_{i \in I}}{\Diamond (g_i)_{i \in I}}.
  \end{equation}
  Then, for all $(f_i)_{i\in I},(g_i)_{i\in I}:\ArXYae{\mu}$, we have
  \begin{equation}
    \label{e:compat-of-almost-ineq-with-op-2}
    (\forall i \in I,\; f_i \leqae{\mu} g_i)
    \IMPLIES
    \eqrelpae{\mu}{\Diamond (f_i)_{i \in I}}{\Diamond (g_i)_{i \in I}}.
  \end{equation}
\end{lemma}

\begin{proof}
  Direct consequence of
  Lemma~\threfc{l:compat-of-almost-bin-rel-with-op}{%
    with $\Eqrel\eqdef\leq$}, and
  \assume{reflexivity of inequality}.
\end{proof}

\begin{remark}
  The previous generic results apply to unary operators~$\Diamond$ such as
  the scalar multiplication, or the absolute value (for which $\card(I)=1$),
  as well as to binary (or $n$-ary) operators~$\Diamond$ such as the
  addition, the multiplication, the maximum, or the minimum (for which
  $\card(I)\geq2$ is finite).
  But it also applies to operators~$\Diamond$ taking a countable number of
  arguments such as the infimum, the supremum, or the limit of a pointwise
  convergent sequence.
\end{remark}

\begin{remark}
  Note that the previous lemma may be useful with~$\Eqrelp$ distinct
  from~$\leq$.
  For instance, when the operator~$\Diamond$ is the scalar multiplication by
  a negative number, we need $\Eqrelp=\geq$.
\end{remark}

\begin{lemma}[definiteness implies almost definiteness]
  \label{l:definiteness-implies-almost-definiteness}
  \mbox{}\\
  Let~$(X,\Sigma,\mu)$ be a measure space.
  Let~$Y$ be a set.
  Assume that $0\in Y$.
  Let~$\Diamond:\ArYY$, lifted into an operator $\ArFXYFXY$.
  Assume that~$\Diamond$ is definite:
  \begin{equation}
    \label{e:definiteness-implies-almost-definiteness-1}
    \forall y \in Y,\quad
    \Diamond (y) = 0
    \IMPLIES
    y = 0.
  \end{equation}
  Then, $\Diamond$~is also ``almost definite'':
  \begin{equation}
    \label{e:definiteness-implies-almost-definiteness-2}
    \forall f \in \FXYae{\mu},\quad
    \Diamond (f) \eqae{\mu} 0
    \IMPLIES
    f \eqae{\mu} 0.
  \end{equation}
\end{lemma}

\begin{proof}
  Direct consequence of
  Lemma~\thref{l:almost-modus-ponens}, and
  Lemma~\thref{l:everywhere-implies-almost-everywhere-for-almost-the-same}.
\end{proof}

\begin{remark}
  Note that Lemma~\ref{l:compat-of-almost-eq-with-op} provides the implication
  in the other direction.
\end{remark}

\clearpage
\section{Uniqueness condition}
\label{s:uniqueness-condition}

\begin{remark}
  The next lemma is useful to establish in Section~\ref{s:the-lebesgue-measure}
  uniqueness of the Lebesgue measure that generalizes the length of bounded
  intervals (see Theorem~\ref{t:caratheodory-lebesgue-meas-on-r}).
\end{remark}

\begin{remark}
  The next proof follows the {\Dplt} scheme
  (see Section~\ref{s:dynkin-p-l-th-monot-class-th-schemes}).
\end{remark}

\begin{lemma}[uniqueness of measures extended from a $\pi$-system]
  \label{l:uniq-of-meas-ext-from-p-syst}
  \mbox{}\\
  Let~$(X,\Sigma)$ be a measurable space.
  Let~$\mu_1$ and~$\mu_2$ be measures on~$(X,\Sigma)$.
  Let~$G\subset\calP(X)$.
  Assume that~$G$ is a $\pi$-system, a generator of~$\Sigma$, contains a
  countable pseudopartition~$(X_n)_{n\in\matN}$ of~$X$ with finite
  measure~$\mu_1$, and that both measures coincide on~$G$.
  Then, we have $\mu_1=\mu_2$.
\end{lemma}

\begin{proof}
  Let~$A\in G$.
  Assume that $\mu_1(A)<\infty$ ({\eg} $A\eqdef X_0$).\\
  Let~$B\in\Sigma$.
  Then, from
  Lemma~\threfc{l:gen-sigma-alg-is-min}{%
    $A\in G\subset\Sigma$}, and
  Lem\-ma~\threfc{l:equiv-def-of-sigma-alg}{%
    closedness under countable intersection with $\card(I)$ equals 2},
  we have $B\cap A\in\Sigma$.
  Let $\calS_A\eqdef\{B\in\Sigma\st\mu_1(B\cap A)=\mu_2(B\cap A)\}$.

  \proofparskip{(1). $G\not=\emptyset\;\Conj\;\Pi_X(G)\subset\calS_A$}\\
  Let~$B\in G$.
  Then, from
  Definition~\threfc{d:p-syst}{closedness under intersection},
  $B\cap A\in G$, {\ie} $B\in\calS_A$.
  Thus, we have $G\subset\calS_A$.
  Hence, from
  Definition~\threfc{d:p-syst}{nonemptiness}, and
  Lemma~\thref{l:p-syst-gen-is-idem},
  we have $G\not=\emptyset$ and $\Pi_X(G)=G\subset\calS_A$.

  \proofparskip{(2). $\calS_A$ is $\lambda$-system}

  From
  \assume{the identity $X\cap A=A$}, and
  Lemma~\threfc{l:equiv-def-of-sigma-alg}{contains the full set},
  we have $X\in\calS_A$.

  Let~$B_1,B_2\in\calS_A$.
  Assume that $B_1\subset B_2$.
  Then, from
  \assume{distributivity of intersection over local complement}, and
  Lemma~\threfc{l:meas-is-monot}{with $\mu_1$ and $\mu_2$},
  we have
  \begin{align*}
    &\mu_2 (B_1 \cap A) = \mu_1 (B_1 \cap A) \leq \mu_1 (A) < \infty,
    \quad\AND\\
    \mu_1 ((B_2 \setminus B_1) \cap A)
    &= \mu_1 (B_2 \cap A \setminus B_1 \cap A)
    = \mu_1 (B_2 \cap A) - \mu_1 (B_1 \cap A)\\
    &= \mu_2 (B_2 \cap A) - \mu_2 (B_1 \cap A)
    = \mu_2 (B_2 \cap A \setminus B_1 \cap A)
    = \mu_2 ((B_2 \setminus B_1) \cap A).
  \end{align*}
  Then, $B_2\setminus B_1\in\calS_A$.
  Thus, $\calS_A$ is closed under local complement.

  Let~$(B_n)_{n\in\matN}\in\calS_A$.
  Assume that for all $n\in\matN$, $B_n\subset B_{n+1}$.
  Then, from
  \assume{distributivity of intersection over union},
  Lemma~\threfc{l:meas-is-cont-from-below}{%
    with $\mu_1$ and $\mu_2$, and nondecreasing $(B_n\cap A)_{n\in\matN}$}, and
  Definition~\thref{d:continuity-from-below},
  we have
  \begin{multline*}
    \mu_1 \left( \bigcup_{n \in \matN} B_n \cap A \right)
    = \mu_1 \left( \bigcup_{n \in \matN} (B_n \cap A) \right)
    = \lim_{n \to \infty} \mu_1 (B_n \cap A)\\
    = \lim_{n \to \infty} \mu_2 (B_n \cap A)
    = \mu_2 \left( \bigcup_{n \in \matN} (B_n \cap A) \right)
    = \mu_2 \left( \bigcup_{n \in \matN} B_n \cap A \right).
  \end{multline*}
  Then, $\bigcup_{n\in\matN}B_n\in\calS_A$.
  Thus, $\calS_A$ is closed under countable monotone union.

  Hence, from
  Lemma~\thref{l:equiv-def-of-l-syst},
  $\calS_A$ is a $\lambda$-system on~$X$.

  \proofparskip{(3). $\calS_A=\Sigma$}
  Direct consequence of~(1), (2), and
  Lemma~\threfc{l:usage-of-dynkin-pi-lambda-th}{%
    with $P$ being the function $B\mapsto\mu_1(B\cap A)=\mu_2(B\cap A)$}.\\
  Hence, for all $B\in\Sigma$, for all $A\in G$ such that $\mu_1(A)<\infty$, we
  have $\mu_1(B\cap A)=\mu_2(B\cap A)$.

  \proofparskip{(4). $\mu_1=\mu_2$}
  Let~$B\in\Sigma$.\\
  For all $n\in\matN$, let $B_n\eqdef B\cap X_n$.
  Then, from
  Definition~\thref{d:pseudopart},
  Lemma~\threfc{l:gen-sigma-alg-is-min}{$X_n\in G\subset\Sigma$},
  Lemma~\threfc{l:equiv-def-of-sigma-alg}{%
    closedness under countable intersection with $\card(I)$ equals 2},
  \assume{compatibility of intersection with pairwise disjunction},
  \assume{left distributivity of intersection over union}, and
  \assume{identity law for intersection},
  $(B_n)_{n\in\matN}\in\Sigma$ is pairwise disjoint, and we have
  $\biguplus_{n\in\matN}B_n=B$. \\
  Let~$n\in\matN$.
  Then, from~(3) (with $B\in\Sigma$, and $A\eqdef X_n\in G$ such that
  $\mu_1(X_n)<\infty$), we have $\mu_1(B_n)=\mu_2(B_n)$.
  Hence, from
  Definition~\threfc{d:meas}{$\mu_1$ and $\mu_2$ are $\sigma$-additive}, and
  Definition~\thref{d:sigma-add-over-meas-space},
  we have
  \begin{equation*}
    \mu_1 (B)
    = \sum_{n \in \matN} \mu_1 (B_n)
    = \sum_{n \in \matN} \mu_2 (B_n)
    = \mu_2 (B).
  \end{equation*}

  \medskip\noindent
  Therefore, we have the equality $\mu_1=\mu_2$.
\end{proof}

\clearpage
\section{Some measures}
\label{s:some-measures}

\begin{lemma}[trivial measure]
  \label{l:trivial-meas}
  \mbox{}\\
  Let~$(X,\Sigma)$ be a measurable space.
  Then, the zero function on~$\Sigma$ is a measure on~$(X,\Sigma)$.

  It is called the {\em trivial measure on~$(X,\Sigma)$}, and $(X,\Sigma,0)$
  is called the {\em trivial measure space}.
\end{lemma}

\begin{proof}
  Direct consequence of
  Definition~\thref{d:meas}, and
  \assume{ordered group properties of~$\matR$}.
\end{proof}

\begin{lemma}[equivalent definition of trivial measure]
  \label{l:equiv-def-of-trivial-meas}
  \mbox{}\\
  Let~$(X,\Sigma,\mu)$ be a measure space.
  Then, $\mu$~is the trivial measure iff
  $\mu(X)=0$.
\end{lemma}

\begin{proof}
  Direct consequence of
  Lemma~\thref{l:trivial-meas},
  Definition~\threfc{d:meas}{$\mu$~is non\-negative}, and
  Lemma~\threfc{l:meas-is-monot}{$\mu$~is nonpositive}.
\end{proof}

\begin{lemma}[counting measure]
  \label{l:count-meas}
  \mbox{}\\
  Let~$(X,\Sigma)$ be a measurable space.
  Let~$Y\subset X$.
  Let~$\delta_Y$ be the function defined by
  \begin{equation}
    \label{e:count-meas}
    \forall A \in \Sigma,\quad
    \delta_Y (A) \eqdef \card (A \cap Y)
  \end{equation}
  with the convention that the cardinality of an infinite set is~$\infty$.
  Then, $\delta_Y$~is a measure on~$(X,\Sigma)$.

  It is called the {\em counting measure (associated with~$Y$)}.
\end{lemma}

\begin{proof}
  From
  \assume{the definition of the cardinality}, and
  \assume{the definition of~$\infty$},
  the function~$\delta_Y$ is nonnegative.
  Moreover, from
  Definition~\thref{d:meas},
  Definition~\threfc{d:measurable-space}{$\Sigma$ is a $\sigma$-algebra},
  Definition~\threfc{d:sigma-alg}{$\emptyset\in\Sigma$}, and
  \assume{the definition of~$\emptyset$},
  we have
  \begin{equation*}
    \delta_Y (\emptyset) = \card (\emptyset) = 0.
  \end{equation*}

  Let~$I\subset\matN$.
  Let~$(A_i)_{i\in I}\in\Sigma$.
  Assume that the~$A_i$'s are pairwise disjoint.
  Then, from
  \assume{distributivity of intersection over disjoint union}, and
  \assume{$\sigma$-additivity of the cardinality (with the convention
    that~$\infty$ is absorbing element for addition in~$\matNbar$)},
  we have
  \begin{equation*}
    \delta_Y \left( \biguplus_{i \in I} A_i \right)
    = \card \left( \biguplus_{i \in I} A_i \cap Y \right)
    = \card \left( \biguplus_{i \in I} (A_i \cap Y) \right)
    = \sum_{i \in I} \card (A_i \cap Y)
    = \sum_{i \in I} \delta_Y (A_i).
  \end{equation*}

  Therefore, from
  Definition~\thref{d:sigma-add-over-meas-space}, and
  Definition~\thref{d:meas},
  $\delta_Y$ is a measure on~$(X,\Sigma)$.
\end{proof}

\begin{lemma}[finiteness of counting measure]
  \label{l:finiteness-of-count-meas}
  \mbox{}\\
  Let~$(X,\Sigma)$ be a measurable space.
  Let~$Y\subset X$. 
  Then, $\delta_Y$~is finite iff
  the set $Y$~is finite.
\end{lemma}

\begin{proof}
  Direct consequence of
  Lemma~\threfc{l:equiv-def-of-sigma-alg}{contains full set},
  Lemma~\threfc{l:count-meas}{$\delta_Y(X)=\card(Y)$}, and
  Definition~\thref{d:finite-meas}.
\end{proof}

\begin{lemma}[$\sigma$-finite counting measure]
  \label{l:sigma-finiteness-of-count-meas}
  \mbox{}\\
  Let~$(X,\Sigma)$ be a measurable space.
  Let~$Y\subset X$.
  Assume that~$\Sigma$ contains all singletons of~$X$.\\
  Then, $\delta_Y$~is $\sigma$-finite implies~$Y$ is countable, and
  $X$~is countable implies $\delta_Y$~is $\sigma$-finite.
\end{lemma}

\begin{proof}
  From
  Definition~\thref{d:meas},
  Definition~\thref{d:measurable-space}, and
  Definition~\threfc{d:sigma-alg}{closedness under countable union},
  the $\sigma$-algebra~$\Sigma$ contains all countable subsets of~$X$.
  Hence, if~$X$ is countable, we have $\Sigma=\calP(X)$.

  \proofparskip{First implication}
  Assume that~$\delta_Y$ is $\sigma$-finite.\\
  Then, from
  Lemma~\thref{l:equiv-def-of-sigma-finite-meas},
  let~$(B_n)_{n\in\matN}\in\Sigma$ such that for all $n\in\matN$,
  $B_n\subset B_{n+1}$, $\delta_Y(B_n)=\card(B_n\cap Y)<\infty$, and
  $X=\bigcup_{n\in\matN}B_n$.
  For all $n\in\matN$, let~$A_n\eqdef B_n\cap Y$.
  Let~$n\in\matN$.
  Then, from
  \assume{De~Morgan's laws}, and
  Lemma~\thref{l:count-meas},
  we have $Y=\bigcup_{n\in\matN}A_n$, and $\card(A_n)=\delta_Y(B_n)<\infty$.
  Hence, from
  \assume{countability of countable union of finite subsets},
  $Y$~is countable.

  \proofparskip{Second implication}
  Assume that~$X$ is countable.
  Then, we have $\Sigma=\calP(X)$.\\
  From
  \assume{the definition of countability},
  let~$\fhi$ be a bijection from~$I\subset\matN$ onto~$X$.
  For all $n\in\matN$, let $A_n\eqdef\fhi([0..n]\cap I)\in\Sigma$.
  Let~$n\in\matN$.
  Then, from
  Lemma~\thref{l:count-meas},
  \assume{monotonicity of the cardinality},
  \assume{preservation of the cardinality by bijection},
  \assume{compatibility of direct image with union}, and
  \assume{De~Morgan's laws},
  we have
  \begin{gather*}
    \delta_Y (A_n)
    = \card(A_n \cap Y)
    \leq \card (A_n)
    = \card ([0..n] \cap I)
    \leq n + 1 < \infty,\\
    \bigcup_{n \in \matN} A_n
    = \fhi \left( \bigcup_{n \in \matN} ([0..n] \cap I) \right)
    = \fhi \left( \bigcup_{n \in \matN} [0..n] \cap I \right)
    = \fhi (I)
    = X.
  \end{gather*}
  Hence, from
  Definition~\thref{d:sigma-finite-meas},
  the counting measure~$\delta_Y$ is $\sigma$-finite.

  \medskip\noindent
  Therefore, we have both implications.
\end{proof}

\begin{remark}
  \label{r:v2-mod4}
  Note that the set~$Y$ is not required to be measurable.
  Note also that the previous lemmas are valid for any
  $\sigma$-algebra~$\Sigma$, including the discrete $\sigma$-algebra
  $\calP(X)$.

  Counting measures are usually considered in the case where $Y=X$ is
  countable, and~$\Sigma$ is the discrete $\sigma$-algebra~$\calP(X)$.
  A typical example is the $\sigma$-finite measure space
  $(\matN,\calP(\matN),\delta_\matN)$.
\end{remark}

\begin{definition}[Dirac measure]
  \label{d:dirac-meas}
  \mbox{}\\
  Let~$(X,\Sigma)$ be a measurable space.
  Let~$a\in X$.
  The counting measure associated with~$\{a\}$ is called
  {\em Dirac measure (at~$a$)};
  it is also denoted $\delta_a\eqdef\delta_{\{a\}}$.
\end{definition}

\begin{lemma}[equivalent definition of Dirac measure]
  \label{l:equiv-def-of-dirac-meas}
  \mbox{}\\
  Let~$(X,\Sigma)$ be a measurable space.
  Let~$a\in X$, and let~$A\in\Sigma$.
  Then, we have $\delta_a(A)=\matUN_A(a)$.
\end{lemma}

\begin{proof}
  Direct consequence of
  Definition~\thref{d:dirac-meas}, and
  Lemma~\threfc{l:count-meas}{%
    $\card(A\cap\{a\})=1$ when $a\in A$ and~0 otherwise}.
\end{proof}

\begin{lemma}[Dirac measure is finite]
  \label{l:dirac-meas-is-finite}
  \mbox{}\\
  Let~$(X,\Sigma)$ be a measurable space.
  Let~$a\in X$.
  Then, $\delta_a$~is finite and $\delta_a(X)=1$.
\end{lemma}

\begin{proof}
  Direct consequence of
  Definition~\thref{d:dirac-meas},
  Lemma~\threfc{l:finiteness-of-count-meas}{$\delta_a$ is finite}, and
  Lemma~\threfc{l:count-meas}{$\card(X\cap\{a\})=1$}.
\end{proof}

\chapter{Measure and numbers}
\label{c:measure-and-numbers}

\minitoc

\section{Negligibility and numbers}
\label{s:negligibility-and-numbers}

\begin{definition}[summability domain]
  \label{d:summability-domain}
  \mbox{}\hfill
  Let~$(X,\Sigma,\mu)$ be a measure space.\\
  Let $f,g:\ArXRb$.
  The {\em summability domain of~$f$ and~$g$} is denoted~$\Dfsum(f,g)$;
  it is defined by
  \begin{equation}
    \label{e:summability-domain}
    \Dfsum (f, g)
    \eqdef [(\{ f = \infty \} \cap \{ g = -\infty \})
    \cup (\{ f = -\infty \} \cap \{ g = \infty \})]^c.
  \end{equation}
\end{definition}

\begin{lemma}[summability on summability domain]
  \label{l:summability-on-summability-domain}
  \mbox{}\\
  Let~$(X,\Sigma,\mu)$ be a measure space.
  Let $f,g\in\calM$.
  Then, $f(x)+g(x)$ is well-defined iff $x\in\Dfsum(f,g)$.
\end{lemma}

\begin{proof}
  Direct consequence of
  Definition~\thref{d:add-in-rbar}, and
  Definition~\thref{d:summability-domain}.
\end{proof}

\begin{lemma}[measurability of summability domain]
  \label{l:meas-of-summability-domain}
  \mbox{}\\
  Let~$(X,\Sigma,\mu)$ be a measure space.
  Let $f,g\in\calM$.
  Then, $\Dfsum(f,g)\in\Sigma$.
\end{lemma}

\begin{proof}
  Direct consequence of
  Lemma~\thref{l:inverse-image-is-meas}, and
  Lemma~\thref{l:equiv-def-of-sigma-alg}{%
    closedness under countable intersection, union and complement}.
\end{proof}

\begin{lemma}[negligibility of summability domain]
  \label{l:negl-of-summability-domain}
  \mbox{}\hfill
  Let~$(X,\Sigma,\mu)$ be a measure space.\\
  Let $f,g\in\calM$.
  Then, $f+g$ is well-defined almost everywhere iff
  $[\Dfsum(f,g)]^c\in\neglset$.
\end{lemma}

\begin{proof}
  Direct consequence of
  Lemma~\thref{l:summability-on-summability-domain}, and
  Definition~\thref{d:prop-almost-satisfied},
\end{proof}

\begin{lemma}[almost sum]
  \label{l:almost-sum}
  \mbox{}\hfill
  Let~$(X,\Sigma,\mu)$ be a measure space.\\
  Let~$f,g\in\calM$.
  Let~$A\eqdef\Dfsum(f,g)$.
  Let $\tf\eqdef f\matUN_A$ and $\tg\eqdef g\matUN_A$.
  Assume that~$f+g$ is well-defined almost everywhere.
  Then, we have $\tf\eqae{\mu}f$, $\tg\eqae{\mu}g$, $\Dfsum(\tf,\tg)=X$, and
  $\tf,\tg,\tf+\tg\in\calM$.

  The sum $\tf+\tg$ is called {\em the almost sum of~$f$ and~$g$};
  it is denoted $f\plusae{\mu}g$.
\end{lemma}

\begin{proof}
  From
  Lemma~\thref{l:negl-of-summability-domain}, and
  Definition~\threfc{d:prop-almost-satisfied}{%
    with $\{\tf\not=f\}=\{\tg\not=g\}=A^c$},
  we have~$\tf\eqae{\mu}f$ and~$\tg\eqae{\mu}g$.
  From
  Lemma~\thref{l:meas-of-summability-domain},
  Lemma~\thref{l:meas-of-indic-fun}, and
  Lemma~\thref{l:m-is-closed-under-mult},
  we have $\tf,\tg\in\calM$.
  Hence, from
  Definition~\thref{d:add-in-rbar}, and
  Lemma~\thref{l:m-is-closed-under-add-when-defined},
  $\tf+\tg$ is well-defined, and we have $\tf+\tg\in\calM$.
\end{proof}

\begin{lemma}[compatibility of almost sum with almost equality]
  \label{l:compat-of-almost-sum-with-almost-eq}
  \mbox{}\\
  Let~$(X,\Sigma,\mu)$ be a measure space.
  Let~$f,g\in\calM$.
  Assume that~$f+g$ is well-defined almost everywhere.
  Let $\fp,\gp\in\calM$.
  Assume that $\fp\eqae{\mu}f$, $\gp\eqae{\mu}g$, and $\Dfsum(\fp,\gp)=X$.\\
  Then, $\fp+\gp\in\calM$, and we have $\fp+\gp\eqae{\mu}\tf\plusae{\mu}\tg$.
\end{lemma}

\begin{proof}
  From
  Lemma~\thref{l:empty-set-is-negl},
  Lemma~\thref{l:negl-of-summability-domain}, and
  Lemma~\thref{l:m-is-closed-under-add-when-defined},
  $\fp+\gp$ is well-defined, and we have $\fp+\gp\in\calM$.
  let $A\eqdef\Dfsum(f,g)$, $\tf\eqdef f\matUN_A$ and $\tg\eqdef g\matUN_A$.
  Therefore, from
  Lemma~\threfc{l:almost-eq-is-equiv-rel}{transitivity},
  Lemma~\threfc{l:compat-of-almost-eq-with-op}{%
    with the binary operator addition}, and
  Lemma~\thref{l:almost-sum},
  we have $\fp\eqae{\mu}\tf$, $\gp\eqae{\mu}\tg$, and
  $\fp+\gp\eqae{\mu}\tf+\tg\eqae{\mu}f\plusae{\mu}g$.
\end{proof}

\begin{remark}
  Note that from Lemma~\ref{l:almost-sum}, such functions~$\fp$ and~$\gp$
  actually exist in~$\calM$.
\end{remark}

\begin{lemma}[almost sum is sum]
  \label{l:almost-sum-is-sum}
  \mbox{}\hfill
  Let~$(X,\Sigma,\mu)$ be a measure space.\\
  Let~$f,g\in\calM$.
  Assume that $\Dfsum(f,g)=X$.
  Then, $f+g\in\calM$, and we have $f\plusae{\mu}g=f+g$.
\end{lemma}

\begin{proof}
  Direct consequence of
  Definition~\threfc{d:summability-domain}{%
    $f+g$ is well-defined every\-where}, and
  Lemma~\threfc{l:almost-sum}{with $\tf=f$ and $\tg=g$}.
\end{proof}

\begin{lemma}[absolute value is almost definite]
  \label{l:abs-is-almost-definite}
  \mbox{}\hfill
  Let~$(X,\Sigma,\mu)$ be a measure space.\\
  Then, the absolute value, lifted into an operator $\ArFXRFXR$, is almost
  definite:
  \begin{equation}
    \label{e:abs-is-almost-definite}
    \forall f \in \FXRae{\mu},\quad
    |f| \eqae{\mu} 0
    \EQUIV
    f \eqae{\mu} 0.
  \end{equation}
\end{lemma}

\begin{proof}
  Direct consequence of
  Lemma~\threfc{l:definiteness-implies-almost-definiteness}{%
    with the unary operator absolute value},
  Lemma~\threfc{l:compat-of-almost-eq-with-op}{%
    with the unary operator absolute value}, and
  Lemma~\thref{l:abs-in-rbar-is-definite}.
\end{proof}

\begin{lemma}[masking almost nowhere]
  \label{l:masking-almost-nowhere}
  \mbox{}\hfill
  Let~$(X,\Sigma,\mu)$ be a measure space.
  Let~$f\in\calM$.
  Let~$A\in\Sigma$.
  Assume that $\mu(A^c)=0$.
  Then, we have $f\eqae{\mu}f\matUN_A$.
\end{lemma}

\begin{proof}
  Direct consequence of
  Definition~\thref{d:meas},
  Definition~\threfc{d:measurable-space}{$\Sigma$ is a $\sigma$-algebra},
  Definition~\threfc{d:sigma-alg}{closedness under complement},
  \assume{the definition of the indicator function
    ($A\subset\{f=f\matUN_A\}$)},
  \assume{monotonicity of complement
    ($\{f=f\matUN_A\}^c$ is included in $A^c$)},
  Definition~\thref{d:negl-subset}. and
  Definition~\thref{d:prop-almost-satisfied}.
\end{proof}

\begin{lemma}[finite nonnegative part]
  \label{l:finite-nonneg-part}
  \mbox{}\hfill
  Let~$(X,\Sigma,\mu)$ be a measure space.
  Let~$f\in\calM$.
  Let~$A\in\Sigma$.
  Let~$A_f\eqdef A\cap f^{-1}(\matRplus)$.
  Let~$\tf_A\eqdef f\matUN_{A_f}$.
  Then, $A_f\in\Sigma$ and $\tf_A\in\calMR\cap\calMplus$.

  Moreover, assume that~$f$ is $\mu$-almost everywhere finite and
  nonnegative, and that~$\mu(A^c)=0$.
  Then, we have $\mu(A_f^c)=0$, {\ie} $f\eqae{\mu}\tf_A$.
\end{lemma}

\begin{proof}
  Let~$B\eqdef f^{-1}(\matRbarplus)$ and~$C\eqdef f^{-1}(\matR)$.
  Then, from
  \assume{compatibility of inverse image with intersection},
  the definition of~$A_f$, and
  \assume{properties of the indicator function},
  we have
  \begin{equation*}
    f^{-1} (\matRplus) = B \cap C,\quad
    A_f = A \cap B \cap C,
    \AND
    \matUN_{A_f} = \matUN_A \matUN_B \matUN_C.
  \end{equation*}
  Hence, from
  Lemma~\thref{l:equiv-def-of-nonneg-and-nonpos-parts}, and
  Definition~\thref{d:finite-part},
  $\tf_A$ is the finite part of $(f\matUN_A)^+$.
  Therefore, from
  Lemma~\threfc{l:meas-and-masking}{$f\matUN_A\in\calM$},
  Lemma~\threfc{l:meas-of-nonneg-and-nonpos-parts}{%
    with $f\matUN_A$}, and
  Lemma~\threfc{l:mplus-is-closed-under-finite-part}{%
    with $(f\matUN_A)^+$},
  we have $\tf_A\in\calMR\cap\calMplus$.

  Assume now that~$f$ is $\mu$-almost everywhere finite and
  nonnegative, and that~$\mu(A^c)=0$.
  Then, from
  Definition~\thref{d:meas},
  Definition~\threfc{d:measurable-space}{$\Sigma$ is a $\sigma$-algebra},
  Definition~\threfc{d:sigma-alg}{closedness under complement},
  Definition~\thref{d:prop-almost-satisfied}, and
  Lemma~\thref{l:negl-of-meas-subset},
  we have
  \begin{equation*}
    B^c, C^c \in \Sigma \AND \mu (B^c) = \mu (C^c) = 0.
  \end{equation*}
  Hence, from
  Definition~\threfc{d:sigma-alg}{closedness under complement},
  \assume{De~Morgan's laws}, and
  Lemma~\thref{l:meas-satisfies-finite-boole-ineq},
  we have
  \begin{equation*}
    A_f^c \in \Sigma
    \AND
    \mu (A_f^c)
    \leq \mu (A^c) + \mu (B^c) + \mu (C^c)
    = 0.
  \end{equation*}
  Therefore, from
  Definition~\threfc{d:meas}{nonnegativeness}, and
  Lemma~\thref{l:masking-almost-nowhere},
  we have $\mu(A_f^c)=0$ and $f\eqae{\mu}\tf_A$.
\end{proof}

\clearpage
\section{The Lebesgue measure}
\label{s:the-lebesgue-measure}

\begin{remark}
  \label{r:v2-new08}
  This section follows {\Cara}'s extension scheme
  (see Section~\ref{s:caratheodory-extension-scheme}).
\end{remark}

\begin{remark}
  We recall the notation~$\lsrbra\cdot,\cdot\rsrbra$ for not specifying open or
  closed bounds for intervals.
\end{remark}

\begin{definition}[length of interval]
  \label{d:len-of-int}
  \mbox{}\\
  Let~$a,b\in\matR$.
  Assume that $a\leq b$.
  The {\em length of interval} from~$a$ to~$b$ is
  $\ell(\lsrbra a,b\rsrbra)\eqdef b-a$.
\end{definition}

\begin{lemma}[length is nonnegative]
  \label{l:len-is-nonneg}
  \mbox{}\hfill
  Let~$a,b\in\matR$,
  Assume that $a\leq b$.
  Then, $\ell((a,b))\geq0$.
\end{lemma}

\begin{proof}
  Direct consequence of
  Definition~\thref{d:len-of-int}.
\end{proof}

\begin{lemma}[length is homogeneous]
  \label{l:len-is-hom}
  \mbox{}\hfill
  We have $\ell(\emptyset)=0$.
\end{lemma}

\begin{proof}
  Let~$a\in\matR$.
  Then, for instance, we have $\emptyset=(a,a)$.
  Therefore, from
  Definition~\thref{d:len-of-int},
  we have $\ell(\emptyset)=0$.
\end{proof}

\begin{lemma}[length of partition]
  \label{l:len-of-partition}
  \mbox{}\hfill
  Let~$a,b,c\in\matR$.
  Assume that $a\leq b$.
  Then, we have
  \begin{equation}
    \label{e:len-of-partition}
    \ell ((a, b) \cap (c, \infty))
    + \ell ((a, b) \cap (c, \infty)^c)
    = \ell ((a, b)).
  \end{equation}
\end{lemma}

\begin{proof}
  Let~$I\eqdef(a,b)$, and $E\eqdef(c,\infty)$.
  We have $E^c=(-\infty,c]$.

  \proofparskip{Case $c\leq a$}
  Then, we have $I\cap E=I$ and $I\cap E^c=\emptyset$.
  Hence, from
  Lemma~\thref{l:len-is-hom},
  we have $\ell(I\cap E)+\ell(I\cap E^c)=\ell(I)$.

  \proofparskip{Case $a<c<b$}
  Then, we have $I\cap E=(a,c)$ and
  $I\cap E^c=[c,b)$.
  Hence, from
  Definition~\thref{d:len-of-int}, and
  \assume{additive abelian group properties of~$\matR$},
  we have
  \begin{equation*}
    \ell (I \cap E) + \ell (I \cap E^c)
    = \ell ((a, c)) + \ell ([c, b))
    = (c - a) + (b - c)
    = b - a
    = \ell (I).
  \end{equation*}

  \proofparskip{Case $b\leq c$}
  Then, we have $I\cap E=\emptyset$ and $I\cap E^c=I$.
  Hence, from
  Lemma~\thref{l:len-is-hom},
  we have $\ell(I\cap E)+\ell(I\cap E^c)=\ell(I)$.

  Therefore, we always have $\ell(I\cap E)+\ell(I\cap E^c)=\ell(I)$.
\end{proof}

\begin{definition}[set of countable cover with bounded open intervals]
  \label{d:set-of-count-cover-with-bounded-open-int}
  \mbox{}\\
  Let~$A\subset\matR$.
  The {\em set of countable covers of~$A$ with bounded open intervals} is
  \begin{equation}
    \label{e:set-of-count-cover-with-bounded-open-int}
    C_A
    \eqdef \left\{
      ((a_n, b_n))_{n \in \matN} \leftst
      (\forall n \in \matN,\; a_n, b_n \in \matR \Conj a_n \leq b_n)
      \CONJ A \subset \bigcup_{n \in \matN} (a_n, b_n) \right.
    \right\}.
  \end{equation}
\end{definition}

\begin{lemma}[set of countable cover with bounded open intervals is nonempty]
  \label{l:set-of-count-cover-with-bounded-open-int-is-nonempty}
  \mbox{}\\
  Let~$A\subset\matR$.
  Then, we have $C_A\not=\emptyset$.
\end{lemma}

\begin{proof}
  For all $n\in\matN$, let $I_n\eqdef(-n,n)$.
  Then, from
  \assume{the Archimedean property of~$\matR$},
  we have $\matR\subset\bigcup_{n\in\matN}I_n$.
  Therefore, from
  Definition~\thref{d:set-of-count-cover-with-bounded-open-int}, and
  \assume{properties of inclusion},
  we have $(I_n)_{n\in\matN}\in C_A$.
\end{proof}

\begin{definition}[$\lambda^\star$, Lebesgue measure candidate]
  \label{d:lambda-star-lebesgue-meas-cand}
  \mbox{}\\
  The {\em Lebesgue measure candidate} is the function
  $\lambda^\star:\ArPRRbp$ defined by
  \begin{equation}
    \label{e:lambda-star-lebesgue-meas-cand}
    \forall A \subset \matR,\quad
    \lambda^\star (A)
    \eqdef \inf \left\{
      \left. \sum_{n \in \matN} \ell (I_n) \rightst
      (I_n)_{n \in \matN} \in C_A
    \right\}.
  \end{equation}
\end{definition}

\begin{remark}
  Note that~$\lambda^\star$ in the previous definition is not $\sigma$-additive
  on~$\calP(\matR)$, {\eg} see~\cite[Ex. 2.28 pp.~86--87]{gh:mip:13}.
  Thus, it is not a measure on~$\calP(\matR)$.
  In fact, there is no measure on~$\calP(\matR)$ that generalizes the
  length of interval, {\eg} see~\cite[Ex. 2.29 pp.~87--88]{gh:mip:13}.
\end{remark}

\begin{lemma}[$\lambda^\star$ is nonnegative]
  \label{l:lambda-star-is-nonneg}
  \mbox{}\hfill
  Let~$A\subset\matR$.
  Then, we have $\lambda^\star(A)\geq0$.
\end{lemma}

\begin{proof}
  Direct consequence of
  Lemma~\thref{l:len-is-nonneg}, and
  \assume{monotonicity of infimum}.
\end{proof}

\begin{lemma}[$\lambda^\star$ is homogeneous]
  \label{l:lambda-star-is-hom}
  \mbox{}\hfill
  We have $\lambda^\star(\emptyset)=0$.
\end{lemma}

\begin{proof}
  From
  Lemma~\thref{l:lambda-star-is-nonneg},
  we have $\lambda(\emptyset)\geq0$.

  For all $n\in\matN$, let $I_n\eqdef(0,0)=\emptyset$.
  Let~$n\in\matN$.
  Then, from
  Definition~\thref{d:len-of-int}, and
  \assume{field properties of~$\matR$},
  we have $\ell(I_n)=0-0=0$.
  Thus, we have $\sum_{n\in\matN}\ell(I_n)=0$.
  Moreover, we have $\emptyset\subset\bigcup_{n\in\matN}I_n$.
  Thus, we have $(I_n)_{n\in\matN}\in C_\emptyset$.
  Hence, from
  Definition~\thref{d:lambda-star-lebesgue-meas-cand}, and
  Definition~\threfc{LM-d:infimum}{lower bound},
  we have $\lambda^\star(\emptyset)\leq0$.

  Therefore $\lambda^\star(\emptyset)=0$.
\end{proof}

\begin{lemma}[$\lambda^\star$ is monotone]
  \label{l:lambda-star-is-monot}
  \mbox{}\hfill
  Let~$A,B\subset\matR$.
  Assume that $A\subset B$.
  Then, $\lambda^\star(A)\leq\lambda^\star(B)$.
\end{lemma}

\begin{proof}
  Let~$(I_n)_{n\in\matN}\in C_B$.
  Then, from
  Definition~\thref{d:set-of-count-cover-with-bounded-open-int},
  and
  \assume{transitivity of inclusion},
  we have $A\subset B\subset\bigcup_{n\in\matN}I_n$.
  Thus, from
  Definition~\thref{d:set-of-count-cover-with-bounded-open-int},
  we have $(I_n)_{n\in\matN}\in C_A$.
  Hence, from
  Definition~\thref{d:lambda-star-lebesgue-meas-cand}, and
  Definition~\threfc{LM-d:infimum}{lower bound, for~$A$},
  we have $\lambda^\star(A)\leq\sum_{n\in\matN}\ell(I_n)$.

  Therefore, from
  Definition~\threfc{LM-d:infimum}{greatest lower bound, for~$B$}, and
  Definition~\thref{d:lambda-star-lebesgue-meas-cand},
  we have $\lambda^\star(A)\leq\lambda^\star(B)$.
\end{proof}

\begin{lemma}[$\lambda^\star$ is $\sigma$-subadditive]
  \label{l:lambda-star-is-sigma-subadd}
  \mbox{}\hfill
  Let~$(A_n)_{n\in\matN}\subset\matR$.
  Then, $\lambda^\star\left(\bigcup_{n\in\matN}A_n\right)
  \leq\sum_{n\in\matN}\lambda^\star(A_n)$.
\end{lemma}

\begin{proof}
  Let~$A\eqdef\bigcup_{n\in\matN}A_n$.

  \proofparskip{Case $\exists n_0\in\matN$ such that
    $\lambda^\star(A_{n_0})=\infty$}
  Then, from
  \assume{totally ordered set properties of~$\matRbarplus$},
  we have
  $\lambda^\star\left(\bigcup_{n\in\matN}A_n\right)\leq\infty=
  \sum_{n\in\matN}\lambda^\star(A_n)$.

  \proofparskip{Case $\forall n\in\matN$, $\lambda^\star(A_n)<\infty$}
  Let~$\eps>0$.
  Let~$n\in\matN$.
  Then, from
  Definition~\thref{d:lambda-star-lebesgue-meas-cand}, and
  Lemma~\thref{LM-l:finite-infimum},
  there exists $(I_{n,m})_{m\in\matN}\in C_{A_n}$ such that
  $\sum_{m\in\matN}\ell(I_{n,m})<\lambda^\star(A_n)+\frac{\eps}{2^{n+1}}$.
  Let~$\fhi:\ArNNxN$ be a bijection.
  Then, from
  Definition~\thref{d:set-of-count-cover-with-bounded-open-int},
  Lemma~\thref{l:double-count-union},
  Lemma~\thref{l:def-of-double-count-union}, and
  \assume{countability of~$\matN^2$},
  we have
  \begin{equation*}
    A
    = \bigcup_{n \in \matN} A_n
    \subset \bigcup_{n \in \matN} \left( \bigcup_{m \in \matN} I_{n,m} \right)
    = \bigcup_{n, m \in \matN} I_{n, m}
    = \bigcup_{p \in \matN} I_{\fhi (p)}.
  \end{equation*}
  Then, from
  Definition~\thref{d:set-of-count-cover-with-bounded-open-int},
  $(I_{\fhi(n)})_{n\in\matN}$ belongs to~$C_A$, and thus, from
  Definition~\thref{d:lambda-star-lebesgue-meas-cand},
  Definition~\threfc{LM-d:infimum}{lower bound},
  Lemma~\thref{l:def-of-double-series-in-rbarplus},
  Lemma~\thref{l:double-series-in-rbarplus}, and
  \assume{additive properties of~$\matRplus$},
  we have
  \begin{equation*}
    \lambda^\star (A)
    \leq \sum_{p \in \matN} \ell (I_{\fhi (p)})
    = \sum_{n, m \in \matN} \ell (I_{n, m})
    = \sum_{n \in \matN} \left(
      \sum_{m \in \matN} \ell (I_{n, m}) \right)
    \leq \sum_{n \in \matN} \lambda^\star (A_n) + \eps.
  \end{equation*}
  Hence, from
  \assume{monotonicity of the limit (when $\eps\to0^+$)},
  we have $\lambda^\star(A)\leq\sum_{n\in\matN}\lambda^\star(A_n)$.

  Therefore, we always have
  $\lambda^\star(A)\leq\sum_{n\in\matN}\lambda^\star(A_n)$.
\end{proof}

\begin{lemma}[$\lambda^\star$ generalizes length of interval]
  \label{l:lambda-star-gen-len-of-int}
  \mbox{}\\
  Let~$a,b\in\matR$.
  Assume that $a\leq b$.
  Then, we have $\lambda^\star(\lsrbra a,b\rsrbra)=b-a$.
\end{lemma}

\begin{proof}
  Let~$a,b\in\matR$.
  Assume that $a\leq b$.

  \proofparskip{(1). Closed interval}
  Let~$\eps>0$.
  Then, we have $[a,b]\subset(a-\frac{\eps}{2},b+\frac{\eps}{2})$.
  Thus, from
  Definition~\thref{d:set-of-count-cover-with-bounded-open-int},
  Definition~\thref{d:lambda-star-lebesgue-meas-cand},
  Definition~\threfc{LM-d:infimum}{lower bound},
  Lemma~\thref{l:lambda-star-is-hom},
  Definition~\thref{d:len-of-int}, and
  \assume{ordered field properties of~$\matR$},
  the sequence $((a-\frac{\eps}{2},b+\frac{\eps}{2}), \emptyset, \ldots)$
  belongs to~$C_{[a,b]}$, and
  $\lambda^\star([a,b])\leq\ell(a-\frac{\eps}{2},b+\frac{\eps}{2})=b-a+\eps$.
  Hence, from
  \assume{monotonicity of the limit (when $\eps\to0^+$)},
  we have $\lambda^\star([a,b])\leq b-a$.

  Let~$((a_n,b_n))_{n\in\matN}\in C_{[a,b]}$ (with $\forall n\in\matN$,
  $a_n\leq b_n$).
  Then, from
  Definition~\thref{d:set-of-count-cover-with-bounded-open-int}, and
  Lemma~\thref{l:finite-cover-of-compact-int},
  there exists $q\in\matN$ and $(i_p)_{p\in[0..q]}\in\matN$ pairwise distinct
  such that $[a,b]\subset\bigcup_{p\in[0..q]}(a_{i_p},b_{i_p})$ with
  $a_{i_0}<a$, $b<b_{i_q}$, and for all $p\in[0..q-1]$, $a_{i_{p+1}}<b_{i_p}$.
  Then, from
  \assume{ordered field properties of~$\matR$},
  Definition~\thref{d:len-of-int}, and
  \assume{totally ordered set properties of~$\matRbarplus$},
  we have
  \begin{align*}
    b - a
    < b_{i_q} - a_{i_0}
    \leq b_{i_q}
    + \sum_{p \in [0..q-1]} (- a_{i_{p + 1}} + b_{i_p}) - a_{i_0}
    & = \sum_{p \in [0..q]} (b_{i_p} - a_{i_p})\\
    & = \sum_{p \in [0..q]} \ell ((a_{i_p}, b_{i_p}))
      \leq \sum_{n \in \matN} \ell ((a_n, b_n)).
  \end{align*}
  Hence, from
  Definition~\thref{d:lambda-star-lebesgue-meas-cand}, and
  Definition~\threfc{LM-d:infimum}{greatest lower bound},
  we have $b-a\leq\lambda^\star([a,b])$.

  Therefore, we have $\lambda^\star([a,b])=b-a$.

  \proofparskip{(2). Open interval}
  Let~$\eps>0$.
  Assume that $\eps<b-a$.
  Then, we have
  \begin{equation*}
    \left[ a + \frac{\eps}{2}, b - \frac{\eps}{2} \right]
    \subset (a, b)
    \subset [a, b].
  \end{equation*}
  Hence, from~(1), and
  Lemma~\thref{l:lambda-star-is-monot},
  we have $b-a-\eps\leq\lambda^\star((a,b))\leq b-a$.
  Therefore, from
  \assume{monotonicity of the limit (when $\eps\to0^+$)},
  we have $\lambda^\star((a,b))=b-a$.

  \proofparskip{(3). Left-open right-closed interval}
  Let~$\eps>0$.
  Assume that $\eps<b-a$.
  Then, we have
  \begin{equation*}
    [a + \eps, b] \subset (a, b] \subset [a, b].
  \end{equation*}
  Hence, from~(1), and
  Lemma~\thref{l:lambda-star-is-monot},
  we have $b-a-\eps\leq\lambda^\star((a,b])\leq b-a$.
  Therefore, from
  \assume{monotonicity of the limit (when $\eps\to0^+$)},
  we have $\lambda^\star((a,b])=b-a$.

  \proofparskip{(4). Left-closed right-open interval}
  Let~$\eps>0$.
  Assume that $\eps<b-a$.
  Then, we have
  \begin{equation*}
    (a, b - \eps] \subset [a, b) \subset [a, b].
  \end{equation*}
  Hence, from~(1), and
  Lemma~\thref{l:lambda-star-is-monot},
  we have $b-a-\eps\leq\lambda^\star([a,b))\leq b-a$.
  Therefore, from
  \assume{monotonicity of the limit (when $\eps\to0^+$)},
  we have $\lambda^\star([a,b))=b-a$.
\end{proof}

\begin{remark}
  \mbox{}\\
  Note that similar proofs using monotonicity provide for all $a\in\matR$,
  $\lambda^\star((-\infty,a\rsrbra)=\lambda^\star(\lsrbra a,\infty))=\infty$.
\end{remark}

\begin{definition}[$\calL$, Lebesgue $\sigma$-algebra]
  \label{d:l-lebesgue-sigma-alg}
  \mbox{}\hfill
  The set~$\calL$ of subsets of~$\matR$ defined by
  \begin{equation}
    \label{e:l-lebesgue-sigma-alg}
    \calL
    \eqdef \left\{
      E \subset \matR \st
      \forall A \subset \matR,\;
      \lambda^\star (A)
      = \lambda^\star (A \cap E) + \lambda^\star (A \cap E^c)
    \right\}
  \end{equation}
  is called the {\em Lebesgue $\sigma$-algebra}.
\end{definition}

\begin{remark}
  The set~$\calL$ is shown below to be a $\sigma$-algebra;
  hence its name.
\end{remark}

\begin{lemma}[equivalent definition of~$\calL$]
  \label{l:equiv-def-of-l}
  \mbox{}\hfill
  We have
  \begin{equation}
    \label{e:equiv-def-of-l}
    \calL
    = \left\{
      E \subset \matR \st
      \forall A \subset \matR,\;
      \lambda^\star (A \cap E) + \lambda^\star (A \cap E^c)
      \leq \lambda^\star (A)
    \right\}.
  \end{equation}
\end{lemma}

\begin{proof}
  Direct consequence of
  Lemma~\threfc{l:compat-of-pseudopart-with-inter}{%
    with $\card(I)$ equals~2},
  Lemma~\thref{l:lambda-star-is-sigma-subadd},
  Lemma~\threfc{l:order-in-rbar-is-total}{antisymmetry}, and
  Definition~\thref{d:l-lebesgue-sigma-alg}.
\end{proof}

\begin{lemma}[$\calL$ is closed under complement]
  \label{l:l-is-closed-under-compl}
  \mbox{}\hfill
  $\calL$ is closed under complement.
\end{lemma}

\begin{proof}
  Let~$E\in\calL$.
  Let~$A\subset\matR$.
  Then, from
  \assume{the double complement law},
  Lemma~\thref{l:add-in-rbarplus-is-comm}, and
  Definition~\thref{d:l-lebesgue-sigma-alg},
  we have
  \begin{equation*}
    \lambda^\star (A \cap E^c) + \lambda^\star (A \cap (E^c)^c)
    = \lambda^\star (A \cap E^c) + \lambda^\star (A \cap E)
    = \lambda^\star (A \cap E) + \lambda^\star (A \cap E^c)
    = \lambda^\star(A).
  \end{equation*}
  Therefore, $E^c\in\calL$.
\end{proof}

\begin{lemma}[$\calL$ is closed under finite union]
  \label{l:l-is-closed-under-finite-union}
  \mbox{}\hfill
  $\calL$~is closed under finite union.
\end{lemma}

\begin{proof}
  For all $n\in[2..\infty)$, let $P(n)$ be the property:
  $\forall(E_i)_{i\in[1..n]}\in\calL$, $\bigcup_{i\in[1..n]}E_i\in\calL$.

  \proofparskip{Induction: $P(2)$}
  Let~$E_1,E_2\in\calL$.
  Let~$E\eqdef E_1\cup E_2$.\\
  Let~$A\subset\matR$.
  Then, from
  \assume{properties of intersection, union and complement},
  we have
  \begin{equation*}
    A \cap E
    = A \cap (E_1 \uplus (E_1^c \cap E_2))
    = (A \cap E_1) \cup (A \cap (E_1^c \cap E_2)).
  \end{equation*}
  Thus, from
  Lemma~\thref{l:lambda-star-is-sigma-subadd},
  \assume{associativity of intersection}, and
  Definition~\threfc{d:l-lebesgue-sigma-alg}{%
    $E_2$, then~$E_1$ belong to~$\calL$},
  we have
  \begin{align*}
    \lambda^\star (A \cap E) + \lambda^\star (A \cap E^c)
    & = \lambda^\star ((A \cap E_1) \cup (A \cap (E_1^c \cap E_2)))
    + \lambda^\star (A \cap (E_1^c \cap E_2^c))\\
    & \leq \lambda^\star (A \cap E_1)
    + \lambda^\star ((A \cap E_1^c) \cap E_2)
    + \lambda^\star ((A \cap E_1^c) \cap E_2^c)\\
    & = \lambda^\star (A \cap E_1) + \lambda^\star (A \cap E_1^c)\\
    & = \lambda^\star (A).
  \end{align*}
  Hence, from
  Lemma~\thref{l:equiv-def-of-l},
  we have $E\in\calL$.

  \proofparskip{Induction; $P(n)$ implies $P(n+1)$}
  Let~$n\in[2..\infty)$.
  Assume that $P(n)$ holds.\\
  Let~$(E_i)_{i\in[1..n+1]}\in\calL$.
  Let~$E\eqdef\bigcup_{i\in[1..n]}E_i$.
  Then, from
  $P(n)$,
  we have $E\in\calL$, and from
  \assume{associativity of union}, and
  $P(2)$,
  we have $\bigcup_{i\in[1..n+1]}E_i=E\cup E_{n+1}\in\calL$.

  Therefore, $P(n)$ holds for all $n\in[2..\infty)$.
\end{proof}

\begin{lemma}[$\calL$ is closed under finite intersection]
  \label{l:l-is-closed-under-finite-inter}
  \mbox{}\hfill
  $\calL$~is closed under finite intersection.
\end{lemma}

\begin{proof}
  Direct consequence of
  \assume{De~Morgan's laws},
  Lemma~\thref{l:l-is-closed-under-compl}, and
  Lemma~\thref{l:l-is-closed-under-finite-union}.
\end{proof}

\begin{lemma}[$\calL$ is set algebra]
  \label{l:l-is-set-alg}
  \mbox{}\hfill
  $\calL$~is a set algebra on~$\matR$.
\end{lemma}

\begin{proof}
  Let~$A\subset\matR$.
  Then, from
  \assume{properties of intersection}, and
  Lemma~\thref{l:lambda-star-is-hom},
  we have
  $\lambda^\star(A\cap\matR)+\lambda^\star(A\cap\matR^c)
  =\lambda^\star(A)+\lambda^\star(\emptyset)
  =\lambda^\star(A)$.
  Hence, $\matR\in\calL$.

  Therefore, from
  Lemma~\thref{l:l-is-closed-under-compl},
  Lemma~\thref{l:l-is-closed-under-finite-inter},
  Lemma~\threfc{l:inter-set-diff-equiv}{%
    $\calL$ is closed under set difference},
  Lemma~\thref{l:other-equiv-def-of-set-alg},
  $\calL$~is a set algebra on~$\matR$.
\end{proof}

\begin{lemma}[$\lambda^\star$ is additive on~$\calL$]
  \label{l:lambda-star-is-add-on-l}
  \mbox{}\hfill
  Let~$n\in[2..\infty)$.
  Let~$(E_i)_{i\in[1..n]}\in\calL$.\\
  Assume that the~$E_i$'s are pairwise disjoint.
  Then, $\lambda^\star$~is additive on~$\calL$:
  \begin{equation}
    \label{e:lambda-star-is-add-on-l}
    \forall A \subset \matR,\quad
    \lambda^\star \left( A \cap \biguplus_{i \in [1..n]} E_i \right)
    = \sum_{i \in [1..n]} \lambda^\star (A \cap E_i).
  \end{equation}
\end{lemma}

\begin{proof}
    For all $n\in[2..\infty)$, let $P(n)$ be the property:
    for all $(E_i)_{i\in[1..n]}\in\calL$,
  \begin{equation*}
    (\forall i, j \in [1..n],\;
    i \not= j \Implies E_i \cap E_j = \emptyset)
    \IMPLIES
    \lambda^\star \left( A \cap \biguplus_{i \in [1..n]} E_i \right)
    = \sum_{i \in [1..n]} \lambda^\star (A \cap E_i).
  \end{equation*}

  \proofparskip{Induction: $P(2)$}
  Let~$E_1,E_2\in\calL$.
  Assume that $E_1\cap E_2=\emptyset$.
  Let~$A\subset\matR$.
  Then, from
  \assume{properties of intersection, union, and complement},
  Lemma~\threfc{l:compat-of-pseudopart-with-inter}{%
    with $\card(I)=2$}, and
  Definition~\threfc{d:l-lebesgue-sigma-alg}{$E_1\in\calL$},
  we have
  \begin{align*}
    \lambda^\star (A \cap (E_1 \uplus E_2))
    & = \lambda^\star ((A \cap E_1) \uplus (A \cap E_2))\\
    & = \lambda^\star ([(A \cap E_1) \uplus (A \cap E_2)] \cap E_1)
    + \lambda^\star ([(A \cap E_1) \uplus (A \cap E_2)] \cap E_1^c)\\
    & = \lambda^\star (A \cap E_1) + \lambda^\star (A \cap E_2).
  \end{align*}

  \proofparskip{Induction: $P(n)$ implies $P(n+1)$}
  Let~$n\in[2..\infty)$.
  Assume that $P(n)$ holds.\\
  Let~$(E_i)_{i\in[1..n+1]}\in\calL$.
  Assume that for all $i,j\in[1..n+1]$, $i\not=j$ implies
  $E_i\cap E_j=\emptyset$.
  Let~$E$ be the disjoint union~$\biguplus_{i\in[1..n]}E_i$.
  Then, from
  \assume{properties of intersection, union, and complement},
  $P(2)$, and
  $P(n)$,
  we have $E\cap E_{n+1}=\emptyset$, and
  \begin{equation*}
    \lambda^\star (A \cap (E \uplus E_{n+1}))
    = \lambda^\star (A \cap E) + \lambda^\star (A \cap E_{n+1})
    = \sum_{i \in [1..n]} \lambda^\star (A \cap E_i)
    + \lambda^\star (A \cap E_{n+1}).
  \end{equation*}

  Therefore, $P(n)$ holds for all $n\in[2..\infty)$.
\end{proof}

\begin{remark}
  The additive property of~$\lambda^\star$ {\perse} actually corresponds to the
  case $A=\matR$.
\end{remark}

\begin{lemma}[$\lambda^\star$ is $\sigma$-additive on~$\calL$]
  \label{l:lambda-star-is-sigma-add-on-l}
  \mbox{}\hfill
  $\lambda^\star$ is $\sigma$-additive on~$(\matR,\calL)$.
\end{lemma}

\begin{proof}
  Let~$(E_n)_{n\in\matN}\in\calL$.
  Assume that for all $i,j\in\matN$, $i\not=j$ implies
  $E_i\cap E_j=\emptyset$.\\
  Let~$E\eqdef\bigcup_{n\in\matN}E_n$.
  Let~$n\in\matN$.
  Then, from
  Lemma~\threfc{l:lambda-star-is-add-on-l}{with $A\eqdef\matR$}, and
  Lemma~\thref{l:lambda-star-is-monot},
  we have
  \begin{equation*}
    \sum_{i \in [0..n]} \lambda^\star (E_i)
    = \lambda^\star \left( \bigcup_{i \in [0..n]} E_i \right)
    \leq \lambda^\star (E).
  \end{equation*}
  Thus, from
  \assume{monotonicity of the limit},
  we have
  \begin{equation*}
    \sum_{n \in \matN} \lambda^\star (E_n) \leq \lambda^\star (E).
  \end{equation*}
  Hence, from
  Lemma~\thref{l:lambda-star-is-sigma-subadd},
  we have
  \begin{equation*}
    \lambda^\star (E) = \sum_{n \in \matN} \lambda^\star (E_n).
  \end{equation*}

  Therefore, from
  Definition~\thref{d:sigma-add-over-meas-space},
  $\lambda^\star$ is $\sigma$-additive over $(\matR,\calL)$.
\end{proof}

\begin{lemma}[partition of countable union in~$\calL$]
  \label{l:part-of-count-union-in-l}
  \mbox{}\\
  Let~$(E_n)_{n\in\matN}\in\calL$.
  Let~$F_0\eqdef E_0$, and for all $n\in\matN$, let
  $F_{n+1}\eqdef E_{n+1}\setminus\bigcup_{i\in[0..n]}F_i$.
  Then,
  \begin{align}
    \label{e:part-of-count-union-in-l-1}
    \forall n \in \matN,\quad & F_n \in \calL,\\
    \label{e:part-of-count-union-in-l-2}
    \forall m, n \in \matN,\quad &
    m \not= n \IMPLIES F_m \cap F_n = \emptyset,\\
    \label{e:part-of-count-union-in-l-3}
    \forall n \in \matN,\quad &
    \bigcup_{i \in [0..n]} E_i = \biguplus_{i \in [0..n]} F_i.
  \end{align}
\end{lemma}

\begin{proof}
  Direct consequence of
  Lemma~\thref{l:l-is-set-alg}, and
  Lemma~\thref{l:part-of-count-union-in-set-alg}.
\end{proof}

\begin{lemma}[$\calL$ is closed under countable union]
  \label{l:l-is-closed-under-count-union}
  \mbox{}\hfill
  $\calL$ is closed under countable union.
\end{lemma}

\begin{proof}
  Let~$I\subset\matN$.
  Let~$(E_i)_{i\in I}\in\calL$.
  Let~$E\eqdef\bigcup_{i\in I}E_i$.

  \proofparskip{Case $\card(I)<\infty$}
  Then, from
  Lemma~\thref{l:l-is-closed-under-finite-union},
  we have $E\in\calL$.

  \proofparskip{Case $\card(I)=\infty$}
  Let~$\fhi:\ArNI$ be a bijection.
  Then, from
  \assume{associativity of union},
  we have $E=\bigcup_{n\in\matN}E_{\fhi(n)}$.
  Let~$F_0\eqdef E_{\fhi(0)}$, and for all $n\in\matN$,
  let $F_{n+1}\eqdef E_{\fhi(n+1)}\setminus\bigcup_{i\in[0..n]}F_i$.

  Let~$n\in\matN$.
  Then, from
  Lemma~\threfc{l:part-of-count-union-in-l}{%
    $\biguplus_{i\in[0..n]}F_i$ is equal to
    $\bigcup_{i\in[0..n]}E_{\fhi(i)}$},
  \assume{properties of union}, and
  \assume{monotonicity of complement},
  we have
  \begin{equation*}
    E^c \subset F^c
    \quad\mbox{where } F \eqdef \biguplus_{i \in [0..n]} F_i.
  \end{equation*}
  Let~$A\subset\matR$.
  Then, from
  Lemma~\threfc{l:lambda-star-is-monot}{%
    with $A\cap E^c\subset A\cap F^c$},
  Lemma~\threfc{l:part-of-count-union-in-l}{%
    the $F_i$'s are disjoint in~$\calL$},
  Lemma~\thref{l:lambda-star-is-add-on-l},
  Lemma~\threfc{l:l-is-closed-under-finite-union}{$F\in\calL$}, and
  Definition~\threfc{d:l-lebesgue-sigma-alg}{with~$F$},
  we have
  \begin{align*}
    \sum_{i \in [0..n]} \lambda^\star (A \cap F_i)
    + \lambda^\star (A \cap E^c)
    & \leq \sum_{i \in [0..n]} \lambda^\star (A \cap F_i)
      + \lambda^\star (A \cap F^c)\\
    & = \lambda^\star (A \cap F) + \lambda^\star (A \cap F^c)
      = \lambda^\star (A).
  \end{align*}
  Thus, from
  Lemma~\thref{l:lambda-star-is-sigma-subadd},
  \assume{distributivity of intersection over union}, and
  \assume{monotonicity of the limit},
  we have
  \begin{equation*}
    \lambda^\star \left( A \cap \bigcup_{n \in \matN} F_n \right)
    + \lambda^\star (A \cap E^c)
    \leq \sum_{n \in \matN} \lambda^\star (A \cap F_n)
    + \lambda^\star (A \cap E^c)
    \leq \lambda^\star (A).
  \end{equation*}
  Hence, since
  \assume{the limit is a function},
  we have $\bigcup_{n\in\matN}F_n=\bigcup_{n\in\matN}E_{\fhi(n)}=E$, and
  \begin{equation*}
    \lambda^\star (A \cap E) + \lambda^\star (A \cap E^c)
    \leq \lambda^\star (A).
  \end{equation*}

  Therefore, from
  Lemma~\thref{l:equiv-def-of-l},
  we have $E\in\calL$.
\end{proof}

\begin{lemma}[rays are Lebesgue-measurable]
  \label{l:rays-are-lebesgue-meas}
  \mbox{}\hfill
  Let $a\in\matR$.
  Then, we have
  \begin{equation}
    \label{e:rays-are-lebesgue-meas}
    (a, \infty),\; (-\infty, a],\; (-\infty, a),\; [a, \infty),\; \in \calL.
  \end{equation}
\end{lemma}

\begin{proof}
  \proofpar{(1). $(a,\infty)\in\calL$}\\
  Let~$A\subset\matR$.
  Let~$\alpha
  \eqdef\lambda^\star(A\cap(a,\infty))+\lambda^\star(A\cap(-\infty,a])$.
  Let us show that $\alpha\leq\lambda^\star(A)$.\\
  \proofpar{Case $\lambda^\star(A)=\infty$} Trivial.
  \proofpar{Case $\lambda^\star(A)$ finite}
  Let~$\eps>0$.
  From
  Definition~\thref{d:lambda-star-lebesgue-meas-cand}, and
  Lemma~\thref{LM-l:finite-infimum},
  let $(I_n)_{n\in\matN}\in C_A$ such that
  \begin{equation*}
    \lambda^\star (A)
    \leq \sum_{n \in \matN} \ell (I_n)
    \leq \lambda^\star (A) + \eps.
  \end{equation*}
  For all $n\in\matN$, let $\Ip_n\eqdef I_n\cap(a,\infty)$ and
  $\Ipp_n\eqdef I_n\cap(-\infty,a]$.
  Let~$n\in\matN$.
  From
  Lemma~\thref{l:int-are-closed-under-finite-inter},
  $\Ip_n$ and~$\Ipp_n$ are intervals (possibly empty).
  Then, from
  Lemma~\thref{l:lambda-star-gen-len-of-int}, and
  since~$\Ip_n$ and~$\Ipp_n$ are contiguous,
  we have
  $\lambda^\star(\Ip_n)+\lambda^\star(\Ipp_n)
  =\ell(\Ip_n)+\ell(\Ipp_n)=\ell(I_n)$.
  Hence, from
  Lemma~\thref{l:lambda-star-is-monot}, and
  Lemma~\thref{l:lambda-star-is-sigma-subadd},
  we have
  \begin{equation*}
    \alpha
    \leq \lambda^\star \left( \bigcup_{n \in \matN} \Ip_n \right)
    + \lambda^\star \left( \bigcup_{n \in \matN} \Ipp_n \right)
    \leq \sum_{n \in \matN} (\lambda^\star (\Ip_n) + \lambda^\star (\Ipp_n))
    = \sum_{n \in \matN} \ell (I_n)
    \leq \lambda^\star (A) + \eps.
  \end{equation*}
  Therefore, from
  \assume{monotonicity of the limit (when $\eps\to0^+$)},
  we have $\alpha\leq\lambda^\star(A)$, and from
  Lemma~\thref{l:equiv-def-of-l},
  we have $(a,\infty)\in\calL$.

  \proofparskip{(2). $(-\infty,a]\in\calL$}\\
  Direct consequence of
  Lemma~\thref{l:l-is-closed-under-compl}, and~(1).

  \proofparskip{(3). $(-\infty,a)\in\calL$}\\
  Direct consequence of
  Lemma~\thref{l:l-is-closed-under-count-union}, and~(2) with
  \begin{equation*}
    (-\infty, a)
    = \bigcup_{n \in \matN} \left( -\infty, a - \frac{1}{n + 1} \right].
  \end{equation*}

  \proofparskip{(4). $[a,\infty)\in\calL$}\\
  Direct consequence of
  Lemma~\thref{l:l-is-closed-under-compl}, and~(3).
\end{proof}

\begin{lemma}[intervals are Lebesgue-measurable]
  \label{l:int-are-lebesgue-meas}
  \mbox{}\\
  Let $a,b\in\matR$.
  Assume that $a\leq b$.
  Then, we have
  \begin{equation}
    \label{e:int-are-lebesgue-meas}
    (a, b),\; [a, b],\; [a, b),\; (a, b]\; \in \calL.
  \end{equation}
\end{lemma}

\begin{proof}
  Direct consequence of
  Lemma~\thref{l:l-is-closed-under-finite-inter}, and
  Lemma~\thref{l:rays-are-lebesgue-meas} with
  \begin{gather*}
    (a, b) = (-\infty, b) \cap (a, \infty),\quad
    [a, b] = (-\infty, b] \cap [a, \infty),\\
    [a, b) = (-\infty, b) \cap [a, \infty),\quad
    (a, b] = (-\infty, b] \cap (a, \infty).
  \end{gather*}
\end{proof}

\begin{lemma}[$\calL$ is $\sigma$-algebra]
  \label{l:l-is-sigma-alg}
  \mbox{}\hfill
  $(\matR,\calL)$ is a measurable space.
\end{lemma}

\begin{proof}
  Direct consequence of
  Lemma~\thref{l:l-is-set-alg},
  Definition~\thref{d:set-alg},
  Lem\-ma~\thref{l:l-is-closed-under-count-union}, and
  Definition~\thref{d:sigma-alg}.
\end{proof}

\begin{lemma}[$\lambda^\star$ is measure on~$\calL$]
  \label{l:lambda-star-is-meas-on-l}
  \mbox{}\hfill
  $(\matR,\calL,\lambda^\star_{|\calL})$ is a measure space.
\end{lemma}

\begin{proof}
  Direct consequence of
  Lemma~\thref{l:l-is-sigma-alg},
  Lemma~\thref{l:lambda-star-is-nonneg},
  Lem\-ma~\thref{l:lambda-star-is-hom},
  Lemma~\thref{l:lambda-star-is-sigma-add-on-l}, and
  Definition~\thref{d:meas}.
\end{proof}

\begin{lemma}[$\calBR$ is sub-$\sigma$-algebra of~$\calL$]
  \label{l:br-is-sub-sigma-alg-of-l}
  \mbox{}\hfill
  We have $\calBR\subset\calL$.
\end{lemma}

\begin{proof}
  Direct consequence of
  Lemma~\thref{l:sigma-alg-gen-is-monot},
  Lemma~\thref{l:borel-sigma-alg-of-r}, and
  Lemma~\thref{l:int-are-lebesgue-meas}.
\end{proof}

\begin{lemma}[$\lambda^\star$ is measure on~$\calBR$]
  \label{l:lambda-star-is-meas-on-br}
  \mbox{}\hfill
  $(\matR,\calBR,\lambda^\star_{|\calBR})$ is a measure space.
\end{lemma}

\begin{proof}
  Direct consequence of
  Lemma~\thref{l:lambda-star-is-meas-on-l}, and
  Lemma~\thref{l:br-is-sub-sigma-alg-of-l}.
\end{proof}

\begin{remark}
  \label{r:v2-new09}
  See the sketch of next proof in
  Section~\ref{s:sketch-of-the-proof-of-caratheodory-ext-th}.
\end{remark}

\begin{theorem}[{\Cara}, Lebesgue measure on~$\matR$]
  \label{t:caratheodory-lebesgue-meas-on-r}
  \mbox{}\\
  There exists a unique measure on~$(\matR,\calBR)$ that generalizes the
  length of bounded open intervals.

  This measure is denoted~$\lambda\eqdef\lambda^\star_{|\calBR}$;
  it is called the {\em {\BoL} measure on Borel subsets (of~$\matR$)}.
\end{theorem}

\begin{proof}
  \proofparskip{Existence}
  Direct consequence of
  Lemma~\thref{l:lambda-star-is-meas-on-br}, and
  Lemma~\threfc{l:lambda-star-gen-len-of-int}{%
    for open intervals}.

  \proofparskip{Uniqueness}\\
  Let~$G\eqdef\{[a,b)\}_{a<b}$.
  Then, we have $G\subset\calP(\matR)$ and $G\not=\emptyset$.
  Hence, from
  Lemma~\thref{l:int-are-closed-under-finite-inter}, and
  Definition~\thref{d:p-syst},
  $G$~is a $\pi$-system.\\
  From
  Lemma~\threfc{l:borel-sigma-alg-of-r}{%
    $\{(a,b)\}_{a<b}$ and $\{[a,b)\}_{a<b}$ generate $\calBR$},
  Lemma~\threfc{l:gen-sigma-alg-is-min}{generators belong to $\calBR$},
  we have
  \begin{equation*}
    \left( \forall a, b \in \matR,\quad
    a < b \IMPLIES (a, b), [a, b) \in \calBR \right)
    \AND
    \Sigma_\matR (G) = \calBR.
  \end{equation*}
  For all $n\in\matN$, let $I_n\eqdef[n,n+1)$.
  Then, $(I_n)_{n\in\matN}$ is obviously pairwise disjoint, and from
  \assume{the Archimedean property of~$\matR$},
  we have $\matR=\biguplus_{n\in\matN}I_n$.
  Let~$n\in\matN$.
  Then, $I_n\in G$, and from
  Lemma~\thref{l:lambda-star-gen-len-of-int},
  we have $\lambda^\star(I_n)=1<\infty$.

  Let~$\mu$ be a measure on~$\calBR$ such that,
  for all $a,b\in\matR$, $a<b$ implies $\mu((a,b))=b-a$.\\
  Let~$a,b\in\matR$.
  Assume that~$a<b$.
  For all~$p\in\matN$, let $A_p\eqdef\left(a-\frac{1}{p+1},b\right)$.
  Then, from
  \assume{the non\-increasing property of $\left(p\mapsto\frac{1}{p+1}\right)$
    in $\matRplus$ and its limit~0 when $p\to\infty$},
  \assume{the definition of inclusion},
  \assume{additive group properties of~$\matR$},
  \assume{linearity of the limit}, and
  Lemma~\threfc{l:meas-is-cont-from-above}{%
    with $\mu_1$ and $(A_p)_{p\in\matN}$},
  the sequence~$(A_p)_{p\in\matN}$ is nonincreasing, for all $p\in\matN$, we
  have $\mu_1(A_p)=b-a+\frac{1}{p+1}<\infty$, and
  \begin{equation*}
    \mu_1 ([a, b))
    = \mu_1 \left( \bigcap_{p \in \matN} A_p \right)
    = \inf_{p \in \matN} \mu_1 (A_p)
    = \lim_{p \to \infty} \left( b - a + \frac{1}{p + 1} \right)
    = b - a.
  \end{equation*}
  Hence, from
  Lemma~\thref{l:lambda-star-gen-len-of-int},
  $\mu$ and~$\lambda^\star_{|\calBR}$ coincide on~$G$.

  Therefore, from
  Lemma~\threfc{l:uniq-of-meas-ext-from-p-syst}{%
    with $X\eqdef\matR$, $\Sigma\eqdef\calBR$, $\mu_1\eqdef\lambda^\star$,
    $\mu_2=\mu$, and $X_n\eqdef I_n$},
  we have, for all $A\in\calBR$, $\mu(A)=\lambda^\star_{|\calBR}(A)$.
\end{proof}

\begin{remark}
  Note that $\neglset(\matR,\calL,\lambda^\star)=
  \neglset(\matR,\calBR,\lambda)\subset\calL$,
  {\eg} see~\cite[Ex. 2.33 pp.~92--94]{gh:mip:13}.
\end{remark}

\begin{lemma}[Lebesgue measure generalizes length of interval]
  \label{l:lebesgue-meas-gen-len-of-int}
  \mbox{}\\
  Let~$a,b\in\matR$.
  Assume that $a\leq b$.
  Then, we have $\lambda(\lsrbra a,b\rsrbra)=b-a$.
\end{lemma}

\begin{proof}
  Direct consequence of
  Theorem~\thref{t:caratheodory-lebesgue-meas-on-r}, and
  Lem\-ma~\thref{l:lambda-star-gen-len-of-int}.
\end{proof}

\begin{remark}
  Note that similarly to~$\lambda^\star$, we can prove for all $a\in\matR$,
  $\lambda((-\infty,a\rsrbra)=\lambda(\lsrbra a,\infty))=\infty$.
\end{remark}

\begin{lemma}[Lebesgue measure is $\sigma$-finite]
  \label{l:lebesgue-meas-is-sigma-finite}
  \mbox{}\hfill
  The measure space~$(\matR,\calBR,\lambda)$ is $\sigma$-finite.
\end{lemma}

\begin{proof}
  Direct consequence of
  Theorem~\thref{t:caratheodory-lebesgue-meas-on-r},
  Lemma~\threfc{l:lambda-star-gen-len-of-int}{%
    with $b=-a\in\matN$},
  \assume{the Archimedean property of~$\matR$}, and
  Definition~\threfc{d:sigma-finite-meas}{with $A_n\eqdef(-n,n)$}.
\end{proof}

\begin{lemma}[Lebesgue measure is diffuse]
  \label{l:lebesgue-meas-is-diffuse}
  \mbox{}\hfill
  The measure space~$(\matR,\calBR,\lambda)$ is diffuse.
\end{lemma}

\begin{proof}
  Direct consequence of
  Theorem~\thref{t:caratheodory-lebesgue-meas-on-r},
  Lemma~\threfc{l:lambda-star-gen-len-of-int}{%
    with $b=a$}, and
  Definition~\threfc{d:diffuse-meas}{$\{a\}=[a,a]$}.
\end{proof}

\chapter{Integration of nonnegative functions}
\label{c:integration-of-nonnegative-functions}

\minitoc

\begin{remark}
  \label{r:v2-new10}
  \mbox{}\\
  The first three sections of this chapter follow steps~1 to~3 of the Lebesgue
  scheme (see Section~\ref{s:lebesgue-scheme}).
\end{remark}

\section{Integration of indicator functions}
\label{s:integration-of-indicator-functions}

\begin{remark}
  In this section, functions take their values in~$\{0,1\}$, and the
  expressions involving integrals are taken in~$\matRbarplus$.
\end{remark}

\begin{definition}[$\calIF$, set of measurable indicator functions]
  \label{d:if-set-of-meas-indic-funs}
  \mbox{}\\
  Let~$(X,\Sigma)$ be a measurable space.
  The {\em set of measurable indicator functions} is denoted
  $\calIF(X,\Sigma)$ (or simply~$\calIF$);
  it is defined by $\calIF(X,\Sigma)\eqdef\{\matUN_A\st A\in\Sigma\}$.
\end{definition}

\begin{lemma}[indicator and support are each other inverse]
  \label{l:indic-and-support-are-each-other-inverse}
  \mbox{}\\
  Let~$(X,\Sigma)$ be a measurable space.
  Let~$A\in\Sigma$, let~$f\in\calIF$.
  Then, we have
  \begin{equation}
    \label{e:indic-and-support-are-each-other-inverse}
    \matUN_A \in \calIF,\quad
    \{ f \not= 0 \} \in \Sigma,\quad
    \{ \matUN_A \not= 0 \} = A
    \AND
    \matUN_{\{ f \not= 0 \}} = f.
  \end{equation}
\end{lemma}

\begin{proof}
  Direct consequence of
  Definition~\thref{d:if-set-of-meas-indic-funs},
  \assume{the definition of the indicator function}, and
  \assume{the definition of inverse image}.
\end{proof}

\begin{lemma}[$\calIF$~is measurable]
  \label{l:if-is-meas}
  \mbox{}\\
  Let~$(X,\Sigma)$ be a measurable space.
  Then, we have $\calIF\subset\calMR\cap\calMplus\subset\calM$.
\end{lemma}

\begin{proof}
  Direct consequence of
  Definition~\thref{d:if-set-of-meas-indic-funs},
  Lem\-ma~\thref{l:meas-of-indic-fun},
  Lemma~\threfc{l:m-and-finite-is-mr}{$\calMR\subset\calM$},
  \assume{nonnegativeness of indicator function}, and
  Definition~\thref{d:mplus-subset-of-nonneg-meas-num-fun}.
\end{proof}

\begin{lemma}[$\calIF$~is $\sigma$-additive]
  \label{l:if-is-sigma-add}
  \mbox{}\hfill
  Let~$(X,\Sigma)$ be a measurable space.
  Let~$I\!\subset\!\matN$.
  Let~$(A_i)_{i\in I}\!\in\!\Sigma$.
  Assume that the~$A_i$'s are pairwise disjoint.
  Then, we have
  $\sum_{i\in I}\matUN_{A_i}=\matUN_{\biguplus_{i\in I}A_i}\in\calIF$.
\end{lemma}

\begin{proof}
  Direct consequence of
  Definition~\thref{d:if-set-of-meas-indic-funs},
  \assume{the formula for the indicator of disjoint union (sum)},
  Definition~\threfc{d:measurable-space}{$\Sigma$ is a $\sigma$-algebra}, and
  Definition~\threfc{d:sigma-alg}{closedness under countable union}.
\end{proof}

\begin{lemma}[$\calIF$~is closed under multiplication]
  \label{l:if-is-closed-under-mult}
  \mbox{}\\
  Let~$(X,\Sigma)$ be a measurable space.
  Let~$A,B\in\Sigma$.
  Then, we have $\matUN_A\,\matUN_B=\matUN_{A\cap B}\in\calIF$.
\end{lemma}

\begin{proof}
  Direct consequence of
  Definition~\thref{d:if-set-of-meas-indic-funs},
  \assume{the formula for the indicator of intersection (product)}, and
  Lemma~\threfc{l:equiv-def-of-sigma-alg}{closedness under intersection}.
\end{proof}

\begin{remark}
  \mbox{}\\
  We recall the notation~$\matUN^V_U$ to denote the indicator function of~$U$
  defined on~$V$ (when~$U\subset V$).
\end{remark}

\begin{lemma}[$\calIF$~is closed under extension by zero]
  \label{l:if-is-closed-under-ext-by-zero}
  \mbox{}\\
  Let~$(X,\Sigma)$ be a measurable space.
  Let~$A\in\Sigma$.
  Let~$Y\subset X$ such that $A\subset Y$.
  Let~$f:\ArYRb$ and~$\hf:\ArXRb$.
  Assume that $\restr{\hf}{Y}=f$ and
  $\restr{f}{A}\in\calIF(A,\Sigma\olcap A)$.
  Then, we have $\hf\,\matUN_A\in\calIF(X,\Sigma)$.
\end{lemma}

\begin{proof}
  Direct consequence of
  Definition~\threfc{d:if-set-of-meas-indic-funs}{%
    $\restr{f}{A}=\matUN^A_B$ with $B\in\Sigma$ such that $B\subset A$}, and
  Lemma~\threfc{l:restr-is-mask}{$\hf\,\matUN_A=\matUN^X_B$}.
\end{proof}

\begin{lemma}[$\calIF$~is closed under restriction]
  \label{l:if-is-closed-under-restr}
  \mbox{}\hfill
  Let~$(X,\Sigma)$ be a measurable space.\\
  Let~$f\in\calIF(X,\Sigma)$.
  Let~$A\in\Sigma$.
  Then, we have $\restr{f}{A}\in\calIF(A,\Sigma\olcap A)$.
\end{lemma}

\begin{proof}
  Direct consequence of
  Definition~\threfc{d:if-set-of-meas-indic-funs}{%
    $f=\matUN^X_B$ with $B\in\Sigma$},
  \assume{the rule for the restriction of an indicator function
    ($\restr{f}{A}=\matUN^A_{B\cap A}$)}, and
  Lemma~\threfc{l:meas-of-meas-subspace}{%
    $B\cap A\in\Sigma\olcap A$}.
\end{proof}

\begin{definition}[integral in~$\calIF$]
  \label{d:int-in-if}
  \mbox{}\hfill
  Let~$(X,\Sigma,\mu)$ be a measure space.
  Let~$f\in\calIF$.\\
  The {\em integral of~$f$ (for the measure~$\mu$)} is denoted
  $\int f\,d\mu$;
  it is defined by
  \begin{equation}
    \label{e:int-in-if}
    \int f \, d\mu \eqdef \mu (\{ f \not= 0 \}) \quad \in \matRbarplus.
  \end{equation}
\end{definition}

\begin{lemma}[equivalent definition of integral in~$\calIF$]
  \label{l:equiv-def-of-int-in-if}
  \mbox{}\\
  Let~$(X,\Sigma,\mu)$ be a measure space.
  Let~$A\in\Sigma$.
  Then, we have $\int\matUN_A\,d\mu=\mu(A)$.
\end{lemma}

\begin{proof}
  Direct consequence of
  Lemma~\threfc{l:indic-and-support-are-each-other-inverse}{%
    $\matUN_A\in\calIF$ and $\{\matUN_A\neq0\}=A\in\Sigma$}, and
  Definition~\thref{d:int-in-if}.
\end{proof}

\begin{lemma}[integral in~$\calIF$ is additive]
  \label{l:int-in-if-is-add}
  \mbox{}\hfill
  Let~$(X,\Sigma,\mu)$ be a measure space.
  Let~$n\in\matN$.
  Let~$(A_i)_{i\in[0..n]}\in\Sigma$.
  Assume that the~$A_i$'s are pairwise disjoint.
  Then, we have
  \begin{equation}
    \label{e:int-in-if-is-add}
    \int \sum_{i \in [0..n]} \matUN_{A_i} \, d\mu
    = \int \matUN_{\biguplus_{i \in [0..n]} A_i} \, d\mu
    = \sum_{i \in [0..n]} \int \matUN_{A_i} \, d\mu.
  \end{equation}
\end{lemma}

\begin{proof}
  Direct consequence of
  Lemma~\thref{l:if-is-sigma-add},
  Lemma~\thref{l:equiv-def-of-int-in-if},
  Lemma~\threfc{l:equiv-def-of-meas}{additivity}, and
  Definition~\thref{d:add-over-meas-space}.
\end{proof}

\begin{lemma}[integral in~$\calIF$ over subset]
  \label{l:int-in-if-over-subset}
  \mbox{}\hfill
  Let~$(X,\Sigma,\mu)$ be a measure space.\\
  Let~$A\in\Sigma$.
  Let~$Y\subset X$ such that $A\subset Y$.
  Let~$f:\ArYRb$.
  Let~$\hf:\ArXRb$.
  Assume that $\restr{\hf}{Y}=f$.
  Then, we have $\restr{f}{A}\in\calIF(A,\Sigma\olcap A)$ iff
  $\hf\,\matUN_A\in\calIF(X,\Sigma)$.

  If so, there exists $B\in\Sigma$ such that $B\subset A$,
  $\restr{f}{A}=\matUN^A_B$, $\hf\,\matUN_A=\matUN^X_B$, and we have
  \begin{equation}
    \label{e:int-in-if-over-subset}
    \int \restr{f}{A} \, d\mu_A = \int \hf \, \matUN_A \, d\mu.
  \end{equation}
  This integral is denoted $\int_Af\,d\mu$;
  it is called {\em integral of~$f$ over~$A$}.
\end{lemma}

\begin{proof}
  \proofpar{Equivalence}
  Direct consequence of
  Lemma~\thref{l:if-is-closed-under-ext-by-zero}, and
  Lemma~\thref{l:if-is-closed-under-restr}.

  \proofparskip{Identity}
  Direct consequence of
  Lemma~\thref{l:restr-is-mask},
  Lemma~\threfc{l:indic-and-support-are-each-other-inverse}{%
    $\restr{f}{A}=\matUN^A_B$ and $\hf\,\matUN_A=\matUN^X_B$ with $B\in\Sigma$
    such that $B\subset A$},
  Lemma~\threfc{l:equiv-def-of-int-in-if}{%
    with $\mu\eqdef\mu_A$, then $A\eqdef B$}, and
  Lemma~\thref{l:trace-meas}.
\end{proof}

\begin{lemma}[integral in~$\calIF$ over subset is additive]
  \label{l:int-in-if-over-subset-is-add}
  \mbox{}\hfill
  Let~$(X,\Sigma,\mu)$ be a measure space.\\
  Let~$n\in\matN$.
  Let~$A,(A_i)_{i\in[0..n]}\in\Sigma$.
  Assume that $(A_i)_{i\in[0..n]}$ is a pseudopartition of~$A$.
  Let~$Y\subset X$ such that $A\subset Y$.
  Let~$f:\ArYRb$.
  Let~$\hf:\ArXRb$.
  Assume that $\restr{\hf}{Y}=f$.\\
  Then, $\hf\,\matUN_A\in\calIF$ iff
  for all $i\in[0..n]$, $\hf\,\matUN_{A_i}\in\calIF$.
  If so, we have
  \begin{equation}
    \label{e:int-in-if-over-subset-is-add}
    \int_A f \, d\mu = \sum_{i \in[0..n]} \int_{A_i} f \, d\mu.
  \end{equation}
\end{lemma}

\begin{proof}
  \proofpar{``Left'' implies ``right''}
  Direct consequence of
  Lemma~\threfc{l:indic-and-support-are-each-other-inverse}{%
    $\hf\,\matUN_A=\matUN_B$ with $B\in\Sigma$ such that $B\subset A$}, and
  Lemma~\threfc{l:if-is-closed-under-mult}{%
    $\hf\,\matUN_{A_i}=\matUN_{B\cap A_i}$}.

  \proofparskip{``Right'' implies ``left''}
  Direct consequence of
  Lemma~\threfc{l:indic-and-support-are-each-other-inverse}{%
    $\hf\,\matUN_{A_i}=\matUN_{B_i}$ with $B_i\in\Sigma$ such that
    $B_i\subset A_i$},
  \assume{monotonicity of intersection},
  Definition~\thref{d:meas},
  Definition~\threfc{d:measurable-space}{$\Sigma$ is a $\sigma$-algebra},
  Definition~\threfc{d:sigma-alg}{%
    $B\eqdef\biguplus_{i\in[0..n]}B_i\in\Sigma$},
  Lemma~\threfc{l:if-is-sigma-add}{with $I\eqdef[0..n]$}, and
  \assume{left distributivity of multiplication over addition in~$\matR$
    ($\hf\,\matUN_A=\matUN_B$)}.

  \medskip\noindent
  Therefore, we have the equivalence.

  \proofparskip{Identity}
  Direct consequence of
  Lemma~\threfc{l:int-in-if-over-subset}{%
    with $A$ and the $A_i$'s},
  Lemma~\threfc{l:if-is-sigma-add}{with $I\eqdef[0..n]$},
  \assume{left distributivity of multiplication over addition
    in~$\matR$}, and
  Lemma~\thref{l:int-in-if-is-add}.
\end{proof}

\begin{remark}
  \mbox{}\\
  In the next lemma, when~$Y$ is uncountable, the sum (of nonnegative values)
  is understood as the supremum for all finite subsets (which is also correct
  of course in the countable case),
  \begin{equation}
    \label{e:remark}
    \sum_{y \in Y} f (y)
    \eqdef \sup_{\substack{Z \subset Y\\\card (Z) < \infty}}
    \sum_{z \in Z} f (z).
  \end{equation}
\end{remark}

\begin{lemma}[integral in~$\calIF$ for counting measure]
  \label{l:int-in-if-for-count-meas}
  \mbox{}\\
  Let~$(X,\Sigma)$ be a measurable space.
  Let~$Y\subset X$.
  Let~$f\in\calIF$.
  Then, we have
  \begin{equation}
    \label{e:int-in-if-for-count-meas}
    \int f \, d\delta_Y = \sum_{y \in Y} f (y).
  \end{equation}
\end{lemma}

\begin{proof}
  Direct consequence of
  Lemma~\thref{l:indic-and-support-are-each-other-inverse},
  Lem\-ma~\threfc{l:equiv-def-of-int-in-if}{%
    with $A=\{f\not=0\}\in\Sigma$},
  Lemma~\thref{l:count-meas},
  \assume{the definition of the cardinality}, and
  \assume{the definition of the indicator function}.
\end{proof}

\clearpage
\section{Integration of nonnegative simple functions}
\label{s:integration-of-nonnegative-simple-functions}

\subsection{Simple function}
\label{ss:simple-function}

\begin{remark}
  In this section, the functions take their values in~$\matR$.
\end{remark}

\begin{definition}[$\calSF$, {\vectorspace} of simple functions]
  \label{d:sf-vector-space-of-simple-funs}
  \mbox{}\\
  Let~$(X,\Sigma)$ be a measurable space.
  The vector subspace of finite linear combinations of indicator functions of
  measurable subsets is called {\em {\vectorspace} of simple functions};
  it is denoted $\calSF(X,\Sigma)$ (or simply~$\calSF$);
  it is defined by $\calSF(X,\Sigma)\eqdef\Span{\calIF(X,\Sigma)}$.
\end{definition}

\begin{lemma}[$\calSF$ simple representation]
  \label{l:sf-simple-repr}
  \mbox{}\\
  Let~$(X,\Sigma)$ be a measurable space.
  Let~$f:\ArXR$.
  Then, we have $f\in\calSF$ iff
  there exists $n\in\matN$, $(a_i)_{i\in[0..n]}\in\matR$ and
  $(A_i)_{i\in[0..n]}\in\Sigma$ such that
  $f=\sum_{i\in[0..n]}a_i\,\matUN_{A_i}$.

  If so, for all $x\in X\setminus\bigcup_{i\in[0..n]}A_i$, we have $f(x)=0$.
\end{lemma}

\begin{proof}
  \proofpar{Equivalence}
  Direct consequence of
  Definition~\thref{d:sf-vector-space-of-simple-funs},
  Definition~\thref{d:if-set-of-meas-indic-funs}, and
  Lemma~\thref{l:indic-and-support-are-each-other-inverse}.

  \proofparskip{Property}\\
  Direct consequence of
  \assume{the definition of the indicator function}, and
  \assume{field properties of~$\matR$ (0~is absorbing element for
    multiplication and identity element for addition)}.
\end{proof}

\begin{remark}
  \label{r:v2-new11}
  Note that in a simple representation, the values~$a_i$'s may not be unique and
  may not be values of the function, and the supports~$A_i$'s may be empty, may not
  be related to preimages of~$a_i$'s, may overlap and thus may not form a
  partition of~$X$.
\end{remark}

\begin{remark}
  We recall that~$f^{-1}(y)$ denotes the subset~$f^{-1}(\{y\})$.
\end{remark}

\begin{lemma}[$\calSF$~canonical representation]
  \label{l:sf-can-repr}
  \mbox{}\\
  Let~$(X,\Sigma)$ be a measurable space.
  Let~$f:\ArXR$.\\
  Then, we have $f\in\calSF$ iff
  $f(X)$~is finite, and for all~$y\in f(X)$, $f^{-1}(y)$ belongs to~$\Sigma$.

  If so, we have the following {\em canonical representation}:
  \begin{equation}
    \label{e:sf-can-repr-1}
    f = \sum_{y \in f (X)} y \, \matUN_{f^{-1} (y)},
  \end{equation}
  {\ie} there exists unique $n\in\matN$, $(a_i)_{i\in[0..n]}\in\matR$, and
  $(A_i)_{i\in[0..n]}\in\Sigma$, such that
  \begin{align}
    \label{e:sf-can-repr-2}
    \forall i \in [0..n - 1],\quad & a_i < a_{i + 1},\\
    \label{e:sf-can-repr-3}
    \forall i \in [0..n],\quad & A_i = f^{-1} (a_i) \not= \emptyset,\\
    \label{e:sf-can-repr-4}
    \forall p, q \in [0..n],\quad
    & p \not= q \IMPLIES A_p \cap A_q = \emptyset,\\
    \label{e:sf-can-repr-5}
    & X = \biguplus_{i \in [0..n]} A_i,\\
    \label{e:sf-can-repr-5bis}
    & f(X) = \{ a_i \st i \in [0..n] \},\\
    \label{e:sf-can-repr-6}
    & f = \sum_{i \in [0..n]} a_i \, \matUN_{A_i}.
  \end{align}
\end{lemma}

\begin{proof}
  \proofpar{``Left'' implies ``right''}
  Assume first that $f\in\calSF$.
  Then, from
  Lemma~\thref{l:sf-simple-repr},
  there exists $n\in\matN$, $(a_i)_{i\in[0..n]}\in\matR$, and
  $(A_i)_{i\in[0..n]}\in\Sigma$, such that
  $f=\sum_{i\in[0..n]}a_i\,\matUN_{A_i}$.
  Thus, from
  \assume{the definition of the indicator function},
  $f$~can only take the values $\sum_{i\in[0..n]}a_i\delta_i$ where
  $(\delta_i)_{i\in[0..n]}\in\{0,1\}$.
  Hence, the cardinality of~$f(X)$ is at most~$2^{n+1}$.

  Let~$(\delta_i)_{i\in[0..n]}\in\{0,1\}$.
  Let~$y\eqdef\sum_{i\in[0..n]}a_i\delta_i\in f(X)$.
  Let~$I_y\subset[0..n]$ such that $i\in I_y$ iff $\delta_i=1$.
  Then, we have\footnote{%
    For instance, the case $I_y=\emptyset$ may correspond to $y=0$ when
    $0\not\in\{a_i\st i\in[0..n]\}$.}
  $y=\sum_{i\in I_y}a_i$.
  Unfortunately, several partial sums of~$a_i$'s may lead to the same
  value~$y$.
  Thus, from
  Lemma~\threfc{l:equiv-def-of-sigma-alg}{%
    closedness under countable intersection and union},
  we have
  \begin{equation*}
    f^{-1} (y)
    = \bigcup_{\substack{%
        I_y \subset \calP ([0..n])\\
        \sum_{i \in I_y} a_i = y}}
    \left( \bigcap_{i \in I_y} A_i \right)
    \in \Sigma.
  \end{equation*}

  From
  \assume{properties of inverse image},
  the collection $(f^{-1}(y))_{y\in f(X)}$ makes a finite partition of the
  whole set~$X$.
  Let~$x\in X$.
  Then, there exists a unique $y\in\matR$ such that $x\in~f^{-1}(y)$.
  Thus, from
  \assume{field properties of~$\matR$},
  we have $f(x)=y=y\times1=y\,\matUN_{f^{-1}(y)}(x)$.
  Hence, Equation~\eqref{e:sf-can-repr-1} holds.
  And Equation~\eqref{e:sf-can-repr-6} is equivalent, up to a nondecreasing
  reordering of the $y\in f(X)$.

  \proofparskip{``Right'' implies ``left''}
  Let~$n\eqdef\card(f(X))-1\in\matN$.
  Let~$(a_i)_{i\in[0..n]}\in\matR$ be the~$n+1$ distinct values taken by the
  function~$f$, {\ie} such that $f(X)=\{a_i\st i\in[0..n]\}$.
  For all $i\in[0..n]$, let
  \begin{equation*}
    A_i \eqdef f^{-1} (a_i) \in \Sigma.
  \end{equation*}
  Then, from
  \assume{the definition and properties of image and inverse image}, and
  \assume{the definition of partition},
  the~$n$ subsets $(A_i)_{i\in[0..n]}$ constitutes a partition of~$X$.
  Thus, from
  \assume{the definition of the indicator function},
  we have $f=\sum_{i\in[0..n]}a_i\,\matUN_{A_i}$.
  Hence, from
  Lemma~\thref{l:sf-simple-repr},
  we have $f\in\calSF$.

  \proofparskip{Uniqueness}
  Direct consequence of~\eqref{e:sf-can-repr-2}--\eqref{e:sf-can-repr-6} since
  $\card(f(X))=n+1$ ({\ie} $n$ is unique),
  $f(X)=\{a_i\st i\in[0..n]\}$ ({\ie} the $a_i$'s are unique), and
  $A_i=f^{-1}(a_i)$ ({\ie} the $A_i$'s are unique).

  \medskip\noindent
  Therefore, we have the equivalence, and the representation is unique.
\end{proof}

\begin{remark}
  \label{r:v2-new12}
  To sum up, in a canonical representation, the values~$a_i$'s are unique and
  are the values of the function, and the supports~$A_i$'s are the nonempty
  preimages of the~$a_i$'s and form a partition of~$X$.
  Hence, the canonical representation of the zero function is~$0\,\matUN_X$.
  Note that Equations~\eqref{e:sf-can-repr-1} and~\eqref{e:sf-can-repr-6} are
  identical, up to an nondecreasing reordering of the $y\in f(X)$.

  Some authors exclude~$y=0$ from the sum over~$f(X)$.
  In this case the partition property expressed
  in~\eqref{e:sf-can-repr-3}--\eqref{e:sf-can-repr-5bis} weakens into the
  pairwise disjunction property~\eqref{e:sf-can-repr-4} and inclusions instead
  of equalities.
  Then, the zero function must be treated differently as its canonical
  representation becomes~$\matUN_\emptyset$.
\end{remark}

\begin{lemma}[$\calSF$~disjoint representation]
  \label{l:sf-disj-repr}
  \mbox{}\\
  Let~$(X,\Sigma)$ be a measurable space.
  Let~$f:\ArXR$.\\
  Then, $f\in\calSF$ iff
  there exists $n\in\matN$, $(a_i)_{i\in[0..n]}\in\matR$, and
  $(A_i)_{i\in[0..n]}\in\Sigma$, such that
  \begin{align}
    \label{e:sf-disj-repr-1}
    \forall i \in [0..n],\quad
    & A_i \subset f^{-1} (a_i),\\
    \label{e:sf-disj-repr-2}
    \forall p, q \in [0..n],\quad
    & p \not= q \IMPLIES A_p \cap A_q = \emptyset,\\
    \label{e:sf-disj-repr-3}
    & X = \biguplus_{i \in [0..n]} A_i,\\
    \label{e:sf-disj-repr-4}
    & f = \sum_{i \in [0..n]} a_i \, \matUN_{A_i}.
  \end{align}

  If so, it is called a {\em disjoint representation (of~$f$)}.
\end{lemma}

\begin{proof}
  Direct consequence of
  Lemma~\threfc{l:sf-can-repr}{``left'' implies ``right''}, and
  Lemma~\threfc{l:sf-simple-repr}{``right'' implies ``left''}.
\end{proof}

\begin{remark}
  \label{r:v2-new13}
  In a disjoint representation, the values~$a_i$'s may not be unique ($a_i=a_j$
  with $i\neq j$ is possible) and may not be the values of the function
  ($A_i=\emptyset$ is possible).
  More precisely, the supports~$A_i$'s may be empty but are subsets of the
  associated preimage of~$a_i$ (thus the values that are not taken by the
  function are associated with an empty support).
  They form a pseudopartition of~$X$ (see Definition~\ref{d:pseudopart}).

  Note that the nonempty~$A_i$'s form a subpartition of
  $(f^{-1}(y))_{y\in f(X)}$ (see next lemma).
\end{remark}

\begin{lemma}[$\calSF$~disjoint representation is subpartition
  of canonical representation]
  \label{l:sf-disj-repr-is-subpart-of-can-repr}
  Let~$(X,\Sigma)$ be a measurable space.
  Let~$f\in\calSF$.
  Let~$n,m\in\matN$, $(a_i)_{i\in[0..m]},(b_j)_{j\in[0..m]}\in\matR$, and
  $(A_i)_{i\in[0..m]},(B_j)_{j\in[0..m]}\in\Sigma$.
  Assume that $f=\sum_{i\in[0..n]}a_i\,\matUN_{A_i}$ is a disjoint
  representation, and that $f=\sum_{j\in[0..m]}b_j\,\matUN_{B_j}$ is the
  canonical representation.\\
  For all~$j\in[0..m]$,
  let $\Ip_j\eqdef\{i\in[0..n]\st\emptyset\not=A_i\subset B_j\}$.
  Let $\Ip\eqdef\{i\in[0..n]\st A_i\not=\emptyset\}$.\\
  Then, we have
  \begin{gather}
    \label{e:sf-disj-repr-is-subpart-of-can-repr-1}
    \forall j \in [0..m],\quad
    (\forall i \in \Ip_j,\; a_i = b_j)
    \CONJ
    B_j = \biguplus_{i \in \Ip_j} A_i,\\
    \label{e:sf-disj-repr-is-subpart-of-can-repr-2}
    (\forall p, q \in [0..m],\;
    p \not= q \;\Implies\; \Ip_p \cap \Ip_q = \emptyset)
    \CONJ
    \biguplus_{j \in [0..m]} \Ip_j = \Ip.
  \end{gather}
\end{lemma}

\begin{proof}
  Let~$i\in[0..n]$ and~$j\in[0..m]$.
  Assume that $A_i\not=\emptyset$.\\
  \proofpar{(0a) $A_i\subset B_j\Implies a_i=b_j$}
  Let $x\in A_i\cap B_j$.
  Then, from
  Lemma~\threfc{l:sf-disj-repr}{$f(x)=a_i$}, and
  Lemma~\threfc{l:sf-can-repr}{$f(x)=b_j$},
  we have $a_i=b_j$.\\
  \proofpar{(0b) $a_i=b_j\Implies A_i\subset B_j$}
  Assume that $a_i=b_j$.
  Then, from
  Lemma~\thref{l:sf-disj-repr}, and
  Lemma~\thref{l:sf-can-repr},
  we have $A_i\subset f^{-1}(a_i)=f^{-1}(b_j)=B_j$.

  \proofparskip{(1)}
  Let~$j\in[0..m]$.\\
  From
  Lemma~\thref{l:sf-disj-repr},
  the $A_i$'s are pairwise disjoint.
  Let $\Ap_j\eqdef\biguplus_{i\in\Ip_j}A_i$.\\
  Let~$x\in B_j$.
  Then, from
  Lemma~\threfc{l:sf-can-repr}{$f(x)=b_j$}, and
  Lemma~\threfc{l:sf-disj-repr}{%
    $X=\biguplus_{i\in[0..n]}A_i$ and $A_i\subset f^{-1}(a_i)$},
  there exists~$i\in[0..n]$ such that $x\in A_i$, {\ie} $A_i\not=\emptyset$ and
  $a_i=f(x)=b_j$.
  Thus, from~(0b), we have $i\in\Ip_j$, and $B_j\subset\Ap_j$.
  Hence, since the other inclusion is obvious, we have equality, and from~(0a),
  property~\eqref{e:sf-disj-repr-is-subpart-of-can-repr-1} holds.

  \proofparskip{(2a)}
  Let~$p,q\in[0..m]$.
  Assume that $p\not=q$.\\
  Let~$i\in\Ip_p\cap\Ip_q$.
  Then, from
  Lemma~\threfc{l:sf-can-repr}{pairwise disjunction},
  we have $\emptyset\not=A_i\subset B_p\cap B_q=\emptyset$, which is
  impossible.
  Hence, we have $\Ip_p\cap\Ip_q=\emptyset$.

  \proofparskip{(2b)}
  Let~$i\in\Ip$, {\ie} $A_i\not=\emptyset$.\\
  Then, from
  Lemma~\threfc{l:sf-disj-repr}{$a_i\in f(X)$}, and
  Lemma~\threfc{l:sf-can-repr}{$f(X)=\{b_j\st j\in[0..m]\}$},
  there exists~$j\in[0..m]$ such that $a_i=b_j$.
  Thus, from~(0b), we have $i\in\Ip_j$, and
  $\Ip\subset\uplus_{j\in[0..m]}\Ip_j$.
  Hence, since the other inclusion is obvious, we have equality, and from~(2a),
  property~\eqref{e:sf-disj-repr-is-subpart-of-can-repr-2} holds.

  \medskip\noindent
  Therefore, both properties hold.
\end{proof}

\begin{lemma}[$\calSF$ is algebra over~$\matR$]
  \label{l:sf-is-alg-over-r}
  \mbox{}\\
  Let~$(X,\Sigma)$ be a measurable space.
  Then, $\calSF$~is a subalgebra of~$\matR^X$.

  Let $f,g\in\calSF$, $n,m\in\matN$,
  $(a_i)_{i\in[0..n]},(b_j)_{j\in[0..m]}\in\matR$,
  and $(A_i)_{i\in[0..n]},(B_j)_{j\in[0..m]}\in\Sigma$ such that
  $f=\sum_{i\in[0..n]}a_i\,\matUN_{A_i}$ and
  $g=\sum_{j\in[0..m]}b_j\,\matUN_{B_j}$.\\
  Let~$N\eqdef nm+n+m$.
  Let~$\fhi:[0..N]\to[0..n]\times[0..m]$ be a bijection.\\
  For all $(i,j)\in[0..n]\times[0..m]$, let~$c^+_{i,j}\eqdef a_i+b_j$,
  $c^*_{i,j}\eqdef a_ib_j$, and $C_{i,j}\eqdef A_i\cap B_j$.\\
  Then, on the one hand, if both representations are disjoint, we have
  \begin{equation}
    \label{e:sf-is-alg-over-r-1}
    f + g = \sum_{k \in [0..N]} c^+_{\fhi (k)} \, \matUN_{C_{\fhi (k)}},
  \end{equation}
  and it is also a disjoint representation.
  On the other hand, we always have
  \begin{equation}
    \label{e:sf-is-alg-over-r-2}
    f g = \sum_{k \in [0..N]} c^*_{\fhi (k)} \, \matUN_{C_{\fhi (k)}},
  \end{equation}
  and it is also a disjoint representation when those of~$f$ and~$g$ are.
\end{lemma}

\begin{proof}
  Since $N+1=(n+1)(m+1)$, such a bijection~$\fhi$ exists.

  \proofparskip{(1)}
  From
  Definition~\thref{d:sf-vector-space-of-simple-funs},
  \assume{the definition of the linear span}, and
  Definition~\thref{LM-d:subspace},
  $\calSF$~is a {\vectorsubspace} of~$\matR^X$.
  From
  \assume{field properties of~$\matR$},
  Lemma~\thref{l:if-is-closed-under-mult}, and
  Lemma~\threfc{l:equiv-def-of-sigma-alg}{closedness under intersection},
  we have Equation~\eqref{e:sf-is-alg-over-r-2} where for all $i\in[0..n]$,
  $j\in[0..m]$, $A_i\cap B_j\in\Sigma$.
  Hence, from
  Lemma~\thref{l:sf-simple-repr},
  we have $fg\in\calSF$.
  Therefore, from
  Lemma~\thref{l:subspace-and-closed-under-mult-is-subalg},
  $\calSF$~is a subalgebra of~$\matR^X$.

  \medskip\noindent
  Assume now that both representations of~$f$ and~$g$ are disjoint.

  \proofparskip{(2a)}
  Then, from
  Lemma~\thref{l:sf-disj-repr},
  \assume{compatibility of intersection with pairwise disjunction},
  \assume{idempotent law for intersection}, and
  \assume{left and right distributivity of intersection over union},
  the~$A_i\cap B_j$'s are pairwise disjoint, and
  \begin{equation*}
    X
    = \biguplus_{i \in [0..n]} A_i \cap \biguplus_{j \in [0..m]} B_j
    = \biguplus_{k \in [0..N]} C_{\fhi (k)}.
  \end{equation*}
  Hence, the~$C_{\fhi(k)}$'s form a pseudopartition of~$X$ (there may be empty
  parts in it, see Remark~\ref{r:v2-new13}).

  \proofparskip{(2b)}
  Let~$i\in[0..n]$.
  Then, from
  \assume{field properties of~$\matR$},
  \assume{the definition of the indicator function ($\matUN_X\equiv1$)},
  Lemma~\threfc{l:sf-disj-repr}{$X=\biguplus_{j\in[0..m]}B_j$},
  Lemma~\threfc{l:if-is-sigma-add}{with $I\eqdef[0..m]$},
  \assume{left distributivity of multiplication over addition in~$\matR$}, and
  Lemma~\thref{l:if-is-closed-under-mult},
  we have
  \begin{equation*}
    \matUN_{A_i}
    = \matUN_{A_i} \, \matUN_{\biguplus_{j \in [0..m]} B_j}
    = \matUN_{A_i} \, \left( \sum_{j \in [0..m]} \matUN_{B_j} \right)
    = \sum_{j \in [0..m]} \left( \matUN_{A_i} \, \matUN_{B_j} \right)
    = \sum_{j \in [0..m]} \matUN_{A_i \cap B_j}.
  \end{equation*}
  Thus, from
  \assume{left distributivity of multiplication over addition in~$\matR$},
  \assume{the definition of intersection ($A_i\cap B_j\subset A_i$)},
  (2a), and
  Lemma~\thref{l:sf-disj-repr},
  we have
  \begin{equation*}
    f
    = \sum_{i \in [0..n]} a_i \,
    \left( \sum_{j \in [0..m]} \matUN_{A_i \cap B_j} \right)
    = \sum_{(i, j) \in [0..n] \times [0..m]} a_i \, \matUN_{A_i \cap B_j},
  \end{equation*}
  and the representation is disjoint.
  Similarly, we obtain the disjoint representation
  \begin{equation*}
    g = \sum_{(i, j) \in [0..n] \times [0..m]} b_j \, \matUN_{A_i \cap B_j}.
  \end{equation*}
  Hence, from
  \assume{additive group properties of~$\matR$},
  \assume{right distributivity of multiplication over addition in~$\matR$}, and
  Lemma~\threfc{l:equiv-def-of-sigma-alg}{closedness under intersection},
  we have Equation~\eqref{e:sf-is-alg-over-r-1} where again, for all
  $k\in[0..N]$, $C_{\fhi(k)}\in\Sigma$.

  \proofparskip{(3a)}
  Let~$a,b\in\matR$.
  Let~$\Diamond$ be a binary operator from~$\matR^2$ to~$\matR$ (using infix
  notation).
  Then, from
  \assume{the definition of inverse image},
  we have
  $(f\Diamond g)(f^{-1}(a)\cap g^{-1}(b))\subset\{a\Diamond b\}$.
  Hence, from
  \assume{identity $A\subset \psi^{-1}(B)\Equiv \psi(A)\subset B$
    (with $\psi\eqdef f\Diamond g$)},
  we have $f^{-1}(a)\cap g^{-1}(b)\subset(f\Diamond g)^{-1}(a\Diamond b)$.

  \proofparskip{(3b)}
  Let~$k\in[0..N]$.
  Let~$(i,j)\eqdef\fhi(k)\in[0..n]\times[0..m]$.
  Then, from
  Lemma~\threfc{l:sf-disj-repr}{%
    $A_i\subset f^{-1}(a_i)$ and $B_j\subset g^{-1}(b_j)$},
  \assume{monotonicity of intersection
    ($A_i\cap B_j$ is a subset of $f^{-1}(a_i)\cap g^{-1}(b_j)$)},
  and~(3a) with the binary operators addition and multiplication,
  we have $C_{\fhi(k)}\subset(f+g)^{-1}(c^+_{\fhi(k)})$ and
  $C_{\fhi(k)}\subset(fg)^{-1}(c^*_{\fhi(k)})$.
  Therefore, from
  Lemma~\thref{l:sf-disj-repr}, and~(2a),
  both Equations~\eqref{e:sf-is-alg-over-r-1} and~\eqref{e:sf-is-alg-over-r-2}
  are disjoint representations.
\end{proof}

\begin{remark}
  \label{r:v2-new14}
  Note that in the previous lemma, formula~\eqref{e:sf-is-alg-over-r-1} is
  wrong when there is nonempty overlapping in the representations of the
  functions.
  Indeed, the values in overlapped parts are counted too many times.

  Note also that even though the representations of~$f$ and~$g$ are canonical,
  the representations of~$f+g$ and~$fg$ in
  Equations~\eqref{e:sf-is-alg-over-r-1} and~\eqref{e:sf-is-alg-over-r-2} are
  only disjoint.
  Indeed, the~$a_i+b_j$'s and~$a_ib_j$'s may not be pairwise distinct, and thus
  the $A_i\cap B_j$'s may no longer be their inverse images.
\end{remark}

\begin{lemma}[$\calSF$~is measurable]
  \label{l:sf-is-meas}
  \mbox{}\\
  Let~$(X,\Sigma)$ be a measurable space.
  Then, we have $\calIF\subset\calSF\subset\calMR\subset\calM$.
\end{lemma}

\begin{proof}
  Direct consequence of
  Definition~\thref{d:sf-vector-space-of-simple-funs},
  Lemma~\thref{l:meas-of-indic-fun},
  Lemma~\thref{l:mr-is-alg}, and
  Lemma~\threfc{l:m-and-finite-is-mr}{$\calMR\subset\calM$}.
\end{proof}

\begin{lemma}[$\calSF$~is closed under extension by zero]
  \label{l:sf-is-closed-under-ext-by-zero}
  \mbox{}\\
  Let~$(X,\Sigma)$ be a measurable space.
  Let~$A\in\Sigma$.
  Let~$Y\subset X$ such that $A\subset Y$.
  Let~$f:\ArYRb$ and~$\hf:\ArXRb$.
  Assume that $\restr{\hf}{Y}=f$ and
  $\restr{f}{A}\in\calSF(A,\Sigma\olcap A)$.
  Then, we have $\hf\,\matUN_A\in\calSF(X,\Sigma)$.
\end{lemma}

\begin{proof}
  Direct consequence of
  Definition~\threfc{d:sf-vector-space-of-simple-funs}{%
    $\restr{f}{A}$ is a linear combination of $f_i$'s in
    $\calIF(A,\Sigma\olcap A)$}, and
  Lemma~\threfc{l:if-is-closed-under-ext-by-zero}{%
    with $\hf_i$ function from $X$ to $\matRbar$ such that
    $\restr{(\hf_i)}{Y}=f_i$}.
\end{proof}

\begin{lemma}[$\calSF$~is closed under restriction]
  \label{l:sf-is-closed-under-restr}
  \mbox{}\hfill
  Let~$(X,\Sigma)$ be a measurable space.\\
  Let~$f\in\calSF(X,\Sigma)$.
  Let~$A\in\Sigma$.
  Then, we have $\restr{f}{A}\in\calSF(A,\Sigma\olcap A)$.
\end{lemma}

\begin{proof}
  Direct consequence of
  Definition~\thref{d:sf-vector-space-of-simple-funs}, and
  Lem\-ma~\thref{l:if-is-closed-under-restr}.
\end{proof}

\subsection{Nonnegative simple function}
\label{ss:nonnegative-simple-function}

\begin{remark}
  From now on, the functions take their values in~$\matRplus$, and the
  expressions involving integrals are taken in~$\matRbarplus$.
\end{remark}

\begin{definition}[$\calSFplus$, subset of nonnegative simple functions]
  \label{d:sfplus-subset-of-nonneg-simple-funs}
  \mbox{}\\
  Let~$(X,\Sigma)$ be a measurable space.
  The {\em subset of nonnegative simple functions} is denoted
  $\calSFplus(X,\Sigma)$ (or simply~$\calSFplus$);
  it is defined by $\calSFplus\eqdef\{f\in\calSF\st f(X)\subset\matRplus\}$.
\end{definition}

\begin{lemma}[$\calSFplus$ disjoint representation]
  \label{l:sfplus-disj-repr}
  \mbox{}\\
  Let~$(X,\Sigma)$ be a measurable space.
  Let~$f:\ArXR$.\\
  Then, $f\in\calSFplus$ iff
  there exists $n\in\matN$, $(a_i)_{i\in[0..n]}\in\matRplus$, and
  $(A_i)_{i\in[0..n]}\in\Sigma$ such that
  \begin{align}
    \label{e:sfplus-disj-repr-1}
    \forall i \in [0..n],\quad
    & A_i \subset f^{-1} (a_i),\\
    \label{e:sfplus-disj-repr-2}
    \forall p, q \in [0..n],\quad
    & p \not= q \IMPLIES A_p \cap A_q = \emptyset,\\
    \label{e:sfplus-disj-repr-3}
    & X = \biguplus_{i \in [0..n]} A_i,\\
    \label{e:sfplus-disj-repr-4}
    & f = \sum_{i \in [0..n]} a_i \, \matUN_{A_i}
  \end{align}
\end{lemma}

\begin{proof}
  \proofpar{``Left'' implies ``right''}
  Direct consequence of
  Definition~\thref{d:sfplus-subset-of-nonneg-simple-funs}, and
  Lemma~\threfc{l:sf-disj-repr}{%
    with $f(A_i)=\{a_i\}\subset\matRplus$}.

  \proofparskip{``Right'' implies ``left''}\\
  Direct consequence of
  Lemma~\threfc{l:sf-disj-repr}{$f\in\calSF$},
  \assume{closedness of addition in~$\matRplus$ ($f\geq0$)}, and
  Definition~\thref{d:sfplus-subset-of-nonneg-simple-funs}.

  \medskip\noindent
  Therefore, we have the equivalence.
\end{proof}

\begin{lemma}[$\calSFplus$ canonical representation]
  \label{l:sfplus-can-repr}
  \mbox{}\\
  Let~$(X,\Sigma)$ be a measurable space.
  Let~$f:\ArXR$.\\
  Then, $f\in\calSFplus$ iff
  there exists unique $n\in\matN$, $(a_i)_{i\in[0..n]}\in\matRplus$, and
  $(A_i)_{i\in[0..n]}\in\Sigma$ such that
  \begin{align}
    \label{e:sfplus-can-repr-2}
    \forall i \in [0..n - 1],\quad & a_i < a_{i + 1},\\
    \label{e:sfplus-can-repr-3}
    \forall i \in [0..n],\quad & A_i = f^{-1} (a_i) \not= \emptyset,\\
    \label{e:sfplus-can-repr-4}
    \forall p, q \in [0..n],\quad
    & p \not= q \IMPLIES A_p \cap A_q = \emptyset,\\
    \label{e:sfplus-can-repr-5}
    & X = \biguplus_{i \in [0..n]} A_i,\\
    \label{e:sfplus-can-repr-6}
    & f = \sum_{i \in [0..n]} a_i \, \matUN_{A_i}
  \end{align}
  which can also be written (with $f^{-1}(y)\in\Sigma$ for all $y\in f(X)$)
  \begin{equation}
    \label{e:sfplus-can-repr-1}
    f = \sum_{y \in f (X)} y \, \matUN_{f^{-1} (y)}.
  \end{equation}
\end{lemma}

\begin{proof}
  \proofpar{``Left'' implies ``right''}
  Direct consequence of
  Definition~\thref{d:sfplus-subset-of-nonneg-simple-funs}, and
  Lemma~\threfc{l:sf-can-repr}{%
    with $a_i\in f(X)\subset\matRplus$}.

  \proofparskip{``Right'' implies ``left''}\\
  Direct consequence of
  Lemma~\threfc{l:sf-can-repr}{$f\in\calSF$},
  \assume{closedness of addition in~$\matRplus$ ($f\geq0$)}, and
  Definition~\thref{d:sfplus-subset-of-nonneg-simple-funs}.

  \medskip\noindent
  Therefore, we have the equivalence.
\end{proof}

\begin{lemma}[$\calSFplus$~disjoint representation is subpartition
  of canonical representation]
  \label{l:sfplus-disj-repr-is-subpart-of-can-repr}
  Let~$(X,\Sigma)$ be a measurable space.
  Let~$f\in\calSFplus$.
  Let~$n,m\in\matN$, $(a_i)_{i\in[0..m]},(b_j)_{j\in[0..m]}\in\matRplus$, and
  $(A_i)_{i\in[0..m]},(B_j)_{j\in[0..m]}\in\Sigma$.
  Assume that $f=\sum_{i\in[0..n]}a_i\,\matUN_{A_i}$ is a disjoint
  representation, and that $f=\sum_{j\in[0..m]}b_j\,\matUN_{B_j}$ is the
  canonical representation.\\
  For all~$j\in[0..m]$,
  let $\Ip_j\eqdef\{i\in[0..n]\st\emptyset\not=A_i\subset B_j\}$.
  Let $\Ip\eqdef\{i\in[0..n]\st A_i\not=\emptyset\}$.\\
  Then, we have
  \begin{gather}
    \label{e:sfplus-disj-repr-is-subpart-of-can-repr-1}
    \forall j \in [0..m],\quad
    B_j = \biguplus_{i \in \Ip_j} A_i
    \CONJ
    (\forall i \in \Ip_j,\; a_i = b_j),\\
    \label{e:sfplus-disj-repr-is-subpart-of-can-repr-2}
    (\forall p, q \in [0..m],\;
    p \not= q \;\Implies\; \Ip_p \cap \Ip_q = \emptyset)
    \CONJ
    \biguplus_{j \in [0..m]} \Ip_j = \Ip.
  \end{gather}
\end{lemma}

\begin{proof}
  Direct consequence of
  Lemma~\thref{l:sf-disj-repr-is-subpart-of-can-repr},
  Lemma~\threfc{l:sfplus-disj-repr}{$a_i\geq0$}, and
  Lemma~\threfc{l:sfplus-can-repr}{$b_i\geq0$}.
\end{proof}

\begin{lemma}[$\calSFplus$ simple representation]
  \label{l:sfplus-simple-repr}
  \mbox{}\\
  Let~$(X,\Sigma)$ be a measurable space.
  Let~$f:\ArXR$.
  Then, we have $f\in\calSFplus$ iff
  there exists $n\in\matN$, $(a_i)_{i\in[0..n]}\in\matRplus$, and
  $(A_i)_{i\in[0..n]}$ in~$\Sigma$ such that,
  $f=\sum_{i\in[0..n]}a_i\,\matUN_{A_i}$.

  If so, for all $x\in X\setminus\bigcup_{i\in[0..n]}A_i$, we have $f(x)=0$.
\end{lemma}

\begin{proof}
  \proofpar{``Left'' implies ``right''}\\
  Direct consequence of
  Lemma~\thref{l:sfplus-can-repr}.

  \proofparskip{``Right'' implies ``left''}\\
  Direct consequence of
  Lemma~\threfc{l:sf-simple-repr}{$f\in\calSF$},
  \assume{closedness of addition in~$\matRplus$ ($f\geq0$)}, and
  Definition~\thref{d:sfplus-subset-of-nonneg-simple-funs}.

  \medskip\noindent
  Therefore, we have the equivalence.

  \proofparskip{Property}
  Direct consequence of
  Lemma~\thref{l:sf-simple-repr}.
\end{proof}

\begin{lemma}[$\calSFplus$ is closed under positive algebra operations]
  \label{l:sfplus-is-closed-under-pos-alg-ops}
  \mbox{}\hfill
  Let~$(X,\Sigma,\mu)$ be a measure space.
  Let~$f,g\in\calSFplus$.
  Let~$a\in\matRplus$.
  Then, we have $f+g,af,fg\in\calSFplus$.

  Let $f,g\in\calSFplus$, $n,m\in\matN$,
  $(a_i)_{i\in[0..n]},(b_j)_{j\in[0..m]}\in\matRplus$,
  and $(A_i)_{i\in[0..n]},(B_j)_{j\in[0..m]}\in\Sigma$ such that
  $f=\sum_{i\in[0..n]}a_i\,\matUN_{A_i}$ and
  $g=\sum_{j\in[0..m]}b_j\,\matUN_{B_j}$.\\
  Let~$N\eqdef nm+n+m$.
  Let~$\fhi$ be a bijection from~$[0..N]$ to~$[0..n]\times[0..m]$.\\
  For all $(i,j)\in[0..n]\times[0..m]$, let~$c^+_{i,j}\eqdef a_i+b_j$,
  $c^*_{i,j}\eqdef a_ib_j$, and $C_{i,j}\eqdef A_i\cap B_j$.\\
  Then, on the one hand, if both representations are disjoint, we have
  \begin{equation}
    \label{e:sfplus-is-closed-under-pos-alg-ops-1}
    f + g = \sum_{k \in [0..N]} c^+_{\fhi (k)} \, \matUN_{C_{\fhi (k)}},
  \end{equation}
  and it is also a disjoint representation.
  On the other hand, we always have
  \begin{equation}
    \label{e:sfplus-is-closed-under-pos-alg-ops-2}
    f g = \sum_{k \in [0..N]} c^*_{\fhi (k)} \, \matUN_{C_{\fhi (k)}},
  \end{equation}
  and it is also a disjoint representation when those of~$f$ and~$g$ are.
\end{lemma}

\begin{proof}
  \proofpar{Closedness}
  Direct consequence of
  Definition~\thref{d:sfplus-subset-of-nonneg-simple-funs},
  Lemma~\thref{l:sf-is-alg-over-r},
  Definition~\thref{d:alg-over-a-field}, and
  \assume{closedness of addition and multiplication in~$\matRplus$}.

  \proofparskip{Identities}
  Direct consequences of
  Lemma~\thref{l:sf-is-alg-over-r},
  \assume{closedness of addition and multiplication in~$\matRplus$},
  Lemma~\thref{l:sfplus-disj-repr}, and
  \assume{associativity and commutativity of intersection}.
\end{proof}

\begin{lemma}[$\calSFplus$~is measurable]
  \label{l:sfplus-is-meas}
  \mbox{}\\
  Let~$(X,\Sigma)$ be a measurable space.
  Then, we have
  $\calIF\subset\calSFplus\subset\calMR\cap\calMplus\subset\calM$.
\end{lemma}

\begin{proof}
  Direct consequence of
  Definition~\thref{d:sfplus-subset-of-nonneg-simple-funs},
  Definition~\thref{d:if-set-of-meas-indic-funs},
  \assume{nonnegativeness of the indicator function},
  Lemma~\thref{l:sf-is-meas},
  Definition~\thref{d:mplus-subset-of-nonneg-meas-num-fun}.
\end{proof}

\subsection{Integration of nonnegative simple function}
\label{ss:integration-of-nonnegative-simple-function}

\begin{lemma}[integral in~$\calSFplus$]
  \label{l:int-in-sfplus}
  \mbox{}\hfill
  Let~$(X,\Sigma,\mu)$ be a measure space.
  Let~$f\in\calSFplus$.\\
  Then, we have $f=\sum_{y\in f(X)}y\,\matUN_{f^{-1}(y)}$, and the sum
  $\sum_{y\in f(X)}y\,\mu(f^{-1}(y))$ is well-defined in~$\matRbarplus$.

  The {\em integral of~$f$ (for the measure~$\mu$)} is still denoted
  $\int f\,d\mu$;
  it is defined by
  \begin{equation}
    \label{e:int-in-sfplus}
    \int f \, d\mu
    \eqdef \sum_{y \in f (X)} y \, \mu \left( f^{-1} (y) \right)
    \quad \in \matRbarplus.
  \end{equation}
\end{lemma}

\begin{proof}
  Direct consequence of
  Lemma~\thref{l:sfplus-can-repr},
  Definition~\threfc{d:meas}{nonnegativeness},
  Lemma~\threfc{l:mult-in-rbarplus-is-closed-mt}{%
    $a_i\in\matRplus\subset\matRbarplus$}, and
  Lemma~\thref{l:add-in-rbarplus-is-closed}.
\end{proof}

\begin{lemma}[integral in~$\calSFplus$ generalizes integral in~$\calIF$]
  \label{l:int-in-sfplus-gen-int-in-if}
  \mbox{}\\
  Let~$(X,\Sigma,\mu)$ be a measure space.
  Let~$f\in\calIF$.
  Then, the values of $\int f\,d\mu$ provided by
  Definition~\thref{d:int-in-if}, and
  Lemma~\thref{l:int-in-sfplus}
  coincide.

  In other terms, for all~$A\in\Sigma$, we have $\matUN_A\in\calSFplus$ and
  $\int\matUN_A\,d\mu=\mu(A)$.
\end{lemma}

\begin{proof}
  Direct consequence of
  Definition~\threfc{d:if-set-of-meas-indic-funs}{$A\in\Sigma$},
  Lemma~\threfc{l:sfplus-simple-repr}{with $n=0$, $a_0=1$, and $A_0=A$},
  Lemma~\thref{l:equiv-def-of-int-in-if}, and
  Lemma~\threfc{l:int-in-sfplus}{%
    with $f(X)=\{0,1\}$, $f^{-1}(0)=A^c$, and $f^{-1}(1)=A$}.
\end{proof}

\begin{lemma}[equivalent definition of the integral in~$\calSFplus$ (disjoint)]
  \label{l:equiv-def-of-int-in-sfplus-disj}
  \mbox{}\\
  Let~$(X,\Sigma,\mu)$ be a measure space.
  Let~$f\in\calSFplus$.
  Let~$n\in\matN$, $(a_i)_{i\in[0..n]}\in\matRplus$, and
  $(A_i)_{i\in[0..n]}\in\Sigma$.
  Assume that $f=\sum_{i\in[0..n]}a_i\,\matUN_{A_i}$ is a disjoint
  representation.
  Then, we have
  \begin{equation}
    \label{e:equiv-def-of-int-in-sfplus-disj}
    \int f \, d\mu = \sum_{i \in [0..n]} a_i \, \mu (A_i).
  \end{equation}
\end{lemma}

\begin{proof}
  From
  Lemma~\thref{l:sfplus-can-repr},
  let~$m\in\matN$, $(b_j)_{j\in[0..m]}\in\matRplus$, and
  $(B_j)_{j\in[0..m]}\in\Sigma$ such that
  $f=\sum_{j\in[0..m]}b_j\,\matUN_{B_j}$ is the canonical representation.\\
  For all~$j\in[0..m]$,
  let $\Ip_j\eqdef\{i\in[0..n]\st\emptyset\not=A_i\subset B_j\}$.
  Let $\Ip\eqdef\{i\in[0..n]\st A_i\not=\emptyset\}$.\\
  Then, from
  Lemma~\thref{l:int-in-sfplus},
  Lemma~\threfc{l:sfplus-disj-repr-is-subpart-of-can-repr}{both properties},
  Definition~\threfc{d:meas}{$\sigma$-additivity and $\mu(\emptyset)=0$},
  Definition~\threfc{d:sigma-add-over-meas-space}{with $I\eqdef \Ip_j$}, and
  \assume{left distributivity of multiplication over addition in~$\matR$},
  we have
  \begin{align*}
    \int f \, d\mu
    & = \sum_{j \in [0..m]} b_j \, \mu (B_j)
    = \sum_{j \in [0..m]} b_j \, \mu \left( \biguplus_{i \in \Ip_j} A_i \right)\\
    & = \sum_{j \in [0..m]} \sum_{i \in \Ip_j} a_i \, \mu (A_i)
    = \sum_{i \in \Ip} a_i \, \mu (A_i)
    = \sum_{i \in [0..n]} a_i \, \mu (A_i).
  \end{align*}
\end{proof}

\begin{remark}
  \label{r:v2-new15}
  The next Lemmas~\ref{l:int-in-sfplus-is-add}
  and~\ref{l:int-in-sfplus-is-add-alt-proof} state the same result (additivity
  of the integral in $\calSFplus$), but their proofs are different: the first
  one relies on the disjoint representation of simple functions, whereas the
  second one sticks to the canonical representation and uses a somewhat tedious
  change of variables.
\end{remark}

\begin{lemma}[integral in~$\calSFplus$ is additive]
  \label{l:int-in-sfplus-is-add}
  \mbox{}\\
  Let~$(X,\Sigma,\mu)$ be a measure space.
  Let~$f,g\in\calSFplus$.
  Then, $f+g\in\calSFplus$, and we have
  \begin{equation}
    \label{e:int-in-sfplus-is-add}
    \int (f + g) \, d\mu = \int f \, d\mu + \int g \, d\mu.
  \end{equation}
\end{lemma}

\begin{proof}
  From
  Lemma~\thref{l:sfplus-is-closed-under-pos-alg-ops},
  we have $f+g\in\calSFplus$.

  From
  Lemma~\thref{l:sfplus-disj-repr},
  let $n,m\in\matN$, $(a_i)_{i\in[0..n]},(b_j)_{j\in[0..m]}\in\matR$ and
  $(A_i)_{i\in[0..n]},(B_j)_{j\in[0..m]}\in\Sigma$ such that
  $f=\sum_{i\in[0..n]}a_i\,\matUN_{A_i}$ and
  $g=\sum_{j\in[0..m]}b_j\,\matUN_{B_j}$ are disjoint representations.
  In particular, we have
  $X=\biguplus_{i\in[0..n]}A_i=\biguplus_{j\in[0..m]}B_j$.

  Let~$N\eqdef nm+n+m$.
  Since $N+1=(n+1)(m+1)$, there exists a
  bijection~$\fhi$ from~$[0..N]$ to~$[0..n]\times[0..m]$.
  For all $(i,j)\in[0..n]\times[0..m]$, let $c_{i,j}\eqdef a_i+b_j$ and
  $C_{i,j}\eqdef A_i\cap B_j$.
  Then, from
  Lemma~\threfc{l:sfplus-is-closed-under-pos-alg-ops}{%
    addition},
  the representation $f+g=\sum_{k\in[0..N]}c_{\fhi(k)}\,\matUN_{C_{\fhi(k)}}$
  is also disjoint.
  Hence, from
  Lemma~\thref{l:equiv-def-of-int-in-sfplus-disj},
  \assume{associativity and commutativity of addition in~$\matR$
    ($\sum_k=\sum_{i,j}$)},
  Lemma~\threfc{l:meas-over-count-pseudopart}{%
    first with $A_i$ and the pseudopartition $(B_j)_{j\in[0..m]}$,
    then with $B_j$ and the pseudopartition $(A_i)_{i\in[0..n]}$}, and
  \assume{left distributivity of multiplication over addition in~$\matR$},
  we have
  \begin{gather*}
    \int (f + g) \, d\mu
    = \sum_{k \in [0..N]} c_{\fhi (k)} \, \mu (C_{\fhi (k)})
    = \sum_{(i, j) \in [0..n] \times [0..m]} c_{i, j} \, \mu (C_{i, j}),\\
    \int f \, d\mu
    = \sum_{i \in [0..n]} a_i \, \mu (A_i)
    = \sum_{(i, j) \in [0..n] \times [0..m]} a_i \, \mu (C_{i, j}),\\
    \int g \, d\mu
    = \sum_{j \in [0..m]} b_j \, \mu (B_j)
    = \sum_{(i, j) \in [0..n] \times [0..m]} b_j \, \mu (C_{i, j}).
  \end{gather*}
  Therefore, from
  \assume{left distributivity of multiplication over addition in~$\matR$},
  we have the equality.
\end{proof}

\begin{lemma}[decomposition of measure in~$\calSFplus$]
  \label{l:decomp-of-meas-in-sfplus}
  \mbox{}\\
  Let~$(X,\Sigma,\mu)$ be a measure space.
  Let~$f,g\in\calSFplus$.
  Let~$y\in f(X)$.
  Then, we have
  \begin{equation}
    \label{e:decomp-of-meas-in-sfplus}
    \mu \left( f^{-1}(y) \right)
    = \sum_{z \in g (X)}
    \mu \left( f^{-1} (y) \cap g^{-1} (z) \right).
  \end{equation}
\end{lemma}

\begin{proof}
  Direct consequence of
  Lemma~\thref{l:sfplus-is-meas},
  Lemma~\threfc{l:some-borel-subsets}{singletons are measurable},
  Lemma~\threfc{l:inverse-image-is-meas-in-r}{%
    $f^{-1}(y),g^{-1}(z)\in\Sigma$},
  Lemma~\threfc{l:sf-can-repr}{%
    $(g^{-1}(z))_{z\in g(X)}$ form a partition of $X$}, and
  Lemma~\threfc{l:meas-over-count-pseudopart}{%
    with $A\eqdef f^{-1}(y)$, $B_i\eqdef g^{-1}(z)$ and $\card(I)$ equals
    $\card(g(X))$}.
\end{proof}

\begin{lemma}[change of variable in sum in~$\calSFplus$]
  \label{l:change-of-variable-in-sum-in-sfplus}
  \mbox{}\\
  Let~$(X,\Sigma,\mu)$ be a measure space.
  Let~$f,g\in\calSFplus$.
  Let~$y\in f(X)$.
  Then, we have
  \begin{equation}
    \label{e:change-of-variable-in-sum-in-sfplus}
    \sum_{z \in g (X)} (y + z) \,
    \mu \left( f^{-1} (y) \cap g^{-1} (z) \right)
    = \sum_{t \in (f + g) (X)} t \,
    \mu \left( f^{-1} (y) \cap (f + g)^{-1} (t) \right).
  \end{equation}
\end{lemma}

\begin{proof}
  For all $z,t\in\matR$, let $A(z)\eqdef f^{-1}(y)\cap g^{-1}(z)$
  and $B(t)\eqdef f^{-1}(y)\cap(f+g)^{-1}(t)$.
  Then, from
  Lemma~\thref{l:sfplus-is-closed-under-pos-alg-ops}, and
  Lemma~\thref{l:sfplus-is-meas},
  we have $f,g,f+g\in\calSFplus\subset\calMplus\subset\calM$.

  Let~$z,t\in\matR$.
  Then, from
  Definition~\thref{d:m-set-of-meas-num-funs},
  Lemma~\threfc{l:inverse-image-is-meas-in-r}{%
    $f^{-1}(y),\, g^{-1}(z),\, (f+g)^{-1}(t) \in\Sigma$}, and
  Lemma~\threfc{l:equiv-def-of-sigma-alg}{%
    closedness under countable intersection},
  we have $A(z),B(t)\in\Sigma$.

  Let~$\calA\eqdef\{z\in g(X)\st\mu(A(z))>0\}$ and
  $\calB\eqdef\{t\in(f+g)(X)\st\mu(B(t))>0\}$.

  \proofparskip{(1). $\forall z\in g(X)$, $A(z)=B(y+z)$}

  Direct consequence of
  the definition of~$A(z)$ and~$B(y+z)$,
  \assume{the definition of the addition of functions to~$\matR$}, and
  \assume{additive abelian group properties of~$\matR$
    (with $y\not=\infty$)}.

  \proofparskip{(2). $\forall t\in(f+g)(X)$, $B(t)=A(-y+t)$}

  Direct consequence of
  the definition of~$B(t)$ and~$A(-y+t)$,
  \assume{the definition of the addition of functions to~$\matR$}, and
  \assume{additive abelian group properties of~$\matR$
    (with $y\not=\infty$)}.

  \proofparskip{(3). $\tau_y=(z\mapsto y+z):\ArAB$ is a bijection}

  From
  \assume{additive abelian group properties of~$\matR$
    (with $y\not=\infty$)},
  the translation~$\tau_y$ is a bijection from~$\matR$ onto itself.

  Let~$z\in\calA$.
  Then, from
  the definition of~$\calA$,
  we have $z\in g(X)$ and $\mu(A(z))>0$.
  Thus, from
  Definition~\threfc{d:meas}{$\mu(\emptyset)=0$ (contrapositive)},
  we have $A(z)\not=\emptyset$.
  Let~$x\in A(z)$.
  Then, from
  the definition of~$A(z)$,
  we have $f(x)=y$ and $g(x)=z$.
  Thus, from
  the definition of the addition of functions to~$\matR$,
  we have $(f+g)(x)=y+z$, {\ie} $y+z\in(f+g)(X)$.
  Moreover, from~(1),
  we have $\mu(B(y+z))=\mu(A(z))>0$.
  Thus, from
  the definition of~$\calB$,
  we have $\tau_y(z)=y+z\in\calB$.
  Hence, we have $\tau_y(\calA)\subset\calB$.

  Conversely, let $t\in\calB$.
  Then, from
  the definition of~$\calB$,
  we have $t\in(f+g)(X)$ and $\mu(B(t))>0$.
  Thus, from
  Definition~\threfc{d:meas}{$\mu(\emptyset)=0$ (contrapositive)},
  we have $B(t)\not=\emptyset$.
  Let~$x\in B(t)$.
  Then, from
  the definition of~$B(t)$,
  we have $f(x)=y$ and $(f+g)(x)=t$.
  Thus, from
  the definition of the addition of functions to~$\matR$, and
  \assume{additive abelian group properties of~$\matR$
    (with $y\not=\infty$)},
  we have $g(x)=-y+t$, {\ie} $-y+t\in g(X)$.
  Moreover, from~(2),
  we have $\mu(A(-y+t))=\mu(B(t))>0$.
  Thus, from
  the definition of~$\calA$,
  we have $\tau_y^{-1}(t)=-y+t\in\calA$.
  Then, from
  \assume{properties of inverse functions},
  we have $t=\tau_y(\tau_y^{-1}(t))\in\tau_y(\calA)$.
  Hence, we have $\calB\subset\tau_y(\calA)$.

  Finally, $\tau_y$~is a bijection from~$\calA$ onto $\tau_y(\calA)=\calB$.

  \medskip
  Therefore, from
  the definitions of~$\calA$ and~$\tau_y$,
  (1), (3), and
  the definition of~$\calB$,
  we have
  \begin{align*}
    \sum_{z \in g (X)} (y + z) \, \mu (A (z))
    & = \sum_{z \in \calA} (y + z) \, \mu (A (z))
      = \sum_{z \in \calA} \tau_y (z) \, \mu (B (\tau_y (z)))\\
    & = \sum_{t \in \tau_y (\calA)} t \, \mu (B (t))
      = \sum_{t \in \calB} t \, \mu (B (t))
      = \sum_{t \in (f + g) (X)} t \, \mu (B (t)).
  \end{align*}
\end{proof}

\begin{remark}
  Note that the previous lemma is still valid when~$f$ and~$g$ are functions
  with possibly changing sign, and $y\in\matR$.
  Moreover, the equalities $A(z)=B(y+z)$ and $B(t)=A(-y+t)$ in the proof are
  still valid for any $z,t\in\matR$, in which case these subsets may be empty.
\end{remark}

\begin{lemma}[integral in~$\calSFplus$ is additive (alternate proof)]
  \label{l:int-in-sfplus-is-add-alt-proof}
  \mbox{}\\
  Let~$(X,\Sigma,\mu)$ be a measure space.
  Let~$f,g\in\calSFplus$.
  Then, $f+g\in\calSFplus$, and we have
  \begin{equation}
    \label{e:int-in-sfplus-is-add-alt-proof}
    \int (f + g) \, d\mu = \int f \, d\mu + \int g \, d\mu.
  \end{equation}
\end{lemma}

\begin{proof}
  From
  Lemma~\thref{l:sfplus-is-closed-under-pos-alg-ops}, and
  Lemma~\thref{l:sfplus-is-meas},
  we have $f,g,f+g\in\calSFplus\subset\calMplus\subset\calM$.

  Let~$y,z,t\in\matR$.
  Then, from
  Definition~\thref{d:m-set-of-meas-num-funs},
  Lem\-ma~\threfc{l:inverse-image-is-meas-in-r}{%
    $f^{-1}(y),g^{-1}(z),(f+g)^{-1}(t)\in\Sigma$}, and
  Lemma~\threfc{l:equiv-def-of-sigma-alg}{%
    closedness under countable intersection},
  we have
  \begin{equation*}
    f^{-1} (y) \cap g^{-1} (z), f^{-1}(y) \cap(f + g)^{-1} (t) \in \Sigma.
  \end{equation*}
  From
  Lemma~\thref{l:int-in-sfplus},
  Lemma~\thref{l:decomp-of-meas-in-sfplus}, and
  Lem\-ma~\thref{l:mult-in-rbarplus-is-distr-over-add-mt},
  we have
  \begin{equation*}
    \int f \, d\mu
    = \sum_{y \in f (X)} y \, \mu \left( f^{-1} (y) \right)
    = \sum_{y \in f (X)} \sum_{z \in g (X)}
    y \, \mu \left( f^{-1} (y) \cap g^{-1} (z) \right).
  \end{equation*}
  In the very same way, we have
  \begin{equation*}
    \int g \, d\mu
    = \sum_{z \in g (X)} z \, \mu \left( g^{-1} (z) \right)
    = \sum_{z \in g (X)} \sum_{y \in f (X)}
    z \, \mu \left( g^{-1} (z) \cap f^{-1} (y) \right).
  \end{equation*}
  Therefore, from
  Lemma~\thref{l:add-in-rbarplus-is-assoc},
  Lemma~\thref{l:add-in-rbarplus-is-comm},
  \assume{commutativity of intersection},
  Lemma~\thref{l:mult-in-rbarplus-is-distr-over-add-mt},
  Lemma~\thref{l:change-of-variable-in-sum-in-sfplus},
  Lemma~\threfc{l:decomp-of-meas-in-sfplus}{%
    with $(f+g)$, $f$ and $t$}, and
  Lemma~\thref{l:int-in-sfplus},
  we have
  \begin{align*}
    \int f \, d\mu + \int g \, d\mu
    & = \sum_{y \in f (X)} \sum_{z \in g (X)} (y + z) \,
      \mu \left( f^{-1} (y) \cap g^{-1} (z) \right)\\
    & = \sum_{y \in f (X)} \sum_{t \in (f + g) (X)} t \,
      \mu \left( f^{-1} (y) \cap (f + g)^{-1} (t) \right)\\
    & = \sum_{t \in (f + g) (X)} t \, \sum_{y \in f (X)}
      \mu \left( (f + g)^{-1} (t) \cap f^{-1} (y) \right)\\
    & = \sum_{t \in (f + g) (X)} t \,
      \mu \left( (f + g)^{-1} (t) \right)\\
    & = \int (f + g) \, d\mu.
  \end{align*}
\end{proof}

\begin{lemma}[integral in~$\calSFplus$ is positive linear]
  \label{l:int-in-sfplus-is-pos-lin}
  \mbox{}\hfill
  Let~$(X,\Sigma,\mu)$ be a measure space.
  Let~$f,g\in\calSFplus$.
  Let~$a\in\matRplus$.
  Then, $f+g,af\in\calSFplus$, and we have
  \begin{equation}
    \label{e:int-in-sfplus-is-pos-lin}
    \int (f + g) \, d\mu = \int f \, d\mu + \int g \, d\mu
    \AND
    \int a f \, d\mu = a \int f \, d\mu.
  \end{equation}
\end{lemma}

\begin{proof}
  \proofpar{Addition}
  Direct consequence of
  Lemma~\thref{l:int-in-sfplus-is-add}, or
  Lemma~\thref{l:int-in-sfplus-is-add-alt-proof}.

  \proofparskip{Nonnegative scalar multiplication}\\
  From
  Lemma~\thref{l:sfplus-is-closed-under-pos-alg-ops},
  we have $af\in\calSFplus$.

  \proofparskip{Case $a=0$}
  Then, from
  \assume{field properties of~$\matR$ ($0f\equiv0$)}, and
  \assume{the definition of the indicator function
    ($\matUN_\emptyset\equiv0$)},
  we have $0f=\matUN_\emptyset$.
  Hence, from
  Lemma~\thref{l:int-in-sfplus-gen-int-in-if},
  Definition~\threfc{d:meas}{homogeneity}, and
  Lemma~\thref{l:zero-prod-prop-in-rbarplus-mt},
  we have
  \begin{equation*}
    \int 0 f \, d\mu
    = \int \matUN_\emptyset \, d\mu
    = \mu (\emptyset)
    = 0
    = 0 \times \int f \, d\mu.
  \end{equation*}

  \proofparskip{Case $a>0$}
  Let~$y\in f(X)$.
  Let~$z=ay\in (af)(X)$.
  Then, from
  Definition~\threfc{d:mult-in-rbar}{%
    for all $a\in\matRplusstar$, $\frac{a}{a}=1$},
  we have $(af)^{-1}(z)=f^{-1}(y)$
  Hence, from
  Lemma~\thref{l:int-in-sfplus},
  we have
  \begin{equation*}
    \int a f \, d\mu
    = \sum_{z \in (af) (X)} z \, \mu ((af)^{-1} (z))
    = \sum_{y \in f (X)} a y \, \mu (f^{-1} (y))
    = a \int f \, d\mu.
  \end{equation*}
\end{proof}

\begin{lemma}[equivalent definition of the integral in~$\calSFplus$ (simple)]
  \label{l:equiv-def-of-int-in-sfplus-simple}
  \mbox{}\\
  Let~$(X,\Sigma,\mu)$ be a measure space.
  Let~$f\in\calSFplus$.
  Let~$n\in\matN$, $(a_i)_{i\in[0..n]}\in\matRplus$, and
  $(A_i)_{i\in[0..n]}\in\Sigma$.
  Assume that $f=\sum_{i\in[0..n]}a_i\,\matUN_{A_i}$ is a simple
  representation.
  Then, we have
  \begin{equation}
    \label{e:equiv-def-of-int-in-sfplus-simple}
    \int f \, d\mu = \sum_{i \in [0..n]} a_i \, \mu (A_i).
  \end{equation}
\end{lemma}

\begin{proof}
  Direct consequence of
  Lemma~\threfc{l:sfplus-simple-repr}{%
    thus such representation exists},
  Lemma~\thref{l:int-in-sfplus-is-pos-lin}, and
  Lemma~\thref{l:int-in-sfplus-gen-int-in-if}.
\end{proof}

\begin{lemma}[integral in~$\calSFplus$ is monotone]
  \label{l:int-in-sfplus-is-monot}
  \mbox{}\\
  Let~$(X,\Sigma,\mu)$ be a measure space.
  Let~$f,g\in\calSFplus$.
  Then, we have
  \begin{equation}
    \label{e:int-in-sfplus-is-monot}
    f \leq g \IMPLIES \int f \, d\mu \leq \int g \, d\mu.
  \end{equation}
\end{lemma}

\begin{proof}
  Assume that $f\leq g$.
  Then, from
  Definition~\thref{d:sfplus-subset-of-nonneg-simple-funs},
  Lemma~\thref{l:sf-is-alg-over-r},
  Definition~\threfc{d:alg-over-a-field}{$\calSF$ is a vector space}, and
  Definition~\threfc{LM-d:space}{$(\calSF,+)$ is an abelian group},
  we have $g=f+(g-f)$ with $f,g-f\in\calSFplus$.
  Therefore, from
  Lemma~\thref{l:int-in-sfplus-is-pos-lin}, and
  Lemma~\threfc{l:int-in-sfplus}{%
    nonnegativeness with $g-f\in\calSFplus$},
  we have
  \begin{equation*}
    \int g \, d\mu
    = \int f \, d\mu + \int (g - f) \, d\mu
    \geq \int f \, d\mu.
  \end{equation*}
\end{proof}

\begin{lemma}[integral in~$\calSFplus$ is continuous]
  \label{l:int-in-sfplus-is-cont}
  \mbox{}\\
  Let~$(X,\Sigma,\mu)$ be a measure space.
  Let~$f\in\calSFplus$.
  Then, we have
  \begin{equation}
    \label{e:int-in-sfplus-is-cont}
    \int f \, d\mu
    = \sup_{\substack{\fhi \in \calSFplus\\\fhi \leq f}} \int \fhi \, d\mu.
  \end{equation}
\end{lemma}

\begin{proof}
  From
  Definition~\threfc{LM-d:supremum}{upper bound}, and
  since $f\in\calSFplus$, we have
  \begin{equation*}
    \int f \, d\mu
    \leq \sup_{\substack{\fhi \in \calSFplus\\\fhi \leq f}} \int \fhi \, d\mu.
  \end{equation*}
  Conversely, let $\fhi\in\calSFplus$ such that $\fhi\leq f$.
  Then, from
  Lemma~\thref{l:int-in-sfplus-is-monot}, and
  Definition~\threfc{LM-d:supremum}{least upper bound},
  we have
  \begin{equation*}
    \sup_{\substack{\fhi \in \calSFplus\\\fhi \leq f}} \int \fhi \, d\mu
    \leq \int f \, d\mu.
  \end{equation*}

  Therefore, we have the equality.
\end{proof}

\begin{lemma}[integral in~$\calSFplus$ over subset]
  \label{l:int-in-sfplus-over-subset}
  \mbox{}\hfill
  Let~$(X,\Sigma,\mu)$ be a measure space.\\
  Let~$A\in\Sigma$.
  Let~$Y\subset X$ such that $A\subset Y$.
  Let~$f:\ArYRb$.
  Let~$\hf:\ArXRb$.
  Assume that $\restr{\hf}{Y}=f$.
  Then, we have $\restr{f}{A}\in\calSFplus(A,\Sigma\olcap A)$ iff
  $\hf\,\matUN_A\in\calSFplus(X,\Sigma)$.
  If so, we have
  \begin{equation}
    \label{e:int-in-sfplus-over-subset}
    \int \restr{f}{A} \, d\mu_A = \int \hf \, \matUN_A \, d\mu.
  \end{equation}
  This integral is still denoted $\int_Af\,d\mu$;
  it is still called {\em integral of~$f$ over~$A$}.
\end{lemma}

\begin{proof}
  \proofpar{Equivalence}
  Direct consequence of
  Lemma~\thref{l:sf-is-closed-under-ext-by-zero}, and
  Lemma~\thref{l:sf-is-closed-under-restr}.

  \proofparskip{Identity}
  Direct consequence of
  Definition~\threfc{d:sf-vector-space-of-simple-funs}{%
    $\restr{f}{A}$ is a linear combination of $f_i$'s in
    $\calIF(A,\Sigma\olcap A)$}
  Lemma~\thref{l:int-in-sfplus-is-pos-lin},
  Lemma~\thref{l:int-in-sfplus-gen-int-in-if}, and
  Lemma~\threfc{l:int-in-if-over-subset}{%
    with $\hf_i:\ArXRb$ such that $\restr{(\hf_i)}{Y}=f_i$}.
\end{proof}

\begin{lemma}[integral in~$\calSFplus$ over subset is additive]
  \label{l:int-in-sfplus-over-subset-is-add}
  \mbox{}\hfill
  Let~$(X,\Sigma,\mu)$ be a measure space.\\
  Let~$n\in\matN$.
  Let~$A,(A_i)_{i\in[0..n]}\in\Sigma$.
  Assume that $(A_i)_{i\in[0..n]}$ is a pseudopartition of~$A$.
  Let~$Y\subset X$ such that $A\subset Y$.
  Let~$f:\ArYRb$.
  Let~$\hf:\ArXRb$.
  Assume that $\restr{\hf}{Y}=f$.\\
  Then, $\hf\,\matUN_A\in\calSFplus$ iff
  for all $i\in[0..n]$, $\hf\,\matUN_{A_i}\in\calSFplus$.
  If so, we have
  \begin{equation}
    \label{e:int-in-sfplus-over-subset-is-add}
    \int_A f \, d\mu = \sum_{i \in[0..n]} \int_{A_i} f \, d\mu.
  \end{equation}
\end{lemma}

\begin{proof}
  \proofpar{(1)}
  Let~$W\in\Sigma$.
  Assume that $W\subset Y$.
  Then, from
  Definition~\thref{d:sf-vector-space-of-simple-funs}, and
  Definition~\thref{d:sfplus-subset-of-nonneg-simple-funs},
  we have $\hf\,\matUN_W\in\calSFplus$ iff
  there exist $p\in\matN$, $(a_j)_{j\in[0..p]}\in\matRplus$, and
  $(f_j)_{j\in[0..p]}\in\calIF$ such that
  \begin{equation*}
    \hf \, \matUN_W = \sum_{j \in [0..p]} a_j f_j.
  \end{equation*}

  \proofparskip{(2)}
  Let~$g\in\calIF$.
  Let~$i\in[0..n]$.
  Then, from
  Definition~\threfc{d:if-set-of-meas-indic-funs}{%
    $g=\matUN_B$ with $B\in\Sigma$}, and
  Lemma~\thref{l:if-is-closed-under-mult},
  we have~$g\,\matUN_{A_i}=\matUN_{B\cap A_i}\in\calIF$.

  \proofparskip{``Left'' implies ``right''}
  Assume that $\hf\,\matUN_A\in\calSFplus$.\\
  Then, from~(1) (with $W\eqdef A$), let $p\in\matN$,
  $(a_j)_{j\in[0..p]}\in\matRplus$, and $(f_j)_{j\in[0..p]}\in\calIF$ such that
  \begin{equation*}
    \hf \, \matUN_A = \sum_{j \in [0..p]} a_j f_j.
  \end{equation*}
  Let~$i\in[0..n]$.
  Let~$j\in[0..p]$.
  Let~$g^i_j\eqdef f_j\,\matUN_{A_i}$.
  Then, from~(2) (with $g\eqdef f_j$),
  we have~$g^i_j\in\calIF$.
  Moreover, since $A_i\subset A$ ({\ie} $A_i=A\cap A_i$), from
  \assume{field properties of~$\matR$}, and
  Lemma~\thref{l:if-is-closed-under-mult}
  we have
  \begin{equation*}
    \hf \, \matUN_{A_i}
    = \hf \, \matUN_{A \cap A_i}
    = \left( \hf \, \matUN_A \right) \, \matUN_{A_i}
    = \sum_{j \in [0..p]} a_j f_j \, \matUN_{A_i}
    = \sum_{j \in [0..p]} a_j g^i_j.
  \end{equation*}
  Hence, from~(1) (with $W\eqdef A_i$), we have
  $\hf\,\matUN_{A_i}\in\calSFplus$.

  \proofparskip{``Right'' implies ``left''}
  Assume that for all $i\in[0..n]$,
  $\hf\,\matUN_{A_i}\in\calSFplus$.\\
  Then, from~(1) (with $W\eqdef A_i$), for all $i\in[0..n]$, let
  $p^i\in\matN$, $(a^i_{j^i})_{j^i\in[0..p^i]}\in\matRplus$, and
  $(f^i_{j^i})_{j^i\in[0..p^i]}$ in~$\calIF$ such that
  $\hf\,\matUN_{A_i}=\sum_{j^i\in[0..p^i]}a^i_{j^i}f^i_j$.
  Let $p \eqdef -1 + \sum_{i \in [0..n]} (1 + p^i)$,
  \begin{equation*}
    (a_j)_{j \in [0..p]}
    \eqdef \left(
      (a^i_{j^i})_{j^i \in [0..p^i]} \right)_{i \in [0..n]}
    \in \matRplus
    \AND
    (f_j)_{j \in [0..p]}
    \eqdef \left(
      (f^i_{j^i} \, \matUN_{A_i})_{j^i \in [0..p^i]} \right)_{i \in [0..n]}
    \in \calIF.
  \end{equation*}
  Then, from
  Lemma~\threfc{l:if-is-sigma-add}{with $I\eqdef[0..n]$}, and
  \assume{field properties of~$\matR$}, and~(2) (with $g\eqdef f^i_{j^i}$),
  we have
  \begin{equation*}
    \hf \, \matUN_A
    = \sum_{i \in [0..n]} \hf \, \matUN_{A_i}
    = \sum_{i \in [0..n]} \sum_{j^i \in [0..p^i]}
    a^i_{j^i} (f^i_j \, \matUN_{A_i})
    = \sum_{j \in [0..p]} a_j f_j.
  \end{equation*}
  Hence, from~(1) (with $W\eqdef A$), we have $\hf\,\matUN_A\in\calSFplus$.

  \medskip\noindent
  Therefore, we have the equivalence.

  \proofparskip{Identity}
  Direct consequence of
  Lemma~\threfc{l:int-in-sfplus-over-subset}{%
    with $A$ and the $A_i$'s}, and
  Lemma~\threfc{l:if-is-sigma-add}{with $I\eqdef[0..n]$}.
\end{proof}

\begin{lemma}[integral in~$\calSFplus$ for counting measure]
  \label{l:int-in-sfplus-for-count-meas}
  \mbox{}\\
  Let~$(X,\Sigma)$ be a measurable space.
  Let~$Y\subset X$.
  Let~$f\in\calSFplus$.
  Then, we have
  \begin{equation}
    \label{e:int-in-sfplus-for-count-meas}
    \int f \, d\delta_Y = \sum_{y \in Y} f (y).
  \end{equation}
\end{lemma}

\begin{proof}
  Direct consequence of
  Lemma~\thref{l:sfplus-simple-repr},
  Lemma~\thref{l:int-in-sfplus-gen-int-in-if},
  Lemma~\thref{l:int-in-if-for-count-meas}, and
  \assume{associativity and commutativity of (possibly uncountable) addition
    in~$\matRplus$}.
\end{proof}

\begin{lemma}[integral in~$\calSFplus$ for counting measure on~$\matN$]
  \label{l:int-in-sfplus-for-count-meas-on-n}
  \mbox{}\\
  Let~$f\in\calSFplus(\matN,\calP(\matN))$.
  Then, we have
  \begin{equation}
    \label{e:int-in-sfplus-for-count-meas-on-n}
    \int f \, d\delta_\matN = \sum_{n \in \matN} f (n).
  \end{equation}
\end{lemma}

\begin{proof}
  Direct consequence of
  Lemma~\threfc{l:int-in-sfplus-for-count-meas}{%
    with $Y\!=\!X\eqdef\matN$ and $\Sigma\eqdef\calP(\matN)$}.
\end{proof}

\begin{lemma}[integral in~$\calSFplus$ for Dirac measure]
  \label{l:int-in-sfplus-for-dirac-meas}
  \mbox{}\\
  Let~$(X,\Sigma)$ be a measurable space.
  Let~$\{a\}\in\Sigma$.
  Let~$f\in\calSFplus$.
  Then, we have
  \begin{equation}
    \label{e:int-in-sfplus-for-dirac-meas}
    \int f \, d\delta_a = f (a).
  \end{equation}
\end{lemma}

\begin{proof}
  Direct consequence of
  Definition~\thref{d:dirac-meas}, and
  Lemma~\threfc{l:int-in-sfplus-for-count-meas}{%
    with $Y\eqdef\{a\}$}.
\end{proof}

\clearpage
\section{Integration of nonnegative measurable functions}
\label{s:integration-of-nonnegative-measurable-functions}

\begin{remark}
  In this section, functions take their values in~$\matRbarplus$, and the
  expressions involving integrals are also taken in~$\matRbarplus$.
\end{remark}

\begin{lemma}[integral in~$\calMplus$]
  \label{l:int-in-mplus}
  \mbox{}\hfill
  Let~$(X,\Sigma,\mu)$ be a measure space.
  Let~$f\in\calMplus$.\\
  Then, $\{\int\fhi\,d\mu\st\fhi\in\calSFplus\Conj\fhi\leq f\}$ admits a
  supremum.

  The {\em integral of~$f$ (for the measure~$\mu$)} is still denoted
  $\int f\,d\mu$;
  it is defined by
  \begin{equation}
    \label{e:int-in-mplus}
    \int f \, d\mu
    \eqdef \sup_{\substack{\fhi \in \calSFplus\\\fhi \leq f}} \int \fhi \, d\mu
    \quad \in \matRbarplus.
  \end{equation}

  A function~$f:\ArXRbp$ is said {\em $\mu$-integrable (in~$\calMplus$)} iff
  $f\in\calMplus$ and $\int f\,d\mu<\infty$.
\end{lemma}

\begin{proof}
  Direct consequence of
  Lemma~\thref{l:int-in-sfplus}, and
  Definition~\threfc{LM-d:supremum}{%
    with the Lebesgue integral in~$\calSFplus$ over
    $\{\fhi\in\calSFplus\st\fhi\leq f\}$},
\end{proof}

\begin{lemma}[integral in~$\calMplus$ generalizes integral in~$\calSFplus$]
  \label{l:int-in-mplus-gen-int-in-sfplus}
  \mbox{}\\
  Let~$(X,\Sigma,\mu)$ be a measure space.
  Let~$f\in\calSFplus$.
  Then, the values of~$\int f\,d\mu$ provided by
  Lemma~\thref{l:int-in-sfplus}, and
  Lemma~\thref{l:int-in-mplus}
  coincide.
\end{lemma}

\begin{proof}
  Direct consequence of
  Lemma~\thref{l:int-in-sfplus},
  Lemma~\thref{l:int-in-sfplus-is-cont}, and
  Lemma~\thref{l:int-in-mplus}.
\end{proof}

\begin{lemma}[integral in~$\calMplus$ of indicator function]
  \label{l:int-in-mplus-of-indic-fun}
  \mbox{}\\
  Let~$(X,\Sigma,\mu)$ be a measure space.
  Let~$A\in\Sigma$.
  Then, we have $\matUN_A\in\calMplus$ and $\int\matUN_A\,d\mu=\mu(A)$.
\end{lemma}

\begin{proof}
  Direct consequence of
  Lemma~\thref{l:sfplus-is-meas},
  Lemma~\thref{l:int-in-mplus-gen-int-in-sfplus}, and
  Lemma~\thref{l:int-in-sfplus-gen-int-in-if}.
\end{proof}

\begin{lemma}[integral in~$\calMplus$ is positive homogeneous]
  \label{l:int-in-mplus-is-pos-hom}
  \mbox{}\\
  Let~$(X,\Sigma,\mu)$ be a measure space.
  Let~$f\in\calMplus$.
  Let~$a\in\matRplus$.
  Then, $af\in\calMplus$, and we have
  \begin{equation}
    \label{e:int-in-mplus-is-pos-hom}
    \int a f \, d\mu = a \int f \, d\mu.
  \end{equation}
\end{lemma}

\begin{proof}
  From
  Lemma~\thref{l:mplus-is-closed-under-nonneg-scalar-mult},
  $af\in\calMplus$.

  \proofparskip{Case $a=0$}
  Then, from
  Definition~\thref{d:sfplus-subset-of-nonneg-simple-funs},
  Definition~\thref{d:sf-vector-space-of-simple-funs}, and
  Definition~\threfc{LM-d:space}{$0\in\calSF$},
  we have
  \begin{equation*}
    0 f = 0 \in \calSFplus \subset \calMplus.
  \end{equation*}
  Hence, from
  Lemma~\thref{l:int-in-mplus},
  Lemma~\thref{l:equiv-def-of-int-in-sfplus-simple},
  Lemma~\thref{l:int-in-sfplus}, and
  Lemma~\thref{l:zero-prod-prop-in-rbarplus-mt},
  we have
  \begin{equation*}
    \int 0 f \, d\mu
    = \int 0 \, d\mu
    = 0
    = 0 \times \int f \, d\mu.
  \end{equation*}

  \proofparskip{Case $a>0$}
  Then, from
  Lemma~\thref{l:mult-in-rbarplus-is-assoc-mt},
  and
  Definition~\threfc{d:mult-in-rbar}{%
    for all $a\in\matRplusstar$, $\frac{a}{a}=1$},
  we have $(\star)\;\forall b\in\matRbarplus,\;\frac{1}{a}(ab)=b$.

  Let~$\fhi\in\calSFplus$ such that $\fhi\leq f$.
  Then, from
  Lemma~\thref{l:sfplus-is-closed-under-pos-alg-ops},
  we have $a\fhi\in\calSFplus$.
  Moreover, from
  \assume{compatibility of multiplication by a positive number with
    order in~$\matRbar$},
  property~($\star$),
  Lemma~\threfc{l:int-in-sfplus-is-pos-lin}{%
    positive homogeneity},
  Lemma~\thref{l:int-in-mplus}, and
  Definition~\threfc{LM-d:supremum}{upper bound},
  we have $a\fhi\leq af$ and
  \begin{equation*}
    \int \fhi \, d\mu
    = \frac{1}{a} \int a \fhi \, d\mu
    \leq \frac{1}{a} \int a f \, d\mu.
  \end{equation*}
  Thus, from
  Lemma~\thref{l:int-in-mplus},
  Definition~\threfc{LM-d:supremum}{least upper bound}, and
  property~($\star$),
  we have
  \begin{equation*}
    a \int f \, d\mu
    = a \sup_{\substack{\fhi \in \calSFplus\\\fhi \leq f}} \int \fhi \, d\mu
    \leq a \frac{1}{a} \int a f \, d\mu
    = \int a f \, d\mu.
  \end{equation*}
  Moreover, the same result also holds for function~$af\in\calMplus$ and
  number~$\frac{1}{a}>0$, and with property~($\star$), we have
  $\int af\,d\mu\leq a\int\frac{1}{a}(af)\,d\mu=a\int f\,d\mu$.
  Hence, we have the equality.

  Therefore, we always have the equality.
\end{proof}

\begin{lemma}[integral in~$\calMplus$ of zero is zero]
  \label{l:int-in-mplus-of-zero-is-zero}
  \mbox{}\\
   Let~$(X,\Sigma,\mu)$ be a measure space.
   Then, $0\in\calMplus$, and we have $\int0\,d\mu=0$.
\end{lemma}

\begin{proof}
  Direct consequence of
  Lemma~\threfc{l:int-in-mplus-is-pos-hom}{%
    with $a\eqdef0$},
  Definition~\thref{d:mult-in-rbar}, and
  Lemma~\thref{l:zero-prod-prop-in-rbarplus-mt}.
\end{proof}

\begin{lemma}[integral in~$\calMplus$ is monotone]
  \label{l:int-in-mplus-is-monot}
  \mbox{}\\
  Let~$(X,\Sigma,\mu)$ be a measure space.
  Let~$f,g\in\calMplus$.
  Then, we have
  \begin{equation}
    \label{e:int-in-mplus-is-monot}
    f \leq g \IMPLIES \int f \, d\mu \leq \int g \, d\mu.
  \end{equation}
\end{lemma}

\begin{proof}
  Assume that $f\leq g$.
  Then,
  $\{\fhi\in\calSFplus\st\fhi\leq f\}
  \subset\{\fhi\in\calSFplus\st\fhi\leq g\}$.
  Therefore, from
  Lemma~\thref{l:int-in-mplus}, and
  \assume{monotonicity of supremum},
  we have
  \begin{equation*}
    \int f \, d\mu \leq \int g \, d\mu.
  \end{equation*}
\end{proof}

\begin{remark}
  \label{r:v2-new16}
  The next proof follows step~3 of the Lebesgue scheme
  (see Section~\ref{s:lebesgue-scheme}).

  See also the sketch of the proof in
  Section~\ref{s:sketch-of-the-proof-of-the-beppo-levi-mono-cv-th}.
\end{remark}

\begin{theorem}[{\BL}, monotone convergence]
  \label{t:beppo-levi-monot-conv}
  \mbox{}\\
  Let~$(X,\Sigma,\mu)$ be a measure space.
  Let~$(f_n)_{n\in\matN}\in\calMplus$.
  Assume that the sequence is pointwise nondecreasing.
  Then, $\lim_{n\to\infty}f_n\in\calMplus$, and we have
  \begin{equation}
    \label{e:beppo-levi-monot-conv}
    \int \lim_{n \to \infty} f_n \, d\mu
    = \lim_{n \to \infty} \int f_n \, d\mu.
  \end{equation}
\end{theorem}

\begin{proof}
  From
  \assume{convergence of monotone sequences in~$\matRbar$}, and
  \assume{completeness of~$\matRbar$},
  the existence of the pointwise limit~$f$ is guaranteed in~$\matRbar$, and
  we have $f=\lim_{n\to\infty}f_n=\sup_{n\in\matN}f_n$.

  From
  Lemma~\thref{l:mplus-is-closed-under-limit-when-pointwise-conv},
  we have $f\in\calMplus$.
  Then, from
  Lemma~\thref{l:int-in-mplus},
  the integral of~$f$ is well-defined.
  Moreover, from
  Lemma~\thref{l:int-in-mplus-is-monot}, and
  \assume{completeness of the extended real numbers},
  the sequence $(\int f_n\,d\mu)_{n\in\matN}$ is nondecreasing, hence
  convergent in~$\matRbar$ towards
  \begin{equation*}
    \lim_{n \to \infty} \int f_n \, d\mu
    = \sup_{n \in \matN} \int f_n \, d\mu.
  \end{equation*}
  Thus, from
  Definition~\threfc{LM-d:supremum}{upper bound}, and
  Lemma~\thref{l:int-in-mplus-is-monot},
  for all $n\in\matN$, we have
  \begin{equation*}
    \int f_n \, d\mu
    \leq \int \sup_{n \in \matN} f_n \, d\mu
    = \int f \, d\mu.
  \end{equation*}
  Hence, from
  Definition~\threfc{LM-d:supremum}{least upper bound},
  we have
  \begin{equation*}
    \lim_{n \to \infty} \int f_n \, d\mu
    = \sup_{n \in \matN} \int f_n \, d\mu
    \leq \int f \, d\mu.
  \end{equation*}

  Let~$a\in(0,1)$.
  Let~$\fhi\in\calSFplus$.
  Assume that $\fhi\leq f$.
  Let~$n\in\matN$.
  Let $A_n\eqdef\{a\fhi\leq f_n\}\subset X$.
  Then, from
  Definition~\threfc{d:sf-vector-space-of-simple-funs}{%
    $-a\fhi\in\calSF$},
  Lemma~\threfc{l:sf-is-meas}{$-a\fhi\in\calM$}, and
  Lemma~\threfc{l:m-is-closed-under-add-when-defined}{%
    $-a\fhi$ takes finite values},
  we have $f_n-a\fhi\in\calM$.
  Thus, from
  Lemma~\threfc{l:meas-of-num-fun}{with $f_n-a\fhi$}, and
  Lemma~\thref{l:int-in-mplus-of-indic-fun},
  we have $A_n\in\Sigma$, and $\matUN_{A_n}\in\calMplus$.
  Moreover, from
  \assume{monotonicity of addition in ~$\matRbarplus$}, and
  \assume{transitivity of order},
  the sequence~$(A_n)_{n\in\matN}$ is nondecreasing.\\
  Let~$x\in X$.
  \proofpar{Case $\fhi(x)=0$}
  Then, for all $n\in\matN$, $x\in A_n$.
  \proofpar{Case $\fhi(x)>0$}
  Then, since $0<a<1$, we have $a\fhi(x)<\fhi(x)\leq f(x)$.
  Thus, from
  \assume{monotonicity of the limit},
  there exists $N\in\matN$ such that for all $n\geq N$,
  $a\fhi(x)\leq f_n(x)$, {\ie} $x\in A_N$.
  Hence, we have $X=\bigcup_{n\in\matN}A_n$.

  Let~$B\in\Sigma$.
  Then, from
  Lemma~\threfc{l:equiv-def-of-sigma-alg}{%
    closedness under countable intersection (with $\card(I)=2$)},
  \assume{monotonicity of intersection}, and
  \assume{distributivity of intersection over union},
  the sequence $(B\cap A_n)_{n\in\matN}$ belongs to $\Sigma$, is
  nondecreasing, and $\bigcup_{n\in\matN}(B\cap A_n)=B$.
  Thus, from
  Lemma~\thref{l:meas-is-cont-from-below}, and
  Definition~\thref{d:continuity-from-below},
  we have
  $\mu(B)=\mu\left(\bigcup_{n\in\matN}(B\cap A_n)\right)
  =\lim_{n\to\infty}\mu(B\cap A_n)$.

  Then, from
  Lemma~\thref{l:equiv-def-of-int-in-sfplus-simple},
  there exists $k\in\matN$, $(a_i)_{i\in[0..k]}\in\matRplus$, and
  $(B_i)_{i\in[0..k]}\in\Sigma$ such that
  \begin{equation*}
    \fhi = \sum_{i \in [0..k]} a_i \, \matUN_{B_i}
    \AND
    \int \fhi \, d\mu = \sum_{i \in [0..k]} a_i \mu (B_i).
  \end{equation*}
  Thus, from
  Lemma~\thref{l:int-in-sfplus-is-pos-lin},
  we have
  \begin{equation*}
    \forall n \in \matN,\;
    \int \fhi \, \matUN_{A_n} \, d\mu
    = \int \sum_{i\in[0..k]} a_i \, \matUN_{B_i \cap A_n} \, d\mu
    = \sum_{i\in[0..k]} a_i \mu (B_i \cap A_n).
  \end{equation*}
  Hence, from
  the previous result on the measure~$\mu$,
  \assume{linearity of the limit}, and
  Lemma~\threfc{l:int-in-sfplus-is-pos-lin}{with $a>0$},
  we have
  \begin{equation*}
    \int \fhi \, d\mu
    = \sum_{i \in [0..k]} a_i \lim_{n \to \infty} \mu (B_i \cap A_n)
    = \lim_{n \to \infty} \int \fhi \, \matUN_{A_n} \, d\mu
    = \frac{1}{a} \lim_{n \to \infty}
    \int a \fhi \, \matUN_{A_n} \, d\mu.
  \end{equation*}
  Let~$n\in\matN$.
  Then, from
  the definition of the~$A_n$'s, and
  \assume{the definition of the indicator function},
  we have $a\fhi\,\matUN_{A_n}\leq f_n\,\matUN_{A_n}\leq f_n$.
  Thus, from
  \assume{compatibility of the multiplication by a positive number with
    order},
  Lemma~\thref{l:int-in-mplus-is-monot}, and
  \assume{monotonicity and linearity of the limit},
  we have
  \begin{equation*}
    \int \fhi \, d\mu
    \leq \frac{1}{a} \lim_{n \to \infty} \int f_n \, d\mu.
  \end{equation*}
  Then, from
  \assume{left continuity of the division at $a=1$}, and
  \assume{monotonicity of the limit},
  \begin{equation*}
    \forall \fhi \in \calSFplus,\;
    \fhi \leq f \Implies
    \int \fhi \, d\mu
    \leq \lim_{n \to \infty} \int f_n \, d\mu.
  \end{equation*}
  Hence, from
  Lemma~\thref{l:int-in-mplus}, and
  Definition~\threfc{LM-d:supremum}{least upper bound},
  \begin{equation*}
    \int f \, d\mu
    = \sup_{\substack{\fhi \in \calSFplus\\\fhi \leq f}} \int \fhi \, d\mu
    \leq \lim_{n \to \infty} \int f_n \, d\mu.
  \end{equation*}

  Therefore, we have
  $\int f\,d\mu=\lim_{n\to\infty}\int f_n\,d\mu$.
\end{proof}

\begin{lemma}[integral in~$\calMplus$ is homogeneous at~$\infty$]
  \label{l:int-in-mplus-is-hom-at-infinity}
  \mbox{}\\
  Let~$(X,\Sigma,\mu)$ be a measure space.
  Let~$f\in\calMplus$.
  Then, $\infty f\in\calMplus$, and we have
  \begin{equation}
    \label{e:int-in-mplus-is-hom-at-infinity}
    \int \infty f \, d\mu = \infty \int f \, d\mu.
  \end{equation}
\end{lemma}

\begin{proof}
  From
  Definition~\threfc{d:mplus-subset-of-nonneg-meas-num-fun}{%
    $f\geq0$},
  \assume{ordered set properties of~$\matRbarplus$},
  Lemma~\thref{l:int-in-mplus-is-pos-hom}, and
  Lemma~\thref{l:mplus-is-closed-under-nonneg-scalar-mult},
  $(nf)_{n\in\matN}$ is a nondecreasing sequence in~$\calMplus$ such that
  $\lim_{n\to\infty}nf=\infty f\in\calMplus$.
  Therefore, from
  Theorem~\thref{t:beppo-levi-monot-conv},
  Lemma~\thref{l:int-in-mplus-is-pos-hom}, and
  \assume{positive homogeneity of the limit in~$\matRbarplus$},
  we have
  \begin{align*}
    \int \infty f \, d\mu
    & = \int \left( \lim_{n \to \infty} n \right) f \, d\mu
      = \int \left( \lim_{n \to \infty} n f \right) \, d\mu\\
    & = \lim_{n \to \infty} \left( \int n f \, d\mu \right)
      = \lim_{n \to \infty} \left( n \int f \, d\mu \right)
      = \left( \lim_{n \to \infty} n \right) \int f \, d\mu
      = \infty \int f \, d\mu.
  \end{align*}
\end{proof}

\begin{definition}[adapted sequence]
  \label{d:adapted-seq}
  \mbox{}\\
  Let~$(X,\Sigma,\mu)$ be a measure space.
  Let~$f\in\calMplus$.
  Let~$(\fhi_n)_{n\in\matN}\in\calSFplus$.
  The sequence $(\fhi_n)_{n\in\matN}$ is called {\em adapted sequence for~$f$}
  iff
  it is nondecreasing and $f=\lim_{n\in\matN}\fhi_n=\sup_{n\in\matN}\fhi_n$.
\end{definition}

\begin{lemma}[adapted sequence in~$\calMplus$]
  \label{l:adapted-seq-in-mplus}
  \mbox{}\\
  Let~$(X,\Sigma,\mu)$ be a measure space.
  Let~$f\in\calMplus$.
  For all $n\in\matN$, let~$\fhi_n:\ArXRp$ defined by
  \begin{equation}
    \label{e:adapted-seq-in-mplus}
    \forall x \in X,\quad
    \fhi_n (x) \eqdef \left\{
      \begin{array}{ll}
        \displaystyle
        \frac{\floor{2^n f (x)}}{2^n} & \mbox{when } f (x) < n,\\
        n & \mbox{otherwise}.
      \end{array}
    \right.
  \end{equation}
  Then, $(\fhi_n)_{n\in\matN}$ is an adapted sequence for~$f$.
\end{lemma}

\begin{proof}
  Let~$n\in\matN$.
  Let~$x\in X$.
  Assume that $f(x)<n$.
  Let~$i\eqdef\floor{2^nf(x)}\in\matN$.
  Then, from
  \assume{the definition of the floor function},
  we have $0\leq i\leq2^nf(x)<i+1\leq n2^n$.
  Thus, for all $i\in\matN$ such that $0\leq i<n2^n$, for all $x\in X$ such
  that $\frac{i}{2^n}\leq f(x)<\frac{i+1}{2^n}$, we have
  $\fhi_n(x)=\frac{i}{2^n}$.
  Hence,
  \begin{equation*}
    \fhi_n
    = \sum_{0 \leq i < n 2^n} \frac{i}{2^n}
    \matUN_{\left\{ \frac{i}{2^n} \leq f < \frac{i + 1}{2^n} \right\}}
    + n \, \matUN_{\{f \geq n\}}.
  \end{equation*}
  Moreover, from
  Lemma~\thref{l:meas-of-num-fun}, and
  Lemma~\threfc{l:equiv-def-of-sigma-alg}{%
    closedness under countable intersection (with $\card(I)=2$)},
  we have
  \begin{equation*}
    \left\{ \frac{i}{2^n} \leq f < \frac{i + 1}{2^n} \right\}
    = \left\{ \frac{i}{2^n} \leq f \right\}
    \cap \left\{ f < \frac{i + 1}{2^n} \right\}
    \in \Sigma
    \AND
    \left\{ f \geq n \right\} \in \Sigma.
  \end{equation*}
  Hence, from
  Definition~\thref{d:sf-vector-space-of-simple-funs}, and
  Definition~\thref{d:sfplus-subset-of-nonneg-simple-funs},
  we have $\fhi_n\in\calSFplus$.

  Let~$n\in\matN$.
  Let~$x\in X$.
  \proofpar{Case $n+1\leq f(x)$}
  Then, we have
  \begin{equation*}
    \fhi_n (x) = n < n + 1 = \fhi_{n+1} (x).
  \end{equation*}
  \proofpar{Case $f(x)<n+1$}
  Let~$i\eqdef\floor{2^{n+1}f(x)}$.
  Then, we have
  \begin{equation*}
    \fhi_{n + 1} (x) = \frac{i}{2^{n + 1}}
    \quad\mbox{with }
    0
    \leq \frac{i}{2^{n + 1}}
    \leq f (x)
    < \frac{i + 1}{2^{n + 1}}.
  \end{equation*}
  \proofpar{Case $\frac{i}{2^{n+1}}\geq n$}
  Then, $f(x)\geq n$ and we have
  \begin{equation*}
    \fhi_n (x) = n
    \leq \frac{i}{2^{n + 1}}
    = \fhi_{n + 1} (x).
  \end{equation*}
  \proofpar{Case $\frac{i}{2^{n+1}}<n$ and $i$ even}
  Let~$j\eqdef\frac{i}{2}\in\matN$.
  Then, from
  \assume{ordered field properties of~$\matR$},
  we have
  \begin{equation*}
    \frac{j}{2^n}
    \leq f (x)
    < \frac{j}{2^n} + \frac{1}{2^{n + 1}}
    < \frac{j + 1}{2^n}.
  \end{equation*}
  Thus, from
  \assume{the definition of the floor function},
  we have $j=\floor{2^nf(x)}$ and
  \begin{equation*}
    \fhi_n (x)
    = \frac{j}{2^n}
    = \frac{i}{2^{n + 1}}
    = \fhi_{n + 1} (x).
  \end{equation*}
  \proofpar{Case $\frac{i}{2^{n+1}}<n$ and $i$ odd}
  Let~$j\eqdef\frac{i-1}{2}\in\matN$.
  Then, from
  \assume{ordered field properties of~$\matR$},
  we have
  \begin{equation*}
    \frac{j}{2^n}
    \leq \frac{j}{2^n} + \frac{1}{2^{n + 1}}
    \leq f (x)
    < \frac{j + 1}{2^n}.
  \end{equation*}
  Thus, from
  \assume{the definition of the floor function},
  we have $j=\floor{2^nf(x)}$ and
  \begin{equation*}
    \fhi_n (x)
    = \frac{j}{2^n}
    = \frac{i - 1}{2^{n + 1}}
    < \frac{i}{2^{n + 1}}
    = \fhi_{n + 1} (x).
  \end{equation*}
  Thus, we always have $\fhi_n(x)\leq \fhi_{n+1}(x)$.
  Hence, the sequence $(\fhi_n)_{n\in\matN}$ is nondecreasing.

  Let~$x\in X$.
  We have $f(x)\in\matRbarplus$.
  \proofpar{Case $f(x)=\infty$}
  Then, for all $n\in\matN$, we have $\fhi_n(x)=n$.
  Hence, $\lim_{n\to\infty}\fhi_n(x)=f(x)$.
  \proofpar{Case $f(x)\in\matRplus$}
  Then, from
  \assume{the Archimedean property of~$\matR$},
  there exists $N\in\matN$ such that $f(x)<N$.
  Let~$n\in\matN$ such that $N\leq n$.
  Then, from
  \assume{the definition of the floor function},
  we have
  \begin{equation*}
    2^n \fhi_n (x)
    = \floor{2^n f (x)}
    \leq 2^n f (x)
    < \floor{2^n f (x)} + 1
    = 2^n \fhi_n (x) + 1.
  \end{equation*}
  Thus, from
  \assume{ordered field properties of~$\matR$ (with $2^n>0$)},
  we have $f(x)-\frac{1}{2^n}<\fhi_n(x)\leq f(x)$.
  Hence, from
  \assume{the squeeze theorem},
  we have $\lim_{n\to\infty}\fhi_n(x)=f(x)$.

  Therefore, from
  Definition~\thref{d:adapted-seq},
  $(\fhi_n)_{n\in\matN}$ is an adapted sequence for~$f$.
\end{proof}

\begin{lemma}[usage of adapted sequences]
  \label{l:usage-of-adapted-seqs}
  \mbox{}\hfill
  Let~$(X,\Sigma,\mu)$ be a measure space.\\
  Let~$f\in\calMplus$.
  Let~$(\fhi_n)_{n\in\matN}\in\calSFplus$ be an adapted sequence of~$f$.
  Then, we have
  \begin{equation}
    \label{e:usage-of-adapted-seqs}
    \int f \, d\mu
    = \int \lim_{n \to \infty} \fhi_n \, d\mu
    = \lim_{n \to \infty} \int \fhi_n \, d\mu.
  \end{equation}
\end{lemma}

\begin{proof}
  Direct consequence of
  Definition~\thref{d:adapted-seq},
  Lemma~\thref{l:adapted-seq-in-mplus},
  Theorem~\thref{t:beppo-levi-monot-conv}. and
  Lemma~\thref{l:int-in-mplus-gen-int-in-sfplus}.
\end{proof}

\begin{lemma}[integral in~$\calMplus$ is additive]
  \label{l:int-in-mplus-is-add}
  \mbox{}\\
  Let~$(X,\Sigma,\mu)$ be a measure space.
  Let~$f,g\in\calMplus$.
  Then, $f+g\in\calMplus$, and we have
  \begin{equation}
    \label{e:int-in-mplus-is-add}
    \int (f + g) \, d\mu = \int f \, d\mu + \int g \, d\mu.
  \end{equation}
\end{lemma}

\begin{proof}
  From
  Lemma~\thref{l:mplus-is-closed-under-add},
  we have $f+g\in\calMplus$.

  From
  Lemma~\thref{l:adapted-seq-in-mplus},
  let~$(\fhi_n)_{n\in\matN},(\psi_n)_{n\in\matN}\in\calSFplus$ be adapted
  sequences for~$f$ and~$g$.
  Then, from
  Lemma~\threfc{l:int-in-sfplus-is-pos-lin}{additivity},
  \assume{monotonicity of addition}, and
  \assume{additivity of the limit},
  $(\fhi_n+\psi_n)_{n\in\matN}\in\calSFplus$ is an adapted sequence for~$f+g$.
  Let~$n\in\matN$.
  Then, from
  Lemma~\threfc{l:int-in-sfplus-is-pos-lin}{additivity},
  we have
  \begin{equation*}
    \int (\fhi_n + \psi_n) \, d\mu = \int \fhi_n \, d\mu + \int \psi_n \, d\mu.
  \end{equation*}
  Therefore, from
  \assume{linearity of the limit when~$n$ goes to infinity}, and
  Lemma~\thref{l:usage-of-adapted-seqs},
  we have
  \begin{equation*}
    \int (f + g) \, d\mu = \int f \, d\mu + \int g \, d\mu.
  \end{equation*}
\end{proof}

\begin{lemma}[integral in~$\calMplus$ is positive linear]
  \label{l:int-in-mplus-is-pos-lin}
  \mbox{}\hfill
  Let~$(X,\Sigma,\mu)$ be a measure space.
  Let~$f,g\in\calMplus$.
  Let~$a\in\matRbarplus$.
  Then, $f+g,af\in\calMplus$, and we have
  \begin{equation}
    \label{e:int-in-mplus-is-pos-lin}
    \int (f + g) \, d\mu = \int f \, d\mu + \int g \, d\mu
    \AND
    \int a f \, d\mu = a \int f \, d\mu.
  \end{equation}
\end{lemma}

\begin{proof}
  Direct consequence of
  Lemma~\thref{l:int-in-mplus-is-add},
  Lemma~\thref{l:int-in-mplus-is-pos-hom}, and
  Lemma~\thref{l:int-in-mplus-is-hom-at-infinity}.
\end{proof}

\begin{lemma}[integral in~$\calMplus$ is $\sigma$-additive]
  \label{l:int-in-mplus-is-sigma-add}
  \mbox{}\\
  Let~$(X,\Sigma,\mu)$ be a measure space.
  Let~$(f_n)_{n\in\matN}\in\calMplus$.
  Then, $\sum_{n\in\matN}f_n\in\calMplus$, and we have
  \begin{equation}
    \label{e:int-in-mplus-is-sigma-add}
    \int \left( \sum_{n \in \matN} f_n \right) \, d\mu
    = \sum_{n \in \matN} \left( \int f_n \, d\mu \right).
  \end{equation}
\end{lemma}

\begin{proof}
  Direct consequence of
  Lemma~\thref{l:mplus-is-closed-under-count-sum}, and
  Theorem~\threfc{t:beppo-levi-monot-conv}{%
    with nondecreasing sequence $(\sum_{i\in[0..n]}f_i)_{n\in\matN}$}.
\end{proof}

\begin{lemma}[integral in~$\calMplus$ of decomposition into nonpositive and
  nonnegative parts]
  \label{l:int-in-mplus-of-decomp-into-nonpos-and-nonneg-parts}
  \mbox{}\hfill
  Let~$(X,\Sigma)$ be a measurable space.
  Let~$f\in\calM$.
  Then, we have
  \begin{equation}
    \label{e:int-in-mplus-of-decomp-into-nonpos-and-nonneg-parts}
    \int | f | \, d\mu = \int f^+ \, d\mu + \int f^- \, d\mu.
  \end{equation}
\end{lemma}

\begin{proof}
  Direct consequence of
  Lemma~\threfc{l:m-is-closed-under-abs}{%
    $|f|$ belongs to $\calMplus$},
  Lemma~\threfc{l:meas-of-nonneg-and-nonpos-parts}{%
    $f^+,f^-\in\calMplus$},
  Lemma~\threfc{l:decomp-into-nonneg-and-nonpos-parts}{%
    $|f|\!=\!f^+\!+\!f^-$}, and
  Lemma~\thref{l:int-in-mplus-is-add}.
\end{proof}

\begin{lemma}[compatibility of integral in~$\calMplus$ with nonpositive and
  nonnegative parts]
  \label{l:compat-of-int-in-mplus-with-nonpos-and-nonneg-parts}
  Let~$(X,\Sigma)$ be a measurable space.
  Let~$f,g\in\calM$ such that $f+g\in\calM$.
  Then, we have
  \begin{equation}
    \label{e:compat-of-int-in-mplus-with-nonpos-and-nonneg-parts}
    \int (f + g)^+ \, d\mu + \int f^- \, d\mu + \int g^- \, d\mu
    = \int (f + g)^- \, d\mu + \int f^+ \, d\mu + \int g^+ \, d\mu.
  \end{equation}
\end{lemma}

\begin{proof}
  Direct consequence of
  Lemma~\thref{l:compat-of-nonpos-and-nonneg-parts-with-add}, and
  Lemma~\thref{l:int-in-mplus-is-add}.
\end{proof}

\begin{lemma}[integral in~$\calMplus$ is almost definite]
  \label{l:int-in-mplus-is-almost-definite}
  \mbox{}\\
  Let~$(X,\Sigma,\mu)$ be a measure space.
  Let~$f$ in~$\calMplus$.
  Then, we have
  \begin{equation}
    \label{e:int-in-mplus-is-almost-definite}
    \int f \, d\mu = 0 \EQUIV f \eqae{\mu} 0.
  \end{equation}
\end{lemma}

\begin{proof}
  Let~$A\eqdef\{f>0\}=f^{-1}(0,\infty]$.
  Then, from
  Lemma~\threfc{l:meas-of-num-fun}{%
    $A\in\Sigma$},
  Lemma~\threfc{l:int-in-mplus-of-indic-fun}{%
    $\matUN_A\in\calMplus$},
  Lemma~\threfc{l:mplus-is-closed-under-nonneg-scalar-mult}{%
    $\infty f,\infty\,\matUN_A\in\calMplus$},
  Lemma~\thref{l:int-in-mplus-is-hom-at-infinity},
  Lemma~\threfc{l:infinity-prod-prop-in-rbarplus-mt}{%
    $\infty f=\infty\,\matUN_A$}, and
  Lemma~\thref{l:int-in-mplus-of-indic-fun}
  we have
  \begin{equation*}
    \infty \int f \, d\mu
    = \int \infty f \, d\mu
    = \int \infty \, \matUN_A \, d\mu
    = \infty \int \, \matUN_A \, d\mu
    = \infty \mu (A).
  \end{equation*}
  Therefore, from
  Lemma~\threfc{l:zero-prod-prop-in-rbarplus-mt}{%
    multiplication by~$\infty$ is definite in~$\matRbarplus$},
  Definition~\threfc{d:mplus-subset-of-nonneg-meas-num-fun}{%
    $\{f=0\}^c=A$},
  Lemma~\thref{l:negl-of-meas-subset}, and
  Definition~\thref{d:prop-almost-satisfied},
  we have
  \begin{align*}
    \int f \, d\mu = 0
    & \EQUIV \infty \int f \, d\mu = 0
      \EQUIV \infty \, \mu (A) = 0\\
    & \EQUIV \mu (A) = 0
      \EQUIV \mu (\{ f = 0 \}^c) = 0
      \EQUIV f \eqae{\mu} 0.
  \end{align*}
\end{proof}

\begin{lemma}[compatibility of integral in~$\calMplus$ with almost binary
  relation]
  \label{l:compat-of-int-in-mplus-with-almost-bin-rel}
  \mbox{}\\
  Let~$(X,\Sigma,\mu)$ be a measure space.
  Let~$\Eqrel$ be a binary relation on~$\calMplus$.
  Let~$\Eqrelp$ be a binary relation on~$\matRbarplus$.
  Assume that we have the properties $\eqrel{0}{0}$ and
  \begin{equation}
    \label{e:compat-of-int-in-mplus-with-almost-bin-rel-1}
    \forall f, g \in \calMplus,\quad
    \eqrel{f}{g}
    \IMPLIES
    \eqrelp{\int f \, d\mu}{\int g \, d\mu}.
  \end{equation}
  Then, we have the property
  \begin{equation}
    \label{e:compat-of-int-in-mplus-with-almost-bin-rel-2}
    \forall f, g \in \calMplus,\quad
    \eqrelae{\mu}{f}{g}
    \IMPLIES
    \eqrelp{\int f \, d\mu}{\int g \, d\mu}.
  \end{equation}
\end{lemma}

\begin{proof}
  Let~$f,g\in\calMplus$.
  Assume that $\eqrelae{\mu}{f}{g}$ holds.

  From
  Definition~\thref{d:almost-bin-rel},
  Definition~\thref{d:prop-almost-satisfied},
  Definition~\thref{d:negl-subset}, and
  \assume{monotonicity of complement},
  let~$A\in\Sigma$ such that $\mu(A)=0$ and $A^c\subset\{\eqrel{f}{g}\}$,
  {\ie} such that $\eqrel{\restr{f}{A^c}}{\restr{g}{A^c}}$ holds.
  Thus, from
  \assume{the definition of the indicator function}, and
  since $\eqrel{0}{0}$, we have $\eqrel{f\,\matUN_{A^c}}{g\,\matUN_{A^c}}$.
  Then, from
  Definition~\thref{d:meas},
  Definition~\threfc{d:measurable-space}{$\Sigma$ is a $\sigma$-algebra},
  Definition~\threfc{d:sigma-alg}{$A^c\in\Sigma$},
  Lemma~\threfc{l:int-in-mplus-of-indic-fun}{%
    $\matUN_A,\matUN_{A^c}\in\calMplus$}, and
  Lemma~\thref{l:mplus-is-closed-under-mult},
  we have
  $f\,\matUN_A,f\,\matUN_{A^c},g\,\matUN_A,g\,\matUN_{A^c}\in\calMplus$.
  Thus, from assumption, the property
  $\eqrelp{\int f\,\matUN_{A^c}\,d\mu}{\int g\,\matUN_{A^c}\,d\mu}$ holds.

  Let~$h\in\{f,g\}$.
  Then, from
  \assume{the definition of the indicator function
    ($\{\matUN_A=0\}^c$ is $A$)},
  Lemma~\threfc{l:negl-of-meas-subset}{%
    $\{\matUN_A=0\}^c\in\neglset$},
  Definition~\threfc{d:prop-almost-satisfied}{%
    $\matUN_A\!\eqae{\mu}\!0$},
  Lemma~\threfc{l:compat-of-almost-eq-with-op}{%
    with the unary operator left multiplication by~$h$}, and
  Lemma~\thref{l:zero-prod-prop-in-rbarplus-mt},
  we have $h\,\matUN_A\eqae{\mu}h0=0$.
  Thus, from
  Lemma~\thref{l:everywhere-implies-almost-everywhere}, and
  Lemma~\threfc{l:almost-eq-is-equiv-rel}{transitivity},
  we have $h\,\matUN_A\eqae{\mu}0$.
  Hence, from
  \assume{properties of the indicator function},
  Lemma~\thref{l:mult-in-rbarplus-is-distr-over-add-mt},
  Lemma~\thref{l:int-in-mplus-is-add},
  Lemma~\thref{l:int-in-mplus-is-almost-definite}, and
  Lemma~\thref{l:zero-is-identity-element-for-add-in-rbar},
  we have for all $h\in\{f,g\}$,
  \begin{equation*}
    \int h \, d\mu
    = \int h (\matUN_A + \matUN_{A^c}) \, d\mu
    = \int h \, \matUN_A \, d\mu + \int h \, \matUN_{A^c} \, d\mu
    = \int h \, \matUN_{A^c} \, d\mu.
  \end{equation*}

  Therefore,
  $\eqrelp{\int f\,d\mu}{\int g\,d\mu}$ also holds.
\end{proof}

\begin{lemma}[compatibility of integral in~$\calMplus$ with almost equality]
  \label{l:compat-of-int-in-mplus-with-almost-eq}
  \mbox{}\\
  Let~$(X,\Sigma,\mu)$ be a measure space.
  Let~$f,g\in\calMplus$.
  Then, we have
  \begin{equation}
    \label{e:compat-of-int-in-mplus-with-almost-eq}
    f \eqae{\mu} g \IMPLIES \int f \, d\mu = \int g \, d\mu.
  \end{equation}
\end{lemma}

\begin{proof}
  Direct consequence of
  Lemma~\threfc{l:compat-of-int-in-mplus-with-almost-bin-rel}{%
    with $\Eqrel=\Eqrelp\eqdef$ equality}, and
  Lemma~\threfc{l:int-in-mplus}{integral is a function}.
\end{proof}

\begin{lemma}[integral in~$\calMplus$ is almost monotone]
  \label{l:int-in-mplus-is-almost-monot}
  \mbox{}\\
  Let~$(X,\Sigma,\mu)$ be a measure space.
  Let~$f,g\in\calMplus$.
  Then, we have
  \begin{equation}
    \label{e:int-in-mplus-is-almost-monot}
    f \leqae{\mu} g \IMPLIES \int f \, d\mu \leq \int g \, d\mu.
  \end{equation}
\end{lemma}

\begin{proof}
  Direct consequence of
  Lemma~\threfc{l:compat-of-int-in-mplus-with-almost-bin-rel}{%
    with $\Eqrel=\Eqrelp\eqdef$ inequality}, and
  Lemma~\thref{l:int-in-mplus-is-monot}.
\end{proof}

\begin{lemma}[Bienaym\'e--Chebyshev inequality]
  \label{l:bienayme-chebyshev-ineq}
  \mbox{}\\
  Let~$(X,\Sigma,\mu)$ be a measure space.
  Let~$f\in\calM$.
  Let~$a\in\matRbarplusstar$.
  Then, we have
  \begin{equation}
    \label{e:bienayme-chebyshev-ineq}
    a \mu \left( \{ | f | \geq a \} \right)
    \leq \int | f | \, d\mu.
  \end{equation}
\end{lemma}

\begin{proof}
  Let~$A\eqdef\{|f|\geq a\}$.
  Then, from
  Lemma~\threfc{l:m-is-closed-under-abs}{%
    $|f|$ belongs to $\calMplus\subset\calM$}, and
  Lemma~\thref{l:meas-of-num-fun},
  we have $A\in\Sigma$.
  Moreover, from
  \assume{the definition of the indicator function}, and
  Lemma~\thref{l:abs-in-rbar-is-nonneg},
  we have $|f|\geq a\,\matUN_A$.
  Therefore, from
  Lemma~\thref{l:meas-of-indic-fun},
  Lemma~\thref{l:mplus-is-closed-under-nonneg-scalar-mult},
  Lemma~\thref{l:int-in-mplus-is-pos-hom},
  Lemma~\thref{l:int-in-mplus-is-hom-at-infinity},
  Lemma~\thref{l:int-in-mplus-is-monot}, and
  Lemma~\thref{l:int-in-mplus-of-indic-fun},
  we have $a\,\matUN_A\in\calMplus$ and
  \begin{equation*}
    \int | f | \, d\mu
    \geq \int a \, \matUN_A \, d\mu
    = a \int \matUN_A \, d\mu
    = a \mu (A).
  \end{equation*}
\end{proof}

\begin{lemma}[integrable in~$\calMplus$ is almost finite]
  \label{l:integrable-in-mplus-is-almost-finite}
  \mbox{}\hfill
  Let~$(X,\Sigma,\mu)$ be a measure space.\\
  Let~$f:\ArXRbp$ be $\mu$-integrable in $\calMplus$.
  Then, we have $\mu(f^{-1}(\infty))=0$, {\ie} $f\ltae{\mu}\infty$.
\end{lemma}

\begin{proof}
  Direct consequence of
  Lemma~\threfc{l:int-in-mplus}{$f\in\calMplus$},
  Definition~\thref{d:mplus-subset-of-nonneg-meas-num-fun},
  Lemma~\threfc{l:bienayme-chebyshev-ineq}{with $a\eqdef\infty$},
  Lemma~\threfc{l:abs-in-rbar-is-nonneg}{$|f|=f$},
  Definition~\threfc{d:ext-real-nums-rbar}{%
    $A\eqdef\{|f|\geq\infty\}=f^{-1}(\infty)$},
  Lemma~\threfc{l:int-in-mplus}{$\int|f|\,d\mu<\infty$},
  \assume{compatibility of multiplication with order in $\matRbar$
    ($\mu(A)\leq0$)},
  Definition~\threfc{d:meas}{nonnegativeness, {\ie} $\mu(A)=0$},
  Lemma~\threfc{l:negl-of-meas-subset}{$A\in\neglset$}, and
  Definition~\threfc{d:prop-almost-satisfied}{$A^c=\{f<\infty\}$},
\end{proof}

\begin{lemma}[bounded by integrable in~$\calMplus$ is integrable]
  \label{l:bounded-by-integrable-in-mplus-is-integrable}
  \mbox{}\\
  Let~$(X,\Sigma,\mu)$ be a measure space.
  Let~$f:\ArXRbp$.
  Then, $f$~is $\mu$-integrable in~$\calMplus$ iff
  there exists $g:\ArXRbp$ such that~$g$ is $\mu$-integrable in~$\calMplus$
  and $f\leq g$.
\end{lemma}

\begin{proof}
  \proofpar{``Left'' implies ``right''}
  Direct consequence of
  Lemma~\threfc{l:order-in-rbar-is-total}{reflexivity, with $g\eqdef f$}.

  \proofparskip{``Right'' implies ``left''}
  Direct consequence of
  Lemma~\threfc{l:int-in-mplus}{$\mu$-integrability},
  Lemma~\thref{l:int-in-mplus-is-monot}, and
  Lemma~\threfc{l:order-in-rbar-is-total}{transitivity}.

  \medskip\noindent
  Therefore, we have the equivalence.
\end{proof}

\begin{lemma}[integral in~$\calMplus$ over subset]
  \label{l:int-in-mplus-over-subset}
  \mbox{}\hfill
  Let~$(X,\Sigma,\mu)$ be a measure space.\\
  Let~$A\in\Sigma$.
  Let~$Y\subset X$ such that $A\subset Y$.
  Let~$f:\ArYRb$.
  Let~$\hf:\ArXRb$.
  Assume that $\restr{\hf}{Y}=f$.
  Then, we have $\restr{f}{A}\in\calMplus(A,\Sigma\olcap A)$ iff
  $\hf\,\matUN_A\in\calMplus(X,\Sigma)$.
  If so, we have
  \begin{equation}
    \label{e:int-in-mplus-over-subset}
    \int \restr{f}{A} \, d\mu_A = \int \hf \, \matUN_A \, d\mu.
  \end{equation}
  This integral is still denoted $\int_Af\,d\mu$;
  it is still called {\em integral of~$f$ over~$A$}.
\end{lemma}

\begin{proof}
  \proofpar{Equivalence}
  Direct consequence of
  Lemma~\thref{l:meas-of-restr},
  Definition~\thref{d:mplus-subset-of-nonneg-meas-num-fun}, and
  \assume{nonnegativeness of the indicator function}.

  \proofparskip{Identity}
  Direct consequence of
  Lemma~\threfc{l:adapted-seq-in-mplus}{%
    let $(\hfhi_n)_{n\in\matN}\in\calSFplus(X,\Sigma)$ be an adapted sequence
    for $\hf\,\matUN_A$},
  Lemma~\threfc{l:sf-is-closed-under-restr}{%
    $\restr{(\hfhi_n)}{A}$ belongs to $\calSFplus(A,\Sigma\olcap A)$},
  \assume{compatibility of restriction of function with monotonicity and
    limit},
  Definition~\threfc{d:adapted-seq}{%
    $(\restr{(\hfhi_n)}{A})_{n\in\matN}$ is an adapted sequences
    for~$\restr{f}{A}$},
  Lemma~\thref{l:usage-of-adapted-seqs}, and
  Lemma~\thref{l:int-in-sfplus-over-subset}.
\end{proof}

\begin{lemma}[integral in~$\calMplus$ over subset is $\sigma$-additive]
  \label{l:int-in-mplus-over-subset-is-sigma-add}
  \mbox{}\hfill
  Let~$(X,\Sigma,\mu)$ be a measure space.\\
  Let~$I\subset\matN$.
  Let~$A,(A_i)_{i\in I}\in\Sigma$.
  Assume that $(A_i)_{i\in I}$ is a pseudopartition of~$A$.
  Let~$Y\subset X$ such that $A\subset Y$.
  Let~$f:\ArYRb$.
  Let~$\hf:\ArXRb$.
  Assume that $\restr{\hf}{Y}=f$.\\
  Then, $\hf\,\matUN_A=\sum_{i\in I}\hf\,\matUN_{A_i}$, and
  $\hf\,\matUN_A\in\calMplus$ iff
  for all $i\in I$, $\hf\,\matUN_{A_i}\in\calMplus$.
  If so, we have
  \begin{equation}
    \label{e:int-in-mplus-over-subset-is-sigma-add}
    \int_A f \, d\mu = \sum_{i \in I} \int_{A_i} f \, d\mu.
  \end{equation}
\end{lemma}

\begin{proof}
  \proofpar{(1). $\hf\,\matUN_A=\sum_{i\in I}\hf\,\matUN_{A_i}$}
  Direct consequence of
  Lemma~\threfc{l:if-is-sigma-add}{with $I\eqdef\matN$}, and
  \assume{left distributivity of multiplication over countable addition
    in~$\matRplus$}.

  \proofparskip{``Left'' implies ``right''}
  Direct consequence of
  Lemma~\threfc{l:if-is-closed-under-mult}{%
    $\matUN_{A_i}=\matUN_{A\cap A_i}=\matUN_A\,\matUN_{A_i}$},
  \assume{associativity of multiplication in~$\matR$},
  Lemma~\thref{l:meas-of-indic-fun},
  Lemma~\threfc{l:m-and-finite-is-mr}{$\calMR\subset\calM$},
  \assume{nonnegativeness of the indicator function},
  Definition~\thref{d:mplus-subset-of-nonneg-meas-num-fun}, and
  Lemma~\thref{l:mplus-is-closed-under-mult}.

  \proofparskip{``Right'' implies ``left''}
  Direct consequence of~(1), and
  Lemma~\thref{l:mplus-is-closed-under-count-sum}.

  \medskip\noindent
  Therefore, we have the equivalence.

  \proofparskip{Identity}
  Direct consequence of
  Lemma~\threfc{l:int-in-mplus-over-subset}{%
    with $A$, then $A_i$}, (1), and
  Lemma~\thref{l:int-in-mplus-is-sigma-add}.
\end{proof}

\begin{lemma}[integral in~$\calMplus$ over singleton]
  \label{l:int-in-mplus-over-singleton}
  \mbox{}\hfill
  Let~$(X,\Sigma,\mu)$ be a measure space.
  Let~$a\in X$.
  Assume that $\{a\}\!\in\!\Sigma$.
  Let~$f:\ArXRbp$.
  Then, we have $f\,\matUN_{\{a\}}=f(a)\,\matUN_{\{a\}}\in\calMplus$, and
  \begin{equation}
    \label{e:int-in-mplus-over-singleton}
    \int_{\{ a \}} f \, d\mu =  f(a) \mu (\{ a \}).
  \end{equation}
\end{lemma}

\begin{proof}
  From
  \assume{the definition of the indicator function},
  Lemma~\thref{l:meas-of-indic-fun}, and
  Lemma~\thref{l:mplus-is-closed-under-nonneg-scalar-mult},
  we have $f\,\matUN_{\{a\}}=f(a)\,\matUN_{\{a\}}\in\calMplus$.
  Then, from
  Lemma~\threfc{l:int-in-mplus-over-subset}{with $A\eqdef\{a\}$, $Y\eqdef X$},
  Lemma~\threfc{l:int-in-mplus-is-pos-lin}{%
    homogeneity in $\matRbarplus$}, and
  Lemma~\thref{l:int-in-mplus-of-indic-fun},
  we have
  \begin{equation*}
    \int_{\{ a \}} f \, d\mu
    = \int f \, \matUN_{\{ a \}} \, d\mu
    = f (a) \int \, \matUN_{\{ a \}} \, d\mu
    = f(a) \mu(\{ a \}).
  \end{equation*}
\end{proof}

\begin{remark}
  \label{r:v2-new17}
  See the sketch of next proof in
  Section~\ref{s:sketch-of-the-proof-of-fatou-lem}.
\end{remark}

\begin{theorem}[{\Fl}]
  \label{t:fatou-lemma}
  \mbox{}\\
  Let~$(X,\Sigma,\mu)$ be a measure space.
  Let~$(f_n)_{n\in\matN}\in\calMplus$.
  Then, $\liminf_{n\to\infty}f_n\in\calMplus$, and we have
  \begin{equation}
    \label{e:fatou-lemma}
    \int \liminf_{n \to \infty} f_n \, d\mu
    \leq \liminf_{n \to \infty} \int f_n \, d\mu.
  \end{equation}
\end{theorem}

\begin{proof}
  From
  Lemma~\threfc{l:liminf-bounded-from-below}{with $m\eqdef0$}, and
  Lemma~\thref{l:m-is-closed-under-liminf},
  we have $\liminf_{n\to\infty}f_n\in\calMplus$.

  Let~$n\in\matN$.
  Let~$g_n\eqdef\inf_{p\in\matN}f_{n+p}$.
  Then, from
  \assume{monotonicity of infimum},
  Lemma~\threfc{l:inf-of-bounded-seq-is-bounded}{with $a\eqdef0$}, and
  Lemma~\thref{l:m-is-closed-under-inf},
  $(g_n)_{n\in\matN}$ is a nondecreasing sequence in~$\calMplus$.
  Thus, from
  Definition~\thref{d:pointwise-conv},
  \assume{properties of nondecreasing sequences in the ordered
    set~$\matRbar$},
  Lemma~\thref{l:mplus-is-closed-under-limit-when-pointwise-conv},
  $(g_n)_{n\in\matN}$ is pointwise convergent in~$\matRbarplus$ towards the
  measurable function \mbox{$g\eqdef\lim_{n\in\matN}g_n$}.
  Moreover, from
  \assume{the nondecreasing property of the sequence in the ordered
    set~$\matRbar$}, and
  Lemma~\thref{l:liminf},
  we have
  \begin{equation*}
    g = \sup_{n \in \matN} g_n = \liminf_{n \to \infty} f_n.
  \end{equation*}
  Hence, from
  Theorem~\threfc{t:beppo-levi-monot-conv}{with $(g_n)_{n\in\matN}$},
  $g$~is measurable and nonnegative, and we have
  \begin{equation*}
    \int \liminf_{n \to \infty} f_n \, d\mu
    = \int g \, d\mu
    = \lim_{n \to \infty} \int g_n \, d\mu.
  \end{equation*}

  Let~$n,p\in\matN$.
  Let~$x\in X$.
  Then, from
  Definition~\threfc{LM-d:infimum}{lower bound},
  we have
  \begin{equation*}
    g_n (x) \leq f_{n + p} (x).
  \end{equation*}
  Then, from
  Lemma~\thref{l:int-in-mplus-is-monot},
  we have
  \begin{equation*}
    \int g_n \, d\mu \leq \int f_{n + p} \, d\mu.
  \end{equation*}
  Hence, from
  Definition~\threfc{LM-d:infimum}{greatest lower bound},
  we have
  \begin{equation*}
    \int g_n \, d\mu \leq \inf_{p \in \matN} \int f_{n + p} \, d\mu.
  \end{equation*}

  Therefore, from
  \assume{monotonicity of the limit in the ordered set~$\matRbar$}, and
  Lemma~\thref{l:liminf},
  we have
  \begin{equation*}
     \int \liminf_{n \to \infty} f_n \, d\mu
    = \lim_{n \to \infty} \int g_n \, d\mu
    \leq \liminf_{n \to \infty} \int f_n \, d\mu.
  \end{equation*}
\end{proof}

\begin{lemma}[integral in~$\calMplus$ of pointwise convergent sequence]
  \label{l:int-in-mplus-of-pointwise-conv-seq}
  \mbox{}\\
  Let~$(X,\Sigma,\mu)$ be a measure space.
  Let~$(f_n)_{n\in\matN}\in\calMplus$.
  Assume that the sequence is pointwise convergent towards~$f$ such that for
  all $n\in\matN$, $f_n\leq f$.
  Then, $f\in\calMplus$, and we have
  \begin{equation}
    \label{e:int-in-mplus-of-pointwise-conv-seq}
    \int \lim_{n \to \infty} f_n \, d\mu
    = \int f \, d\mu
    = \lim_{n \to \infty} \int f_n \, d\mu.
  \end{equation}
\end{lemma}

\begin{proof}
  From
  Lemma~\thref{l:mplus-is-closed-under-limit-when-pointwise-conv},
  we have
  \begin{equation*}
    f = \lim_{n \to \infty} f_n \in \calMplus.
  \end{equation*}
  Moreover, from
  Lemma~\thref{l:int-in-mplus-is-monot},
  we have
  \begin{equation*}
    \forall n \in \matN,\quad \int f_n \, d\mu \leq \int f \, d\mu.
  \end{equation*}
  Thus, from
  Definition~\threfc{LM-d:supremum}{least upper bound},
  Lemma~\thref{l:liminf-and-limsup-of-pointwise-conv},
  and
  Theorem~\thref{t:fatou-lemma},
  we have
  \begin{equation*}
    \limsup_{n \to \infty} \int f_n \, d\mu
    \leq \int f \, d\mu
    = \int \lim_{n \to \infty} f_n \, d\mu
    = \int \liminf_{n \to \infty} f_n \, d\mu
    \leq \liminf_{n \to \infty} \int f_n \, d\mu
  \end{equation*}
  Therefore, from
  Lemma~\thref{l:liminf-limsup-and-pointwise-conv},
  we have
  \begin{equation*}
    \liminf_{n \to \infty} \int f_n \, d\mu
    = \limsup_{n \to \infty} \int f_n \, d\mu
    = \lim_{n \to \infty} \int f_n \, d\mu
    = \int f \, d\mu.
  \end{equation*}
\end{proof}

\begin{lemma}[integral in~$\calMplus$ for counting measure]
  \label{l:int-in-mplus-for-count-meas}
  \mbox{}\\
  Let~$(X,\Sigma)$ be a measurable space.
  Let~$Y\subset X$.
  Let~$f\in\calMplus$.
  Then, we have
  \begin{equation}
    \label{e:int-in-mplus-for-count-meas}
    \int f \, d\delta_Y = \sum_{y \in Y} f (y).
  \end{equation}
\end{lemma}

\begin{proof}
  Direct consequence of
  Lemma~\thref{l:usage-of-adapted-seqs},
  Lemma~\thref{l:int-in-sfplus-for-count-meas}, and
  \assume{compatibility of (possibly uncountable) addition in~$\matRbarplus$
    with limit}.
\end{proof}

\begin{lemma}[integral in~$\calMplus$ for counting measure on~$\matN$]
  \label{l:int-in-mplus-for-count-meas-on-n}
  \mbox{}\\
  Let~$f:\ArNRbp$ be a nonnegative sequence.
  Then, we have
  \begin{equation}
    \label{e:int-in-mplus-for-count-meas-on-n}
    \int f \, d\delta_\matN = \sum_{n \in \matN} f (n).
  \end{equation}
\end{lemma}

\begin{proof}
  Direct consequence of
  Lemma~\threfc{l:int-in-mplus-for-count-meas}{%
    with $Y\!=\!X\eqdef\matN$ and $\Sigma\eqdef\calP(\matN)$}.
\end{proof}

\begin{remark}
  Note that the previous lemma makes nonnegative series be Lebesgue
  integrals for the counting measure on natural numbers.
  Thus, the theory of nonnegative series can be derived from Lebesgue
  integration for nonnegative measurable functions.
\end{remark}

\begin{lemma}[integral in~$\calMplus$ for Dirac measure]
  \label{l:int-in-mplus-for-dirac-meas}
  \mbox{}\\
  Let~$(X,\Sigma)$ be a measurable space.
  Let~$\{a\}\in\Sigma$.
  Let~$f\in\calMplus$.
  Then, we have
  \begin{equation}
    \label{e:int-in-mplus-for-dirac-meas}
    \int f \, d\delta_a = f (a).
  \end{equation}
\end{lemma}

\begin{proof}
  Direct consequence of
  Definition~\thref{d:dirac-meas}, and
  Lemma~\threfc{l:int-in-mplus-for-count-meas}{%
    with $Y\eqdef\{a\}$}.
\end{proof}

\clearpage
\section{Measure and integration over product space}
\label{s:measure-and-integration-over-product-space}

\subsection{Measure over product space}
\label{ss:measure-over-product-space}

\begin{remark}
  For the sake of simplicity, we only present this section in the case of the
  product of two measure spaces.
  When $i\in\{1,2\}$, the complement $\{1,2\}\setminus\{i\}$ is~$\{3-i\}$.

  The definition of sections and some of their properties were presented in
  Section~\ref{s:product-of-measurable-spaces}.
\end{remark}

\begin{lemma}[measure of section]
  \label{l:meas-of-section}
  \mbox{}\hfill
  Let~$(X_1,\Sigma_1,\mu_1)$ and~$(X_2,\Sigma_2,\mu_2)$ be measure spaces.
  Let~$A\in\Sigma_1\otimes\Sigma_2$.
  Let~$i\in\{1,2\}$.
  Let~$j\eqdef3-i$.
  Then, the function~$F^A_i$ defined by
  \begin{equation}
    \label{e:meas-of-section}
    \forall x_i \in X_i,\quad F^A_i (x_i) \eqdef \mu_j (s_i (x_i, A))
  \end{equation}
  is well-defined on~$X_i$, and takes its values in~$\matRbarplus$.
  It is called {\em measure of $i$-th section of~$A$}.
\end{lemma}

\begin{proof}
  Direct consequence of
  Lemma~\thref{l:measurability-of-section}, and
  Definition~\threfc{d:meas}{nonnegativeness}.
\end{proof}

\begin{lemma}[measure of section of product]
  \label{l:meas-of-section-of-prod}
  \mbox{}\hfill
  Let~$(X_1,\Sigma_1,\mu_1)$ and~$(X_2,\Sigma_2,\mu_2)$ be measure spaces.
  Let~$A_1\in\Sigma_1$ and~$A_2\in\Sigma_2$.
  Let~$i\in\{1,2\}$.
  Let~$j\eqdef3-i$.
  Then, we have
  \begin{equation}
    \label{e:meas-of-section-of-prod}
    F^{A_1 \times A_2}_i = \mu_j (A_j) \, \matUN_{A_i}.
  \end{equation}
\end{lemma}

\begin{proof}
  Direct consequence of
  Lemma~\threfc{l:prod-of-meas-subsets-is-meas}{%
    $A_1\times A_2$ belongs to $\Sigma$},
  Lemma~\thref{l:meas-of-section},
  Lemma~\thref{l:section-of-prod},
  Definition~\threfc{d:meas}{%
    $\mu_j(\emptyset)=0$ and $\mu_j(A_j)\in\matRbarplus$},
  Lemma~\thref{l:zero-prod-prop-in-rbarplus-mt}, and
  \assume{the definition of the indicator function}.
\end{proof}

\begin{remark}
  The next proof follows the {\mct} scheme
  (see Section~\ref{s:dynkin-p-l-th-monot-class-th-schemes}).
\end{remark}

\begin{lemma}[measurability of measure of section (finite)]
  \label{l:meas-of-meas-of-section-finite}
  \mbox{}\\
  Let~$(X_1,\Sigma_1,\mu_1)$ and~$(X_2,\Sigma_2,\mu_2)$ be measure spaces.
  Let~$i\in\{1,2\}$.
  Let~$j\eqdef3-i$.
  Assume that~$\mu_j$ is finite.
  Let~$A\in\Sigma_1\otimes\Sigma_2$.
  Then, we have $F^A_i\in\calMplus(X_i,\Sigma_i)$.
\end{lemma}

\begin{proof}
  Let~$X\eqdef X_1\times X_2$.
  From
  Definition~\thref{d:prod-of-subsets-of-parties},
  Definition~\thref{d:tensor-prod-of-sigma-algs}, and
  Definition~\thref{d:gen-set-alg},
  let $\Sigmabar\eqdef\Sigma_1\oltimes\Sigma_2$,
  $\Sigma\eqdef\Sigma_1\otimes\Sigma_2$, and
  $\calA\eqdef\calA_X(\Sigmabar)$.
  Let~$\calS_i\eqdef\{A\in\Sigma\st F^A_i\in\calMplus(X_i,\Sigma_i)\}$.

  \proofparskip{(1). $\Sigmabar\subset\calS_i$}
  Let~$A\in\Sigmabar$.
  Let~$A_1\in\Sigma_1$ and~$A_2\in\Sigma_2$ such that $A=A_1\times A_2$.

  From
  Lemma~\threfc{l:meas-of-section-of-prod}{%
    $F^A_i=\mu_j(A_j)\,\matUN_{A_i}$},
  Lemma~\threfc{l:int-in-mplus-of-indic-fun}{%
    $\matUN_{A_i}\in\calMplus(X_i,\Sigma_i)$},
  Definition~\threfc{d:meas}{$\mu_j(A_j)\in\matRbarplus$}, and
  Lemma~\threfc{l:mplus-is-closed-under-nonneg-scalar-mult}{%
    with $a\eqdef\mu_j(A_j)$},
  $F^A_i$ belongs to~$\calMplus(X_i,\Sigma_i)$.
  Thus, from
  Lemma~\threfc{l:prod-of-meas-subsets-is-meas}{%
    $A\in\Sigma$}, and
  the definition of~$\calS_i$,
  we have $A\in\calS_i$.
  Hence, we have $\Sigmabar\subset\calS_i$.

  \proofparskip{(2). $\calA\subset\calS_i$}
  Let~$A\in\calA$.

  Then, from
  Lemma~\thref{l:gen-set-alg-is-min},
  Definition~\threfc{d:tensor-prod-of-sigma-algs}{%
    $\Sigma=\Sigma_X(\Sigmabar)$},
  Lemma~\threfc{l:sigma-alg-contains-set-alg}{%
    with $G\eqdef\Sigmabar$, thus $\calA\subset\Sigma$},
  Lemma~\thref{l:set-alg-gen-by-prod-of-sigma-algs}, and
  Lemma~\threfc{l:prod-of-meas-subsets-is-meas}{%
    $\Sigmabar\subset\Sigma$},
  we have $A\in\Sigma$, and there exists $n\in\matN$, and
  $(A_p)_{p\in[0..n]}$ in~$\Sigmabar\subset\Sigma$ such that,
  for all $p,q\in[0..n]$, $p\not=q\Implies A_p\cap A_q=\emptyset$, and
  $A=\biguplus_{p\in[0..n]}A_p$.

  Let~$x_i\in X_i$.
  Then, from
  Lemma~\thref{l:meas-of-section},
  Lemma~\threfc{l:count-union-of-sections-is-meas}{with $I\eqdef[0..n]$},
  Lemma~\threfc{l:compat-of-section-with-set-ops}{%
    the $s_i(x_i,A_p)$'s are pairwise disjoint},
  Definition~\threfc{d:meas}{$\mu_j$ is $\sigma$-additive}, and
  Definition~\threfc{d:sigma-add-over-meas-space}{%
    with $I\eqdef[0..n]$ and $\mu\eqdef\mu_j$},
  we have
  \begin{equation*}
    F^A_i (x_i)
    = \mu_j (s_i (x_i, A))
    = \mu_j \left( \biguplus_{p \in [0..n]} s_i (x_i, A_p) \right)
    = \sum_{p \in [0..n]} \mu_j (s_i (x_i, A_p))
    = \sum_{p \in [0..n]} F^{A_p}_i (x_i).
  \end{equation*}
  Thus, from
  (1) (\uline{$A_p\in\calS_i$}),
  the definition of~$\calS_i$, and
  Lemma~\threfc{l:mplus-is-closed-under-count-sum}{%
    with $I\eqdef[0..n]$, thus $F^A_i\in\calMplus(X_i,\Sigma_i)$},
  we have $A\in\calS_i$.
  Hence, we have $\calA\subset\calS_i$.

  \proofparskip{(3). $\calS_i$~is monotone class}
  Let~$(A_n)_{n\in\matN}\in\calS_i\subset\Sigma$.

  Let~$x_i\in X_i$.
  Let~$n\in\matN$.
  Then, from
  Lemma~\thref{l:measurability-of-section},
  Lemma~\thref{l:count-union-of-sections-is-meas}, and
  Lemma~\thref{l:count-inter-of-sections-is-meas},
  we have
  \begin{equation*}
    s_i (x_i, A_n),\quad
    s_i \left( x_i, \bigcup_{n \in \matN} A_n \right)
    = \bigcup_{n \in \matN} s_i (x_i, A_n),\quad
    s_i \left( x_i, \bigcap_{n \in \matN} A_n \right)
    = \bigcap_{n \in \matN} s_i (x_i, A_n)\quad
    \in \Sigma_j.
  \end{equation*}

  Assume first that, for all $n\in\matN$, $A_n\subset A_{n+1}$.
  Let~$A\eqdef\bigcup_{n\in\matN}A_n$.\\
  Let~$x_i\in X_i$.
  Then, from
  Lemma~\thref{l:meas-of-section},
  the definition of~$A$,
  Lemma~\threfc{l:compat-of-section-with-set-ops}{%
    $(s_i(x_i,A_n))_{n\in\matN}$ is nondecreasing}, and
  Lemma~\threfc{l:meas-is-cont-from-below}{with~$\mu_j$},
  we have
  \begin{align*}
    F^A_i (x_i)
    = \mu_j (s_i (x_i, A))
    & = \mu_j \left(
      s_i \left( x_i, \bigcup_{n \in \matN} A_n \right) \right)\\
    & = \mu_j \left( \bigcup_{n \in \matN} s_i (x_i, A_n) \right)
      = \sup_{n \in \matN} \mu_j (s_i (x_i, A_n))
      = \sup_{n \in \matN} F^{A_n}_i (x_i).
  \end{align*}
  Thus, from
  the definition of~$\calS_i$, and
  Lemma~\thref{l:mplus-is-closed-under-sup},
  $F^A_i$~belongs to~$\calMplus(X_i,\Sigma_i)$.
  Hence, from
  the definition of~$\calS_i$,
  we have $A\in\calS_i$, {\ie}~$\calS_i$ is closed under nondecreasing union.

  Assume now that, for all $n\in\matN$, $A_n\supset A_{n+1}$.
  Let~$A\eqdef\bigcap_{n\in\matN}A_n$.\\
  Let~$x_i\in X_i$.
  Then, from
  Lemma~\thref{l:meas-of-section},
  the definition of~$A$,
  Lemma~\threfc{l:compat-of-section-with-set-ops}{%
    $(s_i(x_i,A_n))_{n\in\matN}$ is nonincreasing},
  Definition~\threfc{d:finite-meas}{with~$\mu_j$},
  Lemma~\threfc{l:finite-meas-is-bounded}{%
    $\mu_j(s_i(x_i,A_0))$ is finite}, and
  Lemma~\threfc{l:meas-is-cont-from-above}{%
    with $\mu_j$ and $n_0\eqdef0$},
  we have
  \begin{align*}
    F^A_i (x_i)
    = \mu_j (s_i (x_i, A))
    & = \mu_j \left(
      s_i \left( x_i, \bigcap_{n \in \matN} A_n \right) \right)\\
    & = \mu_j \left( \bigcap_{n \in \matN} s_i (x_i, A_n) \right)
      = \inf_{n \in \matN} \mu_j (s_i (x_i, A_n))
      = \inf_{n \in \matN} F^{A_n}_i (x_i).
  \end{align*}
  Thus, from
  the definition of~$\calS_i$,  and
  Lemma~\thref{l:mplus-is-closed-under-inf},
  $F^A_i\in\calMplus(X_i,\Sigma_i)$.
  Then, from
  the definition of~$\calS_i$,
  we have $A\in\calS_i$, {\ie}~$\calS_i$ is closed under nonincreasing
  intersection.

  Hence, from
  Definition~\thref{d:monot-class},
  $\calS_i$~is a monotone class.

  \medskip\noindent
  Therefore, from
  Definition~\threfc{d:tensor-prod-of-sigma-algs}{%
    $\Sigma=\Sigma_X(\Sigmabar)$}, (2), (3), and
  Lemma~\threfc{l:usage-of-monot-class-th}{with $G\eqdef\Sigmabar$},
  we have for all $A\in\Sigma$, $F^A_i\in\calMplus(X_i,\Sigma_i)$.
\end{proof}

\begin{lemma}[measurability of measure of section]
  \label{l:meas-of-meas-of-section}
  \mbox{}\hfill
  Let~$(X_1,\Sigma_1,\mu_1)$ and~$(X_2,\Sigma_2,\mu_2)$ be $\sigma$-finite
  measure spaces.
  Let~$A\in\Sigma_1\otimes\Sigma_2$.
  Let~$i\in\{1,2\}$.
  Then, we have $F^A_i\in\calMplus(X_i,\Sigma_i)$.
\end{lemma}

\begin{proof}
  Let~$j\eqdef3-i$.
  From
  Lemma~\threfc{l:equiv-def-of-sigma-finite-meas}{%
    with $\mu_j$},
  there exists $(B_{j,n})_{n\in\matN}\in\Sigma_j$ such that, for all
  $n\in\matN$, $B_{j,n}\subset B_{j,n+1}$, $\mu_j(B_{j,n})<\infty$, and
  $X_j=\bigcup_{n\in\matN}B_{j,n}$.

  For all $n\in\matN$, for all $A\in\Sigma_1\otimes\Sigma_2$, from
  Lemma~\threfc{l:restr-meas}{%
    with $Y\eqdef B_{j,n}$}, and
  Lemma~\threfc{l:meas-of-section}{with $\mu_j\eqdef\mu_{j,n}$},
  let~$\mu_{j,n}$ be the restricted measure,
  and~$F^A_{i,n}$ be the measure of $i$-th section of~$A$ defined by
  \begin{equation*}
    \mu_{j, n}
    \eqdef (\mu_j)^\prime_{B_{j, n}}
    = (A_j \in \Sigma_j \longmapsto \mu_j (A_j \cap B_{j, n}))
    \AND
    F^A_{i, n}
    \eqdef (x_i \in X_i \longmapsto \mu_{j, n} (s_i (x_i, A))).
  \end{equation*}
  Let~$n\in\matN$.
  Then, from
  Definition~\threfc{d:finite-meas}{%
    $\mu_{j, n}(X)=\mu_j(B_{j, n})<\infty$},
  $\mu_{j,n}$~is a finite measure on~$(X_j,\Sigma_j)$.
  Moreover, from
  Lemma~\threfc{l:meas-of-meas-of-section-finite}{%
    with $\mu_j\eqdef\mu_{j, n}$ finite},
  we have $F^A_{i,n}\in\calMplus(X_i,\Sigma_i)$.

  Let~$x_i\in X_i$.
  Then, from
  Lemma~\thref{l:measurability-of-section},
  the definition of~$B_{j,n}$,  and
  \assume{distributivity of intersection over union},
  we have
  \begin{equation*}
    \Sigma_j \ni
    s_i (x_i, A)
    = s_i (x_i, A) \cap \bigcup_{n \in \matN} B_{j, n}
    = \bigcup_{n \in \matN} \left( s_i (x_i, A) \cap B_{j, n} \right).
  \end{equation*}
  Thus, from
  Lemma~\threfc{l:equiv-def-of-sigma-alg}{%
    closedness under countable intersection (with $\card(I)=2$)},
  we have $s_i(x_i,A)\cap B_{j,n}\in\Sigma_j$.
  Hence, from
  Lemma~\thref{l:meas-of-section},
  \assume{monotonicity of intersection
    (\uline{$(s_i(x_i,A)\cap B_{j,n})_{n\in\matN}$ is nondecreasing})},
  Lemma~\threfc{l:meas-is-cont-from-below}{%
    with $\mu_j$ and $A_n\eqdef s_i(x_i,A)\cap B_{j,n}$}, and
  the definition of~$\mu_{j,n}$,
  \begin{align*}
    F^A_i (x_i)
    = \mu_j (s_i (x_i, A))
    & = \mu_j \left( s_i (x_i, A) \cap \bigcup_{n \in \matN} B_{j,n} \right)
      = \mu_j \left(
      \bigcup_{n \in \matN} \left(
      s_i (x_i, A) \cap B_{j,n} \right) \right)\\
    & = \sup_{n \in \matN} \mu_j (s_i (x_i, A) \cap B_{j,n})
      = \sup_{n \in \matN} \mu_{j,n} (s_i (x_i, A))
      = \sup_{n \in \matN} F^A_{i,n} (x_i).
  \end{align*}

  Therefore, from
  Lemma~\thref{l:mplus-is-closed-under-sup},
  we have
  \begin{equation*}
    F^A_i = \sup_{n \in \matN} F^A_{i,n} \in \calMplus (X_i, \Sigma_i).
  \end{equation*}
\end{proof}

\begin{definition}[tensor product measure]
  \label{d:tensor-prod-meas}
  \mbox{}\hfill
  Let~$(X_1,\Sigma_1,\mu_1)$ and~$(X_2,\Sigma_2,\mu_2)$ be measure spaces.
  Let~$X\eqdef X_1\times X_2$ and~$\Sigma\eqdef\Sigma_1\otimes\Sigma_2$.
  A measure~$\mu$ on the product measurable space~$(X,\Sigma)$ is called
  {\em tensor product measure (on~$(X,\Sigma)$ relying on~$\mu_1$ and~$\mu_2$)}
  iff
  \begin{equation}
    \label{e:tensor-prod-meas}
    \forall A_1 \in \Sigma_1,\;
    \forall A_2 \in \Sigma_2,\quad
    \mu (A_1 \times A_2)
    = \mu_1 (A_1) \, \mu_2 (A_2).
  \end{equation}
\end{definition}

\begin{definition}[candidate tensor product measure]
  \label{d:cand-tensor-prod-meas}
  \mbox{}\\
  Let~$(X_1,\Sigma_1,\mu_1)$ and~$(X_2,\Sigma_2,\mu_2)$ be measure spaces.
  Let~$i\in\{1,2\}$.
  The function defined by
  \begin{equation}
    \label{e:cand-tensor-prod-meas}
    \forall A \in \Sigma_1 \otimes \Sigma_2,\quad
    (\mu_1 \otimes \mu_2)^i (A)
    \eqdef \int F^A_i \, d\mu_i
  \end{equation}
  is called {\em candidate tensor product measure}.
\end{definition}

\begin{lemma}[candidate tensor product measure is tensor product measure]
  \label{l:cand-tensor-prod-meas-is-tensor-prod-meas}
  \mbox{}\\
  Let~$(X_1,\Sigma_1,\mu_1)$ and~$(X_2,\Sigma_2,\mu_2)$ be $\sigma$-finite
  measure spaces.
  Let~$i\in\{1,2\}$.
  Then, the candidate tensor product measure~$(\mu_1\otimes\mu_2)^i$ is a
  tensor product measure on~$(X_1\times X_2,\Sigma_1\otimes\Sigma_2)$.
\end{lemma}

\begin{proof}
  Let~$X\eqdef X_1\times X_2$ and~$\Sigma\eqdef\Sigma_1\otimes\Sigma_2$.

  \proofparskip{Measure}
  Let~$j\eqdef3-i$.

  Let~$A\in\Sigma$.
  Then, from
  Lemma~\thref{l:meas-of-section},
  Definition~\thref{d:cand-tensor-prod-meas},
  Lemma~\threfc{l:meas-of-meas-of-section}{%
    with the $\sigma$-finite measure $\mu_i$}, and
  Lemma~\threfc{l:int-in-mplus}{%
    nonnegativeness, with $X\eqdef X_i$ and $\Sigma\eqdef\Sigma_i$},
  the value $(\mu_1\otimes\mu_2)^i(A)=\int F^A_i\,d\mu_i$ is well-defined
  in~$\matRbarplus$, and from
  Lemma~\thref{l:meas-of-section},
  it is equal to $\int\mu_j(s_i(x_i,A))\,d\mu_i$.

  From
  Lemma~\threfc{l:compat-of-section-with-set-ops}{$\emptyset$},
  Definition~\threfc{d:meas}{$\mu_j(\emptyset)=0$}, and
  Lemma~\thref{l:int-in-mplus-of-zero-is-zero},
  we have
  \begin{equation*}
    (\mu_1 \otimes \mu_2)^i (\emptyset)
    = \int \mu_j (s_i (x_i, \emptyset)) \, d\mu_i
    = \int \mu_j (\emptyset) \, d\mu_i
    = \int 0 \, d\mu_i
    = 0.
  \end{equation*}

  Let~$(A_n)_{n\in\matN}\in\Sigma$.
  Assume that the~$A_n$'s are pairwise disjoint.
  Let~$A\eqdef\biguplus_{n\in\matN}A_n$.
  Let~$x_i$ be in~$X_i$.
  Let~$n\in\matN$.
  Let~$p,q\in\matN$.
  Assume that $p\not=q$,
  Then, from
  Lemma~\thref{l:measurability-of-section},
  Lemma~\threfc{l:count-union-of-sections-is-meas}{with $I\eqdef\matN$}, and
  Lemma~\threfc{l:compat-of-section-with-set-ops}{%
    with intersection and~$\emptyset$},
  $s_i(x_i,A_n)\in\Sigma_j$, and
  \begin{gather*}
    s_i (x_i, A)
    = s_i \left( x_i, \bigcup_{n \in \matN} A_n \right)
    = \bigcup_{n \in \matN} s_i (x_i, A_n) \in \Sigma_j,\\
    s_i (x_i, A_p) \cap s_i (x_i, A_q)
    = s_i (x_i, A_p \cap A_q)
    = s_i (x_i, \emptyset)
    = \emptyset.
  \end{gather*}
  Thus, from
  Lemma~\thref{l:meas-of-section},
  Definition~\threfc{d:meas}{$\mu_j$~is $\sigma$-additive},
  Definition~\thref{d:sigma-add-over-meas-space},
  we have
  \begin{equation*}
    F^A_i (x_i)
    = \mu_j (s_i (x_i, A))
    = \mu_j \left( \biguplus_{n \in \matN} s_i (x_i, A_n) \right)
    = \sum_{n \in \matN} \mu_j (s_i (x_i, A_n))
    = \sum_{n \in \matN} F^{A_n}_i (x_i).
  \end{equation*}
  Then, from
  Lemma~\thref{l:int-in-mplus-is-sigma-add}, and
  Definition~\thref{d:cand-tensor-prod-meas},
  we have
  \begin{equation*}
    (\mu_1 \otimes \mu_2)^i (A)
    = \int \sum_{n \in \matN} F^{A_n}_i \, d\mu_i
    = \sum_{n \in \matN} \int F^{A_n}_i \, d\mu_i
    = \sum_{n \in \matN} (\mu_1 \otimes \mu_2)^i (A_n).
  \end{equation*}
  Hence, from
  Definition~\thref{d:sigma-add-over-meas-space},
  $(\mu_1\otimes\mu_2)^i$ is $\sigma$-additive.

  Therefore, from
  Definition~\thref{d:meas},
  $(\mu_1\otimes\mu_2)^i$ is a measure on~$(X,\Sigma)$.

  \proofparskip{Identity}
  Let~$A_1\in\Sigma_1$.
  Let~$A_2\in\Sigma_2$.
  Let~$A\eqdef A_1\times A_2$.

  Then, from
  Lemma~\thref{l:prod-of-meas-subsets-is-meas},
  we have $A\in\Sigma$.
  Hence, from
  Definition~\thref{d:cand-tensor-prod-meas},
  Lemma~\thref{l:meas-of-section-of-prod},
  Definition~\threfc{d:meas}{$\mu_j$~is nonnegative},
  Lemma~\threfc{l:int-in-mplus-is-pos-lin}{%
    with $a\eqdef\mu_j (A_j)\in\matRbarplus$},
  Lemma~\threfc{l:int-in-mplus-of-indic-fun}{%
    with $\mu\eqdef\mu_i$ and $A_i\in\Sigma_i$},
  Lemma~\thref{l:mult-in-rbarplus-is-comm-mt}, and
  since $\{1,2\}=\{i,j\}$,
  we have
  \begin{equation*}
    (\mu_1 \otimes \mu_2)^i (A)
    = \int \mu_j (A_j) \, \matUN_{A_i} \, d\mu_i
    = \mu_j (A_j) \, \mu_i (A_i)
    = \mu_1 (A_1) \, \mu_2 (A_2),
  \end{equation*}
  Therefore, from
  Definition~\thref{d:tensor-prod-meas},
  $(\mu_1\otimes\mu_2)^i$ is a tensor product measure.
\end{proof}

\begin{lemma}[tensor product of finite measures]
  \label{l:tensor-prod-of-finite-meas}
  \mbox{}\\
  Let~$(X_1,\Sigma_1,\mu_1)$ and~$(X_2,\Sigma_2,\mu_2)$ be finite measure
  spaces.
  Let~$\mu$ be a tensor product measure
  on~$(X_1\times X_2,\Sigma_1\otimes\Sigma_2)$ relying on~$\mu_1$ and~$\mu_2$.
  Then, $\mu$~is finite.
\end{lemma}

\begin{proof}
  Direct consequence of
  Definition~\thref{d:meas},
  Lemma~\threfc{l:equiv-def-of-sigma-alg}{%
    $X_1\in\Sigma_1$, $X_2\in\Sigma_2$, and
    $X_1\times X_2\in\Sigma_1\otimes\Sigma_2$},
  Definition~\thref{d:tensor-prod-meas},
  Definition~\thref{d:finite-meas}, and
  \assume{closedness of multiplication in~$\matRplus$}.
\end{proof}

\begin{lemma}[tensor product of $\sigma$-finite measures]
  \label{l:tensor-prod-of-sigma-finite-meas}
  \mbox{}\\
  Let~$(X_1,\Sigma_1,\mu_1)$ and~$(X_2,\Sigma_2,\mu_2)$ be $\sigma$-finite
  measure spaces.
  Let~$X\eqdef X_1\times X_2$ and~$\Sigma\eqdef\Sigma_1\otimes\Sigma_2$.
  Let~$\mu$ be a tensor product measure on~$(X,\Sigma)$ relying on~$\mu_1$
  and~$\mu_2$.
  Then, $\mu$~is $\sigma$-finite.

  In particular, let $i\in\{1,2\}$.
  Let $(B_{i,n})_{n\in\matN}\in\Sigma_i$ such that
  \begin{equation}
    \label{e:tensor-prod-of-sigma-finite-meas-1}
    (\forall n \in \matN,\quad
    B_{i, n} \subset B_{i, n + 1}
    \CONJ
    \mu_i (B_{i, n}) < \infty)
    \CONJ
    X_i = \bigcup_{n \in \matN} B_{i, n}.
  \end{equation}
  For all $n\in\matN$, let~$B_n\eqdef B_{1,n}\times B_{2,n}$.
  Then, we have
  \begin{equation}
    \label{e:tensor-prod-of-sigma-finite-meas-2}
    (\forall n \in \matN,\quad
    B_n \in \Sigma
    \CONJ
    B_n \subset B_{n + 1}
    \CONJ
    \mu (B_n) < \infty)
    \CONJ
    X = \bigcup_{n \in \matN} B_n.
  \end{equation}
\end{lemma}

\begin{proof}
  Existence of the~$B_{i,n}$'s comes from
  Lemma~\thref{l:equiv-def-of-sigma-finite-meas}.

  Let~$n\in\matN$.
  Then, from
  the definition of the~$B_n$'s,
  Lemma~\thref{l:prod-of-meas-subsets-is-meas},
  \assume{monotonicity of Cartesian product},
  Definition~\thref{d:tensor-prod-meas}, and
  \assume{closedness of multiplication in~$\matRplus$},
  we have
  \begin{equation*}
    B_n \in \Sigma
    \CONJ
    B_n \subset B_{n + 1}
    \CONJ
    \mu (B_n) = \mu_1 (B_{1, n}) \, \mu_2 (B_{2, n}) < \infty.
  \end{equation*}

  Moreover, from
  \assume{monotonicity of union},
  we have $\bigcup_{n\in\matN}B_n\subset X$.
  Conversely, let~$(x_1,x_2)$ be in~$X$.
  Let~$i\in\{1,2\}$.
  Then, from
  Lemma~\threfc{l:equiv-def-of-sigma-finite-meas}{with~$\mu_i$},
  there exists~$n_i\in\matN$ such that $x_i\in B_{i,n_i}$.
  Thus, from
  monotonicity of the~$B_{i,n}$'s,
  we have $x_i\in B_{i,\max(n_1,n_2)}$.
  Then, from
  the definition of the~$B_n$'s,
  we have $(x_1,x_2)\in B_{\max(n_1,n_2)}$, {\ie}
  $X\subset\bigcup_{n\in\matN}B_n$.
  Hence, we have $X=\bigcup_{n\in\matN}B_n$.

  Therefore, from
  Definition~\thref{d:sigma-finite-meas},
  $\mu$~is $\sigma$-finite.
\end{proof}

\begin{remark}
  The next proof follows the {\mct} scheme
  (see Section~\ref{s:dynkin-p-l-th-monot-class-th-schemes}).
\end{remark}

\begin{lemma}[uniqueness of tensor product measure (finite)]
  \label{l:uniq-of-tensor-prod-meas-finite}
  \mbox{}\\
  Let~$(X_1,\Sigma_1,\mu_1)$ and~$(X_2,\Sigma_2,\mu_2)$ be finite measure
  spaces.
  Then, there exists a unique tensor product measure on the product
  measurable space~$(X_1\times X_2,\Sigma_1\otimes\Sigma_2)$.
\end{lemma}

\begin{proof}
  \proofpar{Existence}
  Direct consequence of
  Lemma~\threfc{l:finite-meas-is-sigma-finite}{%
    thus $\mu_1$ and $\mu_2$ are $\sigma$-finite}, and
  Lemma~\thref{l:cand-tensor-prod-meas-is-tensor-prod-meas}.

  \proofparskip{Uniqueness}
  Let~$X\eqdef X_1\times X_2$, $\Sigmabar\eqdef\Sigma_1\oltimes\Sigma_2$
  and~$\Sigma\eqdef\Sigma_1\otimes\Sigma_2$.
  Let~$m$ and~$\tm$ be tensor product measures on~$(X,\Sigma)$ relying
  on~$\mu_1$ and~$\mu_2$.
  Let~$\calS\eqdef\{A\in\Sigma\st m(A)=\tm(A)\}$.
  Let~$\calA\eqdef\calA_X(\Sigmabar)$.

  \proofparskip{(1). $\Sigmabar\subset\calS$}
  Direct consequence of
  Lemma~\thref{l:prod-of-meas-subsets-is-meas},
  Definition~\threfc{d:tensor-prod-meas}{%
    with $m$ and $\tm$}, and
  the definition of~$\calS$.

  \proofparskip{(2). $\calA\subset\calS$}
  Let~$A\in\calA$.

  Then, from
  Definition~\threfc{d:tensor-prod-of-sigma-algs}{%
    $\Sigma=\Sigma_X(\Sigmabar)$},
  Lemma~\threfc{l:sigma-alg-contains-set-alg}{%
    with $G\eqdef\Sigmabar$, thus $\calA\subset\Sigma$},
  Lemma~\thref{l:set-alg-gen-by-prod-of-sigma-algs}, and
  Lemma~\threfc{l:prod-of-meas-subsets-is-meas}{%
    $\Sigmabar\subset\Sigma$},
  we have $A\in\Sigma$, and there exists $n\in\matN$, and
  $(A_p)_{p\in[0..n]}\in\Sigmabar\subset\Sigma$ such that, for all
  $p,q\in[0..n]$, $p\not=q$ implies emptiness of $A_p\cap A_q$,
  and~$A=\biguplus_{p\in[0..n]}A_p$.
  Thus, from
  Lemma~\threfc{l:equiv-def-of-meas}{additivity for $m$ and $\tm$},
  Definition~\thref{d:add-over-meas-space},
  (1) (\uline{$A_p\in\calS$}), and
  the definition of~$\calS$, and since
  \assume{addition in~$\matRbarplus$ is a function (same arguments yield the
    same result)},
  we have
  \begin{equation*}
    m (A)
    = m \left( \biguplus_{p \in [0..n]} A_p \right)
    = \sum_{p \in [0..n]} m (A_p)
    = \sum_{p \in [0..n]} \tm (A_p)
    = \tm \left( \biguplus_{p \in [0..n]} A_p \right)
    = \tm (A).
  \end{equation*}
  Then, from
  the definition of~$\calS$,
  we have $A\in\calS$.
  Hence, we have $\calA\subset\calS$.

  \proofparskip{(3). $\calS$~is monotone class}
  Let~$(A_n)_{n\in\matN}\in\calS\subset\Sigma$.

  Then, from
  Lemma~\threfc{l:equiv-def-of-sigma-alg}{%
    closedness under countable union and intersection},
  we have $\bigcup_{n\in\matN}A_n,\bigcap_{n\in\matN}A_n\in\Sigma$.

  Assume first that, for all $n\in\matN$, $A_n\subset A_{n+1}$.
  Let~$A\eqdef\bigcup_{n\in\matN}A_n$.\\
  Then, from
  Lemma~\threfc{l:meas-is-cont-from-below}{with~$m$ and~$\tm$},
  the definition of~$\calS$, and since
  \assume{supremum is a function},
  we have
  \begin{equation*}
    m (A)
    = m \left( \bigcup_{n \in \matN} A_n \right)
    = \sup_{n \in \matN} m (A_n)
    = \sup_{n \in \matN} \tm (A_n)
    = \tm \left( \bigcup_{n \in \matN} A_n \right)
    = \tm (A).
  \end{equation*}
  Hence, from
  the definition of~$\calS_i$,
  we have $A\in\calS$, {\ie}~$\calS$ is closed under nondecreasing union.

  Assume now that, for all $n\in\matN$, $A_n\supset A_{n+1}$.
  Let~$A\eqdef\bigcap_{n\in\matN}A_n$.\\
  Then, from
  Lemma~\threfc{l:tensor-prod-of-finite-meas}{%
    with $m$ and $\tm$, both relying on finite measures $\mu_1$ and $\mu_2$},
  Lemma~\threfc{l:finite-meas-is-bounded}{$m(A_0),\tm(A_0)<\infty$},
  Lemma~\threfc{l:meas-is-cont-from-above}{%
    with $\mu\eqdef m,\tm$ and $n_0\eqdef0$},
  the definition of~$\calS$, and since
  \assume{infimum is a function},
  we have
  \begin{equation*}
    m (A)
    = m \left( \bigcap_{n \in \matN} A_n \right)
    = \inf_{n \in \matN} m (A_n)
    = \inf_{n \in \matN} \tm (A_n)
    = \tm \left( \bigcap_{n \in \matN} A_n \right)
    = \tm (A).
  \end{equation*}
  Thus, from
  the definition of~$\calS_i$,
  we have $A\in\calS$, {\ie}~$\calS$ is closed under nonincreasing
  intersection.

  Hence, from
  Definition~\thref{d:monot-class},
  $\calS$~is a monotone class.

  \medskip\noindent
  Therefore, from
  Definition~\threfc{d:tensor-prod-of-sigma-algs}{%
    $\Sigma=\Sigma_X(\Sigmabar)$}, (2), (3), and
  Lemma~\threfc{l:usage-of-monot-class-th}{with $G\eqdef\Sigmabar$},
  we have $m=\tm$.
\end{proof}

\begin{remark}
  \mbox{}\\
  The uniqueness part of the following proof reproduces the schemes of the
  proof of Lemma~\ref{l:meas-of-meas-of-section}.
\end{remark}

\begin{lemma}[uniqueness of tensor product measure]
  \label{l:uniq-of-tensor-prod-meas}
  \mbox{}\\
  Let~$(X_1,\Sigma_1,\mu_1)$ and $(X_2,\Sigma_2,\mu_2)$ be $\sigma$-finite
  measure spaces.
  Let~$X\eqdef X_1\times X_2$ and~$\Sigma\eqdef\Sigma_1\otimes\Sigma_2$.
  Then, there exists a unique tensor product measure on~$(X,\Sigma)$.

  This measure is denoted~$\mu_1\otimes\mu_2$, and for all $i\in\{1,2\}$ with
  $j\eqdef3-i$, we have
  \begin{equation}
    \label{e:uniq-of-tensor-prod-meas}
    \forall A \in \Sigma,\quad
    (\mu_1 \otimes \mu_2) (A)
    = \int \mu_j (s_i (x_i, A)) \, d\mu_i.
  \end{equation}
  It is called the {\em (tensor) product measure of~$\mu_1$ and~$\mu_2$}, and
  $(X,\Sigma,\mu_1\otimes\mu_2)$ is called
  {\em (tensor) product measure space}.
\end{lemma}

\begin{proof}
  \proofpar{Existence}
  Direct consequence of
  Lemma~\thref{l:cand-tensor-prod-meas-is-tensor-prod-meas}.

  \proofparskip{Uniqueness}
  Let~$m$ and~$\tm$ be tensor product measures on~$(X,\Sigma)$ relying
  on~$\mu_1$ and~$\mu_2$.

  From
  Lemma~\thref{l:equiv-def-of-sigma-finite-meas},
  for all $i\in\{1,2\}$, there exists $(B_{i,n})_{n\in\matN}\in\Sigma_i$
  such that, for all $n\in\matN$, $B_{i,n}\subset B_{i,n+1}$,
  $\mu_i(B_{i,n})<\infty$, and $X_i=\bigcup_{n\in\matN}B_{i,n}$.
  For all $n\in\matN$, let $B_n\eqdef B_{1, n}\times B_{2, n}$.
  Then, from
  Lemma~\threfc{l:tensor-prod-of-sigma-finite-meas}{%
    with $\mu\eqdef m,\tm$},
  we have for all $n\in\matN$, $B_n\in\Sigma$,  $B_n\subset B_{n+1}$,
  $m(B_n),\tm(B_n)<\infty$, and $X=\bigcup_{n\in\matN}B_n$.

  For all $i\in\{1,2\}$, for all $n\in\matN$, from
  Lemma~\threfc{l:equiv-def-of-sigma-alg}{%
    ($A\in\Sigma$ implies $A\cap B_n\in\Sigma$)}, and
  Lemma~\threfc{l:restr-meas}{%
    with $Y\eqdef B_{i,n}$ and $\mu\eqdef\mu_i$,
    then $Y\eqdef B_n$ and $\mu\eqdef m,\tm$},
  let~$\mu_{i,n}$, $m_n$ and~$\tm_n$ be the restricted measures defined by
  \begin{gather*}
    \mu_{i, n}
    \eqdef (\mu_i)^\prime_{B_{i, n}}
    = (A_i \in \Sigma_i \longmapsto \mu_i (A_i \cap B_{i,n})),\\
    m_n
    \eqdef m^\prime_{B_n}
    = (A \in \Sigma \longmapsto m (A \cap B_n))
    \AND
    \tm_n
    \eqdef (\tm)^\prime_{B_n}
    = (A \in \Sigma \longmapsto \tm (A \cap B_n)).
  \end{gather*}
  Let~$i\in\{1,2\}$.
  Let~$n\in\matN$.
  Then, from
  Definition~\threfc{d:finite-meas}{%
    with $\mu_{i,n}(X_i)=\mu_i(B_{i,n})$, $m_n(X)=m(B_n)$ and
    $\tm_n(X)=\tm(B_n)$ finite},
  $\mu_{i,n}$, $m_n$ and~$\tm_n$ are finite measures.

  Let~$n\in\matN$.
  Let~$A_1\in\Sigma_1$ and~$A_2\in\Sigma_2$.
  Then, from
  Lemma~\threfc{l:prod-of-meas-subsets-is-meas}{%
    $A_1\times A_2\in\Sigma$},
  the definitions of~$m_n$, $B_n$, $\mu_{1,n}$, $\mu_{2,n}$ and~$\tm_n$,
  \assume{compatibility of intersection with Cartesian product}, and
  Definition~\threfc{d:tensor-prod-meas}{%
    with $\mu\eqdef m,\tm$},
  we have
  \begin{align*}
    m_n (A_1 \times A_2)
    & = m ((A_1 \times A_2) \cap (B_{1, n} \times B_{2, n}))
      = m ((A_1 \cap B_{1, n}) \times (A_2 \cap B_{2, n}))\\
    & = \mu_1 (A_1 \cap B_{1, n}) \, \mu_2 (A_2 \cap B_{2, n})
      = \mu_{1,n} (A_1) \, \mu_{2,n} (A_2)\\
    & = \tm ((A_1 \cap B_{1, n}) \times (A_2 \cap B_{2, n}))
      = \tm ((A_1 \times A_2) \cap (B_{1, n} \times B_{2, n}))
      = \tm_n (A_1 \times A_2).
  \end{align*}
  Thus, from
  Definition~\threfc{d:tensor-prod-meas}{%
    with $\mu_1\eqdef\!\mu_{1,n}$, $\mu_2\eqdef\!\mu_{2,n}$ and
    $\mu\eqdef\!m_n,\tm_n$},
  $m_n$ and~$\tm_n$ are tensor product measures on~$(X,\Sigma)$ both relying on
  finite measures~$\mu_{1,n}$ and~$\mu_{2,n}$.
  Hence, from
  Lemma~\thref{l:uniq-of-tensor-prod-meas-finite},
  we have $m_n=\tm_n$.

  Let~$A\in\Sigma$.
  Then, from
  Lemma~\threfc{l:equiv-def-of-sigma-alg}{%
    $A\cap B_n\in\Sigma$, then $\bigcup_{n\in\matN}(A\cap B_n)\in\Sigma$},
  \assume{distributivity of intersection over union},
  \assume{monotonicity of intersection
    ($(A\cap B_n)_{n\in\matN}$ is nondecreasing)},
  Lemma~\threfc{l:meas-is-cont-from-below}{%
    with~$m$, $\tm$ and $A_n\eqdef A\cap B_n$},
  the definition of~$m_n$ and~$\tm_n$, and since
  \assume{supremum is a function},
  we have
  \begin{align*}
    m (A)
    & = m \left( A \cap \bigcup_{n \in \matN} B_n \right)
      = m \left( \bigcup_{n \in \matN} (A \cap B_n) \right)\\
    & = \sup_{n \in \matN} m (A \cap B_n)
      = \sup_{n \in \matN} m_n (A)
      = \sup_{n \in \matN} \tm_n (A)
      = \sup_{n \in \matN} \tm (A \cap B_n)\\
    & = \tm \left( \bigcup_{n \in \matN} (A \cap B_n) \right)
      = \tm \left( A \cap \bigcup_{n \in \matN} B_n \right)
      = \tm (A).
  \end{align*}

  Therefore, we have $m=\tm$.

  \proofparskip{Identity}
  Direct consequence of the uniqueness result above,
  Lemma~\thref{l:cand-tensor-prod-meas-is-tensor-prod-meas},
  Definition~\thref{d:cand-tensor-prod-meas}, and
  Lemma~\thref{l:meas-of-section}.
\end{proof}

\begin{lemma}[negligibility of measurable section]
  \label{l:negl-of-meas-section}
  \mbox{}\\
  Let~$(X_1,\Sigma_1,\mu_1)$ and $(X_2,\Sigma_2,\mu_2)$ be $\sigma$-finite
  measure spaces.
  Let~$\mu\eqdef\mu_1\otimes\mu_2$.
  Let~$A\in\Sigma_1\otimes\Sigma_2$.
  Let~$i\in\{1,2\}$.
  Let~$j\eqdef3-i$.
  Then, $A$~is $\mu$-negligible iff
  $\mu(A)=0$ iff
  for almost all $x_i\in X_i$, $\mu_j(s_i(x_i,A))=0$ iff
  for almost all $x_i\in X_i$, $s_i(x_i,A)$ is $\mu_j$-negligible.
\end{lemma}

\begin{proof}
  Direct consequence of
  Lemma~\threfc{l:measurability-of-section}{$s_i(x_i,A)\in\Sigma_j$},
  Lemma~\threfc{l:negl-of-meas-subset}{with $A$, then $s_i(x_i,A)$},
  Lemma~\thref{l:uniq-of-tensor-prod-meas},
  Lemma~\threfc{l:int-in-mplus-is-almost-definite}{%
    with $\mu_i$ and $(x_i\mapsto\mu_j(s_i(x_i,A)))$}.
\end{proof}

\subsection{Lebesgue measure over product space}
\label{ss:lebesgue-measure-over-product-space}

\begin{lemma}[Lebesgue measure on~$\matR^2$]
  \label{l:lebesgue-meas-on-r2}
  \mbox{}\\
  There is a unique measure on~$(\matR^2,\calBRtwo)$ that generalizes the
  area of bounded open boxes.

  It is denoted $\lambda^{\otimes2}\eqdef\lambda\otimes\lambda$, and it is
  called the {\em Lebesgue measure on (Borel subsets of)~$\matR^2$}.
\end{lemma}

\begin{proof}
  Direct consequence of
  Theorem~\thref{t:caratheodory-lebesgue-meas-on-r},
  Lemma~\thref{l:lebesgue-meas-is-sigma-finite}, and
  Lemma~\thref{l:uniq-of-tensor-prod-meas}.
\end{proof}

\begin{lemma}[Lebesgue measure on~$\matR^2$ generalizes area of boxes]
  \label{l:lebesgue-meas-on-r2-gen-area-of-boxes}
  \mbox{}\\
  Let~$a_1,b_1,a_2,b_2\in\matR$.
  Assume that $a_1\leq b_1$ and $a_2\leq b_2$.
  Then, we have
  \begin{equation}
    \label{e:lebesgue-meas-on-r2-gen-area-of-boxes}
    \lambda^{\otimes 2} (\lsrbra a_1, b_1\rsrbra \times \lsrbra a_2, b_2\rsrbra)
    = (b_1 - a_1) \, (b_2 - a_2).
  \end{equation}
\end{lemma}

\begin{proof}
  Direct consequence of
  Lemma~\thref{l:lebesgue-meas-on-r2},
  Definition~\thref{d:tensor-prod-meas},
  Lemma~\thref{l:lebesgue-meas-gen-len-of-int}, and
  \assume{the definition of the area of boxes of~$\matR^2$}.
\end{proof}

\begin{lemma}[Lebesgue measure on~$\matR^2$ is zero on lines]
  \label{l:lebesgue-meas-on-r2-is-zero-on-lines}
  \mbox{}\\
  Let~$A\subset\matR^2$ be a box ({\ie} the Cartesian product of two
  intervals).\\
  Then, we have $\lambda^{\otimes2}(A)=0$ iff
  $A$~is a line ({\ie} at least one of the two intervals is a singleton).
\end{lemma}

\begin{proof}
  Direct consequence of
  Lemma~\threfc{l:lebesgue-meas-on-r2-gen-area-of-boxes}{%
    with $a_1=b_1$ or $a_2=b_2$}, and
  Lemma~\thref{l:zero-prod-prop-in-rbarplus-mt}.
\end{proof}

\begin{lemma}[Lebesgue measure on~$\matR^2$ is $\sigma$-finite]
  \label{l:lebesgue-meas-on-r2-is-sigma-finite}
  \mbox{}\\
  The measure space $(\matR^2,\calBRtwo,\lambda^{\otimes2})$ is
  $\sigma$-finite.
\end{lemma}

\begin{proof}
  Direct consequence of
  Lemma~\thref{l:lebesgue-meas-on-r2},
  Lemma~\thref{l:lebesgue-meas-is-sigma-finite}, and
  Lemma~\thref{l:tensor-prod-of-sigma-finite-meas}.
\end{proof}

\begin{lemma}[Lebesgue measure on~$\matR^2$ is diffuse]
  \label{l:lebesgue-meas-on-r2-is-diffuse}
  \mbox{}\\
  The measure space~$(\matR^2,\calBRtwo,\lambda^{\otimes2})$ is diffuse.
\end{lemma}

\begin{proof}
  Direct consequence of
  Lemma~\threfc{l:lebesgue-meas-on-r2-gen-area-of-boxes}{%
    with $a_1=b_1$ and $a_2=b_2$}, and
  Definition~\threfc{d:diffuse-meas}{%
    $\{(a_1,a_2)\}=[a_1,a_1]\times[a_2,a_2]$}.
\end{proof}

\subsection{Integration over product space of nonnegative function}
\label{ss:integration-over-product-space-of-nonnegative-function}

\begin{definition}[partial function of function from product space]
  \label{d:partial-fun-of-fun-from-prod-space}
  \mbox{}\\
  Let~$(X_1,\Sigma_1,\mu_1)$ and~$(X_2,\Sigma_2,\mu_2)$ be measure spaces.
  Let~$f:X_1\times X_2\to\matRbarplus$.\\
  Let~$i\in\{1,2\}$.
  Let~$j\eqdef3-i$.
  Let $\psi\eqdef((x_i,x_j)\mapsto(x_1,x_2))$.\\
  For all $x_i\in X_i$, $f_{x_i}$~denotes the function
  $(x_j\mapsto f\circ\psi(x_i,x_j))$ from~$X_j$ to~$\matRbarplus$.

  Moreover, when~$f_{x_i}\in\calMplus(X_j,\Sigma_j)$, $I_{f,i}$~denotes the
  function $(x_i\mapsto\int f_{x_i}\,d\mu_j)$ from~$X_i$ to~$\matRbarplus$.
\end{definition}

\begin{remark}
  \label{r:v2-new18}
  The next proof follows steps~1 to~3 of the Lebesgue scheme
  (see Section~\ref{s:lebesgue-scheme}).

  See also the sketch of the proof in
  Section~\ref{s:sketch-of-the-proof-of-the-tonelli-th}.
\end{remark}

\begin{theorem}[Tonelli]
  \label{t:tonelli}
  \mbox{}\hfill
  Let~$(X_1,\Sigma_1,\mu_1)$ and~$(X_2,\Sigma_2,\mu_2)$ be $\sigma$-finite
  measure spaces.\\
  Let~$f\in\calMplus(X_1\times X_2,\Sigma_1\otimes\Sigma_2)$.
  Let~$i\in\{1,2\}$.
  Let~$j\eqdef3-i$.\\
  Then, for all $x_i\in X_i$, $f_{x_i}\in\calMplus(X_j,\Sigma_j)$,
  $I_{f,i}\in\calMplus(X_i,\Sigma_i)$, and we have
  \begin{equation}
    \label{e:tonelli}
    \int f \, d(\mu_1 \otimes \mu_2) = \int I_{f, i} \, d\mu_i.
  \end{equation}
\end{theorem}

\begin{proof}
  Let~$X\eqdef X_1\times X_2$, $\Sigma\eqdef\Sigma_1\otimes\Sigma_2$,
  and~$\mu\eqdef\mu_1\otimes\mu_2$.

  \proofparskip{(1). For $f\in\calIF(X,\Sigma)$}\\
  Let~$A\eqdef\{f\not=0\}$.
  Then, from
  Lemma~\thref{l:indic-and-support-are-each-other-inverse},
  we have $A\in\Sigma$ and $f=\matUN_A$.
  Let~$x_i\in X_i$.
  Then, from
  Definition~\thref{d:partial-fun-of-fun-from-prod-space},
  Lemma~\thref{l:indic-of-section},
  Lemma~\threfc{l:measurability-of-section}{$s_i(x_i,A)\in\Sigma_j$}, and
  Lemma~\threfc{l:int-in-mplus-of-indic-fun}{%
    with $A\eqdef s_i(x_i,A)$ and $\mu_j$},
  we have
  \begin{equation*}
    f_{x_i} = \matUN_{s_i (x_i, A)} \in \calMplus (X_j, \Sigma_j)
    \AND
    I_{f,i} = \int \matUN_{s_i (x_i, A)} \, d\mu_j = \mu_j (s_i (x_i, A)).
  \end{equation*}
  Hence, from
  Lemma~\thref{l:meas-of-section}, and
  Lemma~\thref{l:meas-of-meas-of-section},
  we have $I_{f,i}=F^A_i\in\calMplus(X_{i},\Sigma_{i})$.
  Therefore, from
  Lemma~\threfc{l:int-in-mplus-of-indic-fun}{%
    with $\matUN_A\in\calMplus(X,\Sigma)$ and $\mu$}, and
  Lemma~\thref{l:uniq-of-tensor-prod-meas},
  we have
  \begin{equation*}
    \int f \, d\mu
    = \mu (A)
    = \int F^A_i \, d\mu_i
    = \int I_{f, i} \, d\mu_i.
  \end{equation*}

  \proofparskip{(2). For $f\in\calSFplus(X,\Sigma)$}\\
  From
  Lemma~\thref{l:sfplus-simple-repr},
  let~$n\in\matN$, $(a_k)_{k\in[0..n]}\in\matRplus$
  and~$(A_k)_{k\in[0..n]}\in\Sigma$ such that
  $f=\sum_{k\in[0..n]}a_k\,\matUN_{A_k}$.
  Let~$x_i\in X_i$.
  Then, from
  \assume{left linearity of composition}, (1),
  Lemma~\thref{l:mplus-is-closed-under-add},
  Lemma~\thref{l:mplus-is-closed-under-nonneg-scalar-mult}, and
  Lemma~\threfc{l:int-in-mplus-is-pos-lin}{with $\mu_j$},
  we have 
  \begin{equation*}
    f_{x_i} = \sum_{k \in [0..n]} a_k \, (\matUN_{A_k})_{x_i}
    \in \calMplus (X_j, \Sigma_j)
    \AND 
    I_{f, i} = \sum_{k \in [0..n]} a_k \, I_{\matUN_{A_k}, i}
    \in \calMplus (X_i, \Sigma_i).
  \end{equation*}
  Therefore, from
  Lemma~\threfc{l:int-in-mplus-is-pos-lin}{with $\mu$, then $\mu_{i}$},
  and~(1), we have
  \begin{equation*}
    \int f \, d\mu
    = \sum_{k \in [0..n]} a_k \left( \int \matUN_{A_k} \, d\mu \right)
    = \sum_{k \in [0..n]} a_k \left( \int I_{\matUN_{A_k}, i} \, d\mu_i \right)
    = \int I_{f, i} \, d\mu_i.
  \end{equation*}

  \proofparskip{(3). For $f\in\calMplus(X,\Sigma)$}\\
  From
  Lemma~\thref{l:adapted-seq-in-mplus},
  let~$(\fhi_n)_{n\in\matN}\in\calSFplus(X,\Sigma)$ be an adapted sequence
  for~$f$.
  Let~$x_i\in X_i$.
  Then, from
  \assume{compatibility of composition with limit}, (2),
  Lemma~\thref{l:mplus-is-closed-under-limit-when-pointwise-conv}, and
  Lemma~\threfc{l:usage-of-adapted-seqs}{with $\mu_j$},
  we have 
  $f_{x_i}=\lim_{n\to\infty}(\fhi_n)_{x_i}\in\calMplus(X_j,\Sigma_j)$
  and
  $I_{f,i}=\lim_{n\to\infty}I_{\fhi_n,i}\in\calMplus(X_i,\Sigma_i)$.
  Therefore, from
  Lemma~\threfc{l:usage-of-adapted-seqs}{with $\mu$, then $\mu_{i}$},
  we have
  \begin{equation*}
    \int f \, d\mu
    = \lim_{n \to \infty} \int \fhi_n \, d\mu
    = \lim_{n \to \infty} \int I_{\fhi_n, i} \, d\mu_i
    = \int I_{f, i} \, d\mu_i.
  \end{equation*}
\end{proof}

\begin{lemma}[Tonelli over subset]
  \label{l:tonelli-over-subset}
  \mbox{}\hfill
  Let~$(X_1,\Sigma_1,\mu_1)$ and $(X_2,\Sigma_2,\mu_2)$ be $\sigma$-finite
  measure spaces.
  Let~$\Sigma\eqdef\Sigma_1\otimes\Sigma_2$.
  Let~$A\in\Sigma$.
  Let~$Y\subset X_1\times X_2$ such that $A\subset Y$.
  Let~$f:\ArYRbp$.
  Assume that $\restr{f}{A}\in\calMplus(A,\Sigma\olcap A)$.
  Let~$i\in\{1,2\}$ with~$j\eqdef3-i$.
  Let $\psi\eqdef((x_i,x_j)\mapsto(x_1,x_2))$.
  For all $x_i\in X_i$, let $A_{x_i}\eqdef s_i(x_i,A)$, and
  $f^A_{x_i}\eqdef(x_j\in A_{x_i}\mapsto f\circ\psi(x_i,x_j))$.\\
  Let $I^A_{f,i}\eqdef(x_i\in X_i\mapsto\int_{A_{x_i}} f^A_{x_i}\,d\mu_j)$.\\
  Then, for all $x_i\in X_i$,
  $f^A_{x_i}\in\calMplus(A_{x_i},\Sigma_j\olcap A_{x_i})$,
  $I^A_{f,i}\in\calMplus(X_i,\Sigma_i)$, and we have
  \begin{equation}
    \label{e:tonelli-over-subset}
    \int_A f \, d(\mu_1 \otimes \mu_2) = \int I^A_{f, i} \, d\mu_i.
  \end{equation}
\end{lemma}

\begin{proof}
  Let~$X\eqdef X_1\times X_2$.
  Let~$\mu\eqdef\mu_1\otimes\mu_2$.
  Let~$\hf:\ArXRbp$ such that $\restr{\hf}{Y}=f$.
  Then, from
  Lemma~\thref{l:int-in-mplus-over-subset},
  we have $\hf\,\matUN_A\in\calMplus(X,\Sigma)$ and
  $\int_Af\,d\mu=\int\hf\,\matUN_A\,d\mu$.
  Hence, from
  Theorem~\threfc{t:tonelli}{with $f\eqdef\hf\,\matUN_A$}, and
  Definition~\thref{d:partial-fun-of-fun-from-prod-space},
  for all $x_i\in X_i$, $(\hf\matUN_A)_{x_i}\in\calMplus(X_j,\Sigma_j)$,
  $I_{\hf\matUN_A,i}\in\calMplus(X_i,\Sigma_i)$, and we have
  $\int\hf\,\matUN_A\,d\mu=\int I_{\hf\matUN_A,i}\,d\mu_i$.

  Let~$x_i\in X_i$.
  Then, by construction, and from
  Lemma~\thref{l:indic-of-section},
  we have $f^A_{x_i}=(\hf\matUN_A)_{x_i}=\hf_{x_i}\matUN_{A_{x_i}}$.
  Therefore, from
  Lemma~\threfc{l:int-in-mplus-over-subset}{with $\mu_j$},
  we have $f^A_{x_i}\in\calMplus(A_{x_i},\Sigma_j\olcap A_{x_i})$,
  $I^A_{f,i}=\int(\hf\matUN_A)_{x_i}\,d\mu_j=I_{\hf\matUN_A,i}$, {\ie}
  $I^A_{f,i}\in\calMplus(X_i,\Sigma_i)$ and
  $\int_Af\,d\mu=\int I^A_{f,i}\,d\mu_i$.
\end{proof}

\begin{lemma}[Tonelli for tensor product]
  \label{l:tonelli-for-tensor-prod}
  \mbox{}\\
  For all $i\in\{1,2\}$, let~$(X_i,\Sigma_i,\mu_i)$ be a $\sigma$-finite
  measure space, and~$f_i\in\calMplus(X_i,\Sigma_i)$.\\
  Then, $f_1\otimes f_2\in\calMplus(X_1\times X_2,\Sigma_1\otimes\Sigma_2)$,
  and we have
  \begin{equation}
    \label{e:tonelli-for-tensor-prod}
    \int (f_1 \otimes f_2) \, d(\mu_1 \otimes \mu_2)
    = \left( \int f_1 \, d\mu_1 \right)
    \left( \int f_2 \, d\mu_2 \right).
  \end{equation}
\end{lemma}

\begin{proof}
  Let~$X\eqdef X_1\times X_2$, $\Sigma\eqdef\Sigma_1\otimes\Sigma_2$,
  $\mu\eqdef\mu_1\otimes\mu_2$, and~$f\eqdef f_1\otimes f_2$.\\
  Then, from
  Definition~\thref{d:tensor-prod-of-num-funs},
  Lemma~\thref{l:meas-of-tensor-prod-of-num-funs},
  Lemma~\thref{l:mult-in-rbarplus-is-closed-mt}, and
  Definition~\thref{d:mplus-subset-of-nonneg-meas-num-fun},
  $f$~belongs to~$\calMplus(X,\Sigma)$.

  Then, from
  Theorem~\threfc{t:tonelli}{with $i\eqdef1$},
  Definition~\thref{d:partial-fun-of-fun-from-prod-space},
  Definition~\thref{d:tensor-prod-of-num-funs},
  Lemma~\threfc{l:int-in-mplus-is-pos-hom}{%
    with $f\eqdef f_2$ and $a\eqdef f_1(x_1)\in\matRbarplus$}, and
  Lemma~\thref{l:mult-in-rbarplus-is-comm-mt},
  we have
  \begin{equation*}
    \int f \, d\mu
    = \int I_{f, 1} \, d\mu_1
    \quad\mbox{where }
    \forall x_1 \in X_1,\;
    I_{f, 1} (x_1)
    \eqdef \int f_1 (x_1) \, f_2 \, d\mu_2
    = \left( \int f_2 \, d\mu_2 \right) f_1 (x_1).
  \end{equation*}

  Therefore, from
  Lemma~\threfc{l:int-in-mplus-is-pos-hom}{%
    with $a\eqdef\int f_2\,d\mu_2\in\matRbarplus$ and $f\eqdef f_1$}, and
  Lemma~\thref{l:mult-in-rbarplus-is-comm-mt},
  we have
  \begin{equation*}
    \int f \, d\mu
    = \int \left( \int f_2 \, d\mu_2 \right) f_1 \, d\mu_1
    = \left( \int f_1 \, d\mu_1 \right) \left( \int f_2 \, d\mu_2 \right).
  \end{equation*}
\end{proof}

\chapter{Integration of real functions}
\label{c:integration-of-real-functions}

\minitoc

\begin{remark}
  From now on, the expressions involving integrals are taken in~$\matR$, thus
  functions also take almost all their values in~$\matR$.
\end{remark}

\section{Definition of the integral}
\label{s:definition-of-the-integral}

\begin{remark}
  \label{r:v2-new19}
  This section follows step~4 of the Lebesgue scheme
  (see Section~\ref{s:lebesgue-scheme}).
\end{remark}

\begin{definition}[integrability]
  \label{d:integrability}
  \mbox{}\hfill
  Let~$(X,\Sigma,\mu)$ be a measure space.\\
  A function~$f:\ArXRb$ is said
  {\em $\mu$-integrable (in~$\calM$)} iff
  $f^+$ and~$f^-$ are $\mu$-integrable in~$\calMplus$.
\end{definition}

\begin{lemma}[integrable is measurable]
  \label{l:integrable-is-meas}
  \mbox{}\hfill
  Let~$(X,\Sigma,\mu)$ be a measure space.\\
  Let~$f:\ArXRb$.
  Assume that~$f$ is $\mu$-integrable in~$\calM$.
  Then, we have $f\in\calM$.
\end{lemma}

\begin{proof}
  Direct consequence of
  Definition~\thref{d:integrability},
  Lemma~\threfc{l:int-in-mplus}{%
    definition of $\mu$-integrability with $f^+$ and $f^-$}, and
  Lemma~\thref{l:meas-of-nonneg-and-nonpos-parts}.
\end{proof}

\begin{lemma}[equivalent definition of integrability]
  \label{l:equiv-def-of-integrability}
  \mbox{}\hfill
  Let~$(X,\Sigma,\mu)$ be a measure space.
  Let~$f:\ArXRb$.
  Then, $f$~is $\mu$-integrable in~$\calM$ iff
  $f\in\calM$ and $|f|$~is $\mu$-integrable in~$\calMplus$.
\end{lemma}

\begin{proof}
  \proofpar{``Left'' implies ``right''}
  Direct consequence of
  Definition~\thref{d:integrability},
  Lemma~\thref{l:integrable-is-meas},
  Lemma~\thref{l:m-is-closed-under-abs},
  Lemma~\thref{l:int-in-mplus-of-decomp-into-nonpos-and-nonneg-parts},
  Lemma~\thref{l:add-in-rbarplus-is-closed}, and
  Lemma~\threfc{l:int-in-mplus}{%
      $|f|$ is $\mu$-integrable in $\calMplus$}.

  \proofparskip{``Right'' implies ``left''}
  Direct consequence of
  Lemma~\thref{l:meas-of-nonneg-and-nonpos-parts},
  Lemma~\thref{l:int-in-mplus-of-decomp-into-nonpos-and-nonneg-parts},
  Lemma~\threfc{l:infinity-sum-prop-in-rbarplus}{contrapositive},
  Lemma~\threfc{l:int-in-mplus}{%
    $f^+$ and $f^-$ are $\mu$-integrable in $\calMplus$}, and
  Definition~\thref{d:integrability}.

  \medskip\noindent
  Therefore, we have the equivalence.
\end{proof}

\begin{lemma}[compatibility of integrability in~$\calM$ and~$\calMplus$]
  \label{l:compat-of-integrability-in-m-and-mplus}
  \mbox{}\\
  Let~$(X,\Sigma,\mu)$ be a measure space.
  Let~$f:\ArXRb$.
  Assume that~$f$ is nonnegative.\\
  Then, $f$~is $\mu$-integrable in~$\calM$ iff
  $f$~is $\mu$-integrable in~$\calMplus$.
\end{lemma}

\begin{proof}
  Direct consequence of
  Lemma~\thref{l:equiv-def-of-integrability}, and
  Lemma~\threfc{l:equiv-def-of-abs-in-rbar}{$|f|=f$}.
\end{proof}

\begin{lemma}[integrable is almost finite]
  \label{l:integrable-is-almost-finite}
  \mbox{}\hfill
  Let~$(X,\Sigma,\mu)$ be a measure space.\\
  Let~$f:\ArXRb$ be $\mu$-integrable in~$\calM$.
  Then, we have $\mu(f^{-1}(\pm\infty))=0$, {\ie} $|f|\ltae{\mu}\infty$.
\end{lemma}

\begin{proof}
  Direct consequence of
  Lemma~\threfc{l:equiv-def-of-integrability}{%
    $|f|$ is integrable in~$\calMplus$},
  Lemma~\threfc{l:abs-in-rbar-is-even}{%
    $f^{-1}(\pm\infty)=|f|^{-1}(\infty)$}, and
  Lemma~\threfc{l:integrable-in-mplus-is-almost-finite}{with $|f|$}.
\end{proof}

\begin{lemma}[almost bounded by integrable is integrable]
  \label{l:almost-bounded-by-integrable-is-integrable}
  \mbox{}\\
  Let~$(X,\Sigma,\mu)$ be a measure space.
  Let~$f\in\calM$.
  Then, $f$~is $\mu$-integrable in~$\calM$ iff
  there exists $g:\ArXRbp$ such that~$g$ is $\mu$-integrable in~$\calMplus$
  and $|f|\leqae{\mu}g$.
\end{lemma}

\begin{proof}
  \proofpar{``Left'' implies ``right''}
  Direct consequence of
  Lemma~\threfc{l:equiv-def-of-integrability}{with $g\eqdef|f|$},
  Lemma~\threfc{l:almost-order-is-order-rel}{reflexivity}.

  \proofparskip{``Right'' implies ``left''}
  Direct consequence of
  Lemma~\threfc{l:m-is-closed-under-abs}{$|f|\in\calMplus$},
  Lemma~\threfc{l:int-in-mplus-is-almost-monot}{with $|f|$ and $g$},
  Lemma~\threfc{l:order-in-rbar-is-total}{transitivity},
  Lemma~\threfc{l:int-in-mplus}{%
    $|f|$ is $\mu$-integrable in $\calMplus$}, and
  Lemma~\thref{l:equiv-def-of-integrability}.

  \medskip\noindent
  Therefore, we have the equivalence.
\end{proof}

\begin{lemma}[bounded by integrable is integrable]
  \label{l:bounded-by-integrable-is-integrable}
  \mbox{}\\
  Let~$(X,\Sigma,\mu)$ be a measure space.
  Let~$f\in\calM$.\\
  Then, $f$~is $\mu$-integrable in~$\calM$ iff
  there exists $g:\ArXRbp$ $\mu$-integrable in~$\calMplus$ such that
  $|f|\leq g$.
\end{lemma}

\begin{proof}
  \proofpar{``Left'' implies ``right''}
  Direct consequence of
  Lemma~\threfc{l:equiv-def-of-integrability}{with $g\eqdef|f|$}, and
  Lemma~\threfc{l:order-in-rbar-is-total}{reflexivity}.

  \proofparskip{``Right'' implies ``left''}
  Direct consequence of
  Lemma~\thref{l:everywhere-implies-almost-everywhere}, and
  Lemma~\thref{l:almost-bounded-by-integrable-is-integrable}.

  \medskip\noindent
  Therefore, we have the equivalence.
\end{proof}

\begin{definition}[integral]
  \label{d:integral}
  \mbox{}\hfill
  Let~$(X,\Sigma,\mu)$ be a measure space.
  Let~$f$ be $\mu$-integrable in~$\calM$.
  The {\em Lebesgue integral of~$f$ (for the measure~$\mu$)} is still denoted
  $\int f\,d\mu$;
  it is defined by
  \begin{equation}
    \label{e:integral}
    \int f \, d\mu
    \eqdef \int f^+ \, d\mu - \int f^- \, d\mu
    \quad \in \matR.
  \end{equation}
\end{definition}

\begin{lemma}[compatibility of integral in~$\calM$ and~$\calMplus$]
  \label{l:compat-of-int-in-m-and-mplus}
  \mbox{}\hfill
  Let~$(X,\Sigma,\mu)$ be a measure space.
  Let~$f$ be $\mu$-integrable in~$\calM$.
  Assume that~$f$ is nonnegative.
  Then, both
  Lemma~\thref{l:int-in-mplus}, and
  Definition~\thref{d:integral}
  provide the same value for the integral of~$f$.
\end{lemma}

\begin{proof}
  Direct consequence of
  Definition~\threfc{d:nonneg-and-nonpos-parts}{%
    $f^+=f$ and $f^-=0$},
  Definition~\thref{d:integrability},
  Lemma~\threfc{l:int-in-mplus}{%
    $f$~is $\mu$-integrable in~$\calMplus$},
  Definition~\thref{d:integral}, and
  Lemma~\thref{l:int-in-mplus-of-zero-is-zero}.
\end{proof}

\begin{lemma}[integral of zero is zero]
  \label{l:int-of-zero-is-zero}
  \mbox{}\\
  Let~$(X,\Sigma,\mu)$ be a measure space.
  Then, $0$~is $\mu$-integrable in~$\calM$, and $\int0\,d\mu=0$.
\end{lemma}

\begin{proof}
  Direct consequence of
  Lemma~\thref{l:compat-of-int-in-m-and-mplus}, and
  Lemma~\thref{l:int-in-mplus-of-zero-is-zero}.
\end{proof}

\begin{definition}[merge integral in~$\calM$ and~$\calMplus$]
  \label{d:merge-int-in-m-and-mplus}
  \mbox{}\\
  Let~$(X,\Sigma,\mu)$ be a measure space.
  Let~$f\in\calM$.
  The {\em integral of f (for the measure~$\mu$) exists} iff
  $f$~is nonnegative (and $\int f\,d\mu\in\matRbarplus$), or
  $f$~is $\mu$-integrable (and $\int f\,d\mu\in\matR$).
\end{definition}

\begin{lemma}[compatibility of integral with almost equality]
  \label{l:compat-of-int-with-almost-eq}
  \mbox{}\\
  Let~$(X,\Sigma,\mu)$ be a measure space.
  Let~$f$ be $\mu$-integrable in~$\calM$.
  Let~$g\in\calM$.
  Assume that~$g\eqae{\mu}f$.
  Then, $g$~is $\mu$-integrable in~$\calM$, and we have
  \begin{equation}
    \label{e:compat-of-int-with-almost-eq}
    \int g \, d\mu = \int f \, d\mu.
  \end{equation}
\end{lemma}

\begin{proof}
  Direct consequence of
  Lemma~\threfc{l:compat-of-almost-eq-with-op}{%
    with the binary operator $\max$ and the unary operator additive inverse,
    $g^+\eqae{\mu}f^+$ and $g^-\eqae{\mu}f^-$},
  Lemma~\thref{l:compat-of-int-in-mplus-with-almost-eq},
  Definition~\thref{d:integrability}, and
  Definition~\thref{d:integral}.
\end{proof}

\begin{remark}
  From Lemma~\ref{l:integrable-is-almost-finite}, integrable functions can only
  take infinite values on a negligible subset, and from the previous lemma, the
  integral keeps the same value when the function is modified on a negligible
  subset.
  Therefore, we can restrict the study of integrable functions to the sole case
  of real-valued functions.
\end{remark}

\clearpage
\section{Notations for specific cases}
\label{s:notations-for-specific-cases}

\begin{lemma}[integral over subset]
  \label{l:int-over-subset}
  \mbox{}\\
  Let~$(X,\Sigma,\mu)$ be a measure space.
  Let~$A\in\Sigma$.
  Let~$Y\subset X$ such that $A\subset Y$.
  Let~$f:\ArYRb$.
  Let~$\hf:\ArXRb$.
  Assume that $\restr{\hf}{Y}=f$.
  Then, $\restr{f}{A}$~is $\mu_A$-integrable in~$\calM(A,\Sigma\olcap A)$ iff
  $\hf\,\matUN_A$~is $\mu$-integrable in~$\calM(X,\Sigma)$.
  If so, the function~$f$ is said {\em $\mu$-integrable over~$A$}, and we
  have
  \begin{equation}
    \label{e:int-over-subset}
    \int \restr{f}{A} \, d\mu_A = \int \hf \, \matUN_A \, d\mu.
  \end{equation}
  This integral is still denoted $\int_Af\,d\mu$;
  it is still called {\em integral of~$f$ over~$A$}.
\end{lemma}

\begin{proof}
  \proofpar{Equivalence}
  Direct consequence of
  Lemma~\thref{l:compat-of-nonpos-and-nonneg-parts-with-restr},
  Lemma~\threfc{l:compat-of-nonpos-and-nonneg-parts-with-mask}{%
    $(\hf)^\pm$ is an extension of~$f^\pm$ to~$X$},
  Lemma~\threfc{l:int-in-mplus-over-subset}{%
    $\int\restr{f^\pm}{A}\,d\mu_A$ equals $\int (\hf)^\pm\,\matUN_A\,d\mu$ and
    is finite},
  Definition~\thref{d:integrability},
  Lemma~\threfc{l:int-in-mplus}{definition of integrability in~$\calMplus$},
  Lemma~\threfc{l:int-in-mplus-over-subset}{%
    with $\!f^\pm$ and $\hf\!\eqdef\!(\hf)^\pm$}, and
  \assume{the tautology
    \begin{equation*}
      \left\{
        \begin{array}{l}
          P \Equiv Q \Conj R\\
          \hP \Equiv \hQ \Conj \hR\\
          Q \Equiv \hQ\\
          Q \Disj \hQ \Implies (R \Equiv \hR)
        \end{array}
      \right.
      \IMPLIES (P \Equiv \hP).
    \end{equation*}}

  \proofparskip{Identity}
  Direct consequence of
  Definition~\threfc{d:integral}{%
    with $\restr{f}{A}$, then $\hf\,\matUN_A$}, and
  Lemma~\threfc{l:int-in-mplus-over-subset}{%
    with $f\eqdef f^\pm$ and $\hf\eqdef(\hf)^\pm$}.
\end{proof}

\begin{lemma}[integral over subset is $\sigma$-additive]
  \label{l:int-over-subset-is-sigma-add}
  \mbox{}\\
  Let~$(X,\Sigma,\mu)$ be a measure space.
  Let~$I\subset\matN$.
  Let~$A,(A_i)_{i\in I}\in\Sigma$.
  Assume that $A=\biguplus_{i\in I}A_i$.
  Let~$Y\subset X$ such that $A\subset Y$.
  Let~$f:\ArYRb$.
  Let~$\hf:\ArXRb$.
  Assume that $\restr{\hf}{Y}=f$.\\
  Then, $\hf\,\matUN_A\in\calM$ iff for all $i\in I$,
  $\hf\,\matUN_{A_i}\in\calM$.

  Moreover, if $\sum_{i\in I}|\hf\,\matUN_{A_i}|$ is $\mu$-integrable, then
  $f$~is $\mu$-integrable over~$A$, and we have
  \begin{equation}
    \label{e:int-over-subset-is-sigma-add}
    \int_A f \, d\mu = \sum_{i \in I} \int_{A_i} f \, d\mu.
  \end{equation}
\end{lemma}

\begin{proof}
  \proofpar{Equivalence}
  Direct consequence of
  Lemma~\thref{l:equiv-def-of-integrability},
  Lemma~\threfc{l:int-in-mplus-over-subset-is-sigma-add}{%
    equivalence, same proof for $\hf\,\matUN_A\in\calM$ iff for all $i\in I$,
    $\hf\,\matUN_{A_i}\in\calM$ since the countable sum is well-defined}.

  \proofparskip{$f$~is $\mu$-integrable over~$A$}
  Direct consequence of
  \assume{the countable triangle inequality for the absolute value
    in~$\matR$},
  Lemma~\thref{l:int-in-mplus-is-monot}, and
  Lemma~\threfc{l:int-in-mplus-is-sigma-add}{%
    $\int|\hf\,\matUN_A|\,d\mu\leq
    \sum_{i\in I}\int|\hf\,\matUN_{A_i}|\,d\mu$}.

  \proofparskip{Identity}
  Direct consequence of
  Definition~\threfc{d:integral}{with $f\eqdef\hf\,\matUN_A$},
  Lemma~\threfc{l:int-in-mplus-over-subset-is-sigma-add}{%
    with $f^+$ and $f^-$},
  \assume{associativity of addition in~$\matR$
    (both countable sums are finite)}.
\end{proof}

\begin{lemma}[integral over singleton]
  \label{l:int-over-singleton}
  \mbox{}\hfill
  Let~$(X,\Sigma,\mu)$ be a measure space.\\
  Let~$a\in X$.
  Assume that $\{a\}\in\Sigma$.
  Let~$f:\ArXRb$.
  Then, $f$~is $\mu$-integrable over~$\{a\}$ iff
  \begin{equation}
    \label{e:int-over-singleton-1}
    (f (a) \mbox{ and } \mu (\{ a \}) \mbox{ are finite})
    \DISJ f (a) = 0
    \DISJ \mu (\{ a \}) = 0.
  \end{equation}
  If so, we have
  \begin{equation}
    \label{e:int-over-singleton-2}
    \int_{\{ a \}} f \, d\mu = f (a) \mu(\{ a \}).
  \end{equation}
\end{lemma}

\begin{proof}
  From
  Lemma~\threfc{l:int-over-subset}{$\mu$-integrability over subset},
  Lemma~\thref{l:equiv-def-of-integrability},
  Lemma~\threfc{l:int-in-mplus}{$\mu$-integrability in~$\calMplus$},
  \assume{nonnegativeness of the indicator function},
  Lemma~\threfc{l:int-in-mplus-over-singleton}{%
    with $|f|$, $|f\,\matUN_{\{a\}}|$ is equal to
    $|f(a)|\,\matUN_{\{a\}}\in\calMplus$},
  Lemma~\thref{l:finite-prod-prop-in-rbarplus-mt},
  \assume{closedness of absolute value in~$\matR$},
  Definition~\threfc{d:abs-in-rbar}{%
    absolute value is closed in $\{\pm\infty\}$}, and
  Lemma~\thref{l:abs-in-rbar-is-definite},
  we have
  \begin{align*}
    f \mbox{ is $\mu$-integrable over } \{ a \}
    & \EQUIV f \, \matUN_{\{ a \}} \mbox{ is $\mu$-integrable in } \calM\\
    & \EQUIV | f \, \matUN_{\{ a \}} |
      \mbox{ is $\mu$-integrable in } \calMplus\\
    & \EQUIV | f \, \matUN_{\{ a \}} | \in \calMplus
      \Conj \int | f \, \matUN_{\{ a \}} | \, d\mu < \infty\\
    & \EQUIV | f (a) | \mu (\{ a \}) < \infty\\
    & \EQUIV (f (a) \in \matR \CONJ \mu (\{ a \}) \in \matRplus)\\
    & \hphantom{\EQUIV} \DISJ f (a) = 0 \DISJ \mu (\{ a \}) = 0.
  \end{align*}

  Assume that~$f$ is $\mu$-integrable over~$\{a\}$.
  Then, from
  Lemma~\threfc{l:int-over-subset}{%
    $f\,\matUN_{\{a\}}$ is $\mu$-integrable in $\calM$},
  Definition~\thref{d:integrability},
  Definition~\thref{d:nonneg-and-nonpos-parts}, and
  \assume{nonnegativeness of the indicator function},
  \begin{equation*}
    (f \, \matUN_{\{a\}})^+ = f^+ \, \matUN_{\{a\}}
    \mbox{ and }
    (f \, \matUN_{\{a\}})^- = f^- \, \matUN_{\{a\}} \,
    \mbox{ are $\mu$-integrable in } \calMplus.
  \end{equation*}
  Moreover, from
  Lemma~\thref{l:int-over-subset},
  Definition~\thref{d:integral},
  Lemma~\thref{l:int-in-mplus-over-singleton}, and
  Lemma~\threfc{l:decomp-into-nonneg-and-nonpos-parts}{%
    $f(a)= f^+(a)-f^-(a)$},
  we have
  \begin{align*}
    \int_{\{ a \}} f \, d\mu
    & = \int f \, \matUN_{\{ a \}} \, d\mu
      = \int (f \, \matUN_{\{ a \}})^+ \, d\mu
      - \int (f \, \matUN_{\{ a \}})^- \, d\mu\\
    & = f^+ (a) \mu (\{ a \}) - f^- (a) \mu (\{ a \})
      = f (a) \mu (\{ a \}).
  \end{align*}
\end{proof}

\begin{lemma}[integral over interval]
  \label{l:int-over-int}
  \mbox{}\\
  Let~$(\matR,\calBR,\mu)$ be a measure space.
  Assume that~$\mu$ is diffuse.
  Let~$a,b\in\matRbar$ such that $a\leq b$.
  Let~$f$ be a $\mu$-integrable function over $(a,b)$.
  Then, the integral over the interval remains the same when the interval is
  closed at one or both of its finite extremities.

  The integral of~$f$ over an interval with extremities~$a$ and~$b$ is called
  {\em integral of~$f$ from~$a$ to~$b$};
  it is denoted $\int_a^bf(x)\,d\mu(x)$.
  When $a<b$, we assume the convention
  $\int_b^af(x)\,d\mu(x)\eqdef-\int_a^bf(x)\,d\mu(x)$.
\end{lemma}

\begin{proof}
  Direct consequence of
  Definition~\thref{d:diffuse-meas},
  Lemma~\threfc{l:int-over-subset-is-sigma-add}{%
    on $[a,b]=\{a\}\uplus(a,b)\uplus\{b\}$},
  Lemma~\threfc{l:some-borel-subsets}{singletons are measurable},
  Lemma~\thref{l:int-over-singleton}, and
  Definition~\thref{d:diffuse-meas}.
\end{proof}

\begin{lemma}[Chasles relation, integral over split intervals]
  \label{l:chasles-rel-int-over-split-ints}
  \mbox{}\\
  Let~$(\matR,\calBR,\mu)$ be a measure space.
  Assume that~$\mu$ is diffuse.
  Let~$a,b,c\in\matRbar$.
  Let~$f$ be a $\mu$-integrable function over $(\min(a,b,c),\max(a,b,c))$.
  Then, we have
  \begin{equation}
    \label{e:chasles-rel-int-over-split-ints}
    \int_a^b f (x) \, d\mu (x)
    = \int_a^c f (x) \, d\mu (x) + \int_c^b f (x) \, d\mu (x).
  \end{equation}
\end{lemma}

\begin{proof}
  Direct consequence of
  Lemma~\thref{l:int-over-subset-is-sigma-add}.
  Lemma~\threfc{l:int-over-int}{%
    convention for reverse bounds}, and
  \assume{field properties of~$\matR$}.

  It is trivial when $a\leq c\leq b$.
  For instance, when $c<b<a$, we have
  \begin{equation*}
    \int_c^a f (x) \, d\mu (x)
    = \int_c^b f (x) \, d\mu (x) + \int_b^a f (x) \, d\mu (x).
  \end{equation*}
  Thus,
  \begin{align*}
    \int_a^b f (x) \, d\mu (x) & = - \int_b^a f (x) \, d\mu (x)\\
    & = - \int_c^a f (x) \, d\mu (x) + \int_c^b f (x) \, d\mu (x)\\
    & = \int_a^c f (x) \, d\mu (x) + \int_c^b f (x) \, d\mu (x).
  \end{align*}
\end{proof}

\begin{lemma}[integral for counting measure]
  \label{l:int-for-count-meas}
  \mbox{}\\
  Let~$(X,\Sigma)$ be a measurable space.
  Let~$Y\subset X$.
  Let~$f\in\calM$.
  Assume that $\sum_{y\in Y}|f(y)|$ is finite.
  Then, $f$~is $\delta_Y$-integrable in~$\calM$, and we have
  \begin{equation}
    \label{e:int-for-count-meas}
    \int f \, d\delta_Y = \sum_{y \in Y} f (y).
  \end{equation}
\end{lemma}

\begin{proof}
  From
  Lemma~\threfc{l:int-in-mplus-for-count-meas}{%
    $\int|f|\,d\delta_Y$ is finite},
  Lemma~\threfc{l:int-in-mplus}{%
    definition of $\mu$-integrability in $\calMplus$}, and
  Lemma~\thref{l:equiv-def-of-integrability},
  $f$~is $\delta_Y$-integrable in~$\calM$.
  Then, from
  Definition~\thref{d:integral},
  Lemma~\thref{l:int-in-mplus-for-count-meas},
  Lemma~\thref{l:decomp-into-nonneg-and-nonpos-parts}, and
  \assume{associativity and commutativity of (possibly uncountable) addition
    for absolutely convergent sums},
  we have
  \begin{equation*}
    \int f \, d\delta_Y
    = \int f^+ \, d\delta_Y - \int f^- \, d\delta_Y
    = \sum_{y \in Y} f^+ (y) - \sum_{y \in Y} f^- (y)
    = \sum_{y \in Y} f (y).
  \end{equation*}
\end{proof}

\begin{lemma}[integral for counting measure on~$\matN$]
  \label{l:int-for-count-meas-on-n}
  \mbox{}\hfill
  Let~$f:\ArNRb$ be a sequence.\\
  Assume that $\sum_{n\in\matN}|f(n)|$ is finite.
  Then, $f$~is $\delta_Y$-integrable in~$\calM(\matN,\calP(\matN))$, and we
  have
  \begin{equation}
    \label{e:int-for-count-meas-on-n}
    \int f \, d\delta_\matN = \sum_{n \in \matN} f (n).
  \end{equation}
\end{lemma}

\begin{proof}
  Direct consequence of
  Lemma~\threfc{l:int-for-count-meas}{%
    with $Y\!=\!X\eqdef\matN$ and $\Sigma\eqdef\calP(\matN)$}.
\end{proof}

\begin{remark}
  Note that the previous lemma makes absolutely convergent series be
  Lebesgue integrals for the counting measure on natural numbers.
  Thus, the theory of absolutely converging series can be derived from the
  theory of Lebesgue integration.
\end{remark}

\begin{lemma}[integral for Dirac measure]
  \label{l:int-for-dirac-meas}
  \mbox{}\hfill
  Let~$(X,\Sigma)$ be a measurable space.
  Let~$\{a\}\in\Sigma$.
  Let~$f\in\calM$.
  Assume that~$f(a)$ is finite.
  Then, $f$~is $\delta_a$-integrable in~$\calM$, and we have
  \begin{equation}
    \label{e:int-for-dirac-meas}
    \int f \, d\delta_a = f (a).
  \end{equation}
\end{lemma}

\begin{proof}
  Direct consequence of
  Definition~\thref{d:dirac-meas}, and
  Lemma~\threfc{l:int-for-count-meas}{with $Y\eqdef\{a\}$}.
\end{proof}

\begin{definition}[integral for Lebesgue measure on~$\matR$]
  \label{d:int-for-lebesgue-meas-on-r}
  \mbox{}\\
  Let~$a,b\in\matRbar$.
  Let~$Y$ be an interval with extremities~$\min(a,b)$ and~$\max(a,b)$.
  Let~$f:\ArYR$.
  Assume that~$f$ is $\lambda$-integrable over~$Y$ (for the Lebesgue
  measure~$\lambda$).
  The integral of~$f$ from~$a$ to~$b$ is denoted $\int_a^bf(x)\,dx$.
\end{definition}

\clearpage
\section{The seminormed {\vectorspace}~\calLpintitle{1}}
\label{s:the-seminormed-vector-space-llone}

\begin{lemma}[seminorm~$\calLone$]
  \label{l:seminorm-llone}
  \mbox{}\hfill
  Let~$(X,\Sigma,\mu)$ be a measure space.
  The function
  \begin{equation}
    \label{e:seminorm-llone}
    N_1 \eqdef \left(
      f \in \calM \longmapsto \int | f | \, d\mu \in \matRbarplus \right)
  \end{equation}
  is well-defined;
  it is called {\em seminorm~$\calLone$}.
\end{lemma}

\begin{proof}
  Direct consequence of
  Lemma~\threfc{l:m-is-closed-under-abs}{$|f|\in\calMplus$},
  Definition~\thref{d:merge-int-in-m-and-mplus}, and
  Lemma~\threfc{l:int-in-mplus}{nonnegativeness}.
\end{proof}

\begin{remark}
  \mbox{}\\
  The function~$N_1$ is shown below to be a seminormed on the {\vectorspace}
  $\calLone$, hence its name.
\end{remark}

\begin{lemma}[integrable is finite seminorm~$\calLone$]
  \label{l:integrable-is-finite-seminorm-lone}
  \mbox{}\hfill
  Let~$(X,\Sigma,\mu)$ be a measure space.
  Let~$f\in\ArXRb$.
  Then, $f$~is $\mu$-integrable in~$\calM$ iff
  $f\in\calM$ and $N_1(f)<\infty$.
\end{lemma}

\begin{proof}
  \proofpar{``Left'' implies ``right''}
  Direct consequence of
  Lemma~\thref{l:equiv-def-of-integrability},
  Lemma~\threfc{l:int-in-mplus}{definition of $\mu$-integrability}, and
  Lemma~\thref{l:seminorm-llone}.

  \proofparskip{``Right'' implies ``left''}
  Direct consequence of
  Lemma~\threfc{l:m-is-closed-under-abs}{$|f|\in\calMplus$},
  Lemma~\thref{l:seminorm-llone},
  Lemma~\threfc{l:int-in-mplus}{definition of $\mu$-integrability}, and
  Lemma~\thref{l:equiv-def-of-integrability}.
\end{proof}

\begin{lemma}[compatibility of~$N_1$ with almost equality]
  \label{l:compat-of-none-with-almost-eq}
  \mbox{}\hfill
  Let~$(X,\Sigma,\mu)$ be a measure space.\\
  Let $f,g\in\calM$.
  Assume that $f\eqae{\mu}g$.
  Then, we have $N_1(f)=N_1(g)$.
\end{lemma}

\begin{proof}
  Direct consequence of
  Lemma~\thref{l:seminorm-llone},
  Lemma~\threfc{l:compat-of-almost-eq-with-op}{%
    with the unary operator absolute value}, and
  Lemma~\thref{l:compat-of-int-with-almost-eq}.
\end{proof}

\begin{lemma}[$N_1$~is almost definite]
  \label{l:none-is-almost-definite}
  \mbox{}\\
  Let~$(X,\Sigma,\mu)$ be a measure space.
  Then, $N_1$~is almost definite:
  \begin{equation}
    \label{e:none-is-almost-definite}
    \forall f \in \calM,\quad N_1 (f) = 0 \EQUIV f \eqae{\mu} 0.
  \end{equation}
\end{lemma}

\begin{proof}
  Direct consequence of
  Lemma~\thref{l:seminorm-llone},
  Lemma~\thref{l:int-in-mplus-is-almost-definite}, and
  Lemma~\thref{l:abs-is-almost-definite}.
\end{proof}

\begin{lemma}[$N_1$~is absolutely homogeneous]
  \label{l:none-is-abs-hom}
  \mbox{}\\
  Let~$(X,\Sigma,\mu)$ be a measure space.
  Then, $N_1$~is absolutely homogeneous of degree~1:
  \begin{equation}
    \label{e:none-is-abs-hom}
    \forall \lambda \in \matRbar,\;
    \forall f \in \calM,\quad
    N_1 (\lambda f) = | \lambda | N_1 (f).
  \end{equation}
\end{lemma}

\begin{proof}
  Direct consequence of
  Lemma~\thref{l:seminorm-llone},
  \assume{multiplicativity of the absolute value},
  Lemma~\thref{l:int-in-mplus-is-pos-hom}, and
  Lemma~\thref{l:int-in-mplus-is-hom-at-infinity}.
\end{proof}

\begin{lemma}[integral is homogeneous]
  \label{l:int-is-hom}
  \mbox{}\hfill
  Let~$(X,\Sigma,\mu)$ be a measure space.\\
  Let~$f$ be $\mu$-integrable in~$\calM$.
  Let~$a\in\matR$.
  Then, $af$ is $\mu$-integrable in~$\calM$, and we have
  \begin{equation}
    \label{e:int-is-hom}
    \int a f \, d\mu = a \int f \, d\mu.
  \end{equation}
\end{lemma}

\begin{proof}
  From
  Lemma~\threfc{l:integrable-is-meas}{$f\in\calM$},
  Lemma~\threfc{l:m-is-closed-under-scalar-mult}{$af\in\calM$},
  Lemma~\threfc{l:none-is-abs-hom}{%
    $N_1(af)=|a|N_1(f)<\infty$},
  Lemma~\thref{l:seminorm-llone},
  Lemma~\threfc{l:int-in-mplus}{%
    definition of $\mu$-integrability in $\calMplus$}, and
  Lemma~\thref{l:equiv-def-of-integrability},
  $af$~is $\mu$-integrable in~$\calM$.
  Hence, from
  Lemma~\thref{l:meas-of-nonneg-and-nonpos-parts},
  Definition~\thref{d:nonneg-and-nonpos-parts}, and
  \assume{ordered set properties of~$\matRbar$},
  we have $f^\pm\in\calMplus$,
  \begin{equation*}
    (a f)^\pm = a f^\pm \in \calMplus \mbox{ when } a \geq 0,
    \AND
    (a f)^\pm = -a f^\mp \in \calMplus \mbox{ when } a < 0.
  \end{equation*}
  Therefore, from
  Definition~\threfc{d:integral}{with $af$},
  Lemma~\threfc{l:int-in-mplus-is-pos-hom}{%
    with $af^\pm$ or $-af^\mp$},
  \assume{field properties of~$\matR$ (all integrals below are finite)}, and
  Definition~\threfc{d:integral}{with $f$},
  we have in both cases
  \begin{equation*}
    \int a f \, d\mu
    = \int (a f)^+ \, d\mu - \int (a f)^- \, d\mu
    = a \int f^+ \, d\mu - a \int f^- \, d\mu
    = a \int f \, d\mu.
  \end{equation*}
\end{proof}

\begin{remark}
  In the next two lemmas, the summability domain~$\Dfsum(f,g)$ and the almost
  sum $f\plusae{\mu}g$ are respectively defined in
  Definition~\ref{d:summability-domain}, and Lemma~\ref{l:almost-sum}.
\end{remark}

\begin{lemma}[Minkowski inequality in~$\calM$]
  \label{l:minkowski-ineq-in-m}
  \mbox{}\hfill
  Let~$(X,\Sigma,\mu)$ be a measure space.\\
  Let~$f,g\in\calM$.
  Assume that~$f+g$ is well-defined almost everywhere.
  Let $\fp,\gp\in\calM$.
  Assume that $\fp\eqae{\mu}f$, $\gp\eqae{\mu}g$, and $\Dfsum(\fp,\gp)=X$.
  Then, $\fp+\gp,f\plusae{\mu}g\in\calM$, and we have
  \begin{equation}
    \label{e:minkowski-ineq-in-m}
    N_1 (\fp + \gp)
    = N_1 (f \plusae{\mu} g)
    \leq N_1 (f) + N_1 (g).
  \end{equation}
\end{lemma}

\begin{proof}
  From
  Lemma~\thref{l:almost-sum}, and
  Lemma~\thref{l:compat-of-almost-sum-with-almost-eq},
  we have $f\plusae{\mu}g,\fp+\gp\in\calM$ and
  $\fp+\gp\eqae{\mu}f\plusae{\mu}g$.
  Then, from
  Lemma~\thref{l:m-is-closed-under-abs},
  we have $|f|,|g|,|\fp|,|\gp|,|\fp+\gp|\in\calMplus$.
  Therefore, from
  Lemma~\thref{l:compat-of-none-with-almost-eq},
  \assume{symmetry of equality},
  Lemma~\threfc{l:seminorm-llone}{with $\fp+\gp$},
  Lemma~\thref{l:abs-in-rbar-satisfies-triangle-ineq},
  Lemma~\thref{l:int-in-mplus-is-monot},
  Lemma~\thref{l:int-in-mplus-is-add},
  Lemma~\threfc{l:seminorm-llone}{with $\fp$ and $\gp$}, and
  Lemma~\thref{l:compat-of-none-with-almost-eq},
  we have
  \begin{align*}
    N_1 (f \plusae{\mu} g)
    & = N_1 (\fp + \gp)
      = \int | \fp + \gp | \, d\mu\\
    & \leq \int (| \fp | + | \gp |) \, d\mu
      = \int | \fp | \, d\mu + \int | \gp | \, d\mu
      = N_1 (\fp) + N_1 (\gp)
      = N_1 (f) + N_1 (g).
  \end{align*}
\end{proof}

\begin{lemma}[integral is additive]
  \label{l:int-is-add}
  \mbox{}\\
  Let~$(X,\Sigma,\mu)$ be a measure space.
  Let~$f,g$ be $\mu$-integrable in~$\calM$.
  Let $\fp,\gp\in\calM$.
  Assume that $\fp\eqae{\mu}f$, $\gp\eqae{\mu}g$, and $\Dfsum(\fp,\gp)=X$.
  Then, $f+g$ is well-defined almost everywhere, $\fp+\gp$ and $f\plusae{\mu}g$
  are $\mu$-integrable in~$\calM$, and we have
  \begin{equation}
    \label{e:int-is-add}
    \int (\fp + \gp) \, d\mu
    = \int (f \plusae{\mu} g) \, d\mu
    = \int f \, d\mu + \int g \, d\mu.
  \end{equation}
\end{lemma}

\begin{proof}
  From
  Lemma~\thref{l:integrable-is-finite-seminorm-lone},
  $f,g\in\calM$, and $N_1(f),N_1(g)<\infty$.
  From
  Lemma~\thref{l:integrable-is-almost-finite}, and
  Definition~\thref{d:add-in-rbar},
  $f+g$ is well-defined almost everywhere.
  Then, from
  Lemma~\thref{l:minkowski-ineq-in-m}, and
  \assume{closedness of addition in $\matRplus$},
  we have $\fp+\gp,f\plusae{\mu}g\in\calM$, and
  $N_1(\fp+\gp)=N_1(f\plusae{\mu}g)<\infty$.
  Hence, from
  Lemma~\thref{l:integrable-is-finite-seminorm-lone},
  $\fp+\gp$ and $f\plusae{\mu}g$ are $\mu$-integrable in $\calM$.
  Moreover, from
  Lemma~\thref{l:compat-of-int-with-almost-eq},
  $\fp$ and $\gp$ are also $\mu$-integrable in $\calM$.

  From
  Lemma~\thref{l:meas-of-nonneg-and-nonpos-parts},
  we have
  \begin{equation*}
    \fp^\pm, \gp^\pm, (\fp + \gp)^\pm  \in \calMplus.
  \end{equation*}
  Therefore, from
  Lemma~\thref{l:compat-of-int-with-almost-eq},
  \assume{symmetry of equality},
  Definition~\threfc{d:integral}{with $\fp+\gp$},
  Lemma~\threfc{l:compat-of-int-in-mplus-with-nonpos-and-nonneg-parts}{%
    with $\fp$ and $\gp$},
  \assume{field properties of~$\matR$ (all integrals below are finite)},
  Definition~\threfc{d:integral}{with $\fp$ and $\gp$}, and
  Lemma~\threfc{l:compat-of-int-with-almost-eq}{with $f$ and $g$},
  we have
  \begin{align*}
    \int (f \plusae{\mu} g) \, d\mu
    = \int (\fp + \gp) \, d\mu
    & = \int (\fp + \gp)^+ \, d\mu - \int (\fp + \gp)^- \, d\mu\\
    & = \int \fp^+ \, d\mu + \int \gp^+ \, d\mu
      - \left( \int \fp^- \, d\mu + \int \gp^- \, d\mu \right)\\
    & = \left( \int \fp^+ \, d\mu - \int \fp^- \, d\mu \right)
      + \left( \int \gp^+ \, d\mu - \int \gp^- \, d\mu \right)\\
    & = \int \fp \, d\mu + \int \gp \, d\mu
      = \int f \, d\mu + \int g \, d\mu.
  \end{align*}
\end{proof}

\begin{definition}[$\calLone$, {\vectorspace} of integrable functions]
  \label{d:llone-vector-space-of-int-fun}
  \mbox{}\\
  Let~$(X,\Sigma,\mu)$ be a measure space.
  The {\em {\vectorspace} of integrable functions} is denoted
  $\callone{X,\Sigma,\mu}$ (or simply~$\calLone$);
  it is defined by
  \begin{equation}
    \label{e:llp-vector-space-of-int-fun}
    \callone{X, \Sigma, \mu} \eqdef \{ f \in \calMR \st N_1 (f) < \infty \}.
  \end{equation}
\end{definition}

\begin{remark}
  The set~$\calLone$ is shown below to be a (seminormed) {\vectorspace},
  hence its name.
\end{remark}

\begin{lemma}[equivalent definition of $\calLone$]
  \label{l:equiv-def-of-llone}
  \mbox{}\hfill
  Let~$(X,\Sigma,\mu)$ be a measure space.\\
  Let~$f:\ArXRb$.
  Then, $f\in\calLone$ iff
  $f$~is finite and $\mu$-integrable in~$\calM$.
\end{lemma}

\begin{proof}
  \proofpar{``Left'' implies ``right''}
  Direct consequence of
  Definition~\thref{d:llone-vector-space-of-int-fun},
  Lemma~\threfc{l:m-and-finite-is-mr}{$f\in\calM\cap\FXR$},
  Lemma~\thref{l:seminorm-llone},
  Lemma~\threfc{l:int-in-mplus}{%
      $|f|$ is $\mu$-integrable in $\calMplus$}, and
  Lemma~\thref{l:equiv-def-of-integrability}.

  \proofparskip{``Right'' implies ``left''}
  Direct consequence of
  Lemma~\thref{l:equiv-def-of-integrability},
  Lemma~\threfc{l:m-and-finite-is-mr}{$f\in\calMR$},
  Lemma~\thref{l:int-in-mplus},
  Lemma~\thref{l:seminorm-llone}, and
  Definition~\thref{d:llone-vector-space-of-int-fun}.

  \medskip\noindent
  Therefore, we have the equivalence.
\end{proof}

\begin{lemma}[Minkowski inequality in~$\calLone$]
  \label{l:minkowski-ineq-in-llone}
  \mbox{}\\
  Let~$(X,\Sigma,\mu)$ be a measure space.
  Let~$f,g\in\calLone$.
  Then, we have $N_1(f+g)\leq N_1(f)+N_1(g)$.
\end{lemma}

\begin{proof}
  Direct consequence of
  Definition~\threfc{d:llone-vector-space-of-int-fun}{%
    $f$ and $g$ belong to $\calMR\subset\calM$},
  Definition~\threfc{d:summability-domain}{$\Dfsum(f,g)=X$},
  Lemma~\thref{l:almost-sum-is-sum},
  Lemma~\thref{l:everywhere-implies-almost-everywhere}, and
  Lemma~\thref{l:minkowski-ineq-in-m}.
\end{proof}

\begin{lemma}[$\calLone$~is seminormed {\vectorspace}]
  \label{l:llone-is-seminormed-vector-space}
  \mbox{}\\
  Let~$(X,\Sigma,\mu)$ be a measure space.
  Then, $(\calLone,N_1)$ is a seminormed {\vectorspace}.
\end{lemma}

\begin{proof}
  From
  Lemma~\threfc{l:mr-is-vector-space}{with $0:\ArXR$ as zero},
  Definition~\thref{d:llone-vector-space-of-int-fun},
  Lemma~\thref{l:everywhere-implies-almost-everywhere}, and
  Lemma~\threfc{l:none-is-almost-definite}{$N_1(0)=0$},
  $\calLone$ is a subset of the {\vectorspace}~$\calMR$ containing the zero,
  and we have $N_1(\calLone)\subset\matR$.

  Let~$a\in\matR$.
  Let~$f,g\in\calLone\subset\calMR$.
  Then, from
  Definition~\thref{d:llone-vector-space-of-int-fun},
  we have $N_1(f),N_1(g)<\infty$.
  Moreover, from
  Lemma~\thref{l:mr-is-vector-space},
  we have $af,f+g\in\calMR$.
  Thus, from
  Lemma~\thref{l:none-is-abs-hom},
  Lemma~\thref{l:minkowski-ineq-in-llone}, and
  \assume{closedness of multiplication and addition in~$\matR$},
  we have
  \begin{equation*}
    N_1 (a f) = | a | N_1 (f) < \infty
    \AND
    N_1 (f + g) \leq N_1 (f) + N_1 (g) < \infty
  \end{equation*}
  Hence, from
  Definition~\thref{d:llone-vector-space-of-int-fun},
  we have $af,f+g\in\calLone$, $N_1$ is absolutely homogeneous of degree~1,
  and~$N_1$ satisfies the triangle inequality.

  Therefore, from
  Lemma~\threfc{LM-l:closed-under-vector-operations-is-subspace}{%
    $\calLone$~is a {\vectorsubspace} of~$\calMR$},
  Definition~\threfc{LM-d:subspace}{$\calLone$ is a {\vectorspace}}, and
  Definition~\threfc{d:seminorm}{%
    $N_1$ is a seminorm over~$\calLone$},
  $(\calLone,N_1)$ is a seminormed {\vectorspace}.
\end{proof}

\begin{definition}[convergence in~$\calLone$]
  \label{d:conv-in-llone}
  \mbox{}\hfill
  Let~$(X,\Sigma,\mu)$ be a measure space.
  Let~$(f_n)_{n\in\matN},f\in\calLone$.
  The sequence $(f_n)_{n\in\matN}$ is said
  {\em convergent towards~$f$ in~$\calLone$} iff
  $\lim_{n\to\infty}N_1(f_n-f)=0$.
\end{definition}

\begin{lemma}[$\calLone$~is closed under absolute value]
  \label{l:llone-is-closed-under-abs}
  \mbox{}\\
  Let~$(X,\Sigma,\mu)$ be a measure space.
  Let~$f\in\calLone$.
  Then, we have $|f|\in\calLone$.
\end{lemma}

\begin{proof}
  Direct consequence of
  Definition~\thref{d:llone-vector-space-of-int-fun},
  Lemma~\threfc{l:m-is-closed-under-abs}{%
    $|f|\in\calMR$},
  \assume{idempotent law for the absolute value}, and
  Lemma~\threfc{l:seminorm-llone}{$N_1(|f|)=N_1(f)$}.
\end{proof}

\begin{lemma}[bounded by~$\calLone$ is~$\calLone$]
  \label{l:bounded-by-llone-is-llone}
  \mbox{}\\
  Let~$(X,\Sigma,\mu)$ be a measure space.
  Let~$f\in\calM$.
  Then, we have the equivalence
  \begin{equation}
    \label{e:bounded-by-llone-is-llone}
    f \in \calLone \EQUIV \exists g \in \calLone,\quad | f | \leq g.
  \end{equation}
\end{lemma}

\begin{proof}
  \proofpar{``Left'' implies ``right''}
  Direct consequence of
  Lemma~\threfc{l:llone-is-closed-under-abs}{%
    $g\eqdef|f| \in \calLone$}, and
  \assume{reflexivity of order in $\matR$}.

  \proofparskip{``Right'' implies ``left''}
  From
  Lemma~\threfc{l:equiv-def-of-llone}{with $g$},
  Lemma~\thref{l:abs-in-rbar-is-nonneg},
  Lemma~\threfc{l:order-in-rbar-is-total}{%
    transitivity, thus $|f|$ is finite and $g$ is nonnegative},
  Lemma~\thref{l:finite-abs-in-rbar}, and
  Lemma~\threfc{l:compat-of-integrability-in-m-and-mplus}{with $g$},
  we have~$f$ finite and~$g$ $\mu$-integrable in~$\calMplus$.
  Hence, from
  Lemma~\threfc{l:bounded-by-integrable-is-integrable}{%
    $f$ $\mu$-integrable in~$\calM$}, and
  Lemma~\threfc{l:equiv-def-of-llone}{with $f$},
  we have $f\in \calLone$.

  \medskip\noindent
  Therefore, we have the equivalence.
\end{proof}

\begin{lemma}[integral is positive linear form on~$\calLone$]
  \label{l:int-is-pos-lin-form-on-llone}
  \mbox{}\\
  Let~$(X,\Sigma,\mu)$ be a measure space.
  Let~$\calI:\ArLoneR$ be the function defined by for all
  $f\in\calLone$, $\calI(f)\eqdef\int f\,d\mu$.
  Then, $\calI$~is a positive linear form on~$\calLone$;
  hence it is nondecreasing.

  Moreover, for all $f\in\calLone$, we have $|\calI(f)|\leq\calI(|f|)=N_1(f)$.
\end{lemma}

\begin{proof}
  \proofpar{(1). Linearity}
  Direct consequence of
  Lemma~\thref{l:llone-is-seminormed-vector-space},
  Lemma~\thref{l:int-is-hom},
  Definition~\threfc{d:llone-vector-space-of-int-fun}{%
    $\calLone\subset\calMR$},
  Definition~\threfc{d:add-in-rbar}{
    addition is well-defined in $\calLone$}, and
  Lemma~\threfc{l:int-is-add}{with $h\eqdef f+g$}.

  \proofparskip{(2). Nonnegativeness}
  Let~$f\in\calLone$.
  Assume that $f\geq0$.
  Then, from
  Definition~\threfc{d:llone-vector-space-of-int-fun}{$f\in\calMR$},
  Lemma~\threfc{l:m-and-finite-is-mr}{$\calMR\subset\calM$}, and
  Definition~\thref{d:mplus-subset-of-nonneg-meas-num-fun},
  we have $f\in\calMplus$.
  Hence, from
  the definition of~$\calI$, and
  Lemma~\thref{l:int-in-mplus},
  we have $\calI(f)=\int f\,d\mu\geq0$.

  \medskip\noindent
  Therefore, from
  Definition~\thref{LM-d:linear-map}, and
  Definition~\thref{LM-d:linear-form},
  $\calI$~is a nonnegative linear form on~$\calLone$.

  \proofparskip{(3). Monotonicity}
  Let $f,g\in\calLone$.
  Assume that $f\leq g$.
  Then, from
  Lemma~\thref{l:llone-is-seminormed-vector-space},
  Definition~\threfc{LM-d:space}{additive abelian group properties}, and
  \assume{ordered field properties of~$\matR$},
  we have $g=f+(g-f)$ with~$g-f\geq0$.
  Thus, from the definition of~$\calI$ (with $f$), (2), (1), and
  the definition of~$\calI$ (with $g$), we have
  \begin{equation*}
    \calI (f)
    = \int f \, d\mu
    \leq \int f \, d\mu + \int (g - f) \, d\mu
    = \int g \, d\mu
    = \calI (g).
  \end{equation*}
  Hence, $\calI$~is nondecreasing.

  \proofparskip{(4). Inequality}
  Let~$f\in\calLone$.
  Then, from
  Definition~\thref{d:llone-vector-space-of-int-fun},
  Lemma~\thref{l:seminorm-llone},
  Definition~\thref{d:integrability}, and
  Lemma~\thref{l:equiv-def-of-integrability},
  $f^+$, $f^-$ and~$|f|$ are $\mu$-integrable in~$\calMplus$, {\ie} all
  integrals below are finite.
  Hence, from
  the definition of~$\calI$ (with $f$),
  Definition~\thref{d:integral},
  \assume{the triangle inequality for the absolute value in $\matR$},
  Lemma~\thref{l:nonneg-and-nonpos-parts-are-nonneg}, (2),
  Lemma~\thref{l:int-in-mplus-of-decomp-into-nonpos-and-nonneg-parts},
  the definition of~$\calI$ (with $|f|$), and
  Lemma~\thref{l:seminorm-llone},
  we have
  \begin{equation*}
    | \calI (f) |
    = \left| \int f \, d\mu \right |
    = \left| \int f^+ \, d\mu - \int f^- \, d\mu \right|
    \leq \int f^+ \, d\mu + \int f^- \, d\mu
    = \int | f | \, d\mu
    = \calI (| f |)
    = N_1 (f).
  \end{equation*}
\end{proof}

\begin{lemma}[constant function is~$\calLone$]
  \label{l:const-fun-is-llone}
  \mbox{}\\
  Let~$(X,\Sigma,\mu)$ be a finite measure space.
  Let~$a\in\matR$.
  Then, $f\eqdef(x\mapsto a)\in\calLone$, and we have
  \begin{equation}
    \label{e:const-fun-is-llone}
    \int f \, d\mu = a \mu (X).
  \end{equation}
\end{lemma}

\begin{proof}
  Direct consequence of
  Definition~\thref{d:meas},
  Definition~\threfc{d:measurable-space}{$\Sigma$ is a $\sigma$-algebra},
  Definition~\threfc{d:sigma-alg}{$X\in\Sigma$},
  Definition~\threfc{d:sf-vector-space-of-simple-funs}{%
    $|f|=|a|\,\matUN_X\in\calSFplus$},
  Lemma~\threfc{l:int-in-sfplus}{$\int|f|\,d\mu=|a|\mu(X)<\infty$},
  Lemma~\thref{l:int-in-mplus-gen-int-in-sfplus},
  Lemma~\thref{l:compat-of-int-in-m-and-mplus}, and
  Definition~\threfc{d:integral}{%
    $\int f\,d\mu=\sgn(a)\int|f|\,d\mu=a\mu(X)$}.
\end{proof}

\begin{lemma}[first mean value theorem]
  \label{l:first-mean-value-theorem}
  \mbox{}\hfill
  Let~$(X,\Sigma,\mu)$ be a finite measure space.\\
  Assume that~$\mu$ is nonzero.
  Let~$f\in\calM$.
  Assume that~$f$ is bounded.
  Then, $f\in\calLone$, and we have
  \begin{equation}
    \label{e:first-mean-value-theorem}
    \inf (f (X)) \leq \frac{1}{\mu (X)} \int f \, d\mu \leq \sup (f (X)).
  \end{equation}

  Moreover, inequalities are strict iff
  $f$~is not equal to one of its bounds $\mu$-almost everywhere.
\end{lemma}

\begin{proof}
  Let~$m\eqdef\inf(f(X))$ and~$M=\sup(f(X))$.

  \proofpar{Case $0\leq m\leq M$}
  Then, $-m\leq M$ and $|f|\leq M$.
  \proofpar{Case $m<0\leq M$}
  Then, $|f|$~is less than or equal to~$\max(-m,M)$.
  \proofpar{Case $m\leq M<0$}
  Then, $M\leq -m$ and $|f|\leq -m$.
  Thus, in all cases, we have $|f|\leq g$ where $g\eqdef\max(-m,M)$.
  Hence, from
  Lemma~\threfc{l:const-fun-is-llone}{$g\in\calLone$}, and
  Lemma~\thref{l:bounded-by-llone-is-llone},
  we have $f\in\calLone$.
  Therefore, from
  Lemma~\threfc{l:int-is-pos-lin-form-on-llone}{%
    $\calI$~is nondecreasing}, and
  \assume{ordered field properties of~$\matR$ with $0<\mu(X)<\infty$},
  we have
  \begin{equation*}
    m \leq \frac{1}{\mu (X)} \int f \, d\mu \leq M.
  \end{equation*}

  Moreover, from
  Lemma~\threfc{l:const-fun-is-llone}{$(x\mapsto m)\in\calLone$},
  Lemma~\threfc{l:llone-is-seminormed-vector-space}{$f-m\in\calLone$},
  Lemma~\threfc{l:int-is-pos-lin-form-on-llone}{%
    $\int(f-m)\,d\mu$ is zero},
  Lemma~\thref{l:compat-of-int-in-m-and-mplus},
  Lemma~\threfc{l:int-in-mplus-is-almost-definite}{with $f-m\geq0$}, and
  Lemma~\threfc{l:compat-of-almost-eq-with-op}{%
    with the unary operator translation by $m$},
  we have
  \begin{equation*}
    \int f \, d\mu = m \mu (X)
    \EQUIV \int (f - m) \, d\mu = 0
    \EQUIV f - m \eqae{\mu} 0
    \EQUIV f \eqae{\mu} m.
  \end{equation*}
  Similarly, with $M-f\in\calLone\cap\calMplus$, we have
  $\int f\,d\mu=M\mu(X)\Equiv f\eqae{\mu}M$.
  Hence, since
  \assume{$(P\Equiv Q)\Equiv(\neg P\Equiv\neg Q)$},
  we have strict inequalities in~\eqref{e:first-mean-value-theorem}
  iff~$f$ is not equal to~$m$ or~$M$ $\mu$-almost everywhere.
\end{proof}

\begin{lemma}[variant of first mean value theorem]
  \label{l:variant-of-first-mean-value-theorem}
  \mbox{}\\
  Let~$(X,\Sigma,\mu)$ be a finite measure space.
  Assume that~$\mu$ is nonzero.
  Let~$f\in\calM$, $m\eqdef\inf(f(X))$ and~$M\eqdef\sup(f(X))$.
  Assume that~$f$ is bounded, not equal to~$m$ or~$M$ $\mu$-almost
  everywhere, and $(m,M)\subset f(X)$.
  Then, $f\in\calLone$, and we have
  \begin{equation}
    \label{e:variant-of-first-mean-value-theorem}
    \exists x \in X,\quad
   \int f \, d\mu = f (x) \mu (X).
  \end{equation}
\end{lemma}

\begin{proof}
  Direct consequence of
  Lemma~\threfc{l:first-mean-value-theorem}{with strict inequalities},
  \assume{ordered field properties of~$\matR$ with $0<\mu(X)<\infty$}, and
  Definition~\threfc{d:interval}{%
    with $X\eqdef\matR$, there exists $y\in(m,M)$ such that $\int
    f\,d\mu=y\mu(X)$}.
\end{proof}

\begin{remark}
  \label{r:v2-new20}
  See the sketch of next proof in
  Section~\ref{s:sketch-of-the-proof-of-lebesgue-dom-cv-th}.
\end{remark}

\begin{theorem}[Lebesgue, dominated convergence]
  \label{t:lebesgue-dom-conv}
  \mbox{}\\
  Let~$(X,\Sigma,\mu)$ be a measure space.
  Let~$(f_n)_{n\in\matN}\in\calM$.
  Assume that the sequence is pointwise convergent towards~$f$.
  Let~$g\in\calLone$.
  Assume that for all $n\in\matN$, $|f_n|\leq g$.\\
  Then, for all $n\in\matN$, $f_n\in\calLone$, $f\in\calLone$,
  the sequence is convergent towards~$f$ in~$\calLone$, and
  we have
  \begin{equation}
    \label{e:lebesgue-dom-conv}
    \int f \, d\mu = \lim_{n \to \infty} \int f_n \, d\mu.
  \end{equation}
\end{theorem}

\begin{proof}
  From
  Lemma~\thref{l:m-is-closed-under-limit-when-pointwise-conv},
  we have $f\in\calM$.
  Moreover, from
  Lemma~\thref{l:abs-in-rbar-is-cont}, and
  \assume{monotonicity of the limit in~$\matRbar$},
  we have $|f|\leq g$.

  Let~$n\in\matN$.
  Then, from
  Lemma~\thref{l:bounded-by-llone-is-llone},
  we have $f_n,f\in\calLone$.
  Thus, from
  Lemma~\thref{l:llone-is-seminormed-vector-space},
  Definition~\thref{d:seminorm},
  Definition~\threfc{LM-d:space}{$(\calLone,+)$ is an abelian group},
  Lemma~\thref{l:llone-is-closed-under-abs},
  \assume{linearity and compatibility of the limit with the absolute value},
  Lemma~\thref{l:abs-in-rbar-is-nonneg}, and
  Lemma~\thref{l:abs-in-rbar-is-definite},
  we have
  \begin{equation*}
    g_n \eqdef | f_n - f | \in \calLone \cap \calMplus
    \CONJ
    \lim_{n \to \infty} g_n = 0.
  \end{equation*}
  Moreover, from
  \assume{the triangle inequality for the absolute value}, and
  \assume{ordered field properties of~$\matR$},
  we have $g_n\leq|f_n|+|-f|\leq2g$.
  Thus, from
  Lemma~\thref{l:llone-is-seminormed-vector-space},
  Definition~\thref{d:seminorm},
  Definition~\threfc{LM-d:space}{$(\calLone,+)$ is an abelian group},
  Definition~\threfc{d:llone-vector-space-of-int-fun}{%
    $\calLone\subset\calM$}, and
  Definition~\thref{d:mplus-subset-of-nonneg-meas-num-fun},
  we have $2g-g_n\in\calLone\cap\calMplus$, and from
  Lemma~\thref{l:liminf-and-limsup-of-pointwise-conv},
  we have
  $\liminf_{n\to\infty}(2g-g_n)=\lim_{n\to\infty}(2g-g_n)=2g$.

  From
  Theorem~\threfc{t:fatou-lemma}{with $f_n\eqdef 2g-g_n$},
  Lemma~\threfc{l:int-is-pos-lin-form-on-llone}{linearity}, and
  Lemma~\thref{l:duality-liminf-limsup},
  we have
  \begin{align*}
    2 \int g \, d\mu
    = \int \liminf_{n \to \infty} (2g - g_n) \, d\mu
    & \leq \liminf_{n \to \infty} \int (2g - g_n) \, d\mu\\
    & = \liminf_{n \to \infty}
      \left( 2 \int g \, d\mu - \int g_n \, d\mu \right)
      = 2 \int g \, d\mu - \limsup_{n \to \infty} \int g_n \, d\mu.
  \end{align*}
  Thus, from
  \assume{ordered field properties of~$\matR$ (with $\int g\,d\mu$ finite)},
  $\limsup_{n\to\infty}\int g_n\, d\mu\leq0$.
  Hence, from
  Lemma~\threfc{l:int-is-pos-lin-form-on-llone}{%
    $\int g_n\,d\mu\geq0$},
  Lemma~\threfc{l:liminf-bounded-from-below}{%
    $\liminf_{n\to\infty}\int g_n\,d\mu\geq0$},
  Lemma~\threfc{l:liminf-limsup-and-pointwise-conv}{%
    with $\limsup \leq 0 \leq \liminf$},
  the definition of the~$g_n$'s, and
  Lemma~\thref{l:seminorm-llone},
  we have
  \begin{equation*}
    0
    = \liminf_{n \to \infty} \int g_n \, d\mu
    = \limsup_{n \to \infty} \int g_n \, d\mu
    = \lim_{n \to \infty} \int g_n \, d\mu
    = \lim_{n \to \infty} N_1 (f_n - f).
  \end{equation*}
  Therefore, from
  Definition~\thref{d:conv-in-llone},
  $(f_n)_{n\in\matN}$ is convergent towards~$f$ in~$\calLone$.

  Moreover, from
  Lemma~\thref{l:abs-in-rbar-is-nonneg},
  Lemma~\thref{l:int-is-pos-lin-form-on-llone}, and
  the definition of the~$g_n$'s,
  we have
  \begin{equation*}
    0
    \leq \left| \int f_n \, d\mu - \int f \, d\mu \right|
    = \left| \int (f_n - f) \, d\mu \right|
    \leq \int | f_n - f | \, d\mu
    = \int g_n \, d\mu.
  \end{equation*}
  Thus, from
  \assume{the squeeze theorem}, and
  \assume{linearity of the limit},
  we have
  \begin{equation*}
    \lim_{n \to \infty} \int f_n \, d\mu
    = \int f \, d\mu
    = \int \lim_{n \to \infty} f_n \, d\mu.
  \end{equation*}
\end{proof}

\begin{remark}
  \label{r:v2-new21}
  See the sketch of next proof in
  Section~\ref{s:sketch-of-the-proof-of-lebesgue-ext-dom-cv-th}.
\end{remark}

\begin{theorem}[Lebesgue, extended dominated convergence]
  \label{t:lebesgue-ext-dom-conv}
  \mbox{}\\
  Let~$(X,\Sigma,\mu)$ be a measure space.
  Let~$(f_n)_{n\in\matN},f,g\in\calM$.
  Assume that the sequence is $\mu$-almost everywhere pointwise convergent
  towards~$f$, that~$g$ is $\mu$-integrable, and that for all $n\in\matN$, we
  have $|f_n|\leqae{\mu}g$.
  Then, for all $n\in\matN$, $f_n$~is $\mu$-integrable, $f$~is
  $\mu$-integrable,
  the sequence~$(N_1(f_n-f))_{n\in\matN}$ converges towards~0, and we have
  \begin{equation}
    \label{e:lebesgue-ext-dom-conv}
    \int f \, d\mu = \lim_{n \to \infty} \int f_n \, d\mu.
  \end{equation}
\end{theorem}

\begin{proof}
  For all $n\in\matN$, let
  $\tB\eqdef\{f=\liminf_{n\to\infty}f_n\}\cap\{f=\limsup_{n\to\infty}f_n\}$
  and $\tC_n\eqdef\{|f_n|\leq g\}$.

  Let $n\in\matN$.
  Then, from
  Definition~\thref{d:prop-almost-satisfied},
  Definition~\thref{d:negl-subset},
  Definition~\thref{d:meas},
  Definition~\threfc{d:measurable-space}{$\Sigma$ is a $\sigma$-algebra},
  Definition~\threfc{d:sigma-alg}{closedness under complement},
  \assume{monotonicity of complement}, and since
  \assume{involutiveness of complement},
  let $B,C_n\in\Sigma$ such that $B\subset\tB$, $C_n\subset\tC_n$ and
  $\mu(B^c)=\mu(C_n^c)=0$.

  Let $C\eqdef\bigcap_{n\in\matN}C_n$ and $D\eqdef B\cap C$.
  Then, from
  Lemma~\threfc{l:equiv-def-of-sigma-alg}{%
    closedness under countable intersection
    (with $I=\matN$, then $\card(I)=2$)},
  \assume{De~Morgan's laws}, and
  Lemma~\threfc{l:compat-of-null-meas-with-count-union}{%
    with $I=\matN$, then $\card(I)=2$},
  we have $C,D\in\Sigma$, and $\mu(C^c)=\mu(D^c)=0$.
  For all $n\in\matN$, let
  \begin{equation*}
    A \eqdef D \cap g^{-1} (\matRplus),\quad
    \tf_n \eqdef f_n \, \matUN_A,\quad
    \tf \eqdef f \, \matUN_A,\quad
    \tg \eqdef g \, \matUN_A.
  \end{equation*}

  Let $n\in\matN$.
  Then, from
  Lemma~\threfc{l:finite-nonneg-part}{with $g$},
  we have $A\in\Sigma$, $\tg$~belongs to~$\calMR\cap\calMplus$, $\mu(A^c)=0$,
  and $g\eqae{\mu}\tg$.
  Hence, from
  Lemma~\threfc{l:meas-and-masking}{with $f_n$ and $f$}, and
  Lemma~\threfc{l:masking-almost-nowhere}{with $f_n$ and $f$},
  we have $\tf_n,\tf\in\calM$, $f_n\eqae{\mu}\tf_n$ and $f\eqae{\mu}\tf$.

  From
  Lemma~\threfc{l:compat-of-almost-eq-with-op}{%
    with the unary operator absolute value},
  we have $|\tg|\eqae{\mu}|g|$.
  Thus, from
  Lemma~\threfc{l:compat-of-int-with-almost-eq}{%
    with $|g|$ and $|\tg|$},
  Lemma~\threfc{l:equiv-def-of-integrability}{with $g$}, and
  Lemma~\thref{l:seminorm-llone},
  we have $N_1(\tg)=N_1(g)<\infty$.
  Hence, from
  Definition~\thref{d:llone-vector-space-of-int-fun},
  we have $\tg\in\calLone$.

  Let~$x\in X$.
  \proofpar{Case $x\in A$}
  Then, from
  \assume{the definition of the indicator function}, and
  since $A\subset B\subset\tB$, we have
  $\lim_{n\to\infty}\tf_n(x)
  =\lim_{n\to\infty}f_n(x)=f(x)=\tf(x)$.
  Moreover, for all $n\in\matN$, since $A\subset C_n\subset\tC_n$, we have
  $|\tf_n(x)|=|f_n(x)|\leq g(x)=\tg(x)$.
  \proofpar{Case $x\not\in A$}
  Then, from
  \assume{the definition of the indicator function},
  we have for all $n\in\matN$, $\tf_n(x)=\tf(x)=0=\tg(x)$.
  Thus, we have $\lim_{n\to\infty}\tf_n(x)=0=\tf(x)$, and
  $|\tf_n(x)|=0\leq0=\tg(x)$.
  Hence, in all cases, from
  Theorem~\thref{t:lebesgue-dom-conv},
  we have for all $n\in\matN$, $\tf_n,\tf\in\calLone$, and
  \begin{equation*}
    \int \tf \, d\mu = \lim_{n \to \infty} \int \tf_n \, d\mu.
  \end{equation*}

  Therefore, from
  Lemma~\threfc{l:compat-of-almost-eq-with-op}{%
    with the unary operator absolute value, $|\tf_n|\eqae{\mu}|f_n|$ and
    $|\tf|\eqae{\mu}|f|$},
  Lemma~\threfc{l:compat-of-int-with-almost-eq}{%
    with $f_n$ and $\tf_n$, then $f$ and $\tf$},
  Lemma~\threfc{l:equiv-def-of-integrability}{with $f_n$, then $f$},
  Lemma~\thref{l:seminorm-llone}, and
  Definition~\thref{d:llone-vector-space-of-int-fun},
  for all $n\in\matN$, $f_n$ and~$f$ are $\mu$-integrable in~$\calM$, and
  \begin{equation*}
    \int f \, d\mu
    = \int \tf \, d\mu
    = \lim_{n \to \infty} \int \tf_n \, d\mu.
    = \lim_{n \to \infty} \int f \, d\mu.
  \end{equation*}
\end{proof}

\chapter{Conclusions, perspectives}
\label{c:conclusions-perspectives}

We have presented very detailed proofs of the main basic results in measure
theory and Lebesgue integration such as the {\BLt}, {\Fl}, the {\Tont}, and
{\Ldcvt}.

The short-term purpose of this work was to help the formalization in a formal
proof assistant such as {\coq} of the basic concepts in measure theory and
Lebesgue integration and of the proofs of their main properties.
A first milestone towards this is dedicated to the integral of nonnegative
measurable functions~\cite{bol:cfp:21} where special attention was paid to the
formalization of $\sigma$-algebras and of simple functions.

Our mean-term purpose is now  to continue up to the formalization of~$L^p$
Lebesgue spaces and of~$W^{m,p}$ Sobolev spaces as Banach spaces, and in
particular~$L^2$ and~$H^m\eqdef W^{m,2}$ as Hilbert spaces.
This will include parts of the distribution theory.

The long-term purpose of these studies is the formal proof of programs
implementing the {\FEM}.
As a consequence, after having addressed the formalization of the
{\LMT}~\cite{cm:lmt:16,bol:cfp:17}, we will also have to write very detailed
pen-and-paper proofs for the concepts and results of the interpolation and
approximation theory to define the {\FEM} itself.

\chapter*{Acknowledgment}

The authors thank Francis Maisonneuve for having invited them to be in charge
of small classes on this very subject at
\emph{École Nationale Supérieure des Mines de Paris} at the turn of the
millennium, and for more recent fruitful discussions about what was not
detailed in his Lecture Notes.

\tocotherhead{chapter}

\bibliography{biblio_lds}
\bibliographystyle{plainnat}

\appendix
\chapter{Lists of statements}
\label{c:lists-of-statements}

This appendix collects the references (name and number) for all statements
present in Part~\ref{p:detailed-proofs}.
These are split into definitions, lemmas, and theorems.

\tocotherhead{section}

\renewcommand{\listtheoremname}{List of Definitions}
\phantomsection \label{s:list-of-definitions}
\listoftheorems[ignoreall,show=definition]

\clearpage
\renewcommand{\listtheoremname}{List of Lemmas}
\phantomsection \label{s:list-of-lemmas}
\listoftheorems[ignoreall,show=lemma]

\clearpage
\renewcommand{\listtheoremname}{List of Theorems}
\phantomsection \label{s:list-of-theorems}
\listoftheorems[ignoreall,show=theorem]

\clearpage
\begin{sidewaysfigure}
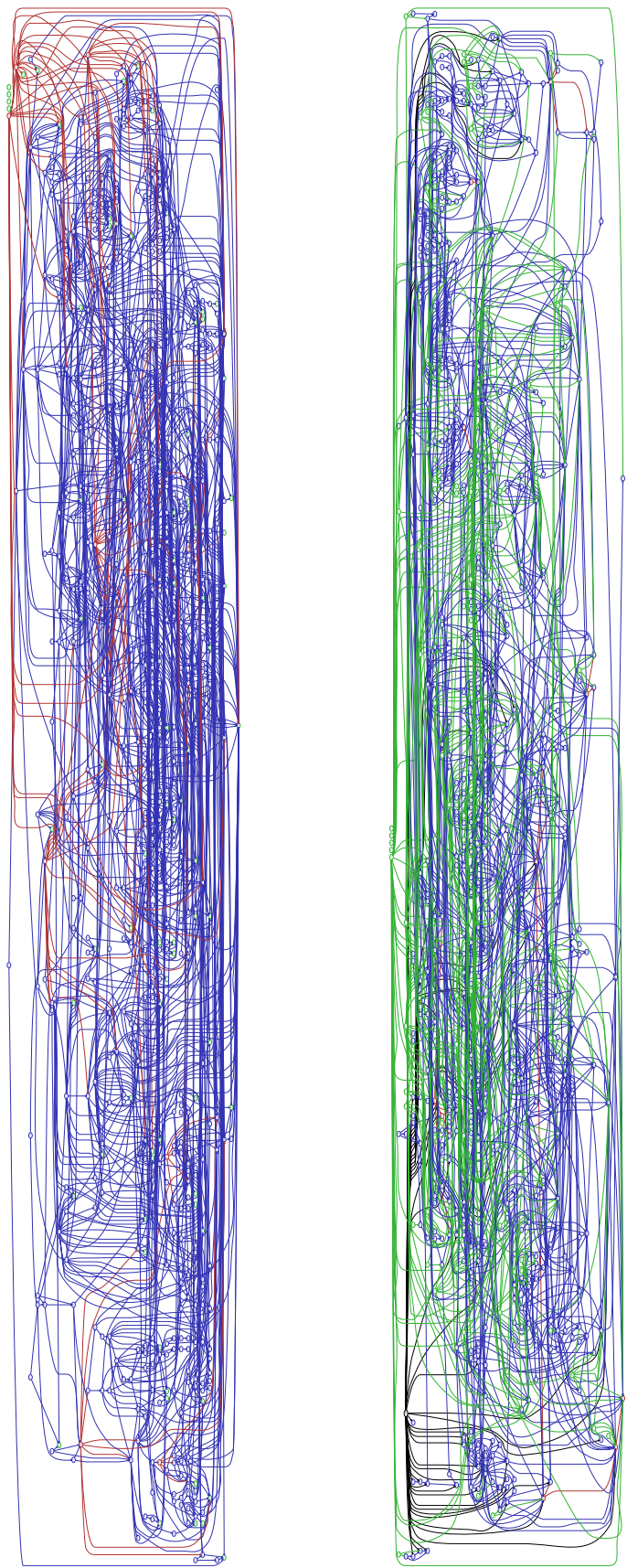

  \centering
  \includegraphics[width=\textheight]{tree1_lds}
  \\[2cm]
  \includegraphics[width=\textheight]{tree2_lds}
  \caption{%
    Dependency graphs.
    Top: the proof cites explicitly\ldots{}
    Bottom: is explicitly cited in the proof of\ldots{}
    Arrows are colored according to the nature of their foot:
    green is for definitions, blue for lemmas, and red for theorems.
    All dependencies are listed in
    Appendices~\ref{c:the-proof-cites-explicitly}
    and~\ref{c:is-explicitly-cited-in-the-proof-of}.
  }
  \label{f:depend-graph}
\end{sidewaysfigure}

\chapter{The proof cites explicitly\ldots}
\label{c:the-proof-cites-explicitly}

This appendix gathers the explicit citations of the statements listed in
Appendix~\ref{c:lists-of-statements} that appear in the proof of each result
(lemmas and theorems).
Statements from~\cite{cm:lmt:16} are anonymized.

The corresponding dependency graph is represented in
Figure~\ref{f:depend-graph} (top).
The dual graph is described in
Appendix~\ref{c:is-explicitly-cited-in-the-proof-of}.

Printing is not advised!

\bigskip

\begin{description}[style=unboxed]

\item[The proof of Lemma~\thref{l:compat-of-pseudopart-with-inter}] \mbox{}\\
  cites explicitly:\\
  Definition~\thref{d:pseudopart}.

\item[The proof of Lemma~\thref{l:technical-inclusion-for-count-union}] \mbox{}\\
  has no explicit citation.

\item[The proof of Lemma~\thref{l:order-is-meaningless-in-count-union}] \mbox{}\\
  cites explicitly:\\
  Lemma~\thref{l:technical-inclusion-for-count-union}.

\item[The proof of Lemma~\thref{l:def-of-double-count-union}] \mbox{}\\
  cites explicitly:\\
  Lemma~\thref{l:order-is-meaningless-in-count-union}.

\item[The proof of Lemma~\thref{l:double-count-union}] \mbox{}\\
  cites explicitly:\\
  Lemma~\thref{l:def-of-double-count-union}.

\item[The proof of Lemma~\thref{l:part-of-count-union}] \mbox{}\\
  has no explicit citation.

\item[The proof of Lemma~\thref{l:restr-is-mask}] \mbox{}\\
  has no explicit citation.

\item[The proof of Lemma~\thref{l:quotient-vector-ops}] \mbox{}\\
  cites explicitly:\\
  Definition~\thref{d:relation-compatible-with-vector-ops}.

\item[The proof of Lemma~\thref{l:quotient-vector-space-equiv-rel}] \mbox{}\\
  cites explicitly:\\
  Statement(s) from~\cite{cm:lmt:16},\\
  Lemma~\thref{l:quotient-vector-ops}.

\item[The proof of Lemma~\thref{l:quotient-vector-space}] \mbox{}\\
  cites explicitly:\\
  Statement(s) from~\cite{cm:lmt:16},\\
  Definition~\thref{d:relation-compatible-with-vector-ops},\\
  Lemma~\thref{l:quotient-vector-space-equiv-rel}.

\item[The proof of Lemma~\thref{l:linear-map-on-quotient-vector-space}] \mbox{}\\
  cites explicitly:\\
  Statement(s) from~\cite{cm:lmt:16},\\
  Lemma~\thref{l:quotient-vector-ops},\\
  Lemma~\thref{l:quotient-vector-space}.

\item[The proof of Lemma~\thref{l:k-is-k-alg}] \mbox{}\\
  cites explicitly:\\
  Definition~\thref{d:alg-over-a-field}.

\item[The proof of Lemma~\thref{l:alg-of-funs-to-alg}] \mbox{}\\
  cites explicitly:\\
  Statement(s) from~\cite{cm:lmt:16},\\
  Definition~\thref{d:alg-over-a-field},\\
  Definition~\thref{d:inherited-alg-ops}.

\item[The proof of Lemma~\thref{l:maps-to-k-is-alg}] \mbox{}\\
  cites explicitly:\\
  Lemma~\thref{l:k-is-k-alg},\\
  Lemma~\thref{l:alg-of-funs-to-alg}.

\item[The proof of Lemma~\thref{l:subspace-and-closed-under-mult-is-subalg}] \mbox{}\\
  cites explicitly:\\
  Statement(s) from~\cite{cm:lmt:16},\\
  Definition~\thref{d:alg-over-a-field},\\
  Definition~\thref{d:subalg}.

\item[The proof of Lemma~\thref{l:closed-under-alg-ops-is-subalg}] \mbox{}\\
  cites explicitly:\\
  Statement(s) from~\cite{cm:lmt:16},\\
  Lemma~\thref{l:subspace-and-closed-under-mult-is-subalg}.

\item[The proof of Lemma~\thref{l:definite-seminorm-is-norm}] \mbox{}\\
  cites explicitly:\\
  Statement(s) from~\cite{cm:lmt:16},\\
  Definition~\thref{d:seminorm}.

\item[The proof of Lemma~\thref{l:empty-open-int}] \mbox{}\\
  cites explicitly:\\
  Definition~\thref{d:interval}.

\item[The proof of Lemma~\thref{l:int-are-closed-under-finite-inter}] \mbox{}\\
  cites explicitly:\\
  Definition~\thref{d:interval}.

\item[The proof of Lemma~\thref{l:empty-inter-of-open-ints}] \mbox{}\\
  cites explicitly:\\
  Lemma~\thref{l:empty-open-int},\\
  Lemma~\thref{l:int-are-closed-under-finite-inter}.

\item[The proof of Lemma~\thref{l:inter-of-topo}] \mbox{}\\
  cites explicitly:\\
  Definition~\thref{d:topological-space}.

\item[The proof of Lemma~\thref{l:gen-topo-is-min}] \mbox{}\\
  cites explicitly:\\
  Lemma~\thref{l:inter-of-topo},\\
  Definition~\thref{d:gen-topo}.

\item[The proof of Lemma~\thref{l:equiv-def-of-gen-topo}] \mbox{}\\
  cites explicitly:\\
  Definition~\thref{d:topological-space},\\
  Lemma~\thref{l:gen-topo-is-min}.

\item[The proof of Lemma~\thref{l:augmented-topo-basis}] \mbox{}\\
  cites explicitly:\\
  Definition~\thref{d:topological-basis}.

\item[The proof of Lemma~\thref{l:topological-basis-of-order-topo}] \mbox{}\\
  cites explicitly:\\
  Lemma~\thref{l:int-are-closed-under-finite-inter},\\
  Lemma~\thref{l:equiv-def-of-gen-topo},\\
  Definition~\thref{d:topological-basis},\\
  Definition~\thref{d:order-topo}.

\item[The proof of Lemma~\thref{l:trace-topo-on-subset}] \mbox{}\\
  cites explicitly:\\
  Definition~\thref{d:trace-of-subsets-of-parties},\\
  Definition~\thref{d:topological-space},\\
  Definition~\thref{d:topological-basis}.

\item[The proof of Lemma~\thref{l:box-topo-on-cartesian-prod}] \mbox{}\\
  cites explicitly:\\
  Definition~\thref{d:prod-of-subsets-of-parties},\\
  Definition~\thref{d:topological-space},\\
  Definition~\thref{d:topological-basis}.

\item[The proof of Lemma~\thref{l:complete-count-topo-basis}] \mbox{}\\
  cites explicitly:\\
  Lemma~\thref{l:augmented-topo-basis},\\
  Definition~\thref{d:second-count}.

\item[The proof of Lemma~\thref{l:compat-of-second-count-with-cartesian-prod}] \mbox{}\\
  cites explicitly:\\
  Definition~\thref{d:prod-of-subsets-of-parties},\\
  Lemma~\thref{l:box-topo-on-cartesian-prod}.

\item[The proof of Lemma~\thref{l:complete-count-topo-basis-of-prod-space}] \mbox{}\\
  cites explicitly:\\
  Definition~\thref{d:prod-of-subsets-of-parties},\\
  Lemma~\thref{l:box-topo-on-cartesian-prod},\\
  Lemma~\thref{l:complete-count-topo-basis}.

\item[The proof of Lemma~\thref{l:equiv-def-of-conv-seq}] \mbox{}\\
  cites explicitly:\\
  Statement(s) from~\cite{cm:lmt:16}.

\item[The proof of Lemma~\thref{l:conv-subseq-of-cauchy-seq}] \mbox{}\\
  cites explicitly:\\
  Statement(s) from~\cite{cm:lmt:16}.

\item[The proof of Lemma~\thref{l:finite-cover-of-compact-int}] \mbox{}\\
  has no explicit citation.

\item[The proof of Lemma~\thref{l:two-is-self-holder-conjugate-in-r}] \mbox{}\\
  cites explicitly:\\
  Definition~\thref{d:holder-conjugates-in-r}.

\item[The proof of Lemma~\thref{l:youngs-ineq-for-prod-in-r}] \mbox{}\\
  cites explicitly:\\
  Definition~\thref{d:holder-conjugates-in-r}.

\item[The proof of Lemma~\thref{l:youngs-ineq-for-prod-in-r-case-p-two}] \mbox{}\\
  cites explicitly:\\
  Lemma~\thref{l:two-is-self-holder-conjugate-in-r},\\
  Lemma~\thref{l:youngs-ineq-for-prod-in-r}.

\item[The proof of Lemma~\thref{l:order-in-rbar-is-total}] \mbox{}\\
  cites explicitly:\\
  Definition~\thref{d:ext-real-nums-rbar}.

\item[The proof of Lemma~\thref{l:zero-is-identity-element-for-add-in-rbar}] \mbox{}\\
  cites explicitly:\\
  Definition~\thref{d:add-in-rbar}.

\item[The proof of Lemma~\thref{l:add-in-rbar-is-assoc-when-defined}] \mbox{}\\
  cites explicitly:\\
  Definition~\thref{d:add-in-rbar}.

\item[The proof of Lemma~\thref{l:add-in-rbar-is-comm-when-defined}] \mbox{}\\
  cites explicitly:\\
  Definition~\thref{d:add-in-rbar}.

\item[The proof of Lemma~\thref{l:infinity-sum-prop-in-rbar}] \mbox{}\\
  cites explicitly:\\
  Definition~\thref{d:add-in-rbar}.

\item[The proof of Lemma~\thref{l:additive-inverse-in-rbar-is-monot}] \mbox{}\\
  cites explicitly:\\
  Definition~\thref{d:ext-real-nums-rbar},\\
  Definition~\thref{d:add-in-rbar}.

\item[The proof of Lemma~\thref{l:mult-in-rbar-is-assoc-when-defined}] \mbox{}\\
  cites explicitly:\\
  Definition~\thref{d:mult-in-rbar}.

\item[The proof of Lemma~\thref{l:mult-in-rbar-is-comm-when-defined}] \mbox{}\\
  cites explicitly:\\
  Definition~\thref{d:mult-in-rbar}.

\item[The proof of Lemma~\thref{l:mult-in-rbar-is-left-distr-over-add-when-defined}] \mbox{}\\
  cites explicitly:\\
  Definition~\thref{d:add-in-rbar},\\
  Definition~\thref{d:mult-in-rbar}.

\item[The proof of Lemma~\thref{l:mult-in-rbar-is-right-distr-over-add-when-defined}] \mbox{}\\
  cites explicitly:\\
  Lemma~\thref{l:mult-in-rbar-is-comm-when-defined},\\
  Lemma~\thref{l:mult-in-rbar-is-left-distr-over-add-when-defined}.

\item[The proof of Lemma~\thref{l:zero-prod-prop-in-rbar}] \mbox{}\\
  cites explicitly:\\
  Definition~\thref{d:mult-in-rbar}.

\item[The proof of Lemma~\thref{l:infinity-prod-prop-in-rbar}] \mbox{}\\
  cites explicitly:\\
  Definition~\thref{d:mult-in-rbar}.

\item[The proof of Lemma~\thref{l:finite-prod-prop-in-rbar}] \mbox{}\\
  cites explicitly:\\
  Definition~\thref{d:mult-in-rbar}.

\item[The proof of Lemma~\thref{l:equiv-def-of-abs-in-rbar}] \mbox{}\\
  cites explicitly:\\
  Definition~\thref{d:abs-in-rbar}.

\item[The proof of Lemma~\thref{l:bounded-abs-in-rbar}] \mbox{}\\
  cites explicitly:\\
  Lemma~\thref{l:additive-inverse-in-rbar-is-monot},\\
  Lemma~\thref{l:equiv-def-of-abs-in-rbar}.

\item[The proof of Lemma~\thref{l:bounded-abs-in-rbar-strict}] \mbox{}\\
  cites explicitly:\\
  Lemma~\thref{l:additive-inverse-in-rbar-is-monot},\\
  Lemma~\thref{l:equiv-def-of-abs-in-rbar}.

\item[The proof of Lemma~\thref{l:finite-abs-in-rbar}] \mbox{}\\
  cites explicitly:\\
  Lemma~\thref{l:bounded-abs-in-rbar-strict}.

\item[The proof of Lemma~\thref{l:abs-in-rbar-is-nonneg}] \mbox{}\\
  cites explicitly:\\
  Definition~\thref{d:abs-in-rbar}.

\item[The proof of Lemma~\thref{l:abs-in-rbar-is-even}] \mbox{}\\
  cites explicitly:\\
  Definition~\thref{d:abs-in-rbar}.

\item[The proof of Lemma~\thref{l:abs-in-rbar-is-definite}] \mbox{}\\
  cites explicitly:\\
  Definition~\thref{d:abs-in-rbar}.

\item[The proof of Lemma~\thref{l:abs-in-rbar-satisfies-triangle-ineq}] \mbox{}\\
  cites explicitly:\\
  Definition~\thref{d:ext-real-nums-rbar},\\
  Definition~\thref{d:add-in-rbar},\\
  Definition~\thref{d:abs-in-rbar}.

\item[The proof of Lemma~\thref{l:exp-and-log-in-rbar-are-inverse}] \mbox{}\\
  cites explicitly:\\
  Definition~\thref{d:exp-and-log-in-rbar}.

\item[The proof of Lemma~\thref{l:exp-in-rbar}] \mbox{}\\
  cites explicitly:\\
  Definition~\thref{d:mult-in-rbar},\\
  Definition~\thref{d:exp-and-log-in-rbar},\\
  Definition~\thref{d:exp-in-rbar}.

\item[The proof of Lemma~\thref{l:topo-of-rbar}] \mbox{}\\
  cites explicitly:\\
  Definition~\thref{d:order-topo},\\
  Lemma~\thref{l:topological-basis-of-order-topo},\\
  Definition~\thref{d:ext-real-nums-rbar}.

\item[The proof of Lemma~\thref{l:trace-topo-on-r}] \mbox{}\\
  cites explicitly:\\
  Definition~\thref{d:topological-basis},\\
  Lemma~\thref{l:topological-basis-of-order-topo},\\
  Lemma~\thref{l:trace-topo-on-subset}.

\item[The proof of Lemma~\thref{l:conv-towards-minus-infinity}] \mbox{}\\
  cites explicitly:\\
  Lemma~\thref{l:topo-of-rbar}.

\item[The proof of Lemma~\thref{l:continuity-of-add-in-rbar}] \mbox{}\\
  cites explicitly:\\
  Definition~\thref{d:add-in-rbar}.

\item[The proof of Lemma~\thref{l:continuity-of-mult-in-rbar}] \mbox{}\\
  cites explicitly:\\
  Definition~\thref{d:mult-in-rbar}.

\item[The proof of Lemma~\thref{l:abs-in-rbar-is-cont}] \mbox{}\\
  cites explicitly:\\
  Definition~\thref{d:abs-in-rbar}.

\item[The proof of Lemma~\thref{l:add-in-rbarplus-is-closed}] \mbox{}\\
  cites explicitly:\\
  Definition~\thref{d:add-in-rbar}.

\item[The proof of Lemma~\thref{l:add-in-rbarplus-is-assoc}] \mbox{}\\
  cites explicitly:\\
  Definition~\thref{d:add-in-rbar},\\
  Lemma~\thref{l:add-in-rbar-is-assoc-when-defined}.

\item[The proof of Lemma~\thref{l:add-in-rbarplus-is-comm}] \mbox{}\\
  cites explicitly:\\
  Definition~\thref{d:add-in-rbar},\\
  Lemma~\thref{l:add-in-rbar-is-comm-when-defined}.

\item[The proof of Lemma~\thref{l:infinity-sum-prop-in-rbarplus}] \mbox{}\\
  cites explicitly:\\
  Lemma~\thref{l:infinity-sum-prop-in-rbar}.

\item[The proof of Lemma~\thref{l:series-are-conv-in-rbarplus}] \mbox{}\\
  has no explicit citation.

\item[The proof of Lemma~\thref{l:technical-upper-bound-in-series-in-rbarplus}] \mbox{}\\
  cites explicitly:\\
  Lemma~\thref{l:series-are-conv-in-rbarplus}.

\item[The proof of Lemma~\thref{l:order-is-meaningless-in-series-in-rbarplus}] \mbox{}\\
  cites explicitly:\\
  Lemma~\thref{l:technical-upper-bound-in-series-in-rbarplus}.

\item[The proof of Lemma~\thref{l:def-of-double-series-in-rbarplus}] \mbox{}\\
  cites explicitly:\\
  Lemma~\thref{l:order-is-meaningless-in-series-in-rbarplus}.

\item[The proof of Lemma~\thref{l:double-series-in-rbarplus}] \mbox{}\\
  cites explicitly:\\
  Lemma~\thref{l:series-are-conv-in-rbarplus},\\
  Lemma~\thref{l:def-of-double-series-in-rbarplus}.

\item[The proof of Lemma~\thref{l:mult-in-rbarplus-is-closed-when-defined}] \mbox{}\\
  cites explicitly:\\
  Definition~\thref{d:mult-in-rbar},\\
  Definition~\thref{d:mult-in-rbarplus}.

\item[The proof of Lemma~\thref{l:zero-prod-prop-in-rbarplus}] \mbox{}\\
  cites explicitly:\\
  Lemma~\thref{l:zero-prod-prop-in-rbar},\\
  Definition~\thref{d:mult-in-rbarplus}.

\item[The proof of Lemma~\thref{l:infinity-prod-prop-in-rbarplus}] \mbox{}\\
  cites explicitly:\\
  Lemma~\thref{l:infinity-prod-prop-in-rbar},\\
  Definition~\thref{d:mult-in-rbarplus}.

\item[The proof of Lemma~\thref{l:finite-prod-prop-in-rbarplus}] \mbox{}\\
  cites explicitly:\\
  Lemma~\thref{l:finite-prod-prop-in-rbar},\\
  Definition~\thref{d:mult-in-rbarplus}.

\item[The proof of Lemma~\thref{l:zero-prod-prop-in-rbar-mt}] \mbox{}\\
  cites explicitly:\\
  Lemma~\thref{l:zero-prod-prop-in-rbar},\\
  Definition~\thref{d:mult-in-rbar-mt}.

\item[The proof of Lemma~\thref{l:infinity-prod-prop-in-rbar-mt}] \mbox{}\\
  cites explicitly:\\
  Lemma~\thref{l:infinity-prod-prop-in-rbar},\\
  Definition~\thref{d:mult-in-rbar-mt}.

\item[The proof of Lemma~\thref{l:finite-prod-prop-in-rbar-mt}] \mbox{}\\
  cites explicitly:\\
  Lemma~\thref{l:infinity-prod-prop-in-rbar-mt}.

\item[The proof of Lemma~\thref{l:mult-in-rbarplus-is-closed-mt}] \mbox{}\\
  cites explicitly:\\
  Definition~\thref{d:mult-in-rbar},\\
  Definition~\thref{d:mult-in-rbarplus},\\
  Definition~\thref{d:mult-in-rbar-mt}.

\item[The proof of Lemma~\thref{l:mult-in-rbarplus-is-assoc-mt}] \mbox{}\\
  cites explicitly:\\
  Lemma~\thref{l:mult-in-rbar-is-assoc-when-defined},\\
  Definition~\thref{d:mult-in-rbar-mt}.

\item[The proof of Lemma~\thref{l:mult-in-rbarplus-is-comm-mt}] \mbox{}\\
  cites explicitly:\\
  Lemma~\thref{l:mult-in-rbar-is-comm-when-defined},\\
  Definition~\thref{d:mult-in-rbar-mt}.

\item[The proof of Lemma~\thref{l:mult-in-rbarplus-is-distr-over-add-mt}] \mbox{}\\
  cites explicitly:\\
  Lemma~\thref{l:mult-in-rbar-is-left-distr-over-add-when-defined},\\
  Lemma~\thref{l:mult-in-rbar-is-right-distr-over-add-when-defined},\\
  Definition~\thref{d:mult-in-rbar-mt}.

\item[The proof of Lemma~\thref{l:zero-prod-prop-in-rbarplus-mt}] \mbox{}\\
  cites explicitly:\\
  Lemma~\thref{l:zero-prod-prop-in-rbarplus},\\
  Definition~\thref{d:mult-in-rbar-mt}.

\item[The proof of Lemma~\thref{l:infinity-prod-prop-in-rbarplus-mt}] \mbox{}\\
  cites explicitly:\\
  Lemma~\thref{l:infinity-prod-prop-in-rbarplus},\\
  Definition~\thref{d:mult-in-rbar-mt}.

\item[The proof of Lemma~\thref{l:finite-prod-prop-in-rbarplus-mt}] \mbox{}\\
  cites explicitly:\\
  Lemma~\thref{l:finite-prod-prop-in-rbar-mt}.

\item[The proof of Lemma~\thref{l:exp-in-rbar-mt}] \mbox{}\\
  cites explicitly:\\
  Definition~\thref{d:exp-in-rbar},\\
  Lemma~\thref{l:exp-in-rbar},\\
  Definition~\thref{d:mult-in-rbar-mt}.

\item[The proof of Lemma~\thref{l:youngs-ineq-for-prod-mt}] \mbox{}\\
  cites explicitly:\\
  Lemma~\thref{l:youngs-ineq-for-prod-in-r},\\
  Definition~\thref{d:add-in-rbar},\\
  Definition~\thref{d:mult-in-rbar},\\
  Lemma~\thref{l:exp-in-rbar},\\
  Lemma~\thref{l:add-in-rbarplus-is-closed},\\
  Lemma~\thref{l:mult-in-rbarplus-is-closed-when-defined},\\
  Lemma~\thref{l:infinity-prod-prop-in-rbarplus},\\
  Lemma~\thref{l:mult-in-rbarplus-is-closed-mt},\\
  Lemma~\thref{l:zero-prod-prop-in-rbarplus-mt},\\
  Lemma~\thref{l:infinity-prod-prop-in-rbarplus-mt}.

\item[The proof of Lemma~\thref{l:youngs-ineq-for-prod-case-p-two-mt}] \mbox{}\\
  cites explicitly:\\
  Lemma~\thref{l:two-is-self-holder-conjugate-in-r},\\
  Lemma~\thref{l:mult-in-rbarplus-is-closed-mt},\\
  Lemma~\thref{l:mult-in-rbarplus-is-assoc-mt},\\
  Lemma~\thref{l:mult-in-rbarplus-is-comm-mt},\\
  Lemma~\thref{l:youngs-ineq-for-prod-mt}.

\item[The proof of Lemma~\thref{l:conn-comp-of-open-subset-of-r-is-open-int}] \mbox{}\\
  cites explicitly:\\
  Statement(s) from~\cite{cm:lmt:16},\\
  Definition~\thref{d:interval},\\
  Definition~\thref{d:topological-space},\\
  Definition~\thref{d:conn-comp-in-r}.

\item[The proof of Lemma~\thref{l:conn-comp-of-open-subset-of-r-is-maximal}] \mbox{}\\
  cites explicitly:\\
  Definition~\thref{d:conn-comp-in-r},\\
  Lemma~\thref{l:conn-comp-of-open-subset-of-r-is-open-int}.

\item[The proof of Lemma~\thref{l:conn-comps-of-open-subset-of-r-equal-or-disj}] \mbox{}\\
  cites explicitly:\\
  Lemma~\thref{l:conn-comp-of-open-subset-of-r-is-maximal}.

\item[The proof of Theorem~\thref{t:count-conn-comps-of-open-subsets-of-r}] \mbox{}\\
  cites explicitly:\\
  Statement(s) from~\cite{cm:lmt:16},\\
  Definition~\thref{d:conn-comp-in-r},\\
  Lemma~\thref{l:conn-comps-of-open-subset-of-r-equal-or-disj}.

\item[The proof of Lemma~\thref{l:rat-approx-of-lower-bound-of-open-int}] \mbox{}\\
  has no explicit citation.

\item[The proof of Lemma~\thref{l:rat-approx-of-upper-bound-of-open-int}] \mbox{}\\
  cites explicitly:\\
  Lemma~\thref{l:rat-approx-of-lower-bound-of-open-int}.

\item[The proof of Lemma~\thref{l:open-int-with-rat-bounds-cover-open-int}] \mbox{}\\
  cites explicitly:\\
  Statement(s) from~\cite{cm:lmt:16},\\
  Definition~\thref{d:topological-space},\\
  Lemma~\thref{l:rat-approx-of-lower-bound-of-open-int},\\
  Lemma~\thref{l:rat-approx-of-upper-bound-of-open-int}.

\item[The proof of Theorem~\thref{t:r-is-second-countable}] \mbox{}\\
  cites explicitly:\\
  Definition~\thref{d:topological-basis},\\
  Definition~\thref{d:second-count},\\
  Theorem~\thref{t:count-conn-comps-of-open-subsets-of-r},\\
  Lemma~\thref{l:open-int-with-rat-bounds-cover-open-int}.

\item[The proof of Lemma~\thref{l:rn-is-second-countable}] \mbox{}\\
  cites explicitly:\\
  Lemma~\thref{l:box-topo-on-cartesian-prod},\\
  Definition~\thref{d:second-count},\\
  Lemma~\thref{l:compat-of-second-count-with-cartesian-prod},\\
  Theorem~\thref{t:r-is-second-countable}.

\item[The proof of Lemma~\thref{l:open-int-with-rat-bounds-cover-open-int-of-rbar}] \mbox{}\\
  cites explicitly:\\
  Lemma~\thref{l:rat-approx-of-lower-bound-of-open-int},\\
  Lemma~\thref{l:rat-approx-of-upper-bound-of-open-int},\\
  Lemma~\thref{l:open-int-with-rat-bounds-cover-open-int}.

\item[The proof of Lemma~\thref{l:rbar-is-second-countable}] \mbox{}\\
  cites explicitly:\\
  Definition~\thref{d:topological-basis},\\
  Definition~\thref{d:second-count},\\
  Lemma~\thref{l:topo-of-rbar},\\
  Theorem~\thref{t:count-conn-comps-of-open-subsets-of-r},\\
  Lemma~\thref{l:open-int-with-rat-bounds-cover-open-int-of-rbar}.

\item[The proof of Lemma~\thref{l:extr-of-const-fun}] \mbox{}\\
  cites explicitly:\\
  Statement(s) from~\cite{cm:lmt:16}.

\item[The proof of Lemma~\thref{l:equiv-def-of-finite-inf}] \mbox{}\\
  cites explicitly:\\
  Statement(s) from~\cite{cm:lmt:16},\\
  Lemma~\thref{l:equiv-def-of-conv-seq}.

\item[The proof of Lemma~\thref{l:equiv-def-of-finite-inf-in-rbar}] \mbox{}\\
  cites explicitly:\\
  Statement(s) from~\cite{cm:lmt:16},\\
  Lemma~\thref{l:extr-of-const-fun},\\
  Lemma~\thref{l:equiv-def-of-finite-inf}.

\item[The proof of Lemma~\thref{l:equiv-def-of-inf}] \mbox{}\\
  cites explicitly:\\
  Statement(s) from~\cite{cm:lmt:16},\\
  Lemma~\thref{l:conv-towards-minus-infinity},\\
  Lemma~\thref{l:equiv-def-of-finite-inf-in-rbar}.

\item[The proof of Lemma~\thref{l:inf-is-smaller-than-sup}] \mbox{}\\
  cites explicitly:\\
  Statement(s) from~\cite{cm:lmt:16},\\
  Lemma~\thref{l:order-in-rbar-is-total}.

\item[The proof of Lemma~\thref{l:inf-is-monot}] \mbox{}\\
  cites explicitly:\\
  Statement(s) from~\cite{cm:lmt:16},\\
  Lemma~\thref{l:order-in-rbar-is-total}.

\item[The proof of Lemma~\thref{l:sup-is-monot}] \mbox{}\\
  cites explicitly:\\
  Statement(s) from~\cite{cm:lmt:16},\\
  Lemma~\thref{l:inf-is-monot}.

\item[The proof of Lemma~\thref{l:compat-of-inf-with-abs}] \mbox{}\\
  cites explicitly:\\
  Statement(s) from~\cite{cm:lmt:16},\\
  Lemma~\thref{l:equiv-def-of-abs-in-rbar},\\
  Lemma~\thref{l:abs-in-rbar-is-even},\\
  Lemma~\thref{l:inf-is-smaller-than-sup},\\
  Lemma~\thref{l:inf-is-monot},\\
  Lemma~\thref{l:sup-is-monot}.

\item[The proof of Lemma~\thref{l:compat-of-sup-with-abs}] \mbox{}\\
  cites explicitly:\\
  Statement(s) from~\cite{cm:lmt:16},\\
  Lemma~\thref{l:abs-in-rbar-is-even},\\
  Lemma~\thref{l:compat-of-inf-with-abs}.

\item[The proof of Lemma~\thref{l:compat-of-translation-with-inf}] \mbox{}\\
  cites explicitly:\\
  Lemma~\thref{l:inf-is-monot}.

\item[The proof of Lemma~\thref{l:compat-of-translation-with-sup}] \mbox{}\\
  cites explicitly:\\
  Lemma~\thref{l:sup-is-monot}.

\item[The proof of Lemma~\thref{l:inf-of-seq-is-monot}] \mbox{}\\
  cites explicitly:\\
  Statement(s) from~\cite{cm:lmt:16},\\
  Lemma~\thref{l:order-in-rbar-is-total}.

\item[The proof of Lemma~\thref{l:sup-of-seq-is-monot}] \mbox{}\\
  cites explicitly:\\
  Statement(s) from~\cite{cm:lmt:16},\\
  Lemma~\thref{l:inf-of-seq-is-monot}.

\item[The proof of Lemma~\thref{l:inf-of-bounded-seq-is-bounded}] \mbox{}\\
  cites explicitly:\\
  Lemma~\thref{l:extr-of-const-fun},\\
  Lemma~\thref{l:inf-of-seq-is-monot}.

\item[The proof of Lemma~\thref{l:sup-of-bounded-seq-is-bounded}] \mbox{}\\
  cites explicitly:\\
  Lemma~\thref{l:extr-of-const-fun},\\
  Lemma~\thref{l:sup-of-seq-is-monot}.

\item[The proof of Lemma~\thref{l:liminf}] \mbox{}\\
  cites explicitly:\\
  Lemma~\thref{l:inf-is-monot}.

\item[The proof of Lemma~\thref{l:liminf-is-inf}] \mbox{}\\
  cites explicitly:\\
  Statement(s) from~\cite{cm:lmt:16},\\
  Lemma~\thref{l:liminf}.

\item[The proof of Lemma~\thref{l:equiv-def-of-liminf}] \mbox{}\\
  cites explicitly:\\
  Statement(s) from~\cite{cm:lmt:16},\\
  Definition~\thref{d:cluster-point},\\
  Lemma~\thref{l:liminf},\\
  Lemma~\thref{l:liminf-is-inf}.

\item[The proof of Lemma~\thref{l:liminf-is-invariant-by-translation}] \mbox{}\\
  cites explicitly:\\
  Definition~\thref{d:cluster-point},\\
  Lemma~\thref{l:equiv-def-of-liminf}.

\item[The proof of Lemma~\thref{l:liminf-is-monot}] \mbox{}\\
  cites explicitly:\\
  Lemma~\thref{l:inf-of-seq-is-monot},\\
  Lemma~\thref{l:sup-of-seq-is-monot},\\
  Lemma~\thref{l:liminf},\\
  Lemma~\thref{l:liminf-is-invariant-by-translation}.

\item[The proof of Lemma~\thref{l:limsup}] \mbox{}\\
  has no explicit citation.

\item[The proof of Lemma~\thref{l:duality-liminf-limsup}] \mbox{}\\
  cites explicitly:\\
  Statement(s) from~\cite{cm:lmt:16},\\
  Lemma~\thref{l:liminf},\\
  Lemma~\thref{l:limsup}.

\item[The proof of Lemma~\thref{l:equiv-def-of-limsup}] \mbox{}\\
  cites explicitly:\\
  Lemma~\thref{l:equiv-def-of-liminf},\\
  Lemma~\thref{l:duality-liminf-limsup}.

\item[The proof of Lemma~\thref{l:liminf-is-smaller-than-limsup}] \mbox{}\\
  cites explicitly:\\
  Lemma~\thref{l:order-in-rbar-is-total},\\
  Lemma~\thref{l:equiv-def-of-liminf},\\
  Lemma~\thref{l:equiv-def-of-limsup}.

\item[The proof of Lemma~\thref{l:limsup-is-monot}] \mbox{}\\
  cites explicitly:\\
  Lemma~\thref{l:liminf-is-monot},\\
  Lemma~\thref{l:duality-liminf-limsup}.

\item[The proof of Lemma~\thref{l:compat-liminf-with-abs}] \mbox{}\\
  cites explicitly:\\
  Lemma~\thref{l:compat-of-inf-with-abs},\\
  Lemma~\thref{l:liminf},\\
  Lemma~\thref{l:limsup}.

\item[The proof of Lemma~\thref{l:compat-limsup-with-abs}] \mbox{}\\
  cites explicitly:\\
  Lemma~\thref{l:abs-in-rbar-is-even},\\
  Lemma~\thref{l:duality-liminf-limsup},\\
  Lemma~\thref{l:compat-liminf-with-abs}.

\item[The proof of Lemma~\thref{l:liminf-and-limsup-of-pointwise-conv}] \mbox{}\\
  cites explicitly:\\
  Lemma~\thref{l:equiv-def-of-liminf},\\
  Lemma~\thref{l:equiv-def-of-limsup},\\
  Definition~\thref{d:pointwise-conv}.

\item[The proof of Lemma~\thref{l:liminf-bounded-from-below}] \mbox{}\\
  cites explicitly:\\
  Statement(s) from~\cite{cm:lmt:16},\\
  Lemma~\thref{l:liminf-is-monot},\\
  Lemma~\thref{l:liminf-and-limsup-of-pointwise-conv}.

\item[The proof of Lemma~\thref{l:liminf-bounded-from-above}] \mbox{}\\
  cites explicitly:\\
  Statement(s) from~\cite{cm:lmt:16},\\
  Lemma~\thref{l:liminf-is-monot},\\
  Lemma~\thref{l:liminf-and-limsup-of-pointwise-conv}.

\item[The proof of Lemma~\thref{l:limsup-bounded-from-below}] \mbox{}\\
  cites explicitly:\\
  Lemma~\thref{l:duality-liminf-limsup},\\
  Lemma~\thref{l:liminf-bounded-from-above}.

\item[The proof of Lemma~\thref{l:limsup-bounded-from-above}] \mbox{}\\
  cites explicitly:\\
  Lemma~\thref{l:duality-liminf-limsup},\\
  Lemma~\thref{l:liminf-bounded-from-below}.

\item[The proof of Lemma~\thref{l:liminf-limsup-and-pointwise-conv}] \mbox{}\\
  cites explicitly:\\
  Statement(s) from~\cite{cm:lmt:16},\\
  Lemma~\thref{l:liminf-is-inf},\\
  Lemma~\thref{l:equiv-def-of-liminf},\\
  Lemma~\thref{l:duality-liminf-limsup},\\
  Lemma~\thref{l:equiv-def-of-limsup},\\
  Lemma~\thref{l:liminf-is-smaller-than-limsup}.

\item[The proof of Lemma~\thref{l:finite-part-is-finite}] \mbox{}\\
  cites explicitly:\\
  Definition~\thref{d:ext-real-nums-rbar},\\
  Definition~\thref{d:finite-part}.

\item[The proof of Lemma~\thref{l:equiv-def-of-nonneg-and-nonpos-parts}] \mbox{}\\
  cites explicitly:\\
  Definition~\thref{d:nonneg-and-nonpos-parts}.

\item[The proof of Lemma~\thref{l:nonneg-and-nonpos-parts-are-nonneg}] \mbox{}\\
  cites explicitly:\\
  Definition~\thref{d:nonneg-and-nonpos-parts}.

\item[The proof of Lemma~\thref{l:nonneg-and-nonpos-parts-are-orthogonal}] \mbox{}\\
  has no explicit citation.

\item[The proof of Lemma~\thref{l:decomp-into-nonneg-and-nonpos-parts}] \mbox{}\\
  cites explicitly:\\
  Definition~\thref{d:add-in-rbar},\\
  Definition~\thref{d:abs-in-rbar},\\
  Lemma~\thref{l:nonneg-and-nonpos-parts-are-orthogonal}.

\item[The proof of Lemma~\thref{l:compat-of-nonpos-and-nonneg-parts-with-add}] \mbox{}\\
  cites explicitly:\\
  Definition~\thref{d:add-in-rbar},\\
  Lemma~\thref{l:add-in-rbarplus-is-assoc},\\
  Lemma~\thref{l:add-in-rbarplus-is-comm},\\
  Lemma~\thref{l:infinity-sum-prop-in-rbarplus},\\
  Definition~\thref{d:nonneg-and-nonpos-parts},\\
  Lemma~\thref{l:nonneg-and-nonpos-parts-are-nonneg},\\
  Lemma~\thref{l:nonneg-and-nonpos-parts-are-orthogonal},\\
  Lemma~\thref{l:decomp-into-nonneg-and-nonpos-parts}.

\item[The proof of Lemma~\thref{l:compat-of-nonpos-and-nonneg-parts-with-mask}] \mbox{}\\
  cites explicitly:\\
  Definition~\thref{d:nonneg-and-nonpos-parts}.

\item[The proof of Lemma~\thref{l:compat-of-nonpos-and-nonneg-parts-with-restr}] \mbox{}\\
  cites explicitly:\\
  Definition~\thref{d:nonneg-and-nonpos-parts}.

\item[The proof of Lemma~\thref{l:nonempty-and-empty-or-full}] \mbox{}\\
  has no explicit citation.

\item[The proof of Lemma~\thref{l:empty-and-full}] \mbox{}\\
  has no explicit citation.

\item[The proof of Lemma~\thref{l:local-compl-and-compl}] \mbox{}\\
  has no explicit citation.

\item[The proof of Lemma~\thref{l:union-disj-local-compl-equiv}] \mbox{}\\
  has no explicit citation.

\item[The proof of Lemma~\thref{l:set-diff-and-local-compl}] \mbox{}\\
  has no explicit citation.

\item[The proof of Lemma~\thref{l:inter-set-diff-equiv}] \mbox{}\\
  has no explicit citation.

\item[The proof of Lemma~\thref{l:union-inter-equiv}] \mbox{}\\
  has no explicit citation.

\item[The proof of Lemma~\thref{l:union-set-diff-equiv}] \mbox{}\\
  cites explicitly:\\
  Lemma~\thref{l:inter-set-diff-equiv},\\
  Lemma~\thref{l:union-inter-equiv}.

\item[The proof of Lemma~\thref{l:finite-ops-equiv}] \mbox{}\\
  has no explicit citation.

\item[The proof of Lemma~\thref{l:finite-union-inter-equiv}] \mbox{}\\
  cites explicitly:\\
  Lemma~\thref{l:union-inter-equiv},\\
  Lemma~\thref{l:finite-ops-equiv}.

\item[The proof of Lemma~\thref{l:count-and-finite-union-disj}] \mbox{}\\
  has no explicit citation.

\item[The proof of Lemma~\thref{l:count-union-disj-local-compl}] \mbox{}\\
  cites explicitly:\\
  Lemma~\thref{l:empty-and-full},\\
  Lemma~\thref{l:union-disj-local-compl-equiv},\\
  Lemma~\thref{l:finite-ops-equiv},\\
  Lemma~\thref{l:count-and-finite-union-disj}.

\item[The proof of Lemma~\thref{l:count-union-inter-equiv}] \mbox{}\\
  has no explicit citation.

\item[The proof of Lemma~\thref{l:count-union-disj-and-monot}] \mbox{}\\
  cites explicitly:\\
  Lemma~\thref{l:part-of-count-union}.

\item[The proof of Lemma~\thref{l:count-union-monot-and-disj}] \mbox{}\\
  has no explicit citation.

\item[The proof of Lemma~\thref{l:count-union-disj-and-union}] \mbox{}\\
  cites explicitly:\\
  Lemma~\thref{l:part-of-count-union},\\
  Lemma~\thref{l:inter-set-diff-equiv},\\
  Lemma~\thref{l:union-inter-equiv},\\
  Lemma~\thref{l:finite-ops-equiv}.

\item[The proof of Lemma~\thref{l:count-union-monot-and-union}] \mbox{}\\
  cites explicitly:\\
  Lemma~\thref{l:set-diff-and-local-compl},\\
  Lemma~\thref{l:inter-set-diff-equiv},\\
  Lemma~\thref{l:union-inter-equiv},\\
  Lemma~\thref{l:count-union-monot-and-disj},\\
  Lemma~\thref{l:count-union-disj-and-union}.

\item[The proof of Lemma~\thref{l:inter-of-p-systs}] \mbox{}\\
  cites explicitly:\\
  Definition~\thref{d:p-syst}.

\item[The proof of Lemma~\thref{l:gen-p-syst-is-min}] \mbox{}\\
  cites explicitly:\\
  Lemma~\thref{l:inter-of-p-systs},\\
  Definition~\thref{d:gen-p-syst}.

\item[The proof of Lemma~\thref{l:p-syst-gen-is-monot}] \mbox{}\\
  cites explicitly:\\
  Lemma~\thref{l:gen-p-syst-is-min}.

\item[The proof of Lemma~\thref{l:p-syst-gen-is-idem}] \mbox{}\\
  cites explicitly:\\
  Lemma~\thref{l:gen-p-syst-is-min}.

\item[The proof of Lemma~\thref{l:equiv-def-of-set-alg}] \mbox{}\\
  cites explicitly:\\
  Lemma~\thref{l:nonempty-and-empty-or-full},\\
  Lemma~\thref{l:empty-and-full},\\
  Lemma~\thref{l:finite-ops-equiv},\\
  Lemma~\thref{l:finite-union-inter-equiv},\\
  Definition~\thref{d:set-alg}.

\item[The proof of Lemma~\thref{l:other-equiv-def-of-set-alg}] \mbox{}\\
  cites explicitly:\\
  Lemma~\thref{l:local-compl-and-compl},\\
  Lemma~\thref{l:set-diff-and-local-compl},\\
  Lemma~\thref{l:inter-set-diff-equiv},\\
  Lemma~\thref{l:equiv-def-of-set-alg}.

\item[The proof of Lemma~\thref{l:set-alg-is-closed-under-local-compl}] \mbox{}\\
  cites explicitly:\\
  Lemma~\thref{l:set-diff-and-local-compl},\\
  Lemma~\thref{l:other-equiv-def-of-set-alg}.

\item[The proof of Lemma~\thref{l:inter-of-set-algs}] \mbox{}\\
  cites explicitly:\\
  Definition~\thref{d:set-alg}.

\item[The proof of Lemma~\thref{l:gen-set-alg-is-min}] \mbox{}\\
  cites explicitly:\\
  Lemma~\thref{l:inter-of-set-algs},\\
  Definition~\thref{d:gen-set-alg}.

\item[The proof of Lemma~\thref{l:set-alg-gen-is-monot}] \mbox{}\\
  cites explicitly:\\
  Lemma~\thref{l:gen-set-alg-is-min}.

\item[The proof of Lemma~\thref{l:set-alg-gen-is-idem}] \mbox{}\\
  cites explicitly:\\
  Lemma~\thref{l:gen-set-alg-is-min}.

\item[The proof of Lemma~\thref{l:part-of-count-union-in-set-alg}] \mbox{}\\
  cites explicitly:\\
  Lemma~\thref{l:part-of-count-union},\\
  Lemma~\thref{l:equiv-def-of-set-alg},\\
  Lemma~\thref{l:other-equiv-def-of-set-alg}.

\item[The proof of Lemma~\thref{l:explicit-set-alg}] \mbox{}\\
  cites explicitly:\\
  Lemma~\thref{l:finite-ops-equiv},\\
  Definition~\thref{d:set-alg},\\
  Lemma~\thref{l:equiv-def-of-set-alg},\\
  Lemma~\thref{l:gen-set-alg-is-min}.

\item[The proof of Lemma~\thref{l:inter-of-monot-classes}] \mbox{}\\
  cites explicitly:\\
  Definition~\thref{d:monot-class}.

\item[The proof of Lemma~\thref{l:gen-monot-class-is-min}] \mbox{}\\
  cites explicitly:\\
  Lemma~\thref{l:inter-of-monot-classes},\\
  Definition~\thref{d:gen-monot-class}.

\item[The proof of Lemma~\thref{l:monot-class-gen-is-monot}] \mbox{}\\
  cites explicitly:\\
  Lemma~\thref{l:gen-monot-class-is-min}.

\item[The proof of Lemma~\thref{l:monot-class-gen-is-idem}] \mbox{}\\
  cites explicitly:\\
  Lemma~\thref{l:gen-monot-class-is-min}.

\item[The proof of Lemma~\thref{l:c-diff-is-symmetric}] \mbox{}\\
  cites explicitly:\\
  Definition~\thref{d:monot-class-and-symm-set-diff}.

\item[The proof of Lemma~\thref{l:c-diff-is-monot-class}] \mbox{}\\
  cites explicitly:\\
  Definition~\thref{d:monot-class},\\
  Definition~\thref{d:monot-class-and-symm-set-diff}.

\item[The proof of Lemma~\thref{l:monot-class-is-closed-under-set-diff}] \mbox{}\\
  cites explicitly:\\
  Lemma~\thref{l:gen-monot-class-is-min},\\
  Definition~\thref{d:monot-class-and-symm-set-diff},\\
  Lemma~\thref{l:c-diff-is-symmetric},\\
  Lemma~\thref{l:c-diff-is-monot-class}.

\item[The proof of Lemma~\thref{l:monot-class-gen-by-set-alg}] \mbox{}\\
  cites explicitly:\\
  Lemma~\thref{l:other-equiv-def-of-set-alg},\\
  Lemma~\thref{l:gen-monot-class-is-min},\\
  Lemma~\thref{l:monot-class-is-closed-under-set-diff}.

\item[The proof of Lemma~\thref{l:equiv-def-of-l-syst}] \mbox{}\\
  cites explicitly:\\
  Lemma~\thref{l:local-compl-and-compl},\\
  Lemma~\thref{l:count-union-disj-local-compl},\\
  Lemma~\thref{l:count-union-disj-and-monot},\\
  Lemma~\thref{l:count-union-monot-and-disj},\\
  Definition~\thref{d:l-syst}.

\item[The proof of Lemma~\thref{l:other-prop-of-l-syst}] \mbox{}\\
  cites explicitly:\\
  Lemma~\thref{l:nonempty-and-empty-or-full},\\
  Lemma~\thref{l:empty-and-full},\\
  Lemma~\thref{l:count-union-inter-equiv},\\
  Definition~\thref{d:l-syst},\\
  Lemma~\thref{l:equiv-def-of-l-syst}.

\item[The proof of Lemma~\thref{l:inter-of-l-systs}] \mbox{}\\
  cites explicitly:\\
  Definition~\thref{d:l-syst}.

\item[The proof of Lemma~\thref{l:gen-l-syst-is-min}] \mbox{}\\
  cites explicitly:\\
  Lemma~\thref{l:inter-of-l-systs},\\
  Definition~\thref{d:gen-l-syst}.

\item[The proof of Lemma~\thref{l:l-syst-gen-is-monot}] \mbox{}\\
  cites explicitly:\\
  Lemma~\thref{l:gen-l-syst-is-min}.

\item[The proof of Lemma~\thref{l:l-syst-gen-is-idem}] \mbox{}\\
  cites explicitly:\\
  Lemma~\thref{l:gen-l-syst-is-min}.

\item[The proof of Lemma~\thref{l:l-inter-is-symmetric}] \mbox{}\\
  cites explicitly:\\
  Definition~\thref{d:l-syst-and-inter}.

\item[The proof of Lemma~\thref{l:l-inter-is-l-syst}] \mbox{}\\
  cites explicitly:\\
  Definition~\thref{d:l-syst},\\
  Lemma~\thref{l:equiv-def-of-l-syst}.

\item[The proof of Lemma~\thref{l:l-syst-with-inter}] \mbox{}\\
  cites explicitly:\\
  Lemma~\thref{l:gen-l-syst-is-min},\\
  Definition~\thref{d:l-syst-and-inter},\\
  Lemma~\thref{l:l-inter-is-symmetric},\\
  Lemma~\thref{l:l-inter-is-l-syst}.

\item[The proof of Lemma~\thref{l:l-syst-is-closed-under-inter}] \mbox{}\\
  cites explicitly:\\
  Definition~\thref{d:l-syst-and-inter},\\
  Lemma~\thref{l:l-syst-with-inter}.

\item[The proof of Lemma~\thref{l:l-syst-gen-by-p-syst}] \mbox{}\\
  cites explicitly:\\
  Lemma~\thref{l:finite-ops-equiv},\\
  Definition~\thref{d:p-syst},\\
  Lemma~\thref{l:gen-l-syst-is-min},\\
  Lemma~\thref{l:l-syst-is-closed-under-inter}.

\item[The proof of Lemma~\thref{l:equiv-def-of-sigma-alg}] \mbox{}\\
  cites explicitly:\\
  Lemma~\thref{l:nonempty-and-empty-or-full},\\
  Lemma~\thref{l:empty-and-full},\\
  Lemma~\thref{l:count-union-inter-equiv},\\
  Definition~\thref{d:sigma-alg}.

\item[The proof of Lemma~\thref{l:sigma-alg-is-set-alg}] \mbox{}\\
  cites explicitly:\\
  Definition~\thref{d:set-alg},\\
  Definition~\thref{d:sigma-alg}.

\item[The proof of Lemma~\thref{l:sigma-alg-is-closed-under-set-diff}] \mbox{}\\
  cites explicitly:\\
  Lemma~\thref{l:other-equiv-def-of-set-alg},\\
  Lemma~\thref{l:set-alg-is-closed-under-local-compl},\\
  Lemma~\thref{l:sigma-alg-is-set-alg}.

\item[The proof of Lemma~\thref{l:other-prop-of-sigma-alg}] \mbox{}\\
  cites explicitly:\\
  Lemma~\thref{l:equiv-def-of-sigma-alg}.

\item[The proof of Lemma~\thref{l:part-of-count-union-in-sigma-alg}] \mbox{}\\
  cites explicitly:\\
  Lemma~\thref{l:part-of-count-union},\\
  Lemma~\thref{l:part-of-count-union-in-set-alg},\\
  Definition~\thref{d:sigma-alg},\\
  Lemma~\thref{l:sigma-alg-is-set-alg}.

\item[The proof of Lemma~\thref{l:inter-of-sigma-algs}] \mbox{}\\
  cites explicitly:\\
  Definition~\thref{d:sigma-alg}.

\item[The proof of Lemma~\thref{l:gen-sigma-alg-is-min}] \mbox{}\\
  cites explicitly:\\
  Lemma~\thref{l:inter-of-sigma-algs},\\
  Definition~\thref{d:gen-sigma-alg}.

\item[The proof of Lemma~\thref{l:sigma-alg-gen-is-monot}] \mbox{}\\
  cites explicitly:\\
  Lemma~\thref{l:gen-sigma-alg-is-min}.

\item[The proof of Lemma~\thref{l:sigma-alg-gen-is-idem}] \mbox{}\\
  cites explicitly:\\
  Lemma~\thref{l:gen-sigma-alg-is-min}.

\item[The proof of Lemma~\thref{l:sigma-alg-is-p-syst}] \mbox{}\\
  cites explicitly:\\
  Definition~\thref{d:p-syst},\\
  Lemma~\thref{l:equiv-def-of-sigma-alg}.

\item[The proof of Lemma~\thref{l:sigma-alg-contains-p-syst}] \mbox{}\\
  cites explicitly:\\
  Lemma~\thref{l:gen-p-syst-is-min},\\
  Lemma~\thref{l:gen-sigma-alg-is-min},\\
  Lemma~\thref{l:sigma-alg-is-p-syst}.

\item[The proof of Lemma~\thref{l:p-syst-contains-sigma-alg}] \mbox{}\\
  cites explicitly:\\
  Lemma~\thref{l:nonempty-and-empty-or-full},\\
  Lemma~\thref{l:count-union-disj-and-union},\\
  Definition~\thref{d:p-syst},\\
  Lemma~\thref{l:gen-p-syst-is-min},\\
  Definition~\thref{d:sigma-alg},\\
  Lemma~\thref{l:gen-sigma-alg-is-min}.

\item[The proof of Lemma~\thref{l:sigma-alg-gen-by-p-syst}] \mbox{}\\
  cites explicitly:\\
  Lemma~\thref{l:gen-p-syst-is-min},\\
  Lemma~\thref{l:gen-sigma-alg-is-min},\\
  Lemma~\thref{l:sigma-alg-gen-is-monot},\\
  Lemma~\thref{l:sigma-alg-contains-p-syst}.

\item[The proof of Lemma~\thref{l:sigma-alg-contains-set-alg}] \mbox{}\\
  cites explicitly:\\
  Lemma~\thref{l:gen-set-alg-is-min},\\
  Lemma~\thref{l:sigma-alg-is-set-alg},\\
  Lemma~\thref{l:gen-sigma-alg-is-min}.

\item[The proof of Lemma~\thref{l:set-alg-contains-sigma-alg}] \mbox{}\\
  cites explicitly:\\
  Lemma~\thref{l:count-union-monot-and-union},\\
  Definition~\thref{d:set-alg},\\
  Lemma~\thref{l:gen-set-alg-is-min},\\
  Definition~\thref{d:sigma-alg},\\
  Lemma~\thref{l:gen-sigma-alg-is-min}.

\item[The proof of Lemma~\thref{l:sigma-alg-gen-by-set-alg}] \mbox{}\\
  cites explicitly:\\
  Lemma~\thref{l:gen-set-alg-is-min},\\
  Lemma~\thref{l:gen-sigma-alg-is-min},\\
  Lemma~\thref{l:sigma-alg-gen-is-monot},\\
  Lemma~\thref{l:sigma-alg-contains-set-alg}.

\item[The proof of Lemma~\thref{l:sigma-alg-is-monot-class}] \mbox{}\\
  cites explicitly:\\
  Definition~\thref{d:monot-class},\\
  Lemma~\thref{l:other-prop-of-sigma-alg}.

\item[The proof of Lemma~\thref{l:sigma-alg-contains-monot-class}] \mbox{}\\
  cites explicitly:\\
  Lemma~\thref{l:gen-monot-class-is-min},\\
  Lemma~\thref{l:gen-sigma-alg-is-min},\\
  Lemma~\thref{l:sigma-alg-is-monot-class}.

\item[The proof of Lemma~\thref{l:monot-class-contains-sigma-alg}] \mbox{}\\
  cites explicitly:\\
  Lemma~\thref{l:count-union-monot-and-union},\\
  Definition~\thref{d:monot-class},\\
  Lemma~\thref{l:gen-monot-class-is-min},\\
  Definition~\thref{d:sigma-alg},\\
  Lemma~\thref{l:gen-sigma-alg-is-min}.

\item[The proof of Lemma~\thref{l:sigma-alg-gen-by-monot-class}] \mbox{}\\
  cites explicitly:\\
  Lemma~\thref{l:gen-monot-class-is-min},\\
  Lemma~\thref{l:gen-sigma-alg-is-min},\\
  Lemma~\thref{l:sigma-alg-gen-is-monot},\\
  Lemma~\thref{l:sigma-alg-contains-monot-class}.

\item[The proof of Lemma~\thref{l:sigma-alg-is-l-syst}] \mbox{}\\
  cites explicitly:\\
  Definition~\thref{d:l-syst},\\
  Lemma~\thref{l:equiv-def-of-sigma-alg},\\
  Lemma~\thref{l:other-prop-of-sigma-alg}.

\item[The proof of Lemma~\thref{l:sigma-alg-contains-l-syst}] \mbox{}\\
  cites explicitly:\\
  Lemma~\thref{l:gen-l-syst-is-min},\\
  Lemma~\thref{l:gen-sigma-alg-is-min},\\
  Lemma~\thref{l:sigma-alg-is-l-syst}.

\item[The proof of Lemma~\thref{l:l-syst-contains-sigma-alg}] \mbox{}\\
  cites explicitly:\\
  Lemma~\thref{l:count-union-disj-and-union},\\
  Definition~\thref{d:l-syst},\\
  Lemma~\thref{l:gen-l-syst-is-min},\\
  Lemma~\thref{l:equiv-def-of-sigma-alg},\\
  Lemma~\thref{l:gen-sigma-alg-is-min}.

\item[The proof of Lemma~\thref{l:sigma-alg-gen-by-l-syst}] \mbox{}\\
  cites explicitly:\\
  Lemma~\thref{l:gen-l-syst-is-min},\\
  Lemma~\thref{l:gen-sigma-alg-is-min},\\
  Lemma~\thref{l:sigma-alg-gen-is-monot},\\
  Lemma~\thref{l:sigma-alg-contains-l-syst}.

\item[The proof of Lemma~\thref{l:other-sigma-alg-gen}] \mbox{}\\
  cites explicitly:\\
  Lemma~\thref{l:sigma-alg-gen-is-monot},\\
  Lemma~\thref{l:sigma-alg-gen-is-idem}.

\item[The proof of Lemma~\thref{l:complete-gen-sigma-alg}] \mbox{}\\
  cites explicitly:\\
  Lemma~\thref{l:gen-sigma-alg-is-min},\\
  Lemma~\thref{l:sigma-alg-gen-is-monot},\\
  Lemma~\thref{l:other-sigma-alg-gen}.

\item[The proof of Lemma~\thref{l:count-sigma-alg-gen}] \mbox{}\\
  cites explicitly:\\
  Definition~\thref{d:sigma-alg},\\
  Lemma~\thref{l:gen-sigma-alg-is-min},\\
  Lemma~\thref{l:other-sigma-alg-gen}.

\item[The proof of Lemma~\thref{l:set-alg-gen-by-prod-of-sigma-algs}] \mbox{}\\
  cites explicitly:\\
  Definition~\thref{d:prod-of-subsets-of-parties},\\
  Lemma~\thref{l:explicit-set-alg},\\
  Lemma~\thref{l:equiv-def-of-sigma-alg}.

\item[The proof of Lemma~\thref{l:p-syst-and-l-syst-is-sigma-alg}] \mbox{}\\
  cites explicitly:\\
  Definition~\thref{d:p-syst},\\
  Lemma~\thref{l:l-syst-gen-is-idem},\\
  Lemma~\thref{l:sigma-alg-gen-is-idem},\\
  Lemma~\thref{l:sigma-alg-contains-l-syst},\\
  Lemma~\thref{l:l-syst-contains-sigma-alg}.

\item[The proof of Theorem~\thref{t:dynkin-pi-lambda-th}] \mbox{}\\
  cites explicitly:\\
  Lemma~\thref{l:gen-l-syst-is-min},\\
  Lemma~\thref{l:l-syst-gen-by-p-syst},\\
  Lemma~\thref{l:sigma-alg-gen-is-idem},\\
  Lemma~\thref{l:sigma-alg-gen-by-l-syst},\\
  Lemma~\thref{l:p-syst-and-l-syst-is-sigma-alg}.

\item[The proof of Lemma~\thref{l:usage-of-dynkin-pi-lambda-th}] \mbox{}\\
  cites explicitly:\\
  Lemma~\thref{l:gen-p-syst-is-min},\\
  Lemma~\thref{l:l-syst-gen-is-monot},\\
  Lemma~\thref{l:l-syst-gen-is-idem},\\
  Definition~\thref{d:gen-sigma-alg},\\
  Lemma~\thref{l:sigma-alg-gen-is-monot},\\
  Theorem~\thref{t:dynkin-pi-lambda-th}.

\item[The proof of Lemma~\thref{l:set-alg-and-monot-class-is-sigma-alg}] \mbox{}\\
  cites explicitly:\\
  Definition~\thref{d:set-alg},\\
  Lemma~\thref{l:monot-class-gen-is-idem},\\
  Lemma~\thref{l:sigma-alg-gen-is-idem},\\
  Lemma~\thref{l:sigma-alg-contains-monot-class},\\
  Lemma~\thref{l:monot-class-contains-sigma-alg}.

\item[The proof of Theorem~\thref{t:monot-class}] \mbox{}\\
  cites explicitly:\\
  Lemma~\thref{l:gen-monot-class-is-min},\\
  Lemma~\thref{l:monot-class-gen-by-set-alg},\\
  Lemma~\thref{l:sigma-alg-gen-is-idem},\\
  Lemma~\thref{l:sigma-alg-gen-by-monot-class},\\
  Lemma~\thref{l:set-alg-and-monot-class-is-sigma-alg}.

\item[The proof of Lemma~\thref{l:usage-of-monot-class-th}] \mbox{}\\
  cites explicitly:\\
  Lemma~\thref{l:gen-set-alg-is-min},\\
  Lemma~\thref{l:monot-class-gen-is-monot},\\
  Lemma~\thref{l:monot-class-gen-is-idem},\\
  Definition~\thref{d:gen-sigma-alg},\\
  Lemma~\thref{l:sigma-alg-gen-is-monot},\\
  Theorem~\thref{t:monot-class}.

\item[The proof of Lemma~\thref{l:some-borel-subsets}] \mbox{}\\
  cites explicitly:\\
  Definition~\thref{d:topological-space},\\
  Definition~\thref{d:sigma-alg},\\
  Lemma~\thref{l:gen-sigma-alg-is-min},\\
  Definition~\thref{d:borel-sigma-alg}.

\item[The proof of Lemma~\thref{l:count-borel-sigma-alg-gen}] \mbox{}\\
  cites explicitly:\\
  Lemma~\thref{l:count-sigma-alg-gen},\\
  Definition~\thref{d:borel-sigma-alg}.

\item[The proof of Lemma~\thref{l:inverse-sigma-alg}] \mbox{}\\
  cites explicitly:\\
  Definition~\thref{d:sigma-alg},\\
  Definition~\thref{d:measurable-space},\\
  Definition~\thref{d:meas-fun}.

\item[The proof of Lemma~\thref{l:image-sigma-alg}] \mbox{}\\
  cites explicitly:\\
  Definition~\thref{d:sigma-alg},\\
  Definition~\thref{d:measurable-space},\\
  Definition~\thref{d:meas-fun}.

\item[The proof of Lemma~\thref{l:identity-fun-is-meas}] \mbox{}\\
  cites explicitly:\\
  Definition~\thref{d:meas-fun}.

\item[The proof of Lemma~\thref{l:const-fun-is-meas}] \mbox{}\\
  cites explicitly:\\
  Lemma~\thref{l:equiv-def-of-sigma-alg}.

\item[The proof of Lemma~\thref{l:inverse-image-of-gen-family}] \mbox{}\\
  cites explicitly:\\
  Lemma~\thref{l:gen-sigma-alg-is-min},\\
  Lemma~\thref{l:sigma-alg-gen-is-monot},\\
  Lemma~\thref{l:sigma-alg-gen-is-idem},\\
  Lemma~\thref{l:inverse-sigma-alg},\\
  Lemma~\thref{l:image-sigma-alg}.

\item[The proof of Lemma~\thref{l:equiv-def-of-meas-fun}] \mbox{}\\
  cites explicitly:\\
  Lemma~\thref{l:sigma-alg-gen-is-monot},\\
  Lemma~\thref{l:sigma-alg-gen-is-idem},\\
  Definition~\thref{d:meas-fun},\\
  Lemma~\thref{l:inverse-image-of-gen-family}.

\item[The proof of Lemma~\thref{l:cont-is-meas}] \mbox{}\\
  cites explicitly:\\
  Lemma~\thref{l:gen-sigma-alg-is-min},\\
  Definition~\thref{d:borel-sigma-alg},\\
  Definition~\thref{d:meas-fun},\\
  Lemma~\thref{l:equiv-def-of-meas-fun}.

\item[The proof of Lemma~\thref{l:compat-of-meas-with-comp}] \mbox{}\\
  cites explicitly:\\
  Definition~\thref{d:meas-fun}.

\item[The proof of Lemma~\thref{l:trace-sigma-alg}] \mbox{}\\
  cites explicitly:\\
  Definition~\thref{d:trace-of-subsets-of-parties},\\
  Definition~\thref{d:sigma-alg},\\
  Definition~\thref{d:measurable-space},\\
  Definition~\thref{d:meas-fun}.

\item[The proof of Lemma~\thref{l:meas-of-meas-subspace}] \mbox{}\\
  cites explicitly:\\
  Definition~\thref{d:trace-of-subsets-of-parties},\\
  Lemma~\thref{l:equiv-def-of-sigma-alg},\\
  Lemma~\thref{l:trace-sigma-alg}.

\item[The proof of Lemma~\thref{l:gen-meas-subspace}] \mbox{}\\
  cites explicitly:\\
  Definition~\thref{d:trace-of-subsets-of-parties},\\
  Lemma~\thref{l:inverse-image-of-gen-family}.

\item[The proof of Lemma~\thref{l:borel-sub-sigma-alg}] \mbox{}\\
  cites explicitly:\\
  Definition~\thref{d:borel-sigma-alg},\\
  Lemma~\thref{l:meas-of-meas-subspace},\\
  Lemma~\thref{l:gen-meas-subspace}.

\item[The proof of Lemma~\thref{l:characterization-of-borel-subsets}] \mbox{}\\
  cites explicitly:\\
  Lemma~\thref{l:compat-of-pseudopart-with-inter},\\
  Definition~\thref{d:sigma-alg},\\
  Lemma~\thref{l:equiv-def-of-sigma-alg},\\
  Definition~\thref{d:borel-sigma-alg},\\
  Lemma~\thref{l:borel-sub-sigma-alg}.

\item[The proof of Lemma~\thref{l:source-restr-of-meas-fun}] \mbox{}\\
  cites explicitly:\\
  Definition~\thref{d:trace-of-subsets-of-parties},\\
  Definition~\thref{d:meas-fun},\\
  Lemma~\thref{l:compat-of-meas-with-comp},\\
  Lemma~\thref{l:trace-sigma-alg}.

\item[The proof of Lemma~\thref{l:destination-restr-of-meas-fun}] \mbox{}\\
  cites explicitly:\\
  Definition~\thref{d:meas-fun}.

\item[The proof of Lemma~\thref{l:meas-of-fun-def-on-pseudopart}] \mbox{}\\
  cites explicitly:\\
  Lemma~\thref{l:equiv-def-of-sigma-alg},\\
  Definition~\thref{d:meas-fun}.

\item[The proof of Lemma~\thref{l:prod-of-meas-subsets-is-meas}] \mbox{}\\
  cites explicitly:\\
  Definition~\thref{d:prod-of-subsets-of-parties},\\
  Lemma~\thref{l:gen-sigma-alg-is-min},\\
  Definition~\thref{d:tensor-prod-of-sigma-algs}.

\item[The proof of Lemma~\thref{l:meas-of-fun-to-prod-space}] \mbox{}\\
  cites explicitly:\\
  Definition~\thref{d:prod-of-subsets-of-parties},\\
  Lemma~\thref{l:equiv-def-of-sigma-alg},\\
  Definition~\thref{d:meas-fun},\\
  Lemma~\thref{l:equiv-def-of-meas-fun},\\
  Definition~\thref{d:tensor-prod-of-sigma-algs},\\
  Lemma~\thref{l:prod-of-meas-subsets-is-meas}.

\item[The proof of Lemma~\thref{l:can-proj-is-meas}] \mbox{}\\
  cites explicitly:\\
  Definition~\thref{d:meas-fun},\\
  Lemma~\thref{l:meas-of-fun-to-prod-space}.

\item[The proof of Lemma~\thref{l:perm-is-meas}] \mbox{}\\
  cites explicitly:\\
  Lemma~\thref{l:meas-of-fun-to-prod-space},\\
  Lemma~\thref{l:can-proj-is-meas}.

\item[The proof of Lemma~\thref{l:gen-prod-meas-space}] \mbox{}\\
  cites explicitly:\\
  Definition~\thref{d:prod-of-subsets-of-parties},\\
  Lemma~\thref{l:gen-sigma-alg-is-min},\\
  Lemma~\thref{l:sigma-alg-gen-is-monot},\\
  Definition~\thref{d:meas-fun},\\
  Lemma~\thref{l:inverse-image-of-gen-family},\\
  Definition~\thref{d:tensor-prod-of-sigma-algs},\\
  Lemma~\thref{l:can-proj-is-meas}.

\item[The proof of Lemma~\thref{l:section-of-prod}] \mbox{}\\
  cites explicitly:\\
  Definition~\thref{d:section-in-cartesian-prod}.

\item[The proof of Lemma~\thref{l:compat-of-section-with-set-ops}] \mbox{}\\
  cites explicitly:\\
  Definition~\thref{d:section-in-cartesian-prod}.

\item[The proof of Lemma~\thref{l:measurability-of-section}] \mbox{}\\
  cites explicitly:\\
  Definition~\thref{d:prod-of-subsets-of-parties},\\
  Definition~\thref{d:sigma-alg},\\
  Lemma~\thref{l:gen-sigma-alg-is-min},\\
  Definition~\thref{d:measurable-space},\\
  Definition~\thref{d:tensor-prod-of-sigma-algs},\\
  Lemma~\thref{l:section-of-prod},\\
  Lemma~\thref{l:compat-of-section-with-set-ops}.

\item[The proof of Lemma~\thref{l:count-union-of-sections-is-meas}] \mbox{}\\
  cites explicitly:\\
  Definition~\thref{d:sigma-alg},\\
  Definition~\thref{d:measurable-space},\\
  Lemma~\thref{l:compat-of-section-with-set-ops},\\
  Lemma~\thref{l:measurability-of-section}.

\item[The proof of Lemma~\thref{l:count-inter-of-sections-is-meas}] \mbox{}\\
  cites explicitly:\\
  Lemma~\thref{l:equiv-def-of-sigma-alg},\\
  Lemma~\thref{l:compat-of-section-with-set-ops},\\
  Lemma~\thref{l:measurability-of-section}.

\item[The proof of Lemma~\thref{l:indic-of-section}] \mbox{}\\
  cites explicitly:\\
  Definition~\thref{d:section-in-cartesian-prod}.

\item[The proof of Lemma~\thref{l:meas-of-fun-from-prod-space}] \mbox{}\\
  cites explicitly:\\
  Definition~\thref{d:meas-fun},\\
  Lemma~\thref{l:measurability-of-section}.

\item[The proof of Lemma~\thref{l:borel-sigma-alg-of-r}] \mbox{}\\
  cites explicitly:\\
  Theorem~\thref{t:count-conn-comps-of-open-subsets-of-r},\\
  Definition~\thref{d:sigma-alg},\\
  Lemma~\thref{l:equiv-def-of-sigma-alg},\\
  Definition~\thref{d:gen-sigma-alg},\\
  Lemma~\thref{l:other-sigma-alg-gen},\\
  Lemma~\thref{l:some-borel-subsets},\\
  Lemma~\thref{l:count-borel-sigma-alg-gen}.

\item[The proof of Lemma~\thref{l:count-gen-of-borel-sigma-alg-of-r}] \mbox{}\\
  cites explicitly:\\
  Theorem~\thref{t:count-conn-comps-of-open-subsets-of-r},\\
  Theorem~\thref{t:r-is-second-countable},\\
  Lemma~\thref{l:borel-sigma-alg-of-r}.

\item[The proof of Lemma~\thref{l:borel-sigma-alg-of-rbar}] \mbox{}\\
  cites explicitly:\\
  Theorem~\thref{t:count-conn-comps-of-open-subsets-of-r},\\
  Lemma~\thref{l:equiv-def-of-sigma-alg},\\
  Definition~\thref{d:gen-sigma-alg},\\
  Lemma~\thref{l:other-sigma-alg-gen},\\
  Lemma~\thref{l:count-borel-sigma-alg-gen}.

\item[The proof of Lemma~\thref{l:borel-subsets-of-rbar-and-r}] \mbox{}\\
  cites explicitly:\\
  Definition~\thref{d:sigma-alg},\\
  Definition~\thref{d:borel-sigma-alg},\\
  Lemma~\thref{l:borel-sub-sigma-alg},\\
  Lemma~\thref{l:characterization-of-borel-subsets}.

\item[The proof of Lemma~\thref{l:borel-sigma-alg-of-rplus}] \mbox{}\\
  cites explicitly:\\
  Definition~\thref{d:sigma-alg},\\
  Lemma~\thref{l:borel-sub-sigma-alg},\\
  Lemma~\thref{l:borel-sigma-alg-of-r}.

\item[The proof of Lemma~\thref{l:borel-sigma-alg-of-rbarplus}] \mbox{}\\
  cites explicitly:\\
  Lemma~\thref{l:gen-meas-subspace},\\
  Lemma~\thref{l:borel-sigma-alg-of-rbar}.

\item[The proof of Lemma~\thref{l:borel-sigma-alg-of-rm}] \mbox{}\\
  cites explicitly:\\
  Lemma~\thref{l:complete-count-topo-basis-of-prod-space},\\
  Lemma~\thref{l:rn-is-second-countable},\\
  Lemma~\thref{l:equiv-def-of-sigma-alg},\\
  Lemma~\thref{l:complete-gen-sigma-alg},\\
  Lemma~\thref{l:count-borel-sigma-alg-gen},\\
  Lemma~\thref{l:gen-prod-meas-space},\\
  Lemma~\thref{l:count-gen-of-borel-sigma-alg-of-r}.

\item[The proof of Lemma~\thref{l:meas-of-indic-fun}] \mbox{}\\
  cites explicitly:\\
  Definition~\thref{d:sigma-alg},\\
  Lemma~\thref{l:equiv-def-of-sigma-alg},\\
  Definition~\thref{d:measurable-space},\\
  Definition~\thref{d:mr-vector-space-of-meas-num-fun-to-r}.

\item[The proof of Lemma~\thref{l:meas-of-num-fun-to-r}] \mbox{}\\
  cites explicitly:\\
  Definition~\thref{d:borel-sigma-alg},\\
  Lemma~\thref{l:some-borel-subsets},\\
  Definition~\thref{d:meas-fun},\\
  Lemma~\thref{l:equiv-def-of-meas-fun},\\
  Lemma~\thref{l:borel-sigma-alg-of-r},\\
  Definition~\thref{d:mr-vector-space-of-meas-num-fun-to-r}.

\item[The proof of Lemma~\thref{l:inverse-image-is-meas-in-r}] \mbox{}\\
  cites explicitly:\\
  Lemma~\thref{l:some-borel-subsets},\\
  Definition~\thref{d:meas-fun},\\
  Definition~\thref{d:mr-vector-space-of-meas-num-fun-to-r}.

\item[The proof of Lemma~\thref{l:mr-is-alg}] \mbox{}\\
  cites explicitly:\\
  Lemma~\thref{l:k-is-k-alg},\\
  Lemma~\thref{l:alg-of-funs-to-alg},\\
  Lemma~\thref{l:closed-under-alg-ops-is-subalg},\\
  Lemma~\thref{l:const-fun-is-meas},\\
  Lemma~\thref{l:cont-is-meas},\\
  Lemma~\thref{l:compat-of-meas-with-comp},\\
  Lemma~\thref{l:meas-of-fun-to-prod-space},\\
  Lemma~\thref{l:borel-sigma-alg-of-rm},\\
  Definition~\thref{d:mr-vector-space-of-meas-num-fun-to-r}.

\item[The proof of Lemma~\thref{l:mr-is-vector-space}] \mbox{}\\
  cites explicitly:\\
  Definition~\thref{d:alg-over-a-field},\\
  Definition~\thref{d:subalg},\\
  Lemma~\thref{l:closed-under-alg-ops-is-subalg},\\
  Lemma~\thref{l:mr-is-alg}.

\item[The proof of Lemma~\thref{l:m-and-finite-is-mr}] \mbox{}\\
  cites explicitly:\\
  Definition~\thref{d:ext-real-nums-rbar},\\
  Definition~\thref{d:meas-fun},\\
  Lemma~\thref{l:borel-subsets-of-rbar-and-r},\\
  Definition~\thref{d:mr-vector-space-of-meas-num-fun-to-r},\\
  Definition~\thref{d:m-set-of-meas-num-funs}.

\item[The proof of Lemma~\thref{l:meas-of-num-fun}] \mbox{}\\
  cites explicitly:\\
  Definition~\thref{d:borel-sigma-alg},\\
  Lemma~\thref{l:some-borel-subsets},\\
  Definition~\thref{d:meas-fun},\\
  Lemma~\thref{l:equiv-def-of-meas-fun},\\
  Lemma~\thref{l:borel-sigma-alg-of-rbar},\\
  Definition~\thref{d:m-set-of-meas-num-funs}.

\item[The proof of Lemma~\thref{l:inverse-image-is-meas}] \mbox{}\\
  cites explicitly:\\
  Lemma~\thref{l:some-borel-subsets},\\
  Definition~\thref{d:meas-fun},\\
  Definition~\thref{d:m-set-of-meas-num-funs}.

\item[The proof of Lemma~\thref{l:m-is-closed-under-finite-part}] \mbox{}\\
  cites explicitly:\\
  Definition~\thref{d:pseudopart},\\
  Definition~\thref{d:ext-real-nums-rbar},\\
  Definition~\thref{d:finite-part},\\
  Lemma~\thref{l:finite-part-is-finite},\\
  Definition~\thref{d:sigma-alg},\\
  Definition~\thref{d:measurable-space},\\
  Lemma~\thref{l:const-fun-is-meas},\\
  Lemma~\thref{l:meas-of-fun-def-on-pseudopart},\\
  Lemma~\thref{l:m-and-finite-is-mr},\\
  Lemma~\thref{l:inverse-image-is-meas}.

\item[The proof of Lemma~\thref{l:m-is-closed-under-add-when-defined}] \mbox{}\\
  cites explicitly:\\
  Definition~\thref{d:pseudopart},\\
  Definition~\thref{d:add-in-rbar},\\
  Definition~\thref{d:finite-part},\\
  Lemma~\thref{l:equiv-def-of-sigma-alg},\\
  Definition~\thref{d:borel-sigma-alg},\\
  Definition~\thref{d:meas-fun},\\
  Lemma~\thref{l:const-fun-is-meas},\\
  Lemma~\thref{l:meas-of-fun-def-on-pseudopart},\\
  Lemma~\thref{l:mr-is-alg},\\
  Definition~\thref{d:m-set-of-meas-num-funs},\\
  Lemma~\thref{l:inverse-image-is-meas},\\
  Lemma~\thref{l:m-is-closed-under-finite-part}.

\item[The proof of Lemma~\thref{l:m-is-closed-under-finite-sum-when-defined}] \mbox{}\\
  cites explicitly:\\
  Lemma~\thref{l:m-is-closed-under-add-when-defined}.

\item[The proof of Lemma~\thref{l:m-is-closed-under-mult}] \mbox{}\\
  cites explicitly:\\
  Definition~\thref{d:mult-in-rbar},\\
  Lemma~\thref{l:infinity-prod-prop-in-rbarplus},\\
  Definition~\thref{d:mult-in-rbar-mt},\\
  Lemma~\thref{l:zero-prod-prop-in-rbarplus-mt},\\
  Definition~\thref{d:finite-part},\\
  Lemma~\thref{l:equiv-def-of-sigma-alg},\\
  Definition~\thref{d:borel-sigma-alg},\\
  Definition~\thref{d:meas-fun},\\
  Lemma~\thref{l:const-fun-is-meas},\\
  Lemma~\thref{l:meas-of-fun-def-on-pseudopart},\\
  Lemma~\thref{l:mr-is-alg},\\
  Definition~\thref{d:m-set-of-meas-num-funs},\\
  Lemma~\thref{l:meas-of-num-fun},\\
  Lemma~\thref{l:inverse-image-is-meas},\\
  Lemma~\thref{l:m-is-closed-under-finite-part}.

\item[The proof of Lemma~\thref{l:m-is-closed-under-finite-prod}] \mbox{}\\
  cites explicitly:\\
  Lemma~\thref{l:m-is-closed-under-mult}.

\item[The proof of Lemma~\thref{l:m-is-closed-under-scalar-mult}] \mbox{}\\
  cites explicitly:\\
  Lemma~\thref{l:const-fun-is-meas},\\
  Lemma~\thref{l:m-is-closed-under-mult}.

\item[The proof of Lemma~\thref{l:m-is-closed-under-inf}] \mbox{}\\
  cites explicitly:\\
  Statement(s) from~\cite{cm:lmt:16},\\
  Lemma~\thref{l:equiv-def-of-sigma-alg},\\
  Lemma~\thref{l:meas-of-num-fun}.

\item[The proof of Lemma~\thref{l:m-is-closed-under-sup}] \mbox{}\\
  cites explicitly:\\
  Statement(s) from~\cite{cm:lmt:16},\\
  Lemma~\thref{l:equiv-def-of-sigma-alg},\\
  Lemma~\thref{l:meas-of-num-fun}.

\item[The proof of Lemma~\thref{l:m-is-closed-under-liminf}] \mbox{}\\
  cites explicitly:\\
  Lemma~\thref{l:liminf},\\
  Lemma~\thref{l:m-is-closed-under-inf},\\
  Lemma~\thref{l:m-is-closed-under-sup}.

\item[The proof of Lemma~\thref{l:m-is-closed-under-limsup}] \mbox{}\\
  cites explicitly:\\
  Lemma~\thref{l:limsup},\\
  Lemma~\thref{l:m-is-closed-under-inf},\\
  Lemma~\thref{l:m-is-closed-under-sup}.

\item[The proof of Lemma~\thref{l:m-is-closed-under-limit-when-pointwise-conv}] \mbox{}\\
  cites explicitly:\\
  Lemma~\thref{l:liminf-limsup-and-pointwise-conv},\\
  Lemma~\thref{l:m-is-closed-under-liminf},\\
  Lemma~\thref{l:m-is-closed-under-limsup}.

\item[The proof of Lemma~\thref{l:meas-and-masking}] \mbox{}\\
  cites explicitly:\\
  Definition~\thref{d:mult-in-rbar},\\
  Definition~\thref{d:sigma-alg},\\
  Definition~\thref{d:measurable-space},\\
  Lemma~\thref{l:meas-of-indic-fun},\\
  Lemma~\thref{l:m-and-finite-is-mr},\\
  Lemma~\thref{l:m-is-closed-under-mult}.

\item[The proof of Lemma~\thref{l:meas-of-restr}] \mbox{}\\
  cites explicitly:\\
  Definition~\thref{d:sigma-alg},\\
  Lemma~\thref{l:equiv-def-of-sigma-alg},\\
  Definition~\thref{d:measurable-space},\\
  Definition~\thref{d:meas-fun},\\
  Definition~\thref{d:m-set-of-meas-num-funs}.

\item[The proof of Lemma~\thref{l:meas-of-nonneg-and-nonpos-parts}] \mbox{}\\
  cites explicitly:\\
  Definition~\thref{d:nonneg-and-nonpos-parts},\\
  Lemma~\thref{l:nonneg-and-nonpos-parts-are-nonneg},\\
  Lemma~\thref{l:decomp-into-nonneg-and-nonpos-parts},\\
  Lemma~\thref{l:const-fun-is-meas},\\
  Lemma~\thref{l:m-is-closed-under-add-when-defined},\\
  Lemma~\thref{l:m-is-closed-under-scalar-mult},\\
  Lemma~\thref{l:m-is-closed-under-sup},\\
  Definition~\thref{d:mplus-subset-of-nonneg-meas-num-fun}.

\item[The proof of Lemma~\thref{l:mplus-is-closed-under-finite-part}] \mbox{}\\
  cites explicitly:\\
  Lemma~\thref{l:m-is-closed-under-finite-part},\\
  Definition~\thref{d:mplus-subset-of-nonneg-meas-num-fun}.

\item[The proof of Lemma~\thref{l:m-is-closed-under-abs}] \mbox{}\\
  cites explicitly:\\
  Lemma~\thref{l:abs-in-rbar-is-nonneg},\\
  Lemma~\thref{l:abs-in-rbar-is-cont},\\
  Lemma~\thref{l:cont-is-meas},\\
  Lemma~\thref{l:compat-of-meas-with-comp},\\
  Definition~\thref{d:mr-vector-space-of-meas-num-fun-to-r},\\
  Definition~\thref{d:m-set-of-meas-num-funs},\\
  Definition~\thref{d:mplus-subset-of-nonneg-meas-num-fun}.

\item[The proof of Lemma~\thref{l:mplus-is-closed-under-add}] \mbox{}\\
  cites explicitly:\\
  Lemma~\thref{l:add-in-rbarplus-is-closed},\\
  Lemma~\thref{l:m-is-closed-under-add-when-defined},\\
  Definition~\thref{d:mplus-subset-of-nonneg-meas-num-fun}.

\item[The proof of Lemma~\thref{l:mplus-is-closed-under-mult}] \mbox{}\\
  cites explicitly:\\
  Lemma~\thref{l:mult-in-rbarplus-is-closed-mt},\\
  Lemma~\thref{l:m-is-closed-under-mult},\\
  Definition~\thref{d:mplus-subset-of-nonneg-meas-num-fun}.

\item[The proof of Lemma~\thref{l:mplus-is-closed-under-nonneg-scalar-mult}] \mbox{}\\
  cites explicitly:\\
  Lemma~\thref{l:mult-in-rbarplus-is-closed-mt},\\
  Lemma~\thref{l:m-is-closed-under-scalar-mult},\\
  Definition~\thref{d:mplus-subset-of-nonneg-meas-num-fun}.

\item[The proof of Lemma~\thref{l:mplus-is-closed-under-inf}] \mbox{}\\
  cites explicitly:\\
  Lemma~\thref{l:inf-of-bounded-seq-is-bounded},\\
  Lemma~\thref{l:m-is-closed-under-inf}.

\item[The proof of Lemma~\thref{l:mplus-is-closed-under-sup}] \mbox{}\\
  cites explicitly:\\
  Lemma~\thref{l:sup-of-bounded-seq-is-bounded},\\
  Lemma~\thref{l:m-is-closed-under-sup}.

\item[The proof of Lemma~\thref{l:mplus-is-closed-under-limit-when-pointwise-conv}] \mbox{}\\
  cites explicitly:\\
  Lemma~\thref{l:m-is-closed-under-limit-when-pointwise-conv},\\
  Definition~\thref{d:mplus-subset-of-nonneg-meas-num-fun}.

\item[The proof of Lemma~\thref{l:mplus-is-closed-under-count-sum}] \mbox{}\\
  cites explicitly:\\
  Lemma~\thref{l:series-are-conv-in-rbarplus},\\
  Lemma~\thref{l:mplus-is-closed-under-add},\\
  Lemma~\thref{l:mplus-is-closed-under-limit-when-pointwise-conv}.

\item[The proof of Lemma~\thref{l:meas-of-tensor-prod-of-num-funs}] \mbox{}\\
  cites explicitly:\\
  Lemma~\thref{l:compat-of-meas-with-comp},\\
  Lemma~\thref{l:can-proj-is-meas},\\
  Definition~\thref{d:m-set-of-meas-num-funs},\\
  Lemma~\thref{l:m-is-closed-under-finite-prod},\\
  Definition~\thref{d:tensor-prod-of-num-funs}.

\item[The proof of Lemma~\thref{l:sigma-add-implies-add}] \mbox{}\\
  cites explicitly:\\
  Definition~\thref{d:add-over-meas-space},\\
  Definition~\thref{d:sigma-add-over-meas-space}.

\item[The proof of Lemma~\thref{l:meas-over-count-pseudopart}] \mbox{}\\
  cites explicitly:\\
  Lemma~\thref{l:compat-of-pseudopart-with-inter},\\
  Lemma~\thref{l:equiv-def-of-sigma-alg},\\
  Definition~\thref{d:sigma-add-over-meas-space},\\
  Definition~\thref{d:meas}.

\item[The proof of Lemma~\thref{l:meas-is-monot}] \mbox{}\\
  cites explicitly:\\
  Definition~\thref{d:add-in-rbar},\\
  Lemma~\thref{l:sigma-alg-is-closed-under-set-diff},\\
  Definition~\thref{d:sigma-add-over-meas-space},\\
  Definition~\thref{d:meas}.

\item[The proof of Lemma~\thref{l:meas-satisfies-finite-boole-ineq}] \mbox{}\\
  cites explicitly:\\
  Definition~\thref{d:sigma-alg},\\
  Lemma~\thref{l:sigma-alg-is-closed-under-set-diff},\\
  Definition~\thref{d:measurable-space},\\
  Definition~\thref{d:sigma-add-over-meas-space},\\
  Definition~\thref{d:meas},\\
  Lemma~\thref{l:meas-is-monot}.

\item[The proof of Lemma~\thref{l:meas-is-cont-from-below}] \mbox{}\\
  cites explicitly:\\
  Lemma~\thref{l:part-of-count-union-in-sigma-alg},\\
  Definition~\thref{d:sigma-add-over-meas-space},\\
  Definition~\thref{d:meas},\\
  Lemma~\thref{l:meas-is-monot},\\
  Definition~\thref{d:continuity-from-below}.

\item[The proof of Lemma~\thref{l:meas-is-cont-from-above}] \mbox{}\\
  cites explicitly:\\
  Lemma~\thref{l:equiv-def-of-sigma-alg},\\
  Lemma~\thref{l:sigma-alg-is-closed-under-set-diff},\\
  Lemma~\thref{l:meas-is-monot},\\
  Definition~\thref{d:continuity-from-below},\\
  Lemma~\thref{l:meas-is-cont-from-below},\\
  Definition~\thref{d:continuity-from-above}.

\item[The proof of Lemma~\thref{l:meas-satisfies-boole-ineq}] \mbox{}\\
  cites explicitly:\\
  Statement(s) from~\cite{cm:lmt:16},\\
  Definition~\thref{d:sigma-alg},\\
  Definition~\thref{d:measurable-space},\\
  Definition~\thref{d:meas},\\
  Lemma~\thref{l:meas-satisfies-finite-boole-ineq},\\
  Lemma~\thref{l:meas-is-cont-from-below}.

\item[The proof of Lemma~\thref{l:equiv-def-of-meas}] \mbox{}\\
  cites explicitly:\\
  Definition~\thref{d:sigma-alg},\\
  Definition~\thref{d:measurable-space},\\
  Definition~\thref{d:add-over-meas-space},\\
  Definition~\thref{d:sigma-add-over-meas-space},\\
  Lemma~\thref{l:sigma-add-implies-add},\\
  Definition~\thref{d:meas},\\
  Definition~\thref{d:continuity-from-below},\\
  Lemma~\thref{l:meas-is-cont-from-below}.

\item[The proof of Lemma~\thref{l:finite-meas-is-bounded}] \mbox{}\\
  cites explicitly:\\
  Lemma~\thref{l:equiv-def-of-sigma-alg},\\
  Lemma~\thref{l:meas-is-monot}.

\item[The proof of Lemma~\thref{l:equiv-def-of-sigma-finite-meas}] \mbox{}\\
  cites explicitly:\\
  Definition~\thref{d:sigma-alg},\\
  Definition~\thref{d:measurable-space},\\
  Definition~\thref{d:meas},\\
  Lemma~\thref{l:meas-satisfies-finite-boole-ineq},\\
  Definition~\thref{d:sigma-finite-meas}.

\item[The proof of Lemma~\thref{l:finite-meas-is-sigma-finite}] \mbox{}\\
  cites explicitly:\\
  Definition~\thref{d:finite-meas},\\
  Definition~\thref{d:sigma-finite-meas}.

\item[The proof of Lemma~\thref{l:trace-meas}] \mbox{}\\
  cites explicitly:\\
  Lemma~\thref{l:trace-sigma-alg},\\
  Lemma~\thref{l:meas-of-meas-subspace},\\
  Definition~\thref{d:meas}.

\item[The proof of Lemma~\thref{l:restr-meas}] \mbox{}\\
  cites explicitly:\\
  Lemma~\thref{l:equiv-def-of-sigma-alg},\\
  Definition~\thref{d:meas}.

\item[The proof of Lemma~\thref{l:equiv-def-of-considerable-subset}] \mbox{}\\
  cites explicitly:\\
  Definition~\thref{d:meas},\\
  Definition~\thref{d:negl-subset},\\
  Definition~\thref{d:considerable-subset}.

\item[The proof of Lemma~\thref{l:negl-of-meas-subset}] \mbox{}\\
  cites explicitly:\\
  Definition~\thref{d:negl-subset}.

\item[The proof of Lemma~\thref{l:empty-set-is-negl}] \mbox{}\\
  cites explicitly:\\
  Definition~\thref{d:sigma-alg},\\
  Definition~\thref{d:measurable-space},\\
  Definition~\thref{d:meas},\\
  Lemma~\thref{l:negl-of-meas-subset}.

\item[The proof of Lemma~\thref{l:compat-of-null-meas-with-count-union}] \mbox{}\\
  cites explicitly:\\
  Statement(s) from~\cite{cm:lmt:16},\\
  Definition~\thref{d:sigma-alg},\\
  Definition~\thref{d:measurable-space},\\
  Definition~\thref{d:meas},\\
  Lemma~\thref{l:meas-satisfies-finite-boole-ineq},\\
  Lemma~\thref{l:meas-satisfies-boole-ineq}.

\item[The proof of Lemma~\thref{l:n-is-closed-under-count-union}] \mbox{}\\
  cites explicitly:\\
  Definition~\thref{d:negl-subset},\\
  Lemma~\thref{l:compat-of-null-meas-with-count-union}.

\item[The proof of Lemma~\thref{l:subset-of-negl-is-negl}] \mbox{}\\
  cites explicitly:\\
  Definition~\thref{d:negl-subset}.

\item[The proof of Lemma~\thref{l:everywhere-implies-almost-everywhere}] \mbox{}\\
  cites explicitly:\\
  Lemma~\thref{l:empty-set-is-negl},\\
  Definition~\thref{d:prop-almost-satisfied}.

\item[The proof of Lemma~\thref{l:everywhere-implies-almost-everywhere-for-almost-the-same}] \mbox{}\\
  cites explicitly:\\
  Lemma~\thref{l:subset-of-negl-is-negl},\\
  Definition~\thref{d:prop-almost-satisfied}.

\item[The proof of Lemma~\thref{l:ext-almost-modus-ponens}] \mbox{}\\
  cites explicitly:\\
  Lemma~\thref{l:n-is-closed-under-count-union},\\
  Lemma~\thref{l:subset-of-negl-is-negl},\\
  Definition~\thref{d:prop-almost-satisfied}.

\item[The proof of Lemma~\thref{l:almost-modus-ponens}] \mbox{}\\
  cites explicitly:\\
  Lemma~\thref{l:everywhere-implies-almost-everywhere},\\
  Lemma~\thref{l:ext-almost-modus-ponens}.

\item[The proof of Lemma~\thref{l:compat-of-almost-bin-rel-with-refl}] \mbox{}\\
  cites explicitly:\\
  Definition~\thref{d:prop-almost-satisfied},\\
  Lemma~\thref{l:everywhere-implies-almost-everywhere},\\
  Lemma~\thref{l:everywhere-implies-almost-everywhere-for-almost-the-same},\\
  Definition~\thref{d:almost-definition},\\
  Definition~\thref{d:almost-bin-rel}.

\item[The proof of Lemma~\thref{l:compat-of-almost-bin-rel-with-sym}] \mbox{}\\
  cites explicitly:\\
  Definition~\thref{d:prop-almost-satisfied},\\
  Lemma~\thref{l:everywhere-implies-almost-everywhere-for-almost-the-same},\\
  Lemma~\thref{l:almost-modus-ponens},\\
  Definition~\thref{d:almost-definition},\\
  Definition~\thref{d:almost-bin-rel}.

\item[The proof of Lemma~\thref{l:compat-of-almost-bin-rel-with-antisym}] \mbox{}\\
  cites explicitly:\\
  Lemma~\thref{l:n-is-closed-under-count-union},\\
  Lemma~\thref{l:subset-of-negl-is-negl},\\
  Definition~\thref{d:prop-almost-satisfied},\\
  Lemma~\thref{l:everywhere-implies-almost-everywhere-for-almost-the-same},\\
  Definition~\thref{d:almost-definition},\\
  Definition~\thref{d:almost-bin-rel}.

\item[The proof of Lemma~\thref{l:compat-of-almost-bin-rel-with-trans}] \mbox{}\\
  cites explicitly:\\
  Lemma~\thref{l:n-is-closed-under-count-union},\\
  Lemma~\thref{l:subset-of-negl-is-negl},\\
  Definition~\thref{d:prop-almost-satisfied},\\
  Lemma~\thref{l:everywhere-implies-almost-everywhere-for-almost-the-same},\\
  Definition~\thref{d:almost-definition},\\
  Definition~\thref{d:almost-bin-rel}.

\item[The proof of Lemma~\thref{l:almost-equiv-is-equiv-rel}] \mbox{}\\
  cites explicitly:\\
  Lemma~\thref{l:compat-of-almost-bin-rel-with-refl},\\
  Lemma~\thref{l:compat-of-almost-bin-rel-with-sym},\\
  Lemma~\thref{l:compat-of-almost-bin-rel-with-trans}.

\item[The proof of Lemma~\thref{l:almost-eq-is-equiv-rel}] \mbox{}\\
  cites explicitly:\\
  Lemma~\thref{l:almost-equiv-is-equiv-rel}.

\item[The proof of Lemma~\thref{l:almost-order-is-order-rel}] \mbox{}\\
  cites explicitly:\\
  Lemma~\thref{l:compat-of-almost-bin-rel-with-refl},\\
  Lemma~\thref{l:compat-of-almost-bin-rel-with-antisym},\\
  Lemma~\thref{l:compat-of-almost-bin-rel-with-trans}.

\item[The proof of Lemma~\thref{l:compat-of-almost-bin-rel-with-op}] \mbox{}\\
  cites explicitly:\\
  Lemma~\thref{l:n-is-closed-under-count-union},\\
  Definition~\thref{d:prop-almost-satisfied},\\
  Lemma~\thref{l:everywhere-implies-almost-everywhere-for-almost-the-same},\\
  Definition~\thref{d:almost-definition},\\
  Definition~\thref{d:almost-bin-rel}.

\item[The proof of Lemma~\thref{l:compat-of-almost-eq-with-op}] \mbox{}\\
  cites explicitly:\\
  Lemma~\thref{l:compat-of-almost-bin-rel-with-op}.

\item[The proof of Lemma~\thref{l:compat-of-almost-ineq-with-op}] \mbox{}\\
  cites explicitly:\\
  Lemma~\thref{l:compat-of-almost-bin-rel-with-op}.

\item[The proof of Lemma~\thref{l:definiteness-implies-almost-definiteness}] \mbox{}\\
  cites explicitly:\\
  Lemma~\thref{l:everywhere-implies-almost-everywhere-for-almost-the-same},\\
  Lemma~\thref{l:almost-modus-ponens}.

\item[The proof of Lemma~\thref{l:uniq-of-meas-ext-from-p-syst}] \mbox{}\\
  cites explicitly:\\
  Definition~\thref{d:pseudopart},\\
  Definition~\thref{d:p-syst},\\
  Lemma~\thref{l:p-syst-gen-is-idem},\\
  Lemma~\thref{l:equiv-def-of-l-syst},\\
  Lemma~\thref{l:equiv-def-of-sigma-alg},\\
  Lemma~\thref{l:gen-sigma-alg-is-min},\\
  Lemma~\thref{l:usage-of-dynkin-pi-lambda-th},\\
  Definition~\thref{d:sigma-add-over-meas-space},\\
  Definition~\thref{d:meas},\\
  Lemma~\thref{l:meas-is-monot},\\
  Definition~\thref{d:continuity-from-below},\\
  Lemma~\thref{l:meas-is-cont-from-below}.

\item[The proof of Lemma~\thref{l:trivial-meas}] \mbox{}\\
  cites explicitly:\\
  Definition~\thref{d:meas}.

\item[The proof of Lemma~\thref{l:equiv-def-of-trivial-meas}] \mbox{}\\
  cites explicitly:\\
  Definition~\thref{d:meas},\\
  Lemma~\thref{l:meas-is-monot},\\
  Lemma~\thref{l:trivial-meas}.

\item[The proof of Lemma~\thref{l:count-meas}] \mbox{}\\
  cites explicitly:\\
  Definition~\thref{d:sigma-alg},\\
  Definition~\thref{d:measurable-space},\\
  Definition~\thref{d:sigma-add-over-meas-space},\\
  Definition~\thref{d:meas}.

\item[The proof of Lemma~\thref{l:finiteness-of-count-meas}] \mbox{}\\
  cites explicitly:\\
  Lemma~\thref{l:equiv-def-of-sigma-alg},\\
  Definition~\thref{d:finite-meas},\\
  Lemma~\thref{l:count-meas}.

\item[The proof of Lemma~\thref{l:sigma-finiteness-of-count-meas}] \mbox{}\\
  cites explicitly:\\
  Definition~\thref{d:sigma-alg},\\
  Definition~\thref{d:measurable-space},\\
  Definition~\thref{d:meas},\\
  Definition~\thref{d:sigma-finite-meas},\\
  Lemma~\thref{l:equiv-def-of-sigma-finite-meas},\\
  Lemma~\thref{l:count-meas}.

\item[The proof of Lemma~\thref{l:equiv-def-of-dirac-meas}] \mbox{}\\
  cites explicitly:\\
  Lemma~\thref{l:count-meas},\\
  Definition~\thref{d:dirac-meas}.

\item[The proof of Lemma~\thref{l:dirac-meas-is-finite}] \mbox{}\\
  cites explicitly:\\
  Lemma~\thref{l:count-meas},\\
  Lemma~\thref{l:finiteness-of-count-meas},\\
  Definition~\thref{d:dirac-meas}.

\item[The proof of Lemma~\thref{l:summability-on-summability-domain}] \mbox{}\\
  cites explicitly:\\
  Definition~\thref{d:add-in-rbar},\\
  Definition~\thref{d:summability-domain}.

\item[The proof of Lemma~\thref{l:meas-of-summability-domain}] \mbox{}\\
  cites explicitly:\\
  Lemma~\thref{l:equiv-def-of-sigma-alg},\\
  Lemma~\thref{l:inverse-image-is-meas}.

\item[The proof of Lemma~\thref{l:negl-of-summability-domain}] \mbox{}\\
  cites explicitly:\\
  Definition~\thref{d:prop-almost-satisfied},\\
  Lemma~\thref{l:summability-on-summability-domain}.

\item[The proof of Lemma~\thref{l:almost-sum}] \mbox{}\\
  cites explicitly:\\
  Definition~\thref{d:add-in-rbar},\\
  Lemma~\thref{l:meas-of-indic-fun},\\
  Lemma~\thref{l:m-is-closed-under-add-when-defined},\\
  Lemma~\thref{l:m-is-closed-under-mult},\\
  Definition~\thref{d:prop-almost-satisfied},\\
  Lemma~\thref{l:meas-of-summability-domain},\\
  Lemma~\thref{l:negl-of-summability-domain}.

\item[The proof of Lemma~\thref{l:compat-of-almost-sum-with-almost-eq}] \mbox{}\\
  cites explicitly:\\
  Lemma~\thref{l:m-is-closed-under-add-when-defined},\\
  Lemma~\thref{l:empty-set-is-negl},\\
  Lemma~\thref{l:almost-eq-is-equiv-rel},\\
  Lemma~\thref{l:compat-of-almost-eq-with-op},\\
  Lemma~\thref{l:negl-of-summability-domain},\\
  Lemma~\thref{l:almost-sum}.

\item[The proof of Lemma~\thref{l:almost-sum-is-sum}] \mbox{}\\
  cites explicitly:\\
  Definition~\thref{d:summability-domain},\\
  Lemma~\thref{l:almost-sum}.

\item[The proof of Lemma~\thref{l:abs-is-almost-definite}] \mbox{}\\
  cites explicitly:\\
  Lemma~\thref{l:abs-in-rbar-is-definite},\\
  Lemma~\thref{l:compat-of-almost-eq-with-op},\\
  Lemma~\thref{l:definiteness-implies-almost-definiteness}.

\item[The proof of Lemma~\thref{l:masking-almost-nowhere}] \mbox{}\\
  cites explicitly:\\
  Definition~\thref{d:sigma-alg},\\
  Definition~\thref{d:measurable-space},\\
  Definition~\thref{d:meas},\\
  Definition~\thref{d:negl-subset},\\
  Definition~\thref{d:prop-almost-satisfied}.

\item[The proof of Lemma~\thref{l:finite-nonneg-part}] \mbox{}\\
  cites explicitly:\\
  Definition~\thref{d:finite-part},\\
  Lemma~\thref{l:equiv-def-of-nonneg-and-nonpos-parts},\\
  Definition~\thref{d:sigma-alg},\\
  Definition~\thref{d:measurable-space},\\
  Lemma~\thref{l:meas-and-masking},\\
  Lemma~\thref{l:meas-of-nonneg-and-nonpos-parts},\\
  Lemma~\thref{l:mplus-is-closed-under-finite-part},\\
  Definition~\thref{d:meas},\\
  Lemma~\thref{l:meas-satisfies-finite-boole-ineq},\\
  Lemma~\thref{l:negl-of-meas-subset},\\
  Definition~\thref{d:prop-almost-satisfied},\\
  Lemma~\thref{l:masking-almost-nowhere}.

\item[The proof of Lemma~\thref{l:len-is-nonneg}] \mbox{}\\
  cites explicitly:\\
  Definition~\thref{d:len-of-int}.

\item[The proof of Lemma~\thref{l:len-is-hom}] \mbox{}\\
  cites explicitly:\\
  Definition~\thref{d:len-of-int}.

\item[The proof of Lemma~\thref{l:len-of-partition}] \mbox{}\\
  cites explicitly:\\
  Definition~\thref{d:len-of-int},\\
  Lemma~\thref{l:len-is-hom}.

\item[The proof of Lemma~\thref{l:set-of-count-cover-with-bounded-open-int-is-nonempty}] \mbox{}\\
  cites explicitly:\\
  Definition~\thref{d:set-of-count-cover-with-bounded-open-int}.

\item[The proof of Lemma~\thref{l:lambda-star-is-nonneg}] \mbox{}\\
  cites explicitly:\\
  Lemma~\thref{l:len-is-nonneg}.

\item[The proof of Lemma~\thref{l:lambda-star-is-hom}] \mbox{}\\
  cites explicitly:\\
  Statement(s) from~\cite{cm:lmt:16},\\
  Definition~\thref{d:len-of-int},\\
  Definition~\thref{d:lambda-star-lebesgue-meas-cand},\\
  Lemma~\thref{l:lambda-star-is-nonneg}.

\item[The proof of Lemma~\thref{l:lambda-star-is-monot}] \mbox{}\\
  cites explicitly:\\
  Statement(s) from~\cite{cm:lmt:16},\\
  Definition~\thref{d:set-of-count-cover-with-bounded-open-int},\\
  Definition~\thref{d:lambda-star-lebesgue-meas-cand}.

\item[The proof of Lemma~\thref{l:lambda-star-is-sigma-subadd}] \mbox{}\\
  cites explicitly:\\
  Statement(s) from~\cite{cm:lmt:16},\\
  Lemma~\thref{l:def-of-double-count-union},\\
  Lemma~\thref{l:double-count-union},\\
  Lemma~\thref{l:def-of-double-series-in-rbarplus},\\
  Lemma~\thref{l:double-series-in-rbarplus},\\
  Definition~\thref{d:set-of-count-cover-with-bounded-open-int},\\
  Definition~\thref{d:lambda-star-lebesgue-meas-cand}.

\item[The proof of Lemma~\thref{l:lambda-star-gen-len-of-int}] \mbox{}\\
  cites explicitly:\\
  Statement(s) from~\cite{cm:lmt:16},\\
  Lemma~\thref{l:finite-cover-of-compact-int},\\
  Definition~\thref{d:len-of-int},\\
  Definition~\thref{d:set-of-count-cover-with-bounded-open-int},\\
  Definition~\thref{d:lambda-star-lebesgue-meas-cand},\\
  Lemma~\thref{l:lambda-star-is-hom},\\
  Lemma~\thref{l:lambda-star-is-monot}.

\item[The proof of Lemma~\thref{l:equiv-def-of-l}] \mbox{}\\
  cites explicitly:\\
  Lemma~\thref{l:compat-of-pseudopart-with-inter},\\
  Lemma~\thref{l:order-in-rbar-is-total},\\
  Lemma~\thref{l:lambda-star-is-sigma-subadd},\\
  Definition~\thref{d:l-lebesgue-sigma-alg}.

\item[The proof of Lemma~\thref{l:l-is-closed-under-compl}] \mbox{}\\
  cites explicitly:\\
  Lemma~\thref{l:add-in-rbarplus-is-comm},\\
  Definition~\thref{d:l-lebesgue-sigma-alg}.

\item[The proof of Lemma~\thref{l:l-is-closed-under-finite-union}] \mbox{}\\
  cites explicitly:\\
  Lemma~\thref{l:lambda-star-is-sigma-subadd},\\
  Definition~\thref{d:l-lebesgue-sigma-alg},\\
  Lemma~\thref{l:equiv-def-of-l}.

\item[The proof of Lemma~\thref{l:l-is-closed-under-finite-inter}] \mbox{}\\
  cites explicitly:\\
  Lemma~\thref{l:l-is-closed-under-compl},\\
  Lemma~\thref{l:l-is-closed-under-finite-union}.

\item[The proof of Lemma~\thref{l:l-is-set-alg}] \mbox{}\\
  cites explicitly:\\
  Lemma~\thref{l:inter-set-diff-equiv},\\
  Lemma~\thref{l:other-equiv-def-of-set-alg},\\
  Lemma~\thref{l:lambda-star-is-hom},\\
  Lemma~\thref{l:l-is-closed-under-compl},\\
  Lemma~\thref{l:l-is-closed-under-finite-inter}.

\item[The proof of Lemma~\thref{l:lambda-star-is-add-on-l}] \mbox{}\\
  cites explicitly:\\
  Lemma~\thref{l:compat-of-pseudopart-with-inter},\\
  Definition~\thref{d:l-lebesgue-sigma-alg}.

\item[The proof of Lemma~\thref{l:lambda-star-is-sigma-add-on-l}] \mbox{}\\
  cites explicitly:\\
  Definition~\thref{d:sigma-add-over-meas-space},\\
  Lemma~\thref{l:lambda-star-is-monot},\\
  Lemma~\thref{l:lambda-star-is-sigma-subadd},\\
  Lemma~\thref{l:lambda-star-is-add-on-l}.

\item[The proof of Lemma~\thref{l:part-of-count-union-in-l}] \mbox{}\\
  cites explicitly:\\
  Lemma~\thref{l:part-of-count-union-in-set-alg},\\
  Lemma~\thref{l:l-is-set-alg}.

\item[The proof of Lemma~\thref{l:l-is-closed-under-count-union}] \mbox{}\\
  cites explicitly:\\
  Lemma~\thref{l:lambda-star-is-monot},\\
  Lemma~\thref{l:lambda-star-is-sigma-subadd},\\
  Definition~\thref{d:l-lebesgue-sigma-alg},\\
  Lemma~\thref{l:equiv-def-of-l},\\
  Lemma~\thref{l:l-is-closed-under-finite-union},\\
  Lemma~\thref{l:lambda-star-is-add-on-l},\\
  Lemma~\thref{l:part-of-count-union-in-l}.

\item[The proof of Lemma~\thref{l:rays-are-lebesgue-meas}] \mbox{}\\
  cites explicitly:\\
  Statement(s) from~\cite{cm:lmt:16},\\
  Lemma~\thref{l:int-are-closed-under-finite-inter},\\
  Definition~\thref{d:lambda-star-lebesgue-meas-cand},\\
  Lemma~\thref{l:lambda-star-is-monot},\\
  Lemma~\thref{l:lambda-star-is-sigma-subadd},\\
  Lemma~\thref{l:lambda-star-gen-len-of-int},\\
  Lemma~\thref{l:equiv-def-of-l},\\
  Lemma~\thref{l:l-is-closed-under-compl},\\
  Lemma~\thref{l:l-is-closed-under-count-union}.

\item[The proof of Lemma~\thref{l:int-are-lebesgue-meas}] \mbox{}\\
  cites explicitly:\\
  Lemma~\thref{l:l-is-closed-under-finite-inter},\\
  Lemma~\thref{l:rays-are-lebesgue-meas}.

\item[The proof of Lemma~\thref{l:l-is-sigma-alg}] \mbox{}\\
  cites explicitly:\\
  Definition~\thref{d:set-alg},\\
  Definition~\thref{d:sigma-alg},\\
  Lemma~\thref{l:l-is-set-alg},\\
  Lemma~\thref{l:l-is-closed-under-count-union}.

\item[The proof of Lemma~\thref{l:lambda-star-is-meas-on-l}] \mbox{}\\
  cites explicitly:\\
  Definition~\thref{d:meas},\\
  Lemma~\thref{l:lambda-star-is-nonneg},\\
  Lemma~\thref{l:lambda-star-is-hom},\\
  Lemma~\thref{l:lambda-star-is-sigma-add-on-l},\\
  Lemma~\thref{l:l-is-sigma-alg}.

\item[The proof of Lemma~\thref{l:br-is-sub-sigma-alg-of-l}] \mbox{}\\
  cites explicitly:\\
  Lemma~\thref{l:sigma-alg-gen-is-monot},\\
  Lemma~\thref{l:borel-sigma-alg-of-r},\\
  Lemma~\thref{l:int-are-lebesgue-meas}.

\item[The proof of Lemma~\thref{l:lambda-star-is-meas-on-br}] \mbox{}\\
  cites explicitly:\\
  Lemma~\thref{l:lambda-star-is-meas-on-l},\\
  Lemma~\thref{l:br-is-sub-sigma-alg-of-l}.

\item[The proof of Theorem~\thref{t:caratheodory-lebesgue-meas-on-r}] \mbox{}\\
  cites explicitly:\\
  Lemma~\thref{l:int-are-closed-under-finite-inter},\\
  Definition~\thref{d:p-syst},\\
  Lemma~\thref{l:gen-sigma-alg-is-min},\\
  Lemma~\thref{l:borel-sigma-alg-of-r},\\
  Lemma~\thref{l:meas-is-cont-from-above},\\
  Lemma~\thref{l:uniq-of-meas-ext-from-p-syst},\\
  Lemma~\thref{l:lambda-star-gen-len-of-int},\\
  Lemma~\thref{l:lambda-star-is-meas-on-br}.

\item[The proof of Lemma~\thref{l:lebesgue-meas-gen-len-of-int}] \mbox{}\\
  cites explicitly:\\
  Lemma~\thref{l:lambda-star-gen-len-of-int},\\
  Theorem~\thref{t:caratheodory-lebesgue-meas-on-r}.

\item[The proof of Lemma~\thref{l:lebesgue-meas-is-sigma-finite}] \mbox{}\\
  cites explicitly:\\
  Definition~\thref{d:sigma-finite-meas},\\
  Lemma~\thref{l:lambda-star-gen-len-of-int},\\
  Theorem~\thref{t:caratheodory-lebesgue-meas-on-r}.

\item[The proof of Lemma~\thref{l:lebesgue-meas-is-diffuse}] \mbox{}\\
  cites explicitly:\\
  Definition~\thref{d:diffuse-meas},\\
  Lemma~\thref{l:lambda-star-gen-len-of-int},\\
  Theorem~\thref{t:caratheodory-lebesgue-meas-on-r}.

\item[The proof of Lemma~\thref{l:indic-and-support-are-each-other-inverse}] \mbox{}\\
  cites explicitly:\\
  Definition~\thref{d:if-set-of-meas-indic-funs}.

\item[The proof of Lemma~\thref{l:if-is-meas}] \mbox{}\\
  cites explicitly:\\
  Lemma~\thref{l:meas-of-indic-fun},\\
  Lemma~\thref{l:m-and-finite-is-mr},\\
  Definition~\thref{d:mplus-subset-of-nonneg-meas-num-fun},\\
  Definition~\thref{d:if-set-of-meas-indic-funs}.

\item[The proof of Lemma~\thref{l:if-is-sigma-add}] \mbox{}\\
  cites explicitly:\\
  Definition~\thref{d:sigma-alg},\\
  Definition~\thref{d:measurable-space},\\
  Definition~\thref{d:if-set-of-meas-indic-funs}.

\item[The proof of Lemma~\thref{l:if-is-closed-under-mult}] \mbox{}\\
  cites explicitly:\\
  Lemma~\thref{l:equiv-def-of-sigma-alg},\\
  Definition~\thref{d:if-set-of-meas-indic-funs}.

\item[The proof of Lemma~\thref{l:if-is-closed-under-ext-by-zero}] \mbox{}\\
  cites explicitly:\\
  Lemma~\thref{l:restr-is-mask},\\
  Definition~\thref{d:if-set-of-meas-indic-funs}.

\item[The proof of Lemma~\thref{l:if-is-closed-under-restr}] \mbox{}\\
  cites explicitly:\\
  Lemma~\thref{l:meas-of-meas-subspace},\\
  Definition~\thref{d:if-set-of-meas-indic-funs}.

\item[The proof of Lemma~\thref{l:equiv-def-of-int-in-if}] \mbox{}\\
  cites explicitly:\\
  Lemma~\thref{l:indic-and-support-are-each-other-inverse},\\
  Definition~\thref{d:int-in-if}.

\item[The proof of Lemma~\thref{l:int-in-if-is-add}] \mbox{}\\
  cites explicitly:\\
  Definition~\thref{d:add-over-meas-space},\\
  Lemma~\thref{l:equiv-def-of-meas},\\
  Lemma~\thref{l:if-is-sigma-add},\\
  Lemma~\thref{l:equiv-def-of-int-in-if}.

\item[The proof of Lemma~\thref{l:int-in-if-over-subset}] \mbox{}\\
  cites explicitly:\\
  Lemma~\thref{l:restr-is-mask},\\
  Lemma~\thref{l:trace-meas},\\
  Lemma~\thref{l:indic-and-support-are-each-other-inverse},\\
  Lemma~\thref{l:if-is-closed-under-ext-by-zero},\\
  Lemma~\thref{l:if-is-closed-under-restr},\\
  Lemma~\thref{l:equiv-def-of-int-in-if}.

\item[The proof of Lemma~\thref{l:int-in-if-over-subset-is-add}] \mbox{}\\
  cites explicitly:\\
  Definition~\thref{d:sigma-alg},\\
  Definition~\thref{d:measurable-space},\\
  Definition~\thref{d:meas},\\
  Lemma~\thref{l:indic-and-support-are-each-other-inverse},\\
  Lemma~\thref{l:if-is-sigma-add},\\
  Lemma~\thref{l:if-is-closed-under-mult},\\
  Lemma~\thref{l:int-in-if-is-add},\\
  Lemma~\thref{l:int-in-if-over-subset}.

\item[The proof of Lemma~\thref{l:int-in-if-for-count-meas}] \mbox{}\\
  cites explicitly:\\
  Lemma~\thref{l:count-meas},\\
  Lemma~\thref{l:indic-and-support-are-each-other-inverse},\\
  Lemma~\thref{l:equiv-def-of-int-in-if}.

\item[The proof of Lemma~\thref{l:sf-simple-repr}] \mbox{}\\
  cites explicitly:\\
  Definition~\thref{d:if-set-of-meas-indic-funs},\\
  Lemma~\thref{l:indic-and-support-are-each-other-inverse},\\
  Definition~\thref{d:sf-vector-space-of-simple-funs}.

\item[The proof of Lemma~\thref{l:sf-can-repr}] \mbox{}\\
  cites explicitly:\\
  Lemma~\thref{l:equiv-def-of-sigma-alg},\\
  Lemma~\thref{l:sf-simple-repr}.

\item[The proof of Lemma~\thref{l:sf-disj-repr}] \mbox{}\\
  cites explicitly:\\
  Lemma~\thref{l:sf-simple-repr},\\
  Lemma~\thref{l:sf-can-repr}.

\item[The proof of Lemma~\thref{l:sf-disj-repr-is-subpart-of-can-repr}] \mbox{}\\
  cites explicitly:\\
  Lemma~\thref{l:sf-can-repr},\\
  Lemma~\thref{l:sf-disj-repr}.

\item[The proof of Lemma~\thref{l:sf-is-alg-over-r}] \mbox{}\\
  cites explicitly:\\
  Statement(s) from~\cite{cm:lmt:16},\\
  Lemma~\thref{l:subspace-and-closed-under-mult-is-subalg},\\
  Lemma~\thref{l:equiv-def-of-sigma-alg},\\
  Lemma~\thref{l:if-is-sigma-add},\\
  Lemma~\thref{l:if-is-closed-under-mult},\\
  Definition~\thref{d:sf-vector-space-of-simple-funs},\\
  Lemma~\thref{l:sf-simple-repr},\\
  Lemma~\thref{l:sf-disj-repr}.

\item[The proof of Lemma~\thref{l:sf-is-meas}] \mbox{}\\
  cites explicitly:\\
  Lemma~\thref{l:meas-of-indic-fun},\\
  Lemma~\thref{l:mr-is-alg},\\
  Lemma~\thref{l:m-and-finite-is-mr},\\
  Definition~\thref{d:sf-vector-space-of-simple-funs}.

\item[The proof of Lemma~\thref{l:sf-is-closed-under-ext-by-zero}] \mbox{}\\
  cites explicitly:\\
  Lemma~\thref{l:if-is-closed-under-ext-by-zero},\\
  Definition~\thref{d:sf-vector-space-of-simple-funs}.

\item[The proof of Lemma~\thref{l:sf-is-closed-under-restr}] \mbox{}\\
  cites explicitly:\\
  Lemma~\thref{l:if-is-closed-under-restr},\\
  Definition~\thref{d:sf-vector-space-of-simple-funs}.

\item[The proof of Lemma~\thref{l:sfplus-disj-repr}] \mbox{}\\
  cites explicitly:\\
  Lemma~\thref{l:sf-disj-repr},\\
  Definition~\thref{d:sfplus-subset-of-nonneg-simple-funs}.

\item[The proof of Lemma~\thref{l:sfplus-can-repr}] \mbox{}\\
  cites explicitly:\\
  Lemma~\thref{l:sf-can-repr},\\
  Definition~\thref{d:sfplus-subset-of-nonneg-simple-funs}.

\item[The proof of Lemma~\thref{l:sfplus-disj-repr-is-subpart-of-can-repr}] \mbox{}\\
  cites explicitly:\\
  Lemma~\thref{l:sf-disj-repr-is-subpart-of-can-repr},\\
  Lemma~\thref{l:sfplus-disj-repr},\\
  Lemma~\thref{l:sfplus-can-repr}.

\item[The proof of Lemma~\thref{l:sfplus-simple-repr}] \mbox{}\\
  cites explicitly:\\
  Lemma~\thref{l:sf-simple-repr},\\
  Definition~\thref{d:sfplus-subset-of-nonneg-simple-funs},\\
  Lemma~\thref{l:sfplus-can-repr}.

\item[The proof of Lemma~\thref{l:sfplus-is-closed-under-pos-alg-ops}] \mbox{}\\
  cites explicitly:\\
  Definition~\thref{d:alg-over-a-field},\\
  Lemma~\thref{l:sf-is-alg-over-r},\\
  Definition~\thref{d:sfplus-subset-of-nonneg-simple-funs},\\
  Lemma~\thref{l:sfplus-disj-repr}.

\item[The proof of Lemma~\thref{l:sfplus-is-meas}] \mbox{}\\
  cites explicitly:\\
  Definition~\thref{d:mplus-subset-of-nonneg-meas-num-fun},\\
  Definition~\thref{d:if-set-of-meas-indic-funs},\\
  Lemma~\thref{l:sf-is-meas},\\
  Definition~\thref{d:sfplus-subset-of-nonneg-simple-funs}.

\item[The proof of Lemma~\thref{l:int-in-sfplus}] \mbox{}\\
  cites explicitly:\\
  Lemma~\thref{l:add-in-rbarplus-is-closed},\\
  Lemma~\thref{l:mult-in-rbarplus-is-closed-mt},\\
  Definition~\thref{d:meas},\\
  Lemma~\thref{l:sfplus-can-repr}.

\item[The proof of Lemma~\thref{l:int-in-sfplus-gen-int-in-if}] \mbox{}\\
  cites explicitly:\\
  Definition~\thref{d:if-set-of-meas-indic-funs},\\
  Lemma~\thref{l:equiv-def-of-int-in-if},\\
  Lemma~\thref{l:sfplus-simple-repr},\\
  Lemma~\thref{l:int-in-sfplus}.

\item[The proof of Lemma~\thref{l:equiv-def-of-int-in-sfplus-disj}] \mbox{}\\
  cites explicitly:\\
  Definition~\thref{d:sigma-add-over-meas-space},\\
  Definition~\thref{d:meas},\\
  Lemma~\thref{l:sfplus-can-repr},\\
  Lemma~\thref{l:sfplus-disj-repr-is-subpart-of-can-repr},\\
  Lemma~\thref{l:int-in-sfplus}.

\item[The proof of Lemma~\thref{l:int-in-sfplus-is-add}] \mbox{}\\
  cites explicitly:\\
  Lemma~\thref{l:meas-over-count-pseudopart},\\
  Lemma~\thref{l:sfplus-disj-repr},\\
  Lemma~\thref{l:sfplus-is-closed-under-pos-alg-ops},\\
  Lemma~\thref{l:equiv-def-of-int-in-sfplus-disj}.

\item[The proof of Lemma~\thref{l:decomp-of-meas-in-sfplus}] \mbox{}\\
  cites explicitly:\\
  Lemma~\thref{l:some-borel-subsets},\\
  Lemma~\thref{l:inverse-image-is-meas-in-r},\\
  Lemma~\thref{l:meas-over-count-pseudopart},\\
  Lemma~\thref{l:sf-can-repr},\\
  Lemma~\thref{l:sfplus-is-meas}.

\item[The proof of Lemma~\thref{l:change-of-variable-in-sum-in-sfplus}] \mbox{}\\
  cites explicitly:\\
  Lemma~\thref{l:equiv-def-of-sigma-alg},\\
  Lemma~\thref{l:inverse-image-is-meas-in-r},\\
  Definition~\thref{d:m-set-of-meas-num-funs},\\
  Definition~\thref{d:meas},\\
  Lemma~\thref{l:sfplus-is-closed-under-pos-alg-ops},\\
  Lemma~\thref{l:sfplus-is-meas}.

\item[The proof of Lemma~\thref{l:int-in-sfplus-is-add-alt-proof}] \mbox{}\\
  cites explicitly:\\
  Lemma~\thref{l:add-in-rbarplus-is-assoc},\\
  Lemma~\thref{l:add-in-rbarplus-is-comm},\\
  Lemma~\thref{l:mult-in-rbarplus-is-distr-over-add-mt},\\
  Lemma~\thref{l:equiv-def-of-sigma-alg},\\
  Lemma~\thref{l:inverse-image-is-meas-in-r},\\
  Definition~\thref{d:m-set-of-meas-num-funs},\\
  Lemma~\thref{l:sfplus-is-closed-under-pos-alg-ops},\\
  Lemma~\thref{l:sfplus-is-meas},\\
  Lemma~\thref{l:int-in-sfplus},\\
  Lemma~\thref{l:decomp-of-meas-in-sfplus},\\
  Lemma~\thref{l:change-of-variable-in-sum-in-sfplus}.

\item[The proof of Lemma~\thref{l:int-in-sfplus-is-pos-lin}] \mbox{}\\
  cites explicitly:\\
  Definition~\thref{d:mult-in-rbar},\\
  Lemma~\thref{l:zero-prod-prop-in-rbarplus-mt},\\
  Definition~\thref{d:meas},\\
  Lemma~\thref{l:sfplus-is-closed-under-pos-alg-ops},\\
  Lemma~\thref{l:int-in-sfplus},\\
  Lemma~\thref{l:int-in-sfplus-gen-int-in-if},\\
  Lemma~\thref{l:int-in-sfplus-is-add},\\
  Lemma~\thref{l:int-in-sfplus-is-add-alt-proof}.

\item[The proof of Lemma~\thref{l:equiv-def-of-int-in-sfplus-simple}] \mbox{}\\
  cites explicitly:\\
  Lemma~\thref{l:sfplus-simple-repr},\\
  Lemma~\thref{l:int-in-sfplus-gen-int-in-if},\\
  Lemma~\thref{l:int-in-sfplus-is-pos-lin}.

\item[The proof of Lemma~\thref{l:int-in-sfplus-is-monot}] \mbox{}\\
  cites explicitly:\\
  Statement(s) from~\cite{cm:lmt:16},\\
  Definition~\thref{d:alg-over-a-field},\\
  Lemma~\thref{l:sf-is-alg-over-r},\\
  Definition~\thref{d:sfplus-subset-of-nonneg-simple-funs},\\
  Lemma~\thref{l:int-in-sfplus},\\
  Lemma~\thref{l:int-in-sfplus-is-pos-lin}.

\item[The proof of Lemma~\thref{l:int-in-sfplus-is-cont}] \mbox{}\\
  cites explicitly:\\
  Statement(s) from~\cite{cm:lmt:16},\\
  Lemma~\thref{l:int-in-sfplus-is-monot}.

\item[The proof of Lemma~\thref{l:int-in-sfplus-over-subset}] \mbox{}\\
  cites explicitly:\\
  Lemma~\thref{l:int-in-if-over-subset},\\
  Definition~\thref{d:sf-vector-space-of-simple-funs},\\
  Lemma~\thref{l:sf-is-closed-under-ext-by-zero},\\
  Lemma~\thref{l:sf-is-closed-under-restr},\\
  Lemma~\thref{l:int-in-sfplus-gen-int-in-if},\\
  Lemma~\thref{l:int-in-sfplus-is-pos-lin}.

\item[The proof of Lemma~\thref{l:int-in-sfplus-over-subset-is-add}] \mbox{}\\
  cites explicitly:\\
  Definition~\thref{d:if-set-of-meas-indic-funs},\\
  Lemma~\thref{l:if-is-sigma-add},\\
  Lemma~\thref{l:if-is-closed-under-mult},\\
  Definition~\thref{d:sf-vector-space-of-simple-funs},\\
  Definition~\thref{d:sfplus-subset-of-nonneg-simple-funs},\\
  Lemma~\thref{l:int-in-sfplus-over-subset}.

\item[The proof of Lemma~\thref{l:int-in-sfplus-for-count-meas}] \mbox{}\\
  cites explicitly:\\
  Lemma~\thref{l:int-in-if-for-count-meas},\\
  Lemma~\thref{l:sfplus-simple-repr},\\
  Lemma~\thref{l:int-in-sfplus-gen-int-in-if}.

\item[The proof of Lemma~\thref{l:int-in-sfplus-for-count-meas-on-n}] \mbox{}\\
  cites explicitly:\\
  Lemma~\thref{l:int-in-sfplus-for-count-meas}.

\item[The proof of Lemma~\thref{l:int-in-sfplus-for-dirac-meas}] \mbox{}\\
  cites explicitly:\\
  Definition~\thref{d:dirac-meas},\\
  Lemma~\thref{l:int-in-sfplus-for-count-meas}.

\item[The proof of Lemma~\thref{l:int-in-mplus}] \mbox{}\\
  cites explicitly:\\
  Statement(s) from~\cite{cm:lmt:16},\\
  Lemma~\thref{l:int-in-sfplus}.

\item[The proof of Lemma~\thref{l:int-in-mplus-gen-int-in-sfplus}] \mbox{}\\
  cites explicitly:\\
  Lemma~\thref{l:int-in-sfplus},\\
  Lemma~\thref{l:int-in-sfplus-is-cont},\\
  Lemma~\thref{l:int-in-mplus}.

\item[The proof of Lemma~\thref{l:int-in-mplus-of-indic-fun}] \mbox{}\\
  cites explicitly:\\
  Lemma~\thref{l:sfplus-is-meas},\\
  Lemma~\thref{l:int-in-sfplus-gen-int-in-if},\\
  Lemma~\thref{l:int-in-mplus-gen-int-in-sfplus}.

\item[The proof of Lemma~\thref{l:int-in-mplus-is-pos-hom}] \mbox{}\\
  cites explicitly:\\
  Statement(s) from~\cite{cm:lmt:16},\\
  Definition~\thref{d:mult-in-rbar},\\
  Lemma~\thref{l:mult-in-rbarplus-is-assoc-mt},\\
  Lemma~\thref{l:zero-prod-prop-in-rbarplus-mt},\\
  Lemma~\thref{l:mplus-is-closed-under-nonneg-scalar-mult},\\
  Definition~\thref{d:sf-vector-space-of-simple-funs},\\
  Definition~\thref{d:sfplus-subset-of-nonneg-simple-funs},\\
  Lemma~\thref{l:sfplus-is-closed-under-pos-alg-ops},\\
  Lemma~\thref{l:int-in-sfplus},\\
  Lemma~\thref{l:int-in-sfplus-is-pos-lin},\\
  Lemma~\thref{l:equiv-def-of-int-in-sfplus-simple},\\
  Lemma~\thref{l:int-in-mplus}.

\item[The proof of Lemma~\thref{l:int-in-mplus-of-zero-is-zero}] \mbox{}\\
  cites explicitly:\\
  Definition~\thref{d:mult-in-rbar},\\
  Lemma~\thref{l:zero-prod-prop-in-rbarplus-mt},\\
  Lemma~\thref{l:int-in-mplus-is-pos-hom}.

\item[The proof of Lemma~\thref{l:int-in-mplus-is-monot}] \mbox{}\\
  cites explicitly:\\
  Lemma~\thref{l:int-in-mplus}.

\item[The proof of Theorem~\thref{t:beppo-levi-monot-conv}] \mbox{}\\
  cites explicitly:\\
  Statement(s) from~\cite{cm:lmt:16},\\
  Lemma~\thref{l:equiv-def-of-sigma-alg},\\
  Lemma~\thref{l:meas-of-num-fun},\\
  Lemma~\thref{l:m-is-closed-under-add-when-defined},\\
  Lemma~\thref{l:mplus-is-closed-under-limit-when-pointwise-conv},\\
  Definition~\thref{d:continuity-from-below},\\
  Lemma~\thref{l:meas-is-cont-from-below},\\
  Definition~\thref{d:sf-vector-space-of-simple-funs},\\
  Lemma~\thref{l:sf-is-meas},\\
  Lemma~\thref{l:int-in-sfplus-is-pos-lin},\\
  Lemma~\thref{l:equiv-def-of-int-in-sfplus-simple},\\
  Lemma~\thref{l:int-in-mplus},\\
  Lemma~\thref{l:int-in-mplus-of-indic-fun},\\
  Lemma~\thref{l:int-in-mplus-is-monot}.

\item[The proof of Lemma~\thref{l:int-in-mplus-is-hom-at-infinity}] \mbox{}\\
  cites explicitly:\\
  Definition~\thref{d:mplus-subset-of-nonneg-meas-num-fun},\\
  Lemma~\thref{l:mplus-is-closed-under-nonneg-scalar-mult},\\
  Lemma~\thref{l:int-in-mplus-is-pos-hom},\\
  Theorem~\thref{t:beppo-levi-monot-conv}.

\item[The proof of Lemma~\thref{l:adapted-seq-in-mplus}] \mbox{}\\
  cites explicitly:\\
  Lemma~\thref{l:equiv-def-of-sigma-alg},\\
  Lemma~\thref{l:meas-of-num-fun},\\
  Definition~\thref{d:sf-vector-space-of-simple-funs},\\
  Definition~\thref{d:sfplus-subset-of-nonneg-simple-funs},\\
  Definition~\thref{d:adapted-seq}.

\item[The proof of Lemma~\thref{l:usage-of-adapted-seqs}] \mbox{}\\
  cites explicitly:\\
  Lemma~\thref{l:int-in-mplus-gen-int-in-sfplus},\\
  Theorem~\thref{t:beppo-levi-monot-conv},\\
  Definition~\thref{d:adapted-seq},\\
  Lemma~\thref{l:adapted-seq-in-mplus}.

\item[The proof of Lemma~\thref{l:int-in-mplus-is-add}] \mbox{}\\
  cites explicitly:\\
  Lemma~\thref{l:mplus-is-closed-under-add},\\
  Lemma~\thref{l:int-in-sfplus-is-pos-lin},\\
  Lemma~\thref{l:adapted-seq-in-mplus},\\
  Lemma~\thref{l:usage-of-adapted-seqs}.

\item[The proof of Lemma~\thref{l:int-in-mplus-is-pos-lin}] \mbox{}\\
  cites explicitly:\\
  Lemma~\thref{l:int-in-mplus-is-pos-hom},\\
  Lemma~\thref{l:int-in-mplus-is-hom-at-infinity},\\
  Lemma~\thref{l:int-in-mplus-is-add}.

\item[The proof of Lemma~\thref{l:int-in-mplus-is-sigma-add}] \mbox{}\\
  cites explicitly:\\
  Lemma~\thref{l:mplus-is-closed-under-count-sum},\\
  Theorem~\thref{t:beppo-levi-monot-conv}.

\item[The proof of Lemma~\thref{l:int-in-mplus-of-decomp-into-nonpos-and-nonneg-parts}] \mbox{}\\
  cites explicitly:\\
  Lemma~\thref{l:decomp-into-nonneg-and-nonpos-parts},\\
  Lemma~\thref{l:meas-of-nonneg-and-nonpos-parts},\\
  Lemma~\thref{l:m-is-closed-under-abs},\\
  Lemma~\thref{l:int-in-mplus-is-add}.

\item[The proof of Lemma~\thref{l:compat-of-int-in-mplus-with-nonpos-and-nonneg-parts}] \mbox{}\\
  cites explicitly:\\
  Lemma~\thref{l:compat-of-nonpos-and-nonneg-parts-with-add},\\
  Lemma~\thref{l:int-in-mplus-is-add}.

\item[The proof of Lemma~\thref{l:int-in-mplus-is-almost-definite}] \mbox{}\\
  cites explicitly:\\
  Lemma~\thref{l:zero-prod-prop-in-rbarplus-mt},\\
  Lemma~\thref{l:infinity-prod-prop-in-rbarplus-mt},\\
  Lemma~\thref{l:meas-of-num-fun},\\
  Definition~\thref{d:mplus-subset-of-nonneg-meas-num-fun},\\
  Lemma~\thref{l:mplus-is-closed-under-nonneg-scalar-mult},\\
  Lemma~\thref{l:negl-of-meas-subset},\\
  Definition~\thref{d:prop-almost-satisfied},\\
  Lemma~\thref{l:int-in-mplus-of-indic-fun},\\
  Lemma~\thref{l:int-in-mplus-is-hom-at-infinity}.

\item[The proof of Lemma~\thref{l:compat-of-int-in-mplus-with-almost-bin-rel}] \mbox{}\\
  cites explicitly:\\
  Lemma~\thref{l:zero-is-identity-element-for-add-in-rbar},\\
  Lemma~\thref{l:mult-in-rbarplus-is-distr-over-add-mt},\\
  Lemma~\thref{l:zero-prod-prop-in-rbarplus-mt},\\
  Definition~\thref{d:sigma-alg},\\
  Definition~\thref{d:measurable-space},\\
  Lemma~\thref{l:mplus-is-closed-under-mult},\\
  Definition~\thref{d:meas},\\
  Definition~\thref{d:negl-subset},\\
  Lemma~\thref{l:negl-of-meas-subset},\\
  Definition~\thref{d:prop-almost-satisfied},\\
  Lemma~\thref{l:everywhere-implies-almost-everywhere},\\
  Definition~\thref{d:almost-bin-rel},\\
  Lemma~\thref{l:almost-eq-is-equiv-rel},\\
  Lemma~\thref{l:compat-of-almost-eq-with-op},\\
  Lemma~\thref{l:int-in-mplus-of-indic-fun},\\
  Lemma~\thref{l:int-in-mplus-is-add},\\
  Lemma~\thref{l:int-in-mplus-is-almost-definite}.

\item[The proof of Lemma~\thref{l:compat-of-int-in-mplus-with-almost-eq}] \mbox{}\\
  cites explicitly:\\
  Lemma~\thref{l:int-in-mplus},\\
  Lemma~\thref{l:compat-of-int-in-mplus-with-almost-bin-rel}.

\item[The proof of Lemma~\thref{l:int-in-mplus-is-almost-monot}] \mbox{}\\
  cites explicitly:\\
  Lemma~\thref{l:int-in-mplus-is-monot},\\
  Lemma~\thref{l:compat-of-int-in-mplus-with-almost-bin-rel}.

\item[The proof of Lemma~\thref{l:bienayme-chebyshev-ineq}] \mbox{}\\
  cites explicitly:\\
  Lemma~\thref{l:abs-in-rbar-is-nonneg},\\
  Lemma~\thref{l:meas-of-indic-fun},\\
  Lemma~\thref{l:meas-of-num-fun},\\
  Lemma~\thref{l:m-is-closed-under-abs},\\
  Lemma~\thref{l:mplus-is-closed-under-nonneg-scalar-mult},\\
  Lemma~\thref{l:int-in-mplus-of-indic-fun},\\
  Lemma~\thref{l:int-in-mplus-is-pos-hom},\\
  Lemma~\thref{l:int-in-mplus-is-monot},\\
  Lemma~\thref{l:int-in-mplus-is-hom-at-infinity}.

\item[The proof of Lemma~\thref{l:integrable-in-mplus-is-almost-finite}] \mbox{}\\
  cites explicitly:\\
  Definition~\thref{d:ext-real-nums-rbar},\\
  Lemma~\thref{l:abs-in-rbar-is-nonneg},\\
  Definition~\thref{d:mplus-subset-of-nonneg-meas-num-fun},\\
  Definition~\thref{d:meas},\\
  Lemma~\thref{l:negl-of-meas-subset},\\
  Definition~\thref{d:prop-almost-satisfied},\\
  Lemma~\thref{l:int-in-mplus},\\
  Lemma~\thref{l:bienayme-chebyshev-ineq}.

\item[The proof of Lemma~\thref{l:bounded-by-integrable-in-mplus-is-integrable}] \mbox{}\\
  cites explicitly:\\
  Lemma~\thref{l:order-in-rbar-is-total},\\
  Lemma~\thref{l:int-in-mplus},\\
  Lemma~\thref{l:int-in-mplus-is-monot}.

\item[The proof of Lemma~\thref{l:int-in-mplus-over-subset}] \mbox{}\\
  cites explicitly:\\
  Lemma~\thref{l:meas-of-restr},\\
  Definition~\thref{d:mplus-subset-of-nonneg-meas-num-fun},\\
  Lemma~\thref{l:sf-is-closed-under-restr},\\
  Lemma~\thref{l:int-in-sfplus-over-subset},\\
  Definition~\thref{d:adapted-seq},\\
  Lemma~\thref{l:adapted-seq-in-mplus},\\
  Lemma~\thref{l:usage-of-adapted-seqs}.

\item[The proof of Lemma~\thref{l:int-in-mplus-over-subset-is-sigma-add}] \mbox{}\\
  cites explicitly:\\
  Lemma~\thref{l:meas-of-indic-fun},\\
  Lemma~\thref{l:m-and-finite-is-mr},\\
  Definition~\thref{d:mplus-subset-of-nonneg-meas-num-fun},\\
  Lemma~\thref{l:mplus-is-closed-under-mult},\\
  Lemma~\thref{l:mplus-is-closed-under-count-sum},\\
  Lemma~\thref{l:if-is-sigma-add},\\
  Lemma~\thref{l:if-is-closed-under-mult},\\
  Lemma~\thref{l:int-in-mplus-is-sigma-add},\\
  Lemma~\thref{l:int-in-mplus-over-subset}.

\item[The proof of Lemma~\thref{l:int-in-mplus-over-singleton}] \mbox{}\\
  cites explicitly:\\
  Lemma~\thref{l:meas-of-indic-fun},\\
  Lemma~\thref{l:mplus-is-closed-under-nonneg-scalar-mult},\\
  Lemma~\thref{l:int-in-mplus-of-indic-fun},\\
  Lemma~\thref{l:int-in-mplus-is-pos-lin},\\
  Lemma~\thref{l:int-in-mplus-over-subset}.

\item[The proof of Theorem~\thref{t:fatou-lemma}] \mbox{}\\
  cites explicitly:\\
  Statement(s) from~\cite{cm:lmt:16},\\
  Lemma~\thref{l:inf-of-bounded-seq-is-bounded},\\
  Lemma~\thref{l:liminf},\\
  Definition~\thref{d:pointwise-conv},\\
  Lemma~\thref{l:liminf-bounded-from-below},\\
  Lemma~\thref{l:m-is-closed-under-inf},\\
  Lemma~\thref{l:m-is-closed-under-liminf},\\
  Lemma~\thref{l:mplus-is-closed-under-limit-when-pointwise-conv},\\
  Lemma~\thref{l:int-in-mplus-is-monot},\\
  Theorem~\thref{t:beppo-levi-monot-conv}.

\item[The proof of Lemma~\thref{l:int-in-mplus-of-pointwise-conv-seq}] \mbox{}\\
  cites explicitly:\\
  Statement(s) from~\cite{cm:lmt:16},\\
  Lemma~\thref{l:liminf-and-limsup-of-pointwise-conv},\\
  Lemma~\thref{l:liminf-limsup-and-pointwise-conv},\\
  Lemma~\thref{l:mplus-is-closed-under-limit-when-pointwise-conv},\\
  Lemma~\thref{l:int-in-mplus-is-monot},\\
  Theorem~\thref{t:fatou-lemma}.

\item[The proof of Lemma~\thref{l:int-in-mplus-for-count-meas}] \mbox{}\\
  cites explicitly:\\
  Lemma~\thref{l:int-in-sfplus-for-count-meas},\\
  Lemma~\thref{l:usage-of-adapted-seqs}.

\item[The proof of Lemma~\thref{l:int-in-mplus-for-count-meas-on-n}] \mbox{}\\
  cites explicitly:\\
  Lemma~\thref{l:int-in-mplus-for-count-meas}.

\item[The proof of Lemma~\thref{l:int-in-mplus-for-dirac-meas}] \mbox{}\\
  cites explicitly:\\
  Definition~\thref{d:dirac-meas},\\
  Lemma~\thref{l:int-in-mplus-for-count-meas}.

\item[The proof of Lemma~\thref{l:meas-of-section}] \mbox{}\\
  cites explicitly:\\
  Lemma~\thref{l:measurability-of-section},\\
  Definition~\thref{d:meas}.

\item[The proof of Lemma~\thref{l:meas-of-section-of-prod}] \mbox{}\\
  cites explicitly:\\
  Lemma~\thref{l:zero-prod-prop-in-rbarplus-mt},\\
  Lemma~\thref{l:prod-of-meas-subsets-is-meas},\\
  Lemma~\thref{l:section-of-prod},\\
  Definition~\thref{d:meas},\\
  Lemma~\thref{l:meas-of-section}.

\item[The proof of Lemma~\thref{l:meas-of-meas-of-section-finite}] \mbox{}\\
  cites explicitly:\\
  Definition~\thref{d:prod-of-subsets-of-parties},\\
  Definition~\thref{d:gen-set-alg},\\
  Lemma~\thref{l:gen-set-alg-is-min},\\
  Definition~\thref{d:monot-class},\\
  Lemma~\thref{l:sigma-alg-contains-set-alg},\\
  Lemma~\thref{l:set-alg-gen-by-prod-of-sigma-algs},\\
  Lemma~\thref{l:usage-of-monot-class-th},\\
  Definition~\thref{d:tensor-prod-of-sigma-algs},\\
  Lemma~\thref{l:prod-of-meas-subsets-is-meas},\\
  Lemma~\thref{l:compat-of-section-with-set-ops},\\
  Lemma~\thref{l:measurability-of-section},\\
  Lemma~\thref{l:count-union-of-sections-is-meas},\\
  Lemma~\thref{l:count-inter-of-sections-is-meas},\\
  Lemma~\thref{l:mplus-is-closed-under-nonneg-scalar-mult},\\
  Lemma~\thref{l:mplus-is-closed-under-inf},\\
  Lemma~\thref{l:mplus-is-closed-under-sup},\\
  Lemma~\thref{l:mplus-is-closed-under-count-sum},\\
  Definition~\thref{d:sigma-add-over-meas-space},\\
  Definition~\thref{d:meas},\\
  Lemma~\thref{l:meas-is-cont-from-below},\\
  Lemma~\thref{l:meas-is-cont-from-above},\\
  Definition~\thref{d:finite-meas},\\
  Lemma~\thref{l:finite-meas-is-bounded},\\
  Lemma~\thref{l:int-in-mplus-of-indic-fun},\\
  Lemma~\thref{l:meas-of-section},\\
  Lemma~\thref{l:meas-of-section-of-prod}.

\item[The proof of Lemma~\thref{l:meas-of-meas-of-section}] \mbox{}\\
  cites explicitly:\\
  Lemma~\thref{l:equiv-def-of-sigma-alg},\\
  Lemma~\thref{l:measurability-of-section},\\
  Lemma~\thref{l:mplus-is-closed-under-sup},\\
  Lemma~\thref{l:meas-is-cont-from-below},\\
  Definition~\thref{d:finite-meas},\\
  Lemma~\thref{l:equiv-def-of-sigma-finite-meas},\\
  Lemma~\thref{l:restr-meas},\\
  Lemma~\thref{l:meas-of-section},\\
  Lemma~\thref{l:meas-of-meas-of-section-finite}.

\item[The proof of Lemma~\thref{l:cand-tensor-prod-meas-is-tensor-prod-meas}] \mbox{}\\
  cites explicitly:\\
  Lemma~\thref{l:mult-in-rbarplus-is-comm-mt},\\
  Lemma~\thref{l:prod-of-meas-subsets-is-meas},\\
  Lemma~\thref{l:compat-of-section-with-set-ops},\\
  Lemma~\thref{l:measurability-of-section},\\
  Lemma~\thref{l:count-union-of-sections-is-meas},\\
  Definition~\thref{d:sigma-add-over-meas-space},\\
  Definition~\thref{d:meas},\\
  Lemma~\thref{l:int-in-mplus},\\
  Lemma~\thref{l:int-in-mplus-of-indic-fun},\\
  Lemma~\thref{l:int-in-mplus-of-zero-is-zero},\\
  Lemma~\thref{l:int-in-mplus-is-pos-lin},\\
  Lemma~\thref{l:int-in-mplus-is-sigma-add},\\
  Lemma~\thref{l:meas-of-section},\\
  Lemma~\thref{l:meas-of-section-of-prod},\\
  Lemma~\thref{l:meas-of-meas-of-section},\\
  Definition~\thref{d:tensor-prod-meas},\\
  Definition~\thref{d:cand-tensor-prod-meas}.

\item[The proof of Lemma~\thref{l:tensor-prod-of-finite-meas}] \mbox{}\\
  cites explicitly:\\
  Lemma~\thref{l:equiv-def-of-sigma-alg},\\
  Definition~\thref{d:meas},\\
  Definition~\thref{d:finite-meas},\\
  Definition~\thref{d:tensor-prod-meas}.

\item[The proof of Lemma~\thref{l:tensor-prod-of-sigma-finite-meas}] \mbox{}\\
  cites explicitly:\\
  Lemma~\thref{l:prod-of-meas-subsets-is-meas},\\
  Definition~\thref{d:sigma-finite-meas},\\
  Lemma~\thref{l:equiv-def-of-sigma-finite-meas},\\
  Definition~\thref{d:tensor-prod-meas}.

\item[The proof of Lemma~\thref{l:uniq-of-tensor-prod-meas-finite}] \mbox{}\\
  cites explicitly:\\
  Definition~\thref{d:monot-class},\\
  Lemma~\thref{l:equiv-def-of-sigma-alg},\\
  Lemma~\thref{l:sigma-alg-contains-set-alg},\\
  Lemma~\thref{l:set-alg-gen-by-prod-of-sigma-algs},\\
  Lemma~\thref{l:usage-of-monot-class-th},\\
  Definition~\thref{d:tensor-prod-of-sigma-algs},\\
  Lemma~\thref{l:prod-of-meas-subsets-is-meas},\\
  Definition~\thref{d:add-over-meas-space},\\
  Lemma~\thref{l:meas-is-cont-from-below},\\
  Lemma~\thref{l:meas-is-cont-from-above},\\
  Lemma~\thref{l:equiv-def-of-meas},\\
  Lemma~\thref{l:finite-meas-is-bounded},\\
  Lemma~\thref{l:finite-meas-is-sigma-finite},\\
  Definition~\thref{d:tensor-prod-meas},\\
  Lemma~\thref{l:cand-tensor-prod-meas-is-tensor-prod-meas},\\
  Lemma~\thref{l:tensor-prod-of-finite-meas}.

\item[The proof of Lemma~\thref{l:uniq-of-tensor-prod-meas}] \mbox{}\\
  cites explicitly:\\
  Lemma~\thref{l:equiv-def-of-sigma-alg},\\
  Lemma~\thref{l:prod-of-meas-subsets-is-meas},\\
  Lemma~\thref{l:meas-is-cont-from-below},\\
  Definition~\thref{d:finite-meas},\\
  Lemma~\thref{l:equiv-def-of-sigma-finite-meas},\\
  Lemma~\thref{l:restr-meas},\\
  Lemma~\thref{l:meas-of-section},\\
  Definition~\thref{d:tensor-prod-meas},\\
  Definition~\thref{d:cand-tensor-prod-meas},\\
  Lemma~\thref{l:cand-tensor-prod-meas-is-tensor-prod-meas},\\
  Lemma~\thref{l:tensor-prod-of-sigma-finite-meas},\\
  Lemma~\thref{l:uniq-of-tensor-prod-meas-finite}.

\item[The proof of Lemma~\thref{l:negl-of-meas-section}] \mbox{}\\
  cites explicitly:\\
  Lemma~\thref{l:measurability-of-section},\\
  Lemma~\thref{l:negl-of-meas-subset},\\
  Lemma~\thref{l:int-in-mplus-is-almost-definite},\\
  Lemma~\thref{l:uniq-of-tensor-prod-meas}.

\item[The proof of Lemma~\thref{l:lebesgue-meas-on-r2}] \mbox{}\\
  cites explicitly:\\
  Theorem~\thref{t:caratheodory-lebesgue-meas-on-r},\\
  Lemma~\thref{l:lebesgue-meas-is-sigma-finite},\\
  Lemma~\thref{l:uniq-of-tensor-prod-meas}.

\item[The proof of Lemma~\thref{l:lebesgue-meas-on-r2-gen-area-of-boxes}] \mbox{}\\
  cites explicitly:\\
  Lemma~\thref{l:lebesgue-meas-gen-len-of-int},\\
  Definition~\thref{d:tensor-prod-meas},\\
  Lemma~\thref{l:lebesgue-meas-on-r2}.

\item[The proof of Lemma~\thref{l:lebesgue-meas-on-r2-is-zero-on-lines}] \mbox{}\\
  cites explicitly:\\
  Lemma~\thref{l:zero-prod-prop-in-rbarplus-mt},\\
  Lemma~\thref{l:lebesgue-meas-on-r2-gen-area-of-boxes}.

\item[The proof of Lemma~\thref{l:lebesgue-meas-on-r2-is-sigma-finite}] \mbox{}\\
  cites explicitly:\\
  Lemma~\thref{l:lebesgue-meas-is-sigma-finite},\\
  Lemma~\thref{l:tensor-prod-of-sigma-finite-meas},\\
  Lemma~\thref{l:lebesgue-meas-on-r2}.

\item[The proof of Lemma~\thref{l:lebesgue-meas-on-r2-is-diffuse}] \mbox{}\\
  cites explicitly:\\
  Definition~\thref{d:diffuse-meas},\\
  Lemma~\thref{l:lebesgue-meas-on-r2-gen-area-of-boxes}.

\item[The proof of Theorem~\thref{t:tonelli}] \mbox{}\\
  cites explicitly:\\
  Lemma~\thref{l:measurability-of-section},\\
  Lemma~\thref{l:indic-of-section},\\
  Lemma~\thref{l:mplus-is-closed-under-add},\\
  Lemma~\thref{l:mplus-is-closed-under-nonneg-scalar-mult},\\
  Lemma~\thref{l:mplus-is-closed-under-limit-when-pointwise-conv},\\
  Lemma~\thref{l:indic-and-support-are-each-other-inverse},\\
  Lemma~\thref{l:sfplus-simple-repr},\\
  Lemma~\thref{l:int-in-mplus-of-indic-fun},\\
  Lemma~\thref{l:adapted-seq-in-mplus},\\
  Lemma~\thref{l:usage-of-adapted-seqs},\\
  Lemma~\thref{l:int-in-mplus-is-pos-lin},\\
  Lemma~\thref{l:meas-of-section},\\
  Lemma~\thref{l:meas-of-meas-of-section},\\
  Lemma~\thref{l:uniq-of-tensor-prod-meas},\\
  Definition~\thref{d:partial-fun-of-fun-from-prod-space}.

\item[The proof of Lemma~\thref{l:tonelli-over-subset}] \mbox{}\\
  cites explicitly:\\
  Lemma~\thref{l:indic-of-section},\\
  Lemma~\thref{l:int-in-mplus-over-subset},\\
  Definition~\thref{d:partial-fun-of-fun-from-prod-space},\\
  Theorem~\thref{t:tonelli}.

\item[The proof of Lemma~\thref{l:tonelli-for-tensor-prod}] \mbox{}\\
  cites explicitly:\\
  Lemma~\thref{l:mult-in-rbarplus-is-closed-mt},\\
  Lemma~\thref{l:mult-in-rbarplus-is-comm-mt},\\
  Definition~\thref{d:mplus-subset-of-nonneg-meas-num-fun},\\
  Definition~\thref{d:tensor-prod-of-num-funs},\\
  Lemma~\thref{l:meas-of-tensor-prod-of-num-funs},\\
  Lemma~\thref{l:int-in-mplus-is-pos-hom},\\
  Definition~\thref{d:partial-fun-of-fun-from-prod-space},\\
  Theorem~\thref{t:tonelli}.

\item[The proof of Lemma~\thref{l:integrable-is-meas}] \mbox{}\\
  cites explicitly:\\
  Lemma~\thref{l:meas-of-nonneg-and-nonpos-parts},\\
  Lemma~\thref{l:int-in-mplus},\\
  Definition~\thref{d:integrability}.

\item[The proof of Lemma~\thref{l:equiv-def-of-integrability}] \mbox{}\\
  cites explicitly:\\
  Lemma~\thref{l:add-in-rbarplus-is-closed},\\
  Lemma~\thref{l:infinity-sum-prop-in-rbarplus},\\
  Lemma~\thref{l:meas-of-nonneg-and-nonpos-parts},\\
  Lemma~\thref{l:m-is-closed-under-abs},\\
  Lemma~\thref{l:int-in-mplus},\\
  Lemma~\thref{l:int-in-mplus-of-decomp-into-nonpos-and-nonneg-parts},\\
  Definition~\thref{d:integrability},\\
  Lemma~\thref{l:integrable-is-meas}.

\item[The proof of Lemma~\thref{l:compat-of-integrability-in-m-and-mplus}] \mbox{}\\
  cites explicitly:\\
  Lemma~\thref{l:equiv-def-of-abs-in-rbar},\\
  Lemma~\thref{l:equiv-def-of-integrability}.

\item[The proof of Lemma~\thref{l:integrable-is-almost-finite}] \mbox{}\\
  cites explicitly:\\
  Lemma~\thref{l:abs-in-rbar-is-even},\\
  Lemma~\thref{l:integrable-in-mplus-is-almost-finite},\\
  Lemma~\thref{l:equiv-def-of-integrability}.

\item[The proof of Lemma~\thref{l:almost-bounded-by-integrable-is-integrable}] \mbox{}\\
  cites explicitly:\\
  Lemma~\thref{l:order-in-rbar-is-total},\\
  Lemma~\thref{l:m-is-closed-under-abs},\\
  Lemma~\thref{l:almost-order-is-order-rel},\\
  Lemma~\thref{l:int-in-mplus},\\
  Lemma~\thref{l:int-in-mplus-is-almost-monot},\\
  Lemma~\thref{l:equiv-def-of-integrability}.

\item[The proof of Lemma~\thref{l:bounded-by-integrable-is-integrable}] \mbox{}\\
  cites explicitly:\\
  Lemma~\thref{l:order-in-rbar-is-total},\\
  Lemma~\thref{l:everywhere-implies-almost-everywhere},\\
  Lemma~\thref{l:equiv-def-of-integrability},\\
  Lemma~\thref{l:almost-bounded-by-integrable-is-integrable}.

\item[The proof of Lemma~\thref{l:compat-of-int-in-m-and-mplus}] \mbox{}\\
  cites explicitly:\\
  Definition~\thref{d:nonneg-and-nonpos-parts},\\
  Lemma~\thref{l:int-in-mplus},\\
  Lemma~\thref{l:int-in-mplus-of-zero-is-zero},\\
  Definition~\thref{d:integrability},\\
  Definition~\thref{d:integral}.

\item[The proof of Lemma~\thref{l:int-of-zero-is-zero}] \mbox{}\\
  cites explicitly:\\
  Lemma~\thref{l:int-in-mplus-of-zero-is-zero},\\
  Lemma~\thref{l:compat-of-int-in-m-and-mplus}.

\item[The proof of Lemma~\thref{l:compat-of-int-with-almost-eq}] \mbox{}\\
  cites explicitly:\\
  Lemma~\thref{l:compat-of-almost-eq-with-op},\\
  Lemma~\thref{l:compat-of-int-in-mplus-with-almost-eq},\\
  Definition~\thref{d:integrability},\\
  Definition~\thref{d:integral}.

\item[The proof of Lemma~\thref{l:int-over-subset}] \mbox{}\\
  cites explicitly:\\
  Lemma~\thref{l:compat-of-nonpos-and-nonneg-parts-with-mask},\\
  Lemma~\thref{l:compat-of-nonpos-and-nonneg-parts-with-restr},\\
  Lemma~\thref{l:int-in-mplus},\\
  Lemma~\thref{l:int-in-mplus-over-subset},\\
  Definition~\thref{d:integrability},\\
  Definition~\thref{d:integral}.

\item[The proof of Lemma~\thref{l:int-over-subset-is-sigma-add}] \mbox{}\\
  cites explicitly:\\
  Lemma~\thref{l:int-in-mplus-is-monot},\\
  Lemma~\thref{l:int-in-mplus-is-sigma-add},\\
  Lemma~\thref{l:int-in-mplus-over-subset-is-sigma-add},\\
  Lemma~\thref{l:equiv-def-of-integrability},\\
  Definition~\thref{d:integral}.

\item[The proof of Lemma~\thref{l:int-over-singleton}] \mbox{}\\
  cites explicitly:\\
  Definition~\thref{d:abs-in-rbar},\\
  Lemma~\thref{l:abs-in-rbar-is-definite},\\
  Lemma~\thref{l:finite-prod-prop-in-rbarplus-mt},\\
  Definition~\thref{d:nonneg-and-nonpos-parts},\\
  Lemma~\thref{l:decomp-into-nonneg-and-nonpos-parts},\\
  Lemma~\thref{l:int-in-mplus},\\
  Lemma~\thref{l:int-in-mplus-over-singleton},\\
  Definition~\thref{d:integrability},\\
  Lemma~\thref{l:equiv-def-of-integrability},\\
  Definition~\thref{d:integral},\\
  Lemma~\thref{l:int-over-subset}.

\item[The proof of Lemma~\thref{l:int-over-int}] \mbox{}\\
  cites explicitly:\\
  Lemma~\thref{l:some-borel-subsets},\\
  Definition~\thref{d:diffuse-meas},\\
  Lemma~\thref{l:int-over-subset-is-sigma-add},\\
  Lemma~\thref{l:int-over-singleton}.

\item[The proof of Lemma~\thref{l:chasles-rel-int-over-split-ints}] \mbox{}\\
  cites explicitly:\\
  Lemma~\thref{l:int-over-subset-is-sigma-add},\\
  Lemma~\thref{l:int-over-int}.

\item[The proof of Lemma~\thref{l:int-for-count-meas}] \mbox{}\\
  cites explicitly:\\
  Lemma~\thref{l:decomp-into-nonneg-and-nonpos-parts},\\
  Lemma~\thref{l:int-in-mplus},\\
  Lemma~\thref{l:int-in-mplus-for-count-meas},\\
  Lemma~\thref{l:equiv-def-of-integrability},\\
  Definition~\thref{d:integral}.

\item[The proof of Lemma~\thref{l:int-for-count-meas-on-n}] \mbox{}\\
  cites explicitly:\\
  Lemma~\thref{l:int-for-count-meas}.

\item[The proof of Lemma~\thref{l:int-for-dirac-meas}] \mbox{}\\
  cites explicitly:\\
  Definition~\thref{d:dirac-meas},\\
  Lemma~\thref{l:int-for-count-meas}.

\item[The proof of Lemma~\thref{l:seminorm-llone}] \mbox{}\\
  cites explicitly:\\
  Lemma~\thref{l:m-is-closed-under-abs},\\
  Lemma~\thref{l:int-in-mplus},\\
  Definition~\thref{d:merge-int-in-m-and-mplus}.

\item[The proof of Lemma~\thref{l:integrable-is-finite-seminorm-lone}] \mbox{}\\
  cites explicitly:\\
  Lemma~\thref{l:m-is-closed-under-abs},\\
  Lemma~\thref{l:int-in-mplus},\\
  Lemma~\thref{l:equiv-def-of-integrability},\\
  Lemma~\thref{l:seminorm-llone}.

\item[The proof of Lemma~\thref{l:compat-of-none-with-almost-eq}] \mbox{}\\
  cites explicitly:\\
  Lemma~\thref{l:compat-of-almost-eq-with-op},\\
  Lemma~\thref{l:compat-of-int-with-almost-eq},\\
  Lemma~\thref{l:seminorm-llone}.

\item[The proof of Lemma~\thref{l:none-is-almost-definite}] \mbox{}\\
  cites explicitly:\\
  Lemma~\thref{l:abs-is-almost-definite},\\
  Lemma~\thref{l:int-in-mplus-is-almost-definite},\\
  Lemma~\thref{l:seminorm-llone}.

\item[The proof of Lemma~\thref{l:none-is-abs-hom}] \mbox{}\\
  cites explicitly:\\
  Lemma~\thref{l:int-in-mplus-is-pos-hom},\\
  Lemma~\thref{l:int-in-mplus-is-hom-at-infinity},\\
  Lemma~\thref{l:seminorm-llone}.

\item[The proof of Lemma~\thref{l:int-is-hom}] \mbox{}\\
  cites explicitly:\\
  Definition~\thref{d:nonneg-and-nonpos-parts},\\
  Lemma~\thref{l:m-is-closed-under-scalar-mult},\\
  Lemma~\thref{l:meas-of-nonneg-and-nonpos-parts},\\
  Lemma~\thref{l:int-in-mplus},\\
  Lemma~\thref{l:int-in-mplus-is-pos-hom},\\
  Lemma~\thref{l:integrable-is-meas},\\
  Lemma~\thref{l:equiv-def-of-integrability},\\
  Definition~\thref{d:integral},\\
  Lemma~\thref{l:seminorm-llone},\\
  Lemma~\thref{l:none-is-abs-hom}.

\item[The proof of Lemma~\thref{l:minkowski-ineq-in-m}] \mbox{}\\
  cites explicitly:\\
  Lemma~\thref{l:abs-in-rbar-satisfies-triangle-ineq},\\
  Lemma~\thref{l:m-is-closed-under-abs},\\
  Lemma~\thref{l:almost-sum},\\
  Lemma~\thref{l:compat-of-almost-sum-with-almost-eq},\\
  Lemma~\thref{l:int-in-mplus-is-monot},\\
  Lemma~\thref{l:int-in-mplus-is-add},\\
  Lemma~\thref{l:seminorm-llone},\\
  Lemma~\thref{l:compat-of-none-with-almost-eq}.

\item[The proof of Lemma~\thref{l:int-is-add}] \mbox{}\\
  cites explicitly:\\
  Definition~\thref{d:add-in-rbar},\\
  Lemma~\thref{l:meas-of-nonneg-and-nonpos-parts},\\
  Lemma~\thref{l:compat-of-int-in-mplus-with-nonpos-and-nonneg-parts},\\
  Lemma~\thref{l:integrable-is-almost-finite},\\
  Definition~\thref{d:integral},\\
  Lemma~\thref{l:compat-of-int-with-almost-eq},\\
  Lemma~\thref{l:integrable-is-finite-seminorm-lone},\\
  Lemma~\thref{l:minkowski-ineq-in-m}.

\item[The proof of Lemma~\thref{l:equiv-def-of-llone}] \mbox{}\\
  cites explicitly:\\
  Lemma~\thref{l:m-and-finite-is-mr},\\
  Lemma~\thref{l:int-in-mplus},\\
  Lemma~\thref{l:equiv-def-of-integrability},\\
  Lemma~\thref{l:seminorm-llone},\\
  Definition~\thref{d:llone-vector-space-of-int-fun}.

\item[The proof of Lemma~\thref{l:minkowski-ineq-in-llone}] \mbox{}\\
  cites explicitly:\\
  Lemma~\thref{l:everywhere-implies-almost-everywhere},\\
  Definition~\thref{d:summability-domain},\\
  Lemma~\thref{l:almost-sum-is-sum},\\
  Lemma~\thref{l:minkowski-ineq-in-m},\\
  Definition~\thref{d:llone-vector-space-of-int-fun}.

\item[The proof of Lemma~\thref{l:llone-is-seminormed-vector-space}] \mbox{}\\
  cites explicitly:\\
  Statement(s) from~\cite{cm:lmt:16},\\
  Definition~\thref{d:seminorm},\\
  Lemma~\thref{l:mr-is-vector-space},\\
  Lemma~\thref{l:everywhere-implies-almost-everywhere},\\
  Lemma~\thref{l:none-is-almost-definite},\\
  Lemma~\thref{l:none-is-abs-hom},\\
  Definition~\thref{d:llone-vector-space-of-int-fun},\\
  Lemma~\thref{l:minkowski-ineq-in-llone}.

\item[The proof of Lemma~\thref{l:llone-is-closed-under-abs}] \mbox{}\\
  cites explicitly:\\
  Lemma~\thref{l:m-is-closed-under-abs},\\
  Lemma~\thref{l:seminorm-llone},\\
  Definition~\thref{d:llone-vector-space-of-int-fun}.

\item[The proof of Lemma~\thref{l:bounded-by-llone-is-llone}] \mbox{}\\
  cites explicitly:\\
  Lemma~\thref{l:order-in-rbar-is-total},\\
  Lemma~\thref{l:finite-abs-in-rbar},\\
  Lemma~\thref{l:abs-in-rbar-is-nonneg},\\
  Lemma~\thref{l:compat-of-integrability-in-m-and-mplus},\\
  Lemma~\thref{l:bounded-by-integrable-is-integrable},\\
  Lemma~\thref{l:equiv-def-of-llone},\\
  Lemma~\thref{l:llone-is-closed-under-abs}.

\item[The proof of Lemma~\thref{l:int-is-pos-lin-form-on-llone}] \mbox{}\\
  cites explicitly:\\
  Statement(s) from~\cite{cm:lmt:16},\\
  Definition~\thref{d:add-in-rbar},\\
  Lemma~\thref{l:nonneg-and-nonpos-parts-are-nonneg},\\
  Lemma~\thref{l:m-and-finite-is-mr},\\
  Definition~\thref{d:mplus-subset-of-nonneg-meas-num-fun},\\
  Lemma~\thref{l:int-in-mplus},\\
  Lemma~\thref{l:int-in-mplus-of-decomp-into-nonpos-and-nonneg-parts},\\
  Definition~\thref{d:integrability},\\
  Lemma~\thref{l:equiv-def-of-integrability},\\
  Definition~\thref{d:integral},\\
  Lemma~\thref{l:seminorm-llone},\\
  Lemma~\thref{l:int-is-hom},\\
  Lemma~\thref{l:int-is-add},\\
  Definition~\thref{d:llone-vector-space-of-int-fun},\\
  Lemma~\thref{l:llone-is-seminormed-vector-space}.

\item[The proof of Lemma~\thref{l:const-fun-is-llone}] \mbox{}\\
  cites explicitly:\\
  Definition~\thref{d:sigma-alg},\\
  Definition~\thref{d:measurable-space},\\
  Definition~\thref{d:meas},\\
  Definition~\thref{d:sf-vector-space-of-simple-funs},\\
  Lemma~\thref{l:int-in-sfplus},\\
  Lemma~\thref{l:int-in-mplus-gen-int-in-sfplus},\\
  Definition~\thref{d:integral},\\
  Lemma~\thref{l:compat-of-int-in-m-and-mplus}.

\item[The proof of Lemma~\thref{l:first-mean-value-theorem}] \mbox{}\\
  cites explicitly:\\
  Lemma~\thref{l:compat-of-almost-eq-with-op},\\
  Lemma~\thref{l:int-in-mplus-is-almost-definite},\\
  Lemma~\thref{l:compat-of-int-in-m-and-mplus},\\
  Lemma~\thref{l:llone-is-seminormed-vector-space},\\
  Lemma~\thref{l:bounded-by-llone-is-llone},\\
  Lemma~\thref{l:int-is-pos-lin-form-on-llone},\\
  Lemma~\thref{l:const-fun-is-llone}.

\item[The proof of Lemma~\thref{l:variant-of-first-mean-value-theorem}] \mbox{}\\
  cites explicitly:\\
  Definition~\thref{d:interval},\\
  Lemma~\thref{l:first-mean-value-theorem}.

\item[The proof of Theorem~\thref{t:lebesgue-dom-conv}] \mbox{}\\
  cites explicitly:\\
  Statement(s) from~\cite{cm:lmt:16},\\
  Definition~\thref{d:seminorm},\\
  Lemma~\thref{l:abs-in-rbar-is-nonneg},\\
  Lemma~\thref{l:abs-in-rbar-is-definite},\\
  Lemma~\thref{l:abs-in-rbar-is-cont},\\
  Lemma~\thref{l:duality-liminf-limsup},\\
  Lemma~\thref{l:liminf-and-limsup-of-pointwise-conv},\\
  Lemma~\thref{l:liminf-bounded-from-below},\\
  Lemma~\thref{l:liminf-limsup-and-pointwise-conv},\\
  Lemma~\thref{l:m-is-closed-under-limit-when-pointwise-conv},\\
  Definition~\thref{d:mplus-subset-of-nonneg-meas-num-fun},\\
  Theorem~\thref{t:fatou-lemma},\\
  Lemma~\thref{l:seminorm-llone},\\
  Definition~\thref{d:llone-vector-space-of-int-fun},\\
  Lemma~\thref{l:llone-is-seminormed-vector-space},\\
  Definition~\thref{d:conv-in-llone},\\
  Lemma~\thref{l:llone-is-closed-under-abs},\\
  Lemma~\thref{l:bounded-by-llone-is-llone},\\
  Lemma~\thref{l:int-is-pos-lin-form-on-llone}.

\item[The proof of Theorem~\thref{t:lebesgue-ext-dom-conv}] \mbox{}\\
  cites explicitly:\\
  Definition~\thref{d:sigma-alg},\\
  Lemma~\thref{l:equiv-def-of-sigma-alg},\\
  Definition~\thref{d:measurable-space},\\
  Lemma~\thref{l:meas-and-masking},\\
  Definition~\thref{d:meas},\\
  Definition~\thref{d:negl-subset},\\
  Lemma~\thref{l:compat-of-null-meas-with-count-union},\\
  Definition~\thref{d:prop-almost-satisfied},\\
  Lemma~\thref{l:compat-of-almost-eq-with-op},\\
  Lemma~\thref{l:masking-almost-nowhere},\\
  Lemma~\thref{l:finite-nonneg-part},\\
  Lemma~\thref{l:equiv-def-of-integrability},\\
  Lemma~\thref{l:compat-of-int-with-almost-eq},\\
  Lemma~\thref{l:seminorm-llone},\\
  Definition~\thref{d:llone-vector-space-of-int-fun},\\
  Theorem~\thref{t:lebesgue-dom-conv}.

\end{description}

\chapter{Is explicitly cited in the proof of\ldots}
\label{c:is-explicitly-cited-in-the-proof-of}

This appendix gathers the explicit citations that appear in the proof of
results (lemmas and theorems) for each statement listed in
Appendix~\ref{c:lists-of-statements}.
Statements from~\cite{cm:lmt:16} are anonymized.

The corresponding dependency graph is represented in
Figure~\ref{f:depend-graph} (bottom).
The dual graph is described in Appendix~\ref{c:the-proof-cites-explicitly}.

Printing is not advised!

\bigskip

\begin{description}[style=unboxed]

\item[Statement(s) from~\cite{cm:lmt:16}] \mbox{}\\
  are explicitly cited in the proof of:\\
  Lemma~\thref{l:quotient-vector-space-equiv-rel},\\
  Lemma~\thref{l:quotient-vector-space},\\
  Lemma~\thref{l:linear-map-on-quotient-vector-space},\\
  Lemma~\thref{l:alg-of-funs-to-alg},\\
  Lemma~\thref{l:subspace-and-closed-under-mult-is-subalg},\\
  Lemma~\thref{l:closed-under-alg-ops-is-subalg},\\
  Lemma~\thref{l:definite-seminorm-is-norm},\\
  Lemma~\thref{l:equiv-def-of-conv-seq},\\
  Lemma~\thref{l:conv-subseq-of-cauchy-seq},\\
  Lemma~\thref{l:conn-comp-of-open-subset-of-r-is-open-int},\\
  Theorem~\thref{t:count-conn-comps-of-open-subsets-of-r},\\
  Lemma~\thref{l:open-int-with-rat-bounds-cover-open-int},\\
  Lemma~\thref{l:extr-of-const-fun},\\
  Lemma~\thref{l:equiv-def-of-finite-inf},\\
  Lemma~\thref{l:equiv-def-of-finite-inf-in-rbar},\\
  Lemma~\thref{l:equiv-def-of-inf},\\
  Lemma~\thref{l:inf-is-smaller-than-sup},\\
  Lemma~\thref{l:inf-is-monot},\\
  Lemma~\thref{l:sup-is-monot},\\
  Lemma~\thref{l:compat-of-inf-with-abs},\\
  Lemma~\thref{l:compat-of-sup-with-abs},\\
  Lemma~\thref{l:inf-of-seq-is-monot},\\
  Lemma~\thref{l:sup-of-seq-is-monot},\\
  Lemma~\thref{l:liminf-is-inf},\\
  Lemma~\thref{l:equiv-def-of-liminf},\\
  Lemma~\thref{l:duality-liminf-limsup},\\
  Lemma~\thref{l:liminf-bounded-from-below},\\
  Lemma~\thref{l:liminf-bounded-from-above},\\
  Lemma~\thref{l:liminf-limsup-and-pointwise-conv},\\
  Lemma~\thref{l:m-is-closed-under-inf},\\
  Lemma~\thref{l:m-is-closed-under-sup},\\
  Lemma~\thref{l:meas-satisfies-boole-ineq},\\
  Lemma~\thref{l:compat-of-null-meas-with-count-union},\\
  Lemma~\thref{l:lambda-star-is-hom},\\
  Lemma~\thref{l:lambda-star-is-monot},\\
  Lemma~\thref{l:lambda-star-is-sigma-subadd},\\
  Lemma~\thref{l:lambda-star-gen-len-of-int},\\
  Lemma~\thref{l:rays-are-lebesgue-meas},\\
  Lemma~\thref{l:sf-is-alg-over-r},\\
  Lemma~\thref{l:int-in-sfplus-is-monot},\\
  Lemma~\thref{l:int-in-sfplus-is-cont},\\
  Lemma~\thref{l:int-in-mplus},\\
  Lemma~\thref{l:int-in-mplus-is-pos-hom},\\
  Theorem~\thref{t:beppo-levi-monot-conv},\\
  Theorem~\thref{t:fatou-lemma},\\
  Lemma~\thref{l:int-in-mplus-of-pointwise-conv-seq},\\
  Lemma~\thref{l:llone-is-seminormed-vector-space},\\
  Lemma~\thref{l:int-is-pos-lin-form-on-llone},\\
  Theorem~\thref{t:lebesgue-dom-conv}.

\item[Definition~\thref{d:pseudopart}] \mbox{}\\
  is explicitly cited in the proof of:\\
  Lemma~\thref{l:compat-of-pseudopart-with-inter},\\
  Lemma~\thref{l:m-is-closed-under-finite-part},\\
  Lemma~\thref{l:m-is-closed-under-add-when-defined},\\
  Lemma~\thref{l:uniq-of-meas-ext-from-p-syst}.

\item[Lemma~\thref{l:compat-of-pseudopart-with-inter}] \mbox{}\\
  is explicitly cited in the proof of:\\
  Lemma~\thref{l:characterization-of-borel-subsets},\\
  Lemma~\thref{l:meas-over-count-pseudopart},\\
  Lemma~\thref{l:equiv-def-of-l},\\
  Lemma~\thref{l:lambda-star-is-add-on-l}.

\item[Lemma~\thref{l:technical-inclusion-for-count-union}] \mbox{}\\
  is explicitly cited in the proof of:\\
  Lemma~\thref{l:order-is-meaningless-in-count-union}.

\item[Lemma~\thref{l:order-is-meaningless-in-count-union}] \mbox{}\\
  is explicitly cited in the proof of:\\
  Lemma~\thref{l:def-of-double-count-union}.

\item[Lemma~\thref{l:def-of-double-count-union}] \mbox{}\\
  is explicitly cited in the proof of:\\
  Lemma~\thref{l:double-count-union},\\
  Lemma~\thref{l:lambda-star-is-sigma-subadd}.

\item[Lemma~\thref{l:double-count-union}] \mbox{}\\
  is explicitly cited in the proof of:\\
  Lemma~\thref{l:lambda-star-is-sigma-subadd}.

\item[Lemma~\thref{l:part-of-count-union}] \mbox{}\\
  is explicitly cited in the proof of:\\
  Lemma~\thref{l:count-union-disj-and-monot},\\
  Lemma~\thref{l:count-union-disj-and-union},\\
  Lemma~\thref{l:part-of-count-union-in-set-alg},\\
  Lemma~\thref{l:part-of-count-union-in-sigma-alg}.

\item[Definition~\thref{d:trace-of-subsets-of-parties}] \mbox{}\\
  is explicitly cited in the proof of:\\
  Lemma~\thref{l:trace-topo-on-subset},\\
  Lemma~\thref{l:trace-sigma-alg},\\
  Lemma~\thref{l:meas-of-meas-subspace},\\
  Lemma~\thref{l:gen-meas-subspace},\\
  Lemma~\thref{l:source-restr-of-meas-fun}.

\item[Definition~\thref{d:prod-of-subsets-of-parties}] \mbox{}\\
  is explicitly cited in the proof of:\\
  Lemma~\thref{l:box-topo-on-cartesian-prod},\\
  Lemma~\thref{l:compat-of-second-count-with-cartesian-prod},\\
  Lemma~\thref{l:complete-count-topo-basis-of-prod-space},\\
  Lemma~\thref{l:set-alg-gen-by-prod-of-sigma-algs},\\
  Lemma~\thref{l:prod-of-meas-subsets-is-meas},\\
  Lemma~\thref{l:meas-of-fun-to-prod-space},\\
  Lemma~\thref{l:gen-prod-meas-space},\\
  Lemma~\thref{l:measurability-of-section},\\
  Lemma~\thref{l:meas-of-meas-of-section-finite}.

\item[Lemma~\thref{l:restr-is-mask}] \mbox{}\\
  is explicitly cited in the proof of:\\
  Lemma~\thref{l:if-is-closed-under-ext-by-zero},\\
  Lemma~\thref{l:int-in-if-over-subset}.

\item[Definition~\thref{d:relation-compatible-with-vector-ops}] \mbox{}\\
  is explicitly cited in the proof of:\\
  Lemma~\thref{l:quotient-vector-ops},\\
  Lemma~\thref{l:quotient-vector-space}.

\item[Lemma~\thref{l:quotient-vector-ops}] \mbox{}\\
  is explicitly cited in the proof of:\\
  Lemma~\thref{l:quotient-vector-space-equiv-rel},\\
  Lemma~\thref{l:linear-map-on-quotient-vector-space}.

\item[Lemma~\thref{l:quotient-vector-space-equiv-rel}] \mbox{}\\
  is explicitly cited in the proof of:\\
  Lemma~\thref{l:quotient-vector-space}.

\item[Lemma~\thref{l:quotient-vector-space}] \mbox{}\\
  is explicitly cited in the proof of:\\
  Lemma~\thref{l:linear-map-on-quotient-vector-space}.

\item[Lemma~\thref{l:linear-map-on-quotient-vector-space}] \mbox{}\\
  is not yet used.

\item[Definition~\thref{d:alg-over-a-field}] \mbox{}\\
  is explicitly cited in the proof of:\\
  Lemma~\thref{l:k-is-k-alg},\\
  Lemma~\thref{l:alg-of-funs-to-alg},\\
  Lemma~\thref{l:subspace-and-closed-under-mult-is-subalg},\\
  Lemma~\thref{l:mr-is-vector-space},\\
  Lemma~\thref{l:sfplus-is-closed-under-pos-alg-ops},\\
  Lemma~\thref{l:int-in-sfplus-is-monot}.

\item[Lemma~\thref{l:k-is-k-alg}] \mbox{}\\
  is explicitly cited in the proof of:\\
  Lemma~\thref{l:maps-to-k-is-alg},\\
  Lemma~\thref{l:mr-is-alg}.

\item[Definition~\thref{d:inherited-alg-ops}] \mbox{}\\
  is explicitly cited in the proof of:\\
  Lemma~\thref{l:alg-of-funs-to-alg}.

\item[Lemma~\thref{l:alg-of-funs-to-alg}] \mbox{}\\
  is explicitly cited in the proof of:\\
  Lemma~\thref{l:maps-to-k-is-alg},\\
  Lemma~\thref{l:mr-is-alg}.

\item[Lemma~\thref{l:maps-to-k-is-alg}] \mbox{}\\
  is not yet used.

\item[Definition~\thref{d:subalg}] \mbox{}\\
  is explicitly cited in the proof of:\\
  Lemma~\thref{l:subspace-and-closed-under-mult-is-subalg},\\
  Lemma~\thref{l:mr-is-vector-space}.

\item[Lemma~\thref{l:subspace-and-closed-under-mult-is-subalg}] \mbox{}\\
  is explicitly cited in the proof of:\\
  Lemma~\thref{l:closed-under-alg-ops-is-subalg},\\
  Lemma~\thref{l:sf-is-alg-over-r}.

\item[Lemma~\thref{l:closed-under-alg-ops-is-subalg}] \mbox{}\\
  is explicitly cited in the proof of:\\
  Lemma~\thref{l:mr-is-alg},\\
  Lemma~\thref{l:mr-is-vector-space}.

\item[Definition~\thref{d:seminorm}] \mbox{}\\
  is explicitly cited in the proof of:\\
  Lemma~\thref{l:definite-seminorm-is-norm},\\
  Lemma~\thref{l:llone-is-seminormed-vector-space},\\
  Theorem~\thref{t:lebesgue-dom-conv}.

\item[Lemma~\thref{l:definite-seminorm-is-norm}] \mbox{}\\
  is not yet used.

\item[Definition~\thref{d:interval}] \mbox{}\\
  is explicitly cited in the proof of:\\
  Lemma~\thref{l:empty-open-int},\\
  Lemma~\thref{l:int-are-closed-under-finite-inter},\\
  Lemma~\thref{l:conn-comp-of-open-subset-of-r-is-open-int},\\
  Lemma~\thref{l:variant-of-first-mean-value-theorem}.

\item[Lemma~\thref{l:empty-open-int}] \mbox{}\\
  is explicitly cited in the proof of:\\
  Lemma~\thref{l:empty-inter-of-open-ints}.

\item[Lemma~\thref{l:int-are-closed-under-finite-inter}] \mbox{}\\
  is explicitly cited in the proof of:\\
  Lemma~\thref{l:empty-inter-of-open-ints},\\
  Lemma~\thref{l:topological-basis-of-order-topo},\\
  Lemma~\thref{l:rays-are-lebesgue-meas},\\
  Theorem~\thref{t:caratheodory-lebesgue-meas-on-r}.

\item[Lemma~\thref{l:empty-inter-of-open-ints}] \mbox{}\\
  is not yet used.

\item[Definition~\thref{d:topological-space}] \mbox{}\\
  is explicitly cited in the proof of:\\
  Lemma~\thref{l:inter-of-topo},\\
  Lemma~\thref{l:equiv-def-of-gen-topo},\\
  Lemma~\thref{l:trace-topo-on-subset},\\
  Lemma~\thref{l:box-topo-on-cartesian-prod},\\
  Lemma~\thref{l:conn-comp-of-open-subset-of-r-is-open-int},\\
  Lemma~\thref{l:open-int-with-rat-bounds-cover-open-int},\\
  Lemma~\thref{l:some-borel-subsets}.

\item[Lemma~\thref{l:inter-of-topo}] \mbox{}\\
  is explicitly cited in the proof of:\\
  Lemma~\thref{l:gen-topo-is-min}.

\item[Definition~\thref{d:gen-topo}] \mbox{}\\
  is explicitly cited in the proof of:\\
  Lemma~\thref{l:gen-topo-is-min}.

\item[Lemma~\thref{l:gen-topo-is-min}] \mbox{}\\
  is explicitly cited in the proof of:\\
  Lemma~\thref{l:equiv-def-of-gen-topo}.

\item[Lemma~\thref{l:equiv-def-of-gen-topo}] \mbox{}\\
  is explicitly cited in the proof of:\\
  Lemma~\thref{l:topological-basis-of-order-topo}.

\item[Definition~\thref{d:topological-basis}] \mbox{}\\
  is explicitly cited in the proof of:\\
  Lemma~\thref{l:augmented-topo-basis},\\
  Lemma~\thref{l:topological-basis-of-order-topo},\\
  Lemma~\thref{l:trace-topo-on-subset},\\
  Lemma~\thref{l:box-topo-on-cartesian-prod},\\
  Lemma~\thref{l:trace-topo-on-r},\\
  Theorem~\thref{t:r-is-second-countable},\\
  Lemma~\thref{l:rbar-is-second-countable}.

\item[Lemma~\thref{l:augmented-topo-basis}] \mbox{}\\
  is explicitly cited in the proof of:\\
  Lemma~\thref{l:complete-count-topo-basis}.

\item[Definition~\thref{d:order-topo}] \mbox{}\\
  is explicitly cited in the proof of:\\
  Lemma~\thref{l:topological-basis-of-order-topo},\\
  Lemma~\thref{l:topo-of-rbar}.

\item[Lemma~\thref{l:topological-basis-of-order-topo}] \mbox{}\\
  is explicitly cited in the proof of:\\
  Lemma~\thref{l:topo-of-rbar},\\
  Lemma~\thref{l:trace-topo-on-r}.

\item[Lemma~\thref{l:trace-topo-on-subset}] \mbox{}\\
  is explicitly cited in the proof of:\\
  Lemma~\thref{l:trace-topo-on-r}.

\item[Lemma~\thref{l:box-topo-on-cartesian-prod}] \mbox{}\\
  is explicitly cited in the proof of:\\
  Lemma~\thref{l:compat-of-second-count-with-cartesian-prod},\\
  Lemma~\thref{l:complete-count-topo-basis-of-prod-space},\\
  Lemma~\thref{l:rn-is-second-countable}.

\item[Definition~\thref{d:second-count}] \mbox{}\\
  is explicitly cited in the proof of:\\
  Lemma~\thref{l:complete-count-topo-basis},\\
  Theorem~\thref{t:r-is-second-countable},\\
  Lemma~\thref{l:rn-is-second-countable},\\
  Lemma~\thref{l:rbar-is-second-countable}.

\item[Lemma~\thref{l:complete-count-topo-basis}] \mbox{}\\
  is explicitly cited in the proof of:\\
  Lemma~\thref{l:complete-count-topo-basis-of-prod-space}.

\item[Lemma~\thref{l:compat-of-second-count-with-cartesian-prod}] \mbox{}\\
  is explicitly cited in the proof of:\\
  Lemma~\thref{l:rn-is-second-countable}.

\item[Lemma~\thref{l:complete-count-topo-basis-of-prod-space}] \mbox{}\\
  is explicitly cited in the proof of:\\
  Lemma~\thref{l:borel-sigma-alg-of-rm}.

\item[Definition~\thref{d:pseudometric}] \mbox{}\\
  is not yet used.

\item[Lemma~\thref{l:equiv-def-of-conv-seq}] \mbox{}\\
  is explicitly cited in the proof of:\\
  Lemma~\thref{l:equiv-def-of-finite-inf}.

\item[Lemma~\thref{l:conv-subseq-of-cauchy-seq}] \mbox{}\\
  is not yet used.

\item[Definition~\thref{d:cluster-point}] \mbox{}\\
  is explicitly cited in the proof of:\\
  Lemma~\thref{l:equiv-def-of-liminf},\\
  Lemma~\thref{l:liminf-is-invariant-by-translation}.

\item[Lemma~\thref{l:finite-cover-of-compact-int}] \mbox{}\\
  is explicitly cited in the proof of:\\
  Lemma~\thref{l:lambda-star-gen-len-of-int}.

\item[Definition~\thref{d:holder-conjugates-in-r}] \mbox{}\\
  is explicitly cited in the proof of:\\
  Lemma~\thref{l:two-is-self-holder-conjugate-in-r},\\
  Lemma~\thref{l:youngs-ineq-for-prod-in-r}.

\item[Lemma~\thref{l:two-is-self-holder-conjugate-in-r}] \mbox{}\\
  is explicitly cited in the proof of:\\
  Lemma~\thref{l:youngs-ineq-for-prod-in-r-case-p-two},\\
  Lemma~\thref{l:youngs-ineq-for-prod-case-p-two-mt}.

\item[Lemma~\thref{l:youngs-ineq-for-prod-in-r}] \mbox{}\\
  is explicitly cited in the proof of:\\
  Lemma~\thref{l:youngs-ineq-for-prod-in-r-case-p-two},\\
  Lemma~\thref{l:youngs-ineq-for-prod-mt}.

\item[Lemma~\thref{l:youngs-ineq-for-prod-in-r-case-p-two}] \mbox{}\\
  is not yet used.

\item[Definition~\thref{d:ext-real-nums-rbar}] \mbox{}\\
  is explicitly cited in the proof of:\\
  Lemma~\thref{l:order-in-rbar-is-total},\\
  Lemma~\thref{l:additive-inverse-in-rbar-is-monot},\\
  Lemma~\thref{l:abs-in-rbar-satisfies-triangle-ineq},\\
  Lemma~\thref{l:topo-of-rbar},\\
  Lemma~\thref{l:finite-part-is-finite},\\
  Lemma~\thref{l:m-and-finite-is-mr},\\
  Lemma~\thref{l:m-is-closed-under-finite-part},\\
  Lemma~\thref{l:integrable-in-mplus-is-almost-finite}.

\item[Lemma~\thref{l:order-in-rbar-is-total}] \mbox{}\\
  is explicitly cited in the proof of:\\
  Lemma~\thref{l:inf-is-smaller-than-sup},\\
  Lemma~\thref{l:inf-is-monot},\\
  Lemma~\thref{l:inf-of-seq-is-monot},\\
  Lemma~\thref{l:liminf-is-smaller-than-limsup},\\
  Lemma~\thref{l:equiv-def-of-l},\\
  Lemma~\thref{l:bounded-by-integrable-in-mplus-is-integrable},\\
  Lemma~\thref{l:almost-bounded-by-integrable-is-integrable},\\
  Lemma~\thref{l:bounded-by-integrable-is-integrable},\\
  Lemma~\thref{l:bounded-by-llone-is-llone}.

\item[Definition~\thref{d:add-in-rbar}] \mbox{}\\
  is explicitly cited in the proof of:\\
  Lemma~\thref{l:zero-is-identity-element-for-add-in-rbar},\\
  Lemma~\thref{l:add-in-rbar-is-assoc-when-defined},\\
  Lemma~\thref{l:add-in-rbar-is-comm-when-defined},\\
  Lemma~\thref{l:infinity-sum-prop-in-rbar},\\
  Lemma~\thref{l:additive-inverse-in-rbar-is-monot},\\
  Lemma~\thref{l:mult-in-rbar-is-left-distr-over-add-when-defined},\\
  Lemma~\thref{l:abs-in-rbar-satisfies-triangle-ineq},\\
  Lemma~\thref{l:continuity-of-add-in-rbar},\\
  Lemma~\thref{l:add-in-rbarplus-is-closed},\\
  Lemma~\thref{l:add-in-rbarplus-is-assoc},\\
  Lemma~\thref{l:add-in-rbarplus-is-comm},\\
  Lemma~\thref{l:youngs-ineq-for-prod-mt},\\
  Lemma~\thref{l:decomp-into-nonneg-and-nonpos-parts},\\
  Lemma~\thref{l:compat-of-nonpos-and-nonneg-parts-with-add},\\
  Lemma~\thref{l:m-is-closed-under-add-when-defined},\\
  Lemma~\thref{l:meas-is-monot},\\
  Lemma~\thref{l:summability-on-summability-domain},\\
  Lemma~\thref{l:almost-sum},\\
  Lemma~\thref{l:int-is-add},\\
  Lemma~\thref{l:int-is-pos-lin-form-on-llone}.

\item[Lemma~\thref{l:zero-is-identity-element-for-add-in-rbar}] \mbox{}\\
  is explicitly cited in the proof of:\\
  Lemma~\thref{l:compat-of-int-in-mplus-with-almost-bin-rel}.

\item[Lemma~\thref{l:add-in-rbar-is-assoc-when-defined}] \mbox{}\\
  is explicitly cited in the proof of:\\
  Lemma~\thref{l:add-in-rbarplus-is-assoc}.

\item[Lemma~\thref{l:add-in-rbar-is-comm-when-defined}] \mbox{}\\
  is explicitly cited in the proof of:\\
  Lemma~\thref{l:add-in-rbarplus-is-comm}.

\item[Lemma~\thref{l:infinity-sum-prop-in-rbar}] \mbox{}\\
  is explicitly cited in the proof of:\\
  Lemma~\thref{l:infinity-sum-prop-in-rbarplus}.

\item[Lemma~\thref{l:additive-inverse-in-rbar-is-monot}] \mbox{}\\
  is explicitly cited in the proof of:\\
  Lemma~\thref{l:bounded-abs-in-rbar},\\
  Lemma~\thref{l:bounded-abs-in-rbar-strict}.

\item[Definition~\thref{d:mult-in-rbar}] \mbox{}\\
  is explicitly cited in the proof of:\\
  Lemma~\thref{l:mult-in-rbar-is-assoc-when-defined},\\
  Lemma~\thref{l:mult-in-rbar-is-comm-when-defined},\\
  Lemma~\thref{l:mult-in-rbar-is-left-distr-over-add-when-defined},\\
  Lemma~\thref{l:zero-prod-prop-in-rbar},\\
  Lemma~\thref{l:infinity-prod-prop-in-rbar},\\
  Lemma~\thref{l:finite-prod-prop-in-rbar},\\
  Lemma~\thref{l:exp-in-rbar},\\
  Lemma~\thref{l:continuity-of-mult-in-rbar},\\
  Lemma~\thref{l:mult-in-rbarplus-is-closed-when-defined},\\
  Lemma~\thref{l:mult-in-rbarplus-is-closed-mt},\\
  Lemma~\thref{l:youngs-ineq-for-prod-mt},\\
  Lemma~\thref{l:m-is-closed-under-mult},\\
  Lemma~\thref{l:meas-and-masking},\\
  Lemma~\thref{l:int-in-sfplus-is-pos-lin},\\
  Lemma~\thref{l:int-in-mplus-is-pos-hom},\\
  Lemma~\thref{l:int-in-mplus-of-zero-is-zero}.

\item[Lemma~\thref{l:mult-in-rbar-is-assoc-when-defined}] \mbox{}\\
  is explicitly cited in the proof of:\\
  Lemma~\thref{l:mult-in-rbarplus-is-assoc-mt}.

\item[Lemma~\thref{l:mult-in-rbar-is-comm-when-defined}] \mbox{}\\
  is explicitly cited in the proof of:\\
  Lemma~\thref{l:mult-in-rbar-is-right-distr-over-add-when-defined},\\
  Lemma~\thref{l:mult-in-rbarplus-is-comm-mt}.

\item[Lemma~\thref{l:mult-in-rbar-is-left-distr-over-add-when-defined}] \mbox{}\\
  is explicitly cited in the proof of:\\
  Lemma~\thref{l:mult-in-rbar-is-right-distr-over-add-when-defined},\\
  Lemma~\thref{l:mult-in-rbarplus-is-distr-over-add-mt}.

\item[Lemma~\thref{l:mult-in-rbar-is-right-distr-over-add-when-defined}] \mbox{}\\
  is explicitly cited in the proof of:\\
  Lemma~\thref{l:mult-in-rbarplus-is-distr-over-add-mt}.

\item[Lemma~\thref{l:zero-prod-prop-in-rbar}] \mbox{}\\
  is explicitly cited in the proof of:\\
  Lemma~\thref{l:zero-prod-prop-in-rbarplus},\\
  Lemma~\thref{l:zero-prod-prop-in-rbar-mt}.

\item[Lemma~\thref{l:infinity-prod-prop-in-rbar}] \mbox{}\\
  is explicitly cited in the proof of:\\
  Lemma~\thref{l:infinity-prod-prop-in-rbarplus},\\
  Lemma~\thref{l:infinity-prod-prop-in-rbar-mt}.

\item[Lemma~\thref{l:finite-prod-prop-in-rbar}] \mbox{}\\
  is explicitly cited in the proof of:\\
  Lemma~\thref{l:finite-prod-prop-in-rbarplus}.

\item[Definition~\thref{d:abs-in-rbar}] \mbox{}\\
  is explicitly cited in the proof of:\\
  Lemma~\thref{l:equiv-def-of-abs-in-rbar},\\
  Lemma~\thref{l:abs-in-rbar-is-nonneg},\\
  Lemma~\thref{l:abs-in-rbar-is-even},\\
  Lemma~\thref{l:abs-in-rbar-is-definite},\\
  Lemma~\thref{l:abs-in-rbar-satisfies-triangle-ineq},\\
  Lemma~\thref{l:abs-in-rbar-is-cont},\\
  Lemma~\thref{l:decomp-into-nonneg-and-nonpos-parts},\\
  Lemma~\thref{l:int-over-singleton}.

\item[Lemma~\thref{l:equiv-def-of-abs-in-rbar}] \mbox{}\\
  is explicitly cited in the proof of:\\
  Lemma~\thref{l:bounded-abs-in-rbar},\\
  Lemma~\thref{l:bounded-abs-in-rbar-strict},\\
  Lemma~\thref{l:compat-of-inf-with-abs},\\
  Lemma~\thref{l:compat-of-integrability-in-m-and-mplus}.

\item[Lemma~\thref{l:bounded-abs-in-rbar}] \mbox{}\\
  is not yet used.

\item[Lemma~\thref{l:bounded-abs-in-rbar-strict}] \mbox{}\\
  is explicitly cited in the proof of:\\
  Lemma~\thref{l:finite-abs-in-rbar}.

\item[Lemma~\thref{l:finite-abs-in-rbar}] \mbox{}\\
  is explicitly cited in the proof of:\\
  Lemma~\thref{l:bounded-by-llone-is-llone}.

\item[Lemma~\thref{l:abs-in-rbar-is-nonneg}] \mbox{}\\
  is explicitly cited in the proof of:\\
  Lemma~\thref{l:m-is-closed-under-abs},\\
  Lemma~\thref{l:bienayme-chebyshev-ineq},\\
  Lemma~\thref{l:integrable-in-mplus-is-almost-finite},\\
  Lemma~\thref{l:bounded-by-llone-is-llone},\\
  Theorem~\thref{t:lebesgue-dom-conv}.

\item[Lemma~\thref{l:abs-in-rbar-is-even}] \mbox{}\\
  is explicitly cited in the proof of:\\
  Lemma~\thref{l:compat-of-inf-with-abs},\\
  Lemma~\thref{l:compat-of-sup-with-abs},\\
  Lemma~\thref{l:compat-limsup-with-abs},\\
  Lemma~\thref{l:integrable-is-almost-finite}.

\item[Lemma~\thref{l:abs-in-rbar-is-definite}] \mbox{}\\
  is explicitly cited in the proof of:\\
  Lemma~\thref{l:abs-is-almost-definite},\\
  Lemma~\thref{l:int-over-singleton},\\
  Theorem~\thref{t:lebesgue-dom-conv}.

\item[Lemma~\thref{l:abs-in-rbar-satisfies-triangle-ineq}] \mbox{}\\
  is explicitly cited in the proof of:\\
  Lemma~\thref{l:minkowski-ineq-in-m}.

\item[Definition~\thref{d:exp-and-log-in-rbar}] \mbox{}\\
  is explicitly cited in the proof of:\\
  Lemma~\thref{l:exp-and-log-in-rbar-are-inverse},\\
  Lemma~\thref{l:exp-in-rbar}.

\item[Lemma~\thref{l:exp-and-log-in-rbar-are-inverse}] \mbox{}\\
  is not yet used.

\item[Definition~\thref{d:exp-in-rbar}] \mbox{}\\
  is explicitly cited in the proof of:\\
  Lemma~\thref{l:exp-in-rbar},\\
  Lemma~\thref{l:exp-in-rbar-mt}.

\item[Lemma~\thref{l:exp-in-rbar}] \mbox{}\\
  is explicitly cited in the proof of:\\
  Lemma~\thref{l:exp-in-rbar-mt},\\
  Lemma~\thref{l:youngs-ineq-for-prod-mt}.

\item[Lemma~\thref{l:topo-of-rbar}] \mbox{}\\
  is explicitly cited in the proof of:\\
  Lemma~\thref{l:conv-towards-minus-infinity},\\
  Lemma~\thref{l:rbar-is-second-countable}.

\item[Lemma~\thref{l:trace-topo-on-r}] \mbox{}\\
  is not yet used.

\item[Lemma~\thref{l:conv-towards-minus-infinity}] \mbox{}\\
  is explicitly cited in the proof of:\\
  Lemma~\thref{l:equiv-def-of-inf}.

\item[Lemma~\thref{l:continuity-of-add-in-rbar}] \mbox{}\\
  is not yet used.

\item[Lemma~\thref{l:continuity-of-mult-in-rbar}] \mbox{}\\
  is not yet used.

\item[Lemma~\thref{l:abs-in-rbar-is-cont}] \mbox{}\\
  is explicitly cited in the proof of:\\
  Lemma~\thref{l:m-is-closed-under-abs},\\
  Theorem~\thref{t:lebesgue-dom-conv}.

\item[Lemma~\thref{l:add-in-rbarplus-is-closed}] \mbox{}\\
  is explicitly cited in the proof of:\\
  Lemma~\thref{l:youngs-ineq-for-prod-mt},\\
  Lemma~\thref{l:mplus-is-closed-under-add},\\
  Lemma~\thref{l:int-in-sfplus},\\
  Lemma~\thref{l:equiv-def-of-integrability}.

\item[Lemma~\thref{l:add-in-rbarplus-is-assoc}] \mbox{}\\
  is explicitly cited in the proof of:\\
  Lemma~\thref{l:compat-of-nonpos-and-nonneg-parts-with-add},\\
  Lemma~\thref{l:int-in-sfplus-is-add-alt-proof}.

\item[Lemma~\thref{l:add-in-rbarplus-is-comm}] \mbox{}\\
  is explicitly cited in the proof of:\\
  Lemma~\thref{l:compat-of-nonpos-and-nonneg-parts-with-add},\\
  Lemma~\thref{l:l-is-closed-under-compl},\\
  Lemma~\thref{l:int-in-sfplus-is-add-alt-proof}.

\item[Lemma~\thref{l:infinity-sum-prop-in-rbarplus}] \mbox{}\\
  is explicitly cited in the proof of:\\
  Lemma~\thref{l:compat-of-nonpos-and-nonneg-parts-with-add},\\
  Lemma~\thref{l:equiv-def-of-integrability}.

\item[Lemma~\thref{l:series-are-conv-in-rbarplus}] \mbox{}\\
  is explicitly cited in the proof of:\\
  Lemma~\thref{l:technical-upper-bound-in-series-in-rbarplus},\\
  Lemma~\thref{l:double-series-in-rbarplus},\\
  Lemma~\thref{l:mplus-is-closed-under-count-sum}.

\item[Lemma~\thref{l:technical-upper-bound-in-series-in-rbarplus}] \mbox{}\\
  is explicitly cited in the proof of:\\
  Lemma~\thref{l:order-is-meaningless-in-series-in-rbarplus}.

\item[Lemma~\thref{l:order-is-meaningless-in-series-in-rbarplus}] \mbox{}\\
  is explicitly cited in the proof of:\\
  Lemma~\thref{l:def-of-double-series-in-rbarplus}.

\item[Lemma~\thref{l:def-of-double-series-in-rbarplus}] \mbox{}\\
  is explicitly cited in the proof of:\\
  Lemma~\thref{l:double-series-in-rbarplus},\\
  Lemma~\thref{l:lambda-star-is-sigma-subadd}.

\item[Lemma~\thref{l:double-series-in-rbarplus}] \mbox{}\\
  is explicitly cited in the proof of:\\
  Lemma~\thref{l:lambda-star-is-sigma-subadd}.

\item[Definition~\thref{d:mult-in-rbarplus}] \mbox{}\\
  is explicitly cited in the proof of:\\
  Lemma~\thref{l:mult-in-rbarplus-is-closed-when-defined},\\
  Lemma~\thref{l:zero-prod-prop-in-rbarplus},\\
  Lemma~\thref{l:infinity-prod-prop-in-rbarplus},\\
  Lemma~\thref{l:finite-prod-prop-in-rbarplus},\\
  Lemma~\thref{l:mult-in-rbarplus-is-closed-mt}.

\item[Lemma~\thref{l:mult-in-rbarplus-is-closed-when-defined}] \mbox{}\\
  is explicitly cited in the proof of:\\
  Lemma~\thref{l:youngs-ineq-for-prod-mt}.

\item[Lemma~\thref{l:zero-prod-prop-in-rbarplus}] \mbox{}\\
  is explicitly cited in the proof of:\\
  Lemma~\thref{l:zero-prod-prop-in-rbarplus-mt}.

\item[Lemma~\thref{l:infinity-prod-prop-in-rbarplus}] \mbox{}\\
  is explicitly cited in the proof of:\\
  Lemma~\thref{l:infinity-prod-prop-in-rbarplus-mt},\\
  Lemma~\thref{l:youngs-ineq-for-prod-mt},\\
  Lemma~\thref{l:m-is-closed-under-mult}.

\item[Lemma~\thref{l:finite-prod-prop-in-rbarplus}] \mbox{}\\
  is not yet used.

\item[Definition~\thref{d:mult-in-rbar-mt}] \mbox{}\\
  is explicitly cited in the proof of:\\
  Lemma~\thref{l:zero-prod-prop-in-rbar-mt},\\
  Lemma~\thref{l:infinity-prod-prop-in-rbar-mt},\\
  Lemma~\thref{l:mult-in-rbarplus-is-closed-mt},\\
  Lemma~\thref{l:mult-in-rbarplus-is-assoc-mt},\\
  Lemma~\thref{l:mult-in-rbarplus-is-comm-mt},\\
  Lemma~\thref{l:mult-in-rbarplus-is-distr-over-add-mt},\\
  Lemma~\thref{l:zero-prod-prop-in-rbarplus-mt},\\
  Lemma~\thref{l:infinity-prod-prop-in-rbarplus-mt},\\
  Lemma~\thref{l:exp-in-rbar-mt},\\
  Lemma~\thref{l:m-is-closed-under-mult}.

\item[Lemma~\thref{l:zero-prod-prop-in-rbar-mt}] \mbox{}\\
  is not yet used.

\item[Lemma~\thref{l:infinity-prod-prop-in-rbar-mt}] \mbox{}\\
  is explicitly cited in the proof of:\\
  Lemma~\thref{l:finite-prod-prop-in-rbar-mt}.

\item[Lemma~\thref{l:finite-prod-prop-in-rbar-mt}] \mbox{}\\
  is explicitly cited in the proof of:\\
  Lemma~\thref{l:finite-prod-prop-in-rbarplus-mt}.

\item[Lemma~\thref{l:mult-in-rbarplus-is-closed-mt}] \mbox{}\\
  is explicitly cited in the proof of:\\
  Lemma~\thref{l:youngs-ineq-for-prod-mt},\\
  Lemma~\thref{l:youngs-ineq-for-prod-case-p-two-mt},\\
  Lemma~\thref{l:mplus-is-closed-under-mult},\\
  Lemma~\thref{l:mplus-is-closed-under-nonneg-scalar-mult},\\
  Lemma~\thref{l:int-in-sfplus},\\
  Lemma~\thref{l:tonelli-for-tensor-prod}.

\item[Lemma~\thref{l:mult-in-rbarplus-is-assoc-mt}] \mbox{}\\
  is explicitly cited in the proof of:\\
  Lemma~\thref{l:youngs-ineq-for-prod-case-p-two-mt},\\
  Lemma~\thref{l:int-in-mplus-is-pos-hom}.

\item[Lemma~\thref{l:mult-in-rbarplus-is-comm-mt}] \mbox{}\\
  is explicitly cited in the proof of:\\
  Lemma~\thref{l:youngs-ineq-for-prod-case-p-two-mt},\\
  Lemma~\thref{l:cand-tensor-prod-meas-is-tensor-prod-meas},\\
  Lemma~\thref{l:tonelli-for-tensor-prod}.

\item[Lemma~\thref{l:mult-in-rbarplus-is-distr-over-add-mt}] \mbox{}\\
  is explicitly cited in the proof of:\\
  Lemma~\thref{l:int-in-sfplus-is-add-alt-proof},\\
  Lemma~\thref{l:compat-of-int-in-mplus-with-almost-bin-rel}.

\item[Lemma~\thref{l:zero-prod-prop-in-rbarplus-mt}] \mbox{}\\
  is explicitly cited in the proof of:\\
  Lemma~\thref{l:youngs-ineq-for-prod-mt},\\
  Lemma~\thref{l:m-is-closed-under-mult},\\
  Lemma~\thref{l:int-in-sfplus-is-pos-lin},\\
  Lemma~\thref{l:int-in-mplus-is-pos-hom},\\
  Lemma~\thref{l:int-in-mplus-of-zero-is-zero},\\
  Lemma~\thref{l:int-in-mplus-is-almost-definite},\\
  Lemma~\thref{l:compat-of-int-in-mplus-with-almost-bin-rel},\\
  Lemma~\thref{l:meas-of-section-of-prod},\\
  Lemma~\thref{l:lebesgue-meas-on-r2-is-zero-on-lines}.

\item[Lemma~\thref{l:infinity-prod-prop-in-rbarplus-mt}] \mbox{}\\
  is explicitly cited in the proof of:\\
  Lemma~\thref{l:youngs-ineq-for-prod-mt},\\
  Lemma~\thref{l:int-in-mplus-is-almost-definite}.

\item[Lemma~\thref{l:finite-prod-prop-in-rbarplus-mt}] \mbox{}\\
  is explicitly cited in the proof of:\\
  Lemma~\thref{l:int-over-singleton}.

\item[Lemma~\thref{l:exp-in-rbar-mt}] \mbox{}\\
  is not yet used.

\item[Definition~\thref{d:holder-conjugates}] \mbox{}\\
  is not yet used.

\item[Lemma~\thref{l:youngs-ineq-for-prod-mt}] \mbox{}\\
  is explicitly cited in the proof of:\\
  Lemma~\thref{l:youngs-ineq-for-prod-case-p-two-mt}.

\item[Lemma~\thref{l:youngs-ineq-for-prod-case-p-two-mt}] \mbox{}\\
  is not yet used.

\item[Definition~\thref{d:conn-comp-in-r}] \mbox{}\\
  is explicitly cited in the proof of:\\
  Lemma~\thref{l:conn-comp-of-open-subset-of-r-is-open-int},\\
  Lemma~\thref{l:conn-comp-of-open-subset-of-r-is-maximal},\\
  Theorem~\thref{t:count-conn-comps-of-open-subsets-of-r}.

\item[Lemma~\thref{l:conn-comp-of-open-subset-of-r-is-open-int}] \mbox{}\\
  is explicitly cited in the proof of:\\
  Lemma~\thref{l:conn-comp-of-open-subset-of-r-is-maximal}.

\item[Lemma~\thref{l:conn-comp-of-open-subset-of-r-is-maximal}] \mbox{}\\
  is explicitly cited in the proof of:\\
  Lemma~\thref{l:conn-comps-of-open-subset-of-r-equal-or-disj}.

\item[Lemma~\thref{l:conn-comps-of-open-subset-of-r-equal-or-disj}] \mbox{}\\
  is explicitly cited in the proof of:\\
  Theorem~\thref{t:count-conn-comps-of-open-subsets-of-r}.

\item[Theorem~\thref{t:count-conn-comps-of-open-subsets-of-r}] \mbox{}\\
  is explicitly cited in the proof of:\\
  Theorem~\thref{t:r-is-second-countable},\\
  Lemma~\thref{l:rbar-is-second-countable},\\
  Lemma~\thref{l:borel-sigma-alg-of-r},\\
  Lemma~\thref{l:count-gen-of-borel-sigma-alg-of-r},\\
  Lemma~\thref{l:borel-sigma-alg-of-rbar}.

\item[Lemma~\thref{l:rat-approx-of-lower-bound-of-open-int}] \mbox{}\\
  is explicitly cited in the proof of:\\
  Lemma~\thref{l:rat-approx-of-upper-bound-of-open-int},\\
  Lemma~\thref{l:open-int-with-rat-bounds-cover-open-int},\\
  Lemma~\thref{l:open-int-with-rat-bounds-cover-open-int-of-rbar}.

\item[Lemma~\thref{l:rat-approx-of-upper-bound-of-open-int}] \mbox{}\\
  is explicitly cited in the proof of:\\
  Lemma~\thref{l:open-int-with-rat-bounds-cover-open-int},\\
  Lemma~\thref{l:open-int-with-rat-bounds-cover-open-int-of-rbar}.

\item[Lemma~\thref{l:open-int-with-rat-bounds-cover-open-int}] \mbox{}\\
  is explicitly cited in the proof of:\\
  Theorem~\thref{t:r-is-second-countable},\\
  Lemma~\thref{l:open-int-with-rat-bounds-cover-open-int-of-rbar}.

\item[Theorem~\thref{t:r-is-second-countable}] \mbox{}\\
  is explicitly cited in the proof of:\\
  Lemma~\thref{l:rn-is-second-countable},\\
  Lemma~\thref{l:count-gen-of-borel-sigma-alg-of-r}.

\item[Lemma~\thref{l:rn-is-second-countable}] \mbox{}\\
  is explicitly cited in the proof of:\\
  Lemma~\thref{l:borel-sigma-alg-of-rm}.

\item[Lemma~\thref{l:open-int-with-rat-bounds-cover-open-int-of-rbar}] \mbox{}\\
  is explicitly cited in the proof of:\\
  Lemma~\thref{l:rbar-is-second-countable}.

\item[Lemma~\thref{l:rbar-is-second-countable}] \mbox{}\\
  is not yet used.

\item[Lemma~\thref{l:extr-of-const-fun}] \mbox{}\\
  is explicitly cited in the proof of:\\
  Lemma~\thref{l:equiv-def-of-finite-inf-in-rbar},\\
  Lemma~\thref{l:inf-of-bounded-seq-is-bounded},\\
  Lemma~\thref{l:sup-of-bounded-seq-is-bounded}.

\item[Lemma~\thref{l:equiv-def-of-finite-inf}] \mbox{}\\
  is explicitly cited in the proof of:\\
  Lemma~\thref{l:equiv-def-of-finite-inf-in-rbar}.

\item[Lemma~\thref{l:equiv-def-of-finite-inf-in-rbar}] \mbox{}\\
  is explicitly cited in the proof of:\\
  Lemma~\thref{l:equiv-def-of-inf}.

\item[Lemma~\thref{l:equiv-def-of-inf}] \mbox{}\\
  is not yet used.

\item[Lemma~\thref{l:inf-is-smaller-than-sup}] \mbox{}\\
  is explicitly cited in the proof of:\\
  Lemma~\thref{l:compat-of-inf-with-abs}.

\item[Lemma~\thref{l:inf-is-monot}] \mbox{}\\
  is explicitly cited in the proof of:\\
  Lemma~\thref{l:sup-is-monot},\\
  Lemma~\thref{l:compat-of-inf-with-abs},\\
  Lemma~\thref{l:compat-of-translation-with-inf},\\
  Lemma~\thref{l:liminf}.

\item[Lemma~\thref{l:sup-is-monot}] \mbox{}\\
  is explicitly cited in the proof of:\\
  Lemma~\thref{l:compat-of-inf-with-abs},\\
  Lemma~\thref{l:compat-of-translation-with-sup}.

\item[Lemma~\thref{l:compat-of-inf-with-abs}] \mbox{}\\
  is explicitly cited in the proof of:\\
  Lemma~\thref{l:compat-of-sup-with-abs},\\
  Lemma~\thref{l:compat-liminf-with-abs}.

\item[Lemma~\thref{l:compat-of-sup-with-abs}] \mbox{}\\
  is not yet used.

\item[Lemma~\thref{l:compat-of-translation-with-inf}] \mbox{}\\
  is not yet used.

\item[Lemma~\thref{l:compat-of-translation-with-sup}] \mbox{}\\
  is not yet used.

\item[Lemma~\thref{l:inf-of-seq-is-monot}] \mbox{}\\
  is explicitly cited in the proof of:\\
  Lemma~\thref{l:sup-of-seq-is-monot},\\
  Lemma~\thref{l:inf-of-bounded-seq-is-bounded},\\
  Lemma~\thref{l:liminf-is-monot}.

\item[Lemma~\thref{l:sup-of-seq-is-monot}] \mbox{}\\
  is explicitly cited in the proof of:\\
  Lemma~\thref{l:sup-of-bounded-seq-is-bounded},\\
  Lemma~\thref{l:liminf-is-monot}.

\item[Lemma~\thref{l:inf-of-bounded-seq-is-bounded}] \mbox{}\\
  is explicitly cited in the proof of:\\
  Lemma~\thref{l:mplus-is-closed-under-inf},\\
  Theorem~\thref{t:fatou-lemma}.

\item[Lemma~\thref{l:sup-of-bounded-seq-is-bounded}] \mbox{}\\
  is explicitly cited in the proof of:\\
  Lemma~\thref{l:mplus-is-closed-under-sup}.

\item[Lemma~\thref{l:liminf}] \mbox{}\\
  is explicitly cited in the proof of:\\
  Lemma~\thref{l:liminf-is-inf},\\
  Lemma~\thref{l:equiv-def-of-liminf},\\
  Lemma~\thref{l:liminf-is-monot},\\
  Lemma~\thref{l:duality-liminf-limsup},\\
  Lemma~\thref{l:compat-liminf-with-abs},\\
  Lemma~\thref{l:m-is-closed-under-liminf},\\
  Theorem~\thref{t:fatou-lemma}.

\item[Lemma~\thref{l:liminf-is-inf}] \mbox{}\\
  is explicitly cited in the proof of:\\
  Lemma~\thref{l:equiv-def-of-liminf},\\
  Lemma~\thref{l:liminf-limsup-and-pointwise-conv}.

\item[Lemma~\thref{l:equiv-def-of-liminf}] \mbox{}\\
  is explicitly cited in the proof of:\\
  Lemma~\thref{l:liminf-is-invariant-by-translation},\\
  Lemma~\thref{l:equiv-def-of-limsup},\\
  Lemma~\thref{l:liminf-is-smaller-than-limsup},\\
  Lemma~\thref{l:liminf-and-limsup-of-pointwise-conv},\\
  Lemma~\thref{l:liminf-limsup-and-pointwise-conv}.

\item[Lemma~\thref{l:liminf-is-invariant-by-translation}] \mbox{}\\
  is explicitly cited in the proof of:\\
  Lemma~\thref{l:liminf-is-monot}.

\item[Lemma~\thref{l:liminf-is-monot}] \mbox{}\\
  is explicitly cited in the proof of:\\
  Lemma~\thref{l:limsup-is-monot},\\
  Lemma~\thref{l:liminf-bounded-from-below},\\
  Lemma~\thref{l:liminf-bounded-from-above}.

\item[Lemma~\thref{l:limsup}] \mbox{}\\
  is explicitly cited in the proof of:\\
  Lemma~\thref{l:duality-liminf-limsup},\\
  Lemma~\thref{l:compat-liminf-with-abs},\\
  Lemma~\thref{l:m-is-closed-under-limsup}.

\item[Lemma~\thref{l:duality-liminf-limsup}] \mbox{}\\
  is explicitly cited in the proof of:\\
  Lemma~\thref{l:equiv-def-of-limsup},\\
  Lemma~\thref{l:limsup-is-monot},\\
  Lemma~\thref{l:compat-limsup-with-abs},\\
  Lemma~\thref{l:limsup-bounded-from-below},\\
  Lemma~\thref{l:limsup-bounded-from-above},\\
  Lemma~\thref{l:liminf-limsup-and-pointwise-conv},\\
  Theorem~\thref{t:lebesgue-dom-conv}.

\item[Lemma~\thref{l:equiv-def-of-limsup}] \mbox{}\\
  is explicitly cited in the proof of:\\
  Lemma~\thref{l:liminf-is-smaller-than-limsup},\\
  Lemma~\thref{l:liminf-and-limsup-of-pointwise-conv},\\
  Lemma~\thref{l:liminf-limsup-and-pointwise-conv}.

\item[Lemma~\thref{l:liminf-is-smaller-than-limsup}] \mbox{}\\
  is explicitly cited in the proof of:\\
  Lemma~\thref{l:liminf-limsup-and-pointwise-conv}.

\item[Lemma~\thref{l:limsup-is-monot}] \mbox{}\\
  is not yet used.

\item[Lemma~\thref{l:compat-liminf-with-abs}] \mbox{}\\
  is explicitly cited in the proof of:\\
  Lemma~\thref{l:compat-limsup-with-abs}.

\item[Lemma~\thref{l:compat-limsup-with-abs}] \mbox{}\\
  is not yet used.

\item[Definition~\thref{d:pointwise-conv}] \mbox{}\\
  is explicitly cited in the proof of:\\
  Lemma~\thref{l:liminf-and-limsup-of-pointwise-conv},\\
  Theorem~\thref{t:fatou-lemma}.

\item[Lemma~\thref{l:liminf-and-limsup-of-pointwise-conv}] \mbox{}\\
  is explicitly cited in the proof of:\\
  Lemma~\thref{l:liminf-bounded-from-below},\\
  Lemma~\thref{l:liminf-bounded-from-above},\\
  Lemma~\thref{l:int-in-mplus-of-pointwise-conv-seq},\\
  Theorem~\thref{t:lebesgue-dom-conv}.

\item[Lemma~\thref{l:liminf-bounded-from-below}] \mbox{}\\
  is explicitly cited in the proof of:\\
  Lemma~\thref{l:limsup-bounded-from-above},\\
  Theorem~\thref{t:fatou-lemma},\\
  Theorem~\thref{t:lebesgue-dom-conv}.

\item[Lemma~\thref{l:liminf-bounded-from-above}] \mbox{}\\
  is explicitly cited in the proof of:\\
  Lemma~\thref{l:limsup-bounded-from-below}.

\item[Lemma~\thref{l:limsup-bounded-from-below}] \mbox{}\\
  is not yet used.

\item[Lemma~\thref{l:limsup-bounded-from-above}] \mbox{}\\
  is not yet used.

\item[Lemma~\thref{l:liminf-limsup-and-pointwise-conv}] \mbox{}\\
  is explicitly cited in the proof of:\\
  Lemma~\thref{l:m-is-closed-under-limit-when-pointwise-conv},\\
  Lemma~\thref{l:int-in-mplus-of-pointwise-conv-seq},\\
  Theorem~\thref{t:lebesgue-dom-conv}.

\item[Definition~\thref{d:finite-part}] \mbox{}\\
  is explicitly cited in the proof of:\\
  Lemma~\thref{l:finite-part-is-finite},\\
  Lemma~\thref{l:m-is-closed-under-finite-part},\\
  Lemma~\thref{l:m-is-closed-under-add-when-defined},\\
  Lemma~\thref{l:m-is-closed-under-mult},\\
  Lemma~\thref{l:finite-nonneg-part}.

\item[Lemma~\thref{l:finite-part-is-finite}] \mbox{}\\
  is explicitly cited in the proof of:\\
  Lemma~\thref{l:m-is-closed-under-finite-part}.

\item[Definition~\thref{d:nonneg-and-nonpos-parts}] \mbox{}\\
  is explicitly cited in the proof of:\\
  Lemma~\thref{l:equiv-def-of-nonneg-and-nonpos-parts},\\
  Lemma~\thref{l:nonneg-and-nonpos-parts-are-nonneg},\\
  Lemma~\thref{l:compat-of-nonpos-and-nonneg-parts-with-add},\\
  Lemma~\thref{l:compat-of-nonpos-and-nonneg-parts-with-mask},\\
  Lemma~\thref{l:compat-of-nonpos-and-nonneg-parts-with-restr},\\
  Lemma~\thref{l:meas-of-nonneg-and-nonpos-parts},\\
  Lemma~\thref{l:compat-of-int-in-m-and-mplus},\\
  Lemma~\thref{l:int-over-singleton},\\
  Lemma~\thref{l:int-is-hom}.

\item[Lemma~\thref{l:equiv-def-of-nonneg-and-nonpos-parts}] \mbox{}\\
  is explicitly cited in the proof of:\\
  Lemma~\thref{l:finite-nonneg-part}.

\item[Lemma~\thref{l:nonneg-and-nonpos-parts-are-nonneg}] \mbox{}\\
  is explicitly cited in the proof of:\\
  Lemma~\thref{l:compat-of-nonpos-and-nonneg-parts-with-add},\\
  Lemma~\thref{l:meas-of-nonneg-and-nonpos-parts},\\
  Lemma~\thref{l:int-is-pos-lin-form-on-llone}.

\item[Lemma~\thref{l:nonneg-and-nonpos-parts-are-orthogonal}] \mbox{}\\
  is explicitly cited in the proof of:\\
  Lemma~\thref{l:decomp-into-nonneg-and-nonpos-parts},\\
  Lemma~\thref{l:compat-of-nonpos-and-nonneg-parts-with-add}.

\item[Lemma~\thref{l:decomp-into-nonneg-and-nonpos-parts}] \mbox{}\\
  is explicitly cited in the proof of:\\
  Lemma~\thref{l:compat-of-nonpos-and-nonneg-parts-with-add},\\
  Lemma~\thref{l:meas-of-nonneg-and-nonpos-parts},\\
  Lemma~\thref{l:int-in-mplus-of-decomp-into-nonpos-and-nonneg-parts},\\
  Lemma~\thref{l:int-over-singleton},\\
  Lemma~\thref{l:int-for-count-meas}.

\item[Lemma~\thref{l:compat-of-nonpos-and-nonneg-parts-with-add}] \mbox{}\\
  is explicitly cited in the proof of:\\
  Lemma~\thref{l:compat-of-int-in-mplus-with-nonpos-and-nonneg-parts}.

\item[Lemma~\thref{l:compat-of-nonpos-and-nonneg-parts-with-mask}] \mbox{}\\
  is explicitly cited in the proof of:\\
  Lemma~\thref{l:int-over-subset}.

\item[Lemma~\thref{l:compat-of-nonpos-and-nonneg-parts-with-restr}] \mbox{}\\
  is explicitly cited in the proof of:\\
  Lemma~\thref{l:int-over-subset}.

\item[Lemma~\thref{l:nonempty-and-empty-or-full}] \mbox{}\\
  is explicitly cited in the proof of:\\
  Lemma~\thref{l:equiv-def-of-set-alg},\\
  Lemma~\thref{l:other-prop-of-l-syst},\\
  Lemma~\thref{l:equiv-def-of-sigma-alg},\\
  Lemma~\thref{l:p-syst-contains-sigma-alg}.

\item[Lemma~\thref{l:empty-and-full}] \mbox{}\\
  is explicitly cited in the proof of:\\
  Lemma~\thref{l:count-union-disj-local-compl},\\
  Lemma~\thref{l:equiv-def-of-set-alg},\\
  Lemma~\thref{l:other-prop-of-l-syst},\\
  Lemma~\thref{l:equiv-def-of-sigma-alg}.

\item[Lemma~\thref{l:local-compl-and-compl}] \mbox{}\\
  is explicitly cited in the proof of:\\
  Lemma~\thref{l:other-equiv-def-of-set-alg},\\
  Lemma~\thref{l:equiv-def-of-l-syst}.

\item[Lemma~\thref{l:union-disj-local-compl-equiv}] \mbox{}\\
  is explicitly cited in the proof of:\\
  Lemma~\thref{l:count-union-disj-local-compl}.

\item[Lemma~\thref{l:set-diff-and-local-compl}] \mbox{}\\
  is explicitly cited in the proof of:\\
  Lemma~\thref{l:count-union-monot-and-union},\\
  Lemma~\thref{l:other-equiv-def-of-set-alg},\\
  Lemma~\thref{l:set-alg-is-closed-under-local-compl}.

\item[Lemma~\thref{l:inter-set-diff-equiv}] \mbox{}\\
  is explicitly cited in the proof of:\\
  Lemma~\thref{l:union-set-diff-equiv},\\
  Lemma~\thref{l:count-union-disj-and-union},\\
  Lemma~\thref{l:count-union-monot-and-union},\\
  Lemma~\thref{l:other-equiv-def-of-set-alg},\\
  Lemma~\thref{l:l-is-set-alg}.

\item[Lemma~\thref{l:union-inter-equiv}] \mbox{}\\
  is explicitly cited in the proof of:\\
  Lemma~\thref{l:union-set-diff-equiv},\\
  Lemma~\thref{l:finite-union-inter-equiv},\\
  Lemma~\thref{l:count-union-disj-and-union},\\
  Lemma~\thref{l:count-union-monot-and-union}.

\item[Lemma~\thref{l:union-set-diff-equiv}] \mbox{}\\
  is not yet used.

\item[Lemma~\thref{l:finite-ops-equiv}] \mbox{}\\
  is explicitly cited in the proof of:\\
  Lemma~\thref{l:finite-union-inter-equiv},\\
  Lemma~\thref{l:count-union-disj-local-compl},\\
  Lemma~\thref{l:count-union-disj-and-union},\\
  Lemma~\thref{l:equiv-def-of-set-alg},\\
  Lemma~\thref{l:explicit-set-alg},\\
  Lemma~\thref{l:l-syst-gen-by-p-syst}.

\item[Lemma~\thref{l:finite-union-inter-equiv}] \mbox{}\\
  is explicitly cited in the proof of:\\
  Lemma~\thref{l:equiv-def-of-set-alg}.

\item[Lemma~\thref{l:count-and-finite-union-disj}] \mbox{}\\
  is explicitly cited in the proof of:\\
  Lemma~\thref{l:count-union-disj-local-compl}.

\item[Lemma~\thref{l:count-union-disj-local-compl}] \mbox{}\\
  is explicitly cited in the proof of:\\
  Lemma~\thref{l:equiv-def-of-l-syst}.

\item[Lemma~\thref{l:count-union-inter-equiv}] \mbox{}\\
  is explicitly cited in the proof of:\\
  Lemma~\thref{l:other-prop-of-l-syst},\\
  Lemma~\thref{l:equiv-def-of-sigma-alg}.

\item[Lemma~\thref{l:count-union-disj-and-monot}] \mbox{}\\
  is explicitly cited in the proof of:\\
  Lemma~\thref{l:equiv-def-of-l-syst}.

\item[Lemma~\thref{l:count-union-monot-and-disj}] \mbox{}\\
  is explicitly cited in the proof of:\\
  Lemma~\thref{l:count-union-monot-and-union},\\
  Lemma~\thref{l:equiv-def-of-l-syst}.

\item[Lemma~\thref{l:count-union-disj-and-union}] \mbox{}\\
  is explicitly cited in the proof of:\\
  Lemma~\thref{l:count-union-monot-and-union},\\
  Lemma~\thref{l:p-syst-contains-sigma-alg},\\
  Lemma~\thref{l:l-syst-contains-sigma-alg}.

\item[Lemma~\thref{l:count-union-monot-and-union}] \mbox{}\\
  is explicitly cited in the proof of:\\
  Lemma~\thref{l:set-alg-contains-sigma-alg},\\
  Lemma~\thref{l:monot-class-contains-sigma-alg}.

\item[Definition~\thref{d:p-syst}] \mbox{}\\
  is explicitly cited in the proof of:\\
  Lemma~\thref{l:inter-of-p-systs},\\
  Lemma~\thref{l:l-syst-gen-by-p-syst},\\
  Lemma~\thref{l:sigma-alg-is-p-syst},\\
  Lemma~\thref{l:p-syst-contains-sigma-alg},\\
  Lemma~\thref{l:p-syst-and-l-syst-is-sigma-alg},\\
  Lemma~\thref{l:uniq-of-meas-ext-from-p-syst},\\
  Theorem~\thref{t:caratheodory-lebesgue-meas-on-r}.

\item[Lemma~\thref{l:inter-of-p-systs}] \mbox{}\\
  is explicitly cited in the proof of:\\
  Lemma~\thref{l:gen-p-syst-is-min}.

\item[Definition~\thref{d:gen-p-syst}] \mbox{}\\
  is explicitly cited in the proof of:\\
  Lemma~\thref{l:gen-p-syst-is-min}.

\item[Lemma~\thref{l:gen-p-syst-is-min}] \mbox{}\\
  is explicitly cited in the proof of:\\
  Lemma~\thref{l:p-syst-gen-is-monot},\\
  Lemma~\thref{l:p-syst-gen-is-idem},\\
  Lemma~\thref{l:sigma-alg-contains-p-syst},\\
  Lemma~\thref{l:p-syst-contains-sigma-alg},\\
  Lemma~\thref{l:sigma-alg-gen-by-p-syst},\\
  Lemma~\thref{l:usage-of-dynkin-pi-lambda-th}.

\item[Lemma~\thref{l:p-syst-gen-is-monot}] \mbox{}\\
  is not yet used.

\item[Lemma~\thref{l:p-syst-gen-is-idem}] \mbox{}\\
  is explicitly cited in the proof of:\\
  Lemma~\thref{l:uniq-of-meas-ext-from-p-syst}.

\item[Definition~\thref{d:set-alg}] \mbox{}\\
  is explicitly cited in the proof of:\\
  Lemma~\thref{l:equiv-def-of-set-alg},\\
  Lemma~\thref{l:inter-of-set-algs},\\
  Lemma~\thref{l:explicit-set-alg},\\
  Lemma~\thref{l:sigma-alg-is-set-alg},\\
  Lemma~\thref{l:set-alg-contains-sigma-alg},\\
  Lemma~\thref{l:set-alg-and-monot-class-is-sigma-alg},\\
  Lemma~\thref{l:l-is-sigma-alg}.

\item[Lemma~\thref{l:equiv-def-of-set-alg}] \mbox{}\\
  is explicitly cited in the proof of:\\
  Lemma~\thref{l:other-equiv-def-of-set-alg},\\
  Lemma~\thref{l:part-of-count-union-in-set-alg},\\
  Lemma~\thref{l:explicit-set-alg}.

\item[Lemma~\thref{l:other-equiv-def-of-set-alg}] \mbox{}\\
  is explicitly cited in the proof of:\\
  Lemma~\thref{l:set-alg-is-closed-under-local-compl},\\
  Lemma~\thref{l:part-of-count-union-in-set-alg},\\
  Lemma~\thref{l:monot-class-gen-by-set-alg},\\
  Lemma~\thref{l:sigma-alg-is-closed-under-set-diff},\\
  Lemma~\thref{l:l-is-set-alg}.

\item[Lemma~\thref{l:set-alg-is-closed-under-local-compl}] \mbox{}\\
  is explicitly cited in the proof of:\\
  Lemma~\thref{l:sigma-alg-is-closed-under-set-diff}.

\item[Lemma~\thref{l:inter-of-set-algs}] \mbox{}\\
  is explicitly cited in the proof of:\\
  Lemma~\thref{l:gen-set-alg-is-min}.

\item[Definition~\thref{d:gen-set-alg}] \mbox{}\\
  is explicitly cited in the proof of:\\
  Lemma~\thref{l:gen-set-alg-is-min},\\
  Lemma~\thref{l:meas-of-meas-of-section-finite}.

\item[Lemma~\thref{l:gen-set-alg-is-min}] \mbox{}\\
  is explicitly cited in the proof of:\\
  Lemma~\thref{l:set-alg-gen-is-monot},\\
  Lemma~\thref{l:set-alg-gen-is-idem},\\
  Lemma~\thref{l:explicit-set-alg},\\
  Lemma~\thref{l:sigma-alg-contains-set-alg},\\
  Lemma~\thref{l:set-alg-contains-sigma-alg},\\
  Lemma~\thref{l:sigma-alg-gen-by-set-alg},\\
  Lemma~\thref{l:usage-of-monot-class-th},\\
  Lemma~\thref{l:meas-of-meas-of-section-finite}.

\item[Lemma~\thref{l:set-alg-gen-is-monot}] \mbox{}\\
  is not yet used.

\item[Lemma~\thref{l:set-alg-gen-is-idem}] \mbox{}\\
  is not yet used.

\item[Lemma~\thref{l:part-of-count-union-in-set-alg}] \mbox{}\\
  is explicitly cited in the proof of:\\
  Lemma~\thref{l:part-of-count-union-in-sigma-alg},\\
  Lemma~\thref{l:part-of-count-union-in-l}.

\item[Lemma~\thref{l:explicit-set-alg}] \mbox{}\\
  is explicitly cited in the proof of:\\
  Lemma~\thref{l:set-alg-gen-by-prod-of-sigma-algs}.

\item[Definition~\thref{d:monot-class}] \mbox{}\\
  is explicitly cited in the proof of:\\
  Lemma~\thref{l:inter-of-monot-classes},\\
  Lemma~\thref{l:c-diff-is-monot-class},\\
  Lemma~\thref{l:sigma-alg-is-monot-class},\\
  Lemma~\thref{l:monot-class-contains-sigma-alg},\\
  Lemma~\thref{l:meas-of-meas-of-section-finite},\\
  Lemma~\thref{l:uniq-of-tensor-prod-meas-finite}.

\item[Lemma~\thref{l:inter-of-monot-classes}] \mbox{}\\
  is explicitly cited in the proof of:\\
  Lemma~\thref{l:gen-monot-class-is-min}.

\item[Definition~\thref{d:gen-monot-class}] \mbox{}\\
  is explicitly cited in the proof of:\\
  Lemma~\thref{l:gen-monot-class-is-min}.

\item[Lemma~\thref{l:gen-monot-class-is-min}] \mbox{}\\
  is explicitly cited in the proof of:\\
  Lemma~\thref{l:monot-class-gen-is-monot},\\
  Lemma~\thref{l:monot-class-gen-is-idem},\\
  Lemma~\thref{l:monot-class-is-closed-under-set-diff},\\
  Lemma~\thref{l:monot-class-gen-by-set-alg},\\
  Lemma~\thref{l:sigma-alg-contains-monot-class},\\
  Lemma~\thref{l:monot-class-contains-sigma-alg},\\
  Lemma~\thref{l:sigma-alg-gen-by-monot-class},\\
  Theorem~\thref{t:monot-class}.

\item[Lemma~\thref{l:monot-class-gen-is-monot}] \mbox{}\\
  is explicitly cited in the proof of:\\
  Lemma~\thref{l:usage-of-monot-class-th}.

\item[Lemma~\thref{l:monot-class-gen-is-idem}] \mbox{}\\
  is explicitly cited in the proof of:\\
  Lemma~\thref{l:set-alg-and-monot-class-is-sigma-alg},\\
  Lemma~\thref{l:usage-of-monot-class-th}.

\item[Definition~\thref{d:monot-class-and-symm-set-diff}] \mbox{}\\
  is explicitly cited in the proof of:\\
  Lemma~\thref{l:c-diff-is-symmetric},\\
  Lemma~\thref{l:c-diff-is-monot-class},\\
  Lemma~\thref{l:monot-class-is-closed-under-set-diff}.

\item[Lemma~\thref{l:c-diff-is-symmetric}] \mbox{}\\
  is explicitly cited in the proof of:\\
  Lemma~\thref{l:monot-class-is-closed-under-set-diff}.

\item[Lemma~\thref{l:c-diff-is-monot-class}] \mbox{}\\
  is explicitly cited in the proof of:\\
  Lemma~\thref{l:monot-class-is-closed-under-set-diff}.

\item[Lemma~\thref{l:monot-class-is-closed-under-set-diff}] \mbox{}\\
  is explicitly cited in the proof of:\\
  Lemma~\thref{l:monot-class-gen-by-set-alg}.

\item[Lemma~\thref{l:monot-class-gen-by-set-alg}] \mbox{}\\
  is explicitly cited in the proof of:\\
  Theorem~\thref{t:monot-class}.

\item[Definition~\thref{d:l-syst}] \mbox{}\\
  is explicitly cited in the proof of:\\
  Lemma~\thref{l:equiv-def-of-l-syst},\\
  Lemma~\thref{l:other-prop-of-l-syst},\\
  Lemma~\thref{l:inter-of-l-systs},\\
  Lemma~\thref{l:l-inter-is-l-syst},\\
  Lemma~\thref{l:sigma-alg-is-l-syst},\\
  Lemma~\thref{l:l-syst-contains-sigma-alg}.

\item[Lemma~\thref{l:equiv-def-of-l-syst}] \mbox{}\\
  is explicitly cited in the proof of:\\
  Lemma~\thref{l:other-prop-of-l-syst},\\
  Lemma~\thref{l:l-inter-is-l-syst},\\
  Lemma~\thref{l:uniq-of-meas-ext-from-p-syst}.

\item[Lemma~\thref{l:other-prop-of-l-syst}] \mbox{}\\
  is not yet used.

\item[Lemma~\thref{l:inter-of-l-systs}] \mbox{}\\
  is explicitly cited in the proof of:\\
  Lemma~\thref{l:gen-l-syst-is-min}.

\item[Definition~\thref{d:gen-l-syst}] \mbox{}\\
  is explicitly cited in the proof of:\\
  Lemma~\thref{l:gen-l-syst-is-min}.

\item[Lemma~\thref{l:gen-l-syst-is-min}] \mbox{}\\
  is explicitly cited in the proof of:\\
  Lemma~\thref{l:l-syst-gen-is-monot},\\
  Lemma~\thref{l:l-syst-gen-is-idem},\\
  Lemma~\thref{l:l-syst-with-inter},\\
  Lemma~\thref{l:l-syst-gen-by-p-syst},\\
  Lemma~\thref{l:sigma-alg-contains-l-syst},\\
  Lemma~\thref{l:l-syst-contains-sigma-alg},\\
  Lemma~\thref{l:sigma-alg-gen-by-l-syst},\\
  Theorem~\thref{t:dynkin-pi-lambda-th}.

\item[Lemma~\thref{l:l-syst-gen-is-monot}] \mbox{}\\
  is explicitly cited in the proof of:\\
  Lemma~\thref{l:usage-of-dynkin-pi-lambda-th}.

\item[Lemma~\thref{l:l-syst-gen-is-idem}] \mbox{}\\
  is explicitly cited in the proof of:\\
  Lemma~\thref{l:p-syst-and-l-syst-is-sigma-alg},\\
  Lemma~\thref{l:usage-of-dynkin-pi-lambda-th}.

\item[Definition~\thref{d:l-syst-and-inter}] \mbox{}\\
  is explicitly cited in the proof of:\\
  Lemma~\thref{l:l-inter-is-symmetric},\\
  Lemma~\thref{l:l-syst-with-inter},\\
  Lemma~\thref{l:l-syst-is-closed-under-inter}.

\item[Lemma~\thref{l:l-inter-is-symmetric}] \mbox{}\\
  is explicitly cited in the proof of:\\
  Lemma~\thref{l:l-syst-with-inter}.

\item[Lemma~\thref{l:l-inter-is-l-syst}] \mbox{}\\
  is explicitly cited in the proof of:\\
  Lemma~\thref{l:l-syst-with-inter}.

\item[Lemma~\thref{l:l-syst-with-inter}] \mbox{}\\
  is explicitly cited in the proof of:\\
  Lemma~\thref{l:l-syst-is-closed-under-inter}.

\item[Lemma~\thref{l:l-syst-is-closed-under-inter}] \mbox{}\\
  is explicitly cited in the proof of:\\
  Lemma~\thref{l:l-syst-gen-by-p-syst}.

\item[Lemma~\thref{l:l-syst-gen-by-p-syst}] \mbox{}\\
  is explicitly cited in the proof of:\\
  Theorem~\thref{t:dynkin-pi-lambda-th}.

\item[Definition~\thref{d:sigma-alg}] \mbox{}\\
  is explicitly cited in the proof of:\\
  Lemma~\thref{l:equiv-def-of-sigma-alg},\\
  Lemma~\thref{l:sigma-alg-is-set-alg},\\
  Lemma~\thref{l:part-of-count-union-in-sigma-alg},\\
  Lemma~\thref{l:inter-of-sigma-algs},\\
  Lemma~\thref{l:p-syst-contains-sigma-alg},\\
  Lemma~\thref{l:set-alg-contains-sigma-alg},\\
  Lemma~\thref{l:monot-class-contains-sigma-alg},\\
  Lemma~\thref{l:count-sigma-alg-gen},\\
  Lemma~\thref{l:some-borel-subsets},\\
  Lemma~\thref{l:inverse-sigma-alg},\\
  Lemma~\thref{l:image-sigma-alg},\\
  Lemma~\thref{l:trace-sigma-alg},\\
  Lemma~\thref{l:characterization-of-borel-subsets},\\
  Lemma~\thref{l:measurability-of-section},\\
  Lemma~\thref{l:count-union-of-sections-is-meas},\\
  Lemma~\thref{l:borel-sigma-alg-of-r},\\
  Lemma~\thref{l:borel-subsets-of-rbar-and-r},\\
  Lemma~\thref{l:borel-sigma-alg-of-rplus},\\
  Lemma~\thref{l:meas-of-indic-fun},\\
  Lemma~\thref{l:m-is-closed-under-finite-part},\\
  Lemma~\thref{l:meas-and-masking},\\
  Lemma~\thref{l:meas-of-restr},\\
  Lemma~\thref{l:meas-satisfies-finite-boole-ineq},\\
  Lemma~\thref{l:meas-satisfies-boole-ineq},\\
  Lemma~\thref{l:equiv-def-of-meas},\\
  Lemma~\thref{l:equiv-def-of-sigma-finite-meas},\\
  Lemma~\thref{l:empty-set-is-negl},\\
  Lemma~\thref{l:compat-of-null-meas-with-count-union},\\
  Lemma~\thref{l:count-meas},\\
  Lemma~\thref{l:sigma-finiteness-of-count-meas},\\
  Lemma~\thref{l:masking-almost-nowhere},\\
  Lemma~\thref{l:finite-nonneg-part},\\
  Lemma~\thref{l:l-is-sigma-alg},\\
  Lemma~\thref{l:if-is-sigma-add},\\
  Lemma~\thref{l:int-in-if-over-subset-is-add},\\
  Lemma~\thref{l:compat-of-int-in-mplus-with-almost-bin-rel},\\
  Lemma~\thref{l:const-fun-is-llone},\\
  Theorem~\thref{t:lebesgue-ext-dom-conv}.

\item[Lemma~\thref{l:equiv-def-of-sigma-alg}] \mbox{}\\
  is explicitly cited in the proof of:\\
  Lemma~\thref{l:other-prop-of-sigma-alg},\\
  Lemma~\thref{l:sigma-alg-is-p-syst},\\
  Lemma~\thref{l:sigma-alg-is-l-syst},\\
  Lemma~\thref{l:l-syst-contains-sigma-alg},\\
  Lemma~\thref{l:set-alg-gen-by-prod-of-sigma-algs},\\
  Lemma~\thref{l:const-fun-is-meas},\\
  Lemma~\thref{l:meas-of-meas-subspace},\\
  Lemma~\thref{l:characterization-of-borel-subsets},\\
  Lemma~\thref{l:meas-of-fun-def-on-pseudopart},\\
  Lemma~\thref{l:meas-of-fun-to-prod-space},\\
  Lemma~\thref{l:count-inter-of-sections-is-meas},\\
  Lemma~\thref{l:borel-sigma-alg-of-r},\\
  Lemma~\thref{l:borel-sigma-alg-of-rbar},\\
  Lemma~\thref{l:borel-sigma-alg-of-rm},\\
  Lemma~\thref{l:meas-of-indic-fun},\\
  Lemma~\thref{l:m-is-closed-under-add-when-defined},\\
  Lemma~\thref{l:m-is-closed-under-mult},\\
  Lemma~\thref{l:m-is-closed-under-inf},\\
  Lemma~\thref{l:m-is-closed-under-sup},\\
  Lemma~\thref{l:meas-of-restr},\\
  Lemma~\thref{l:meas-over-count-pseudopart},\\
  Lemma~\thref{l:meas-is-cont-from-above},\\
  Lemma~\thref{l:finite-meas-is-bounded},\\
  Lemma~\thref{l:restr-meas},\\
  Lemma~\thref{l:uniq-of-meas-ext-from-p-syst},\\
  Lemma~\thref{l:finiteness-of-count-meas},\\
  Lemma~\thref{l:meas-of-summability-domain},\\
  Lemma~\thref{l:if-is-closed-under-mult},\\
  Lemma~\thref{l:sf-can-repr},\\
  Lemma~\thref{l:sf-is-alg-over-r},\\
  Lemma~\thref{l:change-of-variable-in-sum-in-sfplus},\\
  Lemma~\thref{l:int-in-sfplus-is-add-alt-proof},\\
  Theorem~\thref{t:beppo-levi-monot-conv},\\
  Lemma~\thref{l:adapted-seq-in-mplus},\\
  Lemma~\thref{l:meas-of-meas-of-section},\\
  Lemma~\thref{l:tensor-prod-of-finite-meas},\\
  Lemma~\thref{l:uniq-of-tensor-prod-meas-finite},\\
  Lemma~\thref{l:uniq-of-tensor-prod-meas},\\
  Theorem~\thref{t:lebesgue-ext-dom-conv}.

\item[Lemma~\thref{l:sigma-alg-is-set-alg}] \mbox{}\\
  is explicitly cited in the proof of:\\
  Lemma~\thref{l:sigma-alg-is-closed-under-set-diff},\\
  Lemma~\thref{l:part-of-count-union-in-sigma-alg},\\
  Lemma~\thref{l:sigma-alg-contains-set-alg}.

\item[Lemma~\thref{l:sigma-alg-is-closed-under-set-diff}] \mbox{}\\
  is explicitly cited in the proof of:\\
  Lemma~\thref{l:meas-is-monot},\\
  Lemma~\thref{l:meas-satisfies-finite-boole-ineq},\\
  Lemma~\thref{l:meas-is-cont-from-above}.

\item[Lemma~\thref{l:other-prop-of-sigma-alg}] \mbox{}\\
  is explicitly cited in the proof of:\\
  Lemma~\thref{l:sigma-alg-is-monot-class},\\
  Lemma~\thref{l:sigma-alg-is-l-syst}.

\item[Lemma~\thref{l:part-of-count-union-in-sigma-alg}] \mbox{}\\
  is explicitly cited in the proof of:\\
  Lemma~\thref{l:meas-is-cont-from-below}.

\item[Lemma~\thref{l:inter-of-sigma-algs}] \mbox{}\\
  is explicitly cited in the proof of:\\
  Lemma~\thref{l:gen-sigma-alg-is-min}.

\item[Definition~\thref{d:gen-sigma-alg}] \mbox{}\\
  is explicitly cited in the proof of:\\
  Lemma~\thref{l:gen-sigma-alg-is-min},\\
  Lemma~\thref{l:usage-of-dynkin-pi-lambda-th},\\
  Lemma~\thref{l:usage-of-monot-class-th},\\
  Lemma~\thref{l:borel-sigma-alg-of-r},\\
  Lemma~\thref{l:borel-sigma-alg-of-rbar}.

\item[Lemma~\thref{l:gen-sigma-alg-is-min}] \mbox{}\\
  is explicitly cited in the proof of:\\
  Lemma~\thref{l:sigma-alg-gen-is-monot},\\
  Lemma~\thref{l:sigma-alg-gen-is-idem},\\
  Lemma~\thref{l:sigma-alg-contains-p-syst},\\
  Lemma~\thref{l:p-syst-contains-sigma-alg},\\
  Lemma~\thref{l:sigma-alg-gen-by-p-syst},\\
  Lemma~\thref{l:sigma-alg-contains-set-alg},\\
  Lemma~\thref{l:set-alg-contains-sigma-alg},\\
  Lemma~\thref{l:sigma-alg-gen-by-set-alg},\\
  Lemma~\thref{l:sigma-alg-contains-monot-class},\\
  Lemma~\thref{l:monot-class-contains-sigma-alg},\\
  Lemma~\thref{l:sigma-alg-gen-by-monot-class},\\
  Lemma~\thref{l:sigma-alg-contains-l-syst},\\
  Lemma~\thref{l:l-syst-contains-sigma-alg},\\
  Lemma~\thref{l:sigma-alg-gen-by-l-syst},\\
  Lemma~\thref{l:complete-gen-sigma-alg},\\
  Lemma~\thref{l:count-sigma-alg-gen},\\
  Lemma~\thref{l:some-borel-subsets},\\
  Lemma~\thref{l:inverse-image-of-gen-family},\\
  Lemma~\thref{l:cont-is-meas},\\
  Lemma~\thref{l:prod-of-meas-subsets-is-meas},\\
  Lemma~\thref{l:gen-prod-meas-space},\\
  Lemma~\thref{l:measurability-of-section},\\
  Lemma~\thref{l:uniq-of-meas-ext-from-p-syst},\\
  Theorem~\thref{t:caratheodory-lebesgue-meas-on-r}.

\item[Lemma~\thref{l:sigma-alg-gen-is-monot}] \mbox{}\\
  is explicitly cited in the proof of:\\
  Lemma~\thref{l:sigma-alg-gen-by-p-syst},\\
  Lemma~\thref{l:sigma-alg-gen-by-set-alg},\\
  Lemma~\thref{l:sigma-alg-gen-by-monot-class},\\
  Lemma~\thref{l:sigma-alg-gen-by-l-syst},\\
  Lemma~\thref{l:other-sigma-alg-gen},\\
  Lemma~\thref{l:complete-gen-sigma-alg},\\
  Lemma~\thref{l:usage-of-dynkin-pi-lambda-th},\\
  Lemma~\thref{l:usage-of-monot-class-th},\\
  Lemma~\thref{l:inverse-image-of-gen-family},\\
  Lemma~\thref{l:equiv-def-of-meas-fun},\\
  Lemma~\thref{l:gen-prod-meas-space},\\
  Lemma~\thref{l:br-is-sub-sigma-alg-of-l}.

\item[Lemma~\thref{l:sigma-alg-gen-is-idem}] \mbox{}\\
  is explicitly cited in the proof of:\\
  Lemma~\thref{l:other-sigma-alg-gen},\\
  Lemma~\thref{l:p-syst-and-l-syst-is-sigma-alg},\\
  Theorem~\thref{t:dynkin-pi-lambda-th},\\
  Lemma~\thref{l:set-alg-and-monot-class-is-sigma-alg},\\
  Theorem~\thref{t:monot-class},\\
  Lemma~\thref{l:inverse-image-of-gen-family},\\
  Lemma~\thref{l:equiv-def-of-meas-fun}.

\item[Lemma~\thref{l:sigma-alg-is-p-syst}] \mbox{}\\
  is explicitly cited in the proof of:\\
  Lemma~\thref{l:sigma-alg-contains-p-syst}.

\item[Lemma~\thref{l:sigma-alg-contains-p-syst}] \mbox{}\\
  is explicitly cited in the proof of:\\
  Lemma~\thref{l:sigma-alg-gen-by-p-syst}.

\item[Lemma~\thref{l:p-syst-contains-sigma-alg}] \mbox{}\\
  is not yet used.

\item[Lemma~\thref{l:sigma-alg-gen-by-p-syst}] \mbox{}\\
  is not yet used.

\item[Lemma~\thref{l:sigma-alg-contains-set-alg}] \mbox{}\\
  is explicitly cited in the proof of:\\
  Lemma~\thref{l:sigma-alg-gen-by-set-alg},\\
  Lemma~\thref{l:meas-of-meas-of-section-finite},\\
  Lemma~\thref{l:uniq-of-tensor-prod-meas-finite}.

\item[Lemma~\thref{l:set-alg-contains-sigma-alg}] \mbox{}\\
  is not yet used.

\item[Lemma~\thref{l:sigma-alg-gen-by-set-alg}] \mbox{}\\
  is not yet used.

\item[Lemma~\thref{l:sigma-alg-is-monot-class}] \mbox{}\\
  is explicitly cited in the proof of:\\
  Lemma~\thref{l:sigma-alg-contains-monot-class}.

\item[Lemma~\thref{l:sigma-alg-contains-monot-class}] \mbox{}\\
  is explicitly cited in the proof of:\\
  Lemma~\thref{l:sigma-alg-gen-by-monot-class},\\
  Lemma~\thref{l:set-alg-and-monot-class-is-sigma-alg}.

\item[Lemma~\thref{l:monot-class-contains-sigma-alg}] \mbox{}\\
  is explicitly cited in the proof of:\\
  Lemma~\thref{l:set-alg-and-monot-class-is-sigma-alg}.

\item[Lemma~\thref{l:sigma-alg-gen-by-monot-class}] \mbox{}\\
  is explicitly cited in the proof of:\\
  Theorem~\thref{t:monot-class}.

\item[Lemma~\thref{l:sigma-alg-is-l-syst}] \mbox{}\\
  is explicitly cited in the proof of:\\
  Lemma~\thref{l:sigma-alg-contains-l-syst}.

\item[Lemma~\thref{l:sigma-alg-contains-l-syst}] \mbox{}\\
  is explicitly cited in the proof of:\\
  Lemma~\thref{l:sigma-alg-gen-by-l-syst},\\
  Lemma~\thref{l:p-syst-and-l-syst-is-sigma-alg}.

\item[Lemma~\thref{l:l-syst-contains-sigma-alg}] \mbox{}\\
  is explicitly cited in the proof of:\\
  Lemma~\thref{l:p-syst-and-l-syst-is-sigma-alg}.

\item[Lemma~\thref{l:sigma-alg-gen-by-l-syst}] \mbox{}\\
  is explicitly cited in the proof of:\\
  Theorem~\thref{t:dynkin-pi-lambda-th}.

\item[Lemma~\thref{l:other-sigma-alg-gen}] \mbox{}\\
  is explicitly cited in the proof of:\\
  Lemma~\thref{l:complete-gen-sigma-alg},\\
  Lemma~\thref{l:count-sigma-alg-gen},\\
  Lemma~\thref{l:borel-sigma-alg-of-r},\\
  Lemma~\thref{l:borel-sigma-alg-of-rbar}.

\item[Lemma~\thref{l:complete-gen-sigma-alg}] \mbox{}\\
  is explicitly cited in the proof of:\\
  Lemma~\thref{l:borel-sigma-alg-of-rm}.

\item[Lemma~\thref{l:count-sigma-alg-gen}] \mbox{}\\
  is explicitly cited in the proof of:\\
  Lemma~\thref{l:count-borel-sigma-alg-gen}.

\item[Lemma~\thref{l:set-alg-gen-by-prod-of-sigma-algs}] \mbox{}\\
  is explicitly cited in the proof of:\\
  Lemma~\thref{l:meas-of-meas-of-section-finite},\\
  Lemma~\thref{l:uniq-of-tensor-prod-meas-finite}.

\item[Lemma~\thref{l:p-syst-and-l-syst-is-sigma-alg}] \mbox{}\\
  is explicitly cited in the proof of:\\
  Theorem~\thref{t:dynkin-pi-lambda-th}.

\item[Theorem~\thref{t:dynkin-pi-lambda-th}] \mbox{}\\
  is explicitly cited in the proof of:\\
  Lemma~\thref{l:usage-of-dynkin-pi-lambda-th}.

\item[Lemma~\thref{l:usage-of-dynkin-pi-lambda-th}] \mbox{}\\
  is explicitly cited in the proof of:\\
  Lemma~\thref{l:uniq-of-meas-ext-from-p-syst}.

\item[Lemma~\thref{l:set-alg-and-monot-class-is-sigma-alg}] \mbox{}\\
  is explicitly cited in the proof of:\\
  Theorem~\thref{t:monot-class}.

\item[Theorem~\thref{t:monot-class}] \mbox{}\\
  is explicitly cited in the proof of:\\
  Lemma~\thref{l:usage-of-monot-class-th}.

\item[Lemma~\thref{l:usage-of-monot-class-th}] \mbox{}\\
  is explicitly cited in the proof of:\\
  Lemma~\thref{l:meas-of-meas-of-section-finite},\\
  Lemma~\thref{l:uniq-of-tensor-prod-meas-finite}.

\item[Definition~\thref{d:measurable-space}] \mbox{}\\
  is explicitly cited in the proof of:\\
  Lemma~\thref{l:inverse-sigma-alg},\\
  Lemma~\thref{l:image-sigma-alg},\\
  Lemma~\thref{l:trace-sigma-alg},\\
  Lemma~\thref{l:measurability-of-section},\\
  Lemma~\thref{l:count-union-of-sections-is-meas},\\
  Lemma~\thref{l:meas-of-indic-fun},\\
  Lemma~\thref{l:m-is-closed-under-finite-part},\\
  Lemma~\thref{l:meas-and-masking},\\
  Lemma~\thref{l:meas-of-restr},\\
  Lemma~\thref{l:meas-satisfies-finite-boole-ineq},\\
  Lemma~\thref{l:meas-satisfies-boole-ineq},\\
  Lemma~\thref{l:equiv-def-of-meas},\\
  Lemma~\thref{l:equiv-def-of-sigma-finite-meas},\\
  Lemma~\thref{l:empty-set-is-negl},\\
  Lemma~\thref{l:compat-of-null-meas-with-count-union},\\
  Lemma~\thref{l:count-meas},\\
  Lemma~\thref{l:sigma-finiteness-of-count-meas},\\
  Lemma~\thref{l:masking-almost-nowhere},\\
  Lemma~\thref{l:finite-nonneg-part},\\
  Lemma~\thref{l:if-is-sigma-add},\\
  Lemma~\thref{l:int-in-if-over-subset-is-add},\\
  Lemma~\thref{l:compat-of-int-in-mplus-with-almost-bin-rel},\\
  Lemma~\thref{l:const-fun-is-llone},\\
  Theorem~\thref{t:lebesgue-ext-dom-conv}.

\item[Definition~\thref{d:borel-sigma-alg}] \mbox{}\\
  is explicitly cited in the proof of:\\
  Lemma~\thref{l:some-borel-subsets},\\
  Lemma~\thref{l:count-borel-sigma-alg-gen},\\
  Lemma~\thref{l:cont-is-meas},\\
  Lemma~\thref{l:borel-sub-sigma-alg},\\
  Lemma~\thref{l:characterization-of-borel-subsets},\\
  Lemma~\thref{l:borel-subsets-of-rbar-and-r},\\
  Lemma~\thref{l:meas-of-num-fun-to-r},\\
  Lemma~\thref{l:meas-of-num-fun},\\
  Lemma~\thref{l:m-is-closed-under-add-when-defined},\\
  Lemma~\thref{l:m-is-closed-under-mult}.

\item[Lemma~\thref{l:some-borel-subsets}] \mbox{}\\
  is explicitly cited in the proof of:\\
  Lemma~\thref{l:borel-sigma-alg-of-r},\\
  Lemma~\thref{l:meas-of-num-fun-to-r},\\
  Lemma~\thref{l:inverse-image-is-meas-in-r},\\
  Lemma~\thref{l:meas-of-num-fun},\\
  Lemma~\thref{l:inverse-image-is-meas},\\
  Lemma~\thref{l:decomp-of-meas-in-sfplus},\\
  Lemma~\thref{l:int-over-int}.

\item[Lemma~\thref{l:count-borel-sigma-alg-gen}] \mbox{}\\
  is explicitly cited in the proof of:\\
  Lemma~\thref{l:borel-sigma-alg-of-r},\\
  Lemma~\thref{l:borel-sigma-alg-of-rbar},\\
  Lemma~\thref{l:borel-sigma-alg-of-rm}.

\item[Definition~\thref{d:meas-fun}] \mbox{}\\
  is explicitly cited in the proof of:\\
  Lemma~\thref{l:inverse-sigma-alg},\\
  Lemma~\thref{l:image-sigma-alg},\\
  Lemma~\thref{l:identity-fun-is-meas},\\
  Lemma~\thref{l:equiv-def-of-meas-fun},\\
  Lemma~\thref{l:cont-is-meas},\\
  Lemma~\thref{l:compat-of-meas-with-comp},\\
  Lemma~\thref{l:trace-sigma-alg},\\
  Lemma~\thref{l:source-restr-of-meas-fun},\\
  Lemma~\thref{l:destination-restr-of-meas-fun},\\
  Lemma~\thref{l:meas-of-fun-def-on-pseudopart},\\
  Lemma~\thref{l:meas-of-fun-to-prod-space},\\
  Lemma~\thref{l:can-proj-is-meas},\\
  Lemma~\thref{l:gen-prod-meas-space},\\
  Lemma~\thref{l:meas-of-fun-from-prod-space},\\
  Lemma~\thref{l:meas-of-num-fun-to-r},\\
  Lemma~\thref{l:inverse-image-is-meas-in-r},\\
  Lemma~\thref{l:m-and-finite-is-mr},\\
  Lemma~\thref{l:meas-of-num-fun},\\
  Lemma~\thref{l:inverse-image-is-meas},\\
  Lemma~\thref{l:m-is-closed-under-add-when-defined},\\
  Lemma~\thref{l:m-is-closed-under-mult},\\
  Lemma~\thref{l:meas-of-restr}.

\item[Lemma~\thref{l:inverse-sigma-alg}] \mbox{}\\
  is explicitly cited in the proof of:\\
  Lemma~\thref{l:inverse-image-of-gen-family}.

\item[Lemma~\thref{l:image-sigma-alg}] \mbox{}\\
  is explicitly cited in the proof of:\\
  Lemma~\thref{l:inverse-image-of-gen-family}.

\item[Lemma~\thref{l:identity-fun-is-meas}] \mbox{}\\
  is not yet used.

\item[Lemma~\thref{l:const-fun-is-meas}] \mbox{}\\
  is explicitly cited in the proof of:\\
  Lemma~\thref{l:mr-is-alg},\\
  Lemma~\thref{l:m-is-closed-under-finite-part},\\
  Lemma~\thref{l:m-is-closed-under-add-when-defined},\\
  Lemma~\thref{l:m-is-closed-under-mult},\\
  Lemma~\thref{l:m-is-closed-under-scalar-mult},\\
  Lemma~\thref{l:meas-of-nonneg-and-nonpos-parts}.

\item[Lemma~\thref{l:inverse-image-of-gen-family}] \mbox{}\\
  is explicitly cited in the proof of:\\
  Lemma~\thref{l:equiv-def-of-meas-fun},\\
  Lemma~\thref{l:gen-meas-subspace},\\
  Lemma~\thref{l:gen-prod-meas-space}.

\item[Lemma~\thref{l:equiv-def-of-meas-fun}] \mbox{}\\
  is explicitly cited in the proof of:\\
  Lemma~\thref{l:cont-is-meas},\\
  Lemma~\thref{l:meas-of-fun-to-prod-space},\\
  Lemma~\thref{l:meas-of-num-fun-to-r},\\
  Lemma~\thref{l:meas-of-num-fun}.

\item[Lemma~\thref{l:cont-is-meas}] \mbox{}\\
  is explicitly cited in the proof of:\\
  Lemma~\thref{l:mr-is-alg},\\
  Lemma~\thref{l:m-is-closed-under-abs}.

\item[Lemma~\thref{l:compat-of-meas-with-comp}] \mbox{}\\
  is explicitly cited in the proof of:\\
  Lemma~\thref{l:source-restr-of-meas-fun},\\
  Lemma~\thref{l:mr-is-alg},\\
  Lemma~\thref{l:m-is-closed-under-abs},\\
  Lemma~\thref{l:meas-of-tensor-prod-of-num-funs}.

\item[Lemma~\thref{l:trace-sigma-alg}] \mbox{}\\
  is explicitly cited in the proof of:\\
  Lemma~\thref{l:meas-of-meas-subspace},\\
  Lemma~\thref{l:source-restr-of-meas-fun},\\
  Lemma~\thref{l:trace-meas}.

\item[Lemma~\thref{l:meas-of-meas-subspace}] \mbox{}\\
  is explicitly cited in the proof of:\\
  Lemma~\thref{l:borel-sub-sigma-alg},\\
  Lemma~\thref{l:trace-meas},\\
  Lemma~\thref{l:if-is-closed-under-restr}.

\item[Lemma~\thref{l:gen-meas-subspace}] \mbox{}\\
  is explicitly cited in the proof of:\\
  Lemma~\thref{l:borel-sub-sigma-alg},\\
  Lemma~\thref{l:borel-sigma-alg-of-rbarplus}.

\item[Lemma~\thref{l:borel-sub-sigma-alg}] \mbox{}\\
  is explicitly cited in the proof of:\\
  Lemma~\thref{l:characterization-of-borel-subsets},\\
  Lemma~\thref{l:borel-subsets-of-rbar-and-r},\\
  Lemma~\thref{l:borel-sigma-alg-of-rplus}.

\item[Lemma~\thref{l:characterization-of-borel-subsets}] \mbox{}\\
  is explicitly cited in the proof of:\\
  Lemma~\thref{l:borel-subsets-of-rbar-and-r}.

\item[Lemma~\thref{l:source-restr-of-meas-fun}] \mbox{}\\
  is not yet used.

\item[Lemma~\thref{l:destination-restr-of-meas-fun}] \mbox{}\\
  is not yet used.

\item[Lemma~\thref{l:meas-of-fun-def-on-pseudopart}] \mbox{}\\
  is explicitly cited in the proof of:\\
  Lemma~\thref{l:m-is-closed-under-finite-part},\\
  Lemma~\thref{l:m-is-closed-under-add-when-defined},\\
  Lemma~\thref{l:m-is-closed-under-mult}.

\item[Definition~\thref{d:tensor-prod-of-sigma-algs}] \mbox{}\\
  is explicitly cited in the proof of:\\
  Lemma~\thref{l:prod-of-meas-subsets-is-meas},\\
  Lemma~\thref{l:meas-of-fun-to-prod-space},\\
  Lemma~\thref{l:gen-prod-meas-space},\\
  Lemma~\thref{l:measurability-of-section},\\
  Lemma~\thref{l:meas-of-meas-of-section-finite},\\
  Lemma~\thref{l:uniq-of-tensor-prod-meas-finite}.

\item[Lemma~\thref{l:prod-of-meas-subsets-is-meas}] \mbox{}\\
  is explicitly cited in the proof of:\\
  Lemma~\thref{l:meas-of-fun-to-prod-space},\\
  Lemma~\thref{l:meas-of-section-of-prod},\\
  Lemma~\thref{l:meas-of-meas-of-section-finite},\\
  Lemma~\thref{l:cand-tensor-prod-meas-is-tensor-prod-meas},\\
  Lemma~\thref{l:tensor-prod-of-sigma-finite-meas},\\
  Lemma~\thref{l:uniq-of-tensor-prod-meas-finite},\\
  Lemma~\thref{l:uniq-of-tensor-prod-meas}.

\item[Lemma~\thref{l:meas-of-fun-to-prod-space}] \mbox{}\\
  is explicitly cited in the proof of:\\
  Lemma~\thref{l:can-proj-is-meas},\\
  Lemma~\thref{l:perm-is-meas},\\
  Lemma~\thref{l:mr-is-alg}.

\item[Lemma~\thref{l:can-proj-is-meas}] \mbox{}\\
  is explicitly cited in the proof of:\\
  Lemma~\thref{l:perm-is-meas},\\
  Lemma~\thref{l:gen-prod-meas-space},\\
  Lemma~\thref{l:meas-of-tensor-prod-of-num-funs}.

\item[Lemma~\thref{l:perm-is-meas}] \mbox{}\\
  is not yet used.

\item[Lemma~\thref{l:gen-prod-meas-space}] \mbox{}\\
  is explicitly cited in the proof of:\\
  Lemma~\thref{l:borel-sigma-alg-of-rm}.

\item[Definition~\thref{d:section-in-cartesian-prod}] \mbox{}\\
  is explicitly cited in the proof of:\\
  Lemma~\thref{l:section-of-prod},\\
  Lemma~\thref{l:compat-of-section-with-set-ops},\\
  Lemma~\thref{l:indic-of-section}.

\item[Lemma~\thref{l:section-of-prod}] \mbox{}\\
  is explicitly cited in the proof of:\\
  Lemma~\thref{l:measurability-of-section},\\
  Lemma~\thref{l:meas-of-section-of-prod}.

\item[Lemma~\thref{l:compat-of-section-with-set-ops}] \mbox{}\\
  is explicitly cited in the proof of:\\
  Lemma~\thref{l:measurability-of-section},\\
  Lemma~\thref{l:count-union-of-sections-is-meas},\\
  Lemma~\thref{l:count-inter-of-sections-is-meas},\\
  Lemma~\thref{l:meas-of-meas-of-section-finite},\\
  Lemma~\thref{l:cand-tensor-prod-meas-is-tensor-prod-meas}.

\item[Lemma~\thref{l:measurability-of-section}] \mbox{}\\
  is explicitly cited in the proof of:\\
  Lemma~\thref{l:count-union-of-sections-is-meas},\\
  Lemma~\thref{l:count-inter-of-sections-is-meas},\\
  Lemma~\thref{l:meas-of-fun-from-prod-space},\\
  Lemma~\thref{l:meas-of-section},\\
  Lemma~\thref{l:meas-of-meas-of-section-finite},\\
  Lemma~\thref{l:meas-of-meas-of-section},\\
  Lemma~\thref{l:cand-tensor-prod-meas-is-tensor-prod-meas},\\
  Lemma~\thref{l:negl-of-meas-section},\\
  Theorem~\thref{t:tonelli}.

\item[Lemma~\thref{l:count-union-of-sections-is-meas}] \mbox{}\\
  is explicitly cited in the proof of:\\
  Lemma~\thref{l:meas-of-meas-of-section-finite},\\
  Lemma~\thref{l:cand-tensor-prod-meas-is-tensor-prod-meas}.

\item[Lemma~\thref{l:count-inter-of-sections-is-meas}] \mbox{}\\
  is explicitly cited in the proof of:\\
  Lemma~\thref{l:meas-of-meas-of-section-finite}.

\item[Lemma~\thref{l:indic-of-section}] \mbox{}\\
  is explicitly cited in the proof of:\\
  Theorem~\thref{t:tonelli},\\
  Lemma~\thref{l:tonelli-over-subset}.

\item[Lemma~\thref{l:meas-of-fun-from-prod-space}] \mbox{}\\
  is not yet used.

\item[Lemma~\thref{l:borel-sigma-alg-of-r}] \mbox{}\\
  is explicitly cited in the proof of:\\
  Lemma~\thref{l:count-gen-of-borel-sigma-alg-of-r},\\
  Lemma~\thref{l:borel-sigma-alg-of-rplus},\\
  Lemma~\thref{l:meas-of-num-fun-to-r},\\
  Lemma~\thref{l:br-is-sub-sigma-alg-of-l},\\
  Theorem~\thref{t:caratheodory-lebesgue-meas-on-r}.

\item[Lemma~\thref{l:count-gen-of-borel-sigma-alg-of-r}] \mbox{}\\
  is explicitly cited in the proof of:\\
  Lemma~\thref{l:borel-sigma-alg-of-rm}.

\item[Lemma~\thref{l:borel-sigma-alg-of-rbar}] \mbox{}\\
  is explicitly cited in the proof of:\\
  Lemma~\thref{l:borel-sigma-alg-of-rbarplus},\\
  Lemma~\thref{l:meas-of-num-fun}.

\item[Lemma~\thref{l:borel-subsets-of-rbar-and-r}] \mbox{}\\
  is explicitly cited in the proof of:\\
  Lemma~\thref{l:m-and-finite-is-mr}.

\item[Lemma~\thref{l:borel-sigma-alg-of-rplus}] \mbox{}\\
  is not yet used.

\item[Lemma~\thref{l:borel-sigma-alg-of-rbarplus}] \mbox{}\\
  is not yet used.

\item[Lemma~\thref{l:borel-sigma-alg-of-rm}] \mbox{}\\
  is explicitly cited in the proof of:\\
  Lemma~\thref{l:mr-is-alg}.

\item[Definition~\thref{d:mr-vector-space-of-meas-num-fun-to-r}] \mbox{}\\
  is explicitly cited in the proof of:\\
  Lemma~\thref{l:meas-of-indic-fun},\\
  Lemma~\thref{l:meas-of-num-fun-to-r},\\
  Lemma~\thref{l:inverse-image-is-meas-in-r},\\
  Lemma~\thref{l:mr-is-alg},\\
  Lemma~\thref{l:m-and-finite-is-mr},\\
  Lemma~\thref{l:m-is-closed-under-abs}.

\item[Lemma~\thref{l:meas-of-indic-fun}] \mbox{}\\
  is explicitly cited in the proof of:\\
  Lemma~\thref{l:meas-and-masking},\\
  Lemma~\thref{l:almost-sum},\\
  Lemma~\thref{l:if-is-meas},\\
  Lemma~\thref{l:sf-is-meas},\\
  Lemma~\thref{l:bienayme-chebyshev-ineq},\\
  Lemma~\thref{l:int-in-mplus-over-subset-is-sigma-add},\\
  Lemma~\thref{l:int-in-mplus-over-singleton}.

\item[Lemma~\thref{l:meas-of-num-fun-to-r}] \mbox{}\\
  is not yet used.

\item[Lemma~\thref{l:inverse-image-is-meas-in-r}] \mbox{}\\
  is explicitly cited in the proof of:\\
  Lemma~\thref{l:decomp-of-meas-in-sfplus},\\
  Lemma~\thref{l:change-of-variable-in-sum-in-sfplus},\\
  Lemma~\thref{l:int-in-sfplus-is-add-alt-proof}.

\item[Lemma~\thref{l:mr-is-alg}] \mbox{}\\
  is explicitly cited in the proof of:\\
  Lemma~\thref{l:mr-is-vector-space},\\
  Lemma~\thref{l:m-is-closed-under-add-when-defined},\\
  Lemma~\thref{l:m-is-closed-under-mult},\\
  Lemma~\thref{l:sf-is-meas}.

\item[Lemma~\thref{l:mr-is-vector-space}] \mbox{}\\
  is explicitly cited in the proof of:\\
  Lemma~\thref{l:llone-is-seminormed-vector-space}.

\item[Definition~\thref{d:m-set-of-meas-num-funs}] \mbox{}\\
  is explicitly cited in the proof of:\\
  Lemma~\thref{l:m-and-finite-is-mr},\\
  Lemma~\thref{l:meas-of-num-fun},\\
  Lemma~\thref{l:inverse-image-is-meas},\\
  Lemma~\thref{l:m-is-closed-under-add-when-defined},\\
  Lemma~\thref{l:m-is-closed-under-mult},\\
  Lemma~\thref{l:meas-of-restr},\\
  Lemma~\thref{l:m-is-closed-under-abs},\\
  Lemma~\thref{l:meas-of-tensor-prod-of-num-funs},\\
  Lemma~\thref{l:change-of-variable-in-sum-in-sfplus},\\
  Lemma~\thref{l:int-in-sfplus-is-add-alt-proof}.

\item[Lemma~\thref{l:m-and-finite-is-mr}] \mbox{}\\
  is explicitly cited in the proof of:\\
  Lemma~\thref{l:m-is-closed-under-finite-part},\\
  Lemma~\thref{l:meas-and-masking},\\
  Lemma~\thref{l:if-is-meas},\\
  Lemma~\thref{l:sf-is-meas},\\
  Lemma~\thref{l:int-in-mplus-over-subset-is-sigma-add},\\
  Lemma~\thref{l:equiv-def-of-llone},\\
  Lemma~\thref{l:int-is-pos-lin-form-on-llone}.

\item[Lemma~\thref{l:meas-of-num-fun}] \mbox{}\\
  is explicitly cited in the proof of:\\
  Lemma~\thref{l:m-is-closed-under-mult},\\
  Lemma~\thref{l:m-is-closed-under-inf},\\
  Lemma~\thref{l:m-is-closed-under-sup},\\
  Theorem~\thref{t:beppo-levi-monot-conv},\\
  Lemma~\thref{l:adapted-seq-in-mplus},\\
  Lemma~\thref{l:int-in-mplus-is-almost-definite},\\
  Lemma~\thref{l:bienayme-chebyshev-ineq}.

\item[Lemma~\thref{l:inverse-image-is-meas}] \mbox{}\\
  is explicitly cited in the proof of:\\
  Lemma~\thref{l:m-is-closed-under-finite-part},\\
  Lemma~\thref{l:m-is-closed-under-add-when-defined},\\
  Lemma~\thref{l:m-is-closed-under-mult},\\
  Lemma~\thref{l:meas-of-summability-domain}.

\item[Lemma~\thref{l:m-is-closed-under-finite-part}] \mbox{}\\
  is explicitly cited in the proof of:\\
  Lemma~\thref{l:m-is-closed-under-add-when-defined},\\
  Lemma~\thref{l:m-is-closed-under-mult},\\
  Lemma~\thref{l:mplus-is-closed-under-finite-part}.

\item[Lemma~\thref{l:m-is-closed-under-add-when-defined}] \mbox{}\\
  is explicitly cited in the proof of:\\
  Lemma~\thref{l:m-is-closed-under-finite-sum-when-defined},\\
  Lemma~\thref{l:meas-of-nonneg-and-nonpos-parts},\\
  Lemma~\thref{l:mplus-is-closed-under-add},\\
  Lemma~\thref{l:almost-sum},\\
  Lemma~\thref{l:compat-of-almost-sum-with-almost-eq},\\
  Theorem~\thref{t:beppo-levi-monot-conv}.

\item[Lemma~\thref{l:m-is-closed-under-finite-sum-when-defined}] \mbox{}\\
  is not yet used.

\item[Lemma~\thref{l:m-is-closed-under-mult}] \mbox{}\\
  is explicitly cited in the proof of:\\
  Lemma~\thref{l:m-is-closed-under-finite-prod},\\
  Lemma~\thref{l:m-is-closed-under-scalar-mult},\\
  Lemma~\thref{l:meas-and-masking},\\
  Lemma~\thref{l:mplus-is-closed-under-mult},\\
  Lemma~\thref{l:almost-sum}.

\item[Lemma~\thref{l:m-is-closed-under-finite-prod}] \mbox{}\\
  is explicitly cited in the proof of:\\
  Lemma~\thref{l:meas-of-tensor-prod-of-num-funs}.

\item[Lemma~\thref{l:m-is-closed-under-scalar-mult}] \mbox{}\\
  is explicitly cited in the proof of:\\
  Lemma~\thref{l:meas-of-nonneg-and-nonpos-parts},\\
  Lemma~\thref{l:mplus-is-closed-under-nonneg-scalar-mult},\\
  Lemma~\thref{l:int-is-hom}.

\item[Lemma~\thref{l:m-is-closed-under-inf}] \mbox{}\\
  is explicitly cited in the proof of:\\
  Lemma~\thref{l:m-is-closed-under-liminf},\\
  Lemma~\thref{l:m-is-closed-under-limsup},\\
  Lemma~\thref{l:mplus-is-closed-under-inf},\\
  Theorem~\thref{t:fatou-lemma}.

\item[Lemma~\thref{l:m-is-closed-under-sup}] \mbox{}\\
  is explicitly cited in the proof of:\\
  Lemma~\thref{l:m-is-closed-under-liminf},\\
  Lemma~\thref{l:m-is-closed-under-limsup},\\
  Lemma~\thref{l:meas-of-nonneg-and-nonpos-parts},\\
  Lemma~\thref{l:mplus-is-closed-under-sup}.

\item[Lemma~\thref{l:m-is-closed-under-liminf}] \mbox{}\\
  is explicitly cited in the proof of:\\
  Lemma~\thref{l:m-is-closed-under-limit-when-pointwise-conv},\\
  Theorem~\thref{t:fatou-lemma}.

\item[Lemma~\thref{l:m-is-closed-under-limsup}] \mbox{}\\
  is explicitly cited in the proof of:\\
  Lemma~\thref{l:m-is-closed-under-limit-when-pointwise-conv}.

\item[Lemma~\thref{l:m-is-closed-under-limit-when-pointwise-conv}] \mbox{}\\
  is explicitly cited in the proof of:\\
  Lemma~\thref{l:mplus-is-closed-under-limit-when-pointwise-conv},\\
  Theorem~\thref{t:lebesgue-dom-conv}.

\item[Lemma~\thref{l:meas-and-masking}] \mbox{}\\
  is explicitly cited in the proof of:\\
  Lemma~\thref{l:finite-nonneg-part},\\
  Theorem~\thref{t:lebesgue-ext-dom-conv}.

\item[Lemma~\thref{l:meas-of-restr}] \mbox{}\\
  is explicitly cited in the proof of:\\
  Lemma~\thref{l:int-in-mplus-over-subset}.

\item[Definition~\thref{d:mplus-subset-of-nonneg-meas-num-fun}] \mbox{}\\
  is explicitly cited in the proof of:\\
  Lemma~\thref{l:meas-of-nonneg-and-nonpos-parts},\\
  Lemma~\thref{l:mplus-is-closed-under-finite-part},\\
  Lemma~\thref{l:m-is-closed-under-abs},\\
  Lemma~\thref{l:mplus-is-closed-under-add},\\
  Lemma~\thref{l:mplus-is-closed-under-mult},\\
  Lemma~\thref{l:mplus-is-closed-under-nonneg-scalar-mult},\\
  Lemma~\thref{l:mplus-is-closed-under-limit-when-pointwise-conv},\\
  Lemma~\thref{l:if-is-meas},\\
  Lemma~\thref{l:sfplus-is-meas},\\
  Lemma~\thref{l:int-in-mplus-is-hom-at-infinity},\\
  Lemma~\thref{l:int-in-mplus-is-almost-definite},\\
  Lemma~\thref{l:integrable-in-mplus-is-almost-finite},\\
  Lemma~\thref{l:int-in-mplus-over-subset},\\
  Lemma~\thref{l:int-in-mplus-over-subset-is-sigma-add},\\
  Lemma~\thref{l:tonelli-for-tensor-prod},\\
  Lemma~\thref{l:int-is-pos-lin-form-on-llone},\\
  Theorem~\thref{t:lebesgue-dom-conv}.

\item[Lemma~\thref{l:meas-of-nonneg-and-nonpos-parts}] \mbox{}\\
  is explicitly cited in the proof of:\\
  Lemma~\thref{l:finite-nonneg-part},\\
  Lemma~\thref{l:int-in-mplus-of-decomp-into-nonpos-and-nonneg-parts},\\
  Lemma~\thref{l:integrable-is-meas},\\
  Lemma~\thref{l:equiv-def-of-integrability},\\
  Lemma~\thref{l:int-is-hom},\\
  Lemma~\thref{l:int-is-add}.

\item[Lemma~\thref{l:mplus-is-closed-under-finite-part}] \mbox{}\\
  is explicitly cited in the proof of:\\
  Lemma~\thref{l:finite-nonneg-part}.

\item[Lemma~\thref{l:m-is-closed-under-abs}] \mbox{}\\
  is explicitly cited in the proof of:\\
  Lemma~\thref{l:int-in-mplus-of-decomp-into-nonpos-and-nonneg-parts},\\
  Lemma~\thref{l:bienayme-chebyshev-ineq},\\
  Lemma~\thref{l:equiv-def-of-integrability},\\
  Lemma~\thref{l:almost-bounded-by-integrable-is-integrable},\\
  Lemma~\thref{l:seminorm-llone},\\
  Lemma~\thref{l:integrable-is-finite-seminorm-lone},\\
  Lemma~\thref{l:minkowski-ineq-in-m},\\
  Lemma~\thref{l:llone-is-closed-under-abs}.

\item[Lemma~\thref{l:mplus-is-closed-under-add}] \mbox{}\\
  is explicitly cited in the proof of:\\
  Lemma~\thref{l:mplus-is-closed-under-count-sum},\\
  Lemma~\thref{l:int-in-mplus-is-add},\\
  Theorem~\thref{t:tonelli}.

\item[Lemma~\thref{l:mplus-is-closed-under-mult}] \mbox{}\\
  is explicitly cited in the proof of:\\
  Lemma~\thref{l:compat-of-int-in-mplus-with-almost-bin-rel},\\
  Lemma~\thref{l:int-in-mplus-over-subset-is-sigma-add}.

\item[Lemma~\thref{l:mplus-is-closed-under-nonneg-scalar-mult}] \mbox{}\\
  is explicitly cited in the proof of:\\
  Lemma~\thref{l:int-in-mplus-is-pos-hom},\\
  Lemma~\thref{l:int-in-mplus-is-hom-at-infinity},\\
  Lemma~\thref{l:int-in-mplus-is-almost-definite},\\
  Lemma~\thref{l:bienayme-chebyshev-ineq},\\
  Lemma~\thref{l:int-in-mplus-over-singleton},\\
  Lemma~\thref{l:meas-of-meas-of-section-finite},\\
  Theorem~\thref{t:tonelli}.

\item[Lemma~\thref{l:mplus-is-closed-under-inf}] \mbox{}\\
  is explicitly cited in the proof of:\\
  Lemma~\thref{l:meas-of-meas-of-section-finite}.

\item[Lemma~\thref{l:mplus-is-closed-under-sup}] \mbox{}\\
  is explicitly cited in the proof of:\\
  Lemma~\thref{l:meas-of-meas-of-section-finite},\\
  Lemma~\thref{l:meas-of-meas-of-section}.

\item[Lemma~\thref{l:mplus-is-closed-under-limit-when-pointwise-conv}] \mbox{}\\
  is explicitly cited in the proof of:\\
  Lemma~\thref{l:mplus-is-closed-under-count-sum},\\
  Theorem~\thref{t:beppo-levi-monot-conv},\\
  Theorem~\thref{t:fatou-lemma},\\
  Lemma~\thref{l:int-in-mplus-of-pointwise-conv-seq},\\
  Theorem~\thref{t:tonelli}.

\item[Lemma~\thref{l:mplus-is-closed-under-count-sum}] \mbox{}\\
  is explicitly cited in the proof of:\\
  Lemma~\thref{l:int-in-mplus-is-sigma-add},\\
  Lemma~\thref{l:int-in-mplus-over-subset-is-sigma-add},\\
  Lemma~\thref{l:meas-of-meas-of-section-finite}.

\item[Definition~\thref{d:tensor-prod-of-num-funs}] \mbox{}\\
  is explicitly cited in the proof of:\\
  Lemma~\thref{l:meas-of-tensor-prod-of-num-funs},\\
  Lemma~\thref{l:tonelli-for-tensor-prod}.

\item[Lemma~\thref{l:meas-of-tensor-prod-of-num-funs}] \mbox{}\\
  is explicitly cited in the proof of:\\
  Lemma~\thref{l:tonelli-for-tensor-prod}.

\item[Definition~\thref{d:add-over-meas-space}] \mbox{}\\
  is explicitly cited in the proof of:\\
  Lemma~\thref{l:sigma-add-implies-add},\\
  Lemma~\thref{l:equiv-def-of-meas},\\
  Lemma~\thref{l:int-in-if-is-add},\\
  Lemma~\thref{l:uniq-of-tensor-prod-meas-finite}.

\item[Definition~\thref{d:sigma-add-over-meas-space}] \mbox{}\\
  is explicitly cited in the proof of:\\
  Lemma~\thref{l:sigma-add-implies-add},\\
  Lemma~\thref{l:meas-over-count-pseudopart},\\
  Lemma~\thref{l:meas-is-monot},\\
  Lemma~\thref{l:meas-satisfies-finite-boole-ineq},\\
  Lemma~\thref{l:meas-is-cont-from-below},\\
  Lemma~\thref{l:equiv-def-of-meas},\\
  Lemma~\thref{l:uniq-of-meas-ext-from-p-syst},\\
  Lemma~\thref{l:count-meas},\\
  Lemma~\thref{l:lambda-star-is-sigma-add-on-l},\\
  Lemma~\thref{l:equiv-def-of-int-in-sfplus-disj},\\
  Lemma~\thref{l:meas-of-meas-of-section-finite},\\
  Lemma~\thref{l:cand-tensor-prod-meas-is-tensor-prod-meas}.

\item[Lemma~\thref{l:sigma-add-implies-add}] \mbox{}\\
  is explicitly cited in the proof of:\\
  Lemma~\thref{l:equiv-def-of-meas}.

\item[Definition~\thref{d:meas}] \mbox{}\\
  is explicitly cited in the proof of:\\
  Lemma~\thref{l:meas-over-count-pseudopart},\\
  Lemma~\thref{l:meas-is-monot},\\
  Lemma~\thref{l:meas-satisfies-finite-boole-ineq},\\
  Lemma~\thref{l:meas-is-cont-from-below},\\
  Lemma~\thref{l:meas-satisfies-boole-ineq},\\
  Lemma~\thref{l:equiv-def-of-meas},\\
  Lemma~\thref{l:equiv-def-of-sigma-finite-meas},\\
  Lemma~\thref{l:trace-meas},\\
  Lemma~\thref{l:restr-meas},\\
  Lemma~\thref{l:equiv-def-of-considerable-subset},\\
  Lemma~\thref{l:empty-set-is-negl},\\
  Lemma~\thref{l:compat-of-null-meas-with-count-union},\\
  Lemma~\thref{l:uniq-of-meas-ext-from-p-syst},\\
  Lemma~\thref{l:trivial-meas},\\
  Lemma~\thref{l:equiv-def-of-trivial-meas},\\
  Lemma~\thref{l:count-meas},\\
  Lemma~\thref{l:sigma-finiteness-of-count-meas},\\
  Lemma~\thref{l:masking-almost-nowhere},\\
  Lemma~\thref{l:finite-nonneg-part},\\
  Lemma~\thref{l:lambda-star-is-meas-on-l},\\
  Lemma~\thref{l:int-in-if-over-subset-is-add},\\
  Lemma~\thref{l:int-in-sfplus},\\
  Lemma~\thref{l:equiv-def-of-int-in-sfplus-disj},\\
  Lemma~\thref{l:change-of-variable-in-sum-in-sfplus},\\
  Lemma~\thref{l:int-in-sfplus-is-pos-lin},\\
  Lemma~\thref{l:compat-of-int-in-mplus-with-almost-bin-rel},\\
  Lemma~\thref{l:integrable-in-mplus-is-almost-finite},\\
  Lemma~\thref{l:meas-of-section},\\
  Lemma~\thref{l:meas-of-section-of-prod},\\
  Lemma~\thref{l:meas-of-meas-of-section-finite},\\
  Lemma~\thref{l:cand-tensor-prod-meas-is-tensor-prod-meas},\\
  Lemma~\thref{l:tensor-prod-of-finite-meas},\\
  Lemma~\thref{l:const-fun-is-llone},\\
  Theorem~\thref{t:lebesgue-ext-dom-conv}.

\item[Lemma~\thref{l:meas-over-count-pseudopart}] \mbox{}\\
  is explicitly cited in the proof of:\\
  Lemma~\thref{l:int-in-sfplus-is-add},\\
  Lemma~\thref{l:decomp-of-meas-in-sfplus}.

\item[Lemma~\thref{l:meas-is-monot}] \mbox{}\\
  is explicitly cited in the proof of:\\
  Lemma~\thref{l:meas-satisfies-finite-boole-ineq},\\
  Lemma~\thref{l:meas-is-cont-from-below},\\
  Lemma~\thref{l:meas-is-cont-from-above},\\
  Lemma~\thref{l:finite-meas-is-bounded},\\
  Lemma~\thref{l:uniq-of-meas-ext-from-p-syst},\\
  Lemma~\thref{l:equiv-def-of-trivial-meas}.

\item[Lemma~\thref{l:meas-satisfies-finite-boole-ineq}] \mbox{}\\
  is explicitly cited in the proof of:\\
  Lemma~\thref{l:meas-satisfies-boole-ineq},\\
  Lemma~\thref{l:equiv-def-of-sigma-finite-meas},\\
  Lemma~\thref{l:compat-of-null-meas-with-count-union},\\
  Lemma~\thref{l:finite-nonneg-part}.

\item[Definition~\thref{d:continuity-from-below}] \mbox{}\\
  is explicitly cited in the proof of:\\
  Lemma~\thref{l:meas-is-cont-from-below},\\
  Lemma~\thref{l:meas-is-cont-from-above},\\
  Lemma~\thref{l:equiv-def-of-meas},\\
  Lemma~\thref{l:uniq-of-meas-ext-from-p-syst},\\
  Theorem~\thref{t:beppo-levi-monot-conv}.

\item[Lemma~\thref{l:meas-is-cont-from-below}] \mbox{}\\
  is explicitly cited in the proof of:\\
  Lemma~\thref{l:meas-is-cont-from-above},\\
  Lemma~\thref{l:meas-satisfies-boole-ineq},\\
  Lemma~\thref{l:equiv-def-of-meas},\\
  Lemma~\thref{l:uniq-of-meas-ext-from-p-syst},\\
  Theorem~\thref{t:beppo-levi-monot-conv},\\
  Lemma~\thref{l:meas-of-meas-of-section-finite},\\
  Lemma~\thref{l:meas-of-meas-of-section},\\
  Lemma~\thref{l:uniq-of-tensor-prod-meas-finite},\\
  Lemma~\thref{l:uniq-of-tensor-prod-meas}.

\item[Definition~\thref{d:continuity-from-above}] \mbox{}\\
  is explicitly cited in the proof of:\\
  Lemma~\thref{l:meas-is-cont-from-above}.

\item[Lemma~\thref{l:meas-is-cont-from-above}] \mbox{}\\
  is explicitly cited in the proof of:\\
  Theorem~\thref{t:caratheodory-lebesgue-meas-on-r},\\
  Lemma~\thref{l:meas-of-meas-of-section-finite},\\
  Lemma~\thref{l:uniq-of-tensor-prod-meas-finite}.

\item[Lemma~\thref{l:meas-satisfies-boole-ineq}] \mbox{}\\
  is explicitly cited in the proof of:\\
  Lemma~\thref{l:compat-of-null-meas-with-count-union}.

\item[Lemma~\thref{l:equiv-def-of-meas}] \mbox{}\\
  is explicitly cited in the proof of:\\
  Lemma~\thref{l:int-in-if-is-add},\\
  Lemma~\thref{l:uniq-of-tensor-prod-meas-finite}.

\item[Definition~\thref{d:finite-meas}] \mbox{}\\
  is explicitly cited in the proof of:\\
  Lemma~\thref{l:finite-meas-is-sigma-finite},\\
  Lemma~\thref{l:finiteness-of-count-meas},\\
  Lemma~\thref{l:meas-of-meas-of-section-finite},\\
  Lemma~\thref{l:meas-of-meas-of-section},\\
  Lemma~\thref{l:tensor-prod-of-finite-meas},\\
  Lemma~\thref{l:uniq-of-tensor-prod-meas}.

\item[Lemma~\thref{l:finite-meas-is-bounded}] \mbox{}\\
  is explicitly cited in the proof of:\\
  Lemma~\thref{l:meas-of-meas-of-section-finite},\\
  Lemma~\thref{l:uniq-of-tensor-prod-meas-finite}.

\item[Definition~\thref{d:sigma-finite-meas}] \mbox{}\\
  is explicitly cited in the proof of:\\
  Lemma~\thref{l:equiv-def-of-sigma-finite-meas},\\
  Lemma~\thref{l:finite-meas-is-sigma-finite},\\
  Lemma~\thref{l:sigma-finiteness-of-count-meas},\\
  Lemma~\thref{l:lebesgue-meas-is-sigma-finite},\\
  Lemma~\thref{l:tensor-prod-of-sigma-finite-meas}.

\item[Lemma~\thref{l:equiv-def-of-sigma-finite-meas}] \mbox{}\\
  is explicitly cited in the proof of:\\
  Lemma~\thref{l:sigma-finiteness-of-count-meas},\\
  Lemma~\thref{l:meas-of-meas-of-section},\\
  Lemma~\thref{l:tensor-prod-of-sigma-finite-meas},\\
  Lemma~\thref{l:uniq-of-tensor-prod-meas}.

\item[Definition~\thref{d:diffuse-meas}] \mbox{}\\
  is explicitly cited in the proof of:\\
  Lemma~\thref{l:lebesgue-meas-is-diffuse},\\
  Lemma~\thref{l:lebesgue-meas-on-r2-is-diffuse},\\
  Lemma~\thref{l:int-over-int}.

\item[Lemma~\thref{l:finite-meas-is-sigma-finite}] \mbox{}\\
  is explicitly cited in the proof of:\\
  Lemma~\thref{l:uniq-of-tensor-prod-meas-finite}.

\item[Lemma~\thref{l:trace-meas}] \mbox{}\\
  is explicitly cited in the proof of:\\
  Lemma~\thref{l:int-in-if-over-subset}.

\item[Lemma~\thref{l:restr-meas}] \mbox{}\\
  is explicitly cited in the proof of:\\
  Lemma~\thref{l:meas-of-meas-of-section},\\
  Lemma~\thref{l:uniq-of-tensor-prod-meas}.

\item[Definition~\thref{d:negl-subset}] \mbox{}\\
  is explicitly cited in the proof of:\\
  Lemma~\thref{l:equiv-def-of-considerable-subset},\\
  Lemma~\thref{l:negl-of-meas-subset},\\
  Lemma~\thref{l:n-is-closed-under-count-union},\\
  Lemma~\thref{l:subset-of-negl-is-negl},\\
  Lemma~\thref{l:masking-almost-nowhere},\\
  Lemma~\thref{l:compat-of-int-in-mplus-with-almost-bin-rel},\\
  Theorem~\thref{t:lebesgue-ext-dom-conv}.

\item[Definition~\thref{d:complete-meas}] \mbox{}\\
  is not yet used.

\item[Definition~\thref{d:considerable-subset}] \mbox{}\\
  is explicitly cited in the proof of:\\
  Lemma~\thref{l:equiv-def-of-considerable-subset}.

\item[Lemma~\thref{l:equiv-def-of-considerable-subset}] \mbox{}\\
  is not yet used.

\item[Lemma~\thref{l:negl-of-meas-subset}] \mbox{}\\
  is explicitly cited in the proof of:\\
  Lemma~\thref{l:empty-set-is-negl},\\
  Lemma~\thref{l:finite-nonneg-part},\\
  Lemma~\thref{l:int-in-mplus-is-almost-definite},\\
  Lemma~\thref{l:compat-of-int-in-mplus-with-almost-bin-rel},\\
  Lemma~\thref{l:integrable-in-mplus-is-almost-finite},\\
  Lemma~\thref{l:negl-of-meas-section}.

\item[Lemma~\thref{l:empty-set-is-negl}] \mbox{}\\
  is explicitly cited in the proof of:\\
  Lemma~\thref{l:everywhere-implies-almost-everywhere},\\
  Lemma~\thref{l:compat-of-almost-sum-with-almost-eq}.

\item[Lemma~\thref{l:compat-of-null-meas-with-count-union}] \mbox{}\\
  is explicitly cited in the proof of:\\
  Lemma~\thref{l:n-is-closed-under-count-union},\\
  Theorem~\thref{t:lebesgue-ext-dom-conv}.

\item[Lemma~\thref{l:n-is-closed-under-count-union}] \mbox{}\\
  is explicitly cited in the proof of:\\
  Lemma~\thref{l:ext-almost-modus-ponens},\\
  Lemma~\thref{l:compat-of-almost-bin-rel-with-antisym},\\
  Lemma~\thref{l:compat-of-almost-bin-rel-with-trans},\\
  Lemma~\thref{l:compat-of-almost-bin-rel-with-op}.

\item[Lemma~\thref{l:subset-of-negl-is-negl}] \mbox{}\\
  is explicitly cited in the proof of:\\
  Lemma~\thref{l:everywhere-implies-almost-everywhere-for-almost-the-same},\\
  Lemma~\thref{l:ext-almost-modus-ponens},\\
  Lemma~\thref{l:compat-of-almost-bin-rel-with-antisym},\\
  Lemma~\thref{l:compat-of-almost-bin-rel-with-trans}.

\item[Definition~\thref{d:prop-almost-satisfied}] \mbox{}\\
  is explicitly cited in the proof of:\\
  Lemma~\thref{l:everywhere-implies-almost-everywhere},\\
  Lemma~\thref{l:everywhere-implies-almost-everywhere-for-almost-the-same},\\
  Lemma~\thref{l:ext-almost-modus-ponens},\\
  Lemma~\thref{l:compat-of-almost-bin-rel-with-refl},\\
  Lemma~\thref{l:compat-of-almost-bin-rel-with-sym},\\
  Lemma~\thref{l:compat-of-almost-bin-rel-with-antisym},\\
  Lemma~\thref{l:compat-of-almost-bin-rel-with-trans},\\
  Lemma~\thref{l:compat-of-almost-bin-rel-with-op},\\
  Lemma~\thref{l:negl-of-summability-domain},\\
  Lemma~\thref{l:almost-sum},\\
  Lemma~\thref{l:masking-almost-nowhere},\\
  Lemma~\thref{l:finite-nonneg-part},\\
  Lemma~\thref{l:int-in-mplus-is-almost-definite},\\
  Lemma~\thref{l:compat-of-int-in-mplus-with-almost-bin-rel},\\
  Lemma~\thref{l:integrable-in-mplus-is-almost-finite},\\
  Theorem~\thref{t:lebesgue-ext-dom-conv}.

\item[Lemma~\thref{l:everywhere-implies-almost-everywhere}] \mbox{}\\
  is explicitly cited in the proof of:\\
  Lemma~\thref{l:almost-modus-ponens},\\
  Lemma~\thref{l:compat-of-almost-bin-rel-with-refl},\\
  Lemma~\thref{l:compat-of-int-in-mplus-with-almost-bin-rel},\\
  Lemma~\thref{l:bounded-by-integrable-is-integrable},\\
  Lemma~\thref{l:minkowski-ineq-in-llone},\\
  Lemma~\thref{l:llone-is-seminormed-vector-space}.

\item[Lemma~\thref{l:everywhere-implies-almost-everywhere-for-almost-the-same}] \mbox{}\\
  is explicitly cited in the proof of:\\
  Lemma~\thref{l:compat-of-almost-bin-rel-with-refl},\\
  Lemma~\thref{l:compat-of-almost-bin-rel-with-sym},\\
  Lemma~\thref{l:compat-of-almost-bin-rel-with-antisym},\\
  Lemma~\thref{l:compat-of-almost-bin-rel-with-trans},\\
  Lemma~\thref{l:compat-of-almost-bin-rel-with-op},\\
  Lemma~\thref{l:definiteness-implies-almost-definiteness}.

\item[Lemma~\thref{l:ext-almost-modus-ponens}] \mbox{}\\
  is explicitly cited in the proof of:\\
  Lemma~\thref{l:almost-modus-ponens}.

\item[Lemma~\thref{l:almost-modus-ponens}] \mbox{}\\
  is explicitly cited in the proof of:\\
  Lemma~\thref{l:compat-of-almost-bin-rel-with-sym},\\
  Lemma~\thref{l:definiteness-implies-almost-definiteness}.

\item[Definition~\thref{d:almost-definition}] \mbox{}\\
  is explicitly cited in the proof of:\\
  Lemma~\thref{l:compat-of-almost-bin-rel-with-refl},\\
  Lemma~\thref{l:compat-of-almost-bin-rel-with-sym},\\
  Lemma~\thref{l:compat-of-almost-bin-rel-with-antisym},\\
  Lemma~\thref{l:compat-of-almost-bin-rel-with-trans},\\
  Lemma~\thref{l:compat-of-almost-bin-rel-with-op}.

\item[Definition~\thref{d:almost-bin-rel}] \mbox{}\\
  is explicitly cited in the proof of:\\
  Lemma~\thref{l:compat-of-almost-bin-rel-with-refl},\\
  Lemma~\thref{l:compat-of-almost-bin-rel-with-sym},\\
  Lemma~\thref{l:compat-of-almost-bin-rel-with-antisym},\\
  Lemma~\thref{l:compat-of-almost-bin-rel-with-trans},\\
  Lemma~\thref{l:compat-of-almost-bin-rel-with-op},\\
  Lemma~\thref{l:compat-of-int-in-mplus-with-almost-bin-rel}.

\item[Lemma~\thref{l:compat-of-almost-bin-rel-with-refl}] \mbox{}\\
  is explicitly cited in the proof of:\\
  Lemma~\thref{l:almost-equiv-is-equiv-rel},\\
  Lemma~\thref{l:almost-order-is-order-rel}.

\item[Lemma~\thref{l:compat-of-almost-bin-rel-with-sym}] \mbox{}\\
  is explicitly cited in the proof of:\\
  Lemma~\thref{l:almost-equiv-is-equiv-rel}.

\item[Lemma~\thref{l:compat-of-almost-bin-rel-with-antisym}] \mbox{}\\
  is explicitly cited in the proof of:\\
  Lemma~\thref{l:almost-order-is-order-rel}.

\item[Lemma~\thref{l:compat-of-almost-bin-rel-with-trans}] \mbox{}\\
  is explicitly cited in the proof of:\\
  Lemma~\thref{l:almost-equiv-is-equiv-rel},\\
  Lemma~\thref{l:almost-order-is-order-rel}.

\item[Lemma~\thref{l:almost-equiv-is-equiv-rel}] \mbox{}\\
  is explicitly cited in the proof of:\\
  Lemma~\thref{l:almost-eq-is-equiv-rel}.

\item[Lemma~\thref{l:almost-eq-is-equiv-rel}] \mbox{}\\
  is explicitly cited in the proof of:\\
  Lemma~\thref{l:compat-of-almost-sum-with-almost-eq},\\
  Lemma~\thref{l:compat-of-int-in-mplus-with-almost-bin-rel}.

\item[Lemma~\thref{l:almost-order-is-order-rel}] \mbox{}\\
  is explicitly cited in the proof of:\\
  Lemma~\thref{l:almost-bounded-by-integrable-is-integrable}.

\item[Lemma~\thref{l:compat-of-almost-bin-rel-with-op}] \mbox{}\\
  is explicitly cited in the proof of:\\
  Lemma~\thref{l:compat-of-almost-eq-with-op},\\
  Lemma~\thref{l:compat-of-almost-ineq-with-op}.

\item[Lemma~\thref{l:compat-of-almost-eq-with-op}] \mbox{}\\
  is explicitly cited in the proof of:\\
  Lemma~\thref{l:compat-of-almost-sum-with-almost-eq},\\
  Lemma~\thref{l:abs-is-almost-definite},\\
  Lemma~\thref{l:compat-of-int-in-mplus-with-almost-bin-rel},\\
  Lemma~\thref{l:compat-of-int-with-almost-eq},\\
  Lemma~\thref{l:compat-of-none-with-almost-eq},\\
  Lemma~\thref{l:first-mean-value-theorem},\\
  Theorem~\thref{t:lebesgue-ext-dom-conv}.

\item[Lemma~\thref{l:compat-of-almost-ineq-with-op}] \mbox{}\\
  is not yet used.

\item[Lemma~\thref{l:definiteness-implies-almost-definiteness}] \mbox{}\\
  is explicitly cited in the proof of:\\
  Lemma~\thref{l:abs-is-almost-definite}.

\item[Lemma~\thref{l:uniq-of-meas-ext-from-p-syst}] \mbox{}\\
  is explicitly cited in the proof of:\\
  Theorem~\thref{t:caratheodory-lebesgue-meas-on-r}.

\item[Lemma~\thref{l:trivial-meas}] \mbox{}\\
  is explicitly cited in the proof of:\\
  Lemma~\thref{l:equiv-def-of-trivial-meas}.

\item[Lemma~\thref{l:equiv-def-of-trivial-meas}] \mbox{}\\
  is not yet used.

\item[Lemma~\thref{l:count-meas}] \mbox{}\\
  is explicitly cited in the proof of:\\
  Lemma~\thref{l:finiteness-of-count-meas},\\
  Lemma~\thref{l:sigma-finiteness-of-count-meas},\\
  Lemma~\thref{l:equiv-def-of-dirac-meas},\\
  Lemma~\thref{l:dirac-meas-is-finite},\\
  Lemma~\thref{l:int-in-if-for-count-meas}.

\item[Lemma~\thref{l:finiteness-of-count-meas}] \mbox{}\\
  is explicitly cited in the proof of:\\
  Lemma~\thref{l:dirac-meas-is-finite}.

\item[Lemma~\thref{l:sigma-finiteness-of-count-meas}] \mbox{}\\
  is not yet used.

\item[Definition~\thref{d:dirac-meas}] \mbox{}\\
  is explicitly cited in the proof of:\\
  Lemma~\thref{l:equiv-def-of-dirac-meas},\\
  Lemma~\thref{l:dirac-meas-is-finite},\\
  Lemma~\thref{l:int-in-sfplus-for-dirac-meas},\\
  Lemma~\thref{l:int-in-mplus-for-dirac-meas},\\
  Lemma~\thref{l:int-for-dirac-meas}.

\item[Lemma~\thref{l:equiv-def-of-dirac-meas}] \mbox{}\\
  is not yet used.

\item[Lemma~\thref{l:dirac-meas-is-finite}] \mbox{}\\
  is not yet used.

\item[Definition~\thref{d:summability-domain}] \mbox{}\\
  is explicitly cited in the proof of:\\
  Lemma~\thref{l:summability-on-summability-domain},\\
  Lemma~\thref{l:almost-sum-is-sum},\\
  Lemma~\thref{l:minkowski-ineq-in-llone}.

\item[Lemma~\thref{l:summability-on-summability-domain}] \mbox{}\\
  is explicitly cited in the proof of:\\
  Lemma~\thref{l:negl-of-summability-domain}.

\item[Lemma~\thref{l:meas-of-summability-domain}] \mbox{}\\
  is explicitly cited in the proof of:\\
  Lemma~\thref{l:almost-sum}.

\item[Lemma~\thref{l:negl-of-summability-domain}] \mbox{}\\
  is explicitly cited in the proof of:\\
  Lemma~\thref{l:almost-sum},\\
  Lemma~\thref{l:compat-of-almost-sum-with-almost-eq}.

\item[Lemma~\thref{l:almost-sum}] \mbox{}\\
  is explicitly cited in the proof of:\\
  Lemma~\thref{l:compat-of-almost-sum-with-almost-eq},\\
  Lemma~\thref{l:almost-sum-is-sum},\\
  Lemma~\thref{l:minkowski-ineq-in-m}.

\item[Lemma~\thref{l:compat-of-almost-sum-with-almost-eq}] \mbox{}\\
  is explicitly cited in the proof of:\\
  Lemma~\thref{l:minkowski-ineq-in-m}.

\item[Lemma~\thref{l:almost-sum-is-sum}] \mbox{}\\
  is explicitly cited in the proof of:\\
  Lemma~\thref{l:minkowski-ineq-in-llone}.

\item[Lemma~\thref{l:abs-is-almost-definite}] \mbox{}\\
  is explicitly cited in the proof of:\\
  Lemma~\thref{l:none-is-almost-definite}.

\item[Lemma~\thref{l:masking-almost-nowhere}] \mbox{}\\
  is explicitly cited in the proof of:\\
  Lemma~\thref{l:finite-nonneg-part},\\
  Theorem~\thref{t:lebesgue-ext-dom-conv}.

\item[Lemma~\thref{l:finite-nonneg-part}] \mbox{}\\
  is explicitly cited in the proof of:\\
  Theorem~\thref{t:lebesgue-ext-dom-conv}.

\item[Definition~\thref{d:len-of-int}] \mbox{}\\
  is explicitly cited in the proof of:\\
  Lemma~\thref{l:len-is-nonneg},\\
  Lemma~\thref{l:len-is-hom},\\
  Lemma~\thref{l:len-of-partition},\\
  Lemma~\thref{l:lambda-star-is-hom},\\
  Lemma~\thref{l:lambda-star-gen-len-of-int}.

\item[Lemma~\thref{l:len-is-nonneg}] \mbox{}\\
  is explicitly cited in the proof of:\\
  Lemma~\thref{l:lambda-star-is-nonneg}.

\item[Lemma~\thref{l:len-is-hom}] \mbox{}\\
  is explicitly cited in the proof of:\\
  Lemma~\thref{l:len-of-partition}.

\item[Lemma~\thref{l:len-of-partition}] \mbox{}\\
  is not yet used.

\item[Definition~\thref{d:set-of-count-cover-with-bounded-open-int}] \mbox{}\\
  is explicitly cited in the proof of:\\
  Lemma~\thref{l:set-of-count-cover-with-bounded-open-int-is-nonempty},\\
  Lemma~\thref{l:lambda-star-is-monot},\\
  Lemma~\thref{l:lambda-star-is-sigma-subadd},\\
  Lemma~\thref{l:lambda-star-gen-len-of-int}.

\item[Lemma~\thref{l:set-of-count-cover-with-bounded-open-int-is-nonempty}] \mbox{}\\
  is not yet used.

\item[Definition~\thref{d:lambda-star-lebesgue-meas-cand}] \mbox{}\\
  is explicitly cited in the proof of:\\
  Lemma~\thref{l:lambda-star-is-hom},\\
  Lemma~\thref{l:lambda-star-is-monot},\\
  Lemma~\thref{l:lambda-star-is-sigma-subadd},\\
  Lemma~\thref{l:lambda-star-gen-len-of-int},\\
  Lemma~\thref{l:rays-are-lebesgue-meas}.

\item[Lemma~\thref{l:lambda-star-is-nonneg}] \mbox{}\\
  is explicitly cited in the proof of:\\
  Lemma~\thref{l:lambda-star-is-hom},\\
  Lemma~\thref{l:lambda-star-is-meas-on-l}.

\item[Lemma~\thref{l:lambda-star-is-hom}] \mbox{}\\
  is explicitly cited in the proof of:\\
  Lemma~\thref{l:lambda-star-gen-len-of-int},\\
  Lemma~\thref{l:l-is-set-alg},\\
  Lemma~\thref{l:lambda-star-is-meas-on-l}.

\item[Lemma~\thref{l:lambda-star-is-monot}] \mbox{}\\
  is explicitly cited in the proof of:\\
  Lemma~\thref{l:lambda-star-gen-len-of-int},\\
  Lemma~\thref{l:lambda-star-is-sigma-add-on-l},\\
  Lemma~\thref{l:l-is-closed-under-count-union},\\
  Lemma~\thref{l:rays-are-lebesgue-meas}.

\item[Lemma~\thref{l:lambda-star-is-sigma-subadd}] \mbox{}\\
  is explicitly cited in the proof of:\\
  Lemma~\thref{l:equiv-def-of-l},\\
  Lemma~\thref{l:l-is-closed-under-finite-union},\\
  Lemma~\thref{l:lambda-star-is-sigma-add-on-l},\\
  Lemma~\thref{l:l-is-closed-under-count-union},\\
  Lemma~\thref{l:rays-are-lebesgue-meas}.

\item[Lemma~\thref{l:lambda-star-gen-len-of-int}] \mbox{}\\
  is explicitly cited in the proof of:\\
  Lemma~\thref{l:rays-are-lebesgue-meas},\\
  Theorem~\thref{t:caratheodory-lebesgue-meas-on-r},\\
  Lemma~\thref{l:lebesgue-meas-gen-len-of-int},\\
  Lemma~\thref{l:lebesgue-meas-is-sigma-finite},\\
  Lemma~\thref{l:lebesgue-meas-is-diffuse}.

\item[Definition~\thref{d:l-lebesgue-sigma-alg}] \mbox{}\\
  is explicitly cited in the proof of:\\
  Lemma~\thref{l:equiv-def-of-l},\\
  Lemma~\thref{l:l-is-closed-under-compl},\\
  Lemma~\thref{l:l-is-closed-under-finite-union},\\
  Lemma~\thref{l:lambda-star-is-add-on-l},\\
  Lemma~\thref{l:l-is-closed-under-count-union}.

\item[Lemma~\thref{l:equiv-def-of-l}] \mbox{}\\
  is explicitly cited in the proof of:\\
  Lemma~\thref{l:l-is-closed-under-finite-union},\\
  Lemma~\thref{l:l-is-closed-under-count-union},\\
  Lemma~\thref{l:rays-are-lebesgue-meas}.

\item[Lemma~\thref{l:l-is-closed-under-compl}] \mbox{}\\
  is explicitly cited in the proof of:\\
  Lemma~\thref{l:l-is-closed-under-finite-inter},\\
  Lemma~\thref{l:l-is-set-alg},\\
  Lemma~\thref{l:rays-are-lebesgue-meas}.

\item[Lemma~\thref{l:l-is-closed-under-finite-union}] \mbox{}\\
  is explicitly cited in the proof of:\\
  Lemma~\thref{l:l-is-closed-under-finite-inter},\\
  Lemma~\thref{l:l-is-closed-under-count-union}.

\item[Lemma~\thref{l:l-is-closed-under-finite-inter}] \mbox{}\\
  is explicitly cited in the proof of:\\
  Lemma~\thref{l:l-is-set-alg},\\
  Lemma~\thref{l:int-are-lebesgue-meas}.

\item[Lemma~\thref{l:l-is-set-alg}] \mbox{}\\
  is explicitly cited in the proof of:\\
  Lemma~\thref{l:part-of-count-union-in-l},\\
  Lemma~\thref{l:l-is-sigma-alg}.

\item[Lemma~\thref{l:lambda-star-is-add-on-l}] \mbox{}\\
  is explicitly cited in the proof of:\\
  Lemma~\thref{l:lambda-star-is-sigma-add-on-l},\\
  Lemma~\thref{l:l-is-closed-under-count-union}.

\item[Lemma~\thref{l:lambda-star-is-sigma-add-on-l}] \mbox{}\\
  is explicitly cited in the proof of:\\
  Lemma~\thref{l:lambda-star-is-meas-on-l}.

\item[Lemma~\thref{l:part-of-count-union-in-l}] \mbox{}\\
  is explicitly cited in the proof of:\\
  Lemma~\thref{l:l-is-closed-under-count-union}.

\item[Lemma~\thref{l:l-is-closed-under-count-union}] \mbox{}\\
  is explicitly cited in the proof of:\\
  Lemma~\thref{l:rays-are-lebesgue-meas},\\
  Lemma~\thref{l:l-is-sigma-alg}.

\item[Lemma~\thref{l:rays-are-lebesgue-meas}] \mbox{}\\
  is explicitly cited in the proof of:\\
  Lemma~\thref{l:int-are-lebesgue-meas}.

\item[Lemma~\thref{l:int-are-lebesgue-meas}] \mbox{}\\
  is explicitly cited in the proof of:\\
  Lemma~\thref{l:br-is-sub-sigma-alg-of-l}.

\item[Lemma~\thref{l:l-is-sigma-alg}] \mbox{}\\
  is explicitly cited in the proof of:\\
  Lemma~\thref{l:lambda-star-is-meas-on-l}.

\item[Lemma~\thref{l:lambda-star-is-meas-on-l}] \mbox{}\\
  is explicitly cited in the proof of:\\
  Lemma~\thref{l:lambda-star-is-meas-on-br}.

\item[Lemma~\thref{l:br-is-sub-sigma-alg-of-l}] \mbox{}\\
  is explicitly cited in the proof of:\\
  Lemma~\thref{l:lambda-star-is-meas-on-br}.

\item[Lemma~\thref{l:lambda-star-is-meas-on-br}] \mbox{}\\
  is explicitly cited in the proof of:\\
  Theorem~\thref{t:caratheodory-lebesgue-meas-on-r}.

\item[Theorem~\thref{t:caratheodory-lebesgue-meas-on-r}] \mbox{}\\
  is explicitly cited in the proof of:\\
  Lemma~\thref{l:lebesgue-meas-gen-len-of-int},\\
  Lemma~\thref{l:lebesgue-meas-is-sigma-finite},\\
  Lemma~\thref{l:lebesgue-meas-is-diffuse},\\
  Lemma~\thref{l:lebesgue-meas-on-r2}.

\item[Lemma~\thref{l:lebesgue-meas-gen-len-of-int}] \mbox{}\\
  is explicitly cited in the proof of:\\
  Lemma~\thref{l:lebesgue-meas-on-r2-gen-area-of-boxes}.

\item[Lemma~\thref{l:lebesgue-meas-is-sigma-finite}] \mbox{}\\
  is explicitly cited in the proof of:\\
  Lemma~\thref{l:lebesgue-meas-on-r2},\\
  Lemma~\thref{l:lebesgue-meas-on-r2-is-sigma-finite}.

\item[Lemma~\thref{l:lebesgue-meas-is-diffuse}] \mbox{}\\
  is not yet used.

\item[Definition~\thref{d:if-set-of-meas-indic-funs}] \mbox{}\\
  is explicitly cited in the proof of:\\
  Lemma~\thref{l:indic-and-support-are-each-other-inverse},\\
  Lemma~\thref{l:if-is-meas},\\
  Lemma~\thref{l:if-is-sigma-add},\\
  Lemma~\thref{l:if-is-closed-under-mult},\\
  Lemma~\thref{l:if-is-closed-under-ext-by-zero},\\
  Lemma~\thref{l:if-is-closed-under-restr},\\
  Lemma~\thref{l:sf-simple-repr},\\
  Lemma~\thref{l:sfplus-is-meas},\\
  Lemma~\thref{l:int-in-sfplus-gen-int-in-if},\\
  Lemma~\thref{l:int-in-sfplus-over-subset-is-add}.

\item[Lemma~\thref{l:indic-and-support-are-each-other-inverse}] \mbox{}\\
  is explicitly cited in the proof of:\\
  Lemma~\thref{l:equiv-def-of-int-in-if},\\
  Lemma~\thref{l:int-in-if-over-subset},\\
  Lemma~\thref{l:int-in-if-over-subset-is-add},\\
  Lemma~\thref{l:int-in-if-for-count-meas},\\
  Lemma~\thref{l:sf-simple-repr},\\
  Theorem~\thref{t:tonelli}.

\item[Lemma~\thref{l:if-is-meas}] \mbox{}\\
  is not yet used.

\item[Lemma~\thref{l:if-is-sigma-add}] \mbox{}\\
  is explicitly cited in the proof of:\\
  Lemma~\thref{l:int-in-if-is-add},\\
  Lemma~\thref{l:int-in-if-over-subset-is-add},\\
  Lemma~\thref{l:sf-is-alg-over-r},\\
  Lemma~\thref{l:int-in-sfplus-over-subset-is-add},\\
  Lemma~\thref{l:int-in-mplus-over-subset-is-sigma-add}.

\item[Lemma~\thref{l:if-is-closed-under-mult}] \mbox{}\\
  is explicitly cited in the proof of:\\
  Lemma~\thref{l:int-in-if-over-subset-is-add},\\
  Lemma~\thref{l:sf-is-alg-over-r},\\
  Lemma~\thref{l:int-in-sfplus-over-subset-is-add},\\
  Lemma~\thref{l:int-in-mplus-over-subset-is-sigma-add}.

\item[Lemma~\thref{l:if-is-closed-under-ext-by-zero}] \mbox{}\\
  is explicitly cited in the proof of:\\
  Lemma~\thref{l:int-in-if-over-subset},\\
  Lemma~\thref{l:sf-is-closed-under-ext-by-zero}.

\item[Lemma~\thref{l:if-is-closed-under-restr}] \mbox{}\\
  is explicitly cited in the proof of:\\
  Lemma~\thref{l:int-in-if-over-subset},\\
  Lemma~\thref{l:sf-is-closed-under-restr}.

\item[Definition~\thref{d:int-in-if}] \mbox{}\\
  is explicitly cited in the proof of:\\
  Lemma~\thref{l:equiv-def-of-int-in-if}.

\item[Lemma~\thref{l:equiv-def-of-int-in-if}] \mbox{}\\
  is explicitly cited in the proof of:\\
  Lemma~\thref{l:int-in-if-is-add},\\
  Lemma~\thref{l:int-in-if-over-subset},\\
  Lemma~\thref{l:int-in-if-for-count-meas},\\
  Lemma~\thref{l:int-in-sfplus-gen-int-in-if}.

\item[Lemma~\thref{l:int-in-if-is-add}] \mbox{}\\
  is explicitly cited in the proof of:\\
  Lemma~\thref{l:int-in-if-over-subset-is-add}.

\item[Lemma~\thref{l:int-in-if-over-subset}] \mbox{}\\
  is explicitly cited in the proof of:\\
  Lemma~\thref{l:int-in-if-over-subset-is-add},\\
  Lemma~\thref{l:int-in-sfplus-over-subset}.

\item[Lemma~\thref{l:int-in-if-over-subset-is-add}] \mbox{}\\
  is not yet used.

\item[Lemma~\thref{l:int-in-if-for-count-meas}] \mbox{}\\
  is explicitly cited in the proof of:\\
  Lemma~\thref{l:int-in-sfplus-for-count-meas}.

\item[Definition~\thref{d:sf-vector-space-of-simple-funs}] \mbox{}\\
  is explicitly cited in the proof of:\\
  Lemma~\thref{l:sf-simple-repr},\\
  Lemma~\thref{l:sf-is-alg-over-r},\\
  Lemma~\thref{l:sf-is-meas},\\
  Lemma~\thref{l:sf-is-closed-under-ext-by-zero},\\
  Lemma~\thref{l:sf-is-closed-under-restr},\\
  Lemma~\thref{l:int-in-sfplus-over-subset},\\
  Lemma~\thref{l:int-in-sfplus-over-subset-is-add},\\
  Lemma~\thref{l:int-in-mplus-is-pos-hom},\\
  Theorem~\thref{t:beppo-levi-monot-conv},\\
  Lemma~\thref{l:adapted-seq-in-mplus},\\
  Lemma~\thref{l:const-fun-is-llone}.

\item[Lemma~\thref{l:sf-simple-repr}] \mbox{}\\
  is explicitly cited in the proof of:\\
  Lemma~\thref{l:sf-can-repr},\\
  Lemma~\thref{l:sf-disj-repr},\\
  Lemma~\thref{l:sf-is-alg-over-r},\\
  Lemma~\thref{l:sfplus-simple-repr}.

\item[Lemma~\thref{l:sf-can-repr}] \mbox{}\\
  is explicitly cited in the proof of:\\
  Lemma~\thref{l:sf-disj-repr},\\
  Lemma~\thref{l:sf-disj-repr-is-subpart-of-can-repr},\\
  Lemma~\thref{l:sfplus-can-repr},\\
  Lemma~\thref{l:decomp-of-meas-in-sfplus}.

\item[Lemma~\thref{l:sf-disj-repr}] \mbox{}\\
  is explicitly cited in the proof of:\\
  Lemma~\thref{l:sf-disj-repr-is-subpart-of-can-repr},\\
  Lemma~\thref{l:sf-is-alg-over-r},\\
  Lemma~\thref{l:sfplus-disj-repr}.

\item[Lemma~\thref{l:sf-disj-repr-is-subpart-of-can-repr}] \mbox{}\\
  is explicitly cited in the proof of:\\
  Lemma~\thref{l:sfplus-disj-repr-is-subpart-of-can-repr}.

\item[Lemma~\thref{l:sf-is-alg-over-r}] \mbox{}\\
  is explicitly cited in the proof of:\\
  Lemma~\thref{l:sfplus-is-closed-under-pos-alg-ops},\\
  Lemma~\thref{l:int-in-sfplus-is-monot}.

\item[Lemma~\thref{l:sf-is-meas}] \mbox{}\\
  is explicitly cited in the proof of:\\
  Lemma~\thref{l:sfplus-is-meas},\\
  Theorem~\thref{t:beppo-levi-monot-conv}.

\item[Lemma~\thref{l:sf-is-closed-under-ext-by-zero}] \mbox{}\\
  is explicitly cited in the proof of:\\
  Lemma~\thref{l:int-in-sfplus-over-subset}.

\item[Lemma~\thref{l:sf-is-closed-under-restr}] \mbox{}\\
  is explicitly cited in the proof of:\\
  Lemma~\thref{l:int-in-sfplus-over-subset},\\
  Lemma~\thref{l:int-in-mplus-over-subset}.

\item[Definition~\thref{d:sfplus-subset-of-nonneg-simple-funs}] \mbox{}\\
  is explicitly cited in the proof of:\\
  Lemma~\thref{l:sfplus-disj-repr},\\
  Lemma~\thref{l:sfplus-can-repr},\\
  Lemma~\thref{l:sfplus-simple-repr},\\
  Lemma~\thref{l:sfplus-is-closed-under-pos-alg-ops},\\
  Lemma~\thref{l:sfplus-is-meas},\\
  Lemma~\thref{l:int-in-sfplus-is-monot},\\
  Lemma~\thref{l:int-in-sfplus-over-subset-is-add},\\
  Lemma~\thref{l:int-in-mplus-is-pos-hom},\\
  Lemma~\thref{l:adapted-seq-in-mplus}.

\item[Lemma~\thref{l:sfplus-disj-repr}] \mbox{}\\
  is explicitly cited in the proof of:\\
  Lemma~\thref{l:sfplus-disj-repr-is-subpart-of-can-repr},\\
  Lemma~\thref{l:sfplus-is-closed-under-pos-alg-ops},\\
  Lemma~\thref{l:int-in-sfplus-is-add}.

\item[Lemma~\thref{l:sfplus-can-repr}] \mbox{}\\
  is explicitly cited in the proof of:\\
  Lemma~\thref{l:sfplus-disj-repr-is-subpart-of-can-repr},\\
  Lemma~\thref{l:sfplus-simple-repr},\\
  Lemma~\thref{l:int-in-sfplus},\\
  Lemma~\thref{l:equiv-def-of-int-in-sfplus-disj}.

\item[Lemma~\thref{l:sfplus-disj-repr-is-subpart-of-can-repr}] \mbox{}\\
  is explicitly cited in the proof of:\\
  Lemma~\thref{l:equiv-def-of-int-in-sfplus-disj}.

\item[Lemma~\thref{l:sfplus-simple-repr}] \mbox{}\\
  is explicitly cited in the proof of:\\
  Lemma~\thref{l:int-in-sfplus-gen-int-in-if},\\
  Lemma~\thref{l:equiv-def-of-int-in-sfplus-simple},\\
  Lemma~\thref{l:int-in-sfplus-for-count-meas},\\
  Theorem~\thref{t:tonelli}.

\item[Lemma~\thref{l:sfplus-is-closed-under-pos-alg-ops}] \mbox{}\\
  is explicitly cited in the proof of:\\
  Lemma~\thref{l:int-in-sfplus-is-add},\\
  Lemma~\thref{l:change-of-variable-in-sum-in-sfplus},\\
  Lemma~\thref{l:int-in-sfplus-is-add-alt-proof},\\
  Lemma~\thref{l:int-in-sfplus-is-pos-lin},\\
  Lemma~\thref{l:int-in-mplus-is-pos-hom}.

\item[Lemma~\thref{l:sfplus-is-meas}] \mbox{}\\
  is explicitly cited in the proof of:\\
  Lemma~\thref{l:decomp-of-meas-in-sfplus},\\
  Lemma~\thref{l:change-of-variable-in-sum-in-sfplus},\\
  Lemma~\thref{l:int-in-sfplus-is-add-alt-proof},\\
  Lemma~\thref{l:int-in-mplus-of-indic-fun}.

\item[Lemma~\thref{l:int-in-sfplus}] \mbox{}\\
  is explicitly cited in the proof of:\\
  Lemma~\thref{l:int-in-sfplus-gen-int-in-if},\\
  Lemma~\thref{l:equiv-def-of-int-in-sfplus-disj},\\
  Lemma~\thref{l:int-in-sfplus-is-add-alt-proof},\\
  Lemma~\thref{l:int-in-sfplus-is-pos-lin},\\
  Lemma~\thref{l:int-in-sfplus-is-monot},\\
  Lemma~\thref{l:int-in-mplus},\\
  Lemma~\thref{l:int-in-mplus-gen-int-in-sfplus},\\
  Lemma~\thref{l:int-in-mplus-is-pos-hom},\\
  Lemma~\thref{l:const-fun-is-llone}.

\item[Lemma~\thref{l:int-in-sfplus-gen-int-in-if}] \mbox{}\\
  is explicitly cited in the proof of:\\
  Lemma~\thref{l:int-in-sfplus-is-pos-lin},\\
  Lemma~\thref{l:equiv-def-of-int-in-sfplus-simple},\\
  Lemma~\thref{l:int-in-sfplus-over-subset},\\
  Lemma~\thref{l:int-in-sfplus-for-count-meas},\\
  Lemma~\thref{l:int-in-mplus-of-indic-fun}.

\item[Lemma~\thref{l:equiv-def-of-int-in-sfplus-disj}] \mbox{}\\
  is explicitly cited in the proof of:\\
  Lemma~\thref{l:int-in-sfplus-is-add}.

\item[Lemma~\thref{l:int-in-sfplus-is-add}] \mbox{}\\
  is explicitly cited in the proof of:\\
  Lemma~\thref{l:int-in-sfplus-is-pos-lin}.

\item[Lemma~\thref{l:decomp-of-meas-in-sfplus}] \mbox{}\\
  is explicitly cited in the proof of:\\
  Lemma~\thref{l:int-in-sfplus-is-add-alt-proof}.

\item[Lemma~\thref{l:change-of-variable-in-sum-in-sfplus}] \mbox{}\\
  is explicitly cited in the proof of:\\
  Lemma~\thref{l:int-in-sfplus-is-add-alt-proof}.

\item[Lemma~\thref{l:int-in-sfplus-is-add-alt-proof}] \mbox{}\\
  is explicitly cited in the proof of:\\
  Lemma~\thref{l:int-in-sfplus-is-pos-lin}.

\item[Lemma~\thref{l:int-in-sfplus-is-pos-lin}] \mbox{}\\
  is explicitly cited in the proof of:\\
  Lemma~\thref{l:equiv-def-of-int-in-sfplus-simple},\\
  Lemma~\thref{l:int-in-sfplus-is-monot},\\
  Lemma~\thref{l:int-in-sfplus-over-subset},\\
  Lemma~\thref{l:int-in-mplus-is-pos-hom},\\
  Theorem~\thref{t:beppo-levi-monot-conv},\\
  Lemma~\thref{l:int-in-mplus-is-add}.

\item[Lemma~\thref{l:equiv-def-of-int-in-sfplus-simple}] \mbox{}\\
  is explicitly cited in the proof of:\\
  Lemma~\thref{l:int-in-mplus-is-pos-hom},\\
  Theorem~\thref{t:beppo-levi-monot-conv}.

\item[Lemma~\thref{l:int-in-sfplus-is-monot}] \mbox{}\\
  is explicitly cited in the proof of:\\
  Lemma~\thref{l:int-in-sfplus-is-cont}.

\item[Lemma~\thref{l:int-in-sfplus-is-cont}] \mbox{}\\
  is explicitly cited in the proof of:\\
  Lemma~\thref{l:int-in-mplus-gen-int-in-sfplus}.

\item[Lemma~\thref{l:int-in-sfplus-over-subset}] \mbox{}\\
  is explicitly cited in the proof of:\\
  Lemma~\thref{l:int-in-sfplus-over-subset-is-add},\\
  Lemma~\thref{l:int-in-mplus-over-subset}.

\item[Lemma~\thref{l:int-in-sfplus-over-subset-is-add}] \mbox{}\\
  is not yet used.

\item[Lemma~\thref{l:int-in-sfplus-for-count-meas}] \mbox{}\\
  is explicitly cited in the proof of:\\
  Lemma~\thref{l:int-in-sfplus-for-count-meas-on-n},\\
  Lemma~\thref{l:int-in-sfplus-for-dirac-meas},\\
  Lemma~\thref{l:int-in-mplus-for-count-meas}.

\item[Lemma~\thref{l:int-in-sfplus-for-count-meas-on-n}] \mbox{}\\
  is not yet used.

\item[Lemma~\thref{l:int-in-sfplus-for-dirac-meas}] \mbox{}\\
  is not yet used.

\item[Lemma~\thref{l:int-in-mplus}] \mbox{}\\
  is explicitly cited in the proof of:\\
  Lemma~\thref{l:int-in-mplus-gen-int-in-sfplus},\\
  Lemma~\thref{l:int-in-mplus-is-pos-hom},\\
  Lemma~\thref{l:int-in-mplus-is-monot},\\
  Theorem~\thref{t:beppo-levi-monot-conv},\\
  Lemma~\thref{l:compat-of-int-in-mplus-with-almost-eq},\\
  Lemma~\thref{l:integrable-in-mplus-is-almost-finite},\\
  Lemma~\thref{l:bounded-by-integrable-in-mplus-is-integrable},\\
  Lemma~\thref{l:cand-tensor-prod-meas-is-tensor-prod-meas},\\
  Lemma~\thref{l:integrable-is-meas},\\
  Lemma~\thref{l:equiv-def-of-integrability},\\
  Lemma~\thref{l:almost-bounded-by-integrable-is-integrable},\\
  Lemma~\thref{l:compat-of-int-in-m-and-mplus},\\
  Lemma~\thref{l:int-over-subset},\\
  Lemma~\thref{l:int-over-singleton},\\
  Lemma~\thref{l:int-for-count-meas},\\
  Lemma~\thref{l:seminorm-llone},\\
  Lemma~\thref{l:integrable-is-finite-seminorm-lone},\\
  Lemma~\thref{l:int-is-hom},\\
  Lemma~\thref{l:equiv-def-of-llone},\\
  Lemma~\thref{l:int-is-pos-lin-form-on-llone}.

\item[Lemma~\thref{l:int-in-mplus-gen-int-in-sfplus}] \mbox{}\\
  is explicitly cited in the proof of:\\
  Lemma~\thref{l:int-in-mplus-of-indic-fun},\\
  Lemma~\thref{l:usage-of-adapted-seqs},\\
  Lemma~\thref{l:const-fun-is-llone}.

\item[Lemma~\thref{l:int-in-mplus-of-indic-fun}] \mbox{}\\
  is explicitly cited in the proof of:\\
  Theorem~\thref{t:beppo-levi-monot-conv},\\
  Lemma~\thref{l:int-in-mplus-is-almost-definite},\\
  Lemma~\thref{l:compat-of-int-in-mplus-with-almost-bin-rel},\\
  Lemma~\thref{l:bienayme-chebyshev-ineq},\\
  Lemma~\thref{l:int-in-mplus-over-singleton},\\
  Lemma~\thref{l:meas-of-meas-of-section-finite},\\
  Lemma~\thref{l:cand-tensor-prod-meas-is-tensor-prod-meas},\\
  Theorem~\thref{t:tonelli}.

\item[Lemma~\thref{l:int-in-mplus-is-pos-hom}] \mbox{}\\
  is explicitly cited in the proof of:\\
  Lemma~\thref{l:int-in-mplus-of-zero-is-zero},\\
  Lemma~\thref{l:int-in-mplus-is-hom-at-infinity},\\
  Lemma~\thref{l:int-in-mplus-is-pos-lin},\\
  Lemma~\thref{l:bienayme-chebyshev-ineq},\\
  Lemma~\thref{l:tonelli-for-tensor-prod},\\
  Lemma~\thref{l:none-is-abs-hom},\\
  Lemma~\thref{l:int-is-hom}.

\item[Lemma~\thref{l:int-in-mplus-of-zero-is-zero}] \mbox{}\\
  is explicitly cited in the proof of:\\
  Lemma~\thref{l:cand-tensor-prod-meas-is-tensor-prod-meas},\\
  Lemma~\thref{l:compat-of-int-in-m-and-mplus},\\
  Lemma~\thref{l:int-of-zero-is-zero}.

\item[Lemma~\thref{l:int-in-mplus-is-monot}] \mbox{}\\
  is explicitly cited in the proof of:\\
  Theorem~\thref{t:beppo-levi-monot-conv},\\
  Lemma~\thref{l:int-in-mplus-is-almost-monot},\\
  Lemma~\thref{l:bienayme-chebyshev-ineq},\\
  Lemma~\thref{l:bounded-by-integrable-in-mplus-is-integrable},\\
  Theorem~\thref{t:fatou-lemma},\\
  Lemma~\thref{l:int-in-mplus-of-pointwise-conv-seq},\\
  Lemma~\thref{l:int-over-subset-is-sigma-add},\\
  Lemma~\thref{l:minkowski-ineq-in-m}.

\item[Theorem~\thref{t:beppo-levi-monot-conv}] \mbox{}\\
  is explicitly cited in the proof of:\\
  Lemma~\thref{l:int-in-mplus-is-hom-at-infinity},\\
  Lemma~\thref{l:usage-of-adapted-seqs},\\
  Lemma~\thref{l:int-in-mplus-is-sigma-add},\\
  Theorem~\thref{t:fatou-lemma}.

\item[Lemma~\thref{l:int-in-mplus-is-hom-at-infinity}] \mbox{}\\
  is explicitly cited in the proof of:\\
  Lemma~\thref{l:int-in-mplus-is-pos-lin},\\
  Lemma~\thref{l:int-in-mplus-is-almost-definite},\\
  Lemma~\thref{l:bienayme-chebyshev-ineq},\\
  Lemma~\thref{l:none-is-abs-hom}.

\item[Definition~\thref{d:adapted-seq}] \mbox{}\\
  is explicitly cited in the proof of:\\
  Lemma~\thref{l:adapted-seq-in-mplus},\\
  Lemma~\thref{l:usage-of-adapted-seqs},\\
  Lemma~\thref{l:int-in-mplus-over-subset}.

\item[Lemma~\thref{l:adapted-seq-in-mplus}] \mbox{}\\
  is explicitly cited in the proof of:\\
  Lemma~\thref{l:usage-of-adapted-seqs},\\
  Lemma~\thref{l:int-in-mplus-is-add},\\
  Lemma~\thref{l:int-in-mplus-over-subset},\\
  Theorem~\thref{t:tonelli}.

\item[Lemma~\thref{l:usage-of-adapted-seqs}] \mbox{}\\
  is explicitly cited in the proof of:\\
  Lemma~\thref{l:int-in-mplus-is-add},\\
  Lemma~\thref{l:int-in-mplus-over-subset},\\
  Lemma~\thref{l:int-in-mplus-for-count-meas},\\
  Theorem~\thref{t:tonelli}.

\item[Lemma~\thref{l:int-in-mplus-is-add}] \mbox{}\\
  is explicitly cited in the proof of:\\
  Lemma~\thref{l:int-in-mplus-is-pos-lin},\\
  Lemma~\thref{l:int-in-mplus-of-decomp-into-nonpos-and-nonneg-parts},\\
  Lemma~\thref{l:compat-of-int-in-mplus-with-nonpos-and-nonneg-parts},\\
  Lemma~\thref{l:compat-of-int-in-mplus-with-almost-bin-rel},\\
  Lemma~\thref{l:minkowski-ineq-in-m}.

\item[Lemma~\thref{l:int-in-mplus-is-pos-lin}] \mbox{}\\
  is explicitly cited in the proof of:\\
  Lemma~\thref{l:int-in-mplus-over-singleton},\\
  Lemma~\thref{l:cand-tensor-prod-meas-is-tensor-prod-meas},\\
  Theorem~\thref{t:tonelli}.

\item[Lemma~\thref{l:int-in-mplus-is-sigma-add}] \mbox{}\\
  is explicitly cited in the proof of:\\
  Lemma~\thref{l:int-in-mplus-over-subset-is-sigma-add},\\
  Lemma~\thref{l:cand-tensor-prod-meas-is-tensor-prod-meas},\\
  Lemma~\thref{l:int-over-subset-is-sigma-add}.

\item[Lemma~\thref{l:int-in-mplus-of-decomp-into-nonpos-and-nonneg-parts}] \mbox{}\\
  is explicitly cited in the proof of:\\
  Lemma~\thref{l:equiv-def-of-integrability},\\
  Lemma~\thref{l:int-is-pos-lin-form-on-llone}.

\item[Lemma~\thref{l:compat-of-int-in-mplus-with-nonpos-and-nonneg-parts}] \mbox{}\\
  is explicitly cited in the proof of:\\
  Lemma~\thref{l:int-is-add}.

\item[Lemma~\thref{l:int-in-mplus-is-almost-definite}] \mbox{}\\
  is explicitly cited in the proof of:\\
  Lemma~\thref{l:compat-of-int-in-mplus-with-almost-bin-rel},\\
  Lemma~\thref{l:negl-of-meas-section},\\
  Lemma~\thref{l:none-is-almost-definite},\\
  Lemma~\thref{l:first-mean-value-theorem}.

\item[Lemma~\thref{l:compat-of-int-in-mplus-with-almost-bin-rel}] \mbox{}\\
  is explicitly cited in the proof of:\\
  Lemma~\thref{l:compat-of-int-in-mplus-with-almost-eq},\\
  Lemma~\thref{l:int-in-mplus-is-almost-monot}.

\item[Lemma~\thref{l:compat-of-int-in-mplus-with-almost-eq}] \mbox{}\\
  is explicitly cited in the proof of:\\
  Lemma~\thref{l:compat-of-int-with-almost-eq}.

\item[Lemma~\thref{l:int-in-mplus-is-almost-monot}] \mbox{}\\
  is explicitly cited in the proof of:\\
  Lemma~\thref{l:almost-bounded-by-integrable-is-integrable}.

\item[Lemma~\thref{l:bienayme-chebyshev-ineq}] \mbox{}\\
  is explicitly cited in the proof of:\\
  Lemma~\thref{l:integrable-in-mplus-is-almost-finite}.

\item[Lemma~\thref{l:integrable-in-mplus-is-almost-finite}] \mbox{}\\
  is explicitly cited in the proof of:\\
  Lemma~\thref{l:integrable-is-almost-finite}.

\item[Lemma~\thref{l:bounded-by-integrable-in-mplus-is-integrable}] \mbox{}\\
  is not yet used.

\item[Lemma~\thref{l:int-in-mplus-over-subset}] \mbox{}\\
  is explicitly cited in the proof of:\\
  Lemma~\thref{l:int-in-mplus-over-subset-is-sigma-add},\\
  Lemma~\thref{l:int-in-mplus-over-singleton},\\
  Lemma~\thref{l:tonelli-over-subset},\\
  Lemma~\thref{l:int-over-subset}.

\item[Lemma~\thref{l:int-in-mplus-over-subset-is-sigma-add}] \mbox{}\\
  is explicitly cited in the proof of:\\
  Lemma~\thref{l:int-over-subset-is-sigma-add}.

\item[Lemma~\thref{l:int-in-mplus-over-singleton}] \mbox{}\\
  is explicitly cited in the proof of:\\
  Lemma~\thref{l:int-over-singleton}.

\item[Theorem~\thref{t:fatou-lemma}] \mbox{}\\
  is explicitly cited in the proof of:\\
  Lemma~\thref{l:int-in-mplus-of-pointwise-conv-seq},\\
  Theorem~\thref{t:lebesgue-dom-conv}.

\item[Lemma~\thref{l:int-in-mplus-of-pointwise-conv-seq}] \mbox{}\\
  is not yet used.

\item[Lemma~\thref{l:int-in-mplus-for-count-meas}] \mbox{}\\
  is explicitly cited in the proof of:\\
  Lemma~\thref{l:int-in-mplus-for-count-meas-on-n},\\
  Lemma~\thref{l:int-in-mplus-for-dirac-meas},\\
  Lemma~\thref{l:int-for-count-meas}.

\item[Lemma~\thref{l:int-in-mplus-for-count-meas-on-n}] \mbox{}\\
  is not yet used.

\item[Lemma~\thref{l:int-in-mplus-for-dirac-meas}] \mbox{}\\
  is not yet used.

\item[Lemma~\thref{l:meas-of-section}] \mbox{}\\
  is explicitly cited in the proof of:\\
  Lemma~\thref{l:meas-of-section-of-prod},\\
  Lemma~\thref{l:meas-of-meas-of-section-finite},\\
  Lemma~\thref{l:meas-of-meas-of-section},\\
  Lemma~\thref{l:cand-tensor-prod-meas-is-tensor-prod-meas},\\
  Lemma~\thref{l:uniq-of-tensor-prod-meas},\\
  Theorem~\thref{t:tonelli}.

\item[Lemma~\thref{l:meas-of-section-of-prod}] \mbox{}\\
  is explicitly cited in the proof of:\\
  Lemma~\thref{l:meas-of-meas-of-section-finite},\\
  Lemma~\thref{l:cand-tensor-prod-meas-is-tensor-prod-meas}.

\item[Lemma~\thref{l:meas-of-meas-of-section-finite}] \mbox{}\\
  is explicitly cited in the proof of:\\
  Lemma~\thref{l:meas-of-meas-of-section}.

\item[Lemma~\thref{l:meas-of-meas-of-section}] \mbox{}\\
  is explicitly cited in the proof of:\\
  Lemma~\thref{l:cand-tensor-prod-meas-is-tensor-prod-meas},\\
  Theorem~\thref{t:tonelli}.

\item[Definition~\thref{d:tensor-prod-meas}] \mbox{}\\
  is explicitly cited in the proof of:\\
  Lemma~\thref{l:cand-tensor-prod-meas-is-tensor-prod-meas},\\
  Lemma~\thref{l:tensor-prod-of-finite-meas},\\
  Lemma~\thref{l:tensor-prod-of-sigma-finite-meas},\\
  Lemma~\thref{l:uniq-of-tensor-prod-meas-finite},\\
  Lemma~\thref{l:uniq-of-tensor-prod-meas},\\
  Lemma~\thref{l:lebesgue-meas-on-r2-gen-area-of-boxes}.

\item[Definition~\thref{d:cand-tensor-prod-meas}] \mbox{}\\
  is explicitly cited in the proof of:\\
  Lemma~\thref{l:cand-tensor-prod-meas-is-tensor-prod-meas},\\
  Lemma~\thref{l:uniq-of-tensor-prod-meas}.

\item[Lemma~\thref{l:cand-tensor-prod-meas-is-tensor-prod-meas}] \mbox{}\\
  is explicitly cited in the proof of:\\
  Lemma~\thref{l:uniq-of-tensor-prod-meas-finite},\\
  Lemma~\thref{l:uniq-of-tensor-prod-meas}.

\item[Lemma~\thref{l:tensor-prod-of-finite-meas}] \mbox{}\\
  is explicitly cited in the proof of:\\
  Lemma~\thref{l:uniq-of-tensor-prod-meas-finite}.

\item[Lemma~\thref{l:tensor-prod-of-sigma-finite-meas}] \mbox{}\\
  is explicitly cited in the proof of:\\
  Lemma~\thref{l:uniq-of-tensor-prod-meas},\\
  Lemma~\thref{l:lebesgue-meas-on-r2-is-sigma-finite}.

\item[Lemma~\thref{l:uniq-of-tensor-prod-meas-finite}] \mbox{}\\
  is explicitly cited in the proof of:\\
  Lemma~\thref{l:uniq-of-tensor-prod-meas}.

\item[Lemma~\thref{l:uniq-of-tensor-prod-meas}] \mbox{}\\
  is explicitly cited in the proof of:\\
  Lemma~\thref{l:negl-of-meas-section},\\
  Lemma~\thref{l:lebesgue-meas-on-r2},\\
  Theorem~\thref{t:tonelli}.

\item[Lemma~\thref{l:negl-of-meas-section}] \mbox{}\\
  is not yet used.

\item[Lemma~\thref{l:lebesgue-meas-on-r2}] \mbox{}\\
  is explicitly cited in the proof of:\\
  Lemma~\thref{l:lebesgue-meas-on-r2-gen-area-of-boxes},\\
  Lemma~\thref{l:lebesgue-meas-on-r2-is-sigma-finite}.

\item[Lemma~\thref{l:lebesgue-meas-on-r2-gen-area-of-boxes}] \mbox{}\\
  is explicitly cited in the proof of:\\
  Lemma~\thref{l:lebesgue-meas-on-r2-is-zero-on-lines},\\
  Lemma~\thref{l:lebesgue-meas-on-r2-is-diffuse}.

\item[Lemma~\thref{l:lebesgue-meas-on-r2-is-zero-on-lines}] \mbox{}\\
  is not yet used.

\item[Lemma~\thref{l:lebesgue-meas-on-r2-is-sigma-finite}] \mbox{}\\
  is not yet used.

\item[Lemma~\thref{l:lebesgue-meas-on-r2-is-diffuse}] \mbox{}\\
  is not yet used.

\item[Definition~\thref{d:partial-fun-of-fun-from-prod-space}] \mbox{}\\
  is explicitly cited in the proof of:\\
  Theorem~\thref{t:tonelli},\\
  Lemma~\thref{l:tonelli-over-subset},\\
  Lemma~\thref{l:tonelli-for-tensor-prod}.

\item[Theorem~\thref{t:tonelli}] \mbox{}\\
  is explicitly cited in the proof of:\\
  Lemma~\thref{l:tonelli-over-subset},\\
  Lemma~\thref{l:tonelli-for-tensor-prod}.

\item[Lemma~\thref{l:tonelli-over-subset}] \mbox{}\\
  is not yet used.

\item[Lemma~\thref{l:tonelli-for-tensor-prod}] \mbox{}\\
  is not yet used.

\item[Definition~\thref{d:integrability}] \mbox{}\\
  is explicitly cited in the proof of:\\
  Lemma~\thref{l:integrable-is-meas},\\
  Lemma~\thref{l:equiv-def-of-integrability},\\
  Lemma~\thref{l:compat-of-int-in-m-and-mplus},\\
  Lemma~\thref{l:compat-of-int-with-almost-eq},\\
  Lemma~\thref{l:int-over-subset},\\
  Lemma~\thref{l:int-over-singleton},\\
  Lemma~\thref{l:int-is-pos-lin-form-on-llone}.

\item[Lemma~\thref{l:integrable-is-meas}] \mbox{}\\
  is explicitly cited in the proof of:\\
  Lemma~\thref{l:equiv-def-of-integrability},\\
  Lemma~\thref{l:int-is-hom}.

\item[Lemma~\thref{l:equiv-def-of-integrability}] \mbox{}\\
  is explicitly cited in the proof of:\\
  Lemma~\thref{l:compat-of-integrability-in-m-and-mplus},\\
  Lemma~\thref{l:integrable-is-almost-finite},\\
  Lemma~\thref{l:almost-bounded-by-integrable-is-integrable},\\
  Lemma~\thref{l:bounded-by-integrable-is-integrable},\\
  Lemma~\thref{l:int-over-subset-is-sigma-add},\\
  Lemma~\thref{l:int-over-singleton},\\
  Lemma~\thref{l:int-for-count-meas},\\
  Lemma~\thref{l:integrable-is-finite-seminorm-lone},\\
  Lemma~\thref{l:int-is-hom},\\
  Lemma~\thref{l:equiv-def-of-llone},\\
  Lemma~\thref{l:int-is-pos-lin-form-on-llone},\\
  Theorem~\thref{t:lebesgue-ext-dom-conv}.

\item[Lemma~\thref{l:compat-of-integrability-in-m-and-mplus}] \mbox{}\\
  is explicitly cited in the proof of:\\
  Lemma~\thref{l:bounded-by-llone-is-llone}.

\item[Lemma~\thref{l:integrable-is-almost-finite}] \mbox{}\\
  is explicitly cited in the proof of:\\
  Lemma~\thref{l:int-is-add}.

\item[Lemma~\thref{l:almost-bounded-by-integrable-is-integrable}] \mbox{}\\
  is explicitly cited in the proof of:\\
  Lemma~\thref{l:bounded-by-integrable-is-integrable}.

\item[Lemma~\thref{l:bounded-by-integrable-is-integrable}] \mbox{}\\
  is explicitly cited in the proof of:\\
  Lemma~\thref{l:bounded-by-llone-is-llone}.

\item[Definition~\thref{d:integral}] \mbox{}\\
  is explicitly cited in the proof of:\\
  Lemma~\thref{l:compat-of-int-in-m-and-mplus},\\
  Lemma~\thref{l:compat-of-int-with-almost-eq},\\
  Lemma~\thref{l:int-over-subset},\\
  Lemma~\thref{l:int-over-subset-is-sigma-add},\\
  Lemma~\thref{l:int-over-singleton},\\
  Lemma~\thref{l:int-for-count-meas},\\
  Lemma~\thref{l:int-is-hom},\\
  Lemma~\thref{l:int-is-add},\\
  Lemma~\thref{l:int-is-pos-lin-form-on-llone},\\
  Lemma~\thref{l:const-fun-is-llone}.

\item[Lemma~\thref{l:compat-of-int-in-m-and-mplus}] \mbox{}\\
  is explicitly cited in the proof of:\\
  Lemma~\thref{l:int-of-zero-is-zero},\\
  Lemma~\thref{l:const-fun-is-llone},\\
  Lemma~\thref{l:first-mean-value-theorem}.

\item[Lemma~\thref{l:int-of-zero-is-zero}] \mbox{}\\
  is not yet used.

\item[Definition~\thref{d:merge-int-in-m-and-mplus}] \mbox{}\\
  is explicitly cited in the proof of:\\
  Lemma~\thref{l:seminorm-llone}.

\item[Lemma~\thref{l:compat-of-int-with-almost-eq}] \mbox{}\\
  is explicitly cited in the proof of:\\
  Lemma~\thref{l:compat-of-none-with-almost-eq},\\
  Lemma~\thref{l:int-is-add},\\
  Theorem~\thref{t:lebesgue-ext-dom-conv}.

\item[Lemma~\thref{l:int-over-subset}] \mbox{}\\
  is explicitly cited in the proof of:\\
  Lemma~\thref{l:int-over-singleton}.

\item[Lemma~\thref{l:int-over-subset-is-sigma-add}] \mbox{}\\
  is explicitly cited in the proof of:\\
  Lemma~\thref{l:int-over-int},\\
  Lemma~\thref{l:chasles-rel-int-over-split-ints}.

\item[Lemma~\thref{l:int-over-singleton}] \mbox{}\\
  is explicitly cited in the proof of:\\
  Lemma~\thref{l:int-over-int}.

\item[Lemma~\thref{l:int-over-int}] \mbox{}\\
  is explicitly cited in the proof of:\\
  Lemma~\thref{l:chasles-rel-int-over-split-ints}.

\item[Lemma~\thref{l:chasles-rel-int-over-split-ints}] \mbox{}\\
  is not yet used.

\item[Lemma~\thref{l:int-for-count-meas}] \mbox{}\\
  is explicitly cited in the proof of:\\
  Lemma~\thref{l:int-for-count-meas-on-n},\\
  Lemma~\thref{l:int-for-dirac-meas}.

\item[Lemma~\thref{l:int-for-count-meas-on-n}] \mbox{}\\
  is not yet used.

\item[Lemma~\thref{l:int-for-dirac-meas}] \mbox{}\\
  is not yet used.

\item[Definition~\thref{d:int-for-lebesgue-meas-on-r}] \mbox{}\\
  is not yet used.

\item[Lemma~\thref{l:seminorm-llone}] \mbox{}\\
  is explicitly cited in the proof of:\\
  Lemma~\thref{l:integrable-is-finite-seminorm-lone},\\
  Lemma~\thref{l:compat-of-none-with-almost-eq},\\
  Lemma~\thref{l:none-is-almost-definite},\\
  Lemma~\thref{l:none-is-abs-hom},\\
  Lemma~\thref{l:int-is-hom},\\
  Lemma~\thref{l:minkowski-ineq-in-m},\\
  Lemma~\thref{l:equiv-def-of-llone},\\
  Lemma~\thref{l:llone-is-closed-under-abs},\\
  Lemma~\thref{l:int-is-pos-lin-form-on-llone},\\
  Theorem~\thref{t:lebesgue-dom-conv},\\
  Theorem~\thref{t:lebesgue-ext-dom-conv}.

\item[Lemma~\thref{l:integrable-is-finite-seminorm-lone}] \mbox{}\\
  is explicitly cited in the proof of:\\
  Lemma~\thref{l:int-is-add}.

\item[Lemma~\thref{l:compat-of-none-with-almost-eq}] \mbox{}\\
  is explicitly cited in the proof of:\\
  Lemma~\thref{l:minkowski-ineq-in-m}.

\item[Lemma~\thref{l:none-is-almost-definite}] \mbox{}\\
  is explicitly cited in the proof of:\\
  Lemma~\thref{l:llone-is-seminormed-vector-space}.

\item[Lemma~\thref{l:none-is-abs-hom}] \mbox{}\\
  is explicitly cited in the proof of:\\
  Lemma~\thref{l:int-is-hom},\\
  Lemma~\thref{l:llone-is-seminormed-vector-space}.

\item[Lemma~\thref{l:int-is-hom}] \mbox{}\\
  is explicitly cited in the proof of:\\
  Lemma~\thref{l:int-is-pos-lin-form-on-llone}.

\item[Lemma~\thref{l:minkowski-ineq-in-m}] \mbox{}\\
  is explicitly cited in the proof of:\\
  Lemma~\thref{l:int-is-add},\\
  Lemma~\thref{l:minkowski-ineq-in-llone}.

\item[Lemma~\thref{l:int-is-add}] \mbox{}\\
  is explicitly cited in the proof of:\\
  Lemma~\thref{l:int-is-pos-lin-form-on-llone}.

\item[Definition~\thref{d:llone-vector-space-of-int-fun}] \mbox{}\\
  is explicitly cited in the proof of:\\
  Lemma~\thref{l:equiv-def-of-llone},\\
  Lemma~\thref{l:minkowski-ineq-in-llone},\\
  Lemma~\thref{l:llone-is-seminormed-vector-space},\\
  Lemma~\thref{l:llone-is-closed-under-abs},\\
  Lemma~\thref{l:int-is-pos-lin-form-on-llone},\\
  Theorem~\thref{t:lebesgue-dom-conv},\\
  Theorem~\thref{t:lebesgue-ext-dom-conv}.

\item[Lemma~\thref{l:equiv-def-of-llone}] \mbox{}\\
  is explicitly cited in the proof of:\\
  Lemma~\thref{l:bounded-by-llone-is-llone}.

\item[Lemma~\thref{l:minkowski-ineq-in-llone}] \mbox{}\\
  is explicitly cited in the proof of:\\
  Lemma~\thref{l:llone-is-seminormed-vector-space}.

\item[Lemma~\thref{l:llone-is-seminormed-vector-space}] \mbox{}\\
  is explicitly cited in the proof of:\\
  Lemma~\thref{l:int-is-pos-lin-form-on-llone},\\
  Lemma~\thref{l:first-mean-value-theorem},\\
  Theorem~\thref{t:lebesgue-dom-conv}.

\item[Definition~\thref{d:conv-in-llone}] \mbox{}\\
  is explicitly cited in the proof of:\\
  Theorem~\thref{t:lebesgue-dom-conv}.

\item[Lemma~\thref{l:llone-is-closed-under-abs}] \mbox{}\\
  is explicitly cited in the proof of:\\
  Lemma~\thref{l:bounded-by-llone-is-llone},\\
  Theorem~\thref{t:lebesgue-dom-conv}.

\item[Lemma~\thref{l:bounded-by-llone-is-llone}] \mbox{}\\
  is explicitly cited in the proof of:\\
  Lemma~\thref{l:first-mean-value-theorem},\\
  Theorem~\thref{t:lebesgue-dom-conv}.

\item[Lemma~\thref{l:int-is-pos-lin-form-on-llone}] \mbox{}\\
  is explicitly cited in the proof of:\\
  Lemma~\thref{l:first-mean-value-theorem},\\
  Theorem~\thref{t:lebesgue-dom-conv}.

\item[Lemma~\thref{l:const-fun-is-llone}] \mbox{}\\
  is explicitly cited in the proof of:\\
  Lemma~\thref{l:first-mean-value-theorem}.

\item[Lemma~\thref{l:first-mean-value-theorem}] \mbox{}\\
  is explicitly cited in the proof of:\\
  Lemma~\thref{l:variant-of-first-mean-value-theorem}.

\item[Lemma~\thref{l:variant-of-first-mean-value-theorem}] \mbox{}\\
  is not yet used.

\item[Theorem~\thref{t:lebesgue-dom-conv}] \mbox{}\\
  is explicitly cited in the proof of:\\
  Theorem~\thref{t:lebesgue-ext-dom-conv}.

\item[Theorem~\thref{t:lebesgue-ext-dom-conv}]

\end{description}

\end{document}